%% file: HH-white-paper.tex
\documentclass[a4paper,11pt,twoside]{book}
\usepackage[utf8]{inputenc}
\usepackage[english]{babel}
\usepackage{amssymb,amsmath}
\usepackage{fourier}
\usepackage[T1]{fontenc}
\usepackage{setspace}
\usepackage{cite}
\usepackage{graphicx}
\usepackage{import}
\usepackage{slashed}
\usepackage{cancel}
\usepackage{multirow}
\usepackage{placeins}
\usepackage{commath}
\usepackage{enumitem}
\usepackage{xcolor}
\usepackage{enumitem}
\definecolor{darkblue}{RGB}{0, 56, 128}

\usepackage{xspace}
\usepackage{caption,subfig,epsfig}
\usepackage{latexsym}
\usepackage{color}

\usepackage{lineno}

\usepackage{geometry}
\makeatletter
\renewcommand{\@pnumwidth}{2em}
\makeatother

\usepackage{fancyhdr}

\usepackage[toc]{appendix} 

\usepackage{csquotes}

\usepackage[hyperfootnotes=true,bookmarks=false,hypertexnames=true,pagebackref=false,linktocpage=false,hidelinks]{hyperref}
\hypersetup{
   colorlinks=true,%
   linkcolor=darkblue,%
   citecolor=darkblue,%
   filecolor=green,%
   menucolor=cyan,%
   urlcolor=darkblue}

\newcommand*{\ifb}{\,fb$^{-1}$\xspace}

\newcommand*{\instl}{\,cm$^{-2}$s$^{-1}$\xspace}
\newcommand*{\sqrts}{\ensuremath{\sqrt s}\xspace}
\newcommand{\gev}{\,\, \mathrm{GeV}\xspace}
\newcommand{\tev}{\,\, \mathrm{TeV}\xspace}

\newcommand*{\klambda}{\ensuremath{\kappa_{\lambda}}\xspace}

\newcommand*{\W}{\ensuremath{W}\xspace}

\newcommand*{\hh}{\ensuremath{HH}\xspace}
\newcommand{\mH}{\ensuremath{m_H}\xspace}
\newcommand{\mh}{\ensuremath{m_H}\xspace}
\newcommand*{\mhh}{\ensuremath{m_{HH}}\xspace}
\newcommand*{\pTH}{\ensuremath{p^H_{T}}\xspace}
\newcommand*{\pTHH}{\ensuremath{p^{HH}_{T}}\xspace}
\newcommand*{\m}{\ensuremath{m}\xspace}
\newcommand*{\myy}{\ensuremath{\m_{\gamma\gamma}}\xspace}
\newcommand*{\mbb}{\ensuremath{m_{b\bar{b}}}\xspace}
\newcommand*{\mtt}{\ensuremath{\m_{\tau\tau}}\xspace}
\newcommand*{\mjj}{\ensuremath{\m_{jj}}\xspace}
\newcommand*{\hbb}{\ensuremath{H\rightarrow{b}\bar{b}}\xspace}
\newcommand*{\hyy}{\ensuremath{H\rightarrow\gamma\gamma}\xspace}
\newcommand*{\htautau}{\ensuremath{H \rightarrow \tau^+ \tau^-}\xspace}
\newcommand*{\hww}{\ensuremath{H \rightarrow WW^*}\xspace}
\newcommand*{\hzz}{\ensuremath{H \rightarrow ZZ^*}}
\newcommand*{\hgg}{\ensuremath{H\rightarrow\gamma\gamma}\xspace}
\newcommand*{\hhbbbb}{\ensuremath{\hh\rightarrow b\bar{b}b\bar{b}}\xspace}
\newcommand*{\hhbbyy}{\ensuremath{\hh\rightarrow b\bar{b}\gamma\gamma}\xspace}
\newcommand*{\hhbbtt}{\ensuremath{\hh\rightarrow b\bar{b}\tau^{+}\tau^{-}}\xspace}
\newcommand*{\hhbbww}{\ensuremath{\hh\rightarrow b\bar{b}WW^{*}}\xspace}
\newcommand*{\hhbbvv}{\ensuremath{\hh\rightarrow b\bar{b}VV^*}\xspace}
\newcommand*{\hhbbVV}{\ensuremath{\hh\rightarrow b\bar{b}VV^*}\xspace}

\newcommand*{\hhbbZZ}{\ensuremath{\hh\rightarrow b\bar{b}ZZ(4\ell)}\xspace}
\newcommand*{\bbbb}{\ensuremath{b\bar{b}b\bar{b}}\xspace}
\newcommand*{\bbtautau}{\ensuremath{b\bar{b}\tau^+\tau^-}\xspace}
\newcommand*{\bbtt}{\ensuremath{b\bar{b}\tau^+\tau^-}\xspace}
\newcommand*{\bbgg}{\ensuremath{b\bar{b}\gamma\gamma}\xspace}
\newcommand*{\bbyy}{\ensuremath{b\bar{b}\gamma\gamma}\xspace}
\newcommand*{\bbzz}{\ensuremath{b\bar{b}ZZ^{*}}\xspace}
\newcommand*{\bbww}{\ensuremath{b\bar{b}WW^{*}}\xspace}
\newcommand*{\bbvv}{\ensuremath{b\bar{b}VV^*}\xspace}
\newcommand*{\hhwwyy}{\ensuremath{\hh\rightarrow WW^*\gamma\gamma}\xspace}
\newcommand*{\hhwwww}{\ensuremath{\hh\rightarrow WW^{*}WW^{*}}\xspace}
\newcommand*{\wwyy}{\ensuremath{WW^*\gamma\gamma}\xspace}
\newcommand*{\yyww}{\ensuremath{\gamma\gamma WW^*}\xspace}
\newcommand*{\wwww}{\ensuremath{WW^{*}WW^*}\xspace}
\newcommand*{\tautauww}{\ensuremath{\tau^{+}\tau^{-}WW^*}\xspace}
\newcommand*{\tautautautau}{\ensuremath{\tau^{+}\tau^{-}\tau^{+}\tau^{-}}\xspace}
\newcommand*{\Mtilde}{\ensuremath{\widetilde{M}_{X}}\xspace}

\newcommand*{\bb}{\ensuremath{b\bar{b}}\xspace}
\newcommand*{\ttbar}{\ensuremath{t\bar{t}}\xspace}

\newcommand{\pT}{\ensuremath{p_{\mathrm{T}}}\xspace}
\newcommand{\ET}{\ensuremath{E_{\mathrm{T}}}\xspace}

\newcommand*{\bjet}{$b$-jet\xspace}
\newcommand*{\bjets}{$b$-jets\xspace}
\newcommand*{\btag}{$b$-tag\xspace}
\newcommand*{\btags}{$b$-tags\xspace}
\newcommand*{\btagging}{$b$-tagging\xspace}
\newcommand*{\btagged}{$b$-tagged\xspace}

\newcommand*{\ETy}{\ensuremath{E^{\gamma}_{T}}\xspace}
\newcommand*{\pTj}{\ensuremath{p^{j}_{T}}\xspace}

\newcommand*{\pTjj}{\ensuremath{p^{jj}_{T}}\xspace}
\newcommand*{\pTyy}{\ensuremath{p^{\gamma\gamma}_{T}}\xspace}
\newcommand*{\mjjyy}{\ensuremath{\m_{jj\gamma\gamma}}\xspace}

\newcommand*{\contrib}[1]{{\normalfont\normalsize\emph{#1}}}

\DeclareUnicodeCharacter{2009}{\,}

\usepackage{heppennames2}
\usepackage[stable]{footmisc}

\newcommand{\bbWW}{\ensuremath{b \bar{b} W W^*}\xspace}

\def\beq{\begin{equation}}
\def\eeq{\end{equation}}
\def\half{\frac{1}{2}}

\def\Dslash{\not{\hbox{\kern-3pt $D$}}}

\newcommand{\tth}{t\bar{t}H}
\newcommand{\tril}{\lambda_{H^3}}

\newcommand{\dsigmah}{\delta \sigma_{\tril}}
\newcommand{\dBR}{\delta {\rm BR}_{\tril}}

\newcommand{\br}{{\rm BR}}

\newcommand{\kl}{\kappa_{\lambda}}
\newcommand{\kt}{\kappa_{t}}

\newcommand*{\hhAlt}{\ensuremath{HH}\xspace}
\newcommand{\mhAlt}{\ensuremath{m_{H}}\xspace}
\newcommand*{\mhhAlt}{\ensuremath{m_{HH}}\xspace}
\newcommand{\mtAlt}{\ensuremath{m_t}\xspace}
\newcommand*{\ggtohhAlt}{\ensuremath{gg\rightarrow\hhAlt}\xspace}
\newcommand*{\bbbbAlt}{\ensuremath{b\bar{b}b\bar{b}}\xspace}
\newcommand*{\bbttAlt}{\ensuremath{b\bar{b}\tau^+\tau^-}\xspace}
\newcommand*{\bbyyAlt}{\ensuremath{b\bar{b}\gamma\gamma}\xspace}

\newcommand{\lHcube}{\lambda_{H^3}}
\newcommand{\lHcubeSM}{\lHcube^{\rm SM}}

\newcommand{\lHquar}{\lambda_{H^4}}
\newcommand{\lHquarSM}{\lHquar^{\rm SM}}

\newcommand{\nf}{\ensuremath{N_F}\xspace}
\newcommand{\as}{\ensuremath{\alpha_s}\xspace}
\newcommand{\refeq}[1]{Eq.~(\ref{#1})}
\newcommand{\refeqs}[1]{Eqs.~(\ref{#1})}
\newcommand{\reffig}[1]{Fig.~\ref{#1}}
\newcommand{\reffigs}[1]{Figs.~\ref{#1}}
\newcommand{\refta}[1]{Table~\ref{#1}}

\newcommand{\MGAMCNLO}{\textsc{MG5}\_aMC@NLO}
\newcommand{\MCNLO}{MC@NLO}
\newcommand{\pythia}{\textsc{Pythia}}
\newcommand{\sherpa}{\textsc{Sherpa}}
\newcommand{\powheg}{\textsc{Powheg}}
\newcommand{\powhegbox}{\textsc{Powheg-Box}}

\def\beq{\begin{equation}}
\def\eeq#1{\label{#1}\end{equation}}
\def\eeqn{\end{equation}}
\newenvironment{Eqnarray}%
   {\arraycolsep 0.14em\begin{eqnarray}}{\end{eqnarray}}
\def\beqa{\begin{Eqnarray}}
\def\eeqa#1{\label{#1}\end{Eqnarray}}
\def\eeqan{\end{Eqnarray}}
\def\CR{\nonumber \\ }
\def\leqn#1{(\ref{#1})}

\begin{document}

\newgeometry{centering} 

\begin{titlepage}

\begin{center}

\vspace*{1.5cm}
{\Huge \textbf{Higgs boson potential at colliders:} \\
\Huge status and perspectives\\} 

\vspace*{4.5cm}

{\Large
Editors: \\[1ex]
Biagio Di Micco,
Maxime Gouzevitch,\\[1ex]
Javier Mazzitelli
and Caterina Vernieri} \\[2ex]



\end{center}

\vspace*{3cm}
\begin{flushright}
FERMILAB-CONF-19-468-E-T\\
LHCHXSWG-2019-005
\end{flushright}

\end{titlepage}
\restoregeometry

\newpage\mbox{}
\thispagestyle{empty}

\newpage
\pagenumbering{Roman}
\setcounter{page}{1}

\input{authors.tex}

\newpage
\chapter*{Preface}
\input{introduction/preface_new.tex}

\tableofcontents

%
%

\newpage
\pagenumbering{arabic}
\setcounter{page}{1}


\part{Theoretical status}\label{chap:th_status}

\input{theoretical_status.tex}

\part{Status of the measurements at LHC}\label{chap:exp_status}

\input{experimental_status.tex}

\part{Higgs boson potential at future colliders}\label{chap:future}
\input{HH_future.tex}

\chapter*{Acknowledgements}
\input{acknowledgements.tex}



%
%
%

\bibliographystyle{atlasnote}
\bibliography{bibliography.bib,HHinee.bib}

\end{document}

%% file: authors.tex
\begin{flushleft}

{\large {\bf Authors}}

B.~Di Micco$^{1,2}$,
M.~Gouzevitch$^{3}$,
J.~Mazzitelli$^{4}$,
C.~Vernieri$^{5}$ (editors),
J.~Alison$^{6}$,
K.~Androsov$^{7}$,
J.~Baglio$^{8}$,
E.~Bagnaschi$^{9}$,
S.~Banerjee$^{10}$,
P.~Basler$^{11}$
% ...

\begin{itemize}[noitemsep]
\item[$^{1}$] INFN, Sezione di Roma Tre, Italy
\item[$^{2}$] Università degli Studi di Roma Tre, Rome, Italy
\item[$^{3}$] Univ. Lyon, Univ. Claude Bernard Lyon 1, CNRS/IN2P3, IP2I Lyon, F-69622, Villeurbanne, France
\item[$^{4}$] Physik Institut, Universität Zürich, CH-8057 Zürich, Switzerland
\item[$^{5}$] SLAC, Stanford University, Menlo Park, California 94025 USA
\item[$^{6}$] Carnegie Mellon Univerisity, Pittsburgh, USA
\item[$^{7}$] INFN, Sezione di Pisa, Italy
\item[$^{8}$] Institute for Theoretical Physics, Eberhard Karls Universität Tübingen, Tübingen D-72076, Germany
\item[$^{9}$] Paul Scherrer Institut, Villigen PSI CH-5232, Switzerland
\item[$^{10}$] Institute for Particle Physics Phenomenology, Department of Physics, Durham University, Durham DH1 3LE, UK
\item[$^{11}$] Institute for Theoretical Particle Physics (TTP), Karlsruhe Institute of Technology, Karlsruhe D-76128, Germany
\item[$^{12}$] School of Physics and Astronomy, University of Manchester, Manchester, United Kingdom
\item[$^{13}$] Physics Department, Southern Methodist University, Dallas TX, USA
\item[$^{14}$] Institute for Nuclear Physics (IKP), Karlsruhe Institute of Technology, Eggenstein-Leopoldshafen  D-76344, Germany
\item[$^{15}$] LPNHE, Sorbonne Universit\'e, Paris, France
\item[$^{16}$] Dipartimento di Fisica e Astronomia, Università di Bologna and INFN, Sezione di Bologna, 40126 Bologna, Italy
\item[$^{17}$] Department of Physics, Northern Illinois University, DeKalb IL, USA
\item[$^{18}$] Enrico Fermi Institute, University of Chicago, Chicago IL, USA
\item[$^{19}$] Ludwig-Maximilians-Universität München, Fakultät für Physik, Arnold Sommerfeld Center for Theoretical Physics, D–80333 München, Germany
\item[$^{20}$] Theory Division, IFIC, University of Valencia-CSIC, E-46980 Paterna, Valencia, Spain
\item[$^{21}$] Kavli Institute for Cosmological Physics, University of Chicago, Chicago, Illinois, 60637, USA
\item[$^{22}$] Theoretical Physics Department, Fermilab, P.O. Box 500, Batavia, IL 60510, USA
\item[$^{23}$] National Institute of Chemical Physics and Biophysics, Tallinn, Estonia
\item[$^{24}$] ETH Zurich - Institute for Particle Physics and Astrophysics (IPA), Zurich, Switzerland
\item[$^{25}$] Department of Physics, Brookhaven National Laboratory, Upton, N.Y., 11973, U.S.A.
\item[$^{26}$] INFN, Sezione di Bari and Politecnico di Bari, Italy
\item[$^{27}$] INFN, Sezione di Milano, 20133 Milano, Italy
\item[$^{28}$] Department of Theoretical Physics, Jozef Stefan Institute, 1000 Ljubljana, Slovenia
\item[$^{29}$] SUPA, School of Physics and Astronomy, University of Glasgow, Glasgow G12 8QQ, U.K
\item[$^{30}$] Department of Physics and Astronomy, Uppsala University, Uppsala, Sweden
\item[$^{31}$] Lancaster University, Department of Physics, Lancaster, United Kingdom
\item[$^{32}$] Laborat\'orio de Instrumenta\c{c}\~ao e F\'isica Experimental de Part\'iculas, LIP Lisbon, Portugal
\item[$^{33}$] Instituto de Física Teórica UAM/CSIC, Universidad Autónoma de Madrid, 28049, Madrid, Spain
\item[$^{34}$] Institute for Theoretical Physics (ITP), Karlsruhe Institute of Technology, D-76131 Karlsruhe, Germany
\item[$^{35}$] Max-Planck-Institut für Kernphysik Saupfercheckweg 1, 69117 Heidelberg, Germany
\item[$^{36}$] Santa Cruz Institute for Particle Physics, University of California, Santa Cruz, CA 95064, USA
\item[$^{37}$] Institut für Physik, Humboldt-Universität zu Berlin, Berlin D-12489, Germany
\item[$^{38}$] DESY, Hamburg D-22607, Germany
\item[$^{39}$] Max Planck Institute for Physics, München 80805, Germany
\item[$^{40}$] Department of Physics and Astronomy University of Nebraska-Lincoln, Lincoln, NE, 68588
\item[$^{41}$] Theoretical Physics Department, CERN, Geneva, Switzerland
\item[$^{42}$] Ohio State University, Columbus OH, USA
\item[$^{43}$] Institute for Theoretical Physics (ITP), Heidelberg University, Heidelberg D-69120, Germany
\item[$^{44}$] Department of Physics, University of Illinois, Urbana IL, USA
\item[$^{45}$] Amherst Center for Fundamental Interactions, Department of Physics, University of Massachusetts, Amherst, MA 01003
\item[$^{46}$] Department of Physics, Chung-Ang University, Seoul 06974, Korea
\item[$^{47}$] Department of Physics and Astronomy, University of Kansas, Lawrence, Kansas 66045, USA
\item[$^{48}$] Maryland Center for Fundamental Physics, Department of Physics, University of Maryland
\item[$^{49}$] Ottawa-Carleton Institute for Physics, Carleton University, Ottawa, Ontario K1S 5B6, Canada
\item[$^{50}$] Department of Physics and Astronomy, Rice University, Houston, TX, 77005, USA
\item[$^{51}$] Centre for Cosmology, Particle Physics and Phenomenology (CP3), Université Catholique de Louvain, B-1348 Louvain-la-Neuve, Belgium
\item[$^{53}$] Nikhef National Institute for Subatomic Physics and University of Amsterdam, Amsterdam, Netherlands
\item[$^{54}$] RWTH Aachen University, III. Physikalisches Institut A, Aachen, Germany
\item[$^{55}$] Institute for Fundamental Theory, Physics Department, University of Florida, Gainesville, FL 32611, USA
\item[$^{56}$] INFN, Sezione di Torino, Italy
\item[$^{57}$] Rutgers, The State University of New Jersey, Piscataway, USA
\item[$^{58}$] Technische Universität München, Garching D-85748 , Germany
\item[$^{59}$] Università degli Studi di Firenze and INFN Firenze, Sesto Fiorentino (FI), Italy
\item[$^{60}$] Institute for Theoretical Physics Amsterdam and Delta Institute for Theoretical Physics, University of Amsterdam, Amsterdam 1098 XH, The Netherlands
\item[$^{61}$] Nikhef, Theory Group, Amsterdam 1098 XG, The Netherlands
\item[$^{62}$] Institute of Convergence Fundamental Studies and School of Liberal Arts, Seoultech, Seoul 01811, Korea
\item[$^{63}$] Laboratoire de Physique Subatomique et de Cosmologie, Université Grenoble-Alpes, CNRS/IN2P3, 53 Avenue des Martyrs, Grenoble 38026, France
\item[$^{64}$] Départment de Physique Théorique, Université de Genève, Genève 1211, Switzerland
\item[$^{65}$] Theoretical Physics Division, Rudjer Boskovic Institute, Zagreb 10002, Croatia
\item[$^{66}$] CERN, Geneva, Switzerland
\item[$^{67}$] CP3-Origins, University of Southern Denmark, Denmark
\item[$^{68}$] Department of Physics, University of Washington, Seattle WA, USA
\item[$^{69}$] Department of Physics and Astronomy, Wayne State University, Detroit, MI 48201, USA
\item[$^{70}$] Indian Institute of Science Education and Research, Knowledge City, Manauli 140306, Punjab, India
\item[$^{71}$] University of Tokyo, Japan
\item[$^{72}$] IFAE and BIST, Universitat Autónoma de Barcelona, Barcelona, Spain
\item[$^{73}$] Department of Physics and Astronomy, University of Sheffield, United Kingdom
\item[$^{74}$] Leinweber Center for Theoretical Physics, Department of Physics, University of Michigan Ann Arbor, MI 48109 USA
\item[$^{75}$] Department of Physics, University of Florida, Gainesville, FL 32611, USA
\item[$^{76}$] University of Kansas, Lawrence, USA
\item[$^{77}$] Department of Physics and Astronomy, University of Oklahoma, Norman, OK, 73019, USA

\end{itemize}

\end{flushleft}

%% file: introduction/preface_new.tex
With the discovery of the Higgs boson~\cite{Higgs:1964ia,Higgs:1964pj,Higgs:1966ev,Englert:1964et,Guralnik:1964eu,Kibble:1967sv} at the Large Hadron Collider (LHC)~\cite{Aad:2012tfa,Chatrchyan:2012xdj}, all the particles predicted by the Standard Model (SM) of particle physics have now been observed. While for the moment the SM has been able to successfully describe the experimental measurements obtained at particle colliders, many predictions of the model remain to be tested. Furthermore, the quest for a more fundamental description of nature is still ongoing.

The LHC, running underneath the frontier between Switzerland and France from the 2009 to 2018, has measured the Higgs boson couplings to the vector bosons $W$ and $Z$, and the more massive generation of quarks and charged leptons. The most elusive couplings to the first two generations and neutrinos are, however, still wholly untested. Moreover, the energy potential of the Higgs boson field, responsible for the electroweak symmetry breaking (EWSB) mechanism, has not yet been measured by any experiment.

After EWSB, the Higgs boson potential gives rise to cubic and quartic terms in the Higgs boson field, inducing a self-coupling $\lambda$ which, within the SM, is fully parameterised by the Higgs boson mass and the Fermi coupling constant. A measurement of this coupling would, therefore, start shedding light on the actual structure of the potential, whose exact shape can have profound theoretical consequences.

The determination of the Higgs boson self-coupling could have implications on our understanding of fundamental interactions not only at the electroweak scale but also at higher energies. Due to the presence of quantum corrections, the value of the self-coupling runs with the energy. Based on the most precise measurements of the Higgs boson and the top quark mass, the Higgs potential energy function would have a new minimum at very large values of the Higgs field, maybe as large at the GUT scale or even the Planck scale. This property would imply that the vacuum state of the SM, as we see it today, is meta-stable, i.e., that after a period much larger than the foreseeable age of the universe a tunnel transition to a new vacuum state with different laws of physics might be possible. This meta-stability could, for instance, be a potential source of a stochastic background of gravitational waves.
The shape of the Higgs potential, and in particular the value of the Higgs trilinear coupling, controls the dynamics of the electroweak phase transition. As such, it defines the viability of models of electroweak baryogenesis. Moreover, it could also fuel the production of primordial black holes.
In addition, within the current uncertainty on the above mass measurements, the self-coupling $\lambda$ could also tend to zero at the Planck scale, a very peculiar situation which could open the door to new physics scenarios, connecting for instance the Higgs boson field to cosmological inflation. 
While the above is not meant to be a comprehensive list of effects related to the Higgs boson self-coupling, it motivates the quest for an experimental measurement of its value.

Despite the importance of a precise determination of $\lambda$, the SM prediction for this coupling is still far from being tested, since all precision observables show only a  mild dependence on it. Constraints can be derived from perturbative unitarity arguments, but also in these cases, they are quite weak, spanning up to five times the SM predicted value. For a more precise determination of this coupling, the challenging measurement of Higgs boson pair production at colliders is needed.

The target of the present work is to summarise the present theoretical and experimental status of the di-Higgs boson production searches, and of the constraints on the self-coupling $\lambda$ arising from double and single Higgs production measurements at hadron and lepton colliders. The work has started as the proceedings of the Di-Higgs workshop at Colliders, which was held at Fermilab from the 4th to the 9th of September 2018. However, it went beyond the topics discussed at that workshop and included further developments. The editors would like to thank all contributors, the ATLAS and CMS collaborations and the Higgs cross section working group for supporting the editing of this white paper and for providing useful inputs and advice, and in particular Michael Peskin for his careful review of the manuscript and his substantial contribution to the overall shaping of this work.
\vspace*{2cm} \\
\phantom{x}\hspace{10cm}{\Large The  editors}\\
\phantom{x}\hspace{10cm}{\large \textit{B. Di Micco}}\\
\phantom{x}\hspace{10cm}{\large \textit{M. Gouzevitch}}\\
\phantom{x}\hspace{10cm}{\large \textit{J. Mazzitelli}}\\
\phantom{x}\hspace{10cm}{\large \textit{C. Vernieri }}

%% file: theoretical_status.tex
In this Part we aim to provide an overview of the latest theory developments that are relevant to the measurement of the Higgs boson self-coupling and, more in general, to the study of the \hh production process in the context of the SM and beyond.
The theory efforts that are summarised here are vital in order to extract the maximum possible information from the experimental measurements.

In Chapter \ref{chap:HHcxs} we present the latest theoretical predictions for the production cross section of SM Higgs boson pairs in the different production modes, including fixed order results and Monte Carlo generators. We put special focus on the main production mode at the LHC, gluon fusion.
We describe in Chapter \ref{chap:EFT} the developments on the effective field theory approach, crucial for the interpretation of non-resonant deviations from the SM expectations.
We study the impact that a fit of the effective field theory coefficients would have on the Higgs self-coupling determination, both in double and single Higgs final states.
Finally, in Chapter \ref{chap:BSM}, we present specific beyond the SM scenarios that can have sizeable effects in the di-Higgs final state. We mostly focus on signatures coming from new resonant states decaying into Higgs boson pairs, though we also study the impact that new physics contributions might have via loop effects.

\input{HH_cross_section/HH_cross_section_main.tex}
\newpage
\input{EFT/EFTmain.tex}
\newpage
\input{BSMresonance/BSMmain.tex}

%% file: HH_cross_section/HH_cross_section_main.tex
\chapter{HH cross section predictions}\label{chap:HHcxs}
\textbf{Editors: M. Spira, E. Vryonidou}
\vspace{2mm}

\noindent
While the quartic Higgs coupling $\lHquar$ cannot be probed
directly at the LHC due to the small size of the triple-Higgs production
cross section \cite{Plehn:2005nk,Binoth:2006ym,Fuks:2015hna,Maltoni:2014eza}, the
trilinear Higgs coupling can be accessed directly in Higgs pair
production.
At hadron colliders, Higgs boson pairs are dominantly produced in the
loop-induced gluon-fusion mechanism $\ggtohhAlt$, mainly mediated by
top quark loops, similarly to how a single Higgs boson is produced.
An estimate of the dependence of the cross section on the size of the
trilinear coupling is given by the relation
$\Delta\sigma/\sigma \sim -\Delta\lambda/\lambda$ in the vicinity of the
SM value of $\lambda$. This fact clearly illustrates that, in order to determine the trilinear
coupling, the theoretical uncertainties of the corresponding cross
section need to be under control, hence the inclusion of higher-order
corrections in the QCD perturbative expansion becomes indispensable.

In this chapter we will summarise the state of the art of the theoretical
predictions concerning the production of SM Higgs boson pairs at hadron colliders. We start by
describing in Sec.~\ref{sec:production_modes} all the different production modes, then in Sec.~\ref{sec:gluon_fusion} we focus on
the QCD corrections for the main production mode, gluon fusion, and
in Sec.~\ref{sec:xs_vs_lambda} we describe its dependence on the Higgs self-coupling.
Finally, in Sec.~\ref{sec:MC} we review the available Monte Carlo generators.

\section{Overview of production modes}\label{sec:production_modes}

\begin{figure}[t!]
\begin{center}
\includegraphics[width=0.99\textwidth]{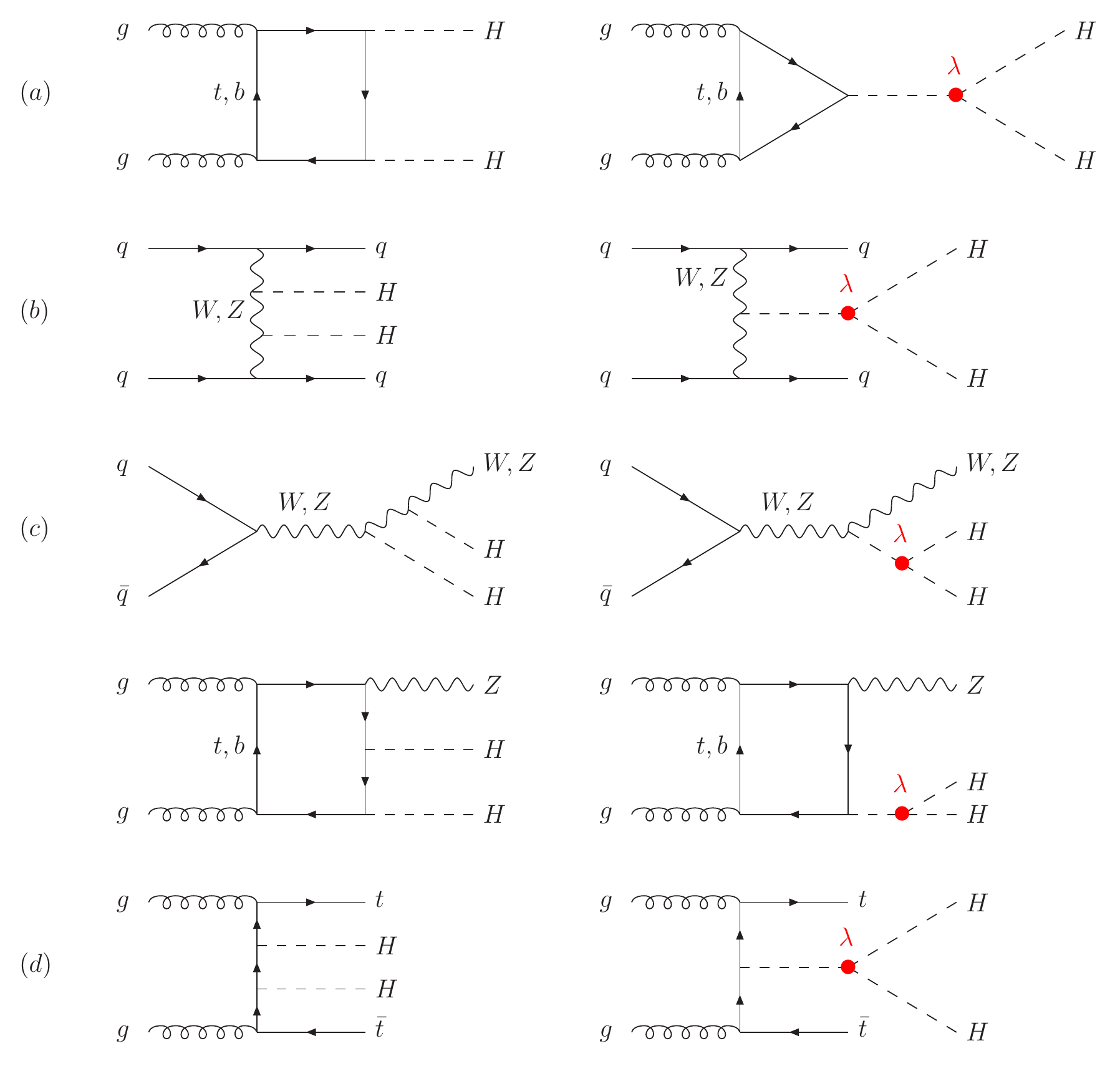}
\end{center}
\caption{Diagrams contributing to Higgs pair
production: (a) gluon fusion, (b) vector-boson fusion, (c) double
Higgs-strahlung and (d) double Higgs bremsstrahlung off top quarks. The
trilinear Higgs coupling contribution is marked in red.}
  \label{fg:pp2hh}
\end{figure}

\begin{figure}
\begin{center}
\includegraphics[width=0.8\textwidth]{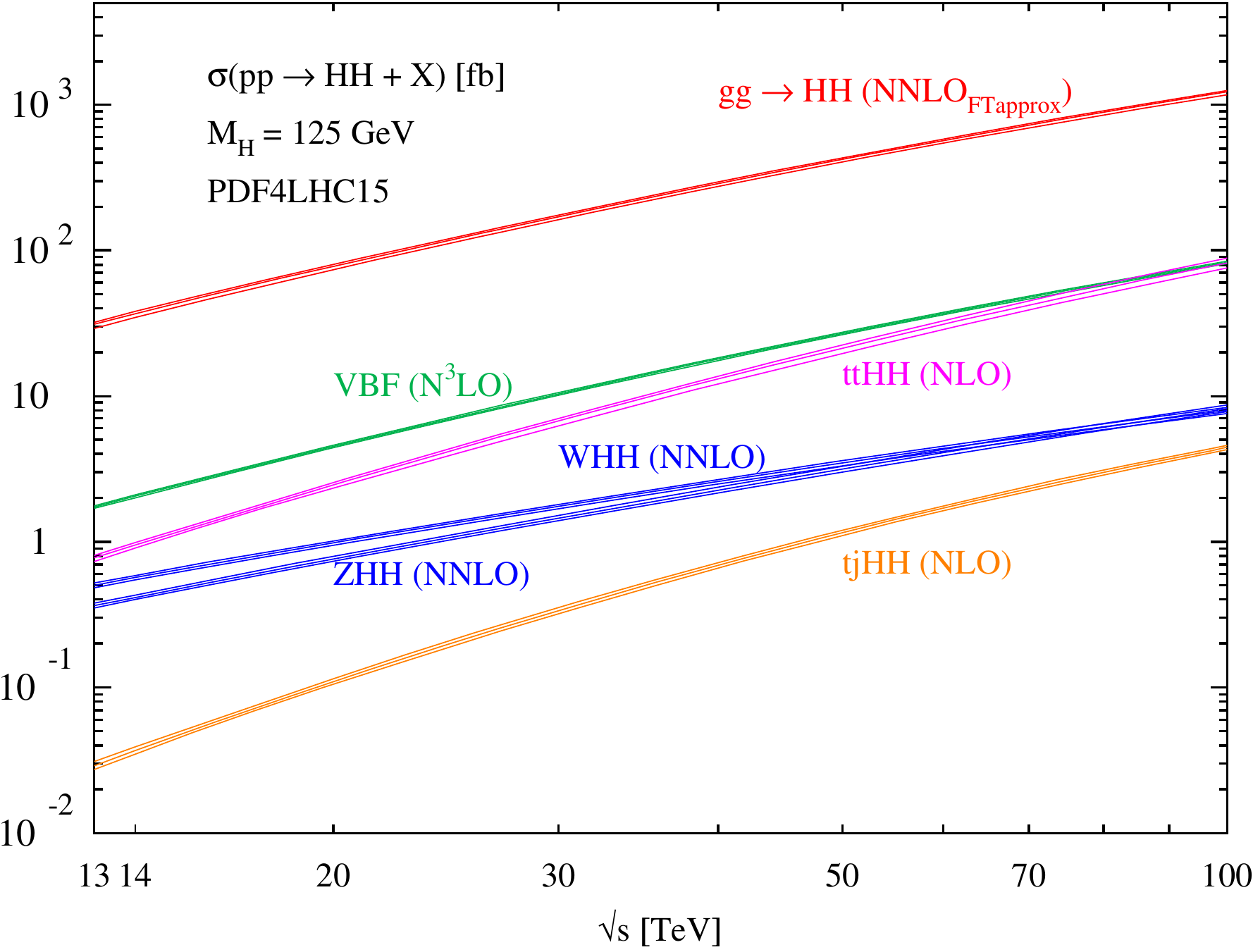}
\vspace*{-0.5cm}
\end{center}
\caption{Total production cross sections for Higgs pairs within the SM
via gluon fusion, vector-boson fusion, double Higgs-strahlung and double
Higgs bremsstrahlung off top quarks. PDF4LHC15 parton densities have
been used with the scale choices according to \refta{table:xsec2}.
The size of the bands shows the total uncertainties originating from the
scale dependence and the PDF+$\as$ uncertainties.}
  \label{fg:HHcxn}
\end{figure}

We individually discuss below the main production modes of Higgs boson pairs
at hadron colliders, briefly summarising the status of the corresponding theoretical
predictions.
Examples of the leading-order (LO) Feynman diagrams are illustrated in \reffig{fg:pp2hh},
a summary plot of the total cross sections -- to the highest available accuracy -- as a function of the collider centre-of-mass energy is shown in \reffig{fg:HHcxn},
and the predictions are also presented in \refta{table:xsec2}.

\begin{table}[t!]
\renewcommand{\arraystretch}{2.0}
\begin{center}
\begin{tabular}{|l|c|c|c|c|}
\hline 
$\sqrt{s}$ & 13 TeV & 14 TeV & 27 TeV & 100 TeV \\ \hline
ggF $\hhAlt$ & $31.05^{+2.2\%}_{-5.0\%}\pm 3.0\%$ & $36.69^{+2.1\%}_{-4.9\%}\pm 3.0\%$ & $139.9^{+1.3\%}_{-3.9\%}\pm 2.5\%$ & $1224^{+0.9\%}_{-3.2\%}\pm 2.4\%$ \\ \hline
VBF $\hhAlt$ & $1.73^{+0.03\%}_{-0.04\%}\pm 2.1\%$ & $2.05^{+0.03\%}_{-0.04\%}\pm 2.1\%$ & $8.40^{+0.11\%}_{-0.04\%} \pm
2.1\%$ & $82.8^{+0.13\%}_{-0.04\%}\pm 2.1\%$ \\ \hline
$Z$\hhAlt & $0.363^{+3.4\%}_{-2.7\%} \pm 1.9\%$ & $0.415^{+3.5\%}_{-2.7\%} \pm 1.8\%$ & $1.23^{+4.1\%}_{-3.3\%} \pm 1.5\%$ & $8.23^{+5.9\%}_{-4.6\%} \pm 1.7\%$ \\ \hline
$W^+\hhAlt$ & $0.329^{+0.32\%}_{-0.41\%}\pm 2.2\%$ & $0.369^{+0.33\%}_{-0.39\%}\pm 2.1\%$ & $0.941^{+0.52\%}_{-0.53\%}\pm 1.8\%$ & $4.70^{+0.90\%}_{-0.96\%}\pm
1.8\%$ \\ \hline
$W^-\hhAlt$ & $0.173^{+1.2\%}_{-1.3\%}\pm 2.8\%$ & $0.198^{+1.2\%}_{-1.3\%}\pm 2.7\%$ & $0.568^{+1.9\%}_{-2.0\%}\pm 2.1\%$ & $3.30^{+3.5\%}_{-4.3\%}\pm 1.9\%$ \\ \hline
${t \bar t} \hhAlt$ & $0.775^{+1.5\%}_{-4.3\%}\pm 3.2\%$ & $0.949^{+1.7\%}_{-4.5\%}\pm 3.1\%$ & $5.24^{+2.9\%}_{-6.4\%}\pm 2.5\%$ & $82.1^{+7.9\%}_{-7.4\%}\pm 1.6\%$ \\ \hline
${t j} \hhAlt$ & $0.0289^{+5.5\%}_{-3.6\%}\pm 4.7\%$ & $0.0367^{+4.2\%}_{-1.8\%}\pm 4.6\%$ & $0.254^{+3.8\%}_{-2.8\%}\pm 3.6\%$ & $4.44^{+2.2\%}_{-2.8\%}\pm 2.4\%$ \\ \hline
\end{tabular} 
\caption{Signal cross sections (in fb) for \hhAlt production including the
available QCD corrections according to the recommendations of the LHC
Higgs Cross Section Working Group \cite{deFlorian:2016spz}. The
renormalization and factorisation scales have been set to $\mhhAlt/2$ for gluon fusion, to the individual virtualities
$Q_{1,2} = \sqrt{-q_{1,2}^2}$ of the t-channel vector-bosons for VBF
(with a lower cut of 1 GeV), to $m_{{HHV}}~({V=W,Z})$ for $HHV$ production, to $m_{t\bar
t}/2$ for ${t \bar t HH}$ and to $\mhhAlt/2$ for ${tjHH}$ production. They have been varied up and down by a
factor of two to obtain the scale uncertainties, indicated as superscript/subscript. PDF4LHC15 parton distributions have been
used to obtain the results, and the corresponding $\as$+PDF uncertainties. The cross sections for ${tjHH}$ involve both
top and anti-top production.
\label{table:xsec2} }  
\end{center} 
\end{table}

\begin{figure}
\begin{center}
\includegraphics[width=0.75\textwidth]{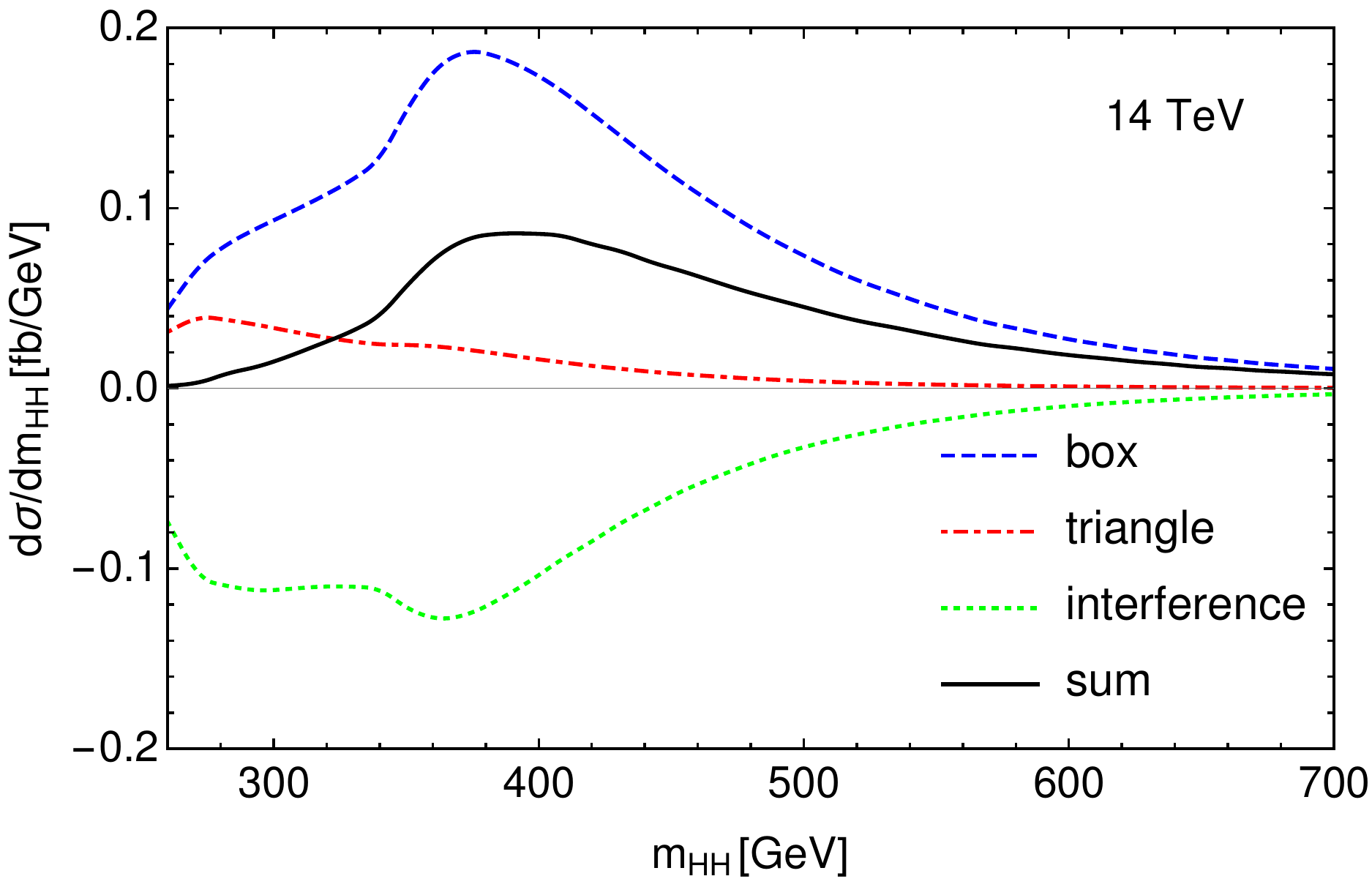}
\end{center}
\vspace*{-5mm}
\caption{Higgs pair invariant mass distribution at leading order for the different contributions to the gluon fusion production mechanism and their interference.}
  \label{fg:LOcontrGGF}
\end{figure}

\paragraph{Gluon fusion.}
Higgs boson pairs are dominantly produced in the
loop-induced gluon-fusion (ggF) mechanism that is mediated by
top quark loops, supplemented by a smaller contribution of bottom quark
loops. There are destructively interfering box (\reffig{fg:pp2hh}a left)
and triangle (\reffig{fg:pp2hh}a right) diagrams, with the latter involving the trilinear Higgs
coupling \cite{Glover:1987nx,Eboli:1987dy,Plehn:1996wb}. 
The relative contribution of these two different pieces, as well as their interference,
 can be observed in the Higgs pair invariant
mass distribution shown in \reffig{fg:LOcontrGGF}.
The effect of the trilinear Higgs self-coupling in the LO total cross section amounts to a reduction of about 50\% with respect to the box-only contribution, due to the large destructive interference.
The QCD corrections are known
up to next-to-leading order (NLO) \cite{Borowka:2016ehy, Borowka:2016ypz,
Baglio:2018lrj}, and at next-to-next-to-leading order (NNLO) in the limit of
heavy top quarks \cite{deFlorian:2013uza, deFlorian:2013jea,
Grigo:2014jma,deFlorian:2016uhr}, including partial finite top quark mass effects \cite{Grazzini:2018bsd}.
Very recently, also the third order corrections have been computed in the heavy top quark limit \cite{Chen:2019lzz}.
The QCD corrections increase the total cross section by about a factor of two with respect to the LO prediction, and they will be discussed in more detail in the following section.

\paragraph{Vector-boson fusion.}
The vector-boson fusion
(VBF) ${qq\to HHqq}$ is the second-largest production mechanism, and it is dominated by $t$-channel $W$ and Z
exchange in analogy to single Higgs production.
It involves continuum diagrams originating from
two Higgs radiations off the virtual $W$ or $Z$ bosons, and diagrams in which a single Higgs boson (off-shell) splits into a Higgs pair (\reffig{fg:pp2hh}b). The QCD corrections are only known in the structure-function
approach, i.e.~where only the $t$-channel $W$ and $Z$ exchange is taken
into account and interference effects for external quarks of the same
flavor are neglected. This approximation is valid at the level of a
percent similar to the single Higgs case. Within this approach the QCD
corrections to the total cross section are known up to N$^3$LO
\cite{Baglio:2012np,Liu-Sheng:2014gxa,Dreyer:2018qbw}, while the
exclusive calculation is available at NNLO \cite{Dreyer:2018rfu}. 
The perturbative corrections alter the total cross section at the level of about 10\%, while they can
be larger for distributions. The moderate size of the QCD corrections
can be traced back to the $t$-channel-diagram dominance, that implies
that the QCD corrections are driven by vertex corrections which can be
obtained from deep inelastic lepton-nucleon scattering (DIS). In turn,
for DIS the residual radiative corrections beyond the proper
implementation of the PDFs at higher orders are moderate; this happens
by construction within the DIS factorization scheme, but holds as well
in the $\overline{\rm MS}$ scheme. 
The NNLO and N$^3$LO corrections range at the per-cent and sub-per-cent level
\cite{Liu-Sheng:2014gxa, Dreyer:2018qbw}.

\paragraph{Double Higgs-strahlung.}
The double Higgs-strahlung's production rate, i.e.~the associated production of Higgs pairs
with a $W$ or $Z$ boson (\reffig{fg:pp2hh}c), is significantly
lower than vector-boson fusion's one. The NLO and NNLO QCD corrections to this process are
known \cite{Baglio:2012np,Li:2016nrr,Li:2017lbf}, and their main component can be
translated from the corresponding calculation of the Drell--Yan process.
These corrections increase the total cross sections by about 30\%.
In the $ZHH$ production channel there is a relevant contamination from the
loop-induced process ${gg\to ZHH}$ adding another $20-30\%$ to $ZHH$
production. The LO contribution of this gluon-induced subprocess is part of
the full NNLO QCD corrections \cite{Baglio:2012np}.

\paragraph{Double Higgs bremsstrahlung off top quarks.}
The associated production of Higgs pairs with top quark pairs (\reffig{fg:pp2hh}d) reaches a cross section value close to the
vector-boson fusion cross section at a 100 TeV
          hadron collider. This is not the case for single Higgs boson production. The NLO
QCD corrections are negative and modify the total cross section at the level of 20\%, and
reduce the residual scale dependence significantly
\cite{Frederix:2014hta}.
In the case of single-top associated production, $tjHH$, the NLO QCD corrections are of a similar size but positive, and scale uncertainties are actually increased with respect to the ones of the LO prediction, the latter not being a reliable estimate of the true perturbative uncertainties~\cite{Frederix:2014hta}.

\section[QCD corrections for gluon fusion]{QCD corrections for gluon fusion \\
\contrib{J.~Baglio, F.~Campanario, P.~Giardino, S.~Glaus, M.~M\"uhlleitner, M.~Spira, J.~Streicher}
}\label{sec:gluon_fusion}

The NLO QCD corrections to the gluon-fusion cross section $\sigma(\ggtohhAlt)$ have first been obtained in the heavy top quark limit (HTL)
\cite{Dawson:1998py} that simplifies the calculation since the top quark loop
contributions reduce to effective couplings between the Higgs boson and gluons, described by
the effective Lagrangian \cite{Ellis:1975ap, Shifman:1979eb,
Inami:1982xt, Spira:1995rr, Kniehl:1995tn}
\begin{equation}
{\cal L}_\mathrm{eff} = \frac{\as}{12\pi} G^{a\mu\nu} G^a _{\mu\nu}
\left(C_1 \frac{H}{v} - C_2 \frac{H^2}{2v^2} \right) \,,
\label{eq:leff}
\end{equation}
with the Wilson coefficients ($L_t = \log \mu_R^2/\mtAlt^2$)
\cite{Chetyrkin:1997iv, Kramer:1996iq, Dawson:1998py, Schroder:2005hy,
Baikov:2016tgj, Grigo:2014jma, Spira:2016zna, Gerlach:2018hen}
\begin{eqnarray}
C_1 & = & 1 + \frac{11}{4} \frac{\as}{\pi} + \left\{ \frac{2777}{288} +
\frac{19}{16} L_t + \nf \left(\frac{L_t}{3}-\frac{67}{69} \right)
\right\} \left(\frac{\as}{\pi} \right)^2 + {\cal O}(\as^3) \,,
\nonumber \\
C_2 & = & C_1 + \left( \frac{35}{24} + \frac{2}{3} \nf \right)
\left(\frac{\as}{\pi} \right)^2 + {\cal O}(\as^3)\,,
\end{eqnarray}
that are presented up to NNLO, but are known up to N$^4$LO
\cite{Schroder:2005hy, Baikov:2016tgj, Spira:2016zna}. Since the top
quark is integrated out, the number of active flavors has to be taken as
$\nf=5$.  Using these effective Higgs couplings to gluons, the
calculation of the NLO QCD corrections is reduced to a one-loop
calculation for the virtual corrections and a tree-level calculation for
the matrix elements of the real corrections.  The NLO final result for
the total gluon-fusion cross section can be decomposed as
\cite{Dawson:1998py}
\begin{equation}
\sigma_{\mathrm{NLO}}({pp \rightarrow H H} + X)  = 
\sigma_{\mathrm{LO}} + \Delta
\sigma_{\mathrm{virt}} + \Delta\sigma_{ gg} + \Delta\sigma_{ gq} +
\Delta\sigma_{ q\bar{q}} \nonumber \,,
\end{equation}
where the different contributions take the following form:
\begin{eqnarray}
\sigma_{\mathrm{LO}} & = & \int_{\tau_0}^1 d\tau~\frac{d{\cal
L}^{ gg}}{d\tau}~
\textcolor[cmyk]{0.91,0.00,0.88,0.20}{\hat\sigma_{\mathrm{LO}}(Q^2 = \tau s)} \,,
\nonumber \\ 
\Delta \sigma_{\mathrm{virt}} & = & \frac{\as(\mu_R)}
{\pi}\int_{\tau_0}^1 d\tau~
\frac{d{\cal L}^{ gg}}{d\tau}~\textcolor[cmyk]{0.91,0.00,0.88,0.20}{\hat
\sigma_{\mathrm{LO}}(Q^2=\tau s)}~\textcolor{red}{C_\mathrm{virt}} \,,
\nonumber \\ 
\Delta \sigma_{ gg} & = & \frac{\alpha_{s}(\mu_R)} {\pi} \int_{\tau_0}^1
d\tau~
\frac{d{\cal L}^{ gg}}{d\tau} \int_{\tau_0/\tau}^1 \frac{dz}{z}~
\textcolor[cmyk]{0.91,0.00,0.88,0.20}{\hat\sigma_{\mathrm{LO}}(Q^2 = z \tau s)}
\left\{ - z P_{ gg} (z) \log \frac{\mu_F^{2}}{\tau s} \right. 
\nonumber \\
& & \left. \hspace*{2.0cm} {} + \textcolor{red}{d_{ gg}(z)} + 6 [1+z^4+(1-z)^4]
\left(\frac{\log (1-z)}{1-z} \right)_+ \right\} \,,
\nonumber \\ 
\Delta \sigma_{ gq} & = & \frac{\alpha_{s}(\mu_R)} {\pi} \int_{\tau_0}^1
d\tau
\sum_{ q,\bar{q}} \frac{d{\cal L}^{ gq}}{d\tau} \int_{\tau_0/\tau}^1
\frac{dz}{z}~
\textcolor[cmyk]{0.91,0.00,0.88,0.20}{\hat \sigma_{\mathrm{LO}}(Q^2 = z \tau s)} \nonumber \\
& & \hspace*{3.0cm} \left\{ -\frac{z}{2} P_{ gq}(z)
\log\frac{\mu_F^{2}}{\tau s(1-z)^2}
+ \textcolor{red}{d_{ gq}(z)}
\vphantom{\frac{\mu_F^{2}}{\tau s(1-z)^2}} \right\} \,,
\nonumber \\ 
\Delta \sigma_{ q\bar q} & = & \frac{\as(\mu_R)}
{\pi} \int_{\tau_0}^1 d\tau
\sum_{ q} \frac{d{\cal L}^{ q\bar q}}{d\tau} \int_{\tau_0/\tau}^1
\frac{dz}{z}~
\textcolor[cmyk]{0.91,0.00,0.88,0.20}{\hat \sigma_{\mathrm{LO}}(Q^2 = z \tau
s)}~\textcolor{red}{d_{ q\bar q}(z)} \,.
\end{eqnarray}
Here $\hat\sigma_{\mathrm{LO}}(Q^2)$ denotes the leading-order partonic cross
section involving the squared invariant mass $Q^2$ of the Higgs boson pair,
$\as(\mu_R)$ the strong coupling constant at the renormalization
scale $\mu_R$, $d{\cal L}^{ij}/d\tau~(i,j={g,q,\bar q})$ the corresponding
parton-parton luminosities at the factorization scale $\mu_F$, and
$P_{ij}(z)~(i,j={g,q,\bar q})$ the individual Altarelli--Parisi splitting
functions \cite{Altarelli:1977zs}. The integration regions are bound by
$\tau_0=4\mhAlt^{2}/s$, with $\mhAlt$ being the Higgs boson mass and $s$ the
square of the hadronic centre-of-mass energy.

\begin{figure}
\begin{center}
\includegraphics[width=1.00\textwidth]{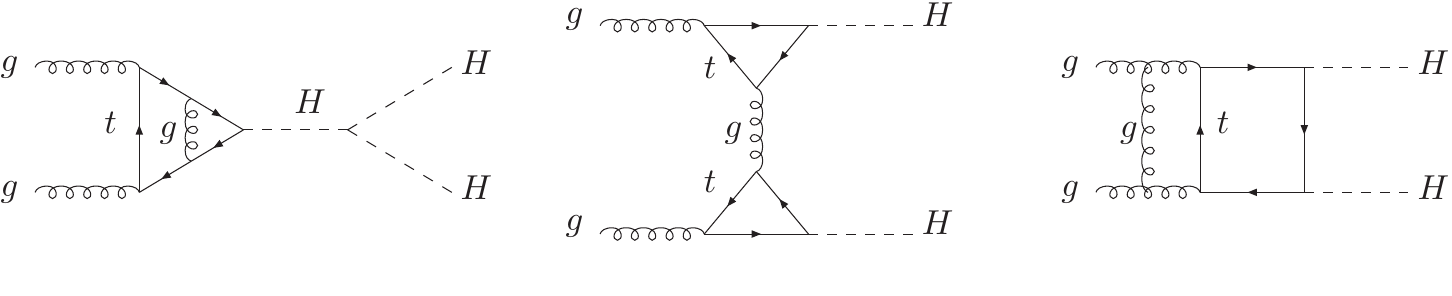}
\vspace*{-1.4cm}
\end{center}
\caption{Examples of two-loop triangle (left), one-particle reducible
(middle) and box (right) diagrams contributing to Higgs pair
production via gluon fusion.}
  \label{fg:gghhvirt}
\end{figure}

The quark-mass dependence is in general encoded in the green factors
$\hat\sigma_{\mathrm{LO}}(Q^2)$ for the LO cross section and the red factors $C_\mathrm{virt}$,
$d_{ij}(z)$ for the virtual and real corrections, respectively. In the
HTL, the latter simplify to
\begin{eqnarray}
 C_\mathrm{virt} & \to & \frac{11}{2} + \pi^2 + C^\infty_{\triangle\triangle} +
\frac{33-2\nf}{6} \log\frac{\mu_R^2}{Q^2}, \nonumber \\
 C_{\triangle\triangle} & = &
\mbox{Re}~\frac{\int_{\hat t_-}^{\hat t_+} d\hat t \left\{ c_1 \left[
(C_\triangle F_\triangle + F_\Box) + \frac{p_T^2}{\hat t} 
G_\Box \right] + (\hat t \leftrightarrow \hat u) \right\}}
{\int_{\hat t_-}^{\hat t_+} d\hat t \left\{ |C_\triangle F_\triangle +
F_\Box |^2 + |G_\Box|^2 \right\}}, \nonumber \\
C^\infty_{\triangle\triangle} & = & \left. C_{\triangle\triangle}
\right|_{c_1 = 2/9}, \nonumber \\
d_{ gg}(z) & \to & - \frac{11}{2} (1-z)^3 \nonumber \\
d_{ gq}(z) & \to & \frac{2}{3} z^2 - (1-z)^2, \nonumber \\
d_{ q\bar q}(z) & \to & \frac{32}{27} (1-z)^3,
\label{eq:coeffvirt}
\end{eqnarray}
where $p_T$ denotes the Higgs transverse momentum, $\hat s,\hat t$ the
partonic Mandelstam variables and $C_{\triangle\triangle}$ is the
contribution of the one-particle reducible diagrams, see
\reffig{fg:gghhvirt}. The integration bounds are given by
\begin{equation}
\hat{t}_\pm = -\frac{1}{2} \left[ Q^2 - 2 \mhAlt^{2} \mp Q^2\sqrt{1-\frac{4
  \mhAlt^2}{Q^2}}\, \right] \, .
\end{equation}
The couplings $C_\triangle$ and $C_\Box$ and the form factors $F_\triangle$, $F_\Box$ and $G_\Box$ in the HTL take the form
\begin{eqnarray}
C_\triangle & = & \lHcube \frac{6 v}{\hat s - m_H^2 + im_H \Gamma_H}, 
\qquad
C_\Box = 1, \nonumber \\ 
F_\triangle & \to & \frac{2}{3}, \qquad 
F_\Box \to -\frac{2}{3}, \nonumber\\[1.5ex]
G_\Box & \to & 0, 
\end{eqnarray}
with the trilinear coupling $\lHcube = m_H^2/(2v^2)$. 
    
For the NLO QCD corrections, the full mass dependence of the LO partonic
cross section has been taken into account first, while treating the
virtual corrections $C_\text{virt}$ and the real corrections $d_{ij}$ in the HTL.
This approach is now called ``Born-improved''. This leads to a reasonable
approximation for invariant Higgs pair masses in the lower range and
approximates the full NLO result for the total cross section within
about 15\% \cite{Borowka:2016ehy, Borowka:2016ypz, Baglio:2018lrj}. The
NLO corrections in the HTL increase the cross section by $80-90\%$
\cite{Dawson:1998py}.
The NNLO QCD corrections have been calculated within the same approximation 
\cite{deFlorian:2013uza, deFlorian:2013jea,Grigo:2014jma,deFlorian:2016uhr}. The main part of these corrections
can be translated from the single Higgs case, since the effective
Lagrangian of \refeq{eq:leff} does not induce a change of kinematics
between single Higgs production and the differential Higgs pair
production with respect to the invariant Higgs pair mass. The new
ingredient of the NNLO calculation is the proper treatment of the
one-particle-reducible contributions $C_{\triangle\triangle}$ of the NLO
corrections. These lead to additional contributions to the NNLO virtual
corrections and the interference of the NLO-real and NLO-virtual
corrections that contributes to the NNLO result, too. Including the full
mass dependence of the LO cross section, the NNLO QCD corrections
increase the total cross section by a more moderate amount of $20-30\%$
\cite{deFlorian:2013jea}. On top of these NNLO QCD corrections, the
soft-gluon resummation (threshold resummation) has been performed at 
next-to-next-to-leading logarithmic accuracy (NNLL) for the total cross
section and invariant mass distribution, resulting in a ${\cal O}(10\%)$ modification of the
total cross section on top of the NNLO result for a central scale $\mu_R = \mu_F = \mhhAlt$,
while the effects are much smaller if the scale $\mu_R = \mu_F = \mhhAlt/2$ is used \cite{Shao:2013bz,
deFlorian:2015moa}.

The calculations in the HTL have been refined by several steps including
mass effects partially at NLO. The inclusion of the full mass effects in
the real correction terms $d_{ij}$ by means of incorporating the full
one-loop real matrix elements for ${gg \to HHg}$, ${gq \to HHq}$, ${q\bar q\to
HHg}$ reduces the Born-improved HTL prediction by about 10\%
\cite{Frederix:2014hta, Maltoni:2014eza}. This improvement is denoted as
``FTapprox'' (for full-theory approximation). This step has been performed by using the \MGAMCNLO{} framework \cite{Alwall:2014hca} for the automatic generation of the  matrix
elements. Another improvement has been accomplished by a systematic
asymptotic large-top-mass expansion of the full NLO corrections at the
integral \cite{Grigo:2013rya} and at the integrand level
\cite{Grigo:2015dia}.  This established sizeable mass effects emerging
from the virtual two-loop corrections.  In addition, the large-top-mass
expansion has been extended to NNLO resulting in expected 5\% mass
effects of the NNLO corrections on top of the NLO result
\cite{Grigo:2015dia}. This situation necessitated the full calculation
of the mass effects at NLO.

The full NLO QCD corrections have been derived by two quite different
methods, both, however, building on a numerical integration of the
two-loop contributions that cannot be integrated analytically with
present state-of-the-art methods. Examples of diagrams of the
NLO virtual corrections are depicted in \reffig{fg:gghhvirt}. It can be
decomposed into triangle, one-particle-reducible and box diagrams. The
triangle diagrams can be obtained from the analogous calculation for
$ gg\to H$ with the Higgs mass replaced by the invariant Higgs pair mass
$Q=\mhhAlt$. The result of the one-particle-reducible diagrams is known
analytically, i.e.~in the notation of \refeq{eq:coeffvirt},
\cite{Degrassi:2016vss, Baglio:2018lrj}
\begin{eqnarray}
c_1 & = & 2 \Big[ I_1(\tau,\lambda_{\hat t})
                -I_2(\tau,\lambda_{\hat t}) \Big]^2, \nonumber \\
I_1(\tau,\lambda) & = & \frac{\tau\lambda}{2(\tau-\lambda)} +
\frac{\tau^2\lambda^2}{2(\tau-\lambda)^2} \left[ f(\tau) - f(\lambda)
\right] + \frac{\tau^2\lambda}{(\tau-\lambda)^2} \left[ g(\tau) -
g(\lambda) \right], \nonumber \\ 
I_2(\tau,\lambda) & = & - \frac{\tau\lambda}{2(\tau-\lambda)}\left[
f(\tau) - f(\lambda) \right],
\end{eqnarray}
with $\tau = 4\mtAlt^2/\mhAlt^2$, $\lambda_{\hat t} = 4 \mtAlt^2/\hat t$ and the
functions
\begin{eqnarray}
f(\tau) & = & \left\{ \begin{array}{ll}
\displaystyle \arcsin^2 \frac{1}{\sqrt{\tau}} & \tau \ge 1 \\
\displaystyle - \frac{1}{4} \left[ \log \frac{1+\sqrt{1-\tau}}
{1-\sqrt{1-\tau}} - i\pi \right]^2 & \tau < 1
\end{array} \right. \, , \nonumber
\\
g(\tau) & = & \left\{ \begin{array}{ll} 
\displaystyle \sqrt{\tau-1} \arcsin \frac{1}{\sqrt{\tau}} & \tau \ge 1
\\
\displaystyle \frac{\sqrt{1-\tau}}{2} \left[ \log \frac{1+\sqrt{1-\tau}}
{1-\sqrt{1-\tau}} - i\pi \right] & \tau < 1 
\end{array} \right. \, .
\end{eqnarray}
This expression has to be inserted in the $C_{\triangle\triangle}$
coefficient of \refeq{eq:coeffvirt}.  The new and cumbersome part of
the full virtual corrections is the calculation of the two-loop box
diagrams that has only been obtained numerically by two different
methods. The full virtual amplitude can be decomposed into two scalar
form factors, one describing the spin-0 component of the full partonic
process and the second the spin-2 one \cite{Glover:1987nx,Plehn:1996wb}.

\begin{figure}
\begin{center}
\includegraphics[width=0.50\textwidth]{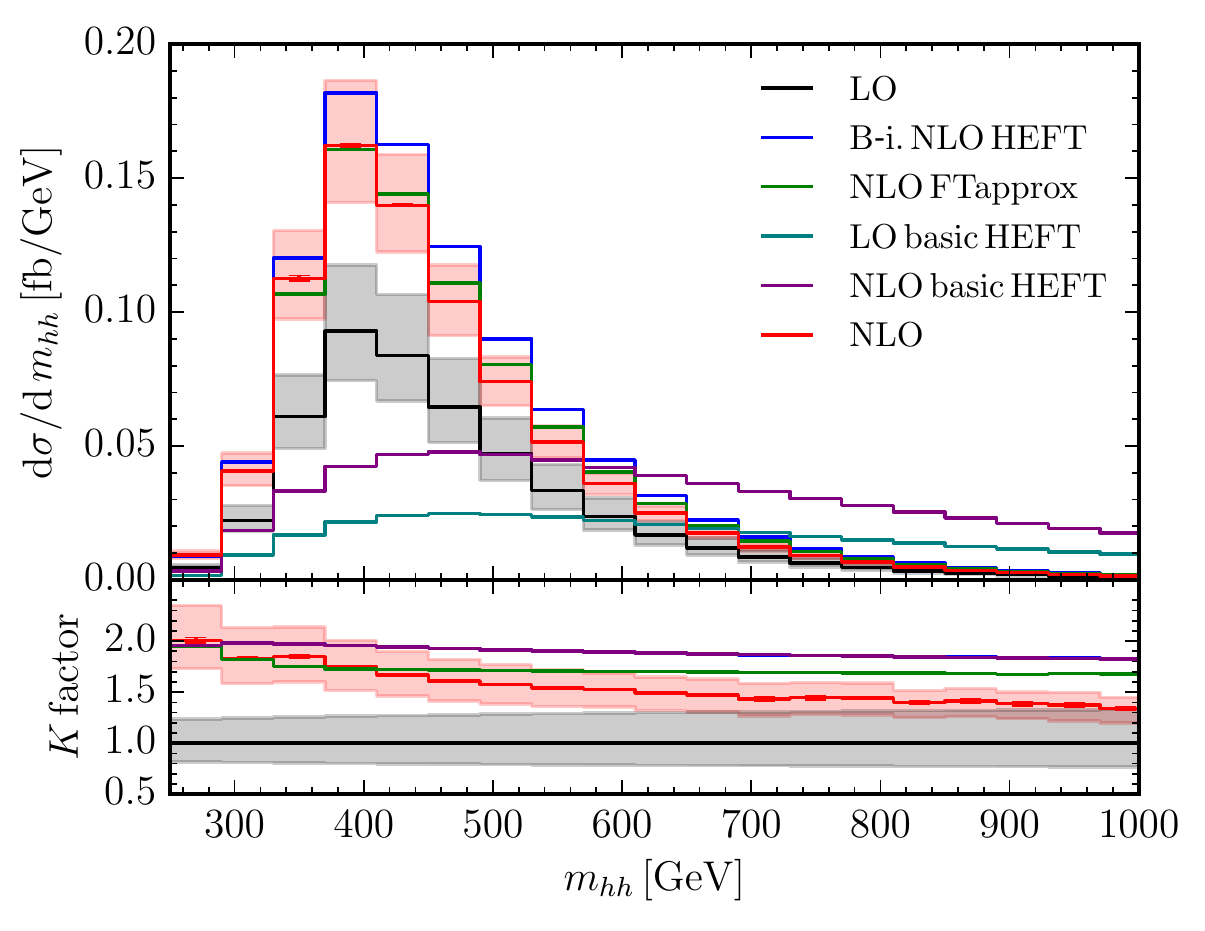}
\hspace*{-0.3cm}
\includegraphics[width=0.50\textwidth]{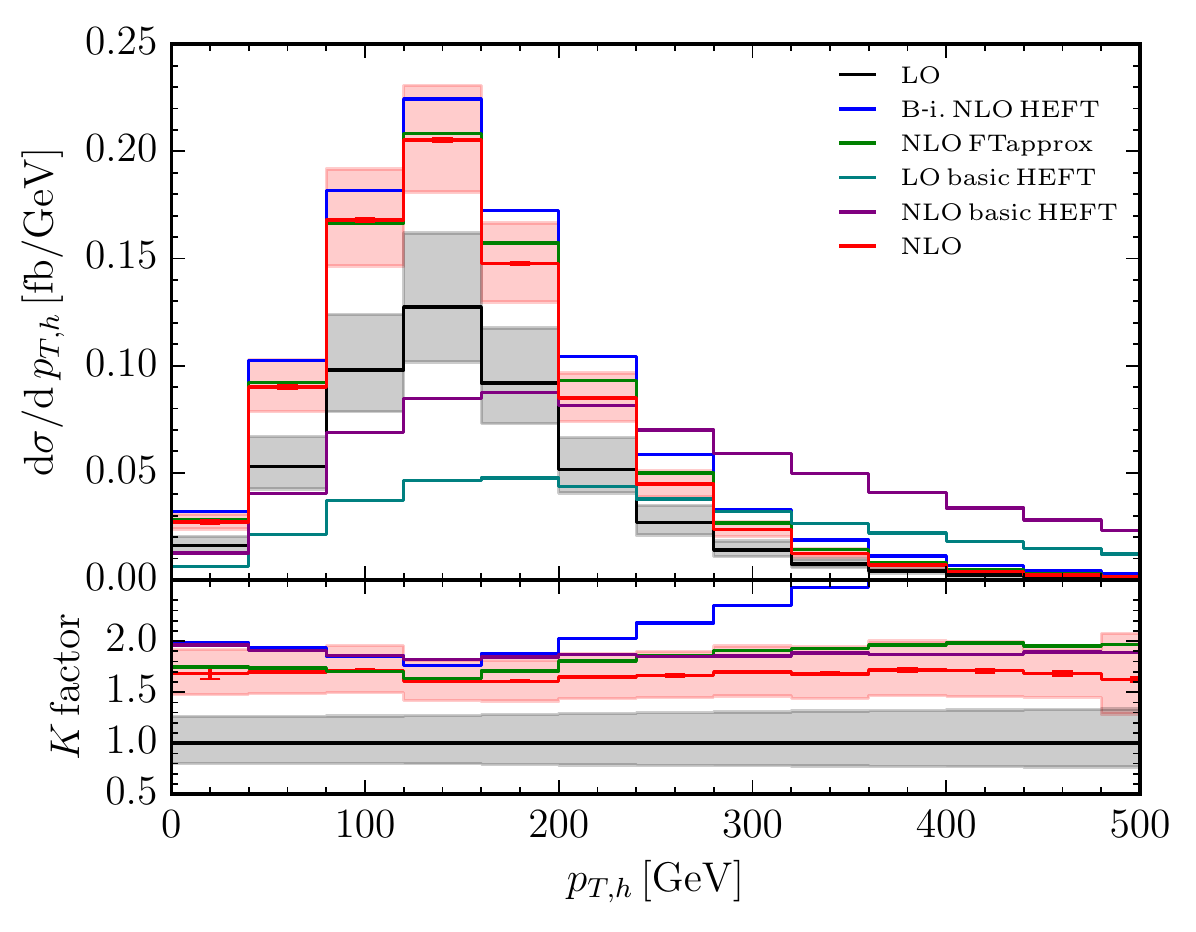}
\vspace*{-0.6cm}
\end{center}
\caption{Higgs pair invariant mass and transverse-momentum
distributions for a center-of-mass energy of 14 TeV in various approximations.
The full NLO results are shown in red. The red bands show the
renormalization and factorization scale dependence obtained from a
7-point scale variation around the central scales
$\mu_R=\mu_F=\mhhAlt/2$~\cite{Borowka:2016ypz}.}
  \label{fg:gghhnlo}
\end{figure}

The first method \cite{Borowka:2016ehy,Borowka:2016ypz} relies on the
reduction of the two-loop form factors to master integrals, Feynman
parameterisation of these master integrals and a sector decomposition to
isolate the ultraviolet and infrared singularities from the two-loop box
integrals. This yields the numerical coefficients of the divergences
that can be checked to cancel against the corresponding ultraviolet
divergences of the counter-terms and the infrared and collinear
singularities of the real corrections numerically. For large invariant
Higgs pair masses, however, the virtual top-antitop pair can become
on-shell so that there are additional threshold singularities inside the
integration region. This has been treated numerically by contour
deformations that exploit the analyticity of the master integrals in the
complex plane and by trading the physical integrals along the real axis
for integrals off the real axis. This procedure leads to numerically
stable results after suitable deformation choices and spending a
sizeable amount of CPU time. The top mass has been renormalized on-shell
and the strong coupling constant $\as$ in the $\overline{\rm
MS}$-scheme with five active flavors. By means of the first method a
grid has been generated for the exclusive calculation of the virtual
corrections to the Higgs pair cross section so that the invariant
Higgs pair mass distribution and the transverse-momentum distribution of
the Higgs bosons in the final state can be obtained. Typical results
after adding the full top-mass corrections are shown in
\reffig{fg:gghhnlo}.  The mass effects induce a reduction of the total
cross section by about 15\% at NLO but turn out to be more sizeable for
large  di-Higgs invariant masses in the differential cross section
\cite{Borowka:2016ehy,Borowka:2016ypz}. The two-dimensional grids in the
 Higgs pair invariant mass and the transverse Higgs momentum are
available and have been included in NLO event generators thus providing
the proper matching to parton showers \cite{Heinrich:2017kxx, Jones:2017giv}.

\begin{figure}
\begin{center}
\includegraphics[width=0.50\textwidth]{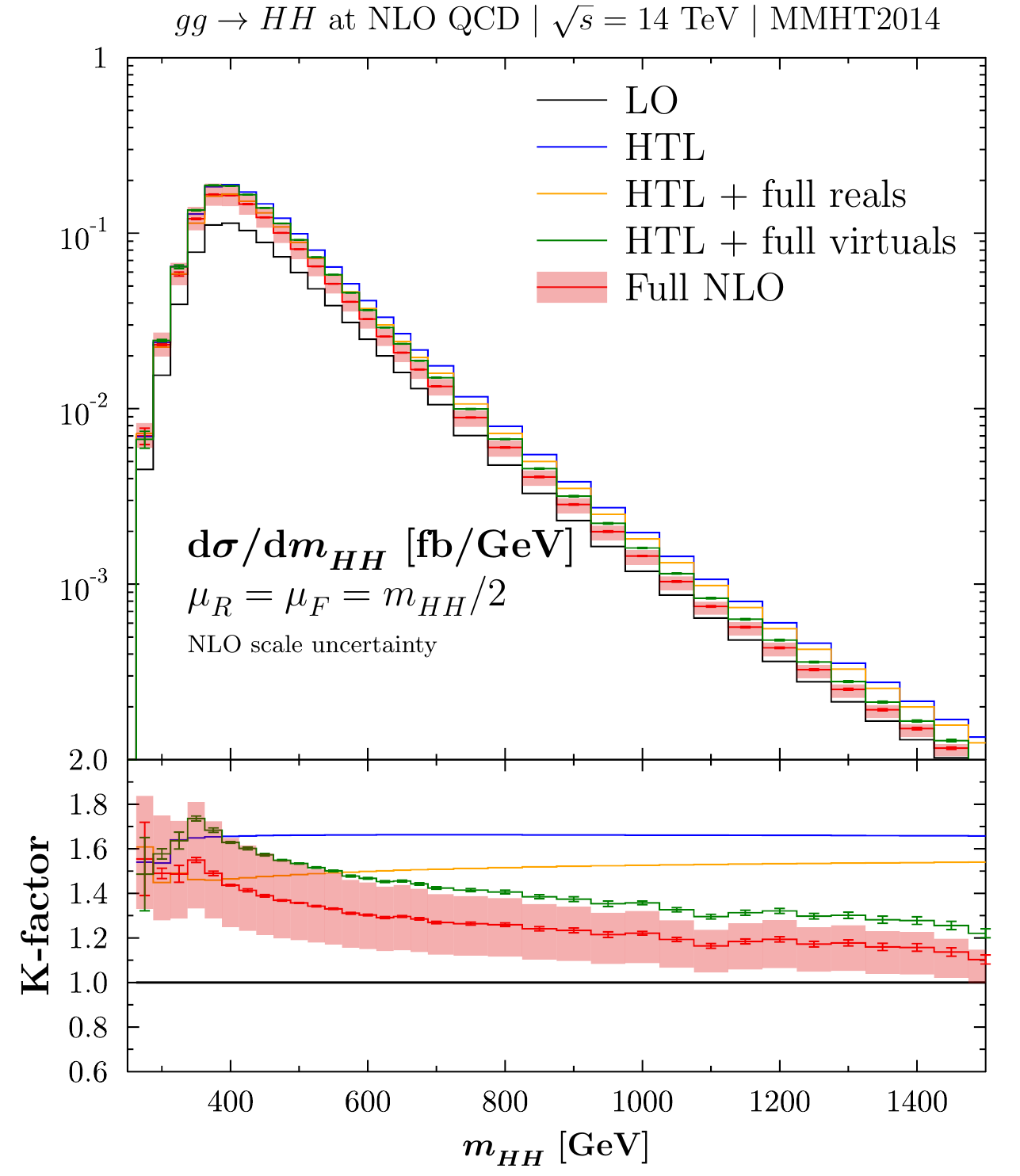}
\hspace*{-0.3cm}
\includegraphics[width=0.50\textwidth]{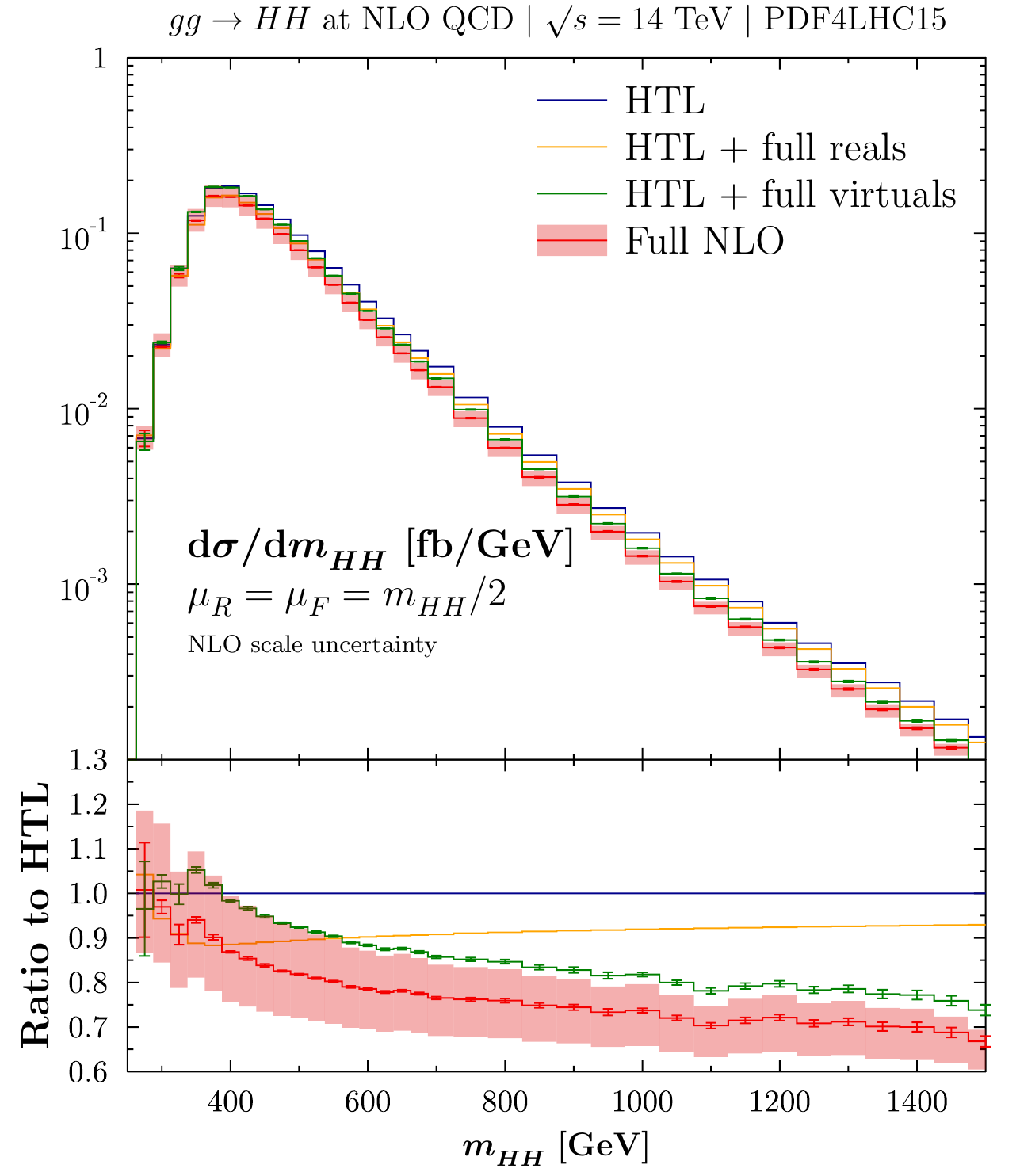}
\vspace*{-0.5cm}
\end{center}
\caption{Higgs pair invariant mass distribution for a collider energy
of 14 TeV in various approximations using MMHT2014 (left) and PDF4LHC15
(right) parton densities. The full NLO results are shown in red. The red
bands show the renormalization and factorization scale dependence
obtained from a 7-point scale variation around the central scales
$\mu_R=\mu_F=\mhhAlt/2$. From Ref.~\cite{Baglio:2018lrj}.}
  \label{fg:gghhnlo2}
\end{figure}

The second method \cite{Baglio:2018lrj} does not perform any tensor
reduction to master integrals, but introduces a Feynman parameterisation
of the full individual two-loop box diagrams. For the isolation of the
ultraviolet divergences end-point subtractions have been made for most
of the diagrams.  However, the diagrams with an external gluon exchange
between the gluons require special subtraction terms for the infrared
singular part. This is also related to the property that these diagrams
develop a second threshold at vanishing  Higgs pair invariant mass in
addition to the threshold at $\mhhAlt= 2\mtAlt$. This method uses special
subtraction terms for these diagrams that cover all singularities and
can be easily integrated analytically over one Feynman parameter. Using
the transformation properties of hypergeometric functions, the infrared
and collinear singularities can be isolated. The numerical stability
of the integrations across the thresholds has been achieved by means of
integration by parts to reduce the power of the singular denominators
and introducing a small imaginary part for the virtual squared top
masses. Since the dependence on this small imaginary part is regular,
i.e.~polynomial, for small values a Richardson extrapolation
\cite{richardson} has been used to obtain the narrow-width approximation
from results at finite values of this imaginary part. The observed
convergence is good and can also be used for a quantitative estimate of
the extrapolation error in addition to the numerical integration error.
In addition to the six-dimensional integration over the Feynman
parameters, the integration over the transverse momentum of the Higgs
bosons in terms of the Mandelstam variable $\hat t$ has been included in
the numerical integration so that the differential cross section in the
invariant Higgs pair mass is obtained directly. Since the $\hat
t$-integration is not finite for individual diagrams, the cancellation of
the divergences in $\hat t$ in the sum of all of them serves as an
additional consistency check of the final result. The numerical
integration together with the Richardson extrapolation requires a huge
amount of CPU time, similar to the other approach. The real corrections
have been calculated by subtracting the corresponding matrix elements in
the HTL for a suitably transformed LO kernel including the full LO mass
dependence from the full real matrix elements. The subtracted pieces
lead to the ``Born-improved" real corrections in the HTL when added back.
Typical final results of this method are displayed in
\reffig{fg:gghhnlo2}, which includes a comparison to the HTL and real
and virtual mass effects individually.

Both methods lead to final results in mutual agreement within their
respective integration errors. The residual small differences are due to
the different top masses chosen in the numerical analysis, $\mtAlt=173$ GeV
for the first method and $\mtAlt=172.5$ GeV (as recommended by the LHC
Higgs Cross Section Working Group) for the second method. The explicit
NLO numbers are collected in \refta{tb:gghhnlo} for several
collider energies.
Note that both calculations have been performed in the narrow-width approximation for the top quark. Finite width effects have been studied at LO and amount to a $\sim-2\%$ for the total cross section \cite{Maltoni:2014eza}.
\begin{table}[tb]
\renewcommand{\arraystretch}{1.8}
\begin{center}
\begin{tabular}{|l|c|c|} \hline
Energy & $\mtAlt=173$ GeV & $\mtAlt=172.5$ GeV \\ \hline
 13 TeV & $27.80(9)^{+13.8\%}_{-12.8\%}$ fb & $27.73(7)^{+13.8\%}_{-12.8\%}$ fb \\ \hline
 14 TeV & $32.91(10)^{+13.6\%}_{-12.6\%}$ fb & $32.78(7)^{+13.5\%}_{-12.5\%}$ fb \\ \hline
 27 TeV & $127.7(2)^{+11.5\%}_{-10.4\%}$ fb  & $127.0(2)^{+11.7\%}_{-10.7\%}$ fb \\ \hline
100 TeV & $1149(2)^{+10.8\%}_{-10.0\%}$ fb   & $1140(2)^{+10.7\%}_{-10.0\%}$ fb \\ \hline
\end{tabular}
\end{center}
\vspace*{-4mm}
\caption{\label{tb:gghhnlo} NLO cross sections for proton colliders
at 13, 14, 27 and 100 TeV center-of-mass energy using PDF4LHC15 parton densities.
The errors in brackets are the numerical integration/extrapolation
errors, while the explicit percentage numbers present the
renormalization and factorization scale dependences. The central scale
choice is $\mu_R=\mu_F=\mhhAlt/2$~\cite{Borowka:2016ehy,Borowka:2016ypz,Baglio:2018lrj}.
}
\renewcommand{\arraystretch}{1.0}
\end{table}

\begin{figure}
\begin{center}
\hspace*{-0.5cm}
\includegraphics[width=0.52\textwidth]{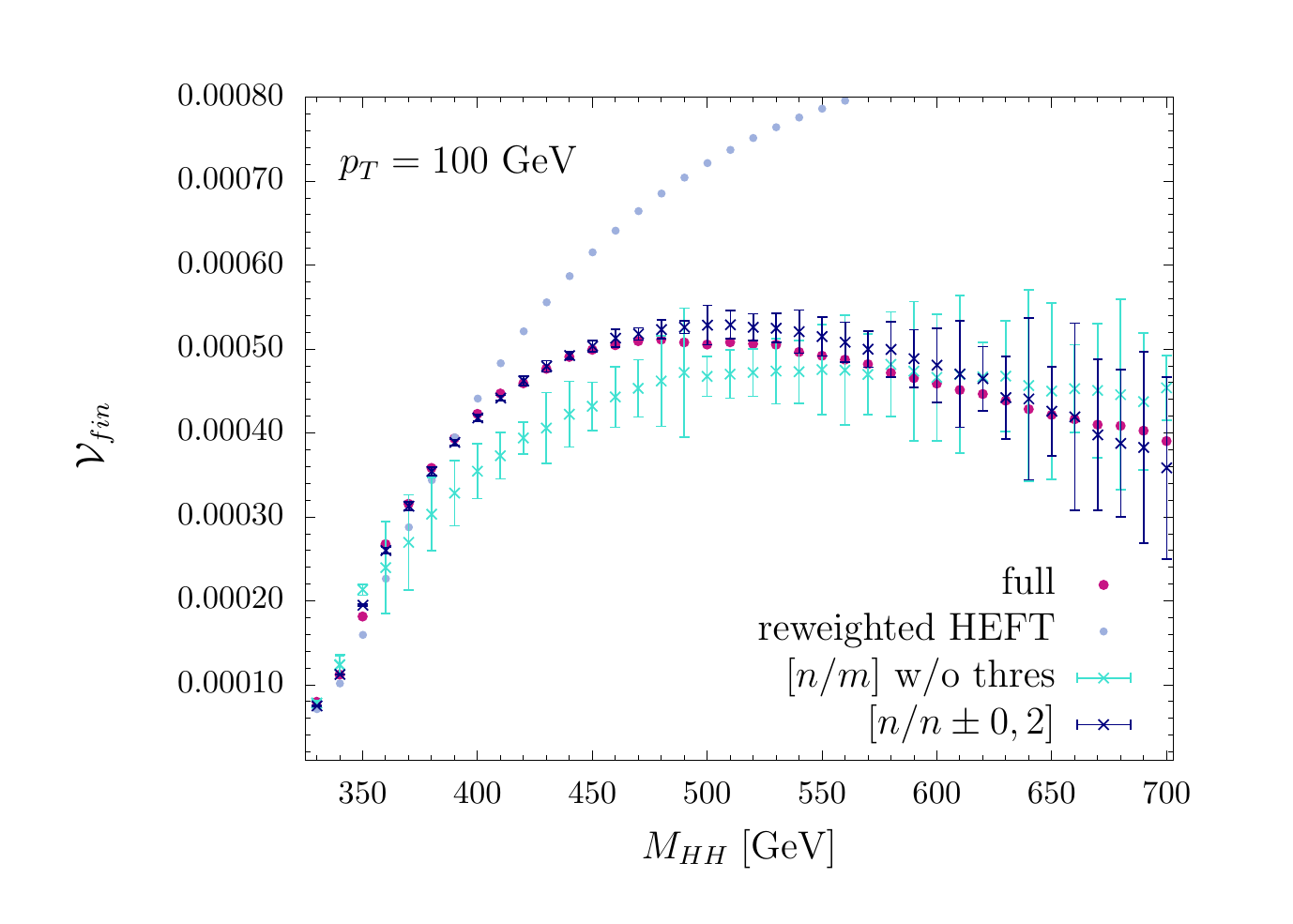}
\hspace*{-1.0cm}
\includegraphics[width=0.52\textwidth]{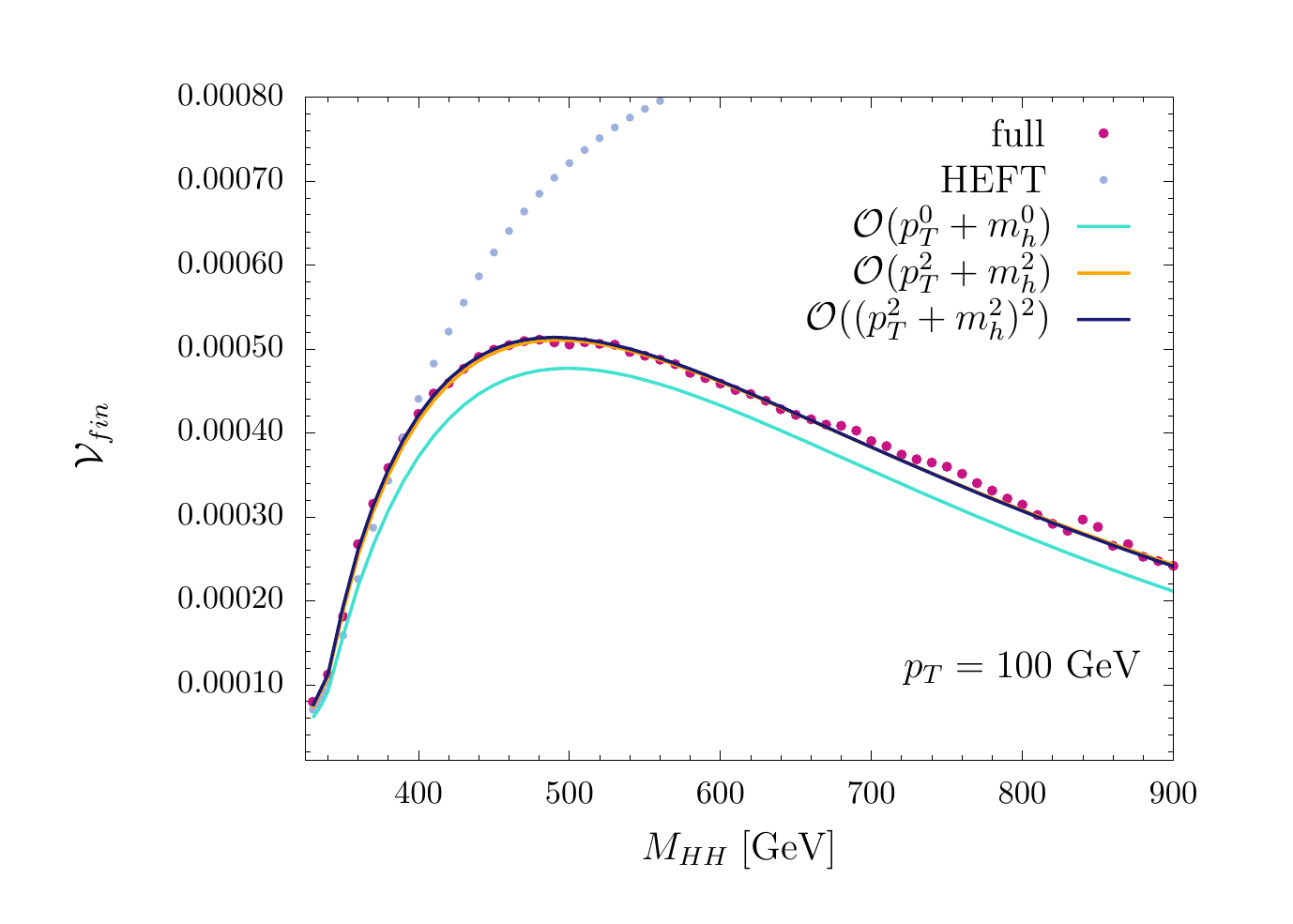}
\hspace*{-1.0cm}
\vspace*{-0.7cm}
\end{center}
\caption{Partonic virtual corrections to the Higgs pair invariant
mass distribution for Pad\'e approximants (left) and the $p_T^2$
expansion (right), for a Higgs transverse momentum $p_T = 100\text{ GeV}$. The full NLO results are shown as red points. The
other curves represent different orders included in the corresponding
expansions~\cite{Grober:2017uho,Bonciani:2018omm}.}
  \label{fg:gghhnloapprox}
\end{figure}

The virtual corrections have also been obtained by expansion methods. A
first approach is based on the large top-mass expansion that is
transformed into a complex polynomial by a suitable conformal mapping
and using Pad\'e approximants above the virtual $ t\bar t$-threshold
\cite{Fleischer:1994ef, Grober:2017uho}. The results are in mutual
agreement with the full numerical integration up to Higgs pair invariant
masses of about $700-800$ GeV \cite{Grober:2017uho}, see
\reffig{fg:gghhnloapprox}. A promising method is provided by a
different expansion in a Lorentz-invariant variable that mainly
corresponds to the transverse momentum of the Higgs bosons. The first
three terms of this expansion result in an excellent agreement with the
full numerical integration for di-Higgs invariant masses up to
$800-900$ GeV \cite{Bonciani:2018omm}, see \reffig{fg:gghhnloapprox}.
Last but not least, the approximate result of the large-$\mhhAlt$
expansion has been obtained analytically, including subleading terms
\cite{Davies:2018ood, Davies:2018qvx}. The latter result exhibits the
detailed logarithmic structure of the full NLO virtual corrections that
may be useful for further improvements. In addition, it provides a first
approach to the contribution of the bottom loops at NLO. Another
proposal for a valuable approximation is provided by a strict expansion
in the Higgs mass while keeping all other kinematical invariants
arbitrary \cite{Xu:2018eos}. However, this last option has not been
worked out completely for $\ggtohhAlt$ yet.

These expansions considerably simplify the problem and allow for an
analytical solution to be found in certain limits. Of course, an
immediate drawback with respect to the numerical calculation is that the
analytical result obtained in this way does not retain the full
dependence on the parameters over which the expansion is performed. On
the other hand, analytical calculations are usually faster and less
computationally intensive than numerical calculations, and can reach
high precision, thus providing a sound alternative in specific
instances and a valid check of the numerical result in the corresponding
limits.

Apart from the renormalization and factorization scale dependence, the
additional uncertainty due to the scale and scheme dependence related to
the top mass has to be taken into account. This has been analyzed in the
framework of the second numerical approach by deriving the full NLO
result not only for the top pole mass, but also for the $\overline{\rm
MS}$ mass at the scales of the top mass itself and in the range of
$\mhhAlt/4$ to $\mhhAlt$ for the scale of the running top mass. The
maximum and minimum of these results have been taken differentially in
$\mhhAlt$. This leads to sizeable additional uncertainties
\cite{Baglio:2018lrj},
\begin{eqnarray}
\frac{d\sigma(\ggtohhAlt)}{dQ}\Big|_{Q=300~{\rm GeV}} & = &
0.02978(7)^{+6\%}_{-34\%}\,
\mathrm{fb/GeV},\nonumber\\
\frac{d\sigma(\ggtohhAlt)}{dQ}\Big|_{Q=400~{\rm GeV}} & = &
0.1609(4)^{+0\%}_{-13\%}\,
\mathrm{fb/GeV},\nonumber\\
\frac{d\sigma(\ggtohhAlt)}{dQ}\Big|_{Q=600~{\rm GeV}} & = &
0.03204(9)^{+0\%}_{-30\%}\,
\mathrm{fb/GeV},\nonumber\\
\frac{d\sigma(\ggtohhAlt)}{dQ}\Big|_{Q=1200~{\rm GeV}} & = &
0.000435(4)^{+0\%}_{-35\%}\,
\mathrm{fb/GeV}
\end{eqnarray}
for the differential cross section at $\sqrt{s}=14$ TeV using PDF4LHC15 parton densities.

\begin{figure}
\begin{center}
\includegraphics[width=0.49\textwidth]{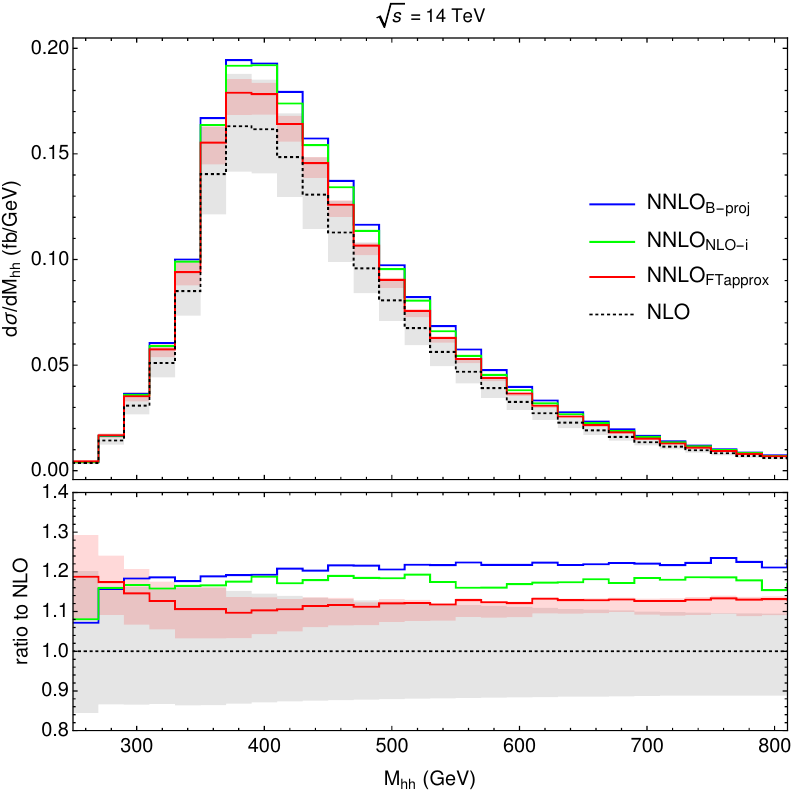}
\hfill
\includegraphics[width=0.487\textwidth]{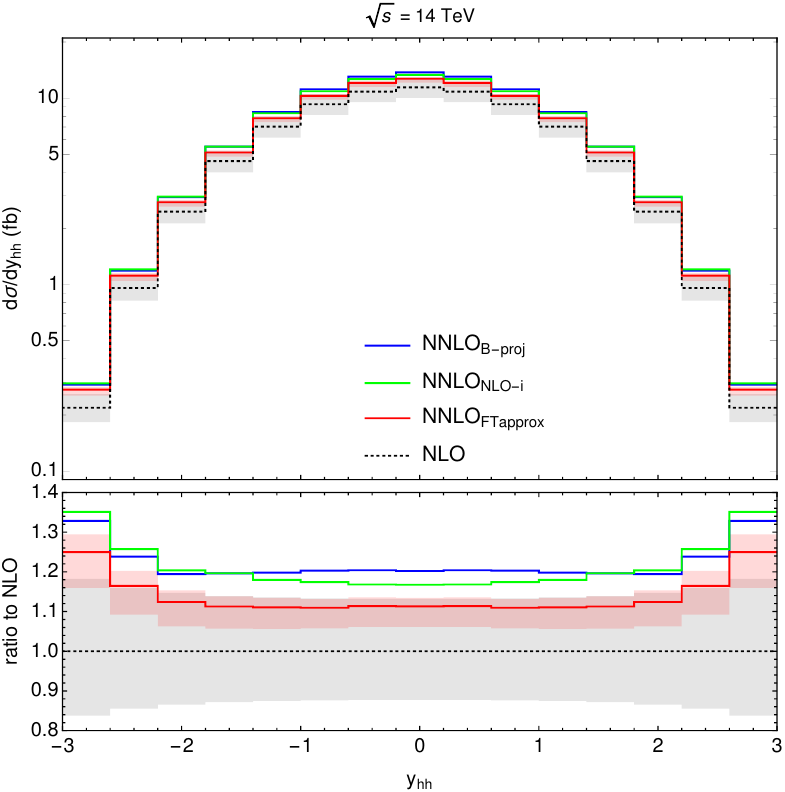}
\vspace*{-0.6cm}
\end{center}
\caption{Higgs pair invariant mass and rapidity
distributions for a collider energy of 14 TeV in various approximations
using PDF4LHC15 parton densities. The NNLO$_{\rm FTapprox}$ results are
shown in red.  The grey and red bands show the scale dependence at NLO
and NNLO~\cite{Grazzini:2018bsd}.}
  \label{fg:gghhnnloex}
\end{figure}

The full NLO corrections have been combined recently with the NNLO QCD
corrections in the HTL, to construct a full NNLO Monte Carlo program for
exclusive Higgs pair production via gluon fusion \cite{Grazzini:2018bsd}.
In this implementation, the second-order corrections have been improved via a reweighting technique to account for partial finite top quark mass effects, in what represents a NNLO extension of the ``FTapprox''.
Within this approach, the NNLO parts of the virtual and real
corrections that are obtained in the HTL at NNLO are rescaled by the ratio between the
corresponding full one-loop (i.e.~LO) amplitudes and the ones obtained in
the HTL for each partonic subprocess individually.
%
The double-real
corrections are added including the full mass dependence, since the
related one-loop amplitudes can be obtained by presently available
automatic tools. This approximation is an
improvement of the previous ``Born-improved'' and ``FTapprox'' approaches used at
NLO, and is expected to deliver more reliable results at NNLO
\cite{Grazzini:2018bsd}. Final predictions at NLO and NNLO are presented in
\refta{tb:gghhnnlo} for different centre-of-mass energies with $\mtAlt = 173$~GeV.
\begin{table}
\renewcommand{\arraystretch}{1.8}
\begin{center}
\begin{tabular}{|l|c|c|c|c|} \hline
Energy & 13 TeV & 14 TeV & 27 TeV & 100 TeV \\ \hline
 NLO                & $27.78^{+13.8\%}_{-12.8\%}$ fb  & $32.88^{+13.5\%}_{-12.5\%}$ fb &
$127.7^{+11.5\%}_{-10.4\%}$ fb & $1147^{+10.7\%}_{-9.9\%}$ fb \\ \hline
 NLO$_{\rm FTapprox}$ & $28.91^{+15.0\%}_{-13.4\%}$ fb& $34.25^{+14.7\%}_{-13.2\%}$ fb &
$134.1^{+12.7\%}_{-11.1\%}$ fb & $1220^{+11.9\%}_{-10.6\%}$ fb \\ \hline
NNLO$_{\rm FTapprox}$ & $31.05^{+2.2\%}_{-5.0\%}$ fb& $36.69^{+2.1\%}_{-4.9\%}$ fb   &
$139.9^{+1.3\%}_{-3.9\%}$ fb   & $1224^{+0.9\%}_{-3.2\%}$ fb \\ \hline
\end{tabular}
\end{center}
\vspace*{-3mm}
\caption{\label{tb:gghhnnlo} NLO and NNLO cross sections for proton
colliders at 13, 14, 27 and 100 TeV centre-of-mass energy using PDF4LHC15 parton
densities. The explicit percentage numbers present the renormalization
and factorization scale dependencies. The central scale choice is
$\mu_R=\mu_F=\mhhAlt/2$~\cite{Grazzini:2018bsd}.}
\renewcommand{\arraystretch}{1.0}
\end{table}
From these values it is visible that the ``FTapprox" method works with an
accuracy of better than 10\% at NLO, so that the NNLO$_{\rm FTapprox}$
results are expected to be more reliable than the left-over uncertainties.
The corresponding NNLO Monte Carlo program can be used to provide NNLO
predictions for exclusive quantities, i.e.~for distributions. Typical
numerical results are shown in \reffig{fg:gghhnnloex}.
In addition, the all-orders resummation of soft-gluon contributions has been performed at NNLL within this approximation, finding -- as it happens in the HTL -- that the effects are very small if the central scale $\mu_R = \mu_F = \mhhAlt/2$ is used, indicating the stability of the perturbative expansion at this order \cite{deFlorian:2018tah}.

\section[Cross section as a function of $\kl$]{Cross section as a function of $\kl$\\
\contrib{G. Heinrich, S.P. Jones, M. Kerner, L. Scyboz}
}\label{sec:xs_vs_lambda}

Non-resonant Higgs boson pair production in gluon fusion is the most promising process to test the trilinear Higgs boson self-coupling at hadron colliders. 
The current constraints at 95\% confidence level from ATLAS and CMS searches, combining various decay channels, are $-5\leq \kl\leq 12.1$~\cite{Aad:2019uzh}
and $-11.8\leq \kl\leq 18.8$~\cite{Sirunyan:2018two}, respectively,
where $\kl=\lHcube/\lHcubeSM$ (see Sec.~\ref{sec:non_res:comb}). 
In order to derive reliable limits on $\kl$ from these searches, it is crucial to have accurate predictions for the cross sections corresponding to non-SM $\lHcube$ values.
The results presented in this section for a generic $\kl$ are NLO-accurate, including the full top quark mass dependence \cite{Heinrich:2019bkc}. They are based on the original calculation of Refs.~\cite{Borowka:2016ehy,Borowka:2016ypz} for the SM cross section, which has been extended to include effects from anomalous couplings in the Higgs sector within a non-linear Effective Field Theory framework in Ref.~\cite{Buchalla:2018yce}.

To obtain a full-fledged NLO generator which also offers the possibility of parton showering, we implemented the calculation in the 
\powhegbox~\cite{Nason:2004rx,Frixione:2007vw,Alioli:2010xd}, building on top of the code presented in Ref.~\cite{Heinrich:2017kxx} for the NLO+PS predictions within the SM;
the code is publicly available in the \powhegbox{\sc-V2} package.\footnote{The code can be found at the website \url{http://powhegbox.mib.infn.it} in the
{\tt User-Processes-V2/ggHH/} directory.}

The results were obtained using the
PDF4LHC15{\tt\_}nlo{\tt\_}30{\tt\_}pdfas~\cite{Butterworth:2015oua,CT14,MMHT14,NNPDF}
parton distribution functions interfaced to the code via
LHAPDF~\cite{Buckley:2014ana}, along with the corresponding value for
$\as$.  The masses of the Higgs boson and the top quark have been
fixed to $\mhAlt=125$\,GeV, $\mtAlt=173$\,GeV, respectively, 
where the pole mass scheme has been employed for the top quark mass.  
The widths $\Gamma_{H}$ and $\Gamma_{t}$ have been set to zero.
Jets are clustered with the
anti-$k_T$ algorithm~\cite{Cacciari:2008gp} as implemented in the
{\tt fastjet} package~\cite{Cacciari:2005hq, Cacciari:2011ma}, with jet
radius $R=0.4$ and a minimum transverse momentum 
$p_{T,\rm{min}}^{\rm{jet}}=20$\,GeV.  The scale uncertainties are
estimated by varying the factorisation and renormalization scales
$\mu_{F}$ and $\mu_{R}$. The scale variation bands result from varying $\mu = \mu_{F} = \mu_{R}$ by a factor of two 
around the central scale $\mu_0 =\mhhAlt/2$. For $\lHcube=\lHcubeSM$, the envelope of the scale variations coincides with the 7-point scale variation band.

\begin{figure}[tb]
  \centering
    \includegraphics[width=0.7\textwidth]{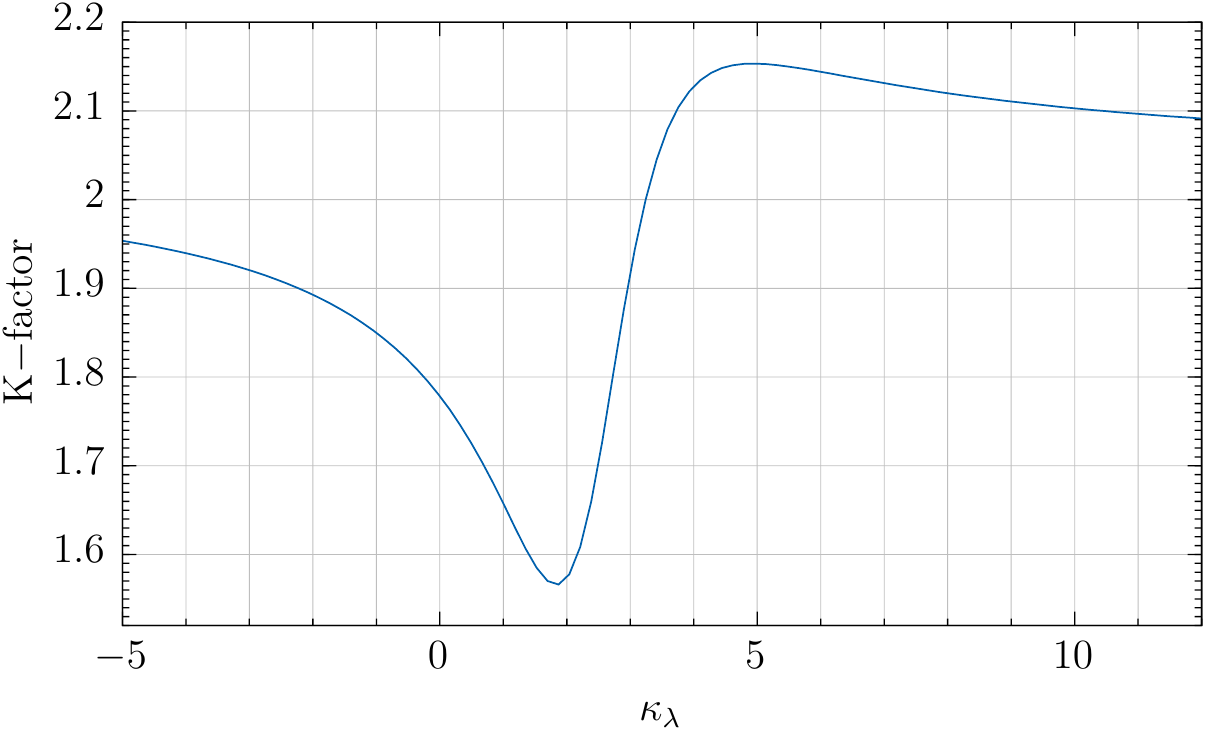}
\caption{Variation of the NLO K-factor with the trilinear coupling for $\sqrt{s}=14$\,TeV \cite{Heinrich:2019bkc}.}
\label{fig:Kfacvariation}
\end{figure}

In \refta{tab:sigmatot} we list total cross sections at 13, 14 and 27\,TeV for various values of the trilinear Higgs coupling. 
\begin{table}[tb]
\renewcommand{\arraystretch}{1.45}
\begin{center}
\begin{tabular}{| c | c | c |c|c|}
\hline
&&&& \\[-2.5ex]
$\;\;\; \kl \;\;\;$ & $\sigma_{\rm{NLO}},\,13 \mathrm{TeV}$\,[fb]& $\sigma_{\rm{NLO}},\,14 \mathrm{TeV}$\,[fb] & $\sigma_{\rm{NLO}},\,27 \mathrm{TeV}$\,[fb] &K-factor, 14TeV\\
&&&& \\[-2.5ex]
\hline 
-1& 116.71$^{+16.4\%}_{-14.3\%}$  & 136.91$^{+16.4\%}_{-13.9\%}$& 504.9$^{+14.1\%}_{-11.8\%}$ & 1.86 \\
\hline
0& 62.51$^{+15.8\%}_{-13.7\%}$ & 73.64$^{+15.4\%}_{-13.4\%}$& 275.29$^{+13.2\%}_{-11.3\%}$ & 1.79 \\
\hline 
1& 27.84$^{+11.6\%}_{-12.9\%}$ & 32.88$^{+13.5\%}_{-12.5\%}$&127.7$^{+11.5\%}_{-10.4\%}$ &1.66 \\
\hline
2 & 12.42$^{+13.1\%}_{-12.0\%}$ & 14.75$^{+12.0\%}_{-11.8\%}$ &  59.10$^{+10.2\%}_{-9.7\%}$ & 1.56\\
\hline
2.4& 11.65$^{+13.9\%}_{-12.7\%}$ & 13.79$^{+13.5\%}_{-12.5\%}$& 53.67$^{+11.4\%}_{-10.3\%}$ & 1.65\\
\hline
3& 16.28$^{+16.2\%}_{-15.3\%}$ & 19.07$^{+17.1\%}_{-14.1\%}$ & 69.84$^{+14.6\%}_{-12.1\%}$ & 1.90 \\
\hline 
5& 81.74$^{+20.0\%}_{-15.6\%}$  & 95.22$^{+19.7\%}_{-11.5\%}$& 330.61$^{+17.4\%}_{-13.6\%}$ & 2.14 \\
\hline 
\end{tabular} 
\end{center}
\vspace*{-4mm}
\caption{Total cross section for Higgs boson pair production at full NLO for different values of $\kl$. The given uncertainties are scale uncertainties, and we use the central value $\mu_R = \mu_F = m_{HH}/2$ \cite{Heinrich:2019bkc}. The K-factors reported for the 14~TeV results are also valid at 13~TeV, with the exception of the $\kl=2$ K-factor which takes the value 1.57.
\label{tab:sigmatot}}
\end{table}
We observe that $\kl=-1$ leads to the largest total cross section of all the considered $\kl$ values.
\refta{tab:sigmatot} also shows that the K-factors vary
substantially as functions of the trilinear coupling, which is
different from the findings in the $\mtAlt\to\infty$ limit~\cite{Grober:2015cwa,deFlorian:2017qfk}.
This fact is illustrated in \reffig{fig:Kfacvariation}, which shows that the K-factor takes values between 1.57 and 2.16
if the trilinear coupling is varied between $-5\leq \kl\leq 12$.

\begin{figure}[tb]
 \includegraphics[width=0.49\textwidth]{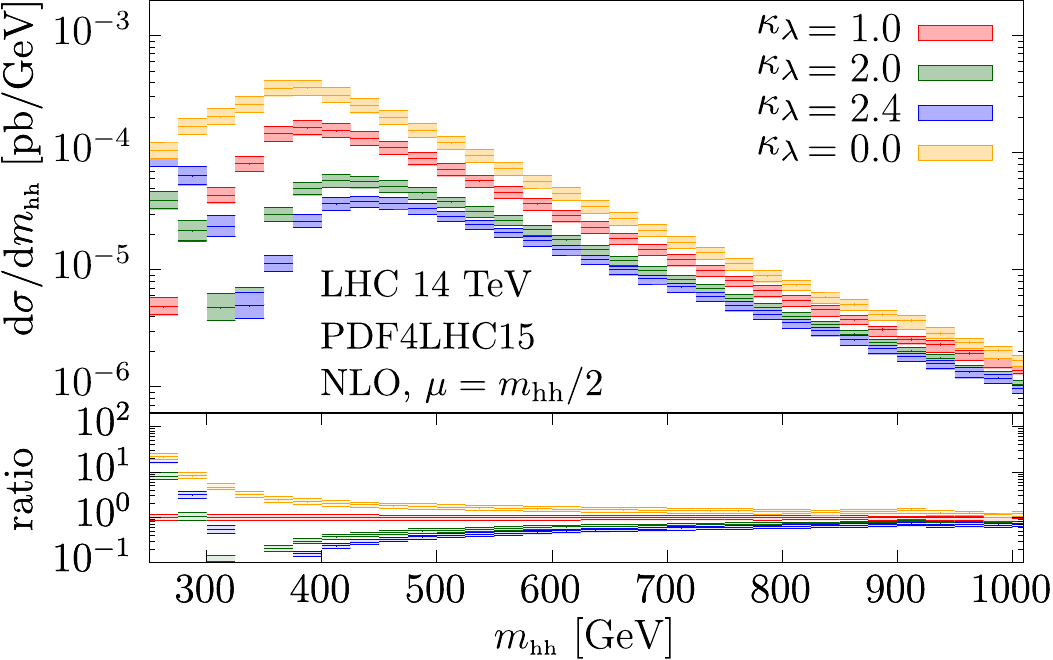}
\label{fig:lambda_small}
\includegraphics[width=0.49\textwidth]{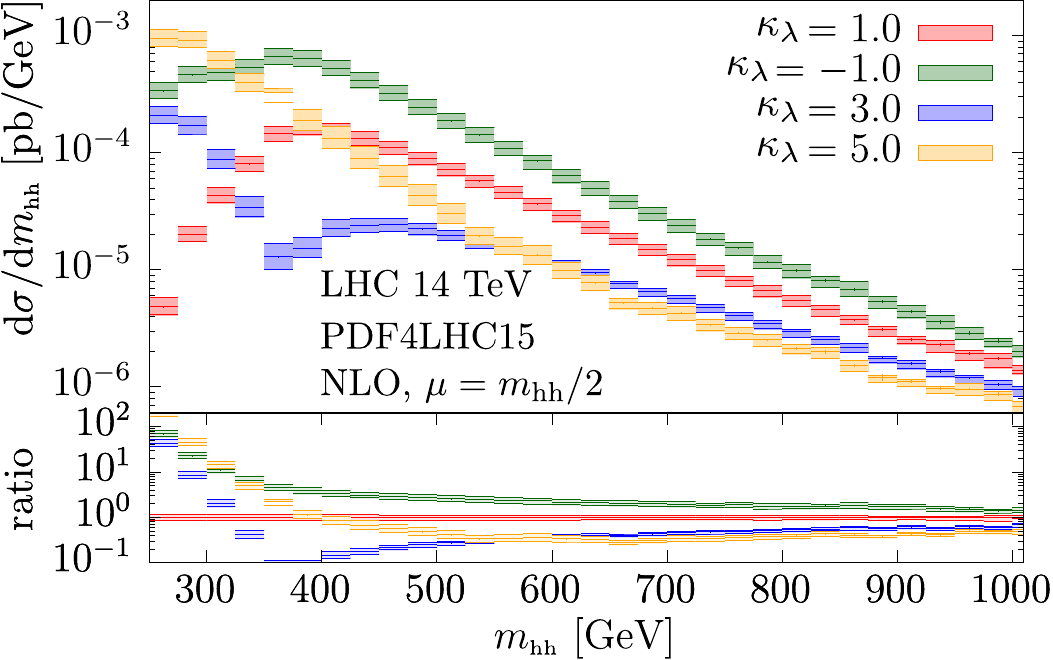}
\label{fig:lambda_large}
\caption{Higgs boson pair invariant mass distributions at 14\,TeV for (left) positive small values of $\kl$ and (right) larger or negative values of $\kl$ \cite{Heinrich:2019bkc}.}
\label{fig:lambdavar14TeV}
\end{figure}

\begin{figure}[tb]
\includegraphics[width=0.49\textwidth]{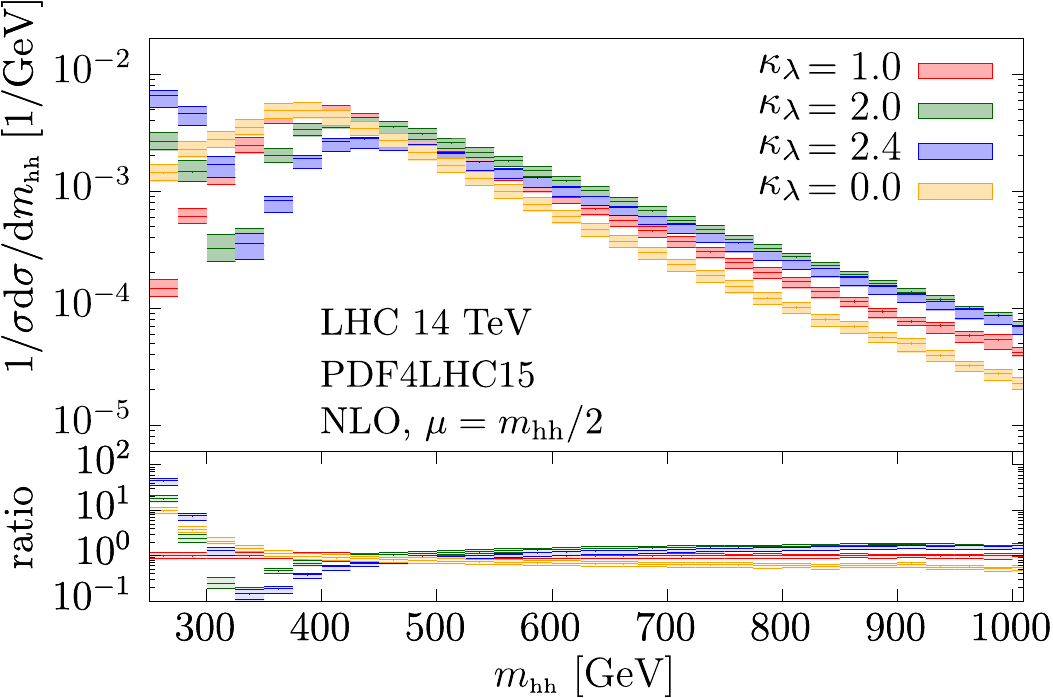}
\label{fig:lambda_small-norm}
\includegraphics[width=0.49\textwidth]{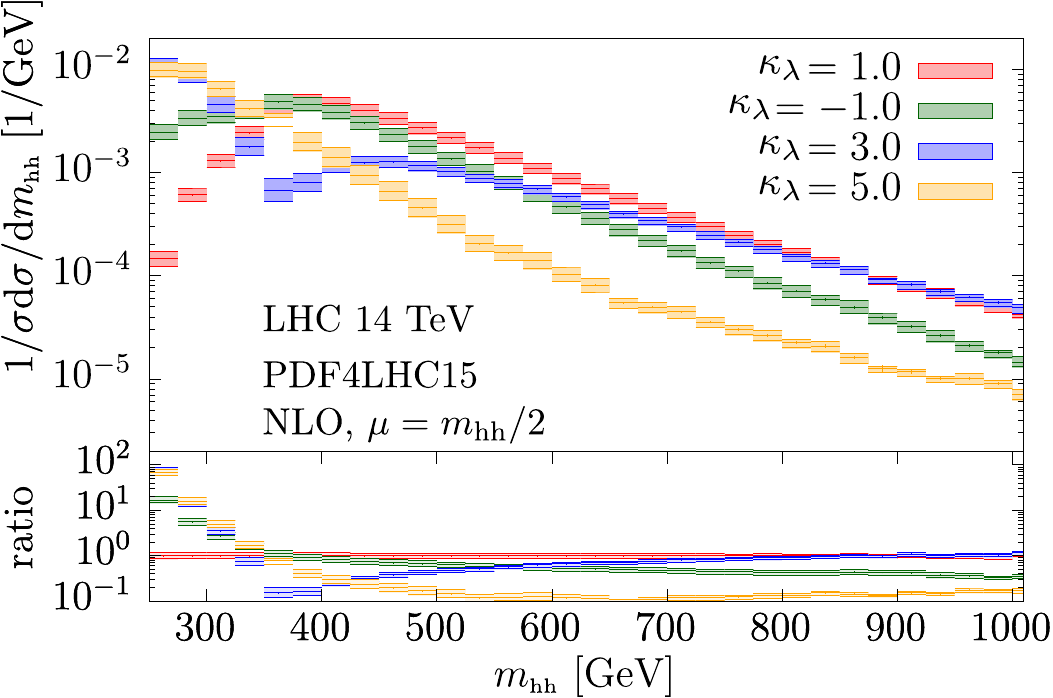}
\label{fig:lambda_large-norm}
\caption{Normalised Higgs boson pair invariant mass distributions at 14\,TeV for (left) positive small values of $\kl$ and (right) larger or negative values of $\kl$ \cite{Heinrich:2019bkc}.}
\label{fig:lambdavar14TeV-norm}
\end{figure}

\begin{figure}[tb]
\includegraphics[width=0.49\textwidth]{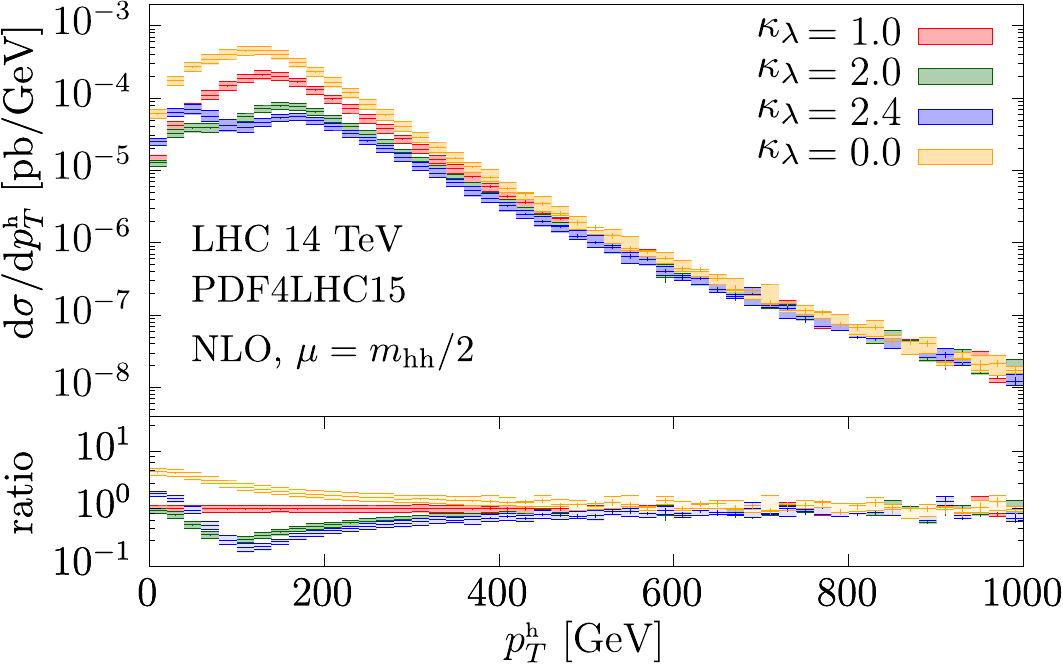}
\label{fig:lambda_small_pTH}
\includegraphics[width=0.49\textwidth]{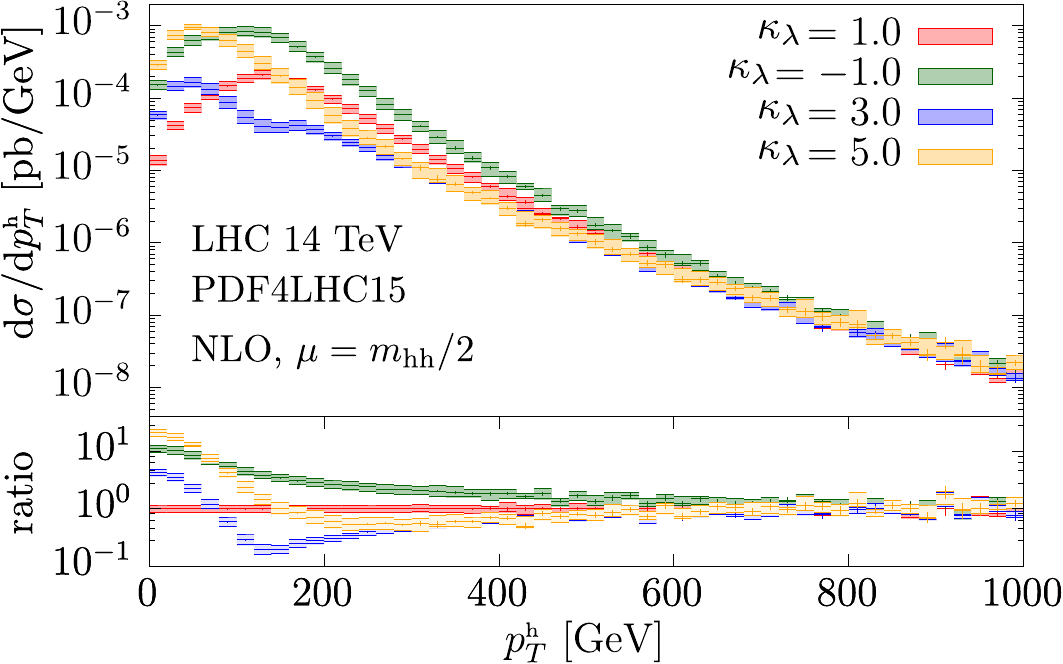}
\label{fig:lambda_large_pTH}
\caption{Higgs boson transverse momentum distributions at 14\,TeV for the considered $\kl$ values \cite{Heinrich:2019bkc}.}
\label{fig:lambdavar14TeV_pTH}
\end{figure}

In \reffigs{fig:lambdavar14TeV} and \ref{fig:lambdavar14TeV-norm} we
show the $\mhhAlt$ distribution for various values of $\kl$. The
results in \reffig{fig:lambdavar14TeV-norm} are distributions
normalised to the total cross section for the corresponding value of $\kl$.
The ratio plots show the ratio to the Standard Model (SM) result. 
A characteristic dip develops in the $\mhhAlt$ distribution around $\kl=2.4$, which is the value of maximal destructive interference between diagrams containing the trilinear coupling (triangle-type contributions) and ``background" diagrams (box-type contributions).
 We provide results for a denser spacing of $\kl$ values around this point. For $\kl < -1$ and $\kl > 5$ the triangle-type contributions dominate increasingly, leading to a shape where the low-$\mhhAlt$ region is more and more enhanced.
 In the transverse momentum distribution of one (any) of the Higgs bosons, shown in \reffig{fig:lambdavar14TeV_pTH}, effects of the destructive interference around $\kl=2.4$ are also visible, however they are less pronounced.

\begin{figure}[tb]
\begin{center}
  \includegraphics[width=0.75\textwidth]{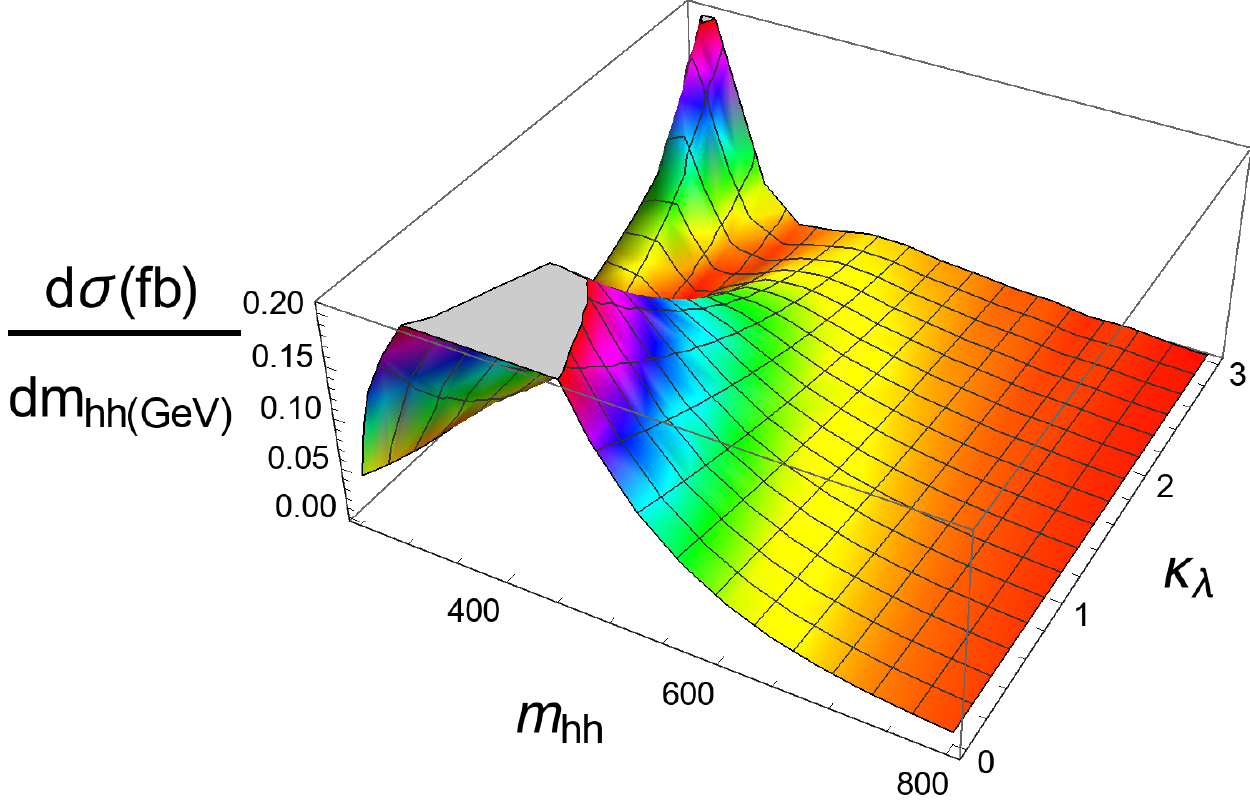}    
\end{center}
\vspace*{-2mm}
\caption{3-dimensional visualisation of the $\mhhAlt$ distribution at
  14\,TeV, as a function of $\kl$ and $\mhhAlt$ \cite{Heinrich:2019bkc}.}
\label{fig:chhh_3D}
\end{figure}
\reffig{fig:chhh_3D} shows the Higgs boson pair invariant mass
distribution at NLO as a function of $\kl$ as a 3-dimensional heat map, where the dip in the $\mhhAlt$ distribution for 
$\kl$ values close to 2.4 is again visible.



To summarise, we have presented in this section full NLO QCD results for Higgs boson pair production 
 for various values of the trilinear Higgs boson coupling.
 We have provided total cross sections for 13, 14 and 27\,TeV, and differential results at 14\,TeV, including scale uncertainties.
 The matrix elements have been implemented in the \powhegbox{\sc-V2} Monte Carlo framework and the corresponding generator is publicly available.

 A combination of the NLO result with full top quark mass dependence presented in this section and the NNLO computed in the (improved) HTL has not yet been done for the case of non-SM $\kl$ values, though work in this direction is in progress \cite{LH_proceedings_inprep}.
 Such a combination would be desirable in order to match the level of accuracy obtained for the SM prediction.
 For the time being, the simplest approach to account for the higher order corrections consists in multiplying the results in Table~\ref{tab:sigmatot} by the SM K-factor, i.e. the ratio of the NNLO$_\text{FTapprox}$ and NLO results in Table~\ref{tb:gghhnnlo}.

\section[Differential predictions and MC generators for gluon fusion]{Differential predictions and MC generators for gluon fusion \\
\contrib{G.~Heinrich, S.P.~Jones, M.~Kerner, S.~Kuttimalai, E.~Vryonidou}
}\label{sec:MC}


The non-resonant production of a pair of Higgs bosons in gluon fusion is available within several public Monte Carlo programs. Currently, the most sophisticated predictions which include a parton shower are based on the NLO matrix-element including a finite top quark mass~\cite{Borowka:2016ehy,Borowka:2016ypz}. The fixed-order result was recently re-calculated and extended to allow also for a running top quark mass~\cite{Baglio:2018lrj}. The NLO calculation was first interfaced to the \powhegbox~\cite{Frixione:2007vw,Alioli:2010xd} and \MGAMCNLO~\cite{Alwall:2014hca,Hirschi:2015iia} in Ref.~\cite{Heinrich:2017kxx}, and to \sherpa~\cite{Gleisberg:2008ta} in Ref.~\cite{Jones:2017giv}.

The matching and parton shower uncertainties have been extensively studied in the literature~\cite{Heinrich:2017kxx,Jones:2017giv,Bendavid:2018nar}, and were found to be large for certain observables. Similar effects have been observed in other processes including the production of a Higgs boson in gluon fusion~\cite{Alioli:2008tz,Bagnaschi:2015bop} and Z-boson pair production in gluon fusion~\cite{Alioli:2016xab}. 

Here, we briefly review the current status of these uncertainties focusing on one of the most sensitive distributions (the $p_{T}$ of the di-Higgs boson system). We will summarise the \MCNLO~\cite{Frixione:2002ik} and \powheg~\cite{Nason:2004rx} matching schemes used in the literature. Results obtained from the \powhegbox, \MGAMCNLO{} and \sherpa{} implementations and via analytic resummation~\cite{Ferrera:2016prr} are compared. The shower uncertainty observed for the \powhegbox{} implementation will also be discussed.

\subsubsection{Parton Shower Matching}
\label{sec:matching}

Already in a pure fixed-order NLO calculation there are
contributions in both the Born phase space $\phi_B$ and in the real
emission phase space $\phi_R=\phi_B\times\phi_1$. In a
parton shower matched calculation, we denote them by $\bar B(\phi_B)$
and $H(\phi_R)$, respectively:
\begin{align}
  \bar B(\phi_B) &= B(\phi_B)+V(\phi_B)+\int D(\phi_R) \Theta(\mu^2_\text{PS}-t(\phi_R))\dif\phi_1,\label{eq:sevent}\\
                   H(\phi_R) &= R(\phi_R) -
                               D(\phi_R)\Theta(\mu^2_\text{PS}-t(\phi_R))\label{eq:hevent}\,.
\end{align}
In \refeqs{eq:sevent} and \eqref{eq:hevent}, $B$ denotes the
leading-order contributions, $V$ the UV-subtracted virtual
corrections, $R$ the real-emission corrections, and $D$ the
differential infrared subtraction terms. The scale $\mu_\text{PS}$ is
the parton shower starting scale and $t(\phi_R)$ is the evolution
variable of the parton shower. Through variations of $\mu_\text{PS}$,
contributions can be shuffled around between $\bar B$ and $H$ while
leaving their sum constant. 

When considering \refeqs{eq:sevent} and \eqref{eq:hevent} by
themselves, real emission configurations are generated only in $H$
events. Furthermore, the emissions are suppressed in the phase space
region $t(\phi_R)<\mu_\text{PS}$ due to the subtraction terms
$D(\phi_R)$. For $t\ll\mu_\text{PS}$, emissions are completely
suppressed since there we have $R\sim D$, and thus $H\sim 0$. These
missing real-emission terms are generated through the first parton
shower emissions off the $\bar B$ events. Taking into account the first
emission, the sum of \refeqs{eq:sevent} and \eqref{eq:hevent} can be
written as
\begin{align}
    \sigma_\text{NLO+PS} = 
    &\int \bar B(\phi_B)
      \left[\Delta(t_0,\mu^2_\text{PS}) + \int
      \Delta(t,\mu^2_\text{PS})\frac{D(\phi_B,
      \phi_1)}{B(\phi_B)}\Theta(\mu_\text{PS}^2-t)\Theta(t-t_0)\dif\phi_1\right]\dif\phi_B\label{eq:sevent_ps}
      \nonumber\\
    +&\int H(\phi_R)\dif\phi_R\,.
\end{align}
Here and in what follows we will assume that the parton shower
splitting kernels are given by $\frac{D}{B}$, i.e. by the kernels that
are also used in the infrared subtraction scheme. The Sudakov form
factor is then given by
$\Delta(t_0,t_1)=\exp\left[
  -\int^{t_1}_{t_0}\frac{D(\phi_R)}{B(\phi_B)}\dif\phi_1 \right]$ and
the infrared cutoff scale of the parton shower is $t_0$. The first
term in the square bracket of \refeq{eq:sevent_ps} corresponds to the
probability of generating no emission above the parton shower cutoff
scale for a $\bar B$ event. The second term represents the probability
of generating an emission somewhere between the starting scale
$\mu_\text{PS}$ and $t_0$. These terms therefore fill the remaining
real-emission phase space region of soft emissions that are subtracted
in $H$ and would otherwise be missing. The scale $\mu_\text{PS}$
therefore separates the real emission phase space in a resummation
region that is populated by the parton shower through the $\bar B$
events and a region that is populated mostly by the fixed-order
real-emission contributions in $H$. Variations of this scale can be
used in order to assess uncertainties associated with this separation.

The \powheg{} method can be understood, in the formulation presented above,
as the limit in which the parton shower starting scale is
set equal to the collider energy $\mu_\text{PS} = \sqrt{s} $ and
$D=R$. This choice leads to $H=0$ and all real emission contributions are therefore
generated by parton shower emission off $\bar B$ events. The choice
$\mu_\text{PS} = \sqrt{s}$ ensures that the full real-emission phase
space is covered. Setting $D=R$ in the first emission ensures
that the fixed-order radiation pattern is recovered in the hard region
where the Sudakov form factor is approximately one. However, setting $D=R$ also 
results in the full real-emission corrections being exponentiated in
the Sudakov form factor. This is in general not justified since $R$
contains hard, non-factorizing contributions. In Ref.~\cite{Alioli:2008tz}
it was instead suggested to use
\begin{align}
D=\frac{h_\mathrm{damp}^2}{p_T^2+h_\mathrm{damp}^2}R\,,
\end{align}
where $p_T$ is the transverse momentum of the Born final state
($p_T=p_{T}^{\hhAlt}$ in the case under consideration). This choice limits the amount 
of hard radiation that gets exponentiated.

\subsubsection{Parton Shower Results}
\label{sec:results}

In the literature, the full NLO di-Higgs boson production calculation has been combined with a parton shower within the \powhegbox, \MGAMCNLO{} and \sherpa{} frameworks. The \powhegbox{} framework relies on the \powheg{} scheme to match the fixed-order calculation with the parton shower, while \MGAMCNLO{} and \sherpa{} use the MC@NLO matching scheme. The \powheg{} and \MGAMCNLO{} implementations generate showered events using a \pythia\,8.2~\cite{Sjostrand:2007gs,Sjostrand:2014zea} shower whilst showered events are generated within \sherpa{} using the built-in Catani-Seymour (CS)~\cite{Schumann:2007mg} or Dire~\cite{Hoche:2015sya} showers.

In general, most distributions were found to be only moderately sensitive to the matching scheme. In particular, matching scheme uncertainties for NLO accurate observables were all found to be within the scale uncertainties~\cite{Heinrich:2017kxx}. However, the impact of the parton shower on the $p_{T}^{\hhAlt}$ (transverse momentum of the di-Higgs boson system), $\Delta\Phi^{\hhAlt}$ (difference in azimuthal angle of the Higgs bosons) and $\Delta R^{\hhAlt}$ (radial separation of the Higgs bosons) was found to be fairly large. The sizeable impact of the parton shower is to be expected as the tails of these distributions are predicted only at the first non-trivial order in the fixed-order calculation. The matching scheme uncertainties for these distributions were also found to be significant and could even become larger than the scale uncertainties.

\begin{figure}
  \centering
  \raisebox{-0.5\height}{\includegraphics[width=.48\linewidth]{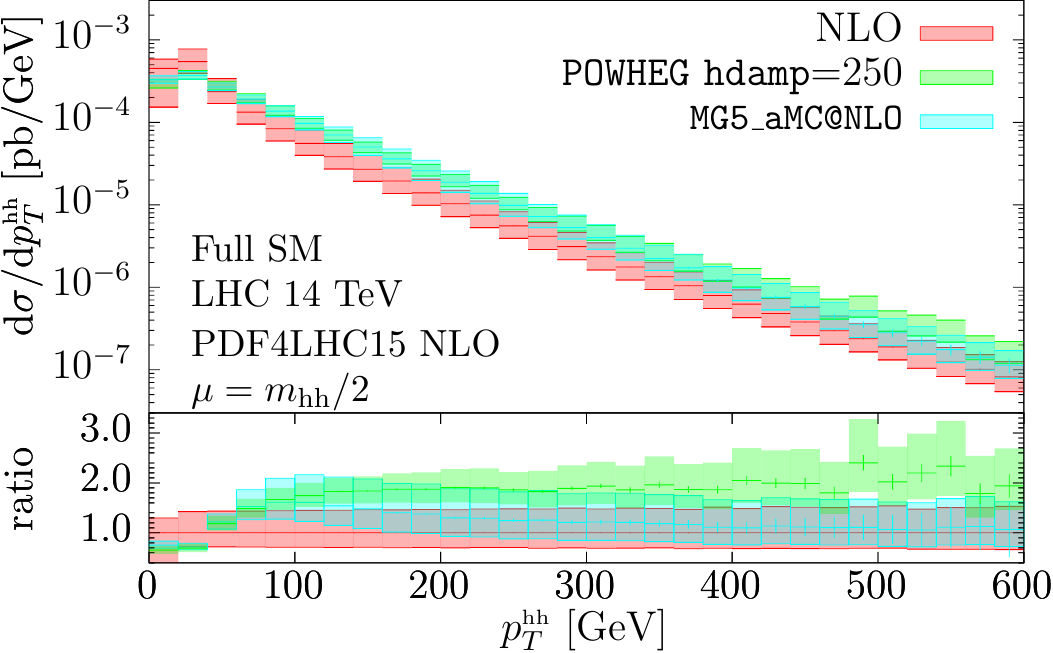}}
  \hfill
  \raisebox{-0.5\height}{\includegraphics[width=.48\linewidth]{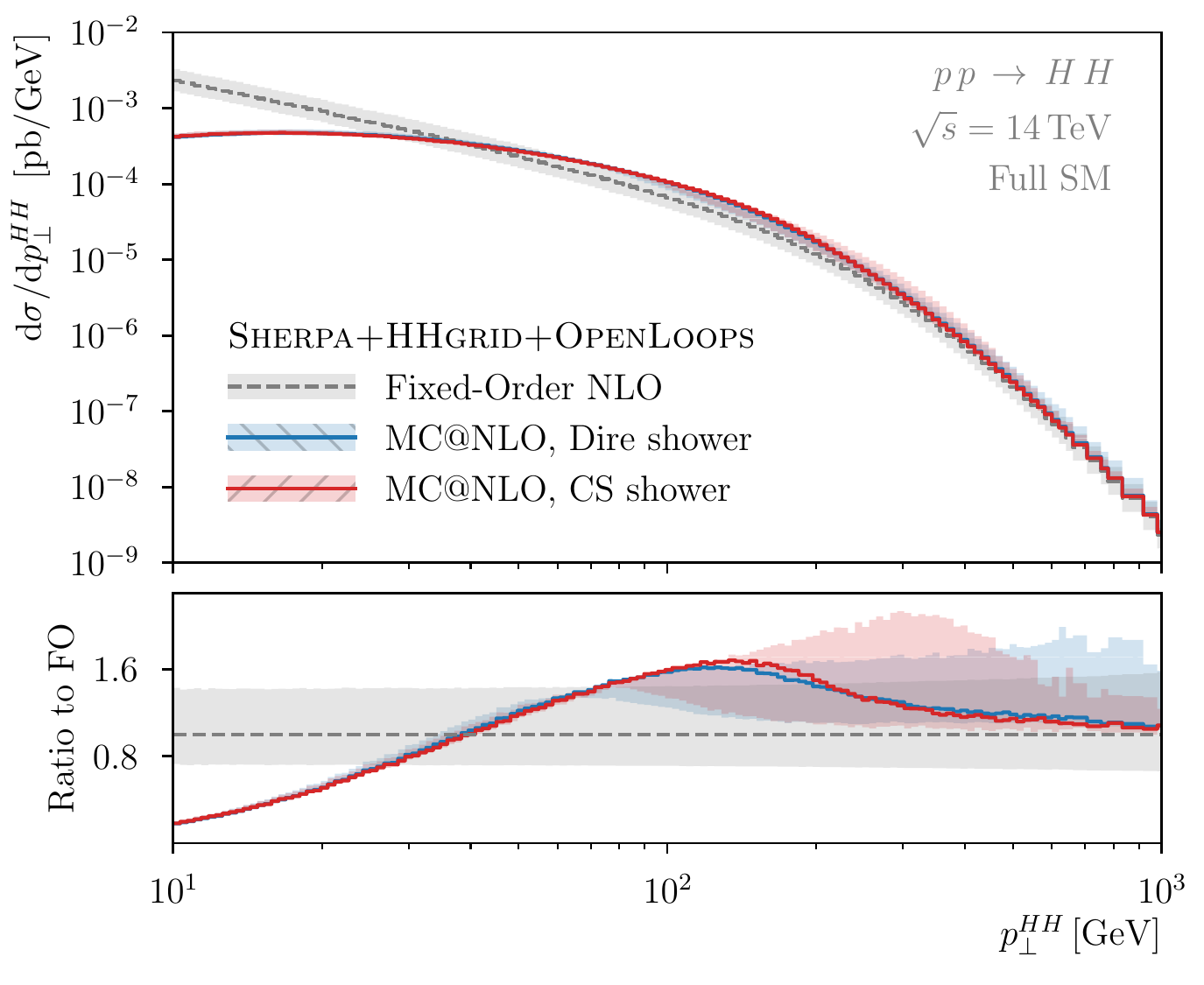}}
  \vspace*{-2mm}
  \caption{
  Left: Comparison between the \powhegbox, \MGAMCNLO{} and NLO fixed-order results for the Higgs boson pair transverse momentum.
  The uncertainty bands were obtained through a 7-point scale variation of the factorization and renormalization scales~\cite{Heinrich:2017kxx}.
  Right: A comparison of the \sherpa{}  and NLO fixed-order results. 
  The \sherpa{} uncertainty bands indicate the shower scale uncertainty obtained by varying $\mu_\mathrm{PS}$. 
  The bands on the fixed-order prediction were obtained by varying $\mu_F$ and $\mu_R$~\cite{Jones:2017giv}.}
  \label{fig:comp-fo}
\end{figure}

In \reffig{fig:comp-fo} the NLO fixed-order result for the $p_{T}$ of the di-Higgs boson system is compared to the showered predictions. The bands displayed on the left plot indicate the scale uncertainty, which is obtained via a 7-point scale variation of the factorization scale $\mu_F$ and renormalization scale $\mu_R$ around the central scale choice $\mu_0 = \mhhAlt/2$, where $\mhhAlt$ is the invariant mass of the Higgs boson pair. In the right plot the grey band indicates the scale uncertainty while the coloured bands display the shower starting scale uncertainty. The \MGAMCNLO{} prediction is produced using a shower starting scale of $\mu_\mathrm{PS} = H_T/2$, where $H_T$ is the sum of the transverse energies of the Higgs bosons. The \powhegbox{} prediction is produced using $h_\mathrm{damp} = 250$ GeV. For the \sherpa{} predictions, the central parton shower starting scale choice in the case of the Dire shower is $\mu_\mathrm{PS} = \mhhAlt/4$, whereas $\mu_\mathrm{PS} = \mhhAlt/2$ is used for the CS shower. The shower scale uncertainty is obtained by varying the parton shower starting scale up and down by a factor of 2. It can be seen that for the central shower starting scale choice the \MGAMCNLO{} and \sherpa{} predictions reproduce the fixed-order result for sufficiently large $p_{T}^{\hhAlt}$, where the fixed-order result can be expected to be reliable to leading order accuracy. On the other hand, the \powhegbox{} result overshoots the fixed-order result by about a factor of 2 for large $p_{T}^{\hhAlt}$. The Dire shower prediction can also significantly overshoot the fixed-order result, but only if the largest shower starting scale is chosen.

\begin{figure}[t]
  \centering
  \raisebox{-0.5\height}{\includegraphics[width=.48\linewidth]{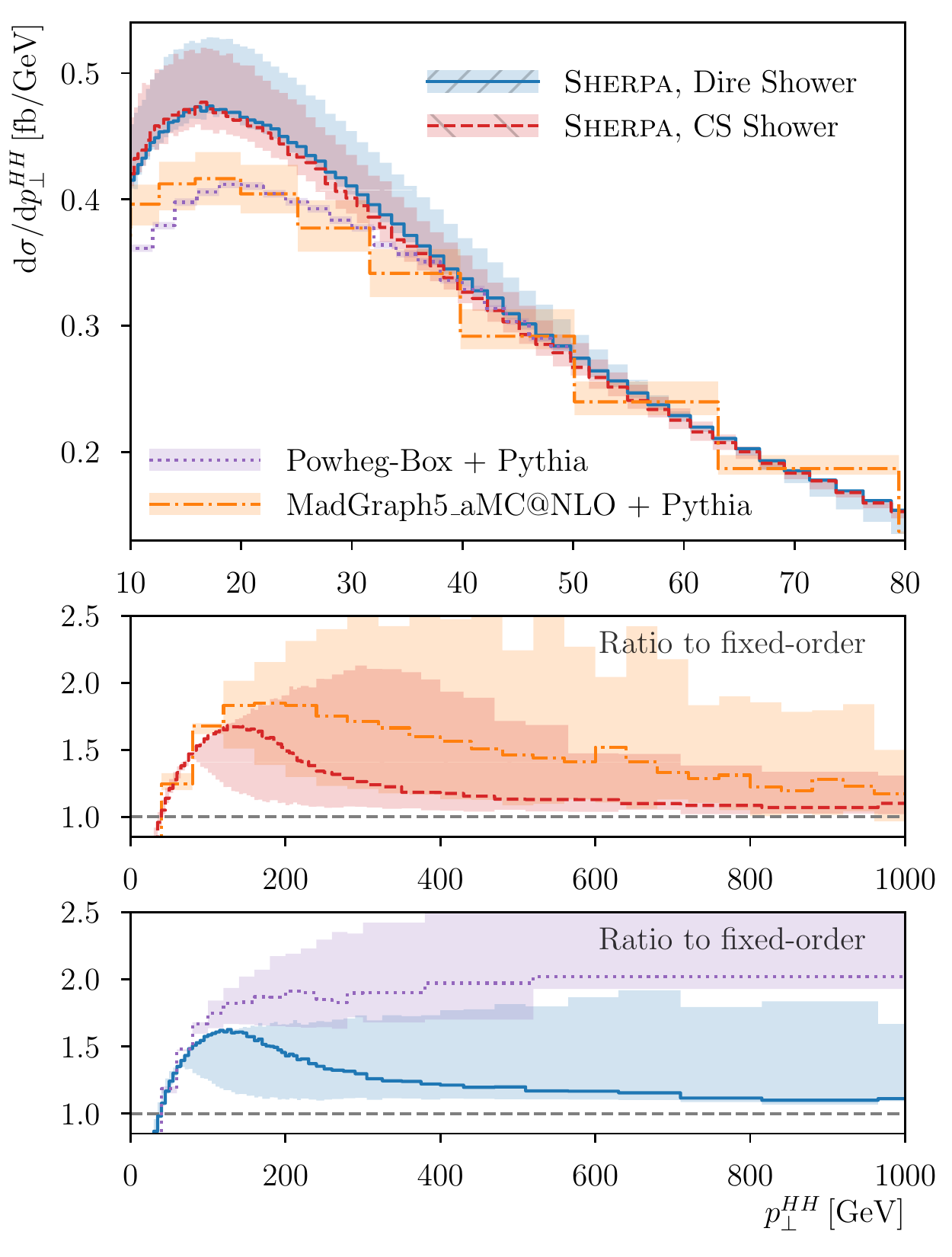}}
  \hfill
  \raisebox{-0.5\height}{\includegraphics[width=.48\linewidth]{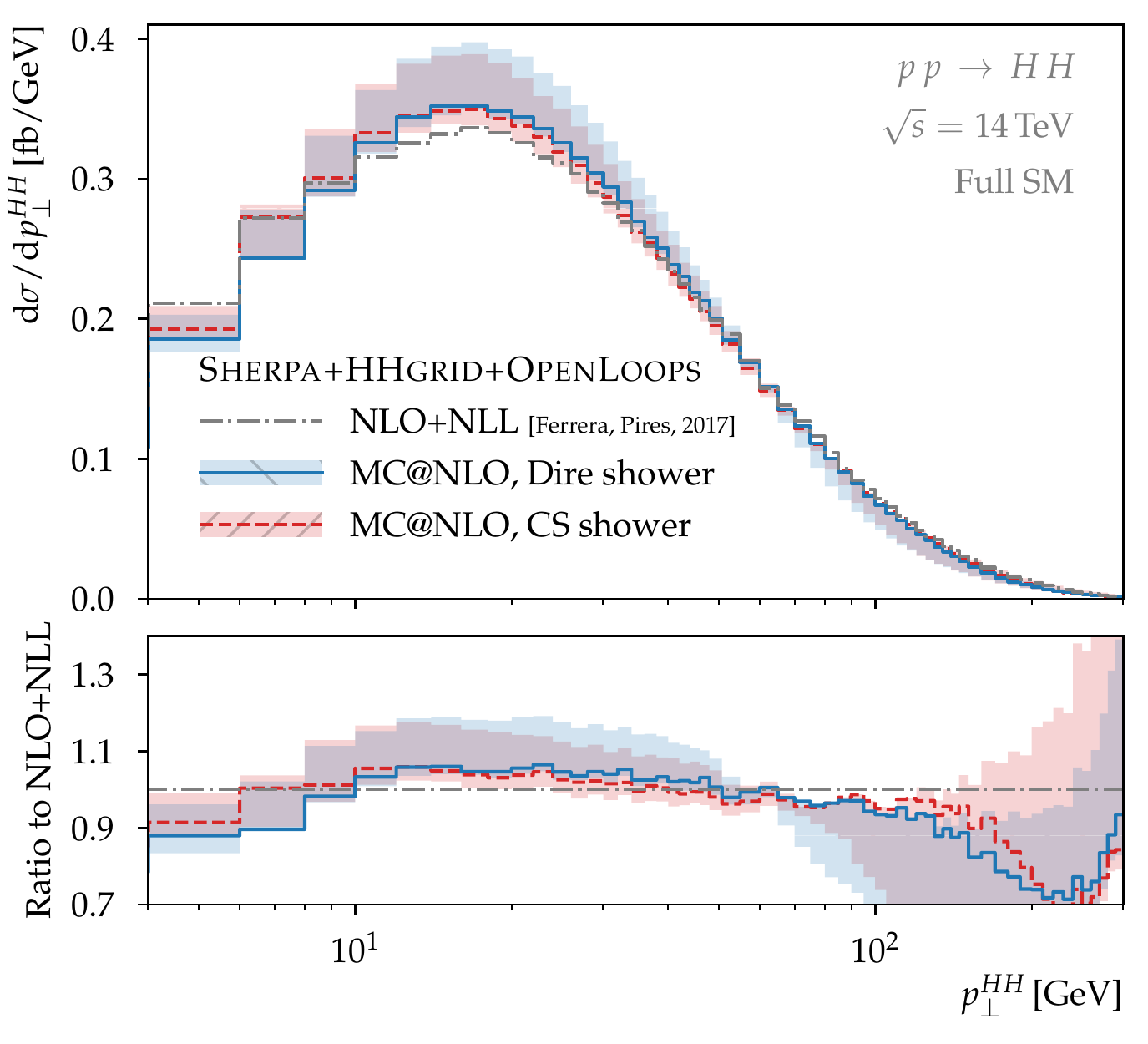}}
  \vspace*{-3mm}
  \caption{Left: Comparison of NLO parton shower matched predictions for the $p_{T}^{\hhAlt}$ spectrum. 
    The lower panels show ratios to the fixed-order
    prediction and cover a wider range of $p_{T}^{\hhAlt}$ than the
    upper panel. The uncertainty bands on the parton shower matched
    predictions were obtained by varying $\mu_\text{PS}$ or $h_{\mathrm{damp}}$ as described in the text.
    Right: Comparison with the NLO+NLL analytic resummation results of Ref.~\cite{Ferrera:2016prr}.
    The uncertainty bands on the \sherpa{} predictions were obtained by varying $\mu_\text{PS}$~\cite{Jones:2017giv}. }
  \label{fig:comp-ps}
\end{figure}

In \reffig{fig:comp-ps} we display a comparison
between the NLO parton shower matched results of \powheg, \MGAMCNLO{} and
\sherpa{} as well as the NLO+NLL result obtained using analytic 
resummation~\cite{Ferrera:2016prr}. 
The bands displayed for the \MGAMCNLO{} and \sherpa{} predictions are
produced by varying the shower starting scale $\mu_\text{PS}$ by a
factor of 2 around their central values. Lacking a natural equivalent
to $\mu_\mathrm{PS}$ in the \powheg{} framework, we display
\powhegbox{} predictions produced with various values of the $h_{\mathrm{damp}}$ parameter.
The nominal (central) \powhegbox{} prediction is produced with $h_{\mathrm{damp}} = 250$ GeV
and the band is produced by varying the $h_{\mathrm{damp}}$ parameter between $150$ GeV and infinity.
We note that although the parton shower predictions differ from each other
significantly less in the low $p_{T}^{\hhAlt}$ region, the shower starting scale 
uncertainty bands do not overlap.  For low $p_{T}^{\hhAlt}$, where the
analytic resummation can be trusted, it is found to be marginally
compatible with the \sherpa{} result and lies between the \sherpa{} result
and that of the other implementations.

Somewhat surprisingly, the matching
uncertainties are very large also at high $p_{T}^{\hhAlt}$. They were investigated in detail in
Refs.~\cite{Jones:2017giv,Bendavid:2018nar}. It was shown
that the large uncertainties are due to the formally sub-leading terms
generated by parton shower emissions off the $\bar B$ terms. Such
contributions are generally restricted to the phase space regions of
soft emissions where $t<\mu_\text{PS}$, as shown in Eq.~\eqref{eq:sevent_ps}. For hard emissions where $t>\mu_\text{PS}$, only
$H$ event contributions remain, which should reproduce the fixed-order
result. If $\mu_\text{PS}$ is sufficiently large, however, the parton
shower emissions in \refeq{eq:sevent_ps} start contributing to the
hard tail of the transverse momentum distribution. These parton shower
emissions do not capture the correct, rapidly falling, fixed-order
spectrum that one observes when the finite top quark mass is accounted
for. They therefore produce the overshoot compared to fixed-order that
we observe in the tails of \reffig{fig:comp-fo}. Varying
$\mu_\text{PS}$ effectively switches this overshoot on and off, thus
generating large uncertainty bands. It is worth noting, however, that
for more moderate choices of the shower starting scale the fixed-order
result is reproduced at large transverse momenta.

\begin{figure}
  \centering
  \includegraphics[width=.48\textwidth]{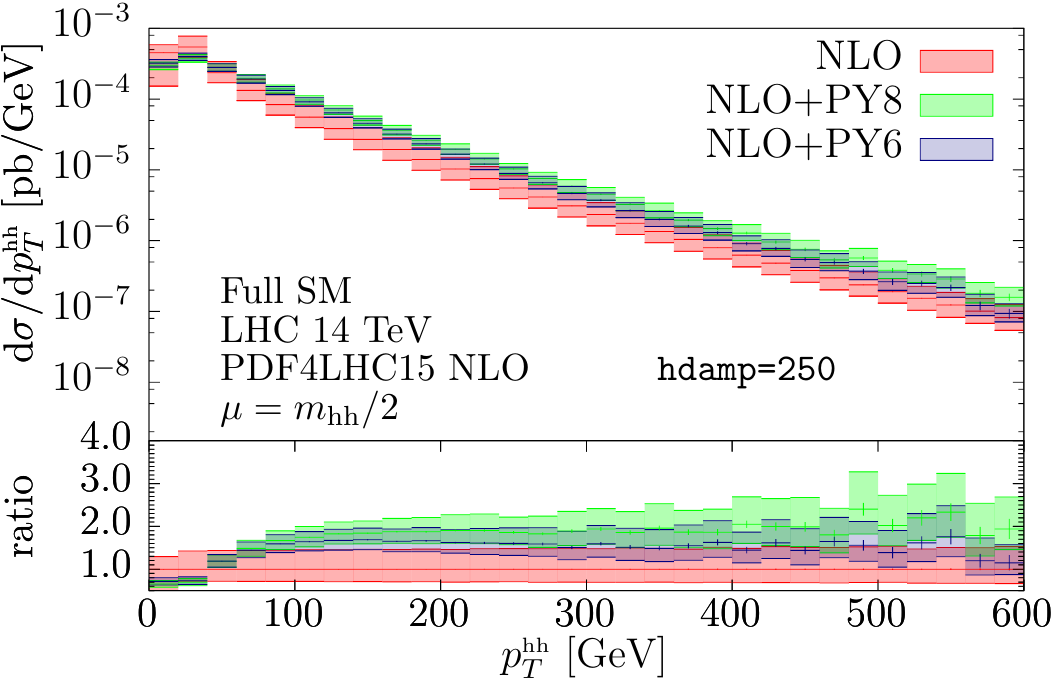}
  \hfill
  \includegraphics[width=.48\textwidth]{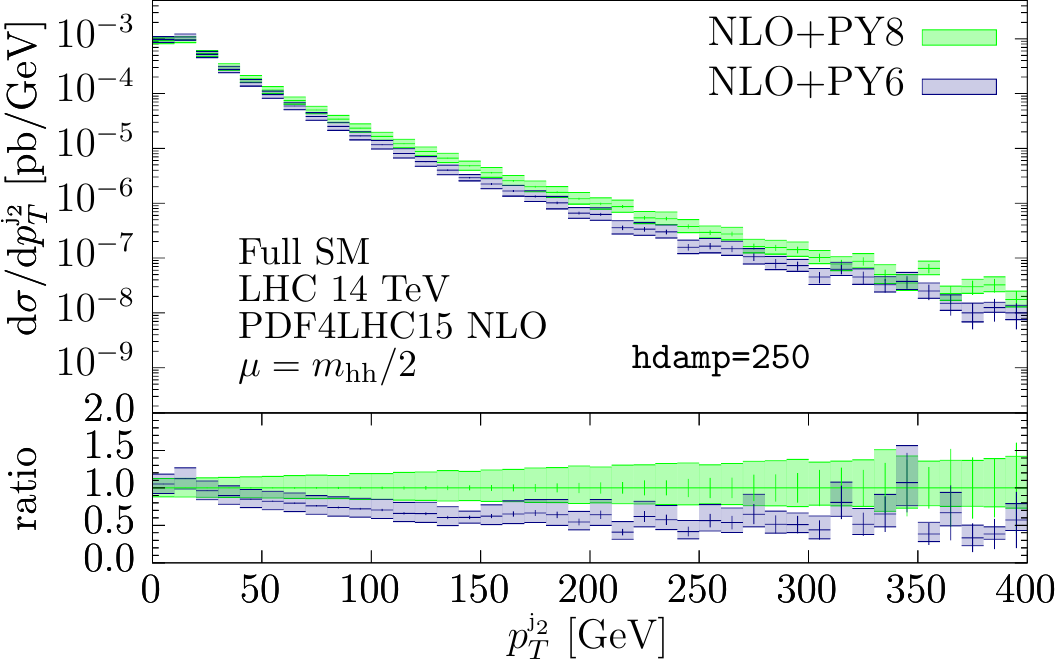}
  \caption{
  Left: The transverse momentum spectrum of the Higgs boson pair obtained using \powhegbox{} in combination with a \pythia\,8.2 and \pythia\,6 shower.
  The scale uncertainty bands represent the variation of $\mu_F$ and $\mu_R$.
  Right: Comparison of the sub-leading jet transverse momentum spectrum generated with \powhegbox{} using a \pythia\,8.2 and \pythia\,6 shower~\cite{Heinrich:2017kxx}.
  }
  \label{fig:comp-pythia}
\end{figure}

Within the \powheg{} matching scheme the $h_{\mathrm{damp}}$ parameter 
can be reduced in order to suppress the overestimated real 
emission at large transverse momenta. In Ref.~\cite{
Heinrich:2017kxx} it was shown that the choice 
$h_\mathrm{damp} = 250 $GeV is sufficient to reproduce the fixed-order 
result at large $p_{T}^{\hhAlt}$ at the Les Houches event Level, i.e. after the 
first hard emission is generated according to the \powheg{} method. 
Nevertheless, as can be seen in both \reffig{fig:comp-fo} and 
\reffig{fig:comp-ps}, the \powhegbox{} result when showered with 
\pythia\,8.2 was still found to be above the fixed-order result even at 
$p_{T}^{\hhAlt} \sim 600$ GeV. In Ref.~\cite{Bendavid:2018nar}, 
one explanation for this behaviour was found to be due to the emission 
of a relatively hard sub-leading jet, which at this order in perturbation 
theory is generated entirely by the parton shower. In \reffig{fig:comp-pythia} 
\powhegbox{} predictions are shown with a \pythia\,8.2 and \pythia\,6 
shower applied. The $p_{T}^{\hhAlt}$ spectrum is considerably softer when the 
\pythia\,6 shower is applied and tends towards the fixed-order prediction 
at large $p_{T}^{\hhAlt}$. In the right panel, the transverse momentum of the 
sub-leading jet, $p_T^{j_2}$, is shown, and we can observe that \pythia\,8.2 predicts a significantly harder jet 
than \pythia\,6. This behaviour is documented elsewhere in the literature~\cite{Corke:2010zj,Corke:2010yf},
and recent developments in \pythia\,8.2 are likely to soften the observed 
behaviour in the tail of the $p_{T}^{\hhAlt}$ distribution~\cite{Cabouat:2017rzi}.


In summary, we have reviewed in this section the studies on
the large matching scheme uncertainties present in Higgs boson 
pair production~\cite{Heinrich:2017kxx,
Jones:2017giv,Bendavid:2018nar}. The origin of the 
uncertainty was found to partly be due to sub-leading terms 
present in the matching procedure, which can lead to a large 
overshoot of the parton shower relative to the fixed order prediction.
There are three factors which play a role: the large K-factor
($\bar{B}-B$), large splitting kernel, and the shower starting scale.
In particular, in the MC@NLO matching scheme the shower starting 
scale must be chosen small enough to prevent the parton shower 
from populating the full phase space, where it will overestimate the
number of hard real emissions. Within the \powheg{} matching 
scheme, a sufficiently low  damping factor 
($h_\mathrm{damp} \lesssim 250$ GeV) or even a hard cut off 
on the hardness of shower emissions ($\mathrm{SCALUP}$) 
must be used to suppress this behaviour.

In the \powhegbox{} implementation it was found that the predictions
for the transverse momentum of the Higgs boson pair differed
significantly depending on whether a \pythia\,8.2 or \pythia\,6 parton 
shower is applied. This was found to be due to the fact that the \pythia\,8.2 
shower generates significantly harder sub-leading jets, which recoil 
against the di-Higgs boson system.

%% file: EFT/EFTmain.tex
\chapter{Effective Field Theory}\label{chap:EFT}
\textbf{Editors: F. Goertz, D. Pagani, G. Panico}
\vspace{2mm}

\section[Introduction to the EFT formalism]{Introduction to the EFT formalism \\
\contrib{G.~Buchalla, C.~Grojean, G.~Heinrich, F.~Maltoni, M.~E.~Peskin, E.~Vryonidou}}
\input{EFT/IntroductionEFT_HH.tex}
\label{subsec:EFTtheory}


\section[EFT vs. complete models: theoretical constraints on $\kl$]{EFT vs. complete models: theoretical constraints on $\kl$ \\
\contrib{L.~Di Luzio, R.~Gr\"ober, S.~Gupta, F.~Maltoni, D.~Pagani, H.~Rzehak, A.~Shivaji, M.~Spannowsky, J.~Wells, X.~Zhao}}
\label{sec:theoretical_constraints_EFT}

\input{EFT/theoreticalconstraints_kappal.tex}


\section{Impact of EFT fit}

\subsection[EFT fit for $HH$ production]{EFT fit for $HH$ production \\
\contrib{F.~Goertz, A.~Papaefstathiou, J.~Zurita}
}

\input{EFT/HH_GPZ.tex}
\label{sec:eft_fit_GPZ}

\subsection[Impact of single Higgs production]{Impact of single Higgs production \\
\contrib{S.~Di~Vita, C.~Grojean, U.~Haisch, F.~Maltoni, D.~Pagani, G.~Panico, M.~Riembau, A.~Shivaji, T.~Vantalon, X. Zhao}
}
\label{tril-single}
\input{EFT/single-WP.tex}



\section[EFT shape benchmarks]{EFT shape benchmarks \\
\contrib{A.~Carvalho, F.~Goertz}
}
\input{EFT/HH_shape_benchmarks.tex}

%% file: EFT/IntroductionEFT_HH.tex
The goal of the study of Higgs pair production, and, more generally, multiple Higgs boson production, is to understand the form of the potential energy function of the Higgs field, and, through this, to understand why the Higgs field acquires a vacuum expectation value, fills the universe, and gives mass to all elementary particles.    The simplest theory incorporating these phenomena is the SM. The SM in fact does not give any insight into these questions.   It is simply a phenomenological model in which all properties of the  Higgs field are input parameters and cannot be explained
within the model.   However, the SM is a tightly constrained structure.  In particular, now that the Higgs boson mass has been measured and the other couplings of the theory are fixed by measurements of particle masses and electroweak (EW) boson couplings, the SM gives precise predictions for the Higgs field potential and other observables.  
Experiments, then, can test whether the SM accurately describes the phenomenon of electroweak symmetry breaking (EWSB), or whether the SM must be replaced by a different, possibly more fundamental or predictive, underlying theory.

To test the SM through Higgs pair production, it is sufficient to work out the cross sections using the SM prediction for the potential and compare these results to experiment.   However, to gain insight into the possibility of alternative theories of
EWSB, it is necessary to understand how these cross sections vary when we go outside the context of the SM.   One way to do this is to compute the relevant pair production cross sections in specific alternative models.
However, it would be good to have a formalism that is not so specific but rather summarises the deviations that might appear in a very wide class of models beyond the SM.

This is the role of Effective Field Theory (EFT).    It is one of the profound ideas of quantum field theory that interactions of arbitrary complexity that act at short distances can be approximated systematically by a Lagrangian with an enumerable set of parameters.  This Lagrangian provides an ``effective'' description of any underlying model in this class.   The EFT  Lagrangian might not be renormalisable in the strictest sense, but it is nevertheless possible to carry out precise calculations that relate the parameters of this Lagrangian to observables~\cite{Weinberg:1978kz}.   For our
purposes, the EFT Lagrangian will be the SM Lagrangian with corrections described by  addition of  local operators.

The EFT formalism addresses the problem of calculating corrections to the
predictions of the SM in a systematic way.
For example, it might seem that the most straightforward way to describe
the effects of new physics on the triple-Higgs coupling is simply to add to the SM
Lagrangian a term
\beq
\Delta {\cal L} = - c h^3 \ .
\eeq{naivehthree}
  This is equivalent to changing the Feynman rules of the SM by multiplying the triple Higgs boson vertex by
$\kappa_\lambda$, with
\beq
\kappa_\lambda =  1 +  2\ c\, v/\mhAlt^2 \ ,
\eeq{kappashift}
where $\mhAlt$ is the Higgs boson mass and $v \simeq 246$~GeV is the Higgs field vacuum expectation value.   We have already seen  calculations in this context in Sec.~\ref{sec:xs_vs_lambda}.
However, the Lagrangian term in \refeq{naivehthree} is consistent with the $SU(2)\times U(1)$ gauge symmetry only if the field $h(x)$ is treated as a gauge singlet.   This requires modifications elsewhere in the Lagrangian.  Alternatively, we can keep $h(x)$ as a component of a complex scalar doublet, as in the SM.  In that case, to have a gauge-invariant Lagrangian, we should recast 
\refeq{naivehthree} as 
\beq
\Delta{\cal L} = - (c/v^2) |\Phi^\dagger \Phi|^3 \ , 
\eeq{betterhthree}
 where $\Phi$ is the SM Higgs doublet field. In both cases, the calculations done in Sec.~\ref{sec:xs_vs_lambda} remain valid to the order at which they were presented.  However, in both cases, the new terms added to  ${\cal L}$ contain additional multi-Higgs vertices. These terms give new contributions to higher-order EW corrections.  It turns out that these terms are needed to  cancel potentially troublesome ultraviolet divergences.  More generally, they allow  us to treat these models with $\kappa_\lambda \neq 1$ in a well-defined way to arbitrary precision. 

The second problem is that a modification of the Higgs self-coupling takes us outside of the SM.  In this context, we might wish to consider the most general set of perturbations due to possible new physics.  Those perturbations will affect the Higgs potential, but they will also modify other interactions that contribute to the Higgs pair production cross sections.  How can we have control over these effects?   The answer is that the possible gauge-invariant terms that we could add to the EFT Lagrangian can be classified according to a systematic expansion parameter, with only a finite number of new terms appearing at each order.  Then we can, order by order, describe the possible ways in which new physics can affect the Higgs pair production cross sections with a finite number of parameters, and constrain them by 
measuring Higgs pair production. But the EFT Lagrangian is \emph{the} Lagrangian, also describing single Higgs boson processes, reactions of the $W$ and $Z$ bosons that do not involve the Higgs boson, and precision EW observables.   This allows us to use data from these other processes to constrain the
new physics parameters and limit their influence on the Higgs pair production cross sections.

There is not a unique way to formulate an EFT description of new physics modifying the SM.   In fact, two different formalisms are used in the literature and, within these, many different approximations are used to simplify the Lagrangians for practical purposes.
In this chapter, we will describe these various approaches and their relation to specific underlying new physics models.

\subsection{Two EFT extensions of the SM}

In the literature, EFT descriptions of new physics beyond the SM are described within two different formalisms, called the HEFT (Higgs Effective Field Theory) and the SMEFT (Standard Model Effective Field Theory).  The HEFT is also referred to as the Electroweak Chiral Lagrangian (EWChL). In relation to the discussion above, the HEFT follows the path of treating the Higgs field $h(x)$ as an $SU(2)\times U(1)$ singlet, while the SMEFT treats $h(x)$ as a component of an $SU(2)\times U(1)$ doublet field $\Phi(x)$.  Both paths lead to self-consistent, gauge-invariant Lagrangians.  The HEFT is the older of the two formalisms.
The SMEFT has come to the fore more recently, specifically motivated by the discovery that the mass of the Higgs boson is not large but, rather, close to the $W$ and $Z$ boson masses. 
In the discussion to follow, we will explain these approaches and some simplifying assumptions used with them in practical calculations.

It is important to emphasise at the start that the HEFT and the SMEFT are different ways to enumerate the same set of operators that can be added to the SM Lagrangian.  In each case, operators are added systematically according to a given scheme of power-counting.  However, the schemes are different in the two cases, so that the same operator might appear at the leading order in one scheme but at a higher order in the other scheme.   In general, the SMEFT is more restrictive and therefore more predictive at a given order in its expansions. 

\subsection{SMEFT}\label{sec:smeft}

The key idea of the SMEFT is to view a model of new physics that extends the SM as being  built from the usual SM fields plus additional fields that act only at short distances or at high energy scales.   We will refer to the mass scale of the new interactions as $M$ in the following discussion.  The fact that the LHC experiments have not yet discovered particles associated with new physics strongly suggests that there is a hierarchy between the mass scale $m_Z$ at which 
SM interactions act and the scale $M$ characteristic of new particle interactions, $M \gg m_Z$.   In this picture,  the Higgs field lives at the scale $m_Z$ and  is described as a full complex doublet of scalar fields $\Phi$, as in the SM.  The SMEFT Lagrangian is taken to be invariant under $SU(2)\times U(1)$.  All of the
fields in the Lagrangian transform {\it linearly}
under gauge transformations. For example, the Higgs field with $I = \frac{1}{2}$, $Y =\frac{1}{2}$ transforms as
\beq
\Phi(x) \to   \exp\left[ - i \alpha^a(x) \frac{\sigma^a}{2} - i \beta(x){\frac{1}{2}} \right] \ \Phi(x)
\ ,  
\eeq{HiggsLTrans}
where $\alpha^a(x)$ and $\beta(x)$ are the $SU(2)$ and $U(1)$ gauge parameters, $\sigma^a$ are the Pauli sigma matrices, and $\half$ is the hypercharge $Y$ of the field $\Phi$. 
The gauge symmetry $SU(2)\times U(1)$ is spontaneously broken when the Higgs field acquires a vacuum expectation value
\beq
\langle \Phi(x) \rangle =  {\frac{1}{\sqrt{2}} }\begin{pmatrix} 0 \cr v\cr \end{pmatrix} \ . 
\eeq{Higgsvev}

The models considered today as the best candidates for a predictive theory of the Higgs potential follow this description.   For example, in the supersymmetric extension of the SM,  the Higgs doublet field is light, with a mass of the order of $m_Z$, while superpartner fields are heavy, with masses  $M = M_\text{SUSY} \gg m_Z$~\cite{Carena:2002es}.  An alternative class of models assumes that the Higgs
field is a multiplet of Goldstone bosons generated by
symmetry breaking at a high mass scale
$M$, with $M$ at  multi-TeV energies.  The potential for the Higgs field is generated at the  much lower scale $v \ll M$ by radiative corrections~\cite{Contino:2010rs,Panico:2015jxa}.

To describe physics at energy scales below the scale $M$, we may integrate out the fields interacting only at high energy.   Then we obtain a Lagrangian that contains only the SM fields, but possibly including operators of higher dimension built from these fields.  In a renormalisable Lagrangian, all terms are operators of dimension 4 or less.  If we add an operator of dimension $d>4$, then, by dimensional analysis, that operator must have  a coefficient proportional to  (mass)$^{4-d}$.  When such operators appear after integrating out heavy fields, their coefficients will be proportional to $M^{4-d}$.   The integration-out preserves the $SU(2)\times U(1)$ gauge symmetry.   Then this procedure will give a Lagrangian of the most general form that can be built from gauge-invariant operators constructed from the SM fields. 

However, it is a property of the SM that the SM Lagrangian is already the most general renormalisable $SU(2)\times U(1)$-invariant Lagrangian (with no strong-CP violation) that can be built from the SM fields. The EFT Lagrangian describing the most general types of new physics at the mass scale $M$ then
takes the form
\beq
  {\cal L} =  {\cal L}_{\rm SM} + \sum_i \frac{\hat c_i}{M^2} {\cal O}_i + \sum_j \frac{\hat d_j}{M^4} {\cal O}_j + \cdots \ ,  
\eeq{EFTLformone}
where the index $i$ runs over dimension-6 operators, the index $j$ runs over dimen\-sion-8 operators, and so on.  There do exist 
operators of dimension $5, 7, \dots$,  but these involve lepton number violation. For example, the dimension 5 operators are neutrino mass terms.   We may ignore them in discussing LHC processes.

Notice that, even though integrating out the new interactions can lead to order-1 modifications in the parameters
of ${\cal L}_{\rm SM}$, those changes are not observable, since in any event the
parameters of ${\cal L}_{\rm SM}$ are determined from experiment.   This means that the observable effects of new physics at the scale $M$ on cross sections at the much lower energy $E$ are at most of size  $E^2/M^2$.   We can use this ratio of energies as an expansion parameter to control the number of new operators that we take under consideration.
In particular,  if $E$ is of order $\mhAlt$, $M$ is of order 1~TeV, and we assume that the coefficients
$\hat c_i$, etc., are of order 1, then the effects of dimension-6 operators are at the level of a few percent while the effects of dimension-8 operators are at the level of $10^{-4}$.  Then it can make sense to drop the terms with operators of dimension-8 and higher and consider only the effects of the dimension-6 operators.   This gives a finite set of parameters describing the most general modification of the SM at short distances. 

If we are considering the experimental implications of one dimension-6 operator, it is very physical to
write the coefficient of this operator in terms of the parameter $M$, which then represents the scale of the new physics 
that gives rise to this operator.  However,  in analyses that involve a large number of 
dimension-6 operators (for example, 6 such operators appear in \refeq{lsmeft} below), it becomes awkward 
to define $M$ in a consistent way. For the rest of this report, then, we will rewrite
\refeq{EFTLformone} using the Higgs field vacuum expectation value as the dimensional parameter. 
This gives a definite, though arbitrary, choice for the dimensional parameter in the EFT coefficients.  Then the EFT Lagrangian will be expanded as
\beq
  {\cal L} =  {\cal L}_{\rm SM} + \sum_i \frac{\overline c_i}{v^2} {\cal O}_i + \sum_j \frac{\overline d_j}{v^4} {\cal O}_j + \cdots \, .  
\eeq{EFTLform}
The statement that $M \gg v$ appears here at the statement that the dimensionless coefficients $\overline c_i$ are much less than 1.   Using the simple  estimation scheme in the previous paragraph, in which we assume that the $M$ is of order 1~TeV and the $\hat c_i$, $\hat d_j$ are of order 1, we would estimate that the $\overline c_i$ are generally of the order of a few
percent, the $\overline d_j$ are generally of the order of $10^{-4}$, etc.   However, this argument is naive and there are important
cases in which the $\overline c_i$ and $\overline d_j$ can be larger. Some of these are relevant to the Higgs self-coupling, as we will see in the next section. 

\subsubsection{How large are the SMEFT parameters?}

Though the number of operators that appear in the SMEFT at dimension~6 is finite, it is very large.  Naive enumeration gives more than 80 operators.  However, linear combinations of operators that vanish by the SM equations of motion do not contribute to S-matrix elements, so we may drop  some operators that appear in these linear combinations in favour of others.  In the literature, there are different choices of which operators to retain and which to drop.   Two commonly used choices are the ``Warsaw basis''~\cite{Grzadkowski:2010es} and the ``SILH basis''~\cite{Giudice:2007fh,Contino:2013kra}.  Comparing these schemes, different operators appear in the descriptions, but the final physics conclusions must be identical.  Still,
even after eliminating as many operators as possible, we are left with an unwieldy number of parameters to work with. 
For one generation of fermions (or assuming the strongest form of flavor universality), there are 59 independent baryon-number-conserving dimension-6 operators that one can build out of SM
fields~\cite{Buchmuller:1985jz,Grzadkowski:2010es}. 

One of these parameters---called $\overline c_6$---multiplies the dimension-6 operator in \refeq{betterhthree} and thus directly induces an $h^3$ vertex that shifts the Higgs self-coupling.   However, other parameters can contribute in the calculation of Higgs pair-production cross sections.  There is another parameter---called $\overline c_H$---that leads to an overall rescaling of all Higgs boson couplings. Other parameters not obviously related to the Higgs self-coupling can also have 
an influence. At the LHC, the Higgs pair production process $gg\to HH$ receives contribution from  triangle and box top quark loop diagrams, with destructive interference.   A change in the value of the top quark Yukawa coupling by 10\%, which can be induced by another  dimension-6 operator,  then turns out to change the extracted value of the Higgs self-coupling by 50\%.   To control effects such as this, we must either argue that
the relevant coefficients $\overline c_i$ are small {\it a priori} or that their values are restricted
by other SM measurements.   In this case, for example, precision measurement of the top quark Yukawa coupling could restrict this source of uncertainty.   It is also possible to use measurements in different regions of phase space to distinguish the effects of different operators.  This strategy has been studied for $gg\to HH$
in Refs.~\cite{Goertz:2014qta,Azatov:2015oxa,Carvalho:2016rys} and for  $e^+e^-\to \nu\bar\nu HH$ in Ref.~\cite{Contino:2013gna}.   More generally, it is possible to combine data from
Higgs pair production with that from other processes affected by dimension-6 perturbations, including precision EW observables, to extract the shift of the Higgs self-coupling through a global fit~\cite{DiVita:2017eyz,DiVita:2017vrr}.

It might also be possible to give {\it a priori} arguments allowing us to ignore
some of the coefficients $\overline c_i$.  In the previous section, we have argued that the $\overline c_i$ might be expected to be only a
few percent in size.   There are some examples in which $\overline c_i$ are known to be smaller. The $S$ and $T$ parameters of precision EW analysis~\cite{Peskin:1990zt} are induced by dimension-6 operator perturbations, and the corresponding $\overline c_i$ coefficients are then bounded by precision EW measurements to be less than $10^{-3}$~\cite{Falkowski:2014tna}.   Constraints from the LHC on the Higgs couplings to $W$, $Z$, and heavy fermions are still at the 10--20\% level~\cite{ATLAS:2018doi,Sirunyan:2018koj},  but the estimate that the corresponding $\overline c_i$ are at the few-percent level neatly explains why no deviations from the SM have yet been observed.   In this report, we will discuss experiments that constrain the Higgs self-coupling at the level of tens of percent.   So perhaps we might even have the opposite problem, that, within the SMEFT, we predict that no deviations of the Higgs self-coupling from the SM will be observable. 

  Fortunately, there are models in which the deviations of the Higgs self-coupling can be of order 1 while the deviations in other parameters remain small.  A variety of such models are studied in Chapter~\ref{chap:BSM}.   We have the possibility for such large deviations when the Higgs field mixes with a SM singlet field that does not directly communicate with the $W$ and $Z$ bosons, or with a new fermion or boson sector that is relatively light compared with the 1~TeV mass scale.   It is typical in these
  models that the same effects that give order-1 shifts of the Higgs potential also give
  few-percent shifts of the $HWW$ and $HZZ$ couplings that can be observed in
  measurements on single Higgs processes.   These couplings, which are measured in single Higgs processes, are already constrained by LHC data, as noted above, and are expected to be measured with much higher accuracy.  So it is possible to bring these pieces of information together to test proposed models.

The largest effects occur in models in which a new boson provides an $s$-channel resonance that can decay to \hh. In such models, the di-Higgs mass spectrum in \hh production can have two distinct peaks, one at high mass corresponding to the resonance and one at  400-500~GeV containing the bulk of the \hh production.   The di-Higgs mass spectrum for a model in this class is shown in Fig.~\ref{fig:phenoshape}.  It should be noted that the EFT description applies only to the lower-energy part of this spectrum, while the resonance at high mass must be  described by a Lagrangian that contains the new particle explicitly. 

    A specific motivation for large modifications of the Higgs potential comes from the idea of EW baryogensis~\cite{Kuzmin:1985mm,Shaposhnikov:1986jp}.  The cosmic excess of baryons over anti-baryons must have been generated in the early universe during a time when the universe was out of equilibrium.   This could have been possible at the EW phase transition, but only if this phase transition was strongly first-order.   In the SM, for $\mhAlt = 125$~GeV, this is not the case.  Altering the Higgs potential to produce a strongly first-order phase transition requires a
   significant change, with a  $h^3$ coefficient about a factor 2 larger than that in the SM~\cite{Grojean:2004xa,Noble:2007kk,Morrissey:2012db,Katz:2014bha,Huang:2016cjm}.   Such a large effect could potentially be observed with high significance in the measurements we will describe.   More information on this point is given in Sec.~\ref{sec:cosmology}. 

Finally, it is also reasonable to take a completely agnostic point of view and ask what
is the maximal allowed value of the Higgs self-coupling.   One possible limit comes from perturbative unitarity, that is, the constraint that tree-level diagrams involving this coupling not violate unitarity bounds.  This gives~\cite{DiVita:2017eyz,Falkowski:2019tft}
\beq
|\kappa_\lambda|< \textrm{Min} \left(600 \xi, 4 \pi \right) \,,
\eeq{firstbound}
where $\xi$ is the typical size of the deviation of the Higgs couplings to other SM particles~\cite{Giudice:2007fh}.   From the LHC measurements quoted above, $\xi$ could be as large as 0.1-0.2. 
The stability of the Higgs potential places a stronger bound on $\kappa_\lambda$~\cite{Falkowski:2019tft},
\beq
| \kappa_\lambda | < 70\, \xi\, .
\eeq{betterbound}
This limit still gives considerable leeway in the search for modifications of the Higgs self-coupling.

Further discussion about the theoretical constraints that can be imposed on $\kl$ from vacuum stability, perturbativity, and by considering specific UV-complete models can be found in Sec.~\ref{sec:theoretical_constraints_EFT}.

\subsubsection{$gg\to HH$ in the SMEFT}

None of the  analyses described in this report confronts the full problem of controlling the dependence of Higgs pair production cross sections on 59 (or more) dimension-6 operators available in the SMEFT.   Most studies restrict themselves either to modification of the $h^3$ coupling only or modifications from a small set of  especially relevant operators.  As we discuss the current analyses, we will clarify for each of them precisely which set of operator contributions is being considered.

If our goal is to extract the Higgs self-coupling at the level of tens of percent, it can make good sense to consider only the subset of operators contributing at the leading order to the process under consideration.  In this section, we describe a
sensible reduction of the operator set for the process $gg\to HH$.

For this process, choosing the ``Warsaw basis'' of dimension-6 operators
defined in Ref.~\cite{Grzadkowski:2010es}, the most important contributions come
from the 6 operators
\beqa
\Delta{\cal L}_6 &=&
\frac{\overline c_H}{v^2}\partial_\mu(\Phi^\dagger\Phi)\partial^\mu(\Phi^\dagger\Phi)
+\frac{\overline c_u}{v^2} (\Phi^\dagger\Phi)\, \bar Q_L\widetilde\Phi\, t_R 
+\frac{\overline c_6}{v^2}\left(\Phi^\dagger\Phi\right)^3\CR
&+&\frac{\overline c_{tG}}{v^2}
\bar Q_L\sigma^{\mu\nu}G_{\mu\nu}\widetilde\Phi t_R 
+\frac{\overline c_{\Phi G}}{v^2}  (\Phi^\dagger\Phi)\, G^a_{\mu\nu}G^{a\,\mu\nu}
+ \frac{\overline c_{\Phi \widetilde G}}{v^2}  (\Phi^\dagger\Phi)\, G^a_{\mu\nu}\widetilde{G}^{a\,\mu\nu}\ .
\eeqa{lsmeft} 
In this  formula, $Q_L$ is the $(t,b)_L$ doublet,
$\widetilde \Phi \equiv i \sigma_2 \Phi$ denotes the charge
conjugate Higgs doublet, and
$\widetilde{G}^{a}_{\mu\nu} \equiv 1/2 \epsilon_{\mu\nu\rho\sigma} G^{a\,\rho\sigma}$.
If we assume CP conservation, the coefficients of the dimension-6 operators $\mathcal{O}_u$ and  $\mathcal{O}_{tG}$ will be  real and the CP-violating operator $\mathcal{O}_{\Phi \widetilde{G}}$ can be ignored.

The operators ${\cal O}_H$ and ${\cal O}_6$ modify the Higgs self interactions. The modifications as a function of the
$\overline c_H$ and $\overline c_6$ coefficients are given by
\beq
\lHcube/\lHcubeSM \equiv \klambda = 1 - \frac{3}{2} c_H + c_6\,,
\qquad\quad
\lHquar/\lHquarSM = 1 - \frac{25}{3} c_H + 6\, c_6\,,
\eeq{eq:klrel}
where $c_H \equiv 2 \overline c_H$ and $c_6 \equiv (2 v^2/\mhAlt^2) \overline c_6$.
The parameter $c_6$ only affects Higgs pair production, but the parameter
$c_H$ also  induces a universal rescaling of single Higgs production cross sections. 

\begin{figure*}[t]
\begin{center}
\includegraphics[width=11cm]{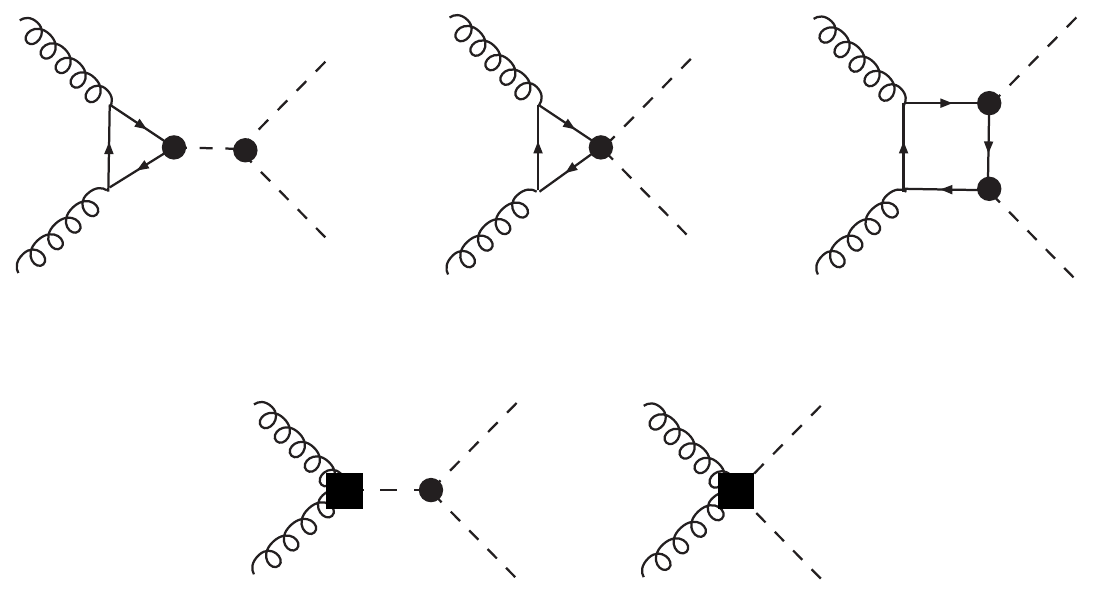}
\end{center}
\caption{Feynman diagrams contributing to $gg \rightarrow \hhAlt$, including SM EFT effects of
$D=6$ operators, whose potential insertions are indicated by black squares and blobs (for simplicity we are neglecting additional diagrams that come from the top dipole operator). See text for details.
These diagrams also correspond to the lowest non-vanishing order (one-loop order in this case) in the non-linear EFT.}
\label{fig:diags}
\end{figure*}

The operators ${\cal O}_u$  and  ${\cal O}_{tG}$  modify the Higgs coupling to the top quark.   The operator ${\cal O}_u$  shifts the top quark Yukawa coupling (relative to the SM relation $ m_t = y_tv/\sqrt{2}$).  The operator ${\cal O}_{tG}$ induces an anomalous colour magnetic dipole for the top quark and a contact interaction
including the Higgs, the gluon and the top quark. These two operators enter
the amplitude for $gg \rightarrow HH$ at the one-loop level. The remaining two
interactions ${\cal O}_{\Phi G}$ and ${\cal O}_{\Phi \widetilde G}$  give contact interactions involving
two Higgs bosons and two gluons. These operators contribute to Higgs pair production already at tree-level. The relevant diagrams for double Higgs production are shown in \reffig{fig:diags}.

Additional SMEFT operators modifying the coupling of the Higgs to the gauge bosons and $b$ quarks become relevant once the decays of the Higgs bosons are taken into account.  Similarly, additional operators will enter once QCD and EW corrections are considered.  The consideration of these operators is beyond the scope of this discussion.

In many SMEFT analyses of Higgs pair production~\cite{Goertz:2014qta,Azatov:2015oxa} a restricted set of dimension-6 operators is used, namely $O_H$, $O_u$, $O_{6}$ and $O_{\Phi G}$. This choice is motivated by theoretical considerations
on the possible origin and size of the effective operators. In a large class of UV theories (including renormalisable,
weakly coupled theories) the dipole operator $O_{tG}$ and the $O_{\Phi G}$ and $O_{\Phi \widetilde G}$ operators
are only induced at loop level, so that their coefficients are expected to be suppressed with respect to the other
dimension-6 operators that can instead be induced at tree level. When this happens, the contributions of the dipole
operator to double Higgs production can be formally considered as two-loop effects, and can be neglected with respect to
the other corrections that arise at one-loop order.
Notice that the $O_{\Phi G}$ operator, although suppressed by a loop factor, contributes at tree-level to double Higgs production.  Therefore it is expected to give corrections comparable to the
$O_H$, $O_u$ and $O_{6}$ operators. On the other hand, if no theory bias is assumed on the origin of the effective operators, a full fit including the dipole operator $O_{tG}$  should be performed.
The contribution of $O_{tG}$ in double Higgs production has been computed in Ref.~\cite{Maltoni:2016yxb}, where its impact on the differential distributions was also studied. 

\begin{figure}[t]
\centering
\includegraphics[width=.7\linewidth]{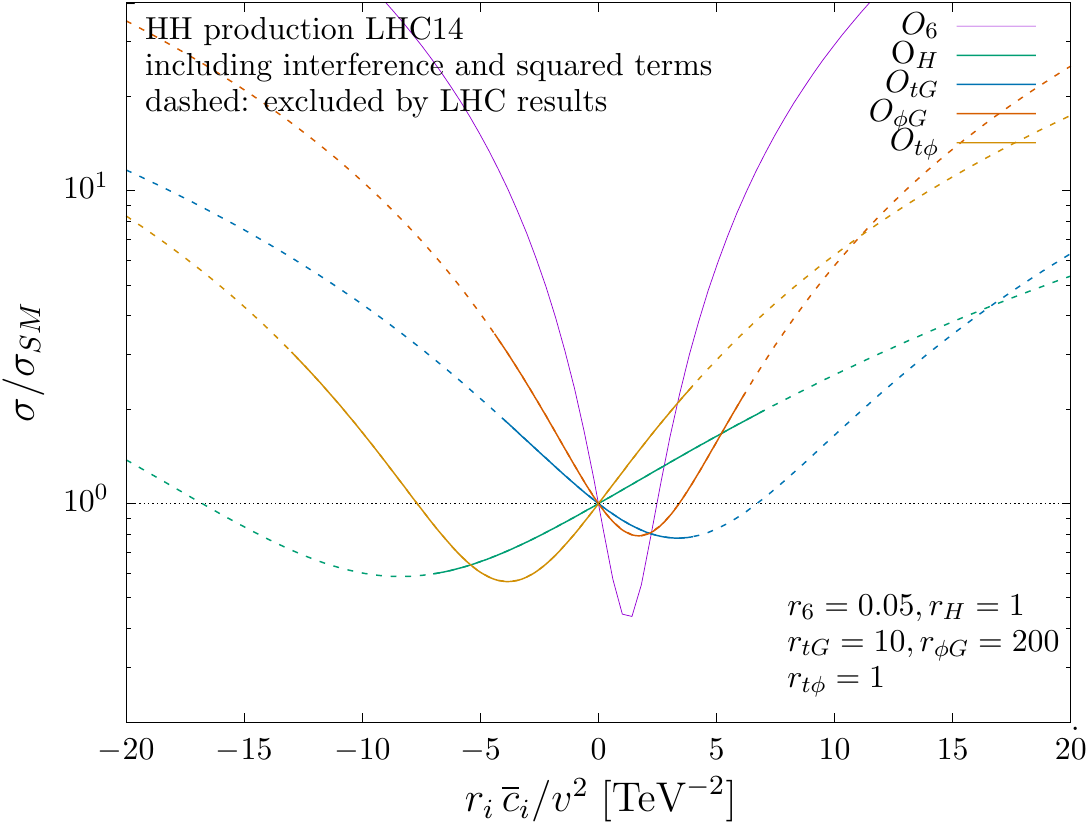}
\caption{Dependence of double Higgs production cross-section on the  Wilson coefficients of the relevant dimension-6 operators. The dashed part of the contours are excluded by LHC Run~1 Higgs and top quark measurements.
Note that each coefficient $\bar{c}_i$ is multiplied by a different factor $r_i$, specified in the figure.
\label{hhplot}}
\end{figure}

In the context of the SMEFT, extracting the triple Higgs coupling from the measurement of the Higgs pair production cross section is more difficult, since
all five operators listed above enter the process. The dependence of the total \hh cross-section on the EFT coefficients of the operators of Eq.~\leqn{lsmeft} is shown in Fig.~\ref{hhplot}.  We see that the total cross section depends rather strongly on all of these coefficients.

A compensating factor is that the coefficients of the operators 
$\mathcal{O}_H$, $\mathcal{O}_u$, $\mathcal{O}_{tG}$ and $\mathcal{O}_{\Phi G}$
can be  constrained by measurements of other processes at the LHC. In particular, top quark  measurements will constrain the dipole operator $\overline c_{tG}$, while
the top Yukawa operator $\overline c_u$ will be constrained by measurements of
$t\bar{t}H$ production and other single Higgs processes.
Similarly, $\overline c_{\Phi G}$ is constrained by measurements of the Higgs production
cross section from gluon fusion, and  $\overline c_H$ can be extracted as a uniform rescaling
of all Higgs couplings.    The current constraints obtained from Run~1 Higgs and
top quark measurements are shown in Fig.~\ref{hhplot} as the points where the
various lines become dashed.
Given these bounds,  only the effect of $\overline c_6$ can lead to deviations
of order 10 in the \hh cross section from the SM predictions. However, to
constrain $\overline c_6$ at levels of order 1, we will need precise
constraints on all of other coefficients that enter the analysis. This demands a global SMEFT interpretation.   We will discuss the impact of a such global fit in Sec.~\ref{sec:eft_fit_GPZ}.

Another aspect to be stressed is the fact that the various effective operators induce different distortions in the double
Higgs invariant mass distribution. A shape analysis can thus help in disentangling the various operators in a global fit.
We will discuss this point in Sec.~\ref{sec:shape_bench}.

\subsubsection{Single Higgs production in the SMEFT}

As we already discussed, another process that is sensitive to modifications of the Higgs trilinear self-coupling
is single Higgs production.
Accessing the self coupling in this way has been entertained in Refs.~\cite{McCullough:2013rea,Gorbahn:2016uoy,Degrassi:2016wml,Bizon:2016wgr,DiVita:2017eyz,Barklow:2017awn,Maltoni:2017ims,DiVita:2017vrr,Maltoni:2018ttu} and is discussed in detail in Sec.~\ref{tril-single}. First experimental results employing this method can be found in Refs.~\cite{CMS:2018rig, ATLAS-PUB-19-009}, and are discussed in Sec.~\ref{singleH_exp} (see also Sec. \ref{sec:singleH_fut}). 
Differently from double Higgs production, in which the Higgs trilinear interaction enters in
LO diagrams, in single Higgs processes such coupling contributes only through NLO corrections and its effects are thus
suppressed by a loop factor. 
In such a situation, model-independent bounds can only be obtained by performing a fit that simultaneously takes into account all the possible deformations of single Higgs interactions that contribute at LO. As discussed in the literature~\cite{DiVita:2017eyz,Gorbahn:2016uoy,Degrassi:2016wml,Bizon:2016wgr} and reviewed in Sec.~\ref{tril-single}, the sensitivity on $\kappa_\lambda$
obtained from single Higgs processes in an exclusive fit (i.e. allowing only $\kappa_\lambda$ to vary and setting all the
other couplings to their SM values) is comparable to the one from double Higgs production. However, once a global fit including deformations in single Higgs couplings is performed, the sensitivity is reduced, especially if no differential information is taken into account (see Fig.~\ref{fig:hllhcchi2}).

The question of identifying a minimal set of effective operators is more complex than what we discussed in the case
of $gg \rightarrow HH$. The difficulty mostly comes from the fact that even small contributions from operators
that enter at LO can easily overshadow the effects due to a modified Higgs trilinear coupling. Here we discuss the
minimal set proposed in Ref.~\cite{DiVita:2017eyz} for an analysis at the high-luminosity LHC. We however stress the fact that a
suitable set of operators can crucially depend not only on the actual collider but also on the sensitivity reached in precision EW
measurements (see for instance Ref.~\cite{DiVita:2017vrr} in the context of future lepton colliders).

Before quoting the operators that we will include, we first discuss our simplifying assumptions. In fact, we will not consider dipole operators (analogous to the one we mentioned in the basis for double Higgs production) and operators that correct the $W$ and $Z$ interactions with the SM fermions. Moreover, we omit four-fermion contact operators (where in particular the ones involving the top quark could be relevant in principle). In all these cases the experimental constraints are weak enough to allow for non-negligible corrections to single Higgs processes.
So, to remove these operators from a global fit, some theoretical assumptions might be needed, which we will rely on in the following.
For instance, as we discussed for $gg \rightarrow HH$, the assumption that dipole operators only arise at loop level makes their contributions
negligible. Moreover, under the assumption of flavor-universality for the new-physics contributions, the corrections
of the $W$ and $Z$ couplings to the SM fermions are constrained at the $10^{-2} - 10^{-3}$ level and can be safely neglected.

A minimal set of operators, following this reasoning, was proposed in Ref~\cite{DiVita:2017eyz} and includes $9$ effective operators in addition to the deformation
of the Higgs trilinear coupling. These operators can be expressed within the SMEFT framework in the ``Higgs basis''~\cite{Falkowski:2001958}
and correspond to
\begin{itemize}
\item 3 for the Yukawa interactions ($\delta y_t,\,\delta y_b,\,\delta y_\tau$),
\item 2 for the contact interactions involving gluons and photons ($c_{gg}\,,c_{\gamma\gamma}$), 
\item 1 for the rescalings of the $HZZ$ and $HWW$ interactions ($\delta c_z$), assuming custodial symmetry is unbroken,
\item 3 for the parameterisation of Higgs interactions with EW bosons featuring non-SM tensor structures
($c_{zz},c_{z\square},c_{z\gamma}$).~\footnote{Since two combinations of these coefficients can also be constrained by di-boson data, the interplay between the gauge and the Higgs sectors cannot be neglected.}
\end{itemize}

The resulting corrections to the Higgs interactions in the unitary gauge are given by
\begin{eqnarray}\label{eq:SMEFT_singleH}
\mathcal{L} &\supset & \, \frac{h}{v} \Bigg[ \delta c_w \frac{g^2 v^2}{2} W_{\mu}^+W^{-\mu} + \delta c_z \frac{(g^2 + g'^2) v^2}{4} Z_\mu Z^\mu \\
&&+c_{ww}\frac{g^2}{2}W_{\mu\nu}^+W^{-\mu\nu} + c_{w\square} g^2\left(W_{\mu}^-\partial_\nu W^{+\mu\nu} + \text{h.c.}\right) + \hat c_{\gamma\gamma}\frac{e^2}{4 \pi^2}A_{\mu\nu}A^{\mu\nu} \nonumber\\
&& +c_{zz}\frac{g^2 + g'^2}{4} Z_{\mu\nu}Z^{\mu\nu}+ \hat c_{z\gamma}\frac{e \sqrt{g^2 + g'^2}}{2 \pi^2}Z_{\mu\nu}A^{\mu\nu}
+c_{z\square}g^2Z_\mu\partial_\nu Z^{\mu\nu}
+c_{\gamma \square}g g' Z_\mu\partial_\nu A^{\mu\nu}
\Bigg]\nonumber\\
&&+ \frac{g_s^2}{48 \pi^2} \left(\hat c_{gg} \frac{h}{v} + \hat c_{gg}^{(2)} \frac{h^2}{2 v^2}\right) G_{\mu\nu} G^{\mu\nu}
-\sum_f \left[ m_f \left(\delta y_f \frac{h}{v} + \delta y_f^{(2)} \frac{h^2}{2v^2}\right) \bar{f}_{R}f_{L}+\text{h.c.}\right]\,,\nonumber
\end{eqnarray}
where the parameters $\delta c_w$, $c_{ww}$, $c_{w\square}$, $c_{\gamma\square}$, $\hat c_{gg}^{(2)}$
and $\delta y_f^{(2)}$ are dependent quantities defined as
\begin{eqnarray}
\delta c_w & =\; & \delta c_z\,, \nonumber\\
c_{ww} & =\; & c_{zz} + 2 \frac{g'^2}{\pi^2 (g^2 + g'^2)} \hat c_{z\gamma} + \frac{g'^4}{\pi^2  (g^2 + g'^2)^2} \hat c_{\gamma \gamma}\,,\nonumber\\
c_{w\square} & =\; & \frac{1}{g^2 - g'^2}\Big[g^2 c_{z\square} + g'^2 c_{zz} - e^2 \frac{g'^2}{\pi^2  (g^2 + g'^2)} \hat c_{\gamma\gamma}
-(g^2-g'^2) \frac{g'^2}{\pi^2 (g^2 + g'^2)} \hat c_{z\gamma} \Big]\,,\nonumber\\
c_{\gamma\square} & =\; & \frac{1}{g^2 - g'^2}\Big[2 g^2 c_{z\square} + \left(g^2 +g'^2\right)c_{zz}
- \frac{e^2}{\pi^2} \hat c_{\gamma\gamma}  - \frac{g^2-g'^2}{\pi^2}\hat c_{z\gamma} \Big]\,,\nonumber\\[1ex]
\hat c_{gg}^{(2)} & =\; & \hat c_{gg}\,,\nonumber\\[1ex]
\delta y_f^{(2)} & =\; & 3 \delta y_f - \delta c_z\,,
\end{eqnarray}
and the relations between the independent couplings and the 
operator coefficients in the Warsaw basis, as appearing in \refeq{lsmeft}, have been worked out in Ref.~\cite{Falkowski:2001958}.
Finally, in the above expressions $g$, $g'$, $g_s$ denote the $SU(2)_L$, $U(1)_Y$ and $SU(3)_c$ gauge couplings
respectively, and $e$ is the electric charge.

 \subsection{HEFT}
 \label{sec:heft}

 The HEFT gives a different way of organising possible operator modifications of the
 SM Lagrangian.   The idea of the HEFT is to describe the low-energy dynamics of
EWSB using a nonlinear realisation of $SU(2)\times U(1)$.
 The SM naturally includes an unbroken global $SU(2)$ symmetry, called ``custodial
 symmetry'', that protects against radiative
 corrections to the relation $m_W = m_Z \cos\theta_w$~\cite{Sikivie:1980hm}.   It is compelling to assume that this custodial symmetry is also present at least approximately  in more general models of EWSB.   Then the pattern of symmetry breaking is $SU(2)\times SU(2)$
 broken to $SU(2)$, the same as the pattern seen in chiral symmetry breaking in the
 QCD
 strong interactions. This suggests taking over the formalism of
 chiral perturbation theory used there to
 successfully describe low energy pion interactions~\cite{Gasser:1983yg,Gasser:1984gg}.

 In this approach, we take the symmetry-breaking field to be a unitary matrix
 of $SU(2)$,
 \beq
 U(x) = \exp[ i \pi^a(x) \sigma^a/v ] \ ,
 \eeq{Udefin}
 where $\pi^a$ are the Goldston boson fields of the SM and $v$ is the SM
 Higgs vacuum expectation value. Global   $SU(2)\times SU(2)$   transformations act on
 $U(x)$ by   $U \to  V_L\,  U(x)\, V_R^\dagger$.  This is a {\it nonlinear} action on the $\pi^a$ fields. The $SU(2)$ gauge symmetry is identified with the left $SU(2)$, and the $U(1)$ gauge symmetry is identified with the rotations about the $\hat 3$ axis in the right $SU(2)$,
 so that an $SU(2)\times U(1)$ gauge transformation is given by 
 \beq
 U(x) \to    \exp\left[ -i\alpha^a(x) \frac{\sigma^a}{2}\right] \ U(x)\  \exp\left[i \beta(x) \frac{\sigma^3}{2}\right]\ .
 \eeq{HiggsNLTrans}
 This transformation law should be contrasted with \refeq{HiggsLTrans}.
 The state of spontaneously broken gauge symmetry is described by
 \beq
 \left\langle U(x) \right\rangle =   {\bf 1} \ .
 \eeq{VEVofU}
 The expectation value leaves invariant the diagonal subgroup of the two original 
 $SU(2)$ symmetries, and this subgroup can be identified with the custodial symmetry.
 The subgroup within this $SU(2)$ of rotations about the $\hat 3$ axis is an
 unbroken gauge symmetry that can be identified with electromagnetism.
 
 A problem with \refeq{HiggsNLTrans} is that it has no place for the
 Higgs boson field $h(x)$.  In this formalism, $h(x)$ must be introduced as an
 $SU(2)\times U(1)$ singlet.   Couplings of the Higgs boson will be introduced into the
 HEFT Lagrangian as polynomials in the dimensionless ratio $h(x)/v$. 

 As in the case of the SMEFT, the HEFT Lagrangian is organised according to
 power-counting rules.  Following the guidance of chiral perturbation theory,
 the Lagrangian can be built up as terms with increasing {\it chiral dimension} $\chi$~\cite{Buchalla:2013rka,Buchalla:2015wfa}.  In this scheme, 
 boson fields are assigned $\chi = 0$, derivatives $\chi = 1$, and fermion bilinears
 $\chi = 1$.   The zeroth order Lagrangian has chiral dimension $\chi = 2$,
 \beq
 {\cal L } = \frac{1}{2} (D_\mu \pi^a)^2 + \half (\partial_\mu h)^2 +  \bar \psi_f (i \Dslash) \psi_f \ ,
 \eeq{zeroL}
 where $D_\mu$ is an appropriate covariant derivative.  It is useful  to think of the effective Lagrangian as being generated perturbatively in successive loop orders $L$,
 with $\chi = 2L+2$.  In this case, a weak coupling constant should also be assigned
 $\chi = 1$.  This counting assigns to  $h(x)/v$ the dimension $\chi=0$ and so
 arbitrary powers of this quantity can appear at each order.   To control this, the HEFT
 Lagrangian should be thought of as a double expansion in $L$ and $h(x)/v$.   A
 systematic expansion of the HEFT Lagrangian and evaluation of constraints on its
 parameters can be found in Ref.~\cite{deBlas:2018tjm}.

 The HEFT approach is well adapted to models of EWSB in which the symmetry breaking has two distinct sources, one from the Higgs field vacuum expectation value, one from strong interaction dynamics at a higher energy.
 Some models in this class are  the low-scale technicolor~\cite{Delgado:2010bb},
 Higgs-dilaton~\cite{Goldberger:2008zz}, composite Higgs~\cite{Agashe:2004rs,Contino:2010rs,Panico:2015jxa}, conformal Higgs~\cite{Haba:2010iv},
 and induced EWSB~\cite{Galloway:2013dma,Chang:2014ida} 
 models.   These models are not yet excluded, but they may be strongly challenged by future more precise measurements of Higgs couplings to fermions.

 In the HEFT, the leading terms in the Lagrangian contributing to a quark mass and fermion-Higgs interactions are
 \beq
 \Delta{\cal L} =  -  m_f \bar  Q_L  \, U \, q_R\  \biggl(1 +  c_f \frac{h}{v} + c_{ff} \frac{h^2}{v^2} \biggr) - \text{h.c.} \,.
 \eeq{HEFTqmass}
 In the SM, $c_f = 1 $ and $c_{ff} = 0$. 
 In the HEFT power-counting, the coefficient $c_f$ is not fixed and can deviate
 from 1 by any amount, and the coefficient $c_{ff}$ also is an independent parameter.
This contrasts strongly with the situation in the SMEFT, in which $c_f = 1 + a v^2/M^2$, where $a$ is a parameter of order 1, and $c_{ff}$ is predicted to be  $c_{ff} = \tfrac{3}{2} a v^2/M^2$ with the same value $a$ (ignoring the contribution from $c_H$), up to corrections of order $1/M^4$.
 At the LHC,
 the Higgs boson couplings are found to be equal to their SM values in terms of the
 particle masses, to an accuracy of 10-20\%.   This is natural in the SMEFT, but it is
 a strong constraint on the parameters of the HEFT.

 With this introduction, we can present the terms in the HEFT Lagrangian most
 relevant to the prediction of the cross section for $gg\to HH$.   These are
\begin{align}
\Delta{\cal L}_\chi=
&-m_t\left(c_t\frac{h}{v}+c_{tt}\frac{h^2}{v^2}\right)\,\bar{t}\,t -
c_{hhh} \frac{\mhAlt^2}{2v} h^3 \nonumber\\
&+\frac{\alpha_s}{8\pi} \left( c_{ggh} \frac{h}{v}+
c_{gghh}\frac{h^2}{v^2}  \right)\, G^a_{\mu \nu} G^{a,\mu \nu}\;.
\label{eq:ewchl}
\end{align}
The couplings $c_t$ and $c_{tt}$ are the
top quark couplings from \refeq{HEFTqmass}. 
To lowest order in the SM, $c_t=c_{hhh}=1$ and $c_{tt}=c_{ggh}=c_{gghh}=0$.
In the HEFT framework, the deviations of the various couplings from their SM
values are not expected to be small.   The relation found in the SMEFT between $c_{t}$ and $c_{tt}$
and the similar SMEFT  relation $c_{gghh}= \half c_{ggh}$ are not present here. However, there is one simplification:
the chromomagnetic operator  does not appear in  \refeq{eq:ewchl} because it 
contributes to $gg\to \hhAlt$ only at 2-loop order in the chiral power counting.

\begin{figure}
\begin{center}
\includegraphics[width=0.48\textwidth]{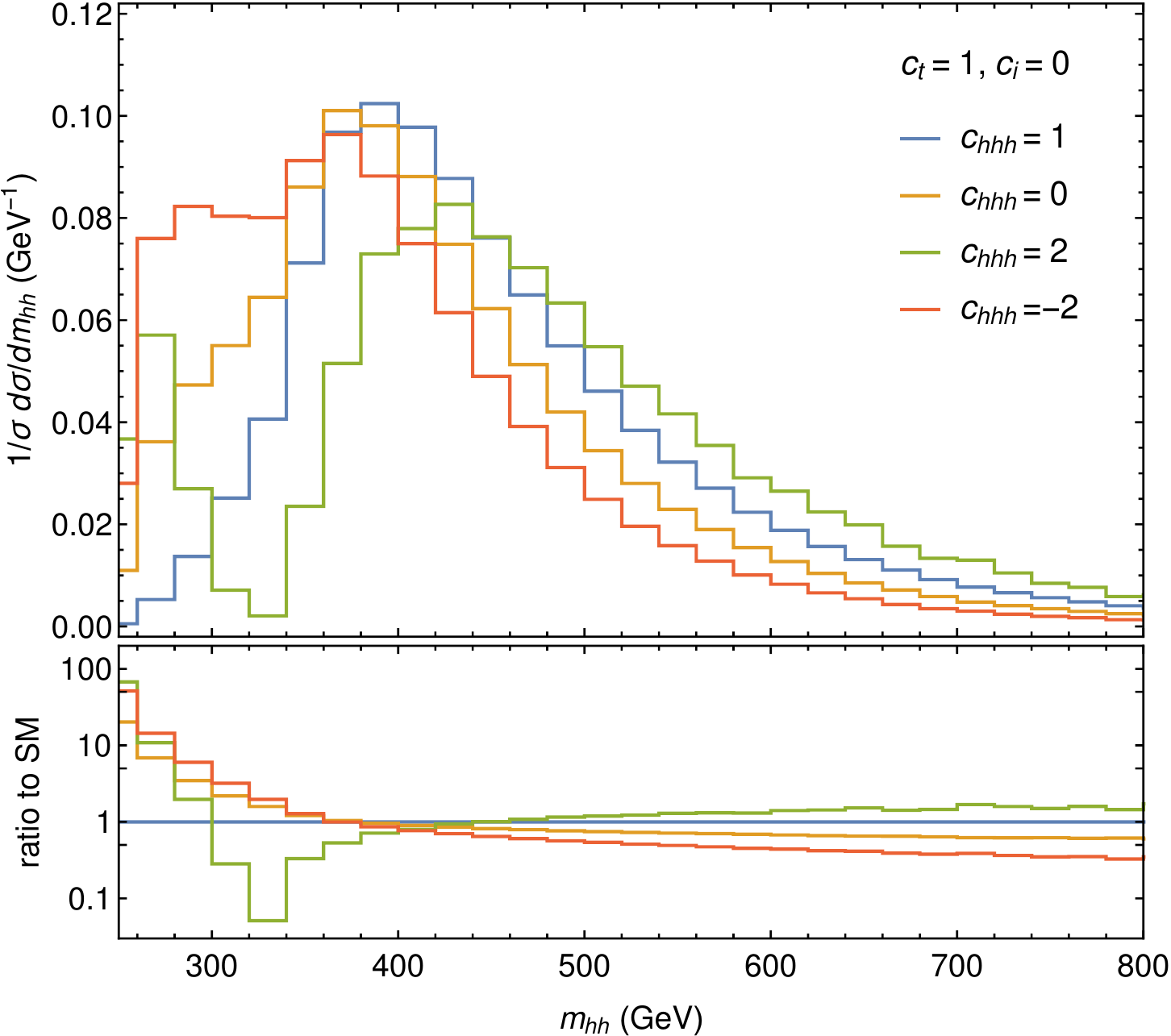}
\hspace*{0.1cm}
\includegraphics[width=0.48\textwidth]{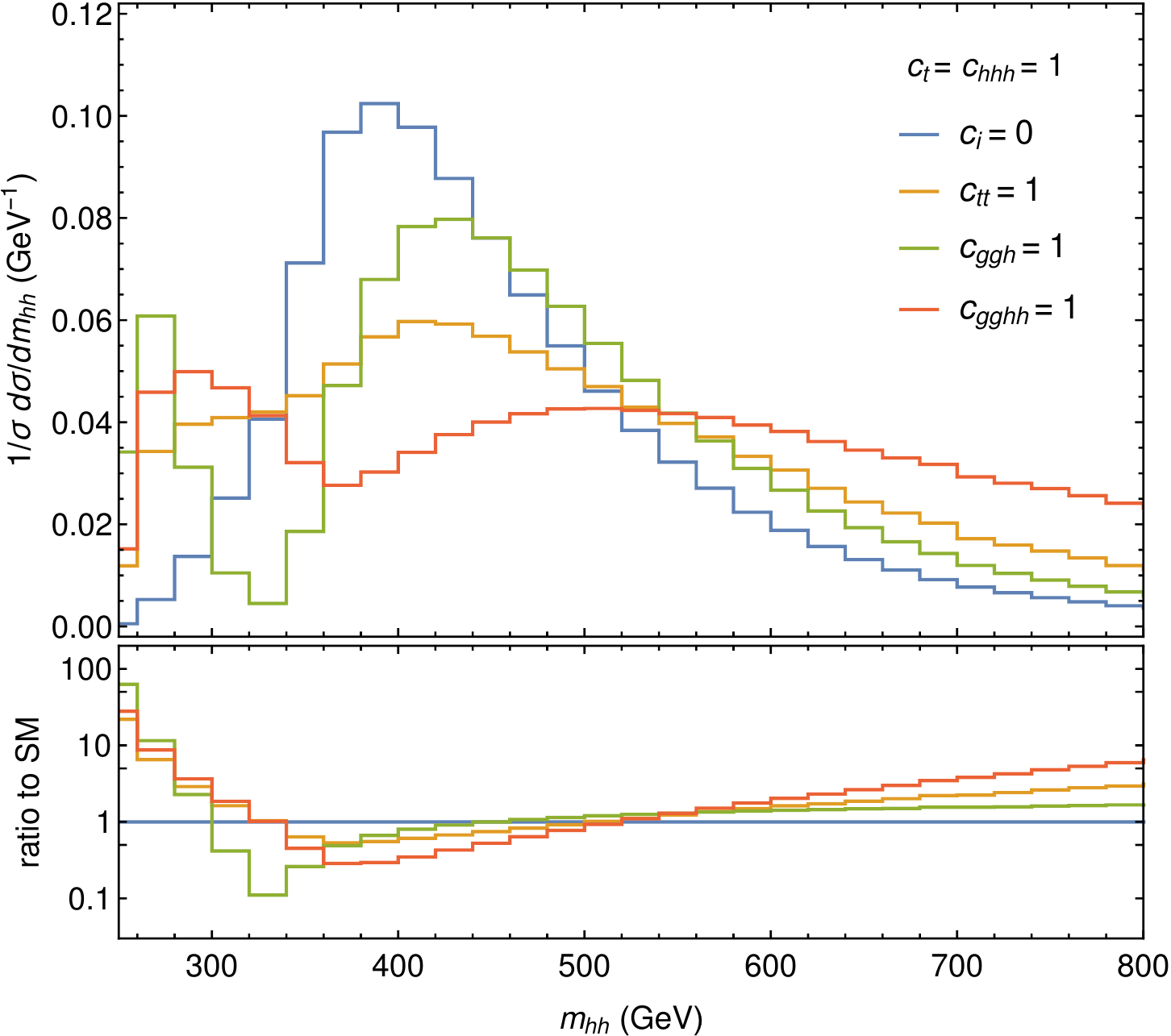}
\vspace*{-0.4cm}
\end{center}
\caption{Normalized Higgs pair invariant mass distribution at 14~TeV for different combinations of the HEFT couplings in \refeq{eq:ewchl}, and the ratio to the SM prediction. Here $c_i=0$ means $i\in \{tt,ggh,gghh\}$, i.e. the blue curve denotes the SM case. The coefficient $c_{hhh}$ is also known as $\kappa_{\lambda}$ in experimental papers. All curves are computed at NLO with full top-quark mass dependence \cite{Buchalla:2018yce}.}
  \label{fig:mhh_eft}
\end{figure}

The contributing diagrams to lowest order are shown in Fig.~\ref{fig:diags}.
All diagrams are at the same order in the chiral power counting
(chiral dimension~4, equivalent to one-loop order).
The diagrams illustrate the interplay between leading-order anomalous couplings
(black dots) within loops, and next-to-leading order terms
(black squares) at tree level. 
In \reffig{fig:mhh_eft} we show, as an illustrative example, the effect that the different operators in \refeq{eq:ewchl} can have on the di-Higgs invariant mass distribution, for several points in the HEFT parameter space. These distributions are computed at NLO in QCD including the full top quark mass dependence, based on the results presented in Ref.~\cite{Buchalla:2018yce}.

As in the case of the SMEFT analysis described earlier, the prediction for the
Higgs pair production cross section depends strongly on all of the parameters in
\refeq{eq:ewchl}.   Thus, to extract the coupling $c_{hhh}$ that determines the
shape of the Higgs potential, we need to constrain the other couplings in
\refeq{eq:ewchl} through a global analysis. This is more difficult in the HEFT formalism compared to the SMEFT because two of the relevant parameters---$c_{tt}$ and $c_{gghh}$ in \refeq{eq:ewchl}---appear only in processes with two Higgs bosons.  Thus it is not possible to use single Higgs data to fix these parameters.  A full analysis in the HEFT thus needs a strategy for constraining the auxiliary HEFT parameters.  For example, analysing $t\bar{t}HH$ production separately from other \hh production processes may allow a determination of $c_{tt}$ independently from $c_{hhh}$.

In conclusion, the HEFT Lagrangian yields a consistent parameterisation
of anomalous Higgs boson properties. Its power-counting encodes the assumption 
that anomalous Higgs couplings are the dominant effects
of new physics in the EW sector. In essence, the HEFT formalism gives
a field-theory basis for the empirical $\kappa$-framework of modified Higgs couplings.

%% file: EFT/theoreticalconstraints_kappal.tex

The current bounds on the trilinear Higgs self-coupling $\lHcube$ are much weaker than those for other Higgs couplings. At the moment, the strongest experimental constraint from double Higgs production has been obtained by the ATLAS collaboration combining  three different analyses \cite{Aaboud:2018knk, Aaboud:2018ftw, Aad:2019uzh, Aaboud:2018sfw}, setting a bound $-5.0 < \kl <12.1 $, where $\kl\equiv\lHcube/\lHcubeSM$. An indirect 
measurement $\kl = 4.0^{+4.3}_{-4.1}$
has also been extracted by the ATLAS collaboration from single Higgs production measurements \cite{ATLAS-PUB-19-009}, following the strategy described in Sec.~\ref{singleH_exp}. No experimental constraints on the quartic Higgs self coupling are available at all.
Given the current situation one can ask the following questions:
\begin{itemize}
\item is there any theoretical argument for constraining the Higgs self-couplings?
\item  how large can Higgs self-couplings be in UV-complete models?
\end{itemize}

In order to address the first question, we will consider both arguments based on
vacuum stability \cite{Cabibbo:1979ay,Buttazzo:2013uya,Degrassi:2012ry}  and perturbativity. Then, we will consider specific UV-complete models for answering the second question.

\subsubsection*{Vacuum stability}
If we consider the modifications induced to the SM potential by dimension-6 operators in \refeq{lsmeft}, in particular the $(\Phi^\dagger\Phi)^3$ operator,
one can distinguish 6 cases for the different sign possibilities for the parameters $\mu^2$ (where $v^2 = -\mu^2/\lambda$), $\lambda$
and $c_6$ \cite{Barger:2003rs}. Two different kinds of instabilities can arise, 
the most obvious one is at large field 
values for $\mu^2>0$, $\lambda>0$, $c_6<0$. 
The other one has to do with the destabilization of the EW minimum against the minimum at 
zero field value, potentially occurring for $\mu^2<0$, $\lambda<0$, $c_6>0$.
In Ref.~\cite{DiLuzio:2017tfn} it has been shown 
that both instabilities cannot be reliably assessed within the EFT, 
so that one cannot infer a model-independent bound on the trilinear Higgs self-coupling 
from stability arguments.  
The instability at large field configurations 
cannot be trusted due to the breakdown of the EFT expansion in the region close to the instability \cite{Burgess:2001tj}, while the occurrence of the low-scale instability requires a rather small value of the cutoff scale, 
making the use of the EFT language questionable.\footnote{However, in a recent study \cite{Falkowski:2019tft} it is argued that vacuum stability argument can still be relevant under some reasonable assumptions about the underlying EFT.}
\par

Also in UV-complete models with modifications of the scalar potential at tree level 
the vacuum instability and the modifications of the trilinear Higgs self-coupling 
are not directly connected, due to the presence of many couplings in the scalar potential 
that decorrelate the two effects. 
Instead, an almost one-to-one correspondence between the two 
is achieved in models where the Higgs self-couplings modifications 
are due to new fermions running in the loop,   
as in the case of right-handed neutrinos.
While in low-scale inverse seesaw models
one can find modifications in the trilinear Higgs self-coupling 
up to 30\% \cite{Baglio:2016bop}, 
the scenarios providing such a large deviation of the trilinear Higgs self-coupling 
drive the Higgs potential into the unstable regime. 
Requesting that this does not occur within one order of magnitude from 
the mass scale of the right-handed neutrinos
(hence not requiring any UV completion below that scale), 
one can bound 
the trilinear Higgs self-coupling modifications 
to be smaller than 
$\kl=0.1\%$ \cite{DiLuzio:2017tfn} via metastability arguments (see e.g. Ref.~\cite{DiLuzio:2015iua}).

\subsubsection*{Perturbativity}
On general grounds, one expects that too large values of the Higgs self-couplings 
will eventually enter the non-perturbative regime. A violation of perturbativity implies that new phenomena such as strong interactions may appear or new massive particles have to be present in the UV-finite model in order to restore perturbativity. On the other hand, non-perturbativity also indicates that LO predictions as well as higher-order corrections cannot be trusted. In view of the following discussion these two complementary aspects have to be kept in mind.
 
\begin{figure}
\begin{center}
\includegraphics[width=0.43\textwidth]{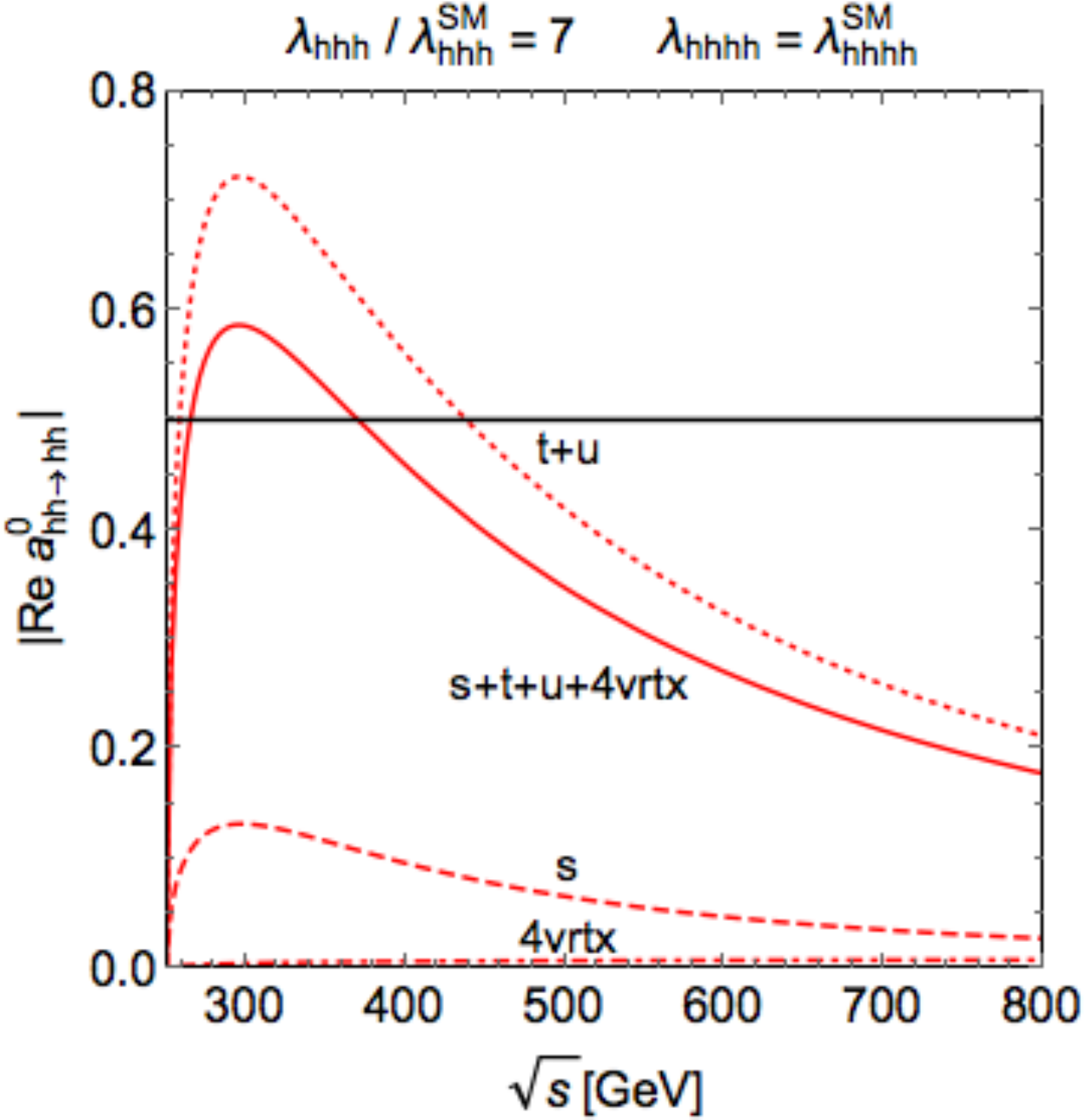}
\hspace*{1cm}
\includegraphics[width=0.43\textwidth]{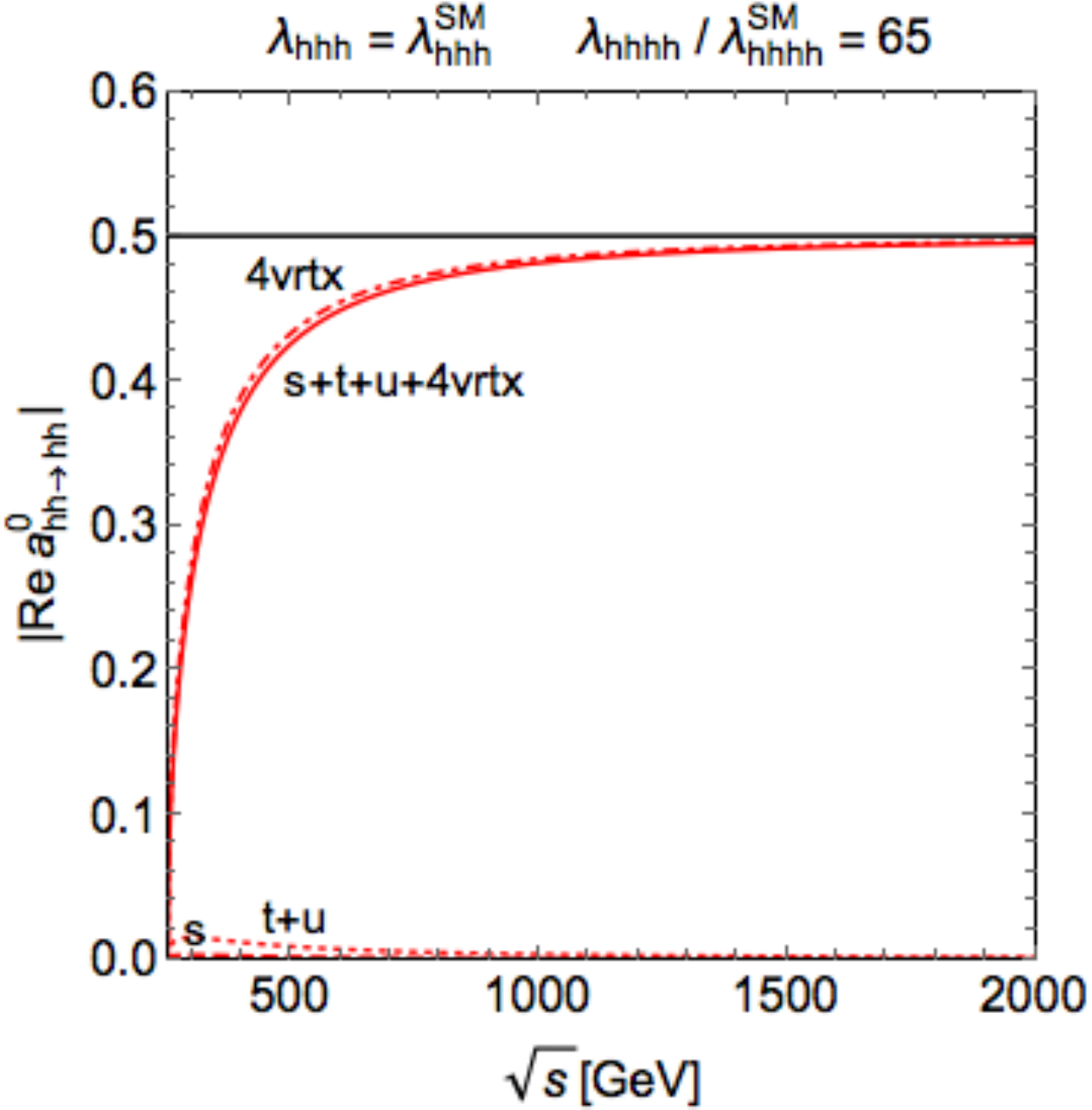}
\caption{Dependence of the $J=0$ partial wave $a^0$ on the centre of mass energy $\sqrt{s}$ for modified trilinear Higgs self-coupling (left) and modified quartic Higgs self-coupling (right). The plots are taken from Ref.~\cite{DiLuzio:2017tfn}. \label{EFT:partialwave}}
\end{center}
\end{figure}

A possible tool to estimate the perturbativity range is based on 
partial wave unitarity. By looking at the $\hhAlt \to \hhAlt$ scattering amplitude 
in the SM broken phase and 
requiring that for the $J=0$ partial wave
$|\text{Re}\,a^{0}|<\frac{1}{2}$,
one finds $|\kl|\lesssim 6.5$ and $|\lambda_{H^4}/\lambda_{H^4}^{\rm SM}| \lesssim 65$. 
Note that the $\kl$ bound is extracted at small $\sqrt{s}$, being the trilinear coupling associated to 
a super-renormalisable operator (cf.~Fig.~\ref{EFT:partialwave} -- left panel), 
while the contribution of the quartic coupling to the partial wave becomes important 
only at large $\sqrt{s}$ (cf.~Fig.~\ref{EFT:partialwave} -- right panel). 
Thanks to this very distinctive kinematic feature, one can
separately set a model-independent bound on the trilinear and quadri-linear 
Higgs self-couplings.
\medskip

An alternative perturbativity criterium is obtained by requiring that 
the loop-corrected trilinear vertex should be smaller than the tree-level vertex, 
or in the case of the quartic that the beta-function satisfies
$|\beta_{\lambda_{H^4}}/ \lambda_{H^4}|<1$ \cite{Goertz:2015nkp, DiLuzio:2016sur}. 
Such criteria lead to bounds very similar to the above mentioned ones, 
leaving us with 
\begin{align}
|\kl&|\lesssim 6\,,\label{EFT:eq:thboundlamhhh}\\ 
|\lambda_{H^4}/\lambda_{H^4}^{\rm SM}| &\lesssim 65  \label{EFT:eq:thboundlamhhhh} \,.
\end{align}

Following the same alternative perturbativity criterion, in Ref.~\cite{Maltoni:2018ttu} a bound equivalent to \refeq{EFT:eq:thboundlamhhh} has been set for  $  c_6\equiv \kl-1$. Parameterising any possible deviation to the SM potential as\footnote{Note that, at variance with \refeq{lsmeft}, the dimension-6 operator is $(\Phi^\dag \Phi -\frac{1}{2}v^2 )^3$, so the coefficients $c_6$ and $\bar{c}_6$ are not simply related by a different normalisation. For more details see Ref.~\cite{Maltoni:2018ttu}. } 
\begin{equation}
V^{\rm NP}(\Phi)\equiv  \sum_{n=3}^{\infty}\frac{ c_{2n} m_H^2}{2 v^{2n-2}}  \left(\Phi^\dag \Phi -\frac{1}{2}v^2 \right)^n \, ,
\label{VNP}
\end{equation}
 one finds that one-loop corrections to the $HHH$ vertex are smaller than its tree-level value only for 
\begin{align}
|\kl-1 &|\lesssim 5\,.\label{EFT:eq:cbsbound} 
\end{align}
It is important to note that the bounds in \refeqs{EFT:eq:thboundlamhhh} and \eqref{EFT:eq:cbsbound} originate from the requirement that the $HHH$ vertex, setting two Higgs bosons on-shell, is perturbative for the full spectrum $\sqrt{\hat s} >2\mhAlt$. The strongest bound arises form the configuration $\mhhAlt\simeq 2\mhAlt$. In general, in other kinematic configurations, the bound is looser. For instance, the $HHH$ vertex enters via one-loop EW corrections the predictions for single-Higgs production and decays modes, however, never with two Higgs bosons on-shell. Therefore, the bound in \refeqs{EFT:eq:thboundlamhhh} and \eqref{EFT:eq:cbsbound}, which indicates where it is sensible to perform a perturbative calculation, does not directly apply to the studies presented in Refs.~\cite{McCullough:2013rea,Gorbahn:2016uoy,Degrassi:2016wml,Bizon:2016wgr,DiVita:2017eyz,Barklow:2017awn,Maltoni:2017ims,DiVita:2017vrr,Maltoni:2018ttu}  and discussed also in Sec.~\ref{tril-single}, where precise predictions for single Higgs production have been proposed as alternative method for the extraction of $\kl$. Trilinear coupling values corresponding to $|\kl|\sim 10$ still lead to reliable perturbative calculations {\it in single Higgs production}, though will lead also to large higher order corrections.\footnote{More details can be found in Refs.~\cite{Degrassi:2016wml,DiLuzio:2016sur}.}

On the other hand, the kinematic configuration corresponding to the most stringent perturbative bounds for $\kl$ corresponds to the threshold region in double Higgs production. This means that if $|\kl|\gtrsim 6$, perturbative predictions for total cross sections in double Higgs production are meaningless. 
 \begin{figure}
 \begin{center}
 \includegraphics[width=9cm]{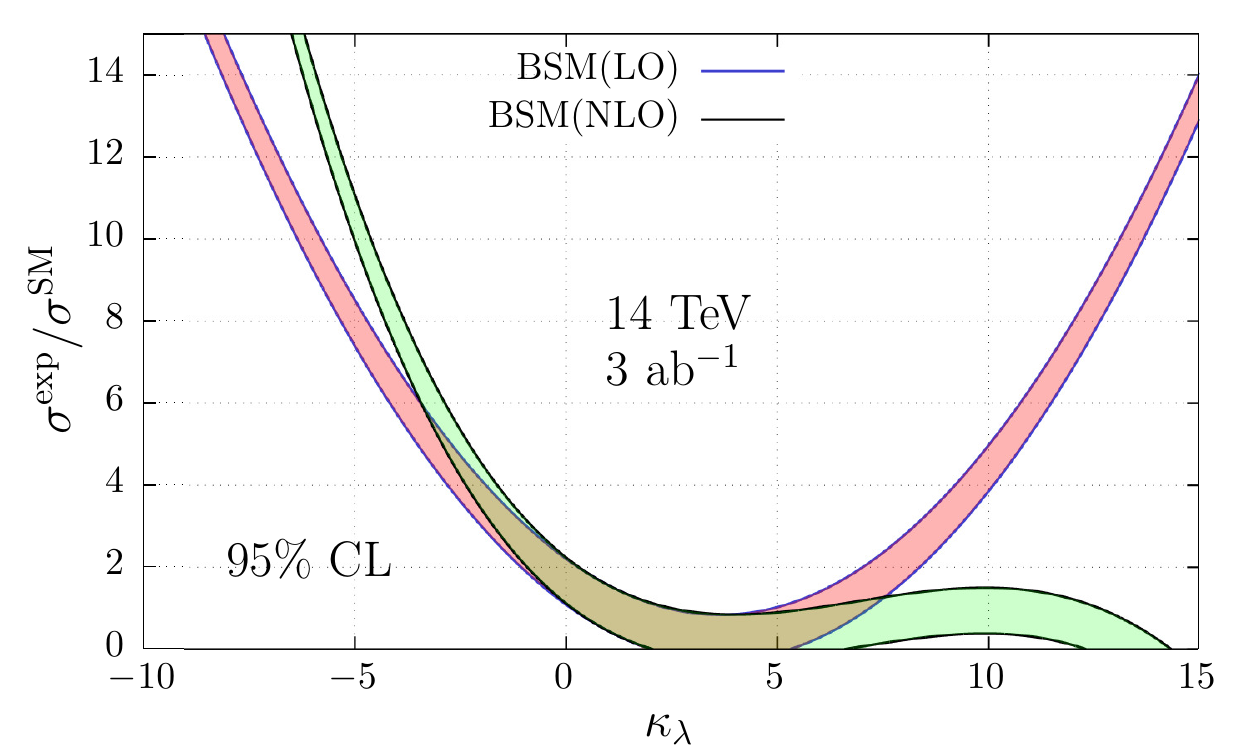}
 \caption{Bounds on $\kl$ that can be set according to the supposedly double Higgs measured cross section, normalised to the corresponding SM prediction. The red band is obtained considering the LO prediction, while the green taking into account one-loop corrections induced by $\kl$. This plot has been taken from Ref.~\cite{Borowka:2018pxx} and adapted for this report. \label{EFT:LOvsNLO}}
 \end{center}
 \end{figure}

In Fig.~\ref{EFT:LOvsNLO} we show a plot taken from Ref.~\cite{Borowka:2018pxx} where the 2$\sigma$ constraints that can be obtained at HL-LHC on ${c}_6$ as function of $\sigma^{\rm exp}/\sigma^{\rm SM}$ are presented. The quantity $\sigma^{\rm exp}$ is the supposedly measured value for the double Higgs cross section, while $\sigma^{\rm SM}$ is the corresponding SM prediction. The constraints are derived using two different approximations: taking into account $\kl=1+{c}_6$ effects only at LO or including also loop-corrections induced by ${c}_6$ itself,\footnote{For a precise definitions of the NLO predictions see Ref.~\cite{Borowka:2018pxx}. Including NLO contributions, it is more convenient to organise the calculation according to $ c_6$ than $\kl$.} {\it i.e.}, at NLO. For $|{c}_6|\gtrsim5$, where perturbativity is violated, NLO and LO  constraints are not compatible. The bottom line is:  when data are fitted via $\sigma_{\rm LO}$ predictions,  $\kl$ or equivalently ${c}_6$ is a parameter of ignorance; only for $|\kl-1|=|{c}_6|\lesssim5$ this parameter coincides with the quantity one is interested in. Moreover, NLO or any higher-order corrections would not improve this situation. Therefore, one can set bounds outside the range   
$|\kl-1|=|{c}_6|\lesssim5$, but only within this region they refer to the parameter in the Lagrangian.

Starting from the parameterisation of BSM effects in \refeq{VNP}, one can derive from the one loop corrections to the $HHH$ vertex perturbative bounds on the coefficient ${c}_8$, which is connected to the quantity $\lambda_{H^4}/\lambda_{H^4}^{\rm SM}$ via the relation  $\lambda_{H^4}/\lambda_{H^4}^{\rm SM}=1+6{c}_6+{c}_8$. Following this strategy, in Ref.~\cite{Borowka:2018pxx}, a bound 
\begin{equation}
|{c}_8|\lesssim 31 \,, \label{EFT:c8bound}
\end{equation}
has been found and, taking into account the bound $|{c}_6|\lesssim 5$, it translates into
\begin{equation}
|\lambda_{H^4}/\lambda_{H^4}^{\rm SM}| \lesssim 61  \label{EFT:l4bound} \,,
\end{equation}
which, although being $\kl$-dependent ({\it e.g.} $|\lambda_{H^4}/\lambda_{H^4}^{\rm SM}| \lesssim 31$ for $\kl\simeq 1$), is in good agreement with \refeq{EFT:eq:thboundlamhhhh}, obtained via a different approach.

Summarising, we find that current limits on the trilinear Higgs self-couplings do not reach the interesting range yet; they are in fact still above the perturbative regime\footnote{The ATLAS and CMS combinations of $HH$ results based on 2016 Run 2 dataset \cite{Aad:2019uzh,Sirunyan:2018two} are discussed in details in Sec. \ref{sec_exp_combination} and a first attempt of combining them is reported in Fig.\ref{fig:comb:ATLAS_CMS}. The strongest constrains results to be $-6.8 < \kl < 14$. Therefore experimental data are currently less constraining than the perturbative conditions. But we expect that the full LHC Run 2 data (a factor three larger than 2016 Run 2 dataset) would reach the perturbative constraints for low $\kl$ values.}. 
Before concluding this subsection, however, we want to mention the results of recent studies \cite{Chang:2019vez,Falkowski:2019tft} that appeared during the writing of this report. In these works, a different approach to the investigation of the possible size of the trilinear has been pursued. In particular, a different question has been posed:
\begin{itemize}
\item if we measure a deviation on the value of the trilinear Higgs self coupling, at which energy scale at least we should  expect new physics?
\end{itemize}
Parameterising the deviation via $\delta_\lambda=\kl-1$ it has been found \cite{Chang:2019vez} that perturbation theory breaks down at the $E_{\rm max}$ scale
\begin{equation}
E_{\rm max}\lesssim \frac{{\rm 13~TeV}}{|\delta_\lambda|}\, ,
\label{EFT:emax}
\end{equation}
regardless of the specific shape of the Higgs potential. Thus, if a
deviation from $\lHcubeSM$ is observed, it would provide a target for the energy to explore at future colliders.

\subsection*{UV-complete models}
We now turn to our second question, namely how large the trilinear Higgs self-coupling can be in renormalisable models. 
As a first step, we need to identify the
class of models with potentially the largest trilinear Higgs self-coupling modifications. 
For simplicity, we restrict ourselves to one particle extensions of the SM and  
focus on the regime where the new states are heavier than the SM ones but not necessarily yet in the EFT regime. 
This is motivated by the fact that we want to concentrate on the case where 
the leading effects in di-Higgs production 
are due to the deviation in the Higgs trilinear. 

The EFT regime can still be very useful in order to classify the SM extensions 
that can potentially yield the largest effects. In fact, we want to select those 
representations that can contribute to the operator $(\Phi^{\dagger}\Phi)^3$ once integrated 
out (see also Ref.~\cite{Goertz:2015dba}).  
\begin{table}
\renewcommand{\arraystretch}{1.4}
\centering
\begin{tabular}{@{} |c|c|
@{}}
\hline
$\phi$ &  $\mathcal{O}_\phi$ 
\\ 
\hline
$(1,1,0)$ & $\phi \Phi \Phi^\dag$ 
\\
$(1,2,\tfrac{1}{2})$ & $\phi \Phi \Phi^\dag \Phi^\dag$ 
\\
$(1,3,0)$ & $\phi \Phi \Phi^\dag$ 
\\
$(1,3,1)$ & $\phi \Phi^\dag \Phi^\dag$ 
\\
$(1,4,\tfrac{1}{2})$ & $\phi \Phi \Phi^\dag \Phi^\dag$ 
\\
$(1,4,\tfrac{3}{2})$ & $\phi \Phi^\dag \Phi^\dag \Phi^\dag$ 
\\
\hline
  \end{tabular}
  \caption{
  List of new scalars $\phi$ 
  inducing a tree-level modification of $\kl$   
  via the tadpole operator $\mathcal{O}_\phi$. The ($SU(3),SU(2),U(1)$) representation is displayed in the left column. \label{EFT:newscalarsHHH} }
\end{table}
In Table~\ref{EFT:newscalarsHHH} we give the complete list of scalar representations 
$\phi$ that introduce a 
tree-level modification to the trilinear Higgs self-coupling in the EFT limit and that 
 are characterized  
by the presence of a tadpole operator $\mathcal{O}_{\phi}$. The $\phi$ states $(1,3,0)$,  $(1,3,1)$, $(1,4,\tfrac{1}{2})$ and $(1,4,\tfrac{3}{2})$ receive a vacuum expectation value that violates custodial symmetry 
and hence these cases are strongly constrained by EW precision measurements, 
while $(1,2,\tfrac{1}{2})$ with the operator $\phi \Phi \Phi^\dag \Phi^\dag$ corresponds to a general two-Higgs doublet model without $Z_2$ symmetry. Such a model leads in general to flavour-changing neutral currents 
and hence requires extra assumptions in the flavour structure. 
We will hence concentrate on the simplest case of a singlet extension $(1,1,0)$, with potential 
\begin{equation}
V(\Phi,\phi) = 
\mu_1^2 |\Phi|^2+\lambda_\Phi |\Phi|^4+\frac{\mu_2^2}{2} \phi^2+\mu_4 |\Phi|^2 \phi 
+\frac{\lambda_3}{2} |\Phi|^2 \phi^2+\frac{\mu_3}{3} \phi^3+\frac{\lambda_2}{4} \phi^4 \, .
\end{equation}
Some of the parameters above can be replaced by phenomenologically more accessible ones, like the mixing angle $\cos\theta$ between the singlet and the doublet fields, the vacuum expectation values and the masses of the Higgs bosons. Choosing as input parameters
$m_1=125\text{ GeV}$, $m_2$, $\theta$, $v_H=246\text{ GeV}$, 
$v_S$, $\lambda_2$, $\lambda_3$, 
we scan them in the range
$800\text{ GeV}< m_2 < 2000 \text{ GeV}$, 
$|v_S|<m_2$, 
$0.9 < \cos\theta < 1$, 
and  in the perturbative regime
$0<\lambda_2<\frac{8}{3} \pi$, 
$|\lambda_3|<16\pi$.
We further check the compatibility with EW precision observables, 
where the strongest bound comes from the measurement of the $W$-boson mass \cite{Lopez-Val:2014jva} 
and a combined fit to the Higgs signal (see also the discussion in Sec.~\ref{sec:BSMspin0}).
The perturbativity bound on $\lambda_{2,3}$ is set by perturbative unitarity, 
while for the dimensional coupling $\mu_3$ we 
require the loop-corrected vertex to be smaller than the tree-level one \cite{DiLuzio:2016sur}. 
In addition, we required the potential to be bounded from below and checked for vacuum stability 
by means of the code {\tt VEVACIOUS} \cite{Camargo-Molina:2013qva}, 
with the model file generated by {\tt SARAH} \cite{Staub:2008uz, Staub:2013tta}.
 \begin{figure}
 \begin{center}
 \includegraphics[width=9cm]{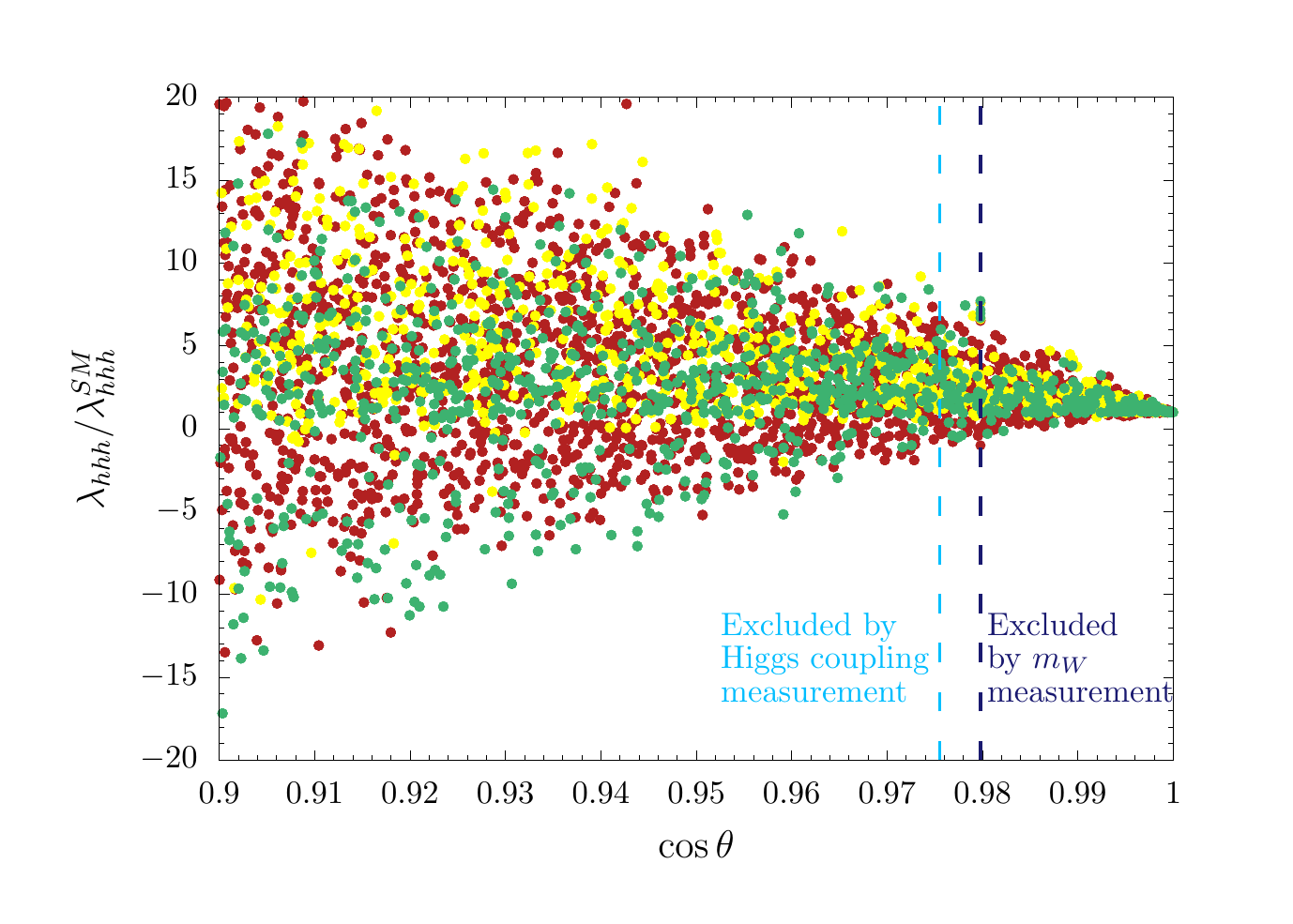}
 \vspace{-0.4cm}
 \caption{Modification of the trilinear Higgs self-coupling obtained from a scan over the singlet model parameters.
  The plot is taken from Ref.~\cite{DiLuzio:2017tfn} and adapted for this report.
 \label{EFT:singletlamhhh}}
 \end{center}
 \end{figure}

The results of the parameter scan can be found in Fig.~\ref{EFT:singletlamhhh}. All points 
on the left of the light blue dashed line are excluded by Higgs coupling measurements, 
while everything on the left of the dark blue line is excluded by the measurement of $m_W$. 
The red, yellow, green points correspond respectively 
to an unstable, metastable, stable EW vacuum. 
As it can be inferred from the figure, the vacuum instability cannot constrain 
the modification of the trilinear Higgs self-coupling. 
The maximal possible deviations allowed in the model are given by
 \begin{equation}
 -1.5<\kl<8.7\,.
 \end{equation}
 
We now discuss the case of the MSSM as an example of a UV-complete model where BSM effects are more complex than in the scenario just considered.
Assuming that at the LHC no further particle related to the EWSB is discovered, in particular no further Higgs bosons, in Ref.~\cite{Gupta:2013zza} the maximal SM deviations of the triple Higgs coupling of the light CP-even Higgs boson was estimated.
 Constraints from the $W$-boson mass have a minimal influence, while viable deviations are mainly constrained by the shape of the discovery potential and the size of the Higgs boson mass.

For a correct determination of the maximal deviations of the triple Higgs coupling, in the MSSM it is crucial that the same approximation is used for the prediction of both the Higgs mass and the triple-Higgs coupling. Also, the input parameters must be the same in order to find
the decoupling behaviour of the MSSM~\cite{Dobado:2002jz}, {\it i.e.}, $\lambda\rightarrow \lambda^{\rm SM}$ for $M_A\rightarrow\infty$.
Taking into account all the corrections given in Ref.~\cite{Carena:1995bx}, which especially includes the  $\mathcal{O}(M_Z^2/v^2 y_t^2)$ terms, the largest deviations were found for $\tan \beta = 5$ and low $M_A$ values, $M_A\sim$ 200 GeV,\footnote{It is important to note that in the region of $\tan \beta = 5$  a relatively light CP-odd Higgs boson of a mass of 200 GeV could be present and still be undiscovered according to the discovery potential assumed in Ref.~\cite{Gupta:2013zza}.} leading to about a 15\% deviation of the SM Higgs triple coupling. Note that the approximation from Ref.~\cite{Carena:1995bx} partly leads to smaller Higgs
mass values and, hence, a wider exclusion of  parts of the parameter points due to a too low Higgs boson mass value w.r.t.~other approximations including further higher-order corrections. In order to account for this effect, a relaxed Higgs boson mass constraint was applied, see Ref.~\cite{Gupta:2013zza} for details.   Instead, for $\tan \beta \geq 10$, the estimated maximal deviation is about 2\%. The latter limit does not change if one assumes that stop quarks are heavier than 2.5~TeV (one should note however that the approximations used to derive the MSSM Higgs mass value and the corresponding triple Higgs coupling have a much larger uncertainty for large stop masses, since large logarithms are not re-summed in this approximation).
On the other hand, the up-to-date results of the searches for heavy Higgs bosons and, in particular, the measurements of the properties of the discovered Higgs boson disfavour such a low value of $M_A$.  For $M_A \gtrsim 350$ GeV, the maximal deviations found are $\lesssim 4\%$.
Thus, it will be very difficult to discover the imprint of the MSSM on the trilinear Higgs self-coupling both at the HL-LHC and at a 100 TeV future collider.

%% file: EFT/HH_GPZ.tex


In terms of the EFT coefficients, as defined in \refeq{lsmeft}, the relevant interaction terms entering 
$\hhAlt$ production in gluon-gluon fusion read \cite{Goertz:2014qta}
\begin{equation}\label{eq:Lgghh}
\begin{split}
\mathcal{L}_{\hhAlt} = &- \frac{ \mhAlt^2 } { 2 v } \left( 1 + c_6 - \frac{3}{2} c_H  \right) h^3 \
-  \Big[ \frac{m_t}{v} \left( 1 + c_t  - \frac{c_H}{2}  \right) \bar{t}_L t_R h 
\\
&
-  \frac{m_t}{v^2}\left( \frac{3 c_t}{2}  -  \frac{ c_H}  { 2 } \right) \bar{t}_L t_R h^2 + \text{h.c.}\Big] \
+ \frac{\alpha_s c_g} {4 \pi } \left(\frac{h^2} {2v^2} + \frac{h}{v} \right) G_{\mu\nu}^a G^{\mu\nu}_a ,
\end{split}
\end{equation}
where we
neglected light fermions, whose impact is in general expected to be small \cite{Goertz:2014qia,Sirunyan:2018koj}, and also neglected effects from the chromomagnetic operator, ${\cal O}_{tG}$.
Moreover, in order to canonically normalise the Higgs kinetic term (after EWSB) and to remove derivative interactions, we employed the field redefinition $h \rightarrow  \left( 1 - \frac{ c_H} {2} \right) h - \frac{ c_H} { 2 v } h^2 - \frac{c_H}{6 v^2} h^3$.

Beyond modifying the trilinear self-coupling and the top-Yukawa coupling (first two terms in Eq.~(\ref{eq:Lgghh})), entering Higgs pair production via the SM-like triangle and box diagrams, as given in the upper panel of Fig.~\ref{fig:diags}, the $D=6$ operators induce new topologies, producing a Higgs pair via 4-point contact interactions with a scalar quark current or with the gluon field strength squared, or finally via a splitting from a contact-like single Higgs production, see the second row of Eq.~(\ref{eq:Lgghh}). The corresponding Feynman diagrams are given in the lower panel of Fig.~\ref{fig:diags}. The resulting differential cross section in the (linear) EFT becomes~\cite{Goertz:2014qta}

\begin{align}\label{eq:diffxsEFT}
\left. \frac{ \mathrm{d} \hat{\sigma} (gg \rightarrow \hhAlt) } { \mathrm{d} \hat{t} } \right|_\mathrm{EFT} &= \frac{ G_F^2 \alpha_s^2 } { 256 (2\pi)^3 } \bigg\{ \Big| ( 1 - 2 c_H + c_t + c_6 )\, \frac{3 \mhAlt^2 } { \hat{s} - \mhAlt^2 }  F_\triangle + (1 - c_H + 2 c_t )\, F_\Box  \nonumber
\\
&\hspace{6em} + (3 c_t - c_H)\, 3 F_\triangle  + 2 c_g\, \Big(1+\frac{3 \mhAlt^2 } { \hat{s} - \mhAlt^2 }\Big) \Big|^2 + \Big|  G_\Box \Big|^2 \bigg\} \,,
\end{align}
where we ordered the various contributions accordingly, and the form factors $F_{\triangle,\Box},G_{\Box}$ take the same form as in the SM \cite{Goertz:2014qta} (see Sec.~\ref{sec:gluon_fusion}) and can be obtained from Ref.~\cite{Plehn:1996wb}. Note that the spin-2 contribution to the box topology, $G_{\Box}$, receives no $D=6$ corrections.

\begin{figure}[t!]
  \begin{center}
  \includegraphics[width=0.5\textwidth]{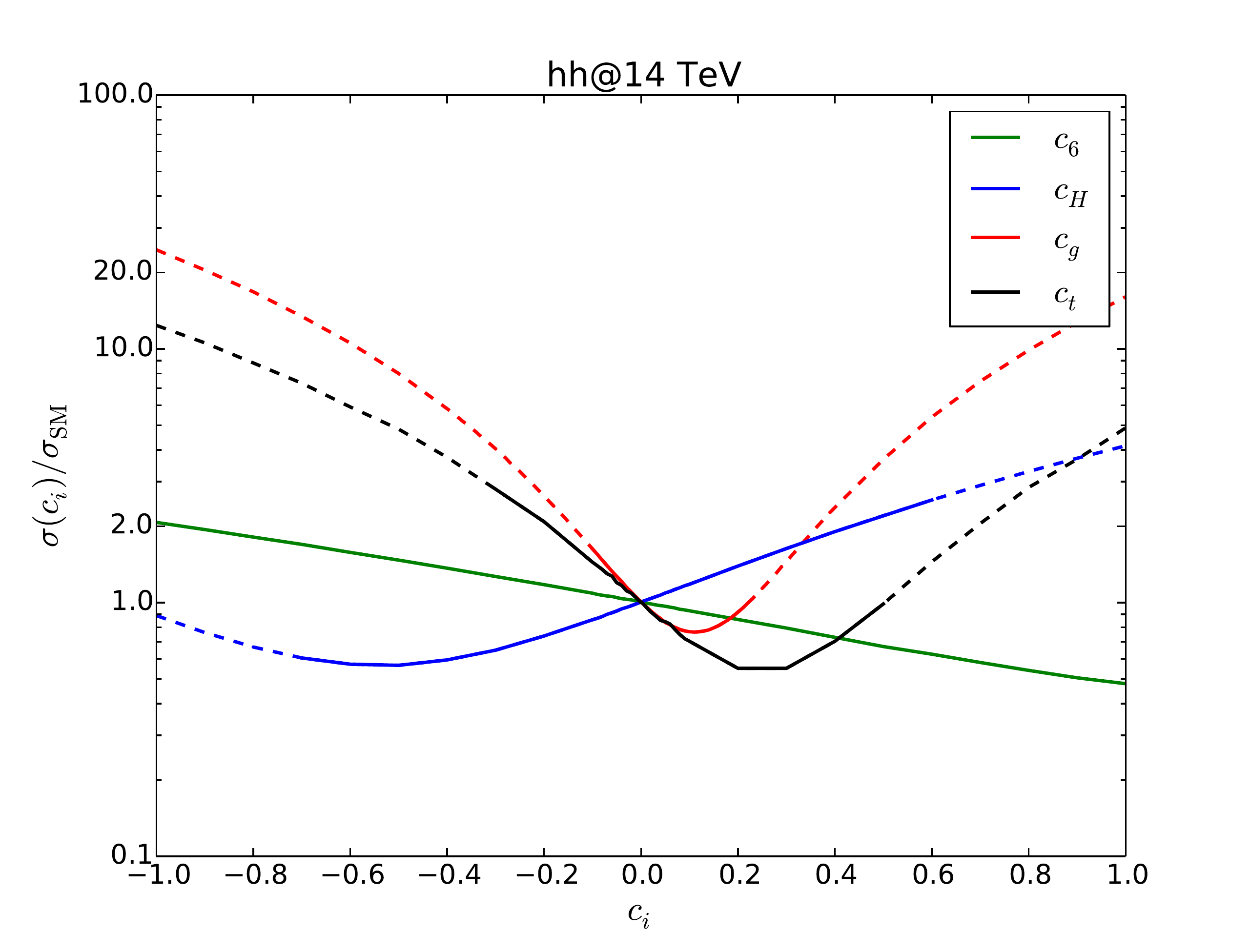}
  \hspace{-0.3cm}
   \includegraphics[width=0.5\textwidth]{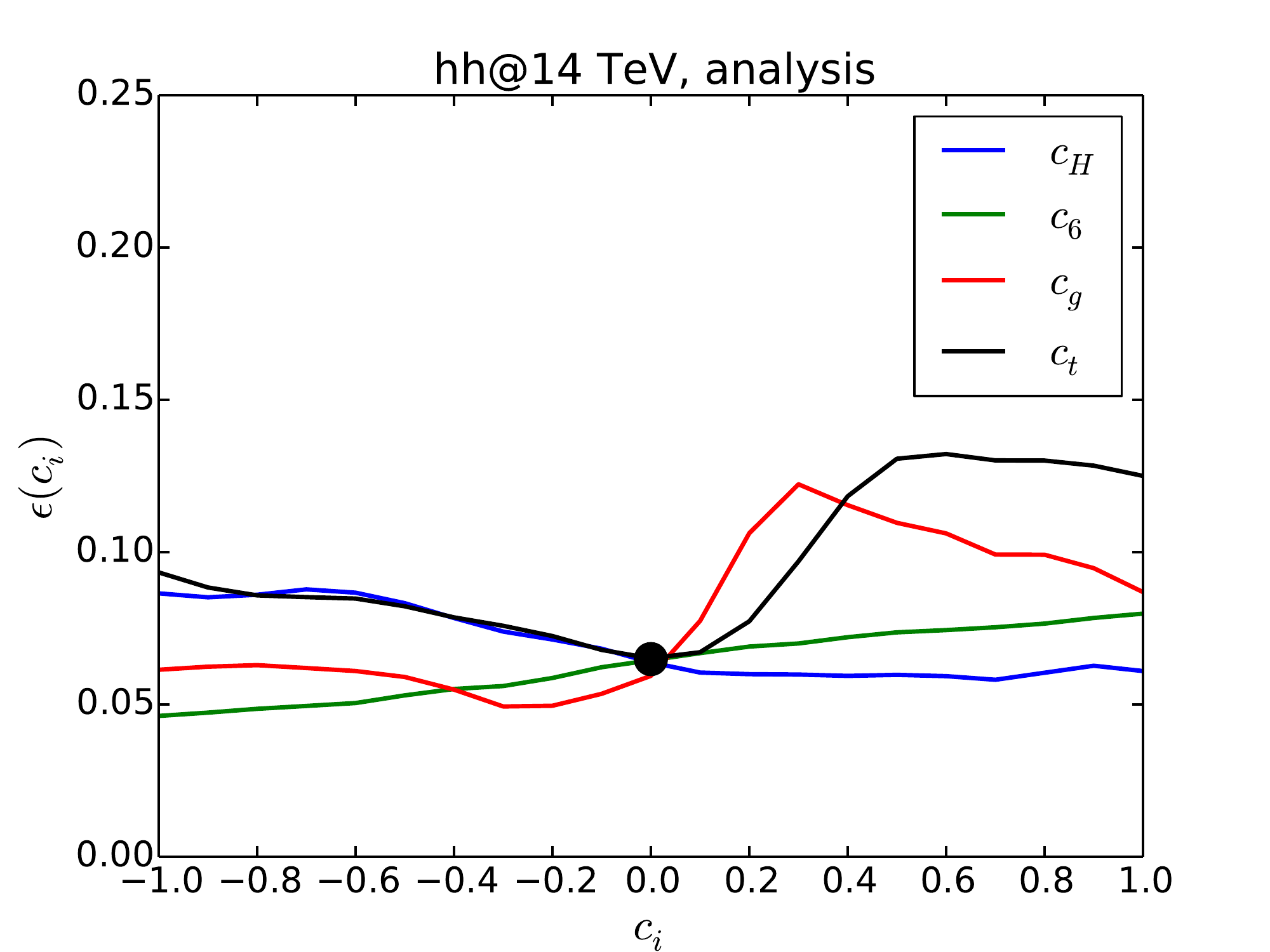}
  \end{center}
  \vspace{-0.5cm}
  \caption{\label{fig:opeffect}
  Left: Relative change in the $\hhAlt$ cross section in dependence on individual operators. The dashed parts of the curves are excluded at 95\% CL from Higgs boson data. Right: Corresponding efficiency of the analysis. See text for details.}
\end{figure}

For our phenomenological analysis we have implemented the Lagrangian in \refeq{eq:Lgghh} into the \texttt{HERWIG++} event generator, which allows to appropriately take into account changes in kinematic distributions that will substantially modify the efficiency of the experimental analysis. To treat higher order QCD corrections, we normalise our results to the NNLO QCD SM calculation of the cross section of $\sim 40$~fb \cite{deFlorian:2013jea} and include a conservative theory uncertainty of $f_\mathrm{th} = 30\%$, comprising scale, PDF plus strong coupling, and $K$-factor uncertainties of $\mathcal{O}(10\%)$ each \cite{deFlorian:2017qfk}.
Regarding the decays of the Higgs pair, we consider the $\hhAlt \rightarrow \bbtt$ final state, where we also include the impact of the $D=6$ operators on the partial widths, via modified Yukawa couplings, as well as the NP effects on the total width, and follow the analysis steps lined out in Ref.~\cite{Dolan:2012rv}. For more details, the reader is referred to Ref.~\cite{Goertz:2014qta}.

In the following, we consider the six-dimensional parameter set $(c_6, c_H, c_g, c_t, c_b, c_\gamma)$, fixing in addition $c_\tau \equiv c_b$ for simplicity. As a first result we present, in the left panel of Fig.~\ref{fig:opeffect}, the impact of varying individual coefficients out of this set on the $\hhAlt$ production cross section relative to the SM value $\sigma(c_i)/\sigma_{\rm SM}$. The dashed parts of the curves represent regions which are excluded at the 95\% CL by  Higgs boson data\footnote{
The 95\% CL limits obtained in Fig.~\ref{fig:opeffect} correspond to the bounds available at the time of the publication of Ref.~\cite{Goertz:2014qta}.
Note the different sign convention in the top-Yukawa coupling $c_t$ with respect to \reffig{hhplot}.
},
employing \texttt{HiggsBounds} and \texttt{HiggsSignals}. We can see a particularly pronounced dependence on the Yukawa-like and gluonic $D=6$ operators, while the negative interference between the triangle and box diagrams leads to a decreased cross section for positive $c_6$. In the right panel of the same figure, we show the efficiency of our analysis, varying the same coefficients, where the non-trivial curves confirm the importance of using Monte Carlo event generation.

\begin{figure}[!t]
\begin{center}
\includegraphics[width=0.31\textwidth]{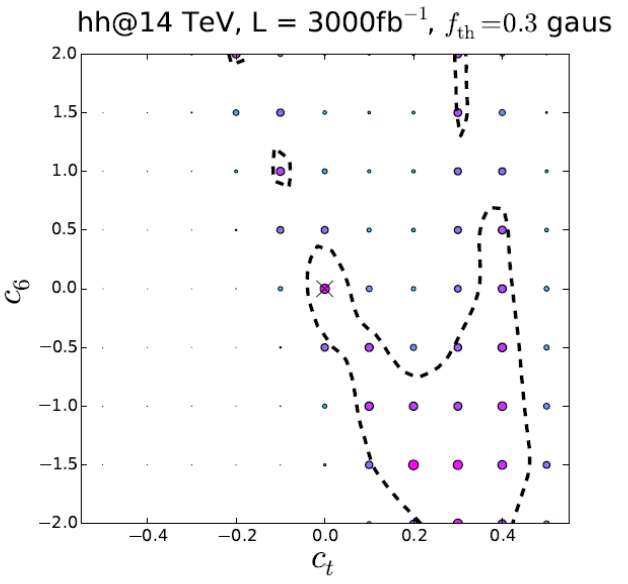}\, \includegraphics[width=0.31\textwidth]{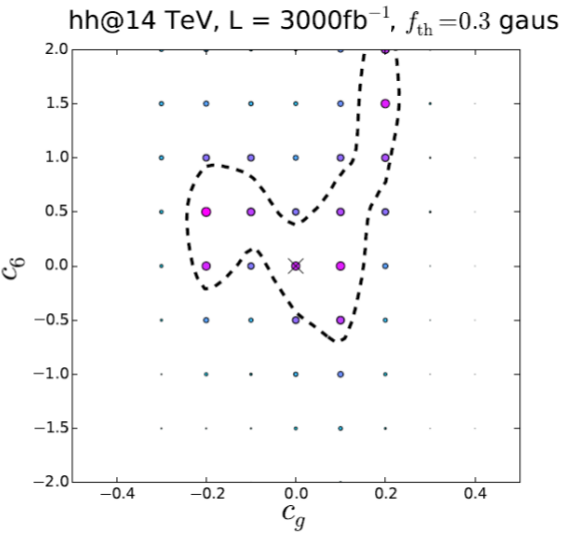}\, \includegraphics[width=0.335\textwidth]{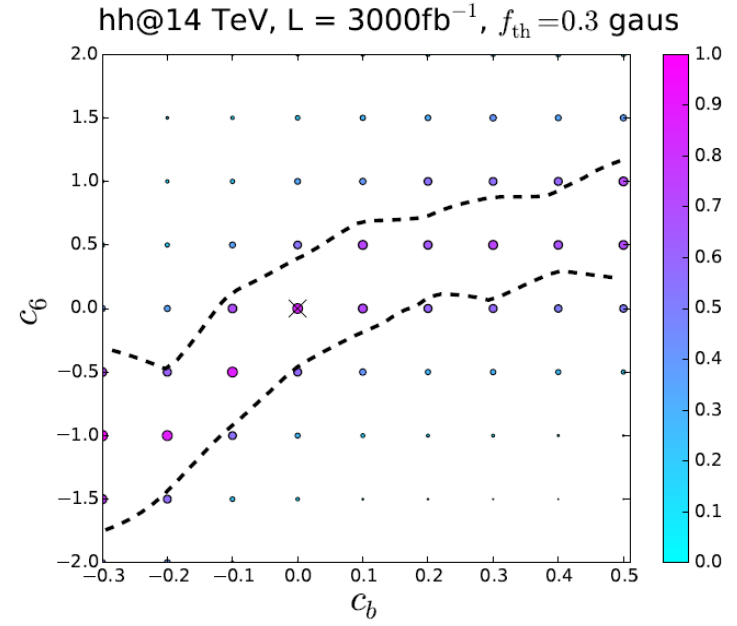}
\end{center}
\vspace*{-0.4cm}
 \caption{Projected exclusions in the $(c_t,c_6)$,
 $(c_g,c_6)$, and $(c_b,c_6)$ planes at the HL-LHC, assuming a theoretical uncertainty of $f_\mathrm{th} = 0.3$. The plots show the $p$-values obtained after marginalisation over the directions orthogonal to the respective planes, including the 1-sigma contours as black dashed lines. See text for details.}
  \label{fig:exc}
\end{figure}

The resulting projected constraints on the EFT coefficients at the HL-LHC with an integrated luminosity of 3000\,fb$^{-1}$, assuming the SM to be true, are presented in Fig.~\ref{fig:exc}. Here, we consider three different two-parameter planes, varying $c_t,c_g$, and $c_b$, each along with $c_6$. We marginalise over those parameters that are not shown, employing a Gaussian weight that corresponds to a projected measurement of the respective (single Higgs) observables at the HL-LHC at the $10\%$ level \cite{Goertz:2014qta}. The plots display the $p$-values obtained for a grid of points in the corresponding planes via a color code and the 1-sigma contours as black dashed lines.

Looking at the $(c_t,c_6)$ plane, shown in the left plot, we see a strong dependence of the self-coupling constraint on $c_t$. In fact, for $c_t \sim 0.2$ the projected bound is significantly shifted compared to $c_t = 0$, since the effects of both coefficients on the production cross section can compensate each other. Beyond that, a similar sensitivity is found with respect to the gluonic contact interactions, entering the $(c_g,c_6)$ plane presented in the middle plot, and on changes in the bottom Yukawa coupling, appearing in the rightmost $(c_b,c_6)$ plane (with $c_b \equiv c_\tau$). In the latter case, a reduction in the production cross section via $c_6>0$ could for example be lifted by an enhanced branching ratio into bottom quarks and $\tau$ leptons.

Finally, we summarise our projected HL-LHC constraints on $c_6$, marginalising over all other coefficients, 
in the following table:
\begin{center}
    \begin{tabular}{|c|c|c|}
    \hline
       full & full ({\bf future}) & $c_6$-only \\
       \hline
       $c_6 \gtrsim -1.2$  & $c_6 \in (-0.6,0.6)$ & $c_6 \in (-0.4,0.4)$  \\
       \hline 
    \end{tabular}\ .
\end{center}    
The result given in the left column corresponds to a marginalisation assuming {\it present} experimental uncertainties for the Higgs couplings, and delivers a very weak projected constraint \cite{Goertz:2014qta}, leaving $c_6$ unbounded from above, which highlights the importance of a combined analysis. Once we consider an improved determination of the Higgs properties via single Higgs production at the $10\%$ level, as discussed above, the projected constraint on $c_6$ improves significantly, reaching the $60\%$ level as presented in the middle column. This is rather close to, but still worse than, the naive bound where only variations in $c_6$ are allowed, leading to a projected $40\%$ determination as given in the last column and agreeing with previous estimates \cite{Goertz:2013kp}.

We close noting that a constraint on the trilinear coupling at the $\lesssim 100\%$ level would in particular mean that one could probe the presence of the only relevant operator in the SMEFT, namely the $\mu^2$ term, whose existence has so far not been established experimentally yet \cite{Goertz:2015dba} and whose absence would lead to a strong decrease of the Higgs pair production cross section of $\sim 70 \%$. It is clear that further studies of Higgs pair production, especially considering its kinematic distributions, would be interesting to still improve constraints and to disentangle different EFT effects, see e.g. Refs.~\cite{Azatov:2015oxa,Carvalho:2017vnu,Borowka:2018pxx}.

%% file: EFT/single-WP.tex
In this section we discuss an alternative strategy for extracting the information on the trilinear Higgs self-coupling: the precise measurement of single Higgs production~\cite{McCullough:2013rea,Gorbahn:2016uoy,Degrassi:2016wml,Bizon:2016wgr,DiVita:2017eyz,Barklow:2017awn,Maltoni:2017ims,DiVita:2017vrr,Maltoni:2018ttu} at the LHC. Indeed, single Higgs production is sensitive to the trilinear Higgs self-coupling via EW one-loop corrections (two loops for gluon-gluon fusion production and $H\rightarrow \gamma\gamma$ decay). Thus, this strategy is based on indirect measurements and it is complementary to the direct measurements via  double Higgs production. In single Higgs production, the effects of a modified Higgs self-coupling are much smaller, but the precision of the experimental measurements is and will be  much better than in the case of double Higgs production. Moreover, many different final states (at the differential level) can be measured, leading to competitive bounds for the trilinear Higgs self-coupling. Even EW precision observables can be helpful for setting these bounds \cite{Degrassi:2017ucl,Kribs:2017znd}.

In this section we recall the most important points of the calculation framework introduced in Refs.~\cite{Gorbahn:2016uoy,Degrassi:2016wml}. Recently, updated numerical results for the effects induced by a modified trilinear Higgs coupling have been presented in Ref.~\cite{Cepeda:2019klc} for inclusive and differential quantities; we do not report them here, but they have been exploited for the projections of the determination of the trilinear Higgs self-coupling that are discussed in this section.  


Assuming on-shell single Higgs production, the signal strength for the  process $i \to H \to f$, {\it i.e.} its rate normalised to the corresponding SM prediction, is $\mu_i ^f =\mu_i \times \mu^f$, where $\mu_i$ and $\mu^f$ are the signal strengths for the production process $i \to H$ and the decay $H \to f$, respectively. Therefore,
 $\mu_i$ and $\mu^f$  can be expressed as
\beq 
\mu_i = 1 + \dsigmah(i) \,,  \qquad 
\mu^f = 1 + \dBR(f) \,,
\eeq{eq:muf}
where $\dsigmah(i)$ and $\dBR(f)$ are the deviations induced by an anomalous interaction, including the case of the trilinear Higgs self-coupling, to the production cross sections and branching ratios, respectively.
This definition can be also extended to the differential level.

In the case of vector boson fusion, $WH$, $ZH$, ${\tth}$ and $tHj$ production, the trilinear Higgs self interactions start to enter only at the one-loop level. On the contrary, gluon-gluon fusion production and the decays H$\to gg, \gamma \gamma$ depend on this coupling only via two-loop EW corrections. It is important to note that in all single Higgs processes the dependence on the quadri-linear Higgs self-coupling is further delayed by one loop order. On the other hand, it is possible in a similar way to probe quartic Higgs self-couplings via EW corrections to double Higgs production \cite{Maltoni:2018ttu,Bizon:2018syu,Borowka:2018pxx}.
Results for future hadron colliders exploiting this strategy to set bounds on the quartic self-coupling are discussed in Sec.~\ref{sec:other_probes}, and in Sec.~\ref{sec_ee:quartic} for the case of $e^+e^-$ machines.



\begin{figure}
\begin{center} 
\includegraphics[width=0.45\textwidth]{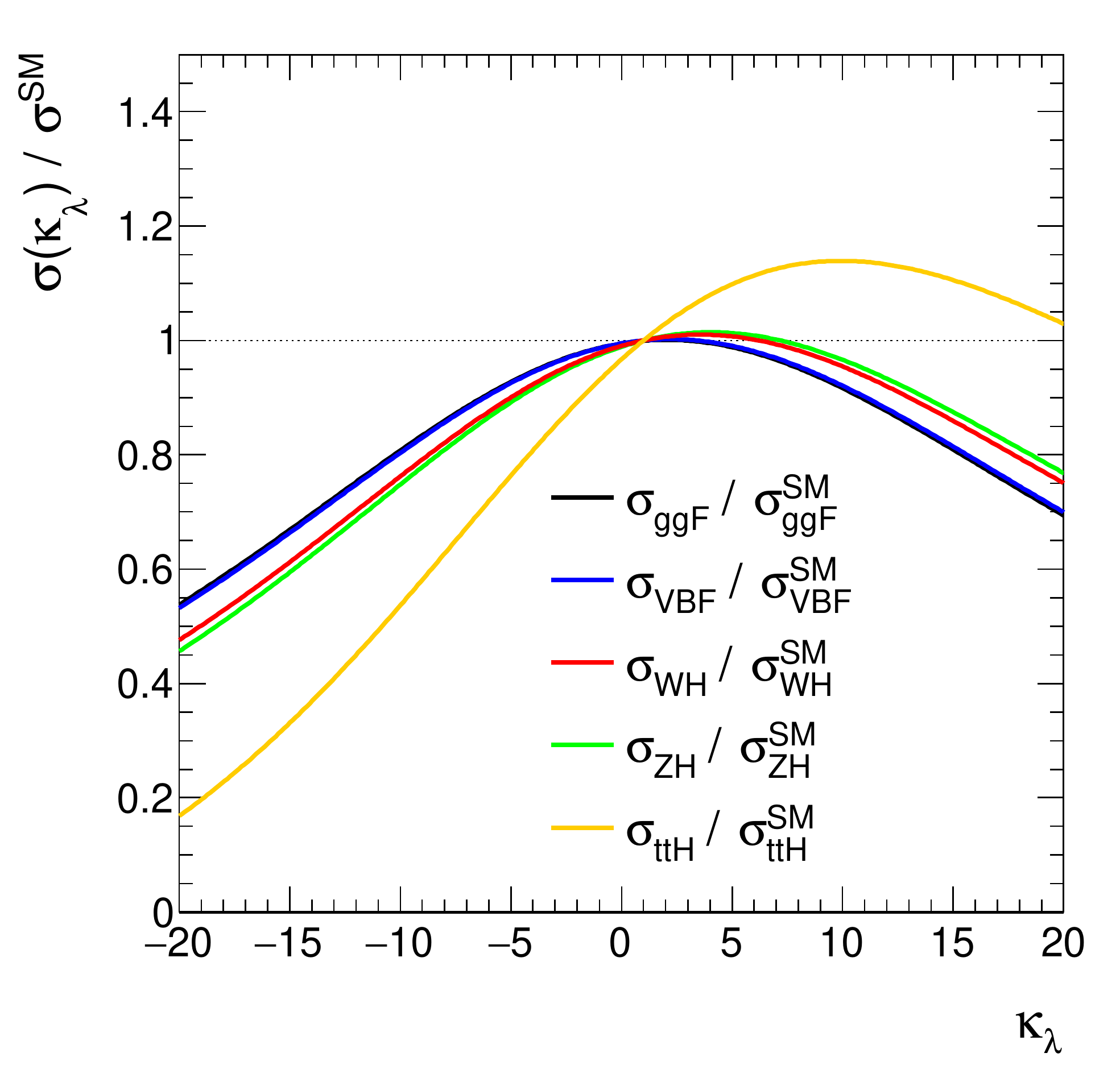}
\includegraphics[width=0.45\textwidth]{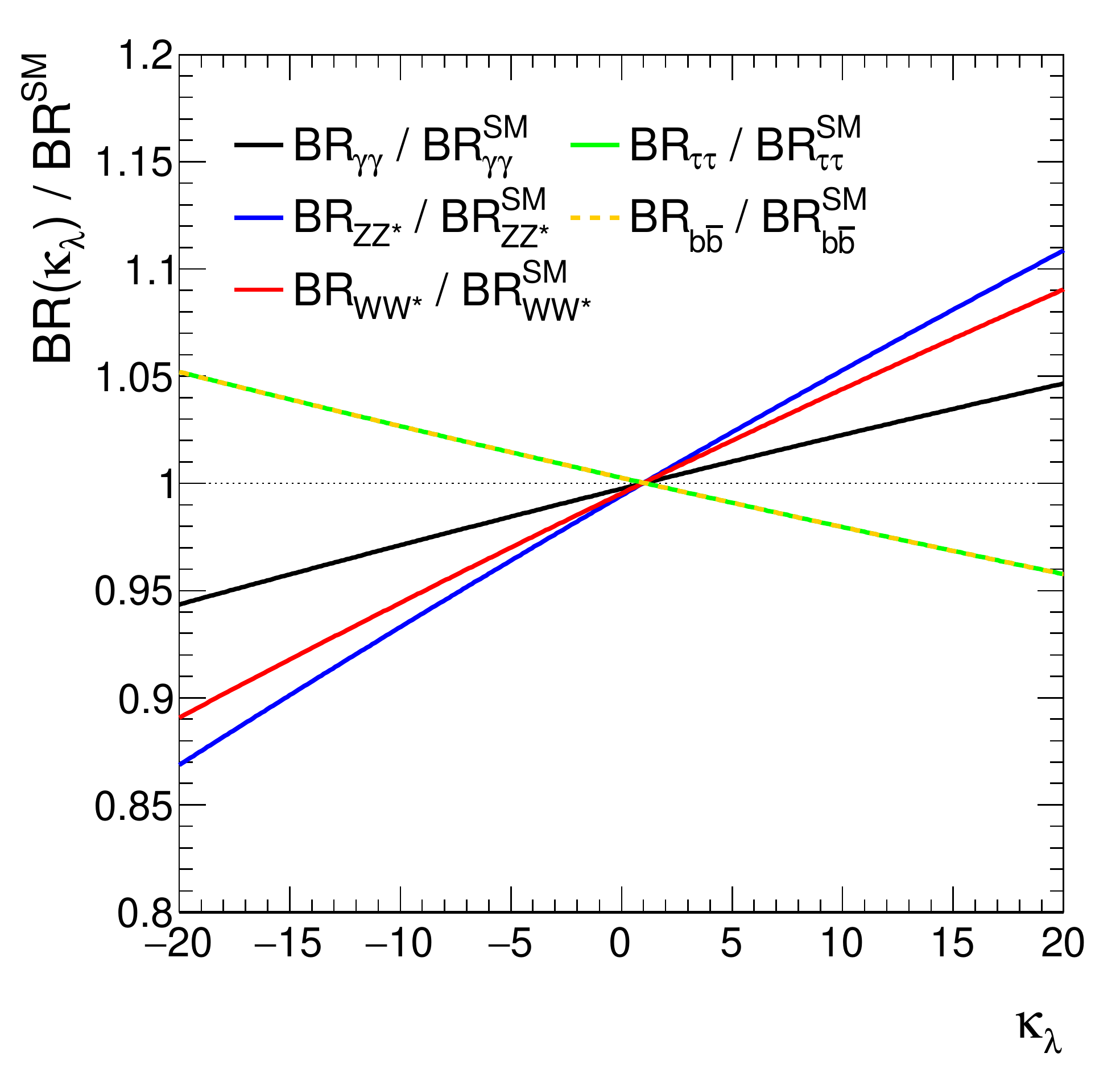}
\vspace*{-0.3cm}
 \caption{Variation of the cross sections (left) and branching fractions (right) as a function of the trilinear coupling modifier~\cite{ATLAS-PUB-19-009}.}
 \label{fig:singleWP1} 
 \end{center} 
 \end{figure}

The anomalous trilinear Higgs self interactions can be parameterized via $\kl$ (see \eqref{eq:klrel}).
Each single Higgs production or decay channel receives two different kinds of $\tril$-dependent contributions ~\cite{Gorbahn:2016uoy,Degrassi:2016wml}. First, a process and kinematic dependent contribution, denoted in the literature as $C_1$, which parameterises the linear dependence on $\kl$. Second, a universal contribution that is associated to the renormalization of the Higgs wave function and induces a quadratic dependence on $\kl$. On the contrary, in the case of the decays only a linear dependence on $\kl$ is present, due to the cancellation of the effects associated to the Higgs wave function renormalization.
Specifically, the signal strength $\mu_i$ for the production process $i 
\to H$ can be written in the following way,
\begin{eqnarray}
\label{eq:mui}
\mu_{i}(\kappa_\lambda)=\frac{\sigma^{\mathrm{BSM}}(i)}{\sigma^{\mathrm{SM}}(i)}= 
1+\delta \mu_i(\kappa_\lambda) + Z_H^\mathrm{BSM} (\kappa_i^2-1)\, ,
\end{eqnarray}
where $Z_{H}^{\textrm{BSM}}\left(\kappa_\lambda\right)$ is defined as:
\begin{equation}
Z_H^{\textrm{BSM}}\left(\kappa_\lambda\right) = 
\frac{1}{1-(\kappa_\lambda^2-1)\delta Z_H} \quad
\textrm{with} \quad \delta Z_H = -1.536 \times 10^{-3} \, ;
\label{eq:ZH}
\end{equation}
$\kappa_i^2 
=\sigma^{\textrm{BSM}}_{\textrm{LO}}(i)/\sigma^{\textrm{SM}}_{\textrm{LO}}(i)$ 
takes into account additional variations of Higgs boson couplings to 
other particles (\textit{e.g.} fermions, vector bosons) or it can be 
taken equal to one when variations of the trilinear-coupling only are 
considered. Assuming that (NLO) QCD corrections factorise anomalous 
$\kappa_\lambda$ effects and taking into account also NLO EW corrections in the SM and on top of $Z_{H}^{\textrm{BSM}}$, the quantity 
$\delta\mu_i(\kappa_\lambda)$ is defined as
\begin{equation}
\label{eq:muf3}
      \delta\mu_i(\kappa_\lambda) = 
\frac{\sigma_{\mathrm{NLO}}^{\mathrm{BSM}}(i)}{\sigma_{\mathrm{NLO}}^{\mathrm{SM}}(i)} 
-1
         =  Z_H^{\mathrm{BSM}} \left[ 1 + \frac{ (\kappa_\lambda-1)C_1^i 
}{ K_{\mathrm{EW}}(i)} \right]-1\, ,
\end{equation}
where $K_{\mathrm{EW}}(i) \equiv 
\sigma_{\mathrm{NLO_{EW}}}^{\mathrm{SM}}(i) / \sigma_{\mathrm{LO}}(i) $ 
is the NLO EW $K$-factor in the SM, which therefore includes also the 
Higgs self-coupling one-loop corrections in the SM.
The values of $K_{\mathrm{EW}}(i)$ and $C_1^i$ for the different 
production mechanisms can be found in Ref.~\cite{Maltoni:2017ims} for 
the inclusive case, and also the differential values have been presented therein (see Figs.~2-10). It is worth to 
note that, although the size of $K_{\mathrm{EW}}(i)$ is quite sizeable  
and has a non negligible impact on the prediction of the (differential) cross sections $\sigma$, in the case of $\delta\mu_i(\kappa_\lambda)$ 
its impact is very small~\cite{Maltoni:2017ims}.
Finally, each decay process $H \to f$ is scaled by the signal strength
\begin{equation}
\label{eq:muf2}
\mu^f(\kappa_\lambda) \simeq  \frac{\kappa_f^2  +(\kappa_\lambda - 1) 
C_1^f }{\sum_j {\mathrm{BR}}^{\mathrm{SM}}(j) [\kappa_j^2 
+(\kappa_\lambda - 1) C_1^j]  } \, ,
\end{equation}
where $\sum_{j}$ runs over all the Higgs boson decay channels and 
$\kappa_j$ is the branching fraction modifier for the $j$ final state, 
$\kappa_j^2 = 
\textrm{BR}^{\textrm{BSM}}_{\textrm{LO}}(j)/\textrm{BR}^{\textrm{SM}}_{\textrm{LO}}(j)
$.
The dependence of the production cross sections and branching fractions with $\kappa_\lambda$ is shown in Fig.~\ref{fig:singleWP1}.

\begin{figure}[tb]
\begin{center}
\includegraphics[width=0.55\textwidth]{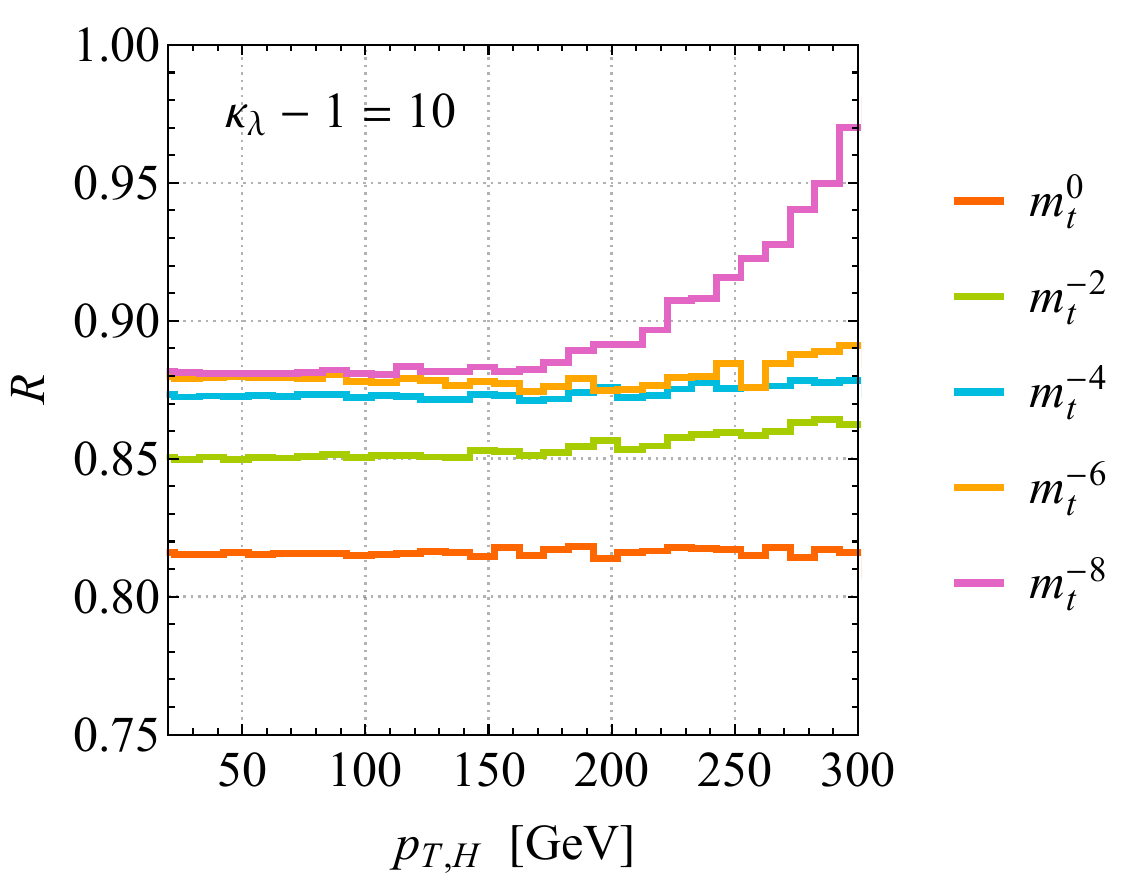} 
\vspace*{-2mm}
\caption{\label{fig:Rpth10} Effect of $\kappa_\lambda$ corrections  on  the $p_{T,H}$ spectrum in $pp \to H+{\rm jet}$ production. As indicated, the curves correspond to different orders in the asymptotic expansion in the top-quark mass $m_t$, and all show the ratio between the new-physics and the SM prediction for the choice $\kappa_\lambda - 1 = 10$. } 
\end{center}
\end{figure}

The processes ${WH}$, ${ZH}$, and especially ${\tth}$, entail a larger linear dependence on $\tril$ with respect to the other processes.
Moreover, also a stronger kinematic dependence is present, with larger values associated to the threshold region \cite{Degrassi:2016wml,Bizon:2016wgr,Maltoni:2017ims}. In the case of VBF, the kinematic dependence is instead rather flat \cite{Degrassi:2016wml,Bizon:2016wgr,Maltoni:2017ims}.
Fully differential results for these production mechanisms can be obtained with the code presented in Ref.~\cite{Maltoni:2017ims}.
%
The calculation of differential effects for gluon-gluon fusion would be desirable, but it is not yet available due to its higher complexity, as it involves the evaluation of two-loop EW diagrams for the process $pp \to H+\text{jet}$.
The calculation of the relevant amplitudes in an asymptotic expansion near the limit of infinitely heavy top quark has been performed for a generic $\kappa_\lambda$ in Ref.~\cite{Gorbahn:2019lwq}. The corresponding numerical results indicate that the effect of $\kappa_\lambda$ variations in the $p_{T,H}$ spectrum are almost flat within the range of validity of the expansion (i.e.~$p_{T,h} < m_t \simeq 173 \, {\rm GeV}$). This feature is illustrated in Fig.~\ref{fig:Rpth10} for the  choice $\kappa_\lambda - 1 = 10$. Above the top threshold, distortions of the $p_{T,H}$ distribution due to the $\kappa_\lambda$ corrections are, however, expected.

Since single Higgs production processes have already been measured, constraints on $\tril$ can be set following this strategy. Especially, since $C_1$ is different for any production and decay channel, a fit involving different measurements can be very powerful for the determination of a single parameter. 
Based on the results presented in Ref.~\cite{Khachatryan:2016vau}, which do not exploit differential information, assuming the only deviations from the SM are associated to $\tril$, the following $2\sigma$ bounds can be set ~\cite{Degrassi:2016wml}:
\begin{equation}
 -9.4<\kl<17.0 \qquad {\rm at ~8 ~TeV}\label{k38TeV}
\end{equation}
and following the same approach, based on the results presented in Ref.~\cite{Sirunyan:2018koj}, 
\begin{equation}
-4.7< \kl <12.6 \qquad {\rm at ~13 ~TeV}\, .\label{k313TeV}
\end{equation} 
Notably, bounds in \refeq{k313TeV} are competitive with the currently strongest bounds from double Higgs production measurements \cite{Aad:2019uzh}. Very recently, the first experimental results obtained following this strategy have been presented by ATLAS \cite{ATLAS-PUB-19-009}. This measurement is in good agreement with the estimate in \refeq{k313TeV} and is discussed in detail in Sec.~\ref{singleH_exp}.


The aforementioned limits, however, assume a very peculiar BSM scenario, in which the only relevant effects originate from the trilinear Higgs coupling, allowing for $\mathcal{O}(1)$ deviations without any effect on other Higgs couplings. In fact, these limits critically depend on other aspects~\cite{Maltoni:2017ims, DiVita:2017eyz}. First, the number of additional parameters, which are related to other anomalous interactions, and the number of independent measurements that are taken into account in the fit. Second, the inclusion or not of the information from differential distributions. Third, the fit assumptions on the size of the theoretical and experimental uncertainties. Also for these reasons, ATLAS and CMS analyses with a full-fledged treatment of all the correlations and with different assumptions on the  the number of BSM parameters are essential. The first of these kind of analyses, which has been presented in Ref.~\cite{ATLAS-PUB-19-009} and it is also discussed in Sec.~\ref{singleH_exp}, is supporting the validity of this strategy. 

As shown in Ref.~\cite{Maltoni:2017ims}, assuming only deviations on the Yukawa coupling of the top quark ($\kappa_t$) and/or a common rescaling of the Higgs gauge interactions ($\kappa_V$), limits are mildly affected. On the other hand, in general, a new dynamic affecting the Higgs self-coupling would leave a more complex imprint on the other Higgs interactions and can have a strong impact on the bound on $\tril$~\cite{DiVita:2017eyz}. 
Adopting the EFT framework described in Ref.~\cite{DiVita:2017eyz}, nine additional coefficients parameterise the possible deviations in single Higgs production (see \refeq{eq:SMEFT_singleH} and related discussion):
\begin{equation}
    \delta y_t,\,\delta y_b,\,\delta y_\tau, \,
    c_{gg},\,c_{\gamma\gamma}, \,
    \delta c_z, \,
    c_{zz},\,c_{z\square},\,c_{z\gamma} .
\label{eq:ninepara}
\end{equation}


For the determination of $\tril$, a global fit is important not only because it involves different processes that entail a different dependence on $\tril$, but also because it allows to assess the robustness of bounds such as those in \refeqs{k38TeV} and \eqref{k313TeV}, where only $\tril$ variations are considered. 
For example, a global fit using only inclusive single Higgs observables such as those presented in Ref.~\cite{DiVita:2017eyz}, which is based on only nine independent measurements, and taking into account the additional nine EFT deviations listed above, suffers from a flat direction. Therefore, $\tril$ remains unconstrained under these assumptions. On the other hand, its presence in the fit decreases the accuracy in the determination of some of the other nine coefficients. In order to lift this degeneracy, it is possible to include data from differential measurements. Indeed $\delta\kl\equiv \kl-1$ has a non-flat effect on single Higgs distributions.

\begin{figure}
	\centering
	\includegraphics[width=0.45\linewidth]{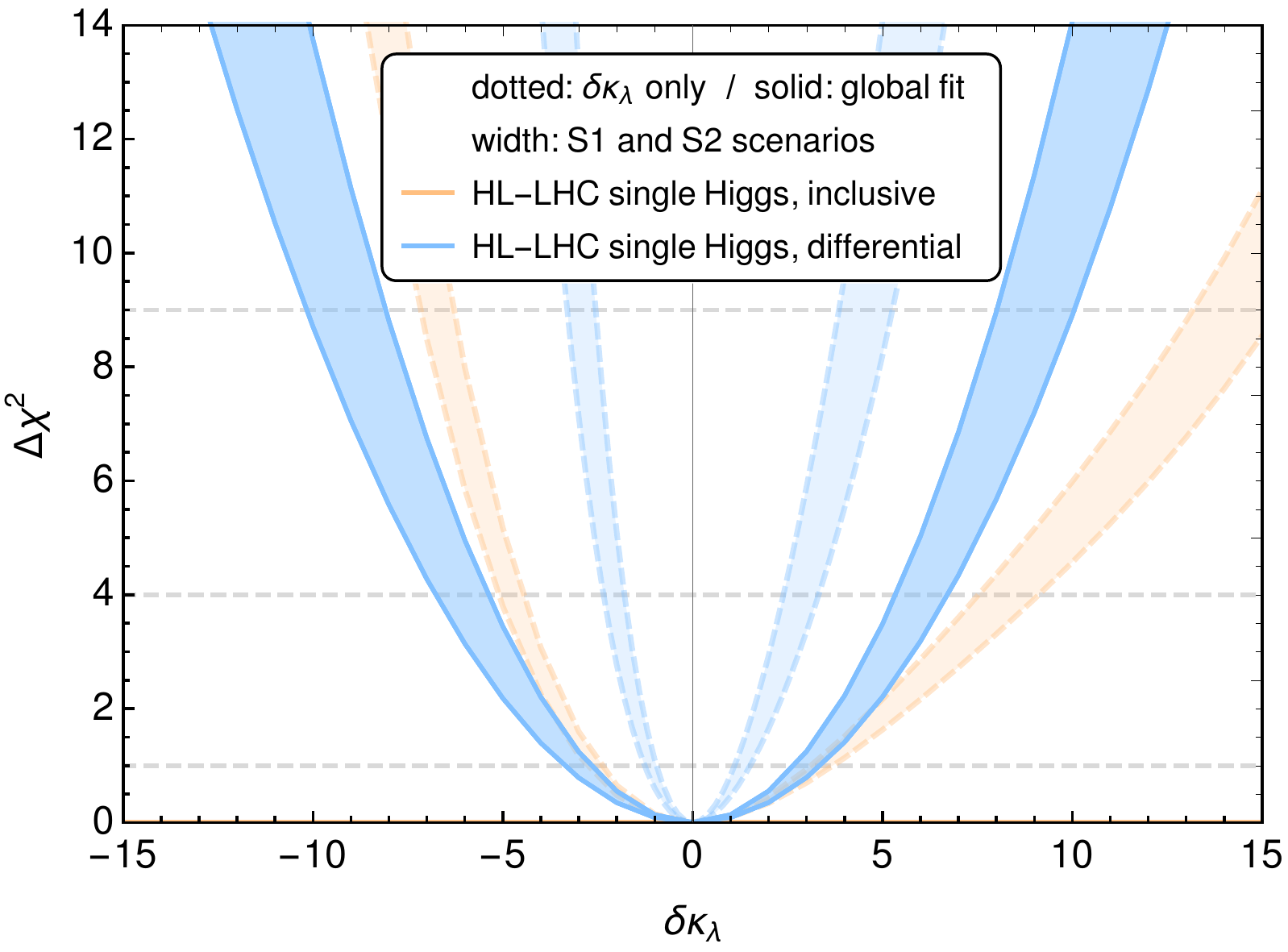}\hfill
	\includegraphics[width=0.45\linewidth]{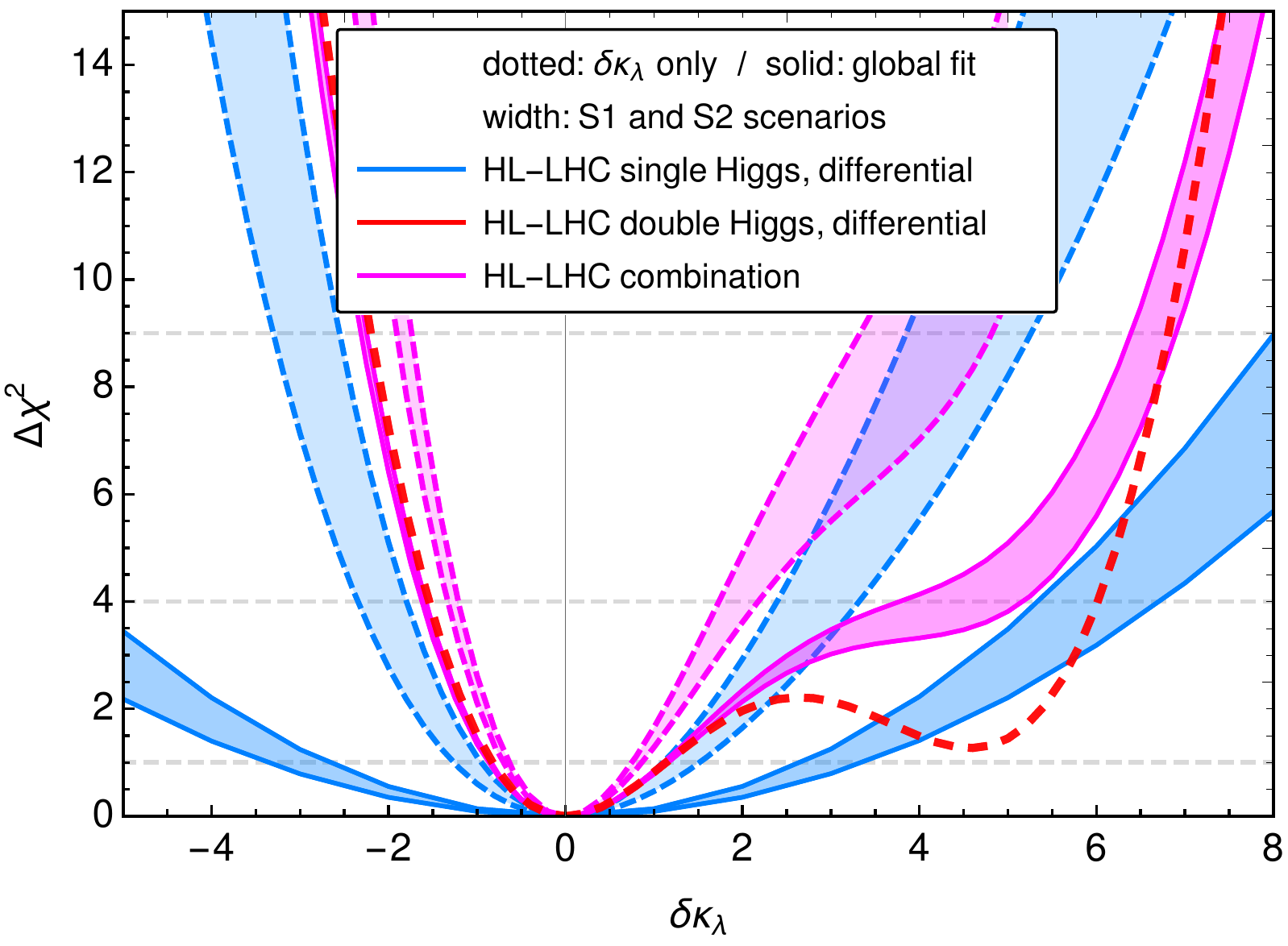}
	\caption{HL-LHC at 13 TeV and 3\,ab$^{-1}$. Left: Single Higgs with only inclusive measurements (orange) and including differential information (blue) with only $\kl$ (pale colour) or marginalising over the nine EFT coefficients (strong colour).  Right: Constraints from differential single Higgs (blue), differential double Higgs (dashed red) and their combination (pink).}
	\label{fig:hllhcchi2}
\end{figure}

We summarise the global fit for the HL-LHC in Fig.~\ref{fig:hllhcchi2}.
The width of the bands represent the results obtained assuming two different uncertainty scenarios, S1 and S2, which correspond to the projected uncertainties on the inclusive signal strengths recommended by the ATLAS and CMS collaborations for the different production and branching ratio\footnote{The first scenario (S1) assumes the same uncertainties as those used in the published in ATLAS and CMS Run 2 analyses~\cite{Aad:2019uzh, Sirunyan:2018two}. The second scenario (S2) features a reduction of the systematic uncertainties due to the improvements expected to be reached at the end of HL-LHC program~\cite{Cepeda:2019klc}.}.
In the case of differential distributions, as a first step, the projections of the uncertainties are estimated by rescaling the statistical uncertainties bin by bin. However, this is a very conservative estimate, because it assumes the background to be flat, while this one is typically larger at lower energies. Therefore, following the CMS analysis on $t\bar{t}H$ production with $H \to \gamma\gamma$ \cite{CMS:2018rig} as a template, we have tilted the background accordingly\footnote{With this procedure a good agreement with the CMS analysis is found, for this channel only. As a simple guess, we use it for the rest of the uncertainties}.
In the left plot, we show the $\Delta\chi^2$ for single Higgs projections including differential information (blue), both assuming only $\delta\kappa$ effects (pale colour between dotted lines) and  profiling over the other nine parameters (strong colour between solid lines). Since the lines are not very separated, we can understand that constraints are mostly  dominated by statistics. In the case of orange bands, we do not include the differential information and we show only the case in which only $\delta\klambda$ effects are present. 
As can we see form Fig.~\ref{fig:hllhcchi2}, including the nine EFT parameters, the constraints on the trilinear coupling are weaker due to correlations. The strongest effects are due to the correlations between $\delta y_t$ and  $c_{gg}$, and also between $\delta y_b$ and $\delta c_z$. On the other hand, the differential information  partially removes flat directions. In the right plot we compare and combine the constraints, including differential observables (blue), with those achievable via double Higgs production, according to Ref.~\cite{Azatov:2015oxa} (red). Their combination is depicted in pink. Allowing non-negligible effects from all the nine EFT parameters, double Higgs is leading to much stronger constraints. Nevertheless, single Higgs data are expected to be relevant and help in lifting the degenerate minima around $\delta \kappa_\lambda\sim 5$.

{
In conclusion, the indirect bounds on the Higgs self-coupling arising from single Higgs production are competitive to those obtained from double Higgs measurements in the case of exclusive $\kl$ variations, and while they become weaker in a fit that includes all the relevant EFT operators, they can still help to improve the bounds obtained from double Higgs production if differential information is included.
At the LHC both direct and indirect constraints on $\kl$ have been independently derived, mostly by considering only $\kl$ variations.
In some cases also a restricted set of additional operators was considered, though still not including potentially large effects from others  (e.g. simultaneous $\kl$-$\delta y_t$ fits without including $c_{gg}$, which can produce effects of a similar size as $\delta y_t$ in gluon fusion).
The natural step forward is the simultaneous fit of both direct and indirect constraints, and the gradual inclusion of the relevant EFT operators in the analysis, which becomes even more important as the experimental sensitivity increases.
}

%% file: EFT/HH_shape_benchmarks.tex
\label{sec:shape_bench}

The differential distributions for the non-resonant \hhAlt signal depend critically on the Higgs boson anomalous couplings to the SM particles. This happens in all the di-Higgs production modes, however due the quantity of possible free parameters
and a cancellation between some of the diagrams contributing to the process, the effect is stronger in the ggF production mode, the dominant production process of Higgs boson pairs in the most reasonable parts of the EFT parameter space.  The dependence of the signal on $\mhhAlt$ 
can vary as much as being localised around $\mhhAlt =$ 250~GeV, to contain dips around $\mhhAlt =$ 400~GeV and/or to contain a non-negligible tail of events that could extend up to 800~GeV or even beyond 1~TeV, when Higgs anomalous couplings are allowed to vary on a theoretically reasonable range (see for instance \reffig{fig:mhh_eft}). 

In order to construct an analysis aiming to find new physics effects, it is very useful to have a finite set of benchmarks that cover the most typical kinematic scenarios for the signal, especially if we expect a small signal rate on top of a sizeable background. On the experimental side, benchmarks are used for very practical reasons: they define a finite number of simulations to be done with optimal coverage of signal possibilities; those simulations are primary used to check the sanity of the data analysis on different phase space regions, and eventually for specific selections, optimization and/or design of subcategories.  In a scenario where small variations of continuous parameters lead to non-negligible changes on signal shapes, it is not obvious how to construct a finite set of benchmarks that would be comprehensive on most of the possible signal shapes based solely on theoretical  principles. To maximise the potential of an early LHC discovery to anomalous di-Higgs production, it seems natural to define the benchmarks based on kinematic features. As by construction only kinematic features are used to define the benchmark points, those are referred as {\it shape benchmarks}~\cite{Carvalho:2015ttv}.

To define the shape benchmarks, a Monte Carlo is used to simulate the possible signal shapes on a large portion of the theory parameter space. 
At LO in ggF, the di-Higgs system can be fully described with two kinematic variables: $\mhhAlt$  and the angle between one of the bosons and the beam pipe measured in the di-Higgs centre of mass reference frame ($\cos\theta^*$). A large Monte Carlo sampling  (1507 samples) populating the parameter space of Higgs anomalous couplings with a range of variations slightly larger than the reasonable theory and experimental limits, provided a rich sampling of possible distributions on the $(\mhhAlt , \cos\theta^*_{\rm HH})$ plane. The parameters used for this scan are the ones described in \refeq{eq:ewchl}. In the scan $c_{hhh}$ is allowed to vary between -15 and 15, $c_t$ between 0.5 and 2.5, while $c_{ggh}$ and $2c_{gghh}$ range between -1 and 1 and $c_{tt}$ between -3 and 3. A more detailed description of the input grid can be found in Ref.~~\cite{Carvalho:2015ttv} (note the different normalisation of the EFT parameters).

\begin{figure}
  \begin{center}
  \includegraphics[width=0.99\textwidth]{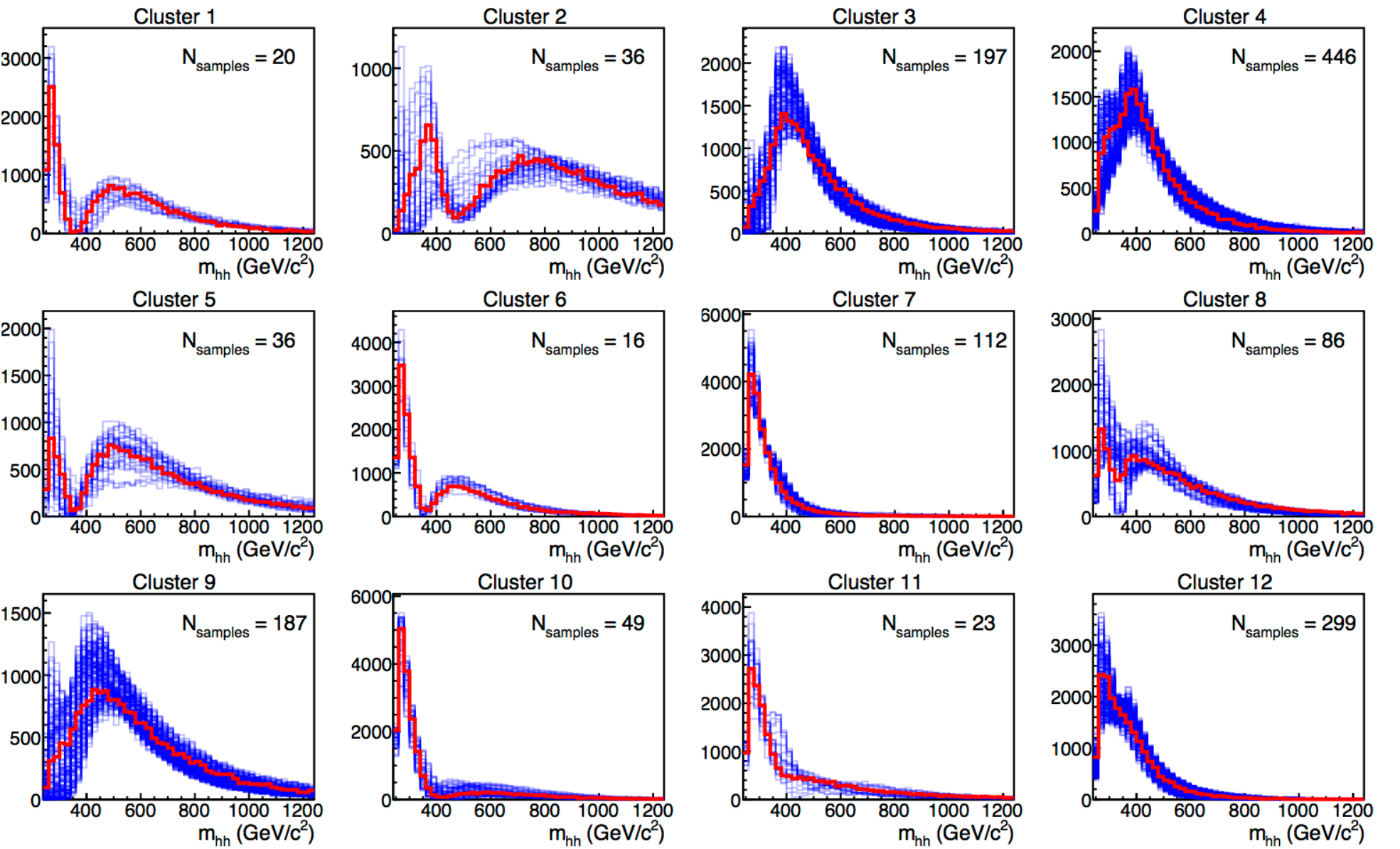}
  \caption{Generation-level distributions for the di-Higgs invariant mass \mhh. The red lines correspond to the benchmark of each cluster, while the blue lines describe the other members of each cluster~\cite{Carvalho:2015ttv}.}\label{fig:EFTbenchmhh}
  \end{center}
\end{figure}

A statistical Two-Sample test (TS-test) based on binned distributions on the $\mhhAlt$ and $\cos\theta^*$ variables is then used as an order parameter to group the large input sample on a smaller set of clusters, such that on each of these clusters the members are the most similar between themselves. The shape benchmark is defined as the element most similar to all the other samples of the cluster (according to the TS-test).
%
The final number of clusters, $N_\text{clus}$, and therefore the number of shape benchmarks, was chosen such that a reasonable trade-off between homogeneity and multiplicity of the clusters is achieved. To this end, the value $N_\text{clus}=12$ was found to be optimal: one cluster less would result on a too heterogeneous cluster, while one more would define a redundant subset of shape benchmarks.
The values of the EFT coefficients for each of the benchmarks are listed in Table~\ref{tab:EFTpnt},
and the \mhh distribution for each of the clusters (and benchmarks) is presented in Fig.~\ref{fig:EFTbenchmhh}.

A large variability in the kinematic topologies is related to the local minima of the total cross sections (where the largest cancellations among the different contributions occur). As a result of this connection, the points in a given cluster are usually distributed in a couple of simply connected regions of couplings (see Figs.~7-9 in Ref.~\cite{Carvalho:2015ttv}), demonstrating the robustness of the method on consistently separating regions of the theory parameter space. The same strategy can be applied to the other di-Higgs processes, as for example VBF, considering that more kinematic variables are necessary to feed the TS-test, as more variables are necessary to describe the process.

\begin{table}
\begin{center}
{
\renewcommand{\arraystretch}{1.2}
\begin{tabular}{ c | c  c  c  c  c }
\hline
Benchmark & $c_{hhh}$ & $c_t$ & $c_{tt}$ & $c_{ggh}$ & $c_{gghh}$ \\
\hline
1 & 7.5 & 1.0 & $-1.0$ & 0.0 & 0.0 \\
\hline
2 & 1.0 & 1.0 & 0.5 & $-\frac{1.6}{3}$ & $-0.2$ \\
\hline
3 & 1.0 & 1.0 & $-1.5$ & 0.0 & $\frac{0.8}{3}$  \\
\hline
4 & $-3.5$ & 1.5 & $-3.0$ & 0.0 & 0.0 \\ 
\hline
5 & 1.0 & 1.0 & 0.0 &   $\frac{1.6}{3}$ & $\frac{1.0}{3}$\\
\hline
6 & 2.4 & 1.0 & 0.0 & $\frac{0.4}{3}$ & $\frac{0.2}{3}$  \\
\hline
7 & 5.0 & 1.0 & 0.0 & $\frac{0.4}{3}$ & $\frac{0.2}{3}$  \\
\hline
8 & 15.0 & 1.0 & 0.0 & $-\frac{2.0}{3}$ & $-\frac{1.0}{3}$\\
\hline
9 & 1.0 & 1.0 & 1.0 &  $-0.4$ &  $-0.2$ \\
\hline
10 & 10.0 & 1.5 & $-1.0$ & 0.0 & 0.0 \\
\hline
11 & 2.4 & 1.0 & 0.0 & $\frac{2.0}{3}$ & $\frac{1.0}{3}$ \\
\hline
12 & 15.0 & 1.0 & 1.0 & 0.0 & 0.0 \\
\hline
SM & 1.0 & 1.0 & 0.0 & 0.0 & 0.0 \\
\hline
\end{tabular}
}
\end{center}
\vspace*{-0.5cm}
\caption{ Parameter values of the twelve benchmarks~\cite{Carvalho:2015ttv}. The SM reference is also shown. \label{tab:EFTpnt}}
\end{table}

The comparison of the constraints obtained by a given experimental analysis on each of these shape benchmarks can provide a useful insight, since they summarise the typical distortions on the signal distributions, therefore allowing to understand to which portion of the phase space each analysis is more/less sensitive. CMS results interpreted in terms of these shape benchmarks are presented in Sec.~\ref{sec_exp_combination}; the observed upper limits on the di-Higgs production cross section can vary up to two orders of magnitude across different shape benchmarks.
 It is important to highlight, however, that the usage of shape benchmarks and its presence on a set of final results does not substitute the need for other ways of producing and  presenting results.
If we want to have precise limits on Higgs anomalous couplings, a possible approach is to propose sets of 1D and 2D scans to be directly produced by the collaborations.
%
However, it is not practical to generate large grids only for interpretation purposes. In this sense, to have the MC generated on terms of shape benchmarks can be useful as basis for MC re-weighting, as they by construction contain events that populate all parts of the possible phase space~\cite{Carvalho:2017vnu}.

Even with a re-weighting method that allows to produce results in several kinds of parameter scans without large computing resources, it is not possible for the experiments to cover with a finite set of scans all the dimensionality of the EFT considering any possible correlation between anomalous couplings of all possible EFT UV completions .
The design of a format for the results that allows an easy reinterpretation is imperative for the long-term usage of the huge experimental work that is invested on di-Higgs searches. Some ideas are discussed in Sec.~\ref{chap:res_pres}.
As long as such a method is not defined and implemented,
an alternative for a first coarse estimation of the effects of anomalous couplings in specific portions of the EFT is to use the TS-test to find which is the shape benchmark that these specific points are more similar to. The limit for this point can be estimated to be equal to the one of the most similar shape benchmark. An exercise of such procedure on the Run 2 $HH\to b\bar{b}\gamma\gamma$ ATLAS and CMS analyses can be found in Ref.~\cite{Carvalho:2017vnu}. As the mapping between a new investigated point and the most similar shape benchmark is based only on kinematic information, this technique also allows to estimate the limits beyond the EFT domain, as for example the case of interference with resonances when a localised peak is not obvious.

The shape benchmarks were also used to study the typical shape modifications that can occur at NLO level, including the full top-quark mass dependence \cite{Buchalla:2018yce}. While the main qualitative features of the distributions (the position of peaks, dips, and the presence or not of sizeable high mass tails) are mostly unchanged from LO to NLO, it was found that the NLO corrections are important for a correct experimental assessment of the Higgs anomalous couplings, with K-factors that can present large variations across the $\mhhAlt$ range, 
and whose size also depends considerably on the value of the anomalous couplings (see e.g. Fig.~\ref{fig:Kfacvariation} for the case of $c_{hhh}$).
Based on these results, it is clear that an extension of the shape benchmarks to NLO is desirable. However, having in mind that the main goal of these benchmarks is to provide a first assessment of the sensitivity of a given analysis to shape modifications and their connection to the EFT parameter space, and not to provide a precise description that includes all the richness present in a full EFT scan, their present LO formulation
might be sufficiently accurate for their purpose, though NLO studies (in the lines of Ref.~\cite{Buchalla:2018yce}) to confirm this statement are in order.

%% file: BSMresonance/BSMmain.tex
\chapter{New Physics in Higgs pair production}\label{chap:BSM}
\textbf{Editors: R.~Gr{\"o}ber, I.~M.~Lewis, Z.~Liu}
\vspace{2mm}

\noindent
While model-independent approaches in effective field theory are usually applicable
for heavy new physics, spectacular new physics signatures can show up in Higgs boson pair production
in the presence of new degrees of freedom, with masses below the validity 
range of an effective field theory. 
Strongly enhanced cross sections for Higgs pair production are typical in the presence of
a light new resonance decaying to a Higgs boson pair. This is a common
feature of models with extended scalar sectors with sizeable couplings of the new resonance to
a pair of Higgs bosons.

The simplest extension providing such a new
scalar resonance is the SM augmented with a new scalar that is a singlet under the SM gauge groups, 
which we discuss in Sec.~\ref{sec:BSMspin0}. We put a particular emphasis
on the interference effects with the background of the box diagrams
and the triangle diagram with SM Higgs boson exchange in Sec.~\ref{sec:inteference}. 
In Sec.~\ref{sec:BSMmodels}, we turn to various other models with extended Higgs sectors,
the complex two-Higgs doublet model (C2HDM), the singlet extension 
of the 2HDM, the next-to-minimal supersymmetric extension of the SM (NMSSM) 
and the Georgi-Machacek model. In the context of these models, we provide benchmarks
for resonant production of a SM-like Higgs boson pair and
final states with different Higgs bosons. We will shortly comment on spin-2 resonances decaying to 
a Higgs pair in Sec.~\ref{sec:spin2}.

The Higgs boson pair production cross section can also be modified by the presence of new colored 
particles in the gluon-induced loop. Prime examples are scalar particles, as for instance
top squarks in supersymmetry \cite{BarrientosBendezu:2001di,Batell:2015koa, Agostini:2016vze}, or new vector-like fermions, as they would appear for 
instance in Composite Higgs Models \cite{Gillioz:2012se, Grober:2016wmf}. Double Higgs production allows to break the degeneracy present in
single Higgs production between a shift in the top Yukawa coupling and new physics 
in the gluon fusion loop \cite{Gillioz:2012se, Azatov:2015oxa, Azatov:2016xik}. We will discuss the impact of new particles in the 
loop in Sec.~\ref{sec:particleinloop}.

In Sec.~\ref{sec:cosmology}, we address the impact of a measurement of the Higgs pair production 
cross section on cosmology. Since in successful models of electroweak baryogenesis
a deformed potential with respect to the SM one is required, 
a measurement of the trilinear Higgs self-coupling has a direct impact on 
possible explanations of the matter-antimatter asymmetry of the universe.
Finally, in Sec.~\ref{sec:DM}, we will show that searches for final states with Higgs pairs and missing energy
have the potential to uncover a dark sector that could provide a dark matter candidate.

\input{BSMresonance/BSMspin0.tex}
\input{BSMresonance/BSMinterference.tex}

\input{BSMresonance/BSMspin2.tex}
\section{BSM Models and Benchmarks \label{sec:BSMmodels}}
\input{BSMresonance/BSMC2HDM.tex}

\input{BSMresonance/BSM1S2HDM.tex}

\input{BSMresonance/BSMhMSSM.tex}
\input{BSMresonance/BSMNMSSM.tex}

\input{BSMresonance/BSMGM.tex}

\input{BSMresonance/BSMparticleinloop.tex}

\input{BSMresonance/BSMMSSM.tex}

\input{BSMresonance/BSMGWCosmo.tex}

\input{BSMresonance/BSMDarkMatter.tex}

\input{BSMresonance/BSMSummary.tex}

%% file: BSMresonance/BSMspin0.tex
\section[Spin-0 models]{Spin-0 models \\
\contrib{S.~Dawson, I.~M.~Lewis, T.~Robens, T.~Stefaniak, M.~Sullivan}
}
\label{sec:BSMspin0}

Resonant double Higgs production is one of the most spectacular signatures to look for in Higgs physics.  The simplest extension of the SM, the addition of a real gauge singlet scalar~\cite{Davoudiasl:2004be,Schabinger:2005ei,Patt:2006fw, Barger:2007im, Bowen:2007ia}, can result in resonant double Higgs production~\cite{Bowen:2007ia,Dolan:2012ac,Pruna:2013bma,Cooper:2013kia,No:2013wsa,Chen:2014ask,Martin-Lozano:2015dja,Dawson:2015haa,Godunov:2015nea,Lu:2015qqa,Robens:2015gla,Bojarski:2015kra,Robens:2016xkb,Nakamura:2017irk,Huang:2017jws,Chang:2017niy,Lewis:2017dme,Dawson:2017jja,Alves:2018jsw,
Alves:2018oct,Carena:2018vpt}.
The most general renormalizable scalar potential can be expressed (using the parametrization of Ref.~\cite{Chen:2014ask}) in the following way~{\cite{OConnell:2006rsp,Barger:2007im}} 
\begin{eqnarray}
V(\Phi,S)&=&-\mu^2\Phi^\dagger \Phi +\lambda\left(\Phi^\dagger\Phi\right)^2+\frac{a_1}{2}\Phi^\dagger\Phi S+\frac{a_2}{2}\Phi^\dagger\Phi S^2\nonumber\\
&&+b_1 S+\frac{b_2}{2}S^2+\frac{b_3}{3}S^3+\frac{b_4}{4}S^4,\label{eq:SingletPot}
\end{eqnarray}
where $S=(v_S+s)/\sqrt{2}$ is a gauge singlet scalar, $\Phi=\left(0,v+h\right)^{\rm T}/\sqrt{2}$ is the Higgs doublet, $v_S$ is the $S$ vacuum expectation value (vev), $v$ is the Higgs vev, $h$ is the SM Higgs boson, and $s$ is a new scalar boson. At the renormalizable level, \refeq{eq:SingletPot} contains {all possible} interactions between $S$ and the SM particles.\footnote{At dimension-5 in an effective field theory, $S$ can have additional couplings to SM particles~\cite{Dawson:2016ugw,Bauer:2016hcu,Chala:2017sjk} and has a qualitatively different phenomenology, which we neglect here.}  After electroweak symmetry breaking, $S$ and $h$ mix{, resulting in} two mass eigenstates $h_{1,2}$ with masses {$m_{1,2}$, where} {by definition} {$m_2\,\geq\,m_1$}. Here we concentrate on the case that $m_1=125$~GeV.\footnote{The case $m_2\,=\,125\,{\rm GeV}$ is also {viable, see e.g.~Refs.}~\cite{Robens:2015gla,Robens:2016xkb,Ilnicka:2018def}.} The $h-s$ mixing angle $\theta$ is defined as
\begin{eqnarray}
\begin{pmatrix} h_1\\ h_2\end{pmatrix}=\begin{pmatrix} \cos\theta & \sin\theta \\ -\sin\theta & \cos\theta \end{pmatrix}\begin{pmatrix}h\\ s\end{pmatrix}
\end{eqnarray}
{Due to this mixing, the couplings of $h_1$ ($h_2$) to SM fermions and gauge bosons are universally suppressed by $\cos\theta$ ($\sin\theta$), relative to the SM Higgs couplings}.
Hence, the production cross section for $h_2$ is given by the SM Higgs production rate at a mass of $m_2$ suppressed by $\sin^2\theta$, and the observed Higgs boson $h_1$ rates are suppressed by $\cos^2\theta$ relative to the SM. 
 
 If $m_2>2\,m_1$, \refeq{eq:SingletPot} allows for on-shell $h_2\rightarrow h_1h_1$ decays.  
The branching ratios of the decays of the heavy {scalar $h_2$ to a Higgs boson pair, $h_1 h_1$, and to final states with SM particles (collectively denoted by `SM')} are given by
\begin{eqnarray}
\text{BR}_{h_2\rightarrow h_1 h_1}&=&\frac{\Gamma_{h_2\rightarrow h_1 h_1}}{\sin^2\theta\,\Gamma_\text{SM, tot}+\Gamma_{h_2\rightarrow h_1h_1}},\;
\text{BR}_{h_2\rightarrow \text{SM}}\,=\, \frac{\sin^2\theta\,\Gamma_{\text{SM}, h_2\rightarrow\text{SM}}}{\sin^2\theta\,\Gamma_\text{SM, tot}+\Gamma_{h_2\rightarrow h_1h_1}},\nonumber\\
&&\label{eq:brdefs}
\end{eqnarray}
{respectively,} where $\Gamma_{h_2\rightarrow h_1 h_1}$ is the partial width of the $h_2\rightarrow h_1h_1$ decay, $\Gamma_{\text{SM},\,h_2\rightarrow\text{SM}}$ is the partial width of the SM Higgs boson at mass $m_2$ {decaying to a SM particle final state}, and $\Gamma_\text{SM, tot}$ denotes the total width of the SM Higgs boson with mass $m_2$.\footnote{Electroweak higher-order corrections to the $h_2\,\rightarrow\,h_1 h_1$ decay width have e.g.~been presented in Ref.~\cite{Bojarski:2015kra} and can amount to up to $10\%$. We neglect these effects in the remainder of our discussion.}

{Imposing further symmetries decreases the number of free parameters.  For example, a $Z_2$ symmetry with the transformation properties $S\rightarrow-S,\;{\Phi \rightarrow \Phi},\;\text{SM}\rightarrow\text{SM}$ requires that $a_1=b_1=b_3=0$}. {If $S$ acquires a vev, $v_S\ne 0$, the $Z_2$ symmetry becomes softly broken.} {We will discuss both the non-$Z_2$ and softly broken $Z_2$ case below.}

\subsubsection{Constraints}
\label{Sec:constraints}

{Detailed discussions of experimental and theoretical constraints on the model can be found in Refs.~\cite{Pruna:2013bma,Robens:2015gla,Robens:2016xkb,Ilnicka:2018def}. Here we briefly summarize these constraints, which are obeyed by the benchmark scenarios proposed here, and refer the reader to the literature for more details.}

{The theoretical constraints that we consider are vacuum stability (both at the low and high scale, $\mu \sim 10^{10}\,{\rm GeV}$),
perturbative unitarity, as well as 
perturbativity of the couplings in the scalar potential (at the low and high scale, $\mu \sim 10^{10}\,{\rm GeV}$). 
The experimental constraints are the
agreement with electroweak precision observables~\cite{Baak:2014ora}, 
with the observed $W$ boson mass~\cite{Schael:2013ita,Aaltonen:2013vwa,Aaboud:2017svj,Tanabashi:2018oca}, $M_W = 80.379 \pm 0.012~{\rm GeV}$ following Ref.~\cite{Lopez-Val:2014jva},
with null-results from LHC Higgs searches (using  \texttt{HiggsBounds}, version 5.4.0beta~\cite{Bechtle:2008jh,Bechtle:2011sb,Bechtle:2013gu,Bechtle:2013wla,Bechtle:2015pma}),
and with Run-1 and Run-2 Higgs boson rate measurements (using \texttt{HiggsSignals}, version 2.2.3beta~\cite{Bechtle:2013xfa}).}\footnote{{With respect to the most recent literature~\cite{Ilnicka:2018def}, this work contains updated LHC Higgs search limits, in particular for $h_2\,\rightarrow\,h_1\,h_1$ signatures~\cite{Aaboud:2018ewm,Aaboud:2018sfw,Aaboud:2018bun,Sirunyan:2018two}, updated Higgs boson signal rate measurements from LHC Run~2, as well as an updated $W$ boson mass value.}}

\subsubsection{$Z_2$}

In the {softly-broken} $Z_2$-symmetric scenario, the  {scalar sector is described by five parameters after electroweak symmetry breaking}, namely, $m_{1}$, $m_2$, $v$, $\sin\theta$, and $\tan\beta \equiv\tfrac{v}{v_s}$. {Two of these parameters, $v \approx 246~{\rm GeV}$ and $m_1 \approx 125~{\rm GeV}$, are fixed by experimental measurements, leaving only three free model parameters.} {The analytic expression for the partial decay width for $h_2 \rightarrow h_1 h_1$ at leading order can be found in Refs.~\cite{Schabinger:2005ei, Bowen:2007ia, Robens:2015gla, Robens:2016xkb}.}
Note that the specific choice of $\tan\beta\,=\,\cot\theta$ leads to $\Gamma_{h_2\rightarrow h_1 h_1}\,=\,0$.

\begin{figure}[ht!]
\begin{center}
\includegraphics[width=0.7\textwidth]{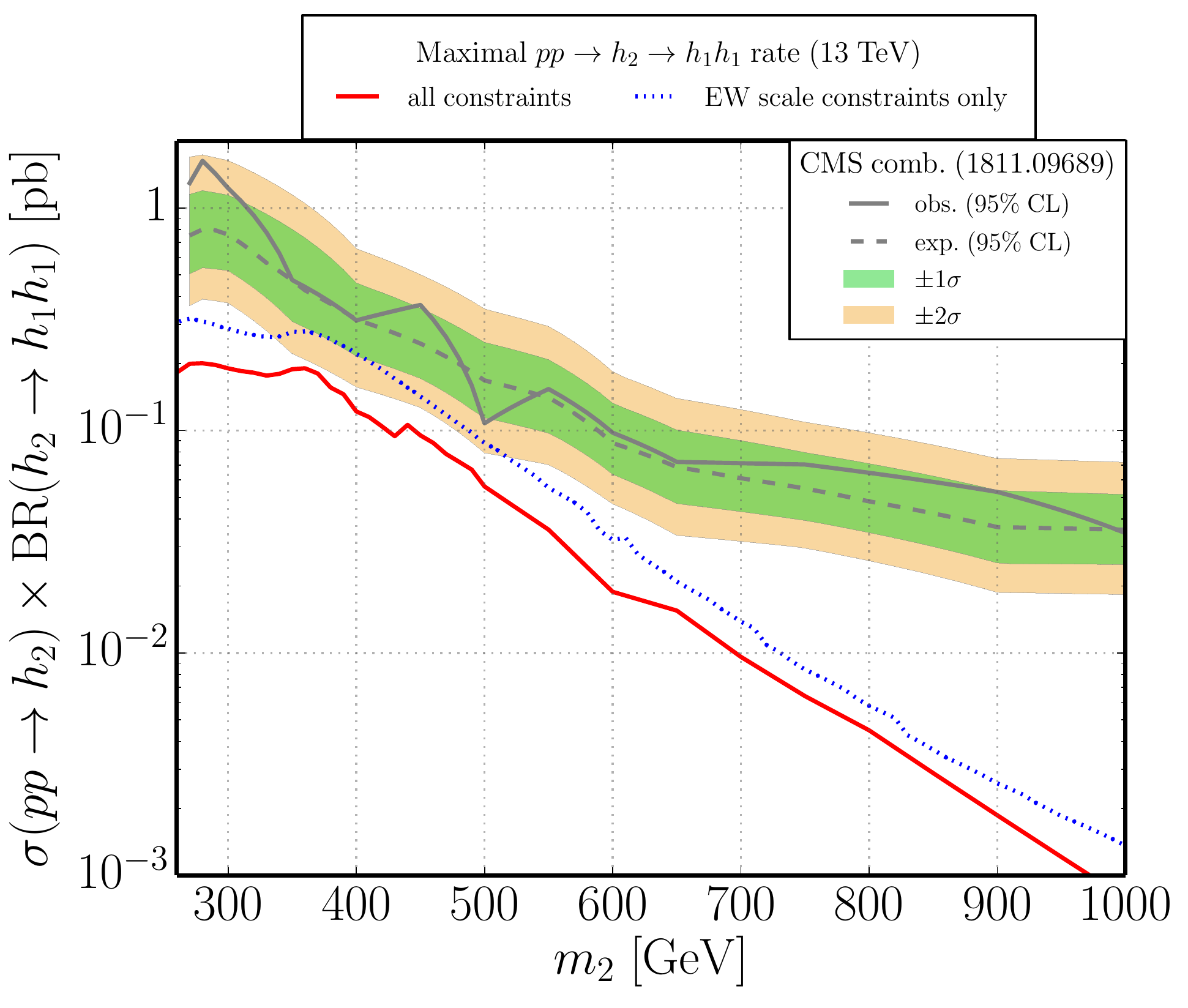}
\caption{\label{fig:xshigh_hh} {Maximal allowed $pp \to h_2 \to h_1h_1$ signal rate at the 13 {\rm TeV}~LHC in the softly-broken $Z_2$-symmetric case. Shown are values after applying (red solid) all constraints and (blue dotted) only constraints at the EW scale. The corresponding $\text{BR}_{\text{max}}^{h_2\to h_1h_1}$ values are given in \refta{tab:highm2}. For comparison we include the current strongest cross section limit (at $95\%~\mathrm{CL}$), obtained from the combination of various CMS $h_2\to h_1h_1$ searches at $13~\mathrm{TeV}$ with up to $36~\mathrm{fb}^{-1}$ of data~\cite{Sirunyan:2018two}. }}
\end{center}
\end{figure}

Given the constraints in Sec.~\ref{Sec:constraints}, \refta{tab:highm2} lists the allowed values of $\sin\theta$ and $\mathrm{BR}_{h_2\to h_1h_1}$ for scenarios with $m_2\geq 2m_1$. The maximal allowed signal rate (i.e.~production cross section times branching ratio) for $pp\to h_2\to h_1h_1$ at the $13~{\rm TeV}$ LHC is shown in \reffig{fig:xshigh_hh}, {in direct comparison with the current strongest upper cross section limit from the CMS combination of $h_2\to h_1h_1$ searches~\cite{Sirunyan:2018two}}.\footnote{We rescaled the NNLO+NNLL gluon fusion cross section of the SM Higgs boson~\cite{deFlorian:2016spz} by $\sin^2\theta$.} {Both \refta{tab:highm2} and \reffig{fig:xshigh_hh} present maximal $\mathrm{BR}_{h_2\to h_1h_1}$ values after applying all constraints, as well as after applying EW-scale constraints only. The latter includes tests of vacuum stability and perturbativity at the EW-scale, but does not require perturbitivity and vacuum stability at a higher scale $\mu\sim 10^{10}~\mathrm{GeV}$.}

\begin{table}[ht!]
\begin{center}
\begin{tabular}{|c|c|c|c|c|}
\hline
$m_2 [{\rm GeV}]$&$|\sin\theta|_\text{max}$&$\text{BR}^{h_2
\rightarrow h_1 h_1}_\text{min}$&$\text{BR}^{h_2
\rightarrow h_1 h_1}_\text{max}$ & 
$\text{BR}^{h_2 \rightarrow h_1 h_1}_\text{max}$ \\ 
 & & & (\emph{all constraints})  & (\emph{EW-scale constraints}) \\
\hline
255 &0.22	& 0.13 &0.26  &0.47 \\
260&	0.22 & 0.17 & 0.32 & 0.54\\
265&	0.22&0.20  &0.35 &0.57 \\
280&	0.22& 0.23&0.39 &0.60\\
290&	0.22&0.24&0.40 &0.61\\ 
305&	0.22&0.25&0.40 &0.60\\
325&	0.22& 0.26 & 0.40 & 0.58\\
345&	0.22&0.26&0.39 &0.56\\
365&	0.22& 0.24 & 0.36 & 0.53\\
395&	0.20& 0.23 &0.33 &0.49\\
430&0.20&0.23&0.30 &0.45\\
470&0.22&0.21 & 0.28 & 0.42 \\
520&0.21&0.21 & 0.27 & 0.39 \\
590&0.20& 0.22 &0.26 &0.36 \\
665&0.21& 0.21& {0.26} & {0.35} \\
770&0.20& 0.22&0.25 &0.33\\
875&0.19&0.22&0.25 &0.31\\
920&0.18&0.23&0.25 &0.31\\
975&0.17&0.23&0.25 &0.31\\
1000&0.16&0.23&0.25 &0.31\\
\hline
\end{tabular}
\end{center}
\vspace*{-0.3cm}
\caption{\label{tab:highm2} {Maximal and minimal allowed branching ratios of the decay $h_2
\rightarrow h_1 h_1$, evaluated at the maximal allowed value of $|\sin\theta|$. Note that minimal values for the $\text{BR}(h_2
\rightarrow h_1 h_1)$ stem from $\sin\theta\,\geq\,0$. For the maximal $\text{BR}(h_2
\rightarrow h_1 h_1)$ we give the values obtained after applying all constraints as well as after applying only EW-scale constraints, i.e.~requiring perturbative couplings and vacuum stability at the EW scale but not up to a high scale $\mu\sim 10^{10}~\mathrm{GeV}$.
The numbers supersede those presented in Table V of Ref.~\cite{Robens:2016xkb}.}}
\end{table}

\subsubsection{Non-$Z_2$}
In the non-$Z_2$ limit, all parameters in \refeq{eq:SingletPot} are allowed.  Since there is no symmetry associated with the scalar $S$, its vev is non-physical and we are allowed to set it to zero: $v_S=0$~\cite{Chen:2014ask,Lewis:2017dme}.  There are now five physical parameters: the Higgs doublet vev $v=246$~GeV, the scalar singlet vev $v_S=0$, the observed Higgs boson mass $m_1=125$~GeV, the heavy scalar mass $m_2$ assumed to be $m_2>2\,m_1$, and the $h-s$ mixing angle $\theta$.  Hence, 5 of the potential parameters $\mu^2,b_1,a_1,b_2,\lambda$ can be solved for~\cite{Chen:2014ask,Lewis:2017dme}: 
\begin{gather}
\mu^2=\lambda\,v^2,\quad b_1=-\frac{v^2}{4}a_1,\quad a_1=\sqrt{2}\frac{m_1^2-m_2^2}{v}\sin\,2\,\theta\nonumber\\
b_2=2\,m_1^2\sin^2\theta+2\,m_2^2\cos^2\theta-\frac{a_2}{2}v^2,\quad \lambda=\frac{m_1^2\cos^2\theta+m_2^2\sin^2\theta}{2\,v^2}.
\end{gather}
This leaves the additional potential parameters $a_2,b_3,b_4$ free. The free parameters of the model are then:
\begin{eqnarray}
m_1=125~{\rm GeV},\,m_2,\,v=246~{\rm GeV},\,v_S=0,\,\theta,\,a_2,\,b_3,\,b_4\,.
\end{eqnarray}

 This situation differs from the $Z_2$ limit, where all potential parameters can be solved for in terms of masses, vevs, and the mixing angle.  This additonal freedom leads to a more complex vacuum structure.   Indeed, with the additional freedom in the non-$Z_2$ model, there are six potential extrema of the potential in \refeq{eq:SingletPot} with Higgs vevs that are not $246$~GeV.  The new scalar is a gauge singlet and its vev cannot contribute to the $W$ and $Z$ masses.  Hence, to get the observed electroweak symmetry breaking pattern, we demand that $(v,v_S)=(246~{\rm GeV},0)$ is the global minimum.  This puts stringent constraints on the potential parameters $a_2$ and $b_3$ which contribute to the $h_2-h_1-h_1$ coupling relevant for $h_2\rightarrow h_1h_1$ decays:
\begin{eqnarray}
V(h_1,h_2)& \supset & \frac{\lambda_{211}}{2}h_2 h_1^2\nonumber\\
\lambda_{211}&=&\frac{b_3}{\sqrt{2}}\,\sin^2\theta\cos\theta+\frac{a_1}{2\,\sqrt{2}}\cos\theta\left(\cos^2\theta-2\,\sin^2\theta\right)\nonumber\\
&&+\frac{a_2}{2}\,v\,\sin\theta\,(2\,\cos^2\theta-\sin^2\theta)-6\,\lambda\,v\,\sin\theta\,\cos^2\theta.
\end{eqnarray}
This limits how large the $h_2\rightarrow h_1h_1$ branching ratios and $pp\rightarrow h_1h_1$ production cross section can be~\cite{Chen:2014ask,Lewis:2017dme}.

\begin{figure}
\begin{center}
\includegraphics[width=0.46\textwidth]{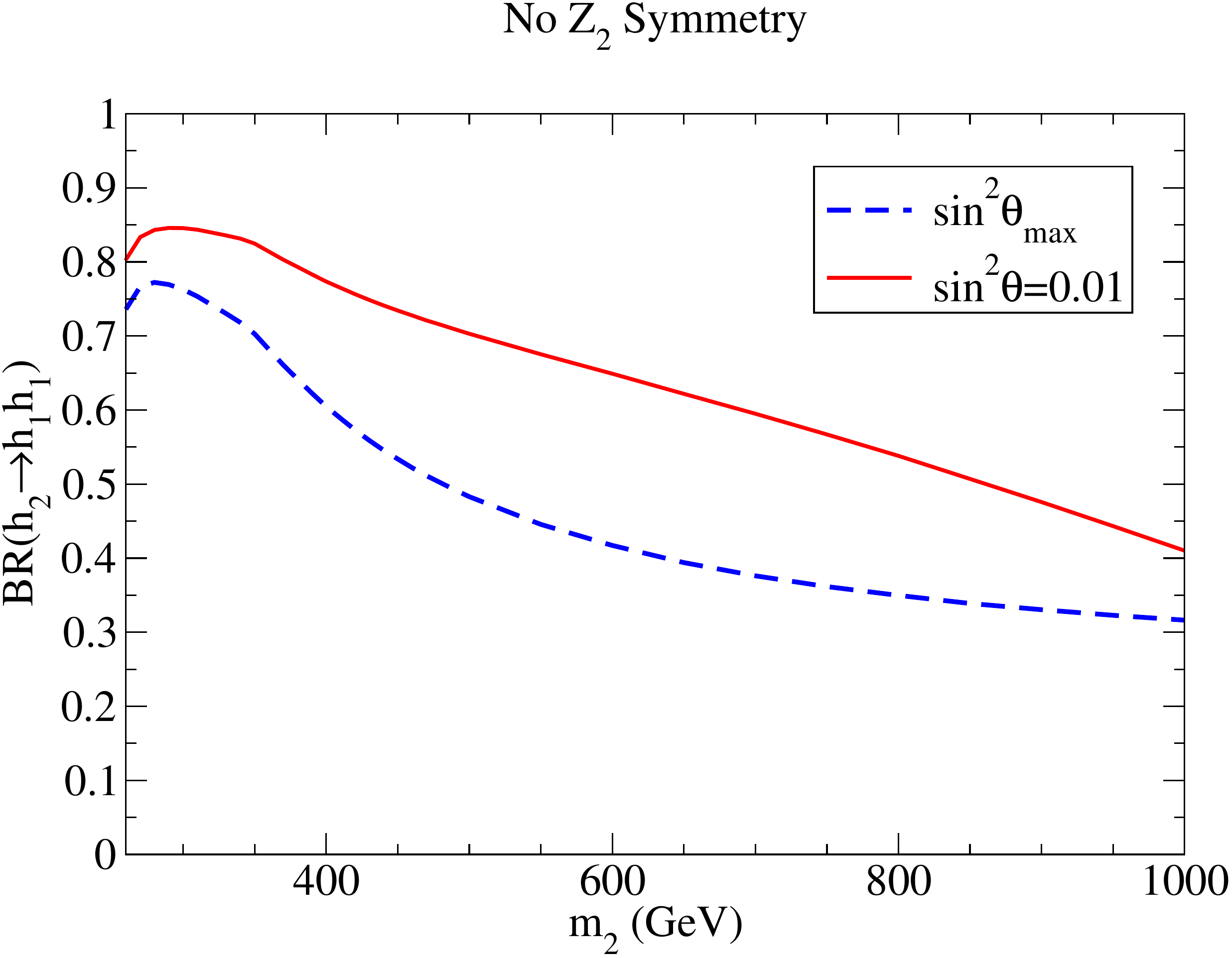}
\hspace*{0.2cm}
\includegraphics[width=0.46\textwidth]{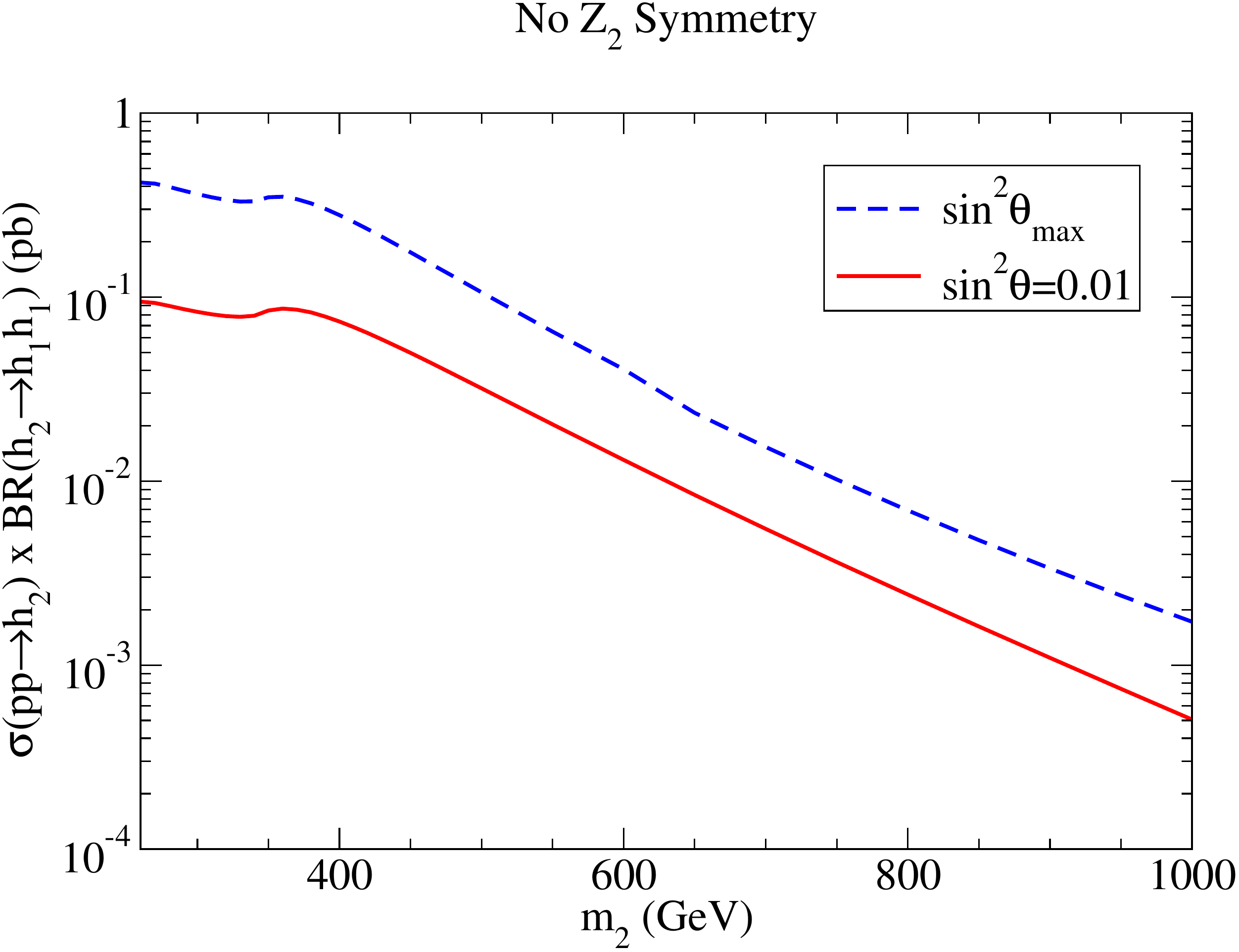}
\end{center}
\vspace{-0.3cm}
\caption{Maximum (left) branching ratios for $h_2\rightarrow h_1h_1$ and (right) resonant double Higgs production normalized to the SM rate in the non-$Z_2$ singlet model for various $h-s$ mixing angles~\cite{Lewis:2017dme}.}
  \label{fig:SingletNonZ2}
\end{figure}

In \reffig{fig:SingletNonZ2} we show the largest $h_2\rightarrow h_1h_1$ branching ratios (left) and $pp\rightarrow h_1h_1$ production cross sections (right) allowed under the constraint the the global minimum correctly breaks electroweak symmetry~\cite{Lewis:2017dme}.  The $S^4$ potential parameter $b_4$ is set to the upper limit consistent with perturbative unitarity~\cite{Chen:2014ask,Lewis:2017dme}. {The lines $\sin\theta_{\text{max}}$ correspond to the maximal allowed mixing angle from Higgs precision measurements and $W$ mass measurements: $|\sin\theta|\lesssim 0.22$ for $250~{\rm GeV}\lesssim m_2 \lesssim 622~{\rm GeV}$ and $|\sin\theta|\lesssim 0.21$ for $m_2\gtrsim 622~{\rm GeV}$.  We do not take into account perturbative limits from RGE running up to a scale of $\sim4\times 10^{10}$ GeV.}  {The mixing angle, and hence $h_2$ production rate, for $\sin^2\theta=0.01$ is smaller than any mixing angle considered in \reffig{fig:xshigh_hh}.  Hence, even though BR($h_2\rightarrow h_1h_1$) is largest for $\sin^2\theta=0.01$, it is still allowed due to the suppressed production rates of $h_2$.}  Even with these constraints, the $h_2\rightarrow h_1h_1$ branching ratio can be above $80\%$, and the $pp\rightarrow h_1h_1$ production rate can be one order of magnitude larger than the SM prediction.

%% file: BSMresonance/BSMinterference.tex
\section{Interference Effects}
\label{sec:inteference}

It has been noted recently that the gluon-induced production of Higgs boson pairs via a heavy scalar resonance often has large and non-trivial interference effects with the continuum SM Higgs pair production
process, which can be considered in this case a background to the resonant BSM signal.
Furthermore, in realistic models, the interfering non-resonant processes are often modified as well with respect to the SM amplitude due to modifications to the Higgs couplings. In this section, we examine these subtle interference effects. First, we discuss the interference effects of the underlying scalar resonance. Second, we show the importance of the on-shell interference effect driven by the dynamical phase generated by the loop diagrams. At the end of this section, a general parameterisation of the effective interactions and a general picture of the overall interference effects are shown.

\subsection[Off-shell Interference]{Off-shell Interference \\
\contrib{I.~M.~Lewis}
}
\input{BSMresonance/BSMinterference_offshell.tex}
\subsection[On-shell Interference]{On-shell Interference \\
\contrib{M.~Carena, Z.~Liu, M.~Riembau}
}
\input{BSMresonance/BSMinterference_onshell.tex}
\subsection[Overall Interference]{Overall Interference \\
\contrib{E.~Bagnaschi, A.~Carvalho, R.~Gr\"ober, S.~Liebler, J.~Quevillon}
}
\input{BSMresonance/BSMinterference_overall.tex}

%% file: BSMresonance/BSMinterference_offshell.tex
In the narrow-width approximation (NWA), it is usually assumed that the interference effects near a resonance scale as $\Gamma/M$, where $\Gamma$ is the width of the resonance and $M$ its mass.  In the limit $\Gamma/M\ll1$, these effects are negligible.  Indeed, in the narrow width approximation only the resonance makes an important contribution to the process  (see the next section for when these arguments fail).  However, away from the resonance peak interference effects can be sizeable.  This is especially true for the Higgs boson and other scalars, where there are non-decoupling effects due to couplings being proportional to masses.  To illustrate this, we consider the $Z_2$ symmetric singlet model introduced in Sec.~\ref{sec:BSMspin0}.  

\begin{figure}[t]
\begin{center}
\includegraphics[width=0.46\textwidth,clip]{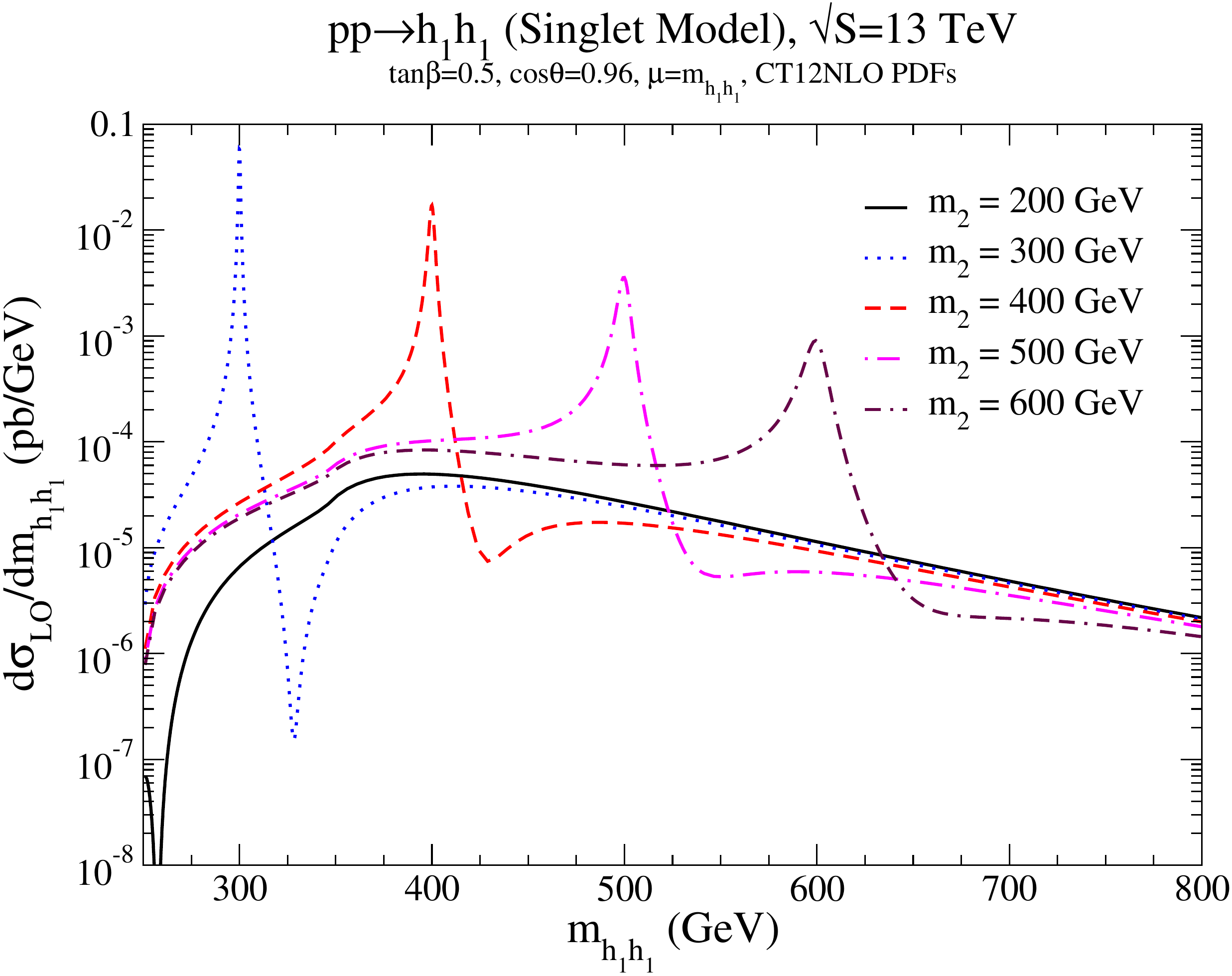}
\hspace*{0.2cm}
\includegraphics[width=0.46\textwidth,clip]{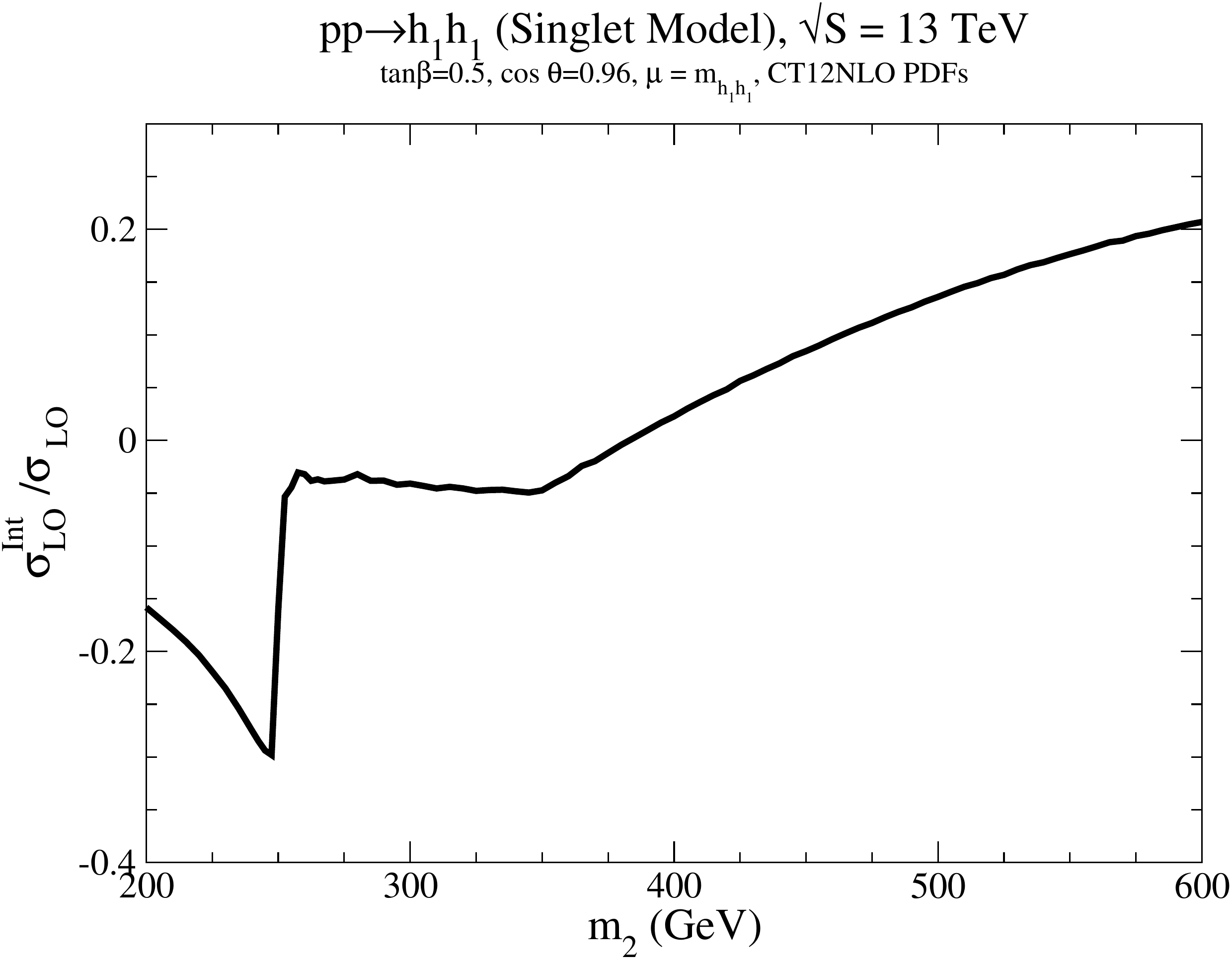}
\end{center}
\vspace{-0.3cm}
\caption{\label{fig:OffShellInt} Left: Invariant mass distributions of di-Higgs final state in the $Z_2$ symmetric singlet model for various singlet masses.  Right:  Fractional contribution of the interference of $h_2$ s-channel contribution with $h_1$ s-channel and box diagram contributions~Ref.~\cite{Dawson:2015haa}. }
\end{figure}

The di-Higgs invariant mass ($m_{h_1h_1}$) distributions for $pp\rightarrow h_1h_1$ and various heavy scalar masses are shown on the left of \reffig{fig:OffShellInt}.  We can observe that, if $m_{h_1 h_1}\ll m_{2}$, the invariant mass becomes independent of the resonance mass.  There are three contributions to $gg\rightarrow h_1h_1$: a top quark triangle with $s$-channel $h_1$, a top quark triangle with $s$-channel $h_2$, and a top quark box diagram.  The box diagram is independent of $h_2$, although there is a uniform suppression from the scalar coupling, as discussed in Sec.~\ref{sec:BSMspin0}.  However, both $s$-channel diagrams depend on the trilinear scalar couplings which are altered from SM predictions.

The trilinear scalar couplings are defined in the scalar potential as
\begin{eqnarray}
V(h_1,h_2)\supset \frac{\lambda_{111}}{3!}h_1^3+\frac{\lambda_{112}}{2}h_1 h_2^2,
\end{eqnarray}
and in the $Z_2$ singlet extension of the SM they are
\begin{eqnarray}
\lambda_{111}&=&\frac{3\,m_1^2}{v}\left(\cos^3\theta+\tan\beta\,\sin^3\theta\right)\,,\nonumber\\
\lambda_{112}&=&-\frac{m_2^2}{2\,v}\sin\,2\theta\left(\cos\theta-\tan\beta\,\sin\theta\right)\left(1+\frac{2\,m_1^2}{m_2^2}\right)\,.
\end{eqnarray}
The $s$-channel contribution to the leading order amplitude is then
\begin{eqnarray}
F_1^{\rm tri}=m_{h_1h_1}^2\left(\frac{\lambda_{111}\,v\,\cos\,\theta}{m_{h_1h_1}^2-m_1^2+i\,m_1\,\Gamma_{h_1}}-\frac{\lambda_{112}\,v\,\sin\theta}{m_{h_1h_1}^2-m_2^2+i\,m_2\,\Gamma_{h_2}}\right)F_{\Delta},\label{eq:Ftri}
\end{eqnarray}
where $\Gamma_{h_j}$ is the total width of $h_j$, $s$ is the center or momentum energy squared, and $F_{\Delta}$ is a form factor for the triangle top loop~\cite{Plehn:1996wb,Glover:1987nx} normalized according to Ref.~\cite{Dawson:2015haa}.
In the limit $m_1,m_{h_1h_1}\ll m_2$, \refeq{eq:Ftri} becomes
\begin{eqnarray}
F_1^{\rm tri}\xrightarrow[m_1,m_{h_1h_1}\ll m_2]{} &&m_{h_1h_1}^2\left(\frac{3\,m_1^2\,\left(\cos^3\theta+\tan\beta\,\sin^3\theta\right)\cos\,\theta}{m_{h_1h_1}^2-m_1^2+i\,m_1\,\Gamma_{h_1}}\right.\nonumber\\
&&\left.-\frac{1}{2}\sin\,2\theta\,\sin\theta\left(\cos\theta-\tan\beta\,\sin\theta\right)\right)F_{\Delta}.
\end{eqnarray}
Hence, the amplitude has no explicit dependence on $m_2$ and the distribution is independent of $m_2$ for $s\ll m_2^2$ as shown in the left of \reffig{fig:OffShellInt}~\cite{Dawson:2015haa}.

Since the di-Higgs invariant mass distribution is independent of $m_2$, the interference between the $s$-channel $h_2$ resonance and other contributions are independent of $h_2$ for $m_1,m_{h_1h_1}\ll m_2$.  As $m_2$ increases, the area of of the distribution satisfying $m_{h_1h_1}\ll m_2$ increases.  Hence, the size of the interference between the $h_2$ resonance and other contributions becomes increasingly large~\cite{Dawson:2015haa}.  To illustrate this effect, the fractional contribution of the interference of the $h_2$ s-channel with $h_1$ s-channel and box diagram contributions is shown on the right hand side of \reffig{fig:OffShellInt}.  The leading order interference cross section is labelled as $\sigma_\text{LO}^\text{Int}$ while the total cross section with all contributions is labelled as $\sigma_\text{LO}$.  As can be seen, the interference can contribute upwards of $20\%$ to the total cross section for $m_2\sim 600$~GeV and the interference contribution increases as $m_2$ increases.

  It should be noted this is effect is due to the $h_1-h_2-h_2$ couplings being proportional to the $h_2$ mass squared.  The mass dependence of the coupling then cancels the mass dependence in the propagator.  Since the Higgs is at the very least a major contributor to fundamental mass, this effect is relatively generic and interference effects are important in Higgs physics.\footnote{Interference effects between scalar contributions to $gg\rightarrow VV$, where $V=W^\pm,Z$, and continuum SM contributions can also be important~\cite{Maina:2015ela,Kauer:2015hia,Kauer:2019qei}.}

%% file: BSMresonance/BSMinterference_onshell.tex
In the case of a singlet resonance,  constraints from SM precision measurements  make these searches more challenging. From one side, precision measurements  imply that  the singlet-doublet mixing parameter is constrained to be small over a large region of parameter space.  From the other side, the singlet only couples to SM particles through mixing with the SM Higgs doublet. This results in a reduction of the di-Higgs production via singlet resonance decays. In particular, the singlet resonance amplitude  becomes of the same order as the SM  triangle  and box diagram amplitudes. Most important, in this work we show that a large relative phase between the SM box diagram and the singlet triangle diagram becomes important. This special on-shell interference effect has important phenomenological implications.

We will consider the simplest extension of the SM that can assist the scalar potential to induce a strongly first-order electroweak phase transition, consisting of an additional real scalar singlet with a $Z_2$ symmetry. Detailed relations between the bare parameters and physical parameters can be found in Ref.~\cite{Carena:2018vpt}.

The on-shell interference effect may enhance or suppress the conventional Breit-Wigner resonance production. 
Examples in Higgs physics known in the literature, such as $gg\to h\to\gamma\gamma$~\cite{Campbell:2017rke} and $gg\to H\to t\bar t$~\cite{Carena:2016npr}, are both destructive.
We discuss in detail in this section the on-shell interference effect between the resonant singlet amplitude and the SM di-Higgs box diagram. We show that in the singlet extension of the SM considered in this paper, the on-shell interference effect is generically constructive and could be large in magnitude, thus enhances the signal production rate.

The interference effect between two generic amplitudes can be denoted as non-resonant amplitude $A_{nr}$ and resonant amplitude $A_{res}$.
The resonant amplitude $A_{res}$, defined as
\begin{equation}
A_{res} = a_{res} \frac {\hat s} {\hat s - m^2 + i \Gamma m},
\end{equation}
has a pole in the region of interest and 
we parameterize it as the product of a fast varying piece containing its propagator and a slowly varying piece $a_{res}$ that generically is a product of couplings and loop-functions. The general interference effect can then be parameterised as~\cite{Carena:2016npr,Campbell:2017rke},
\begin{eqnarray}
|\mathcal{M}|_{int}^2 &=& 2\mbox{Re}(A_{res}\times A_{nr}^*)\,=\,2\left(\mathcal{I}_{int} + \mathcal{R}_{int}\right),\nonumber \\
\mathcal{R}_{int} &\equiv& |A_{nr}||a_{res}|\frac {\hat s (\hat s - m^2)} {(\hat s - m^2)^2+\Gamma^2 m^2} \cos(\delta_{res}-\delta_{nr})\nonumber \\
\mathcal{I}_{int} &\equiv& |A_{nr}||a_{res}|\frac {\hat s \Gamma m} {(\hat s - m^2)^2+\Gamma^2 m^2} \sin(\delta_{res}-\delta_{nr}),
\label{eq:decomposition}
\end{eqnarray}
where $\delta_{res}$ and $\delta_{nr}$ denote the complex phases of $a_{res}$ and $A_{nr}$, respectively.

The special interference effect $\mathcal{I}_{int}$ only appears between the singlet resonant diagram and the SM box diagram. This interference effect is proportional to the relative phase between the loop functions $\sin(\delta_\vartriangleright-\delta_\square)$ and the imaginary part of the scalar propagator which is sizeable near the scalar mass pole. 

\begin{figure}[t]  
  \centering
  \includegraphics[width=.60\textwidth]{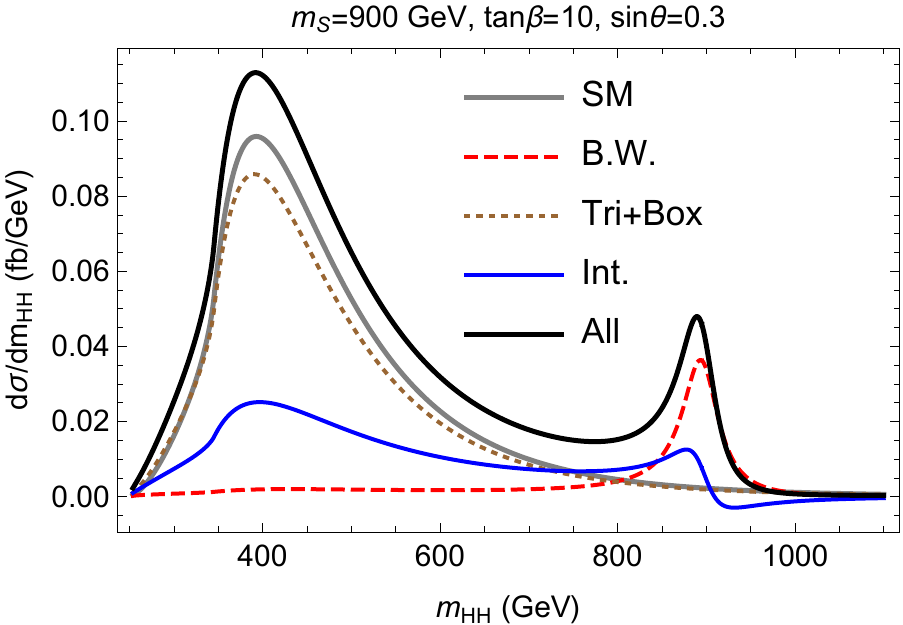}
  \caption{
  The differential di-Higgs distribution for a benchmark point of the singlet extension of the SM shown in linear scale and over a broad range of the di-Higgs invariant mass. The full results for the SM and the singlet SM extension  are shown by the grey and black curves, respectively. In the singlet extension of the SM, the contributions from the resonant singlet diagram, the non-resonant diagram and the interference between them are shown in red (dashed), brown (dotted) and blue curves, respectively \cite{Carena:2018vpt}.  
  }
  \label{fig:phenoshape} 
\end{figure}

In \reffig{fig:phenoshape} we display the differential cross section as a function of the Higgs pair invariant mass for a benchmark point with a heavy scalar mass of 900 GeV, mixing angle $\sin\theta=0.3$ and $\tan\beta=10$. 
The differential cross section is shown in linear scale for a broad range of di-Higgs invariant masses,  including the low invariant mass regime favoured by parton distribution functions at hadron colliders.

This particular choice for the benchmark shows well the separation of the scalar resonance peak and the threshold enhancement peak above the $t\bar t$-threshold. The SM Higgs pair invariant mass distribution is given by the grey curve while the black curve depicts the di-Higgs invariant mass distribution from the singlet extension of the SM. 
It is informative to present all three pieces that contribute to the full result of the di-Higgs production, namely (i) the resonance contribution (red, dashed curve), (ii) the SM non-resonance contribution (box and triangle diagrams given by the brown, dotted curve), and (iii) the interference between them (blue curve). 
Note that the small difference between the ``Tri+Box'' and the ``SM'' line shapes is caused by the doublet-singlet scalar mixing, which leads to a $\cos\theta$ suppression of the SM-like Higgs coupling to top quarks as well as a modified SM-like Higgs trilinear coupling $\lambda_{HHH}$.
We observe that the full results show an important enhancement in the di-Higgs production across a  large range of invariant masses. This behaviour is anticipated from the decomposition analysis in the previous section. There is a clear net effect from  the interference curve shown in blue.  Close to the the scalar mass pole at 900 GeV, the on-shell interference effect enhances the Breit-Wigner resonances peak (red, dashed curve) by about 25\%. Off-the resonance peak, and especially at the threshold peak, the interference term (blue curve) gives a sizeable enhancement to the cross section as well. Hence, a combined differential analysis in the Higgs pair invariant mass is crucial in probing the singlet extension of the SM. 

The interference pattern between the resonant heavy scalar contribution and the SM non-resonant triangle and box contributions show interesting features. 
We highlight the constructive on-shell interference effect that uniquely arises between the heavy scalar resonance diagram and the SM box diagram, due to a large relative phase between the loop functions involved.
We observe that the on-shell interference effect can be as large as 40\% of the Breit-Wigner resonance contribution and enhances notably the total signal strength, making it necessary taking into account in heavy singlet searches. Detailed parametric dependence of the on-shell interference on the model parameters can be found in Ref.~\cite{Carena:2018vpt}.

%% file: BSMresonance/BSMinterference_overall.tex
The search for a heavy Higgs boson resonance in the di-Higgs final states is accompanied
by interferences between the resonant signal and the di-Higgs continuum background,
where the latter includes the SM-like Higgs boson $s$-channel contribution.
In this section we summarize a model-independent study on such interference
effects, see also Ref.~\cite{Brooijmans:2018xbu}. 

We introduce an effective coupling of the heavy Higgs boson $H$ to gluons,
\begin{equation}
{\cal L} \supset \frac{\as}{12\pi v}\,c_H\,H\,G^a_{\mu\nu}G^{a,\mu\nu}\,.
\end{equation}
The Wilson coefficient $c_H$ can in general be complex number, parameterized as
\begin{equation}
c_H=|c_H| e^{i \theta_H}\,.
\end{equation}
This effective interaction accounts for particles $P$ coupling the new Higgs boson $H$ to gluons, for which the threshold $2m_P$ of
the corresponding loop can be either lighter or heavier than the Higgs boson mass $m_H$;
in the former case the effective loop-induced coupling becomes complex.  Note that while formally the description through 
an effective operator is not valid for $2 m_P\le m_{H}$, 
we do not restrict ourselves to a specific model and can hence condense the amplitude to the given form. 
However, we assume the Wilson coefficient and its phase to be constant, whereas for a concrete model realization
the loop-induced coupling can inherit a dependence on the final-state invariant mass $m_{hh}$, where
$h$ denotes the SM-like Higgs boson. This is particularly true for large width $\Gamma_H$ and
in the vicinity of the threshold region $m_{hh}\sim 2m_P$.

In addition to $c_H$,
we also choose the mass of $H$, the width $\Gamma_H$ and the trilinear Higgs-boson self-coupling of the SM-like Higgs boson $\lambda_{hhh}$ (normalised to its SM value) as free input parameters. 
Any effect
in the Higgs-boson self-coupling $\lambda_{Hhh}$ can be absorbed into $c_H$, which is why we keep $\lambda_{Hhh}$ fixed.
In summary, we vary the following parameters freely
\begin{equation}
|c_H|, \; \theta_H, \;\Gamma_H,\;m_H,\;\lambda_{hhh}\,.
\end{equation}

For our analysis we use the code {\tt HPAIR} \cite{hpair}, which incorporates the $s$-channel resonance by a Breit-Wigner propagator
of the heavy Higgs boson $H$
\begin{equation}
\frac{1}{m_{hh}^2-m_H^2+ i m_H \Gamma_H}\,,
\end{equation}
with $m_{hh}$ again denoting the invariant mass of the Higgs-boson pair.
In order to classify the interferences we split the differential cross section in three contributions
\begin{equation}
\frac{d\sigma}{dm_{hh}}=\frac{d\sigma_S}{dm_{hh}}+\frac{d\sigma_I}{dm_{hh}}+\frac{d\sigma_B}{dm_{hh}}\,.
\end{equation}
The signal cross section $\sigma_S$ contains the $s$-channel exchange of a heavy Higgs boson $gg\to H\to hh$ only, while
the background cross section $\sigma_B$ contains all non-resonant diagrams with final state $hh$,
namely the triangle and box diagrams equivalent to the ones of the SM process. The interference cross section $\sigma_I$
is proportional to $2\text{Re}(A_S A_B^*)$, where $A_S$ denotes the signal amplitude and $A_B$ the background amplitude.

As a measure of the interference effects we introduce the following parameters
\begin{align}
 \eta&=\left.\int_{m_H-10\Gamma_H}^{m_H+10\Gamma_H}dm_{hh}\left(\frac{d\sigma_S}{dm_{hh}}+\frac{d\sigma_I}{dm_{hh}}\right)\right/
 \int_{m_H-10\Gamma_H}^{m_H+10\Gamma_H}dm_{hh}\left(\frac{d\sigma_S}{dm_{hh}}\right)\,, \nonumber 
 \\
 \eta_-&=\left.\int_{m_H-10\Gamma_H}^{m_{hh}^I}dm_{hh}\left(\frac{d\sigma_S}{dm_{hh}}+\frac{d\sigma_I}{dm_{hh}}\right)\right/
 \int_{m_H-10\Gamma_H}^{m_{hh}^I}dm_{hh}\left(\frac{d\sigma_S}{dm_{hh}}\right)\,,\nonumber
 \\
 \eta_+&=\left.\int_{m_{hh}^I}^{m_H+10\Gamma_H}dm_{hh}\left(\frac{d\sigma_S}{dm_{hh}}+\frac{d\sigma_I}{dm_{hh}}\right)\right/
 \int_{m_{hh}^I}^{m_H+10\Gamma_H}dm_{hh}\left(\frac{d\sigma_S}{dm_{hh}}\right)\,. 
\end{align}
The first parameter $\eta$ yields, if multiplied with the signal cross section, the overall change of the signal cross section due to the interference effects.
Instead $\eta_+$ and $\eta_-$ measure the interference effects if the peak structure is distorted. For instance, typically there could be a
peak-dip structure. In this case the two curves $d\sigma_S/dm_{hh}$ and $d(\sigma_S+\sigma_I)/dm_{hh}$
intersect in $m_{hh}^I$  and we can define $\eta_+$ and $\eta_-$ as the measures of the peak distortion.
The boundaries in the definition of the $\eta$'s, $m_H\pm 10\Gamma_H$, capture the majority of the peak structure.
We refrain from using very large widths or an even larger boundary of the integration,
since (i) the crossing of the top threshold for the background diagrams at $m_{hh}\sim 2 m_t$
would require a more thorough analysis of the background effects for this peculiar case,
and (ii) we choose the Wilson coefficient to be constant.

For the scan over the parameter space we consider
\begin{equation}
\begin{split}
|c_{H}|\in [0.001, 5]\,, \hspace*{0.2cm} \theta_{H} \in \{0,\frac{\pi}{4},\frac{\pi}{2}\}, \hspace*{0.2cm} m_{H}\in [0.3,1.4]\,\text{TeV}, \hspace*{0.2cm}\Gamma_{H}/m_{H}\in [10^{-4},0.2] \,.
\end{split}
\end{equation}
For the trilinear Higgs-boson self-coupling we use the values $\lambda_{hhh}\in({0,1,2)}\lambda_{hhh}^{SM}$ while keeping $\lambda_{Hhh}=\lambda_{hhh}^{SM}$.
\begin{figure}
\begin{center}
\includegraphics[width=0.6\textwidth]{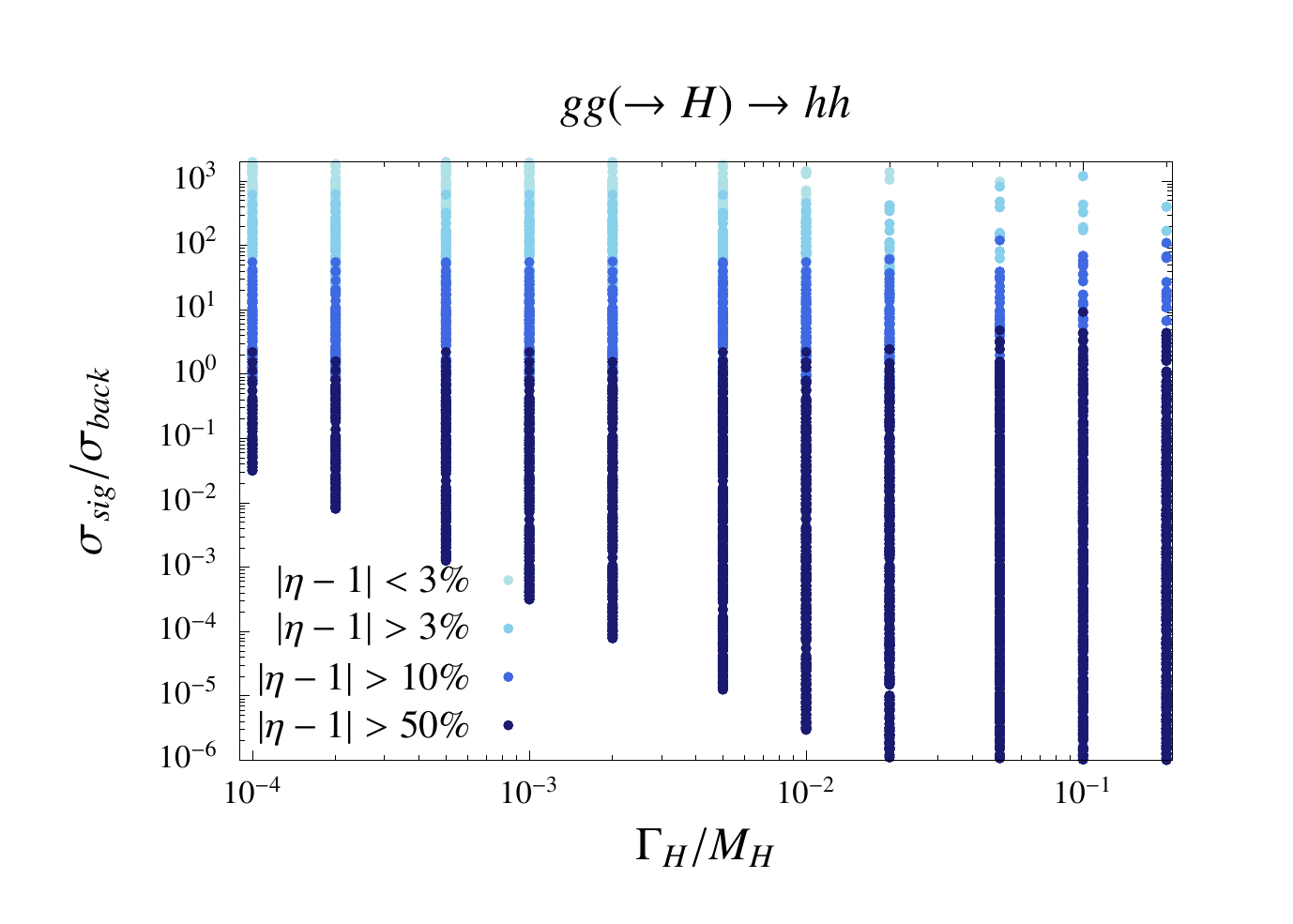}
\end{center}
\vspace{-0.7cm}
\caption{Relative difference (in percentage) of the interference factor $\eta$ from 1 in the ($\Gamma_H/m_H,\sigma_{sig}/\sigma_{back}$) plane. The scan was performed as indicated in the main text~\cite{Brooijmans:2018xbu}. \label{MIresonance:fig1}}
\end{figure}
In \reffig{MIresonance:fig1} we show the dependence of our measure~$\eta$ on $\Gamma_H/m_H$ and $\sigma_{sig}/\sigma_{back}$.
The latter ratio is defined through
\begin{equation}
\sigma_{sig}=\int_{m_{H}-10 \Gamma_{H}}^{m_{H}+10 \Gamma_{H}}dm_{hh} \frac{d\sigma_{S}}{dm_{hh}} \qquad\text{and}\qquad
\sigma_{back}=\int_{m_{H}-10 \Gamma_{H}}^{m_{H}+10 \Gamma_{H}}dm_{hh} \frac{d\sigma_{B}}{dm_{hh}}\,.
\end{equation}
The different colors in the figure indicate if $\eta$ differs from $1$ by less than $3$\%, between $3$--$10$\%, between $10$--$50$\% or by more than $50$\%.
It turns out that the interference effects mostly depend on the size of the ratio of the signal over background cross section.
Instead the interference shows little dependence on the width of the heavy Higgs boson, $\Gamma_H$,
as long as we consider masses $m_H> 2 m_t$, i.e. the region in which the background process develops
an imaginary part. Though, the largest values in $\eta$ are obtained for large width only.
\begin{figure}
\begin{center}
\includegraphics[width=0.6\textwidth]{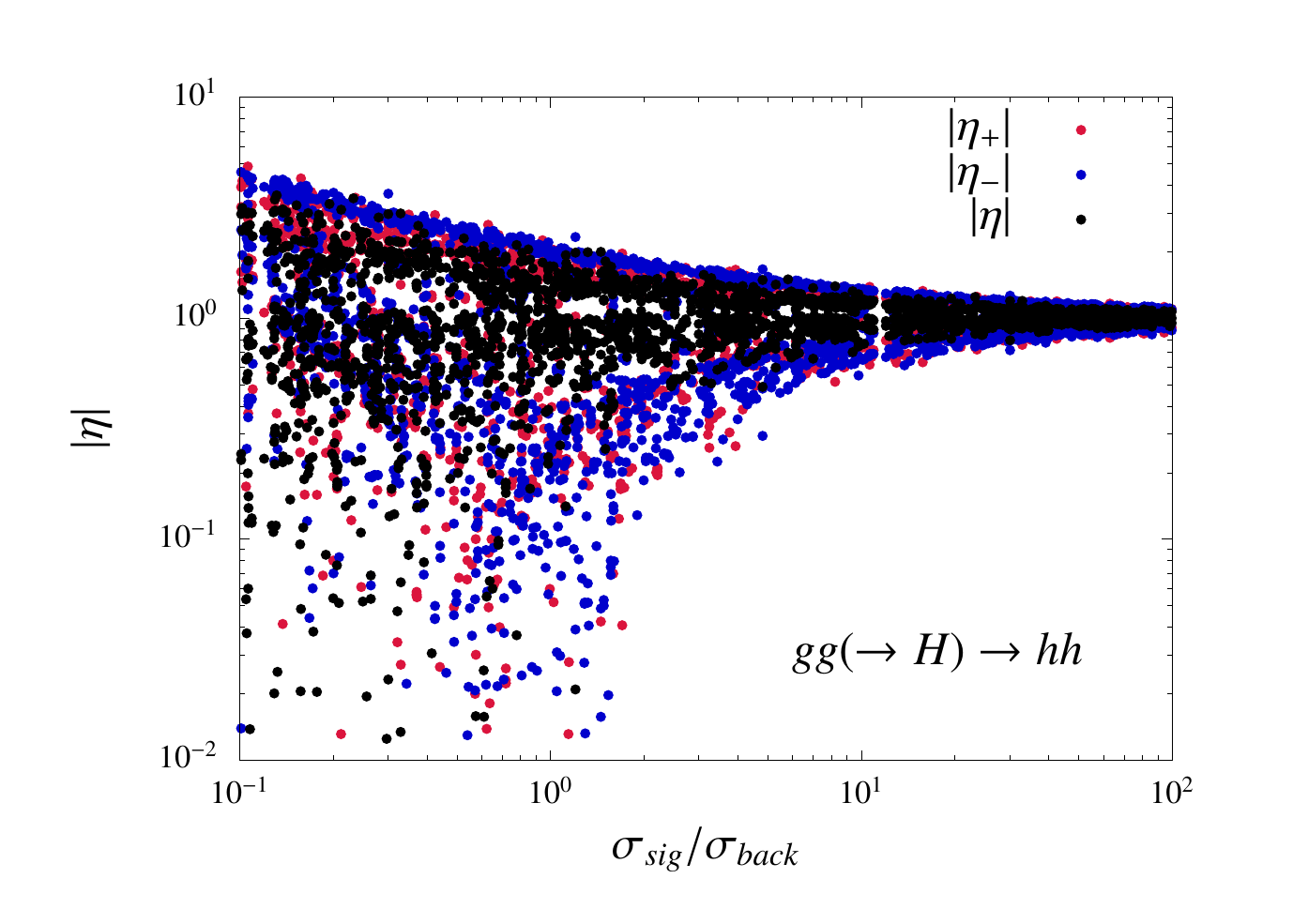}
\end{center}
\vspace{-0.7cm}
\caption{Interference factors $\eta$, $\eta_+$ and $\eta_-$ as a function of $\sigma_{sig}/\sigma_{back}$ for $gg \to H \to hh$~\cite{Brooijmans:2018xbu}.
\label{MIresonance:fig2}}
\end{figure}
In order to emphasize the dependence on  $\sigma_{sig}/\sigma_{back}$ we show $\eta$ (black points), $\eta_+$ (red points) and $\eta_-$ (blue points)
over $\sigma_{sig}/\sigma_{back}$ in \reffig{MIresonance:fig2}. We see that for $\sigma_{sig}/\sigma_{back} \approx 10$ the interference effects can already increase the cross 
section by a factor of $1.5$, and therefore should be definitively taken into account in order to obtain accurate predictions.

In conclusion, we find that interference effects should be taken into account once the LHC reaches sensitivity of $10$ times the SM di-Higgs background process.
We parameterised
the increase in the signal cross section due to interference effects by a parameter~$\eta$. To get a handle on the possible peak distortion 
we introduced the parameters~$\eta_{\pm}$. Note that neither of the parameters accounts for a possible peak shift. Further work should assess whether a peak distortion can be resolved experimentally and how the proposed general parameterisation compares to concrete model realizations.

%% file: BSMresonance/BSMspin2.tex
\section[Spin-2 models]{Spin-2 models \\
\contrib{B.~Dillon, H.~M.~Lee}
}
\label{sec:spin2}

\noindent Extra-dimensional models provide ideal benchmark scenarios for spin-2 resonances decaying to a pair of Higgs bosons. 
Warped extra dimensional models in particular are very well motivated extensions to the SM, providing a natural solution to the electroweak hierarchy problem and an explanation of the hierarchies in the flavour sector.
In addition to this, they are intimately connected with strongly coupled extensions of the SM such as composite Higgs models through the AdS/CFT correspondence.
Metric fluctuations in an extra dimension give rise to massive towers of spin-2 states, the Kaluza-Klein (KK) gravitons, and a light spin-0 state, the radion.
The couplings of these states are determined by the wavefunction overlaps between the SM particles and the metric fluctuations.
Most phenomenoligical studies of the KK gravitons and the radion assume a warped extra dimension described by the Randall-Sundrum (RS) metric \cite{Randall:1999ee}.
The masses of the KK gravitons are typically above $1$ TeV, while the radion may take a much lighter mass due to it being generated from backreaction on the metric \cite{Csaki:2000zn}.
In this RS scenario the electroweak hierarchy is solved by localising the wavefunction of the Higgs field near the IR brane, where the overlaps with the metric fluctuation wavefunctions are large.
For this reason channels with a resonant di-Higgs production are important to probe the KK graviton and radion.

We assume that only the lightest KK graviton, which we denote as $X_{\mu\nu}(x)$,  is accessible  in the experiment.  This state couples to the  SM particles through the energy-momentum tensor as ${\cal L}_X=-\frac{c_i}{\Lambda}\, X_{\mu\nu} T^{\mu\nu}_i$, and we can write the partial decay widths to SM particles as ~\cite{Lee:2013bua,Falkowski:2016glr}
\begin{align}
    \Gamma(X\rightarrow gg)=& \frac{c_g^2m_X^3}{10\pi \Lambda^2}  \,, \;\;   ~~\Gamma(X\rightarrow \gamma\gamma)=\frac{c_{\gamma\gamma}^2m_X^3}{80\pi \Lambda^2} \,,  \nonumber \\
    \Gamma(X\rightarrow hh)=&\frac{c_h^2m_X^3}{960\pi \Lambda^2}(1-4r_h)^{5/2} \,,  \nonumber \\
    \Gamma(X\rightarrow rr)=&\frac{c_{r}^2m_X^3}{960\pi \Lambda^2}(1-4r_r)^{5/2} \,,  \nonumber \\
    \Gamma(X\rightarrow f\bar{f})=&\frac{N_c(c_{fl}^2+c_{fr}^2)m_X^3}{320\pi \Lambda^2} (1-4r_f)^{3/2}(1+8r_f/3) \,, \nonumber \\
    \Gamma(X\rightarrow ZZ)=&\frac{m_X^3}{80\pi \Lambda^2}\sqrt{1-4r_Z}\left(c_{ZZ}^2+\frac{c_h^2}{12}+\frac{r_Z}{3}\left(3c_h^2-20c_hc_{ZZ}-9c_{ZZ}^2\right) \right.\nonumber \\
    &\left.+2\frac{r_Z^2}{3}\left(7c_h^2+10c_hc_{ZZ}+9c_{ZZ}^2\right)\right) \,,  \nonumber \\
    \Gamma(X\rightarrow WW)=&  \frac{m_X^3}{40\pi \Lambda^2}\sqrt{1-4r_W}\left(c_{W}^2+\frac{c_h^2}{12}+\frac{r_W}{3}\left(3c_h^2-20c_hc_{W}-9c_{W}^2\right) \right.\nonumber \\
    &\left.+2\frac{r_W^2}{3}\left(7c_h^2+10c_hc_{W}+9c_{W}^2\right)\right) \,,  \nonumber \\
    \Gamma(X\rightarrow Z\gamma)=& \frac{c_{Z\gamma}^2m_X^3}{40\pi \Lambda^2}(1-r_Z)^3\left(1+\frac{r_Z}{2}+\frac{r_Z^2}{6}\right)  \,,
\end{align}
where $\Lambda$ is the IR scale determining the KK graviton interactions, the $c_i$ coefficients are determined by integrals over the wavefunctions of the states, $r_i=(m_i/m_X)^2$, and $m_X$ is the lightest KK graviton mass.
The decay width to two radion states is also included here with the radion field denoted by $r(x)$, where we assume that the radion mass is lower than $m_X/2$.
We also use the following relations: $c_{\gamma\gamma}=s_{\theta}^2c_W+c_{\theta}^2c_B$, $c_{ZZ}=c_{\theta}^2c_W+s_{\theta}^2c_B$, $c_{Z\gamma}=s_{\theta}c_{\theta}(c_W-c_B)$.
The decay to $Z\gamma$ is only non-zero when brane-kinetic terms for the gauge fields are present, since it is only this that can break the degeneracy between $c_W$ and $c_B$.
Without fine-tuning we can assume that these brane-kinetic terms are negligible.

\begin{figure}[tb]
  \begin{center}
\includegraphics[scale=0.4]{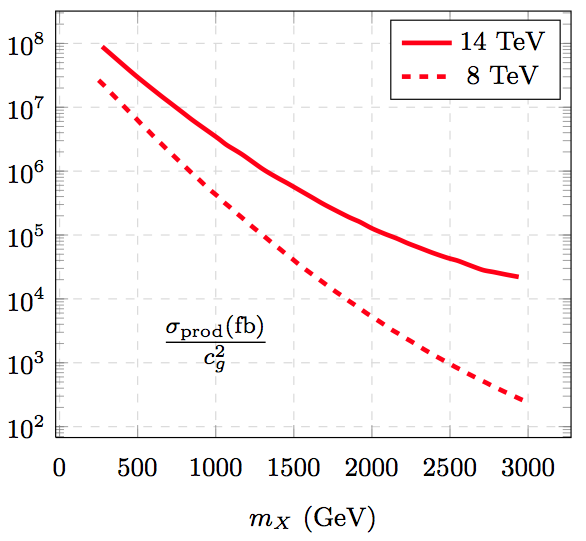}
    \caption{Production cross-section of the KK graviton via gluon fusion at centre of mass energies of $8$ and $14$ TeV, figure taken from Ref.~\cite{Dillon:2016fgw}.  The dependence of this quantity on the graviton coupling to gluons has been factored out. \label{Gxsec}}
  \end{center}
\end{figure}

There are important differences that occur between different incarnations of RS models regarding the nature of the SM fields.
In the most basic set-up, all SM fields are placed on the IR brane.
This is problematic since large couplings between light fermions and the KK states are induced leading to unacceptable levels of flavour violating processes for KK masses in the TeV range.
The solution is then to allow all SM fields to propagate in the bulk of the extra dimension, with the lighter fermions localised near the UV region such that the overlap of their wavefunctions and those of the KK gravitons is significantly reduced.
This also leads to an elegant description of fermion mass generation, whereby the Yukawa couplings of light fermions receive natural exponential suppression due to wavefunction overlaps.
With the Higgs field and the top quark being localised in the IR region and the lighter fermions towards the UV region, it has been shown that the flavour hierarchy structure of the SM can be achieved with $\mathcal{O}(1)$ 5D Yukawa couplings \cite{Huber:2000ie}.
The gauge fields in the extra dimension are restricted to having flat profiles by gauge invariance.
The same mechanism that gives rise to hierarchical Yukawa couplings also gives rise to hierarchical KK graviton couplings, since its wavefunction is localised in the IR region of the extra dimension.
Therefore the most important decay modes of this state will be to the heavier particles of the SM: $t\bar{t}$, $HH$, $W^+W^-$, and $ZZ$. 
The dominant production mechanism for the lightest KK graviton is via gluon fusion, and in Fig.~\ref{Gxsec} the corresponding cross-section is shown as a function of the coupling $c_g$ for both the $8$ TeV and $14$ TeV centre of mass energies.
While VBF typically has a production cross-section an order of magnitude lower than gluon fusion, it has been shown that VBF searches can sometimes provide better sensitivity due to an enhanced background rejection \cite{Brooijmans:2014eja,Aaboud:2017itg}.
More work is required for a detailed study of the phenomenology of the KK graviton produced via VBF.

\begin{figure}[tb]
  \begin{center}
\includegraphics[scale=0.42]{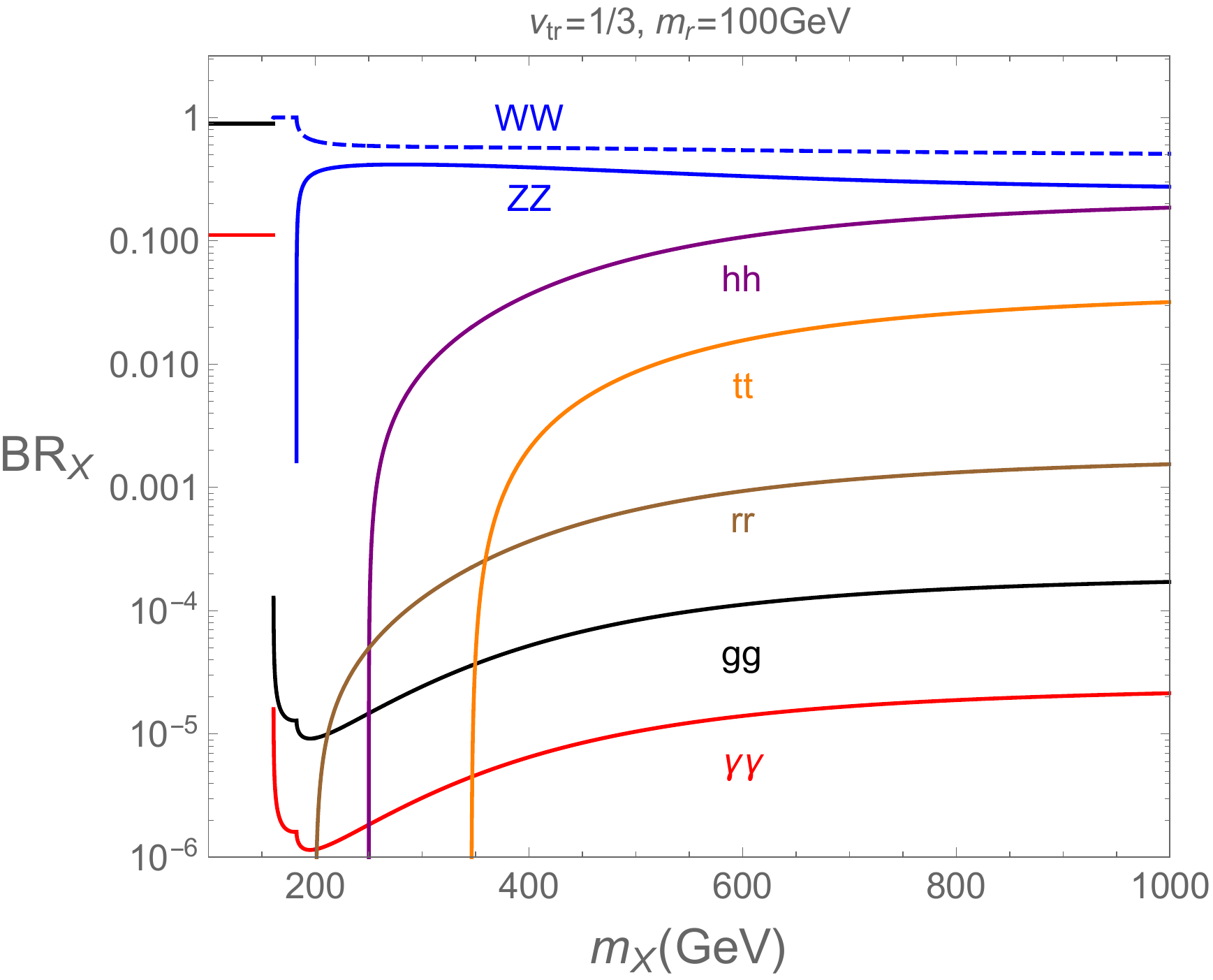} 
\includegraphics[scale=0.42]{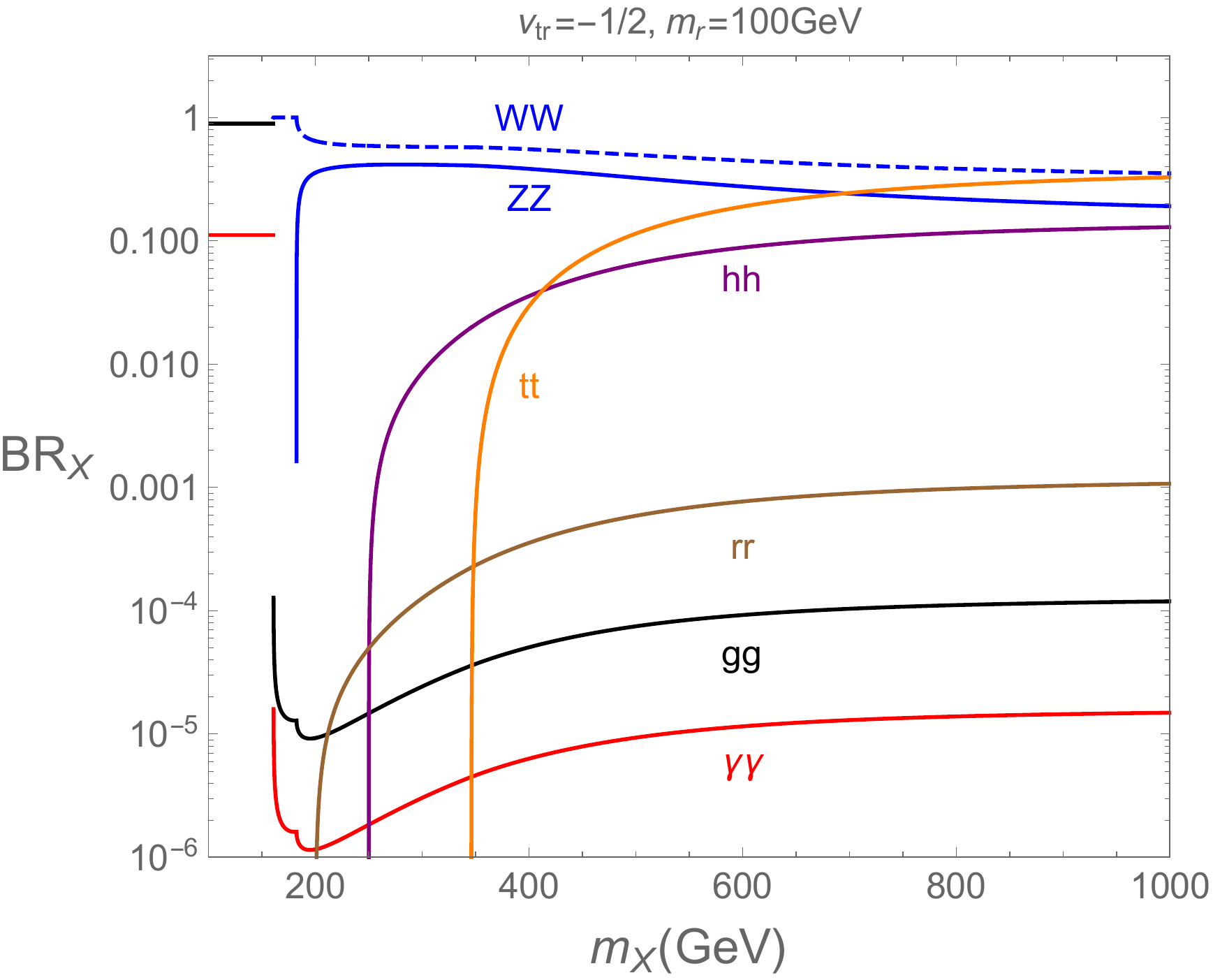}
    \caption{Branching ratios of KK graviton for the bulk RS benchmark model with $\nu_{tr}=-\frac{1}{3}$ (left) and $\nu_{tr}=-\frac{1}{2}$ (right). }
      \label{BR}
  \end{center}
\end{figure}

We consider a benchmark scenario for resonant di-Higgs production due to the decay of a KK graviton with the Higgs field localised on the IR brane, while the SM gauge fields and fermions propagate in bulk. In this case, we have $c_g=c_W=c_Z=[\log(M_P/\Lambda)]^{-1}\sim 0.03$ and $c_h=1$.
The localization of the top quark is controlled by the bulk mass parameter, ${\cal L}_{5D}\supset -{\rm sgn}(y)\nu_{ti} k\, {\bar t}_i t_i$ with $i=l,r$ for the bulk fermions containing either left-handed or right-handed top quarks. For $\nu_{tl}=-\frac{1}{2}$ (that is, a flat left-handed top) and arbitrary $\nu_{tr}$, we have the couplings of the KK graviton to the top quark \cite{Fitzpatrick:2007qr,Lee:2013bua} as
\begin{align}
c^2_{tl}=0, \qquad c^2_{tr}=\frac{1-2\nu_{tr}}{1-e^{-kL(1-2\nu_{tr})}}\, \int^1_0 dy\, y^{2-2\nu_{tr}}\, \frac{J_2(3.83 y)}{J_2(3.83)},
\end{align}
where $J_2$ represents the second order Bessel function.
Finally, the radion coupling to KK graviton is determined geometrically \cite{Dillon:2016tqp} by
\begin{align}
c_r=\frac{1}{3}\int^1_0 dy\, y^3\, \frac{J_2(3.83 y)}{J_2(3.83)} = 0.09.
\end{align}
In Fig.~\ref{BR}, we depict the branching ratios of the KK graviton as a function of the KK graviton mass for the bulk RS model with $\nu_{tr}=\frac{1}{3}$ on the left and $\nu_{tr}=-\frac{1}{2}$ on the right. We have chosen the radion mass to $m_r=100\,{\rm GeV}$ in both plots. 
The branching ratio to the di-Higgs channel is as large as $10\%$ or so, while the $t{\bar t}$ channel can be comparable to $WW$ and $ZZ$ channels when the top quark is localised near the IR brane. 
The branching ratio to the di-radion channel is at the level of $0.1\%$ but it can be also interesting, depending on its decay modes.  

We also remark on the case where the SM particles are localised away from the IR brane and dark matter is localised on the IR brane. This is the so-called the dark brane scenario considered in Ref.~\cite{Lee:2013bua}. In this case, the KK graviton can be regarded as a mediator 
for dark matter \cite{Lee:2014caa,Han:2015cty,Carrillo-Monteverde:2018phy,Dillon:2016tqp,Rueter:2017nbk}, leading to a sizable invisible decay rate of the KK graviton into a dark matter pair.  Furthermore, the KK graviton can decay sizably into a radion pair, which then decays into SM particles \cite{Dillon:2016tqp}.  For a relatively heavy radion, the radion decays into massive particles such as top quarks, di-Higgs, etc., with comparable branching ratios as those for massive particles, even if there is a volume suppression for the SM particles delocalized from the IR brane. Therefore, in the dark brane scenario there are interesting signatures from multiple Higgs production due to the cascade decay of the KK graviton. 

Recent results from the ATLAS collaboration place a lower bound of $3.4$ TeV on the mass of a KK graviton resonance decaying to an all hadronic final state via $t\bar{t}$ \cite{Aaboud:2019roo}.
Searches for a KK graviton decaying via $t\bar{t}$ to lepton-plus-jets final states can constrain the lightest state to lay above $3.8$~TeV for a $15\%$ width \cite{Aaboud:2018mjh}.
Weaker constraints have been obtained through consideration of the VBF production mode \cite{Aaboud:2017itg,Aaboud:2017fgj}, however these will be interesting for future study.
Direct production of a radion can also give rise to di-Higgs signatures, however this is covered by the sections on scalar mediators, and for a recent in depth analysis of radion phenomenology we refer the reader to Ref.~\cite{Ahmed:2015uqt}.
Interesting effects on both the radion and KK graviton phenomenology has been observed in the presence of brane-localised kinetic terms \cite{Davoudiasl:2008hx,McDonald:2008ss,Dillon:2016fgw,Dillon:2016bsb,Dillon:2017ctw}, most notably resulting in a lowering of the KK graviton mass with respect to the scale of other resonances.
Interesting and detailed studies of KK graviton effects in di-Higgs production have been studied in Refs.~\cite{Gouzevitch:2013qca,Zhang:2015mnh}.
In Ref.~\cite{Gouzevitch:2013qca} the authors studied both the scenario with the SM on the IR brane and the SM in the bulk.
They developed a strategy to search for resonant di-Higgs production via a KK graviton in the $b\bar{b}b\bar{b}$ final state and showed that a large range of the parameter space can be explored.
Lastly, it is noteworthy that these techniques and results are equally applicable to the search for spin-2 composite resonances arising in composite Higgs models, thus expanding the theoretical motivation to search for spin-2 resonances in di-Higgs production.

%% file: BSMresonance/BSMC2HDM.tex
\subsection[Complex 2-Higgs-Doublet Model]{Complex 2-Higgs-Doublet Model \\
\contrib{P. Basler, S. Dawson, C. Englert, M. M\"uhlleitner}
}

The 2-Higgs-Doublet Model (2HDM) \cite{Gunion:1989we,Branco:2011iw} 
contains 2 $SU(2)_L$ doublets, $\Phi_1$ and $\Phi_2$.  
Assuming no flavor changing neutral currents and a softly broken $Z_2$ symmetry,  there are four
 different types of 2HDMs, which are defined by the Higgs doublet that couples to each kind of fermions, and are summarised in
\refta{tab:2hdmtypes}.\footnote{Phenomenological implications of the 2HDM model on the Higgs self-coupling are not discussed in this document,
the interested reader can find a study within the Gildener-Weinberg models in Ref.~\cite{Lane:2019dbc}.}
\begin{table}
\begin{center}
\begin{tabular}{|r|ccc|} \hline
& $u$-type & $d$-type & leptons \\ \hline
type I (T1) & $\Phi_2$ & $\Phi_2$ & $\Phi_2$ \\
type II (T2) & $\Phi_2$ & $\Phi_1$ & $\Phi_1$ \\
lepton-specific & $\Phi_2$ & $\Phi_2$ & $\Phi_1$ \\
flipped & $\Phi_2$ & $\Phi_1$ & $\Phi_2$ \\ \hline
\end{tabular}
\caption{The four Yukawa types of the ${Z}_2$-symmetric 2HDM,
  defined by the Higgs doublet that couples to each kind of fermions. \label{tab:2hdmtypes}}
\end{center}
\end{table}
The  complex or CP-violating
2HDM (C2HDM), described in Ref.~\cite{Fontes:2014xva}, allows for two complex phases, and the C2HDM has nine independent parameters \cite{ElKaffas:2007rq},
\begin{equation}
v \;, \quad t_\beta \;, \quad \alpha_{1,2,3}
\;, \quad m_{H_i} \;, \quad m_{H_j} \;, \quad m_{H^\pm} \;, \quad 
\mbox{Re}(m_{12}^2) \; ,
\label{eq:2hdminputset}
\end{equation}
where $  t_\beta= \frac{v_2}{v_1}$, $\alpha_i$ are the angles that diagonalise the Higgs mass matrix, and
$m_{H_i}$ and $m_{H_j}$ are  any two of the three neutral Higgs
boson mass eignestates. The third mass is calculated from the
other parameters~\cite{ElKaffas:2007rq}. 

Within the C2HDM it is possible to produce final states with two different Higgs bosons.
Compared to the SM di-Higgs production rate, in the C2HDM the cross sections can be enhanced in  the case of
resonant production of a heavy Higgs boson that decays
into a pair of lighter Higgs bosons, or due to
Higgs self-couplings that differ from the SM value.

CP-conserving 2HDM benchmarks for double Higgs production can
be found in Refs. \cite{Haber:2015pua,Baglio:2014nea}.  Here we summarise  the benchmark points for 
the C2HDM model presented in Ref. \cite{Basler:2018dac}.  
 \refta{tab:maxcxn} gives the maximum cross
section values for Higgs pair production
that are compatible with all present experimental and theoretical
constraints.
 The SM-like Higgs boson is $h$, the lighter 
of the non-SM-like neutral Higgs bosons is  $H_\downarrow$,  and the
heavier one is $H_\uparrow$. 
\begin{table}[b!]
\centering
\vspace*{0.2cm}
\begin{tabular}{|c|c|c|} \hline
$H_iH_j$/model & T1 & T2 \\ \hline
$hh$ & 794 & 63.2 \\
$h H_\downarrow$ & 49.17 & 11.38 \\
$h H_\uparrow$ & 17.65 & 13.50 \\
$H_\downarrow H_\downarrow$ & 3196 & 0.31 \\
$H_\downarrow H_\uparrow$ & 12.58 & 0.31 \\
$H_\uparrow H_\uparrow$ & 7.10 & 0.23 \\ \hline
\end{tabular}
\caption{Maximum cross section values at $\sqrt{s}=14\text{ TeV}$  in fb for LO gluon fusion into
  Higgs pairs, $\sigma( gg \to H_i H_j)$, in the C2HDM T1 and T2 scenarios, with
  an exclusion luminosity $\ge 64$~fb$^{-1}$ that satisfy all
  theoretical and experimental constraints~\cite{Basler:2018dac,Fontes:2017zfn}.
 \label{tab:maxcxn}}
\end{table}
\begin{figure*}[t!]
  \centering
  \includegraphics[width=0.47\linewidth]{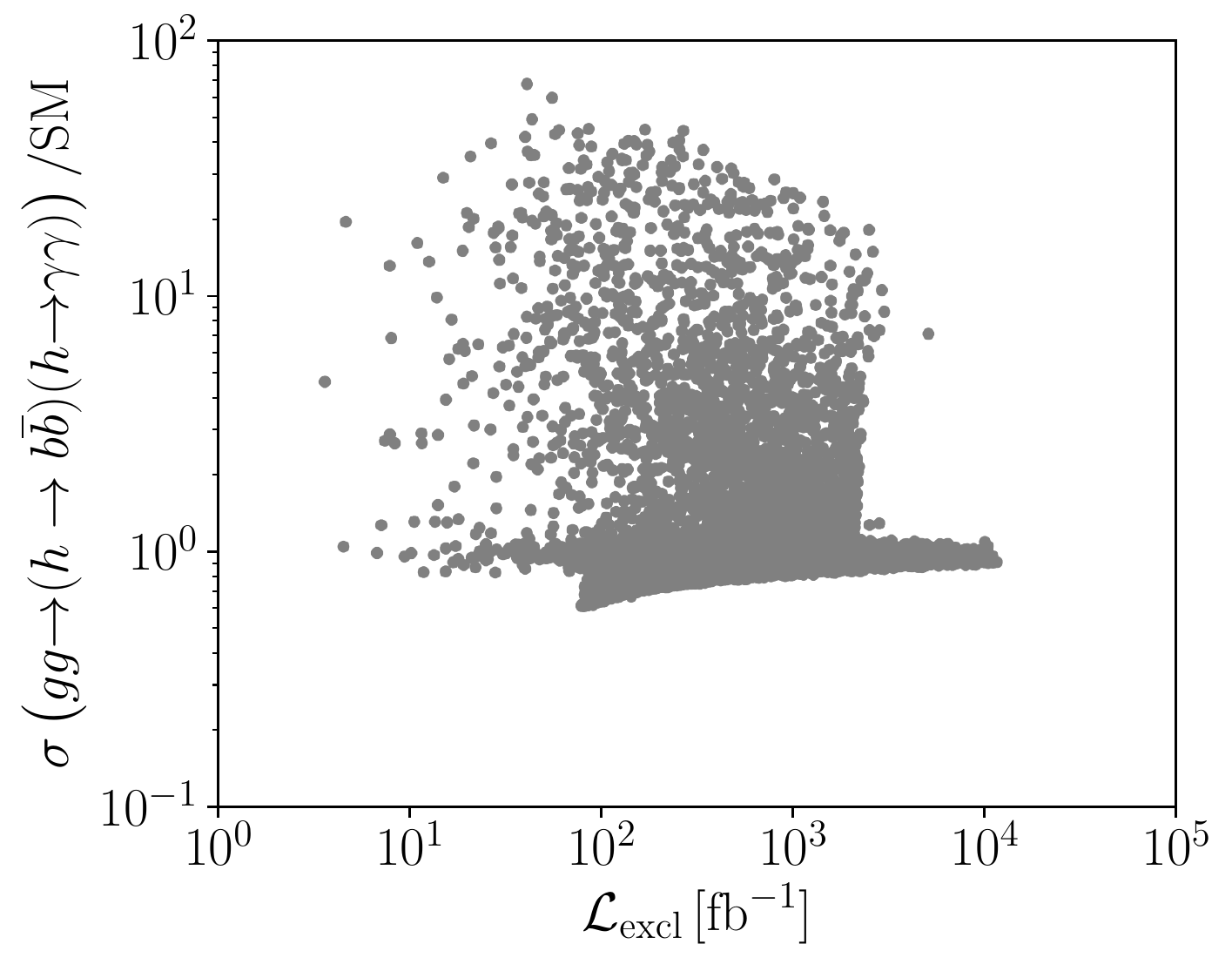}
  \hspace*{0.2cm}
  \includegraphics[width=0.47\linewidth]{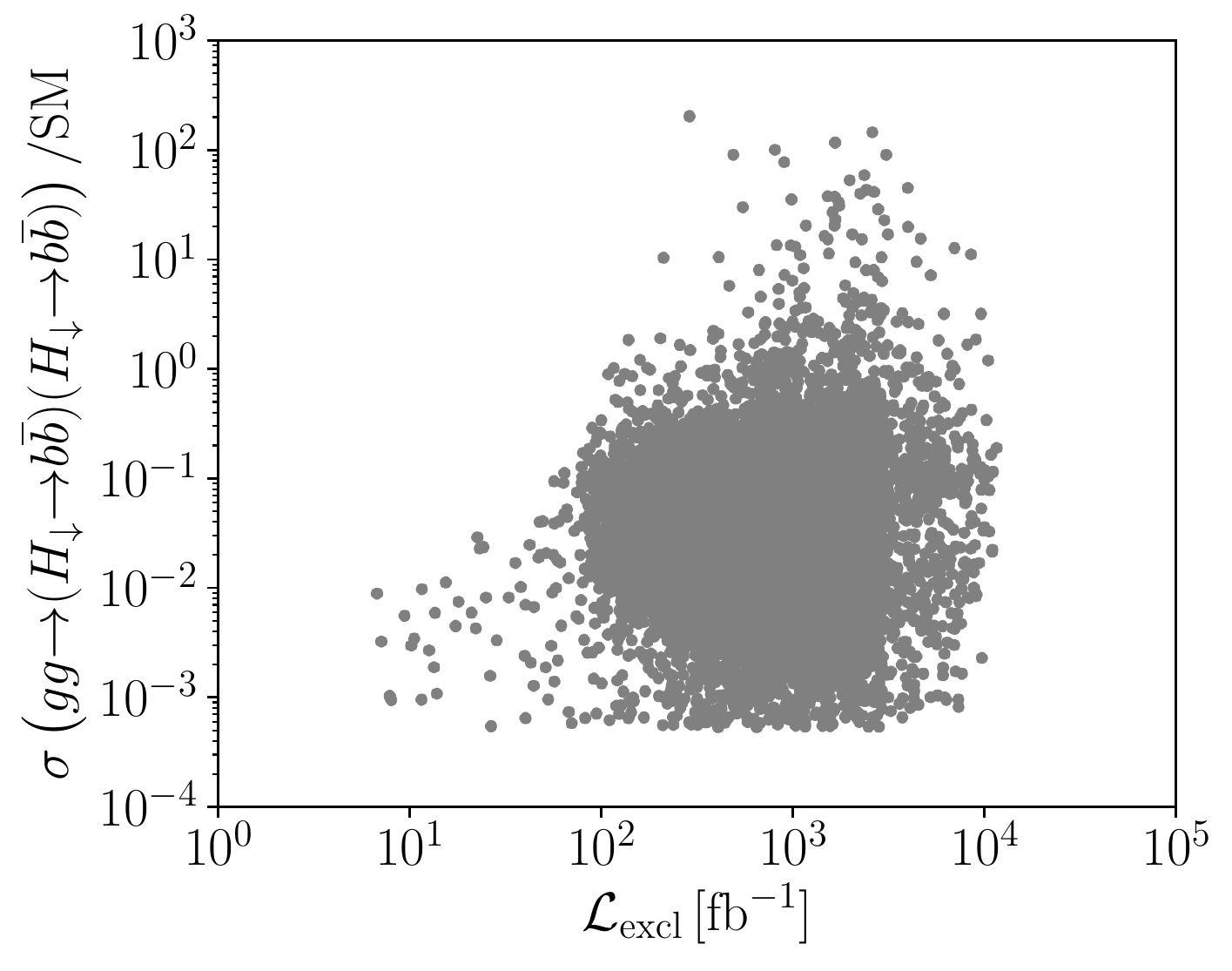}
  \caption{    Higgs pair production cross sections normalized to
    the SM value for SM-like Higgs pairs decaying into
    \bbyyAlt (left) and light-non-SM-like Higgs pairs
    decaying into \bbbbAlt (right) as a function of the
    exclusion luminosity, for the C2HDM T1 scenarios passing our applied constraints~\cite{Basler:2018dac,Fontes:2017zfn}. \label{fig:finalversuslumi}}
\end{figure*}
 \refta{tab:maxcxn} demonstrates that
 in  both  the T1 and T2 scenarios the maximum cross section
for $hh$ production can exceed the SM value: in T1 by a factor of
about 40 and in T2 by a factor of about 3.2. 
The large enhancements are due to
 the resonant production
of an $H_\downarrow$ or $H_\uparrow$  that decays
into a pair of SM-like Higgs bosons.  The reason for the  smaller enhancement
in $hh$ production in T2 compared to T1 is the overall heavier Higgs
spectrum.

Based on extrapolations from current searches, we study the exclusion luminosity, i.e., the
integrated luminosity at which a parameter point could be excluded experimentally.
The most promising final states are \bbyyAlt~\cite{Baur:2003gp}, \bbttAlt~\cite{Baur:2003gpa,Dolan:2012rv,Barr:2013tda} and
\bbbbAlt~\cite{Baur:2003gpa,deLima:2014dta,Wardrope:2014kya}. 
In \reffig{fig:finalversuslumi} we  show (for all the parameter points that pass the theoretical
and experimental constraints), 
the cross section values for $hh$ production in the T1 scenario normalized to
the  SM Higgs pair production in the
\bbyyAlt final state (left) and for $H_\downarrow
H_\downarrow$ production in the \bbbbAlt final state
as a function of the exclusion luminosity. 
 \reffig{fig:finalversuslumi} shows that
 the production of
a SM-like Higgs pair decaying to
\bbyyAlt can exceed the SM rate by up to a factor of
68. This maximum enhancement factor is  roughly the same for all final states.

In the $H_\downarrow H_\downarrow$ final state with both
$H_\downarrow$'s decaying into bottom quarks 
or to \bbttAlt, the enhancement can
be up to a factor of about 200.   Due to a smaller
branching ratio into photons, however, the maximum allowed enhancement
in the \bbyyAlt final state is a
factor  of 40.

The remaining di-Higgs production processes are less promising. 
The enhancement factor for $h H_\downarrow$ production is less than  3
in the \bbbbAlt and \bbttAlt final states.
The $h\rightarrow \bb, H_\downarrow\rightarrow \gamma\gamma$ rate
is below the SM rate, while  $h\rightarrow \gamma\gamma, H_\downarrow\rightarrow 
\bb$ has an enhancement factor around 3.
 Other final states have rates below the SM values.  

The situation is less promising in the C2HDM T2. The
maximum enhancement over the SM rate for $hh$
production with the  decay into the \bbttAlt, \bbbbAlt, or \bbyyAlt final states is around 4.5. All
other final states lead to smaller rates than in the SM. 

We conclude that there are promising
di-Higgs signatures with large rates in the C2HDM T1  for SM-like
Higgs pair production and  also for final states with non-SM-like Higgs
bosons. The new neutral Higgs bosons appear in SM-like final states,
however, with different kinematic correlations due to different
masses. 
The stringent constraints on the di-Higgs production rates present in the T2 scenario could exclude it if signatures much larger than in the SM were to be found.

%% file: BSMresonance/BSM1S2HDM.tex
\subsection[Singlet extensions of 2HDM]{Singlet extensions of 2HDM \\
\contrib{N.~R.~Shah}
}

The extension of a 2HDM by a complex singlet $S$ gives rise to the generic Higgs potential~\cite{Carena:2015moc, Baum:2018zhf}:
\begin{equation}\label{Vall}
    V = V_{2{\rm HDM}} + V_S \;,
\end{equation}
where
\begin{equation} \begin{split} \label{eq:V2HDM}
	V_{\rm 2HDM} &= m_{11}^2 \Phi_1^\dagger \Phi_1 + m_{22}^2 \Phi_2^\dagger \Phi_2 - \left( m_{12}^2 \Phi_1^\dagger \Phi_2 + {\rm h.c.} \right) \\
	& \qquad+ \frac{\lambda_1}{2} \left( \Phi_1^\dagger \Phi_1 \right)^2 + \frac{\lambda_2}{2} \left( \Phi_2^\dagger \Phi_2 \right)^2 + \lambda_3 \left( \Phi_1^\dagger \Phi_1 \right) \left( \Phi_2^\dagger \Phi_2 \right) + \lambda_4 \left( \Phi_1^\dagger \Phi_2 \right) \left( \Phi_2^\dagger \Phi_1 \right) \\
	& \qquad+ \left[ \frac{\lambda_5}{2} \left( \Phi_1^\dagger \Phi_2 \right)^2 + \lambda_6 \left( \Phi_1^\dagger \Phi_1 \right) \left( \Phi_1^\dagger \Phi_2 \right) + \lambda_7 \left( \Phi_2^\dagger \Phi_2 \right) \left( \Phi_1^\dagger \Phi_2 \right) + \rm{h.c.} \right],
\end{split} \end{equation}
and
\begin{equation} \begin{split} \label{eq:VS}
	V_S &= \left( \xi S + {\rm h.c.} \right) + m_S^2 S^\dagger S + \left( \frac{m_S'^2}{2} S^2 + {\rm h.c.} \right) \\
	& \qquad+ \left( \frac{\mu_{S1}}{6} S^3 + {\rm h.c.} \right) + \left( \frac{\mu_{S2}}{2} S S^\dagger S + {\rm h.c.} \right) \\
	& \qquad+ \left( \frac{\lambda''_1}{24} S^4 + {\rm h.c.} \right) + \left( \frac{\lambda''_2}{6} S^2 S^\dagger S + {\rm h.c.} \right) + \frac{\lambda''_3}{4} \left( S^\dagger S \right)^2 \\
	& \qquad+ \left[ S \left( \mu_{11} \Phi_1^\dagger \Phi_1 + \mu_{22} \Phi_2^\dagger \Phi_2 + \mu_{12} \Phi_1^\dagger \Phi_2 + \mu_{21} \Phi_2^\dagger \Phi_1 \right) + {\rm h.c.} \right] \\
	& \qquad+ S^\dagger S \left[ \lambda'_1 \Phi_1^\dagger \Phi_1 + \lambda'_2 \Phi_2^\dagger \Phi_2 + \left( \lambda'_3 \Phi_1^\dagger \Phi_2 + {\rm h.c.} \right) \right] \\ 
	& \qquad+ \left[ S^2 \left( \lambda'_4 \Phi_1^\dagger \Phi_1 + \lambda'_5 \Phi_2^\dagger \Phi_2 + \lambda'_6 \Phi_1^\dagger \Phi_2 + \lambda'_7 \Phi_2^\dagger \Phi_1 \right) + {\rm h.c.} \right].
\end{split} \end{equation}
$\Phi_1$, $\Phi_2$ are $SU(2)$ doublets with hypercharge $Y=1/2$. The $m_{ij}^2$ parameters have dimension mass squared, while the $\lambda_i$ are dimensionless. The parameter $\xi$ has dimensions mass cubed, the parameters $\{\mu_{Si}, \mu_{ij}\}$ have dimension mass, and the $\{\lambda'_i, \lambda''_i\}$ are dimensionless. In the CP-conserving case, all parameters can be chosen manifestly real. As customary, after minimisation, we define
$v_1 \equiv \left\langle \Phi_1\right\rangle, v_2 \equiv \left\langle \Phi_2\right\rangle,  v_S \equiv \left\langle S\right\rangle, v \equiv \sqrt{v_1^2 + v_2^2} $ and $\tan\beta \equiv v_1/v_2 \;.$
The observed mass of the $Z$ boson $m_Z = 91.2\,$GeV is obtained for $v = 174\,$GeV. A similar structure for the Higgs potential is also obtained for the general NMSSM~\cite{Ellwanger:2009dp}, and a mapping is provided to both the general and the $Z_3$ invariant NMSSM in Ref.~\cite{Baum:2018zhf}. 

The potential given above is described by 27 arbitrary parameters and at first glance appears difficult to analyze. However, the 125 GeV Higgs mass and its SM-like couplings enable us to constrain these significantly. In particular, most of the relevant phenomenology can be mostly parameterized in terms of  physical parameters like masses and mixing angles.\footnote{The mapping from the physical parameters to the parameters in the potential can be found in Ref.~\cite{Baum:2018zhf}.} To see this, it is useful to rotate the Higgs fields to the {\it extended Higgs basis}~\cite{Georgi:1978ri, Donoghue:1978cj, Gunion:1989we, Lavoura:1994fv, Botella:1994cs, Branco99, Gunion:2002zf,Carena:2015moc}\footnote{Note that there are different conventions in the literature for the Higgs basis differing by an overall sign of $H^{\rm NSM}$ and $A^{\rm NSM}$. Taking these into account, the potential in \refeq{Vall} and couplings for the 2HDM+S can be mapped directly to the potential and couplings given in the appendices of Ref.~\cite{Carena:2015moc}.}
\begin{align} \label{eq:Hbasis1}
	\begin{bmatrix} G^+ \\ \frac{1}{\sqrt{2}} \left(H^{\rm SM} + i G^0\right) \end{bmatrix} &= \sin\beta \Phi_1 + \cos\beta \Phi_2 \;, \\
	\begin{bmatrix} H^+ \\ \frac{1}{\sqrt{2}} \left(H^{\rm NSM} + i A^{\rm NSM}\right) \end{bmatrix} &= \cos\beta \Phi_1 - \sin\beta \Phi_2 \;, \\
	\frac{1}{\sqrt{2}} \left( H^{\rm S} + i A^{\rm S} \right) &= S \;,
	\label{eq:Hbasis-1}
\end{align} 
where $\{H^{\rm SM}, H^{\rm NSM}, H^{\rm S}\}$ and $\{A^{\rm NSM}, A^{\rm S}\}$ are the neutral CP-even and CP-odd real Higgs basis interaction states, $G^0$ ($G^\pm$) is the neutral (charged) Goldstone mode, and NSM stands for Non-SM. In this basis, of the states coming from the doublets, only $\left\langle H^{\rm SM}\right\rangle = \sqrt{2} v$ acquires a vev, and it is straightforward to work out the coupling of SM fermions to the Higgs basis states. For concreteness, in the following a Type~II Yukawa structure is assumed. However, the results shown will in general hold  for a different Yukawa structure. Some quantitative details may change due to the change in the Yukawa enhancement or suppression of the fermion couplings, but such modifications will be small since mostly  the low $\tan\beta = \mathcal{O}(1)$ regime is considered. 

The three CP-even mass eigenstates are denoted
\begin{equation}
h_i = \{ h_{125}, H, h\}\;, 
\end{equation}
where $h_{125}$ is identified with the $m_{h_{125}} \approx 125\,$GeV SM-like state observed at the LHC, and $H$ and $h$ are ordered by masses, $m_H > m_h$. Each mass eigenstate is an admixture of the extended Higgs basis interaction states,
\begin{equation}
	h_i = S_{h_i}^{\rm SM} H^{\rm SM} + S_{h_i}^{\rm NSM} H^{\rm NSM} + S_{h_i}^{\rm S} H^{\rm S} \;,
\end{equation}
where $S_{h_i}^{j}$ with $j = \{$SM, NSM, S$\}$ denotes the components of the mass eigenstates in terms of the interaction basis. Likewise, the two CP-odd mass eigenstates are denoted 
\begin{equation}
a_i = \{A, a\}\;,
\end{equation}
where again $m_A > m_a$, and
\begin{equation}
	a_i = P_{a_i}^{\rm NSM} A^{\rm NSM} + P_{a_i}^{\rm S} A^{\rm S} \;,
\end{equation}
where the components are similarly denoted by $P_{a_i}^{j}$. The observed SM-like nature of $h_{125}$ implies that \begin{equation}
 S_{h_{125}}^{\rm SM} \approx 1 \;, \qquad \{ (S_{h_{125}}^{\rm NSM})^2, (S_{h_{125}}^{\rm S})^2 \} \ll 1\;,
\end{equation} 
or, in other words, $h_{125}$ mass eigenstate must approximately be {\it aligned} with the $H^{\rm SM}$ interaction state.

First a few conditions that alignment imposes on the phenomenology are highlighted. The most important point is that alignment forbids the NSM or S-like CP-even Higgs bosons from coupling to pairs of $h_{125}$ or vector bosons~($W$ or $Z$). Additionally the CP-odd state couplings to $h_{125}$ and $Z$ are also forbidden. Instead, there can be interesting {\it Higgs cascade decays} of the heavy Higgs bosons to final states involving only one $h_{125}$ or a $Z$ such as $(H^{\rm{NSM}} \to H^{\rm{S}} H^{\rm{SM}})$ or $(A^{\rm{NSM}} \to H^{\rm{S}} Z)$. The singlets couple only to  SM particles via their mixing with the other states, or to a possible Dark Matter~(DM) state $\chi_1$. Hence depending on the mixing angles and the arbitrary coupling to  DM, such decays could result in $h_{125}$ or $Z$ plus visible or invisible signatures.  

\begin{figure}
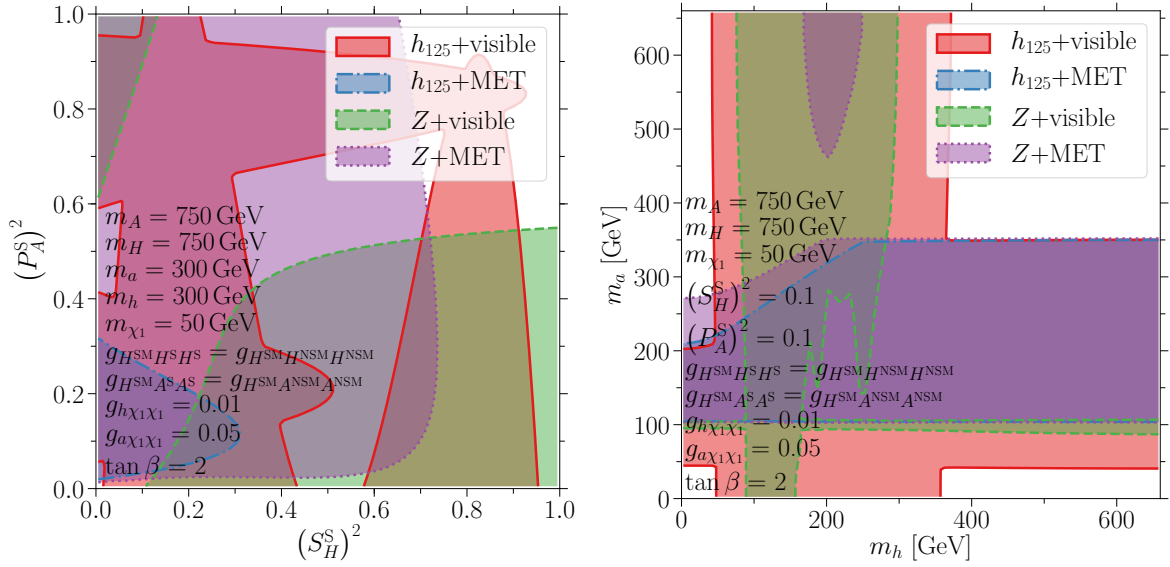

	\includegraphics[width=0.49\linewidth]{{{BSMresonance/BSMfigs/2HDMS_reach_mixing}}}
	\includegraphics[width=0.49\linewidth]{{{BSMresonance/BSMfigs/2HDMS_reach_masses}}}
	\caption{Regions of 2HDM+S parameter space within the future reach of the different Higgs cascade search modes as indicated in the legend at the LHC with ${\cal L} = 3000\,{\rm fb}^{-1}$ of data. The left panel shows the accessible regions in the plane of the singlet fraction of the parent Higgs bosons $(S_H^{\rm S})^2$ vs $(P_A^{\rm S})^2$. The right panel shows the reach in the plane of the masses of the daughter Higgs bosons produced in the Higgs cascades, $m_h$ vs $m_a$. The remaining parameters are fixed to the values indicated in the labels~\cite{Baum:2018zhf}.}
	\label{fig:2HDMSreach}
\end{figure}

We collected all the current search results and projections available for the relevant decays, as well as performed detailed collider simulations where needed, to obtain the projection for the reach at the LHC with 3000 fb$^{-1}$ of data~\cite{Baum:2018zhf}. \reffig{fig:2HDMSreach} presents an example of the reach we obtain for benchmark scenarios. While the mass of the parent Higgs bosons is fixed at 750 GeV in \reffig{fig:2HDMSreach}, and  perfect alignment is assumed, the effect of varying these quantities is easy to deduce. First, different masses for the parent Higgs bosons would primarily affect the gluon fusion production cross section, whose scaling with mass is well known. The affect of misalignment would be quantitatively negligible on the reach of the Higgs cascades discussed here. However, various decay chains not considered in the above analysis, such as $(H\to h_{125} h_{125})$ or $(A\to Z h_{125})$, would be present. Such decays are  suppressed by either the NSM or S component of $h_{125}$ compared to decays into $h$. The reach for such decays can be extrapolated from those presented by the convolution of the relevant decay widths with the misalignment  of $h_{125}$ and identifying $m_h =125$ GeV in the right panel of \reffig{fig:2HDMSreach}.  Observe that the results presented can also be mapped to the case of the decoupled singlet, i.e. an effective 2HDM, with non-degenerate CP-odd and CP-even NSM-like Higgs bosons, by appropriately choosing the NSM and S components for the parent and daughter Higgs bosons in the decay chain of interest.

Finally, note that \reffig{fig:2HDMSreach} is meant to summarise only the prospects of exploring the 2HDM+S parameter space using Higgs cascades. The regions displayed do not take into account existing bounds from searches for additional Higgs boson beyond $h_{125}$. In particular, the charged Higgs does not play a role in any of the searches shown. Existing constraints on charged Higgs bosons, e.g. from flavor physics observables, can be satisfied by choosing a sufficiently large mass of the charged Higgs. Recall that the masses of the physical Higgs bosons are treated as free parameters. Even without considering effects of mixing, mass splittings of order of a few 100\,GeV between the mostly doublet-like pseudo-scalar and the charged Higgs are easily achievable. Furthermore, in more complete models with larger particle content than the 2HDM+S considered here, such as the NMSSM, indirect observables such as those from flavor physics receive additional contributions beyond those from the charged Higgs which may loosen the bounds on the mass of the charged Higgs, cf. Refs.~\cite{Altmannshofer:2012ks,Carena:2018nlf}.

In summary, as evident from \reffig{fig:2HDMSreach}, there are large regions of parameter space in reach of the different Higgs cascade search modes. In particular,  Higgs cascades enable the LHC to probe regions of parameter space challenging to access with traditional searches for the direct decays of additional Higgs states: singlet-like light states are difficult to directly produce due to the small couplings to pairs of SM particles. On the other hand, doublet-like states are readily produced, but if their mass is above the kinematic threshold allowing for decays into pairs of top quarks, for low $\tan\beta$, $\Phi \to t\bar{t}$ decays will dominate over the decays into other SM states. Pairs of top quarks produced from an $s$-channel resonance are very difficult to detect at the LHC due to interference effects with the QCD background, which makes the $m_\Phi \gtrsim 350\,$GeV, low $\tan\beta$ region extremely challenging to probe at the LHC through direct Higgs decays with current search strategies~\cite{Dicus:1994bm,Barcelo:2010bm,Barger:2011pu,Bai:2014fkl,Jung:2015gta,Craig:2015jba,Gori:2016zto,Carena:2016npr}.
Detailed LHC benchmarks optimising qualitative features of the different cascade decay signals discussed here, including alignment suppressed decays into pairs of $h_{125}$, are presented in Ref.~\cite{Baum:2019pqc}.

%% file: BSMresonance/BSMhMSSM.tex
\subsection[hMSSM]{hMSSM \\
\contrib{S.~Liebler, M.~M\"uhlleitner, M.~Spira}
}
\label{sec:hmssm}

An effective approximation of the MSSM for scenarios of large
SUSY particle masses but small and moderate values of the Higgsino mass
parameter $\mu$ relative to the stop masses is provided by the hMSSM
\cite{Djouadi:2013uqa, Djouadi:2013vqa, Maiani:2013hud,
Djouadi:2015jea}. This approach starts from the scalar Higgs mass matrix
including radiative corrections,
\begin{eqnarray}
{\cal M}^2 = &M_{Z}^2
\left(
\begin{array}{cc}
  c^2_\beta & -s_\beta c_\beta \\
 -s_\beta c_\beta & s^2_\beta \\
\end{array}
\right)
+ M_{A}^2
\left(
\begin{array}{cc}
 s^2_\beta & -s_\beta c_\beta \\
 -s_\beta c_\beta& c^2_\beta \\
\end{array}
\right)
+
\left(
\begin{array}{cc}
 \Delta {\cal M}_{11}^2~~ &  \Delta {\cal M}_{12}^2 \\
 \Delta {\cal M}_{12}^2~~ &\Delta {\cal M}_{22}^2 \\
\end{array}
\right) \,,
\end{eqnarray}
where we use the short--hand notation $s_\beta \equiv \sin\beta$, etc.,
and introduce the radiative corrections through the general matrix
elements $\Delta {\cal M}_{ij}^2$. The hMSSM approach starts by neglecting diagonal and off-diagonal entries of the radiative corrections,
\begin{equation}
\Delta {\cal M}_{11}^2 \sim \Delta {\cal M}_{12}^2 \sim 0 \,,
\end{equation}
and just keeping $\Delta {\cal M}_{22}^2$. Since the off-diagonal entries
are proportional to the $\mu$ parameter, this approximation restricts $\mu$
to small or moderate values in comparison with the other SUSY masses, thus
excluding large $\mu$ parameters \cite{Djouadi:2015jea, Bagnaschi:2015hka,
Lee:2015uza}. In this way, all radiative corrections can
be described by the parameter $\epsilon$,
\begin{equation}
\epsilon = \Delta \mathcal{M}_{22}^2 = \frac{M_h^2(M_A^2+M_Z^2-M_h^2)-M_A^2M_Z^2c^2_{2\beta}}{M_Z^2c_\beta^2+M_A^2s_\beta^2-M_h^2} \,,
\end{equation}
which is related to the pseudo-scalar mass $M_A$, the parameter $\tan\beta =
v_2/v_1$ that is determined by the vevs of the two neutral CP-even Higgs
fields, and the light scalar Higgs mass $M_h$. In this way the radiative
corrections are traced back to the known light scalar Higgs mass that is
identified with the mass of the discovered SM-like Higgs boson $M_h=125$
GeV. The remaining parameters of the MSSM Higgs sector are given by
\begin{eqnarray}
M_{H}^2 & = & \frac{(M_{A}^2+M_{Z}^2-M_{h}^2)(M_{Z}^2 c^{2}_{\beta}
+M_{A}^2
s^{2}_{\beta}) - M_{A}^2 M_{Z}^2 c^{2}_{2\beta} } {M_{Z}^2 c^{2}_{\beta}+M_{A}^2
s^{2}_{\beta} - M_{h}^2}\,, \nonumber \\
\alpha & = & -\arctan\left(\displaystyle\frac{ (M_{Z}^2+M_{A}^2)
c_{\beta} s_{\beta}} {M_{Z}^2
c^{2}_{\beta}+M_{A}^2 s^{2}_{\beta} - M_{h}^2}\right) \,, \nonumber \\
M_{H^\pm}^2 & = & M_A^2 + M_W^2 \, ,
\end{eqnarray}
where $M_H$ denotes the heavy scalar Higgs mass, $\alpha$ the CP-even
mixing angle, $M_{H^\pm}$ the charged Higgs mass and $M_W$ the $W$ mass.
The upper bound on the light scalar Higgs mass is lifted to
\begin{equation}
M_h^2 \le M_Z^2 \cos^2 2\beta + \epsilon \sin^2 \beta \, .
\end{equation}

The hMSSM determines all Higgs masses and mixing angles by three input
parameters, $M_A$, $M_h$ and $\tan\beta$. The hMSSM approach, however, can be
understood as an approximation to a low-energy 2HDM with heavy SUSY
particles being integrated out \cite{Chalons:2017wnz, Liebler:2018zul}.
The parameter $\epsilon$ then plays the role of the matching of the
low-energy 2HDM to the full MSSM. This point of view allows to extend
the simplified $\epsilon$ approximation to the Higgs self-couplings,
too. However, an explicit analysis revealed additional contributions to
the effective $\epsilon$ parameter for the Higgs self-couplings that are
determined by the top mass alone \cite{Liebler:2018zul},
\begin{equation}
\overline\epsilon=\epsilon - \frac{24\sqrt{2} G_F
m_t^4}{(4\pi)^2s_\beta^2}\frac{2}{3} \, .
\end{equation}
The trilinear Higgs self-couplings induce Higgs pair production processes at
the LHC and are given, in terms of this modified $\overline\epsilon$ parameter,
\begin{align}
\lambda_{hhh}^{\overline\epsilon}  &=  \lambda_{hhh} + \frac{3c_\alpha^3}{vs_\beta} \overline\epsilon\,,\qquad
&\lambda_{Hhh}^{\overline\epsilon}  &=  \lambda_{Hhh} + \frac{3s_\alpha c_\alpha^2}{vs_\beta} \overline\epsilon\,,
\nonumber \\
\lambda_{HHh}^{\overline\epsilon}  &=  \lambda_{HHh} + \frac{3 s_\alpha^2c_\alpha}{vs_\beta}\overline\epsilon\,,\qquad
&\lambda_{HHH}^{\overline\epsilon}  &=  \lambda_{HHH} +
\frac{3s_\alpha^3}{vs_\beta}\overline\epsilon\,,
\nonumber \\
\lambda_{hAA}^{\overline\epsilon}  &=  \lambda_{hAA} + \frac{c_\alpha c_\beta^2}{vs_\beta}\overline\epsilon\,,\qquad
&\lambda_{HAA}^{\overline\epsilon}  &=  \lambda_{HAA} + \frac{s_\alpha c_\beta^2}{vs_\beta}\overline\epsilon\,,
\end{align}
where the tree-level couplings are given (in terms of the radiatively
corrected mixing angle $\alpha$) by
\begin{align}
\label{eq:LOlambdas}
\lambda_{hhh}  &=  3 \frac{M_Z^2}{v}c_{2\alpha} s_{\alpha+\beta}\,,\qquad
&\lambda_{Hhh}  &=  \frac{M_Z^2}{v} ( 2 s_{2\alpha}s_{\alpha+\beta} - c_{2\alpha} c_{\alpha+\beta})\,,\nonumber \\
\lambda_{HHh}  &=  \frac{M_Z^2}{v} (- 2 s_{2\alpha} c_{\alpha+\beta} - c_{2\alpha} s_{\alpha+\beta})\,,\qquad
&\lambda_{HHH}  &=  3\frac{M_Z^2}{v} c_{2\alpha} c_{\alpha+\beta}\,, \nonumber \\
\lambda_{hAA}  &=  \frac{M_Z^2}{v} c_{2\beta} s_{\alpha+\beta}\,,\qquad
&\lambda_{HAA}  &=  - \frac{M_Z^2}{v} c_{2\beta} c_{\alpha+\beta}\,.
\end{align}
The vacuum expectation value $v$ is related to the Fermi constant by
$v=1/\sqrt{\sqrt{2} G_F}$. Using these radiatively corrected Higgs
masses, mixing angles and trilinear Higgs couplings, the corresponding
Higgs pair production processes can be investigated involving the
dominant radiative corrections within the hMSSM approximation.

%% file: BSMresonance/BSMNMSSM.tex
\subsection[NMSSM]{NMSSM \\
\contrib{P. Basler, S. Dawson, C. Englert,  M. M\"uhlleitner}
}

The NMSSM contains a complex gauge singlet superfield, $\hat{S}$, along with the $SU(2)_L$ doublet
superfields $\hat{H}_u$ and $\hat{H}_d$  of the MSSM
~\cite{Ellwanger:2009dp,Maniatis:2009re}. The additional
contribution to the superpotential due to $\hat{S}$ is,
\begin{eqnarray}
\Delta {\cal W} &=& 
- \epsilon_{ij} \lambda \hat{S} \hat{H}_d^i \hat{H}_u^j +
\frac{\kappa}{3} \, \widehat{S}^3 \, ,
\label{eq:superpotential}
\end{eqnarray}
with the $SU(2)_L$ indices $i,j=1,2$ and the totally antisymmetric tensor $\epsilon_{ij}$ where $\epsilon_{12}=\epsilon^{12}=1$.
The scalar component of the singlet, $S$, contributes 
 the trilinear soft SUSY breaking
interactions,
\begin{eqnarray}
\label{eq:Trilmass}
-{\cal L}_\text{tril}&=&  
- \epsilon_{ij} \lambda A_\lambda S H_d^i H_u^j
+ \frac{1}{3}
\kappa  A_\kappa S^3 \;.
\end{eqnarray}
 The set of six parameters describing the
tree-level NMSSM Higgs sector is
\begin{equation}
\lambda\ , \ \kappa\ , \ A_{\lambda} \ , \ A_{\kappa}, \
\tan \beta =v_u/ v_d \ , \
\mu_\mathrm{eff} = \lambda v_s/\sqrt{2}\; .
\end{equation}
The sign conventions are such that $\lambda$ and $\tan\beta$ are
positive, while $\kappa,A_\lambda,A_\kappa$ and $\mu_{\text{eff}}$ can
take both signs. Diagonalizing the scalar mass matrix gives  three CP-even mass eigenstates, $h$, $H_\downarrow$, and $H_\uparrow$, two
CP-odd mass eigenstates $A_\downarrow$ and $A_\uparrow$,  and
two charged Higgs bosons.
Here, $h$ denotes the SM-like Higgs boson and $H_{\downarrow}$ ($H_\uparrow$) the lighter (heavier) non-SM-like CP-even Higgs boson, $A_\downarrow$ and $A_\uparrow$ denote the lighter and heavier pseudoscalar, respectively. Note that we restrict ourselves to the CP-conserving NMSSM.
The Higgs boson masses are calculated from the input parameters. 

We  scan  the NMSSM parameter space \cite{Basler:2018dac} \footnote{In contrast to \cite{Basler:2018dac} HiggsSignals v.2.2.2 \cite{Bechtle:2013xfa,Bechtle:2008jh,Bechtle:2011sb,Bechtle:2013gu,Bechtle:2013wla} is used for the following part and the lower bound on the chargino mass has been relaxed to the LEP limit of $94$ GeV.} and require  that all experimental constraints from Higgs production, LHC SUSY searches, and  dark matter
limits
are satisfied. Within the allowed parameter space we are interested in double Higgs production with non-SM like signatures.
After satisfying the constraints, we  define an approximate ``exclusion luminosity''  at which single Higgs measurements would  become sensitive to a particular scenario. This allows
us to directly compare the discovery potential of  double-Higgs production to  single Higgs measurements and to identify interesting regions of the NMSSM parameter space.

The enlarged Higgs sector of the NMSSM leads to processes with two different Higgs bosons in the final state. Also the production of two pseudoscalars in the final state is possible. The cross sections can be enhanced relative to the SM \hh rate in the case of the resonant production of a heavy Higgs boson that decays into a pair of lighter Higgs bosons. Note, however, that due to supersymmetry, the Higgs self-coupling are determined in terms of the gauge couplings,  restricting large deviations from the SM Higgs self-coupling. The Higgs bosons can additionally decay into non-SM final states such as e.g. neutralinos, giving signatures with new and interesting features.


Scanning over NMSSM parameter points that meet all criteria and that have exclusion luminosities above $64~\text{fb}^{-1}$, the maximum  
 enhancement of the gluon fusion rate to $hh$ pairs is found to be slightly less than a factor of two. 
The $H_\downarrow H_\downarrow$  cross section
can become very large mainly because of the allowed smallness of the $H_\downarrow$ mass,
$m_{H_\downarrow} \sim 38$~GeV. The maximum value\footnote{All values are for a center of mass energy of $14$ TeV and at LO.  The inclusion of NLO QCD corrections roughly adds a factor of 2.} of $70~\text{fb}$
in $h A_\downarrow$ production is due to the rather small mass, $m_{A_\downarrow}=69$~ 
GeV, in combination with resonant 
$A_\uparrow$ production followed by the decay 
to $h A_\downarrow$.
 Finally, the enhancement in $A_\downarrow
A_\downarrow$ production  with a production cross section of $70~\text{fb}$ is due to the  smallness of the
$A_\downarrow$ mass of $m_{A_\downarrow}= 69$~GeV combined with
the resonant $H_\downarrow$ production 
decaying subsequently into 
$A_\downarrow A_\downarrow$. (The branching ratio
of $H_\uparrow\rightarrow A_\downarrow
A_\downarrow$ is very small for this parameter point.)

\begin{figure*}[t]
  \centering 
  \includegraphics[width=0.47\linewidth]{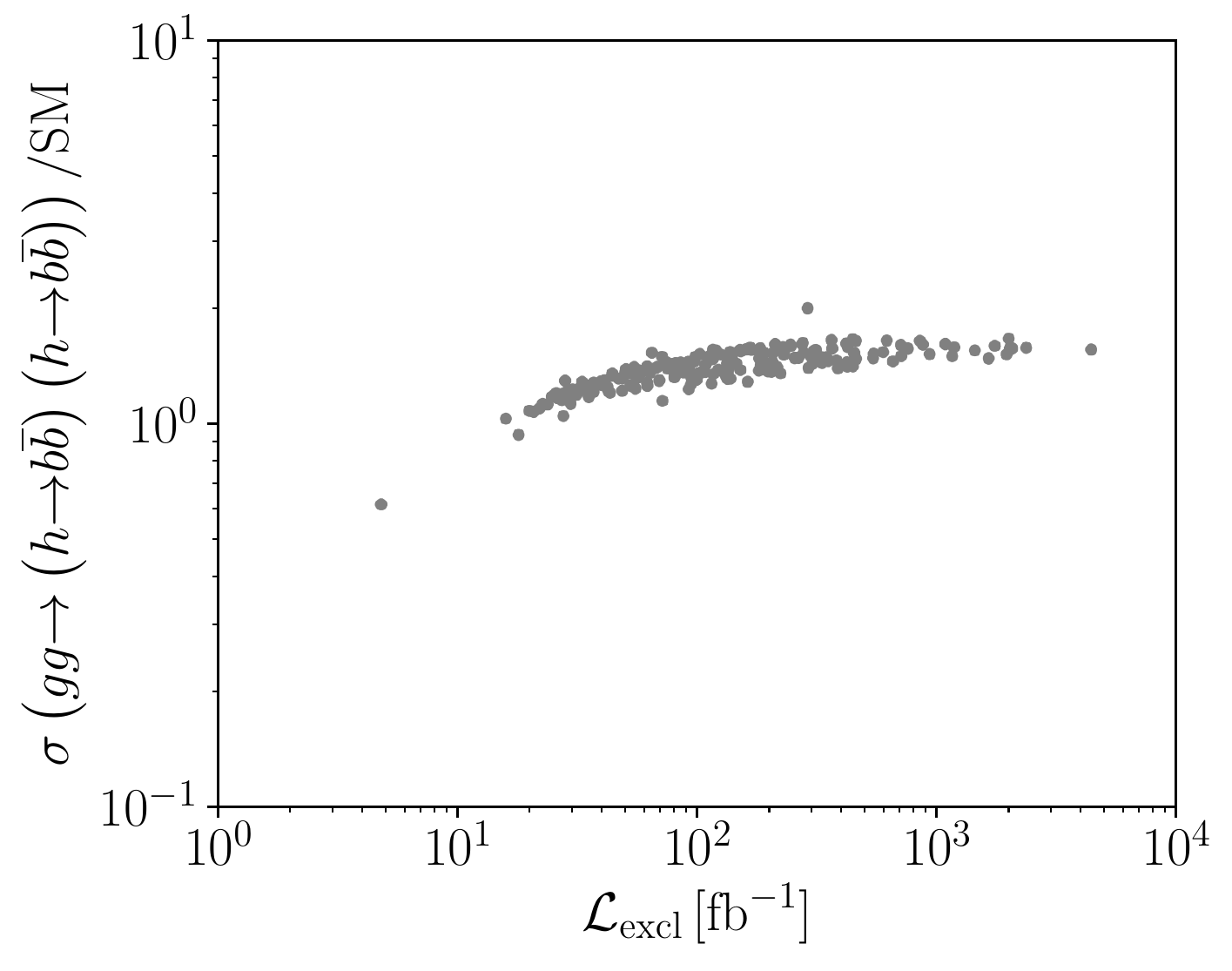}
  \hspace*{0.2cm}
  \includegraphics[width=0.47\linewidth]{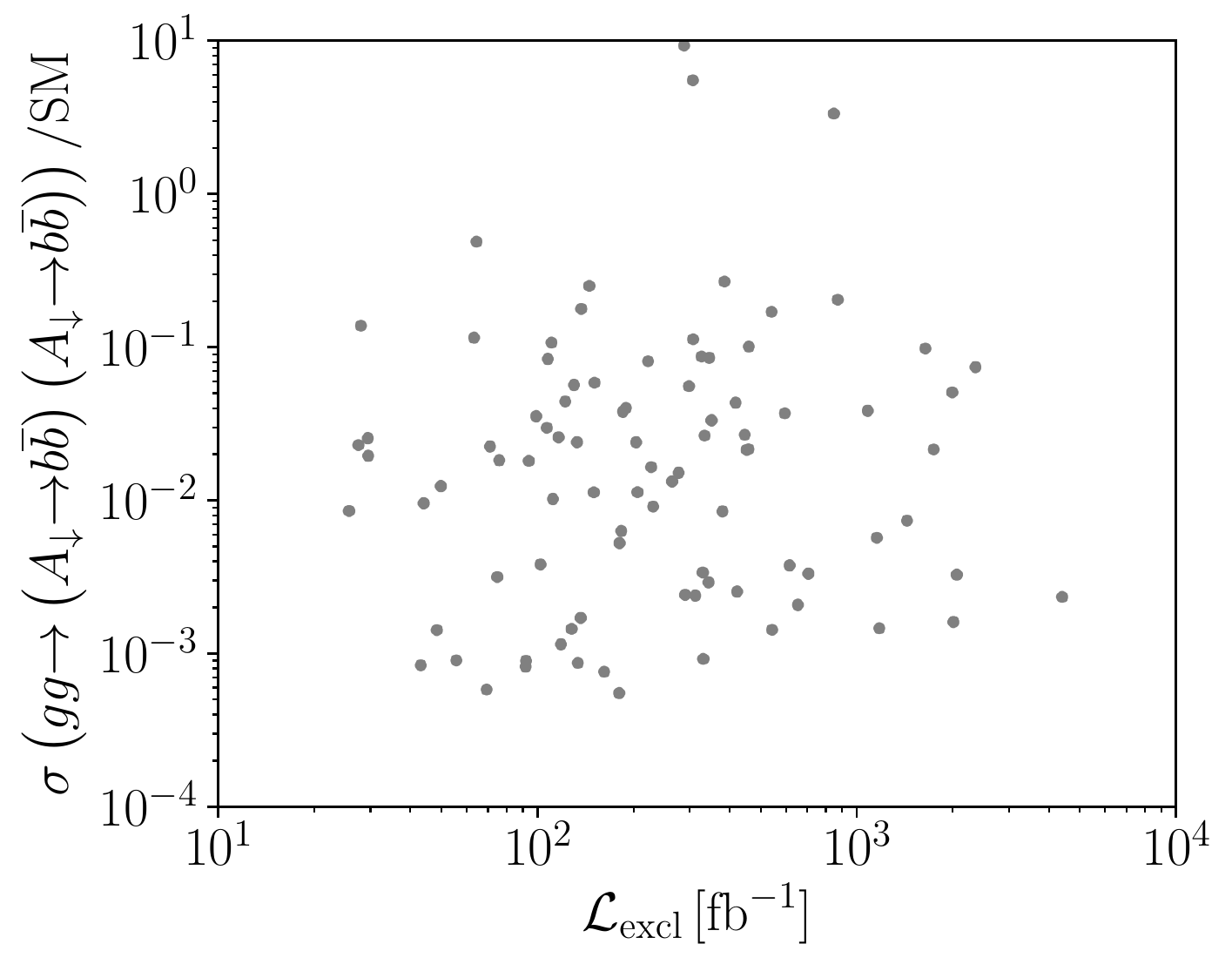}
  \caption{Scatter plots for NMSSM scenarios passing the applied
    experimental constraints: Higgs pair production cross sections normalized to
    the SM value for SM-like Higgs pairs decaying into
    \bbbbAlt (left) and $A_\downarrow A_\downarrow$ Higgs pairs
    decaying into \bbbbAlt (right) as a function of the
    exclusion luminosity~\cite{Basler:2018dac}. \label{fig:nmssmfinalversuslumi}} 
\end{figure*}

\begin{figure}[t]
  \centering
  \includegraphics[width=0.5\linewidth]{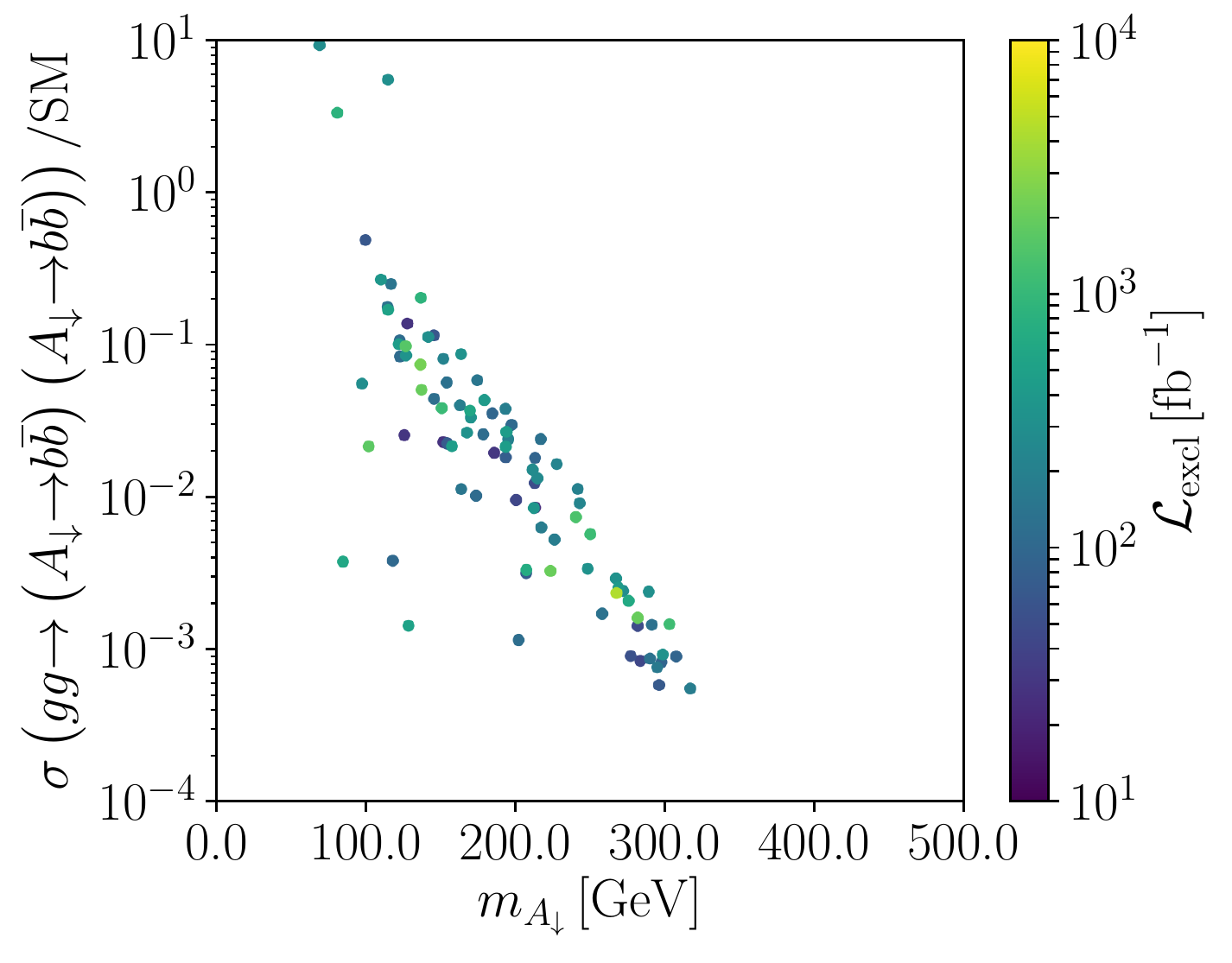}
  \caption{NMSSM: Scatter plots for \bbbbAlt  final state rates from
    $A_\downarrow A_\downarrow$ production normalized to the SM rate
    as a function of $m_{A_\downarrow}$. The colour code denotes the
    exclusion luminosity~\cite{Basler:2018dac}. \label{fig:diadownarrow}}
\end{figure}

In \reffig{fig:nmssmfinalversuslumi} we show the NMSSM cross sections for $hh$ pair production in the \bbbbAlt  final state (left) and for $A_\downarrow A_\downarrow$ production in the \bbbbAlt  final state (right) normalized to the corresponding SM values 
as a function of the exclusion luminosity and for all parameter points that pass the experimental restrictions.
As seen in \reffig{fig:nmssmfinalversuslumi}~(left),
the \bbbbAlt  final state rates from SM-like Higgs pair production 
exceed the SM $HH$ rate by at most a factor of 2 and  only for
higher exclusion luminosities. 
As can be inferred from Fig. \ref{fig:diadownarrow}~(right), the maximum value for $A_\downarrow A_\downarrow$ production with subsequent decay into \bbbbAlt is 9.3 compared to the SM value, at an exclusion luminosity of 
$287~\text{fb}^{-1}$.  The maximum value found for the production of a SM like Higgs boson $h$ together with $H_{\downarrow}$ and subsequent decay into \bbbbAlt 
(not shown here), has an enhancement of~$\sim$~4.6 at an exclusion luminosity of $449~\text{fb}^{-1}$.

Because the 
 light pseudoscalar, $A_\downarrow$, can be
relatively light and decays dominantly into $b\bar{b}$,  
the enhancement factors
can be up to $\sim$ 5- 10 in these processes.
\reffig{fig:diadownarrow}   shows the production of $A_\downarrow
A_\downarrow$ with  decay into the \bbbbAlt final state normalized
to   SM  di-Higgs $HH$
production decaying into the \bbbbAlt final state, as a function of the mass of the light pseudoscalar. The
color code denotes the exclusion luminosity. For masses
below 125~GeV, the rates are  enhanced because of the large
di-Higgs production cross sections.
Above the top-pair threshold, the exclusion luminosities are on average lower than below the threshold due to the exclusion limits in the top-pair final state.
 For masses below the SM-like Higgs
mass, however, there are parameter points where the exclusion
luminosities can exceed $100~\text{fb}^{-1}$ 
up to about $900~\text{fb}^{-1}$ while
still featuring enhanced rates. The reason that these points are not 
excluded from single Higgs searches is that light Higgs states with
dominant decays into $b\bar{b}$ final states are difficult to
probe. On the other hand this enhancement combined with the large
di-Higgs production cross section implies significant  \bbbbAlt final state rates that
may be tested at the high luminosities. This is an example of the
interplay between difficult single-Higgs searches and large exotic
di-Higgs rates, where new physics may be found. 

%% file: BSMresonance/BSMGM.tex
\subsection[Georgi-Machacek Model]{Georgi-Machacek Model \\
\contrib{H.~E.~Logan}
}

The Georgi-Machacek (GM) model~\cite{Georgi:1985nv,Chanowitz:1985ug} is an extended Higgs model whose scalar sector consists of the usual complex isospin doublet $(\phi^+, \phi^0)$ with hypercharge $Y = 1/2$, a real triplet $(\xi^+, \xi^0, \xi^0)$ with $Y = 0$, and a complex triplet $(\chi^{++}, \chi^+, \chi^0)$ with $Y = 1$.  The scalar potential is constructed to preserve a global SU(2)$_L \times$SU(2)$_R$ symmetry so that custodial symmetry is preserved after electroweak symmetry breaking, ensuring that the electroweak $\rho$ parameter is equal to one at tree level.  Additional details can be found in Sec.~IV.4.4 of Ref.~\cite{deFlorian:2016spz}.

In addition to a light custodial-singlet scalar $h$, usually identified with the SM-like 125~GeV Higgs boson, the physical spectrum contains a custodial fiveplet $(H_5^{++}$, $H_5^+$, $H_5^0$, $H_5^-$, $H_5^{--})$ with common mass $m_5$, a custodial triplet $(H_3^+$, $H_3^0$, $H_3^-)$ with common mass $m_3$, and a heavier custodial singlet $H$ with mass $m_H$.  Here we focus on the decay $H \to hh$.  Custodial symmetry forbids $H_3^0$ and $H_5^0$ from decaying into $hh$.

$H$ can be produced at the LHC via the same processes as a heavy SM Higgs boson.  For $H$ masses above $2 m_h$, the only relevant production modes are gluon fusion and vector boson fusion.  We compute the signal cross sections for the 13~TeV LHC as follows, focusing on the \hhbbbb final state:
\begin{eqnarray}
	\sigma(gg \to H \to hh \to \bbbbAlt) &=& \sigma(gg \to H_{\rm SM}) \times (\kappa_f^H)^2 
		\times {\rm BR}(H \to hh) \nonumber \\
		&& \times [{\rm BR}(h \to \bb)]^2, 
		\label{eq:BSMGM1} \\
	\sigma({\rm VBF} \to H \to hh \to \bbbbAlt) &=& \sigma({\rm VBF} \to H_{\rm SM}) \times (\kappa_V^H)^2 
		\times {\rm BR}(H \to hh) \nonumber \\
		&& \times [{\rm BR}(h \to \bb)]^2.
		\label{eq:BSMGM2}
\end{eqnarray}
We take the SM cross sections $\sigma(gg \to H_{\rm SM})$ and $\sigma({\rm VBF} \to H_{\rm SM})$ from Ref.~\cite{deFlorian:2016spz}, where for the gluon fusion process we use the cross sections computed to NNLO+NNLL QCD accuracy.  The remaining factors are computed using the public code GMCALC~1.4.1~\cite{Hartling:2014xma}.  $\kappa_f^H$ and $\kappa_V^H$ are the coupling modification factors for $H$ couplings to fermion pairs and vector boson pairs, respectively.  The branching ratio of $H \to hh$ depends on a combination of the parameters of the scalar potential.  The branching ratio of $h \to b \bar b$ depends mainly on the custodial-singlet Higgs boson mixing angle and the triplet scalar vacuum expectation value, and is constrained by LHC Higgs signal strength measurements to be close to its SM value.

\begin{figure}
\resizebox{0.5\textwidth}{!}{\includegraphics{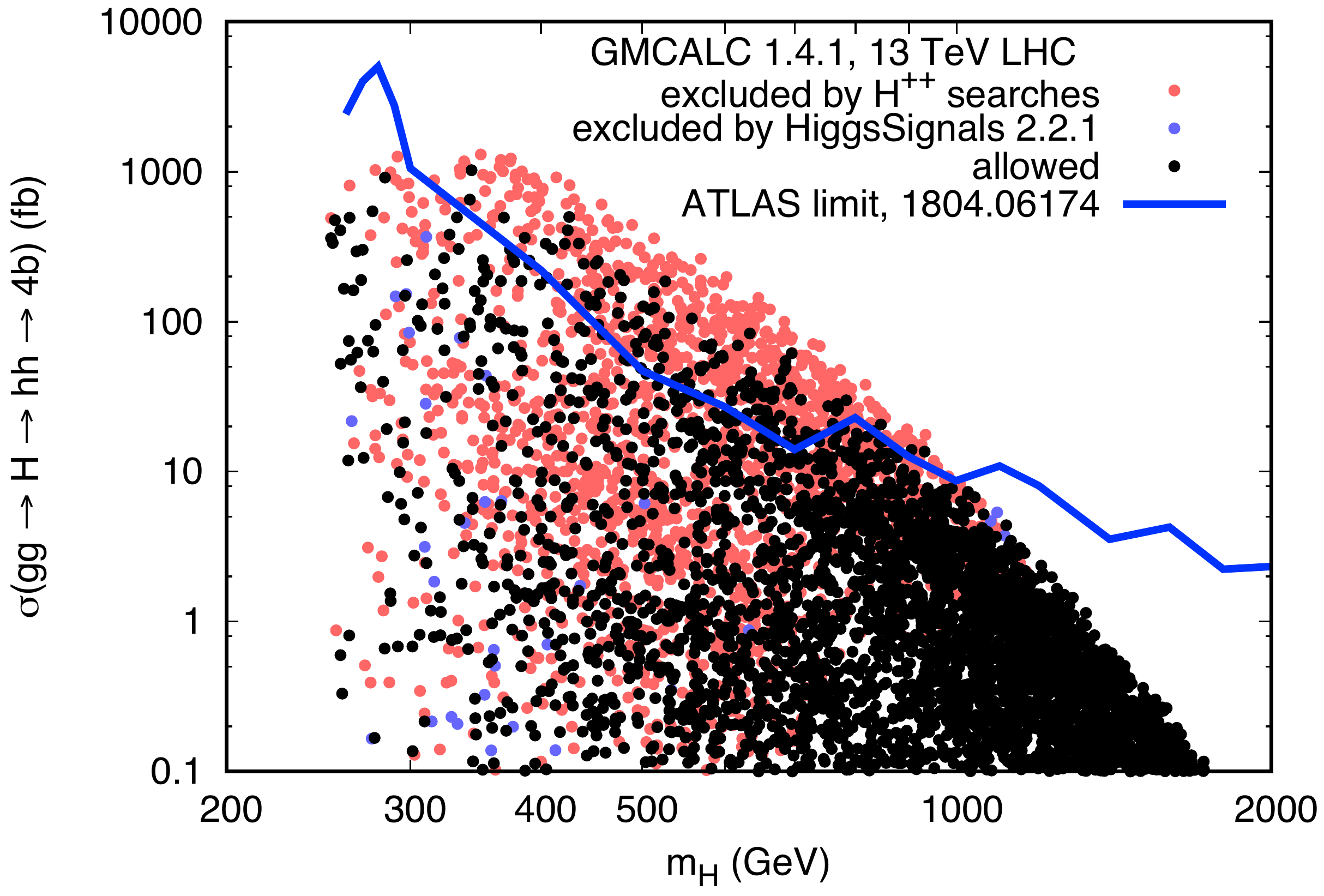}}%
\resizebox{0.5\textwidth}{!}{\includegraphics{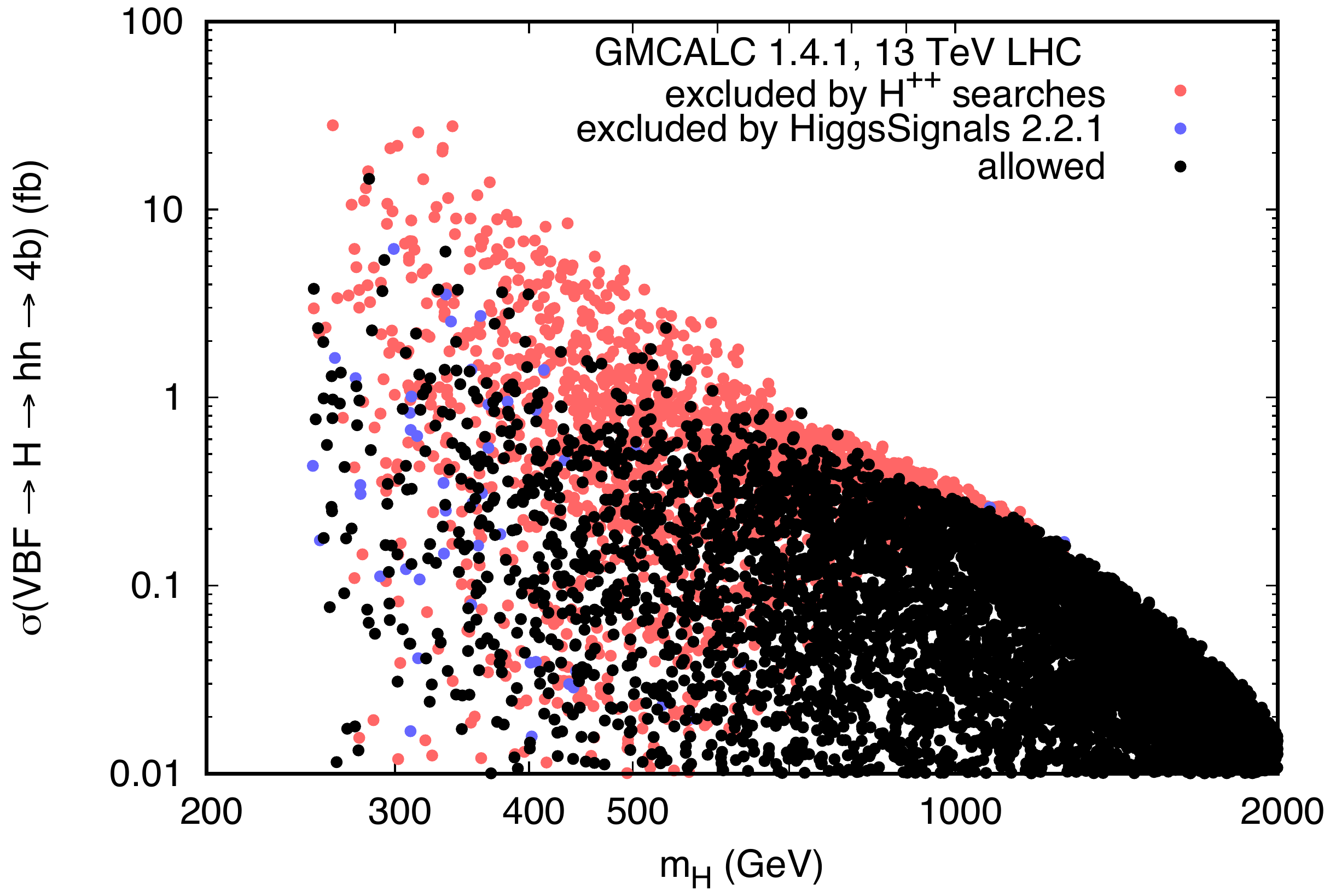}}
\caption{Cross sections for $H \to hh \to \bbbbAlt$ in 13~TeV $pp$ collisions produced via gluon fusion (left) or vector boson fusion (right) as a function of the $H$ mass in the Georgi-Machacek model.  The points represent a scan over the full model parameter space imposing only theoretical constraints.  Red and violet points are already excluded by other searches.}
\label{fig:BSMGM-scans}
\end{figure}

We scan over the full GM model parameter space, requiring that the scalar quartic couplings satisfy perturbative unitarity constraints~\cite{Aoki:2007ah,Hartling:2014zca} and that the potential is bounded from below and has no deeper minima than the desired vacuum~\cite{Hartling:2014zca}.  The resulting signal cross sections are shown in \reffig{fig:BSMGM-scans} as a function of the $H$ mass.  We include only the resonant processes of \refeqs{eq:BSMGM1} and (\ref{eq:BSMGM2}), and do not consider interference with the non-resonant SM-like $pp \to h^* \to hh \to \bbbbAlt$ process (for comparison, the \emph{total} SM cross sections times branching ratios for the non-resonant $gg \to hh \to \bbbbAlt$ and VBF$\to hh \to \bbbbAlt$ processes are 10~fb~\cite{Grazzini:2018bsd} and 0.55~fb~\cite{deFlorian:2016spz}, respectively).

In red we indicate the scan points that are excluded by existing LHC searches other than $H \to hh$.  The most stringent of these is a CMS search for doubly-charged scalar production in VBF with decays to like-sign $W$ bosons using 35.9~fb$^{-1}$ of $pp$ data at 13~TeV~\cite{Sirunyan:2017ret}, which sets an upper bound on the production cross section of $H_5^{\pm\pm}$ as a function of its mass.  

In violet we indicate the scan points that are allowed by direct searches but excluded by measurements of the 125~GeV Higgs boson properties.  We apply the constraint by using HiggsSignals 2.2.1~\cite{Bechtle:2013xfa} to compute a p-value, which we require to be larger than 0.05 for the point to be allowed at the 95\% confidence level.  Because we want to apply the constraint separately for each scan point, we take the number of free model parameters to be zero in the calculation of the p-value.  This maximises the p-value and (conservatively) excludes the smallest number of points.
The black points in \reffig{fig:BSMGM-scans} are still allowed after applying these constraints.  

The thick blue line in the left panel of \reffig{fig:BSMGM-scans} shows the current ATLAS limit on $\sigma(pp \to {\rm Scalar} \to hh \to \bbbbAlt)$ using 27.5--36.1~fb$^{-1}$ of $pp$ data at 13~TeV~\cite{Aaboud:2018knk}.  This search already excludes new parameter space in the GM model for $m_H$ between about 300~GeV and 1~TeV that is not otherwise constrained by previous searches. The model therefore serves as a useful benchmark for interpreting $H \to hh$ searches that will be performed using the full LHC Run 2 dataset. 

%% file: BSMresonance/BSMparticleinloop.tex
\section[New Particles in the Loop]{New Particles in the Loop \\
\contrib{S.~Dawson, I.~M.~Lewis}
} \label{sec:particleinloop}

BSM physics can contribute to di-Higgs production through
 new colored scalars \cite{Belyaev:1999mx,BarrientosBendezu:2001di,Asakawa:2010xj,Kribs:2012kz,Enkhbat:2013oba,Heng:2013cya} or fermonic \cite{Liu:2004pv,Dib:2005re,Pierce:2006dh,Ma:2009kj,Han:2009zp,Dawson:2012mk,Edelhaeuser:2015zra,Chen:2014xra} particles contributing to the loop amplitudes.  If new particles get their masses from a different source than the Higgs, the contributions to single and double Higgs production can be different~\cite{Pierce:2006dh}. These new particles can then significantly change the rates as well as the kinematic distributions in double production and keep single Higgs production close to the SM prediction \cite{Asakawa:2010xj,Batell:2015koa,Huang:2017nnw}.
\subsubsection{Heavy VLQs}
To be consistent with single Higgs rates, new heavy quarks cannot get all their mass from the Higgs mechanism and must be vector-like~\cite{Kribs:2007nz}.  We focus on two cases:  $SU(2)_L$ singlet up-type vector-like quark (VLQ) $U$ and a full generation of up- and down-type VLQs.
\begin{eqnarray}
Q=\begin{pmatrix} T\\ B\end{pmatrix},\,U,\,D,
\end{eqnarray}
where $Q$ is a vector-like $SU(2)_L$ doublet, $U$ is an up-type vector-like $SU(2)_L$ singlet, and $D$ is a down-type vector-like $SU(2)_L$ singlet.  For simplicity and to avoid low energy constraints, we only consider mixing with the third generation SM quarks:
\begin{eqnarray}
q_L=\begin{pmatrix} t_L\\ b_L\end{pmatrix}, t_R, b_R.
\end{eqnarray}

\subsubsection{Singlet VLQ}
The singlet VLQ, third generation quarks, and Higgs boson couple via
\begin{eqnarray}
\mathcal{L}=-\lambda_b \overline{q}_L\Phi b_R-\lambda_t \overline{q}_L\widetilde{\Phi}t_R-\lambda_1\overline{q}_L\widetilde{\Phi}U_R+M_1\,\overline{U}_Lt_R+M_2\,\overline{U}_LU_R+{\rm h.c.}\label{eq:VLQsinglet}
\end{eqnarray}
Since $U_R$ and $t_R$ have the same quantum numbers, $M_1$ can be rotated via a field redefinition between $U_R$ and $t_R$~\cite{Dawson:2012mk}.  Hence, there are four physical free parameters: the bottom quark mass $m_b$, the observed top quark mass $m_t=173$~GeV, the heavy top partner mass $m_T$, and the left-handed mixing angle between the top quark and top partner $\theta_L$.  The right-handed mixing angle $\theta_R$ can be determined by the Higgs vev, $m_t$, $m_T$, and $\theta_L$~\cite{Dawson:2012mk}.  Electroweak precision constraints constrain $\sin\theta_L\lesssim 0.16-0.12$ for $m_T\sim1-2$~TeV~\cite{Chen:2017hak,Dawson:2012di,Aguilar-Saavedra:2013qpa}.

\begin{figure}
\begin{center}
\includegraphics[width=0.438\textwidth]{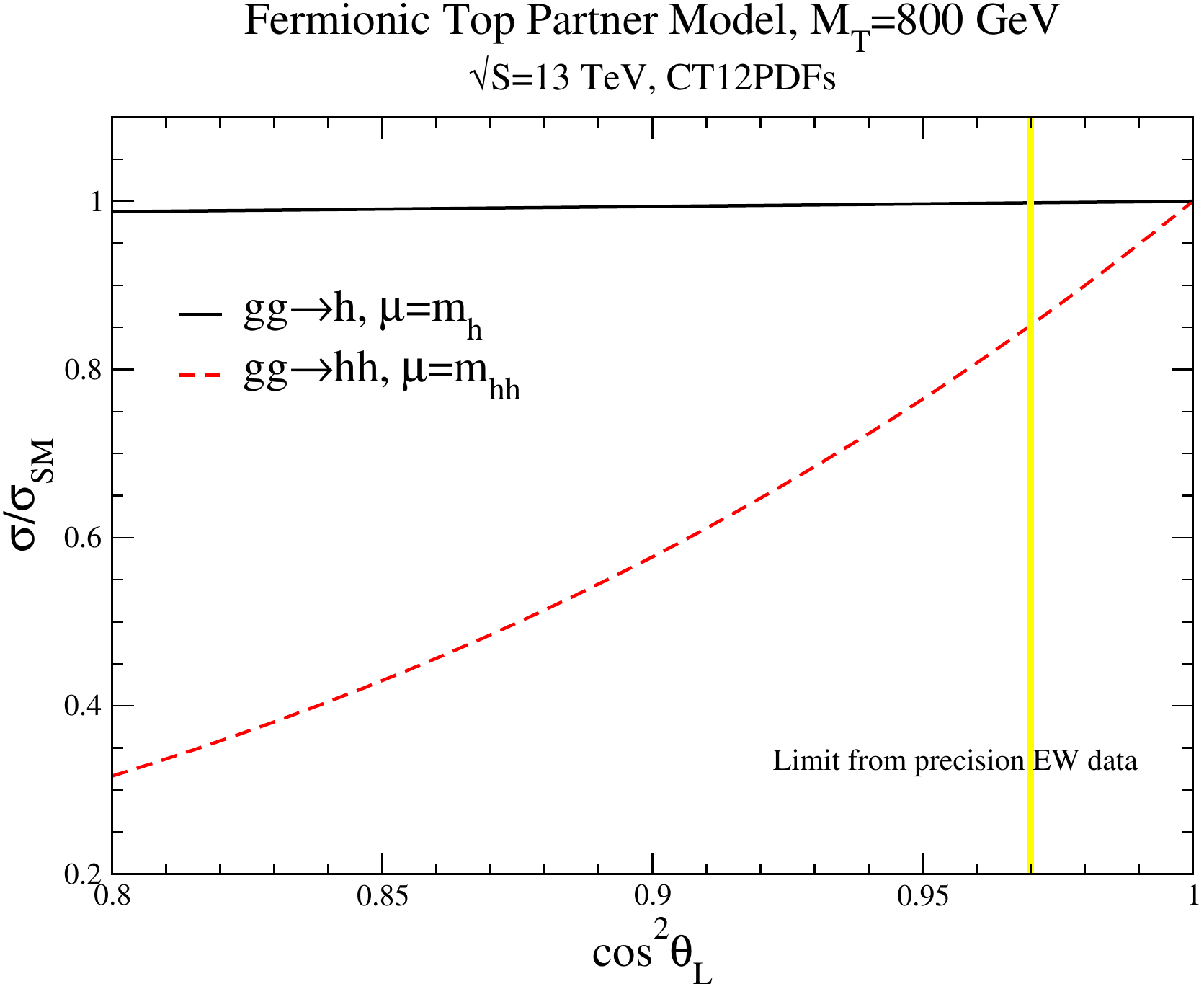}
\hspace*{0.3cm}
\includegraphics[width=0.47\textwidth]{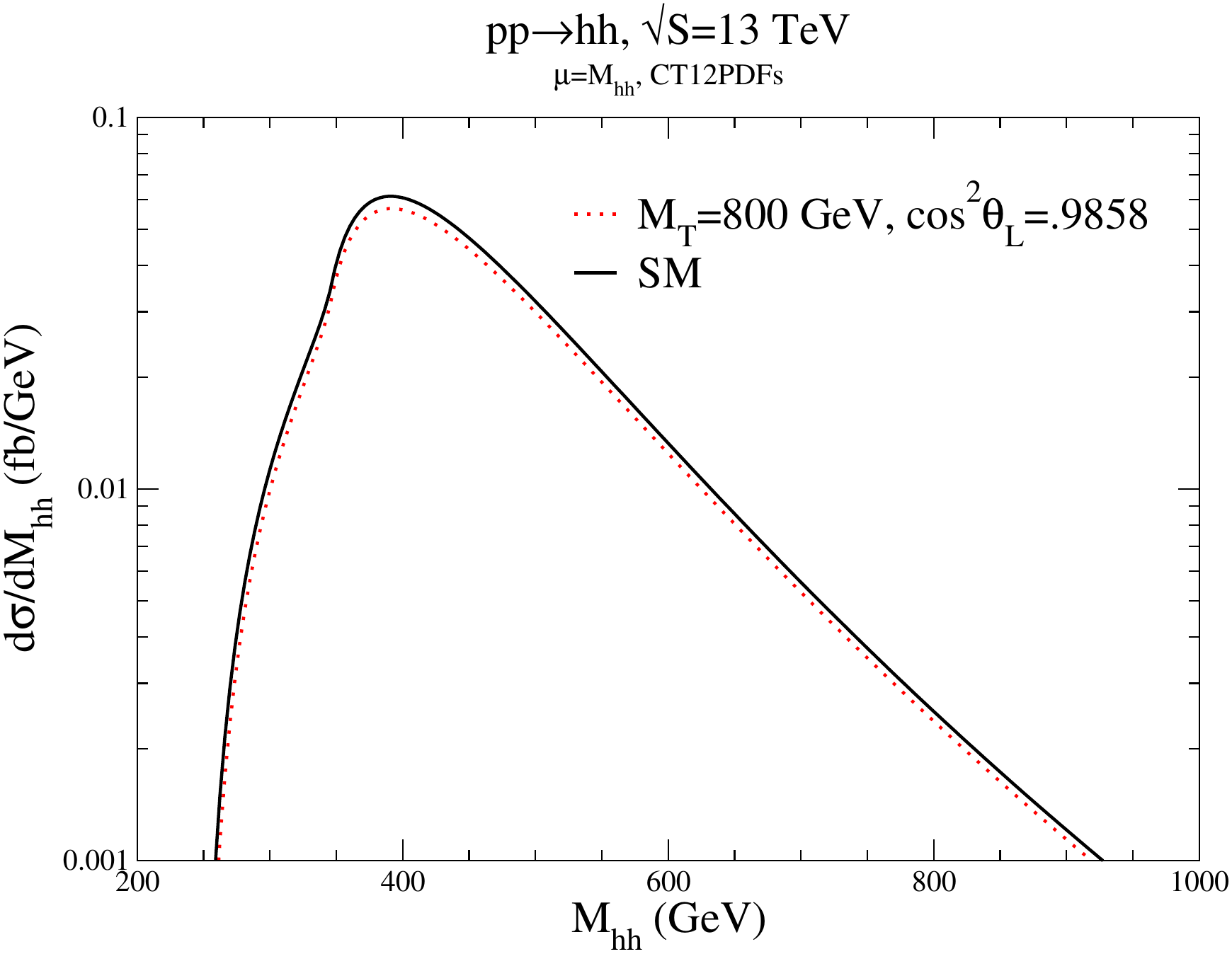}
\end{center}
\vspace{-0.5cm}
\caption{Left: Deviations in (solid black) single Higgs and (red dashed) double Higgs production away from SM predictions in the singlet VLQ model. Vertical yellow band indicates EW precision constraints on $\theta_L$ and allowed values are to the right.  Right: Di-Higgs invariant mass distributions in (black) SM and (red dashed) singlet VLQ model in \refeq{eq:VLQsinglet}~\cite{Dawson:2015oha}.}
  \label{fig:VLQ1}
\end{figure}

In left hand side of \reffig{fig:VLQ1} we show deviations away from SM predictions for single and double Higgs production in the singlet VLQ model as a function of the mixing angle with $M_T=800$~GeV~\cite{Chen:2014xra}.  The single Higgs production is always nearly SM-like, while substantial deviations in double Higgs production are possible.  These deviations are always a suppression~\cite{Dawson:2012mk,Chen:2014xra}.  However, once (yellow solid) EW precision measurements are taken into account, double Higgs production is forced to be within $\sim15\%$ the SM value.    The kinematic distributions are nearly SM like, as shown in the right hand side of \reffig{fig:VLQ1}.  For higher top partner masses, EW precision constraints on $\theta_L$ become more stringent~\cite{Chen:2017hak,Dawson:2012di,Aguilar-Saavedra:2013qpa}, the effects of the top partner decouple more, and invariant mass distributions continue to be SM like~\cite{Chen:2014xra}.

\subsubsection{Full VLQ Generation}
A more complicated scenario is to assume a full generation of VLQs.   The couplings with the third generation quarks and Higgs are~\cite{Chen:2014xra}
\begin{eqnarray}
\mathcal{L}&=&-\lambda_b\overline{q}_L \Phi b_R-\lambda_t\overline{q}_L\widetilde{\Phi}t_R-M\overline{Q}_LQ_R-M_U\overline{U}_LU_R-M_D\overline{D}_LD_R\nonumber\\
&&-\lambda_1\overline{Q}_L\widetilde{\Phi}U_R-\lambda_2\overline{Q}_L\Phi D_R-\lambda_3\overline{Q}_R\widetilde{\Phi}U_L-M_4\overline{q}_LQ_R-M_5 \overline{U}_Lt_R\nonumber\\
&&-M_6\overline{D}_Lb_R-\lambda_{7}\overline{q}_L\widetilde{\Phi}U_R-\lambda_8\overline{q}_L\Phi D_R-\lambda_9\overline{Q}_L\widetilde{\Phi}t_R-\lambda_{10}\overline{Q}_L\Phi b_R\nonumber\\
&&-\lambda_{11}\overline{Q}_R\Phi D_L+{\rm h.c.}\label{eq:VLQgen}
\end{eqnarray}
There are the top quark, the bottom quark, two top partners $T_{1,2}$, and two bottom quark partners $B_{1,2}$ mass eigenstates after electroweak symmetry breaking.  The free parameters are the top quark mass, the bottom quark mass, the two top partner masses $M_{T_{1,2}}$, two bottom partner masses $M_{B_{1,2}}$, and twelve mixing angles.

\begin{figure}
\begin{center}
\includegraphics[width=0.5\textwidth]{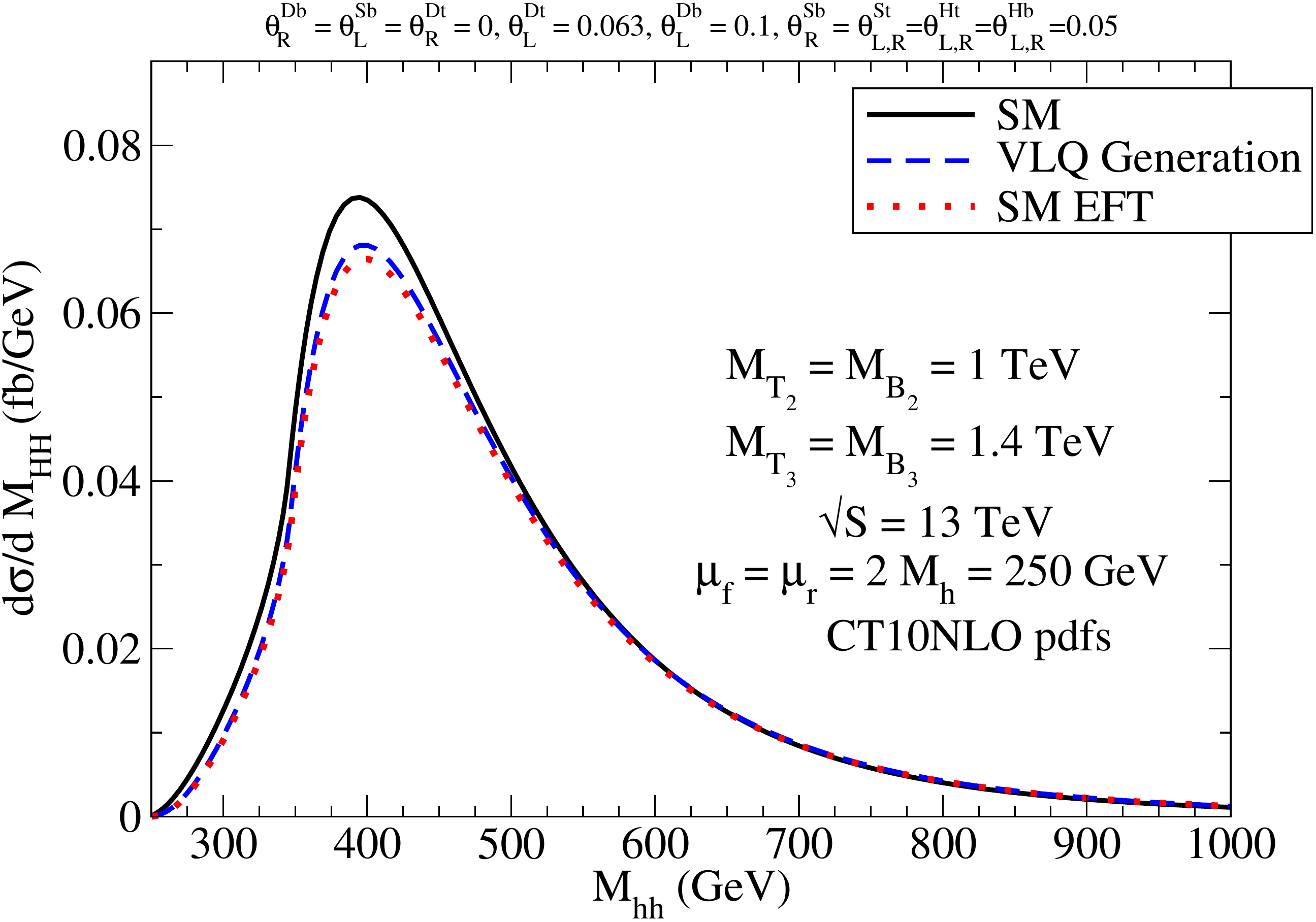}
\end{center}
\vspace*{-0.5cm}
\caption{Double Higgs invariant mass distribution for the (solid) SM, (blue dashed) VLQ generation described in \refeq{eq:VLQgen}, and (red dotted) the VLQs integrated out and matched onto the SM EFT~\cite{Chen:2014xra}.}
  \label{fig:VLQ2}
\end{figure}

The invariant mass distributions for this model are shown in \reffig{fig:VLQ2}.  The pattern of the top and bottom partner masses and values of mixing angles are chosen to be consistent with electroweak precision data, single Higgs rates, and maximize deviations in double Higgs rates~\cite{Chen:2014xra}.  After all constraints are taken into account, the (black solid) SM and (blue dashed) VLQ distributions are very similar.  

The red dotted line in \reffig{fig:VLQ2} shows the distribution calculated by integrating out the heavy top and bottom partners and matching on the SM EFT.  This introduces new point-like interactions between the gluons and Higgs boson and new four point interactions $t-t-h-h$~\cite{Chen:2014xra}.  The EFT agrees very well with the VLQ model in the region of validity $M_{hh}< M_{T_{1,2}},M_{B,{1,2}}$.  In the EFT it is clear that the single and double Higgs rates depend on the same parameters. Hence, the two rates are tightly related, and indeed de-correlated~\cite{Chen:2014xra}.  That is, if the single Higgs rate increases the double Higgs rate is decreased.  Since single and double Higgs rates are bound together, we must go to a region of parameters space where new particles are light and the EFT is not valid~\cite{Asakawa:2010xj,Batell:2015koa,Huang:2017nnw}.

\subsubsection{Colored scalars}
We consider  an $SU(2)_L$ singlet,  $SU(3)_c$ complex scalar, $s$,
\begin{eqnarray}
{\cal L}_{s,c}&=&(D_\mu s)^*(D^\mu s)-m_0^2 s^*s -\frac{\lambda_s}{2} (s^*s)^2-\kappa s^*s \left|H^\dagger H\right|+{\cal L}_{SM}
\, ,
\label{eq:lsc}
\end{eqnarray}
where $H$ is the SM $SU(2)_L$ doublet with $\langle H\rangle =(0, v/\sqrt{2})^T$. If the scalar, $s$, is real,
\begin{eqnarray}
{\cal L}_{s,r}&=&\frac{1}{2} (D_\mu s)(D^\mu s)
-\frac{m_0^2}{2}s^2 -\frac{\lambda_s}{4} s^4
-\frac{\kappa}{2}s^2 \left|H^\dagger H\right|+{\cal L}_{SM}\, .
\end{eqnarray}
The physical mass for either a real or complex scalar is, 
$m_s^2=m_0^2+\frac{\kappa v^2}{2}$ and 
  $m_0=0$ is the limit where the scalar gets all of its mass from electroweak symmetry breaking.

 \subsubsection{Scalar Top Partners}
\begin{figure}[t]
\begin{center}
\includegraphics[scale=0.7]{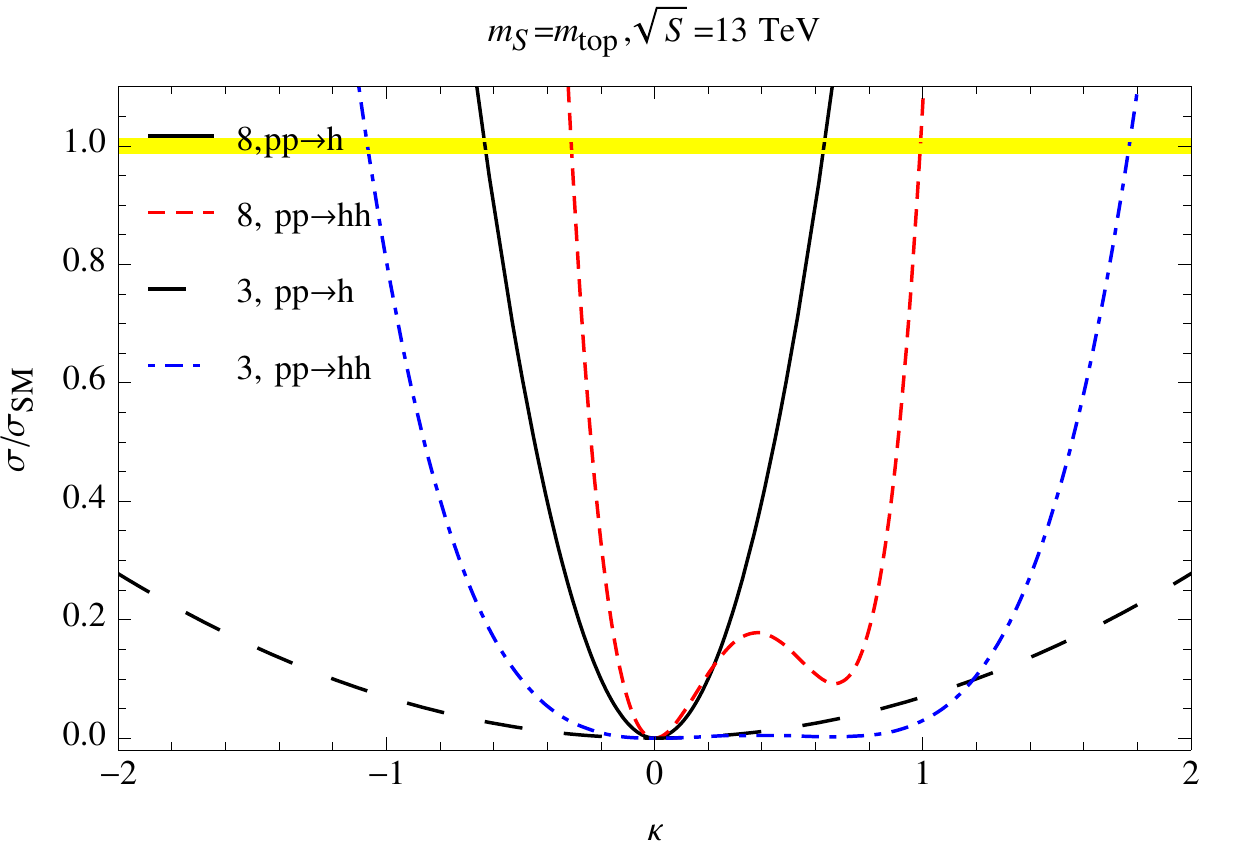}
\end{center}
\vspace{-0.5cm}
\caption{Comparison of $1h$ (dashed black) and $2h$ production (blue dot-dash) to the SM rate, when the SM top quark is replaced by a color
triplet scalar with mass, $m_s=173 $ GeV.  The solid black (red dashed) curves correspond to the ratios  to the SM predictions for $1h$ and $2h$ with a color
octet 
scalar replacing the top quark~\cite{Dawson:2015oha}. 
}
  \label{fig:scal1}
\end{figure}

\begin{figure}[t]
\begin{center}
\includegraphics[scale=0.7]{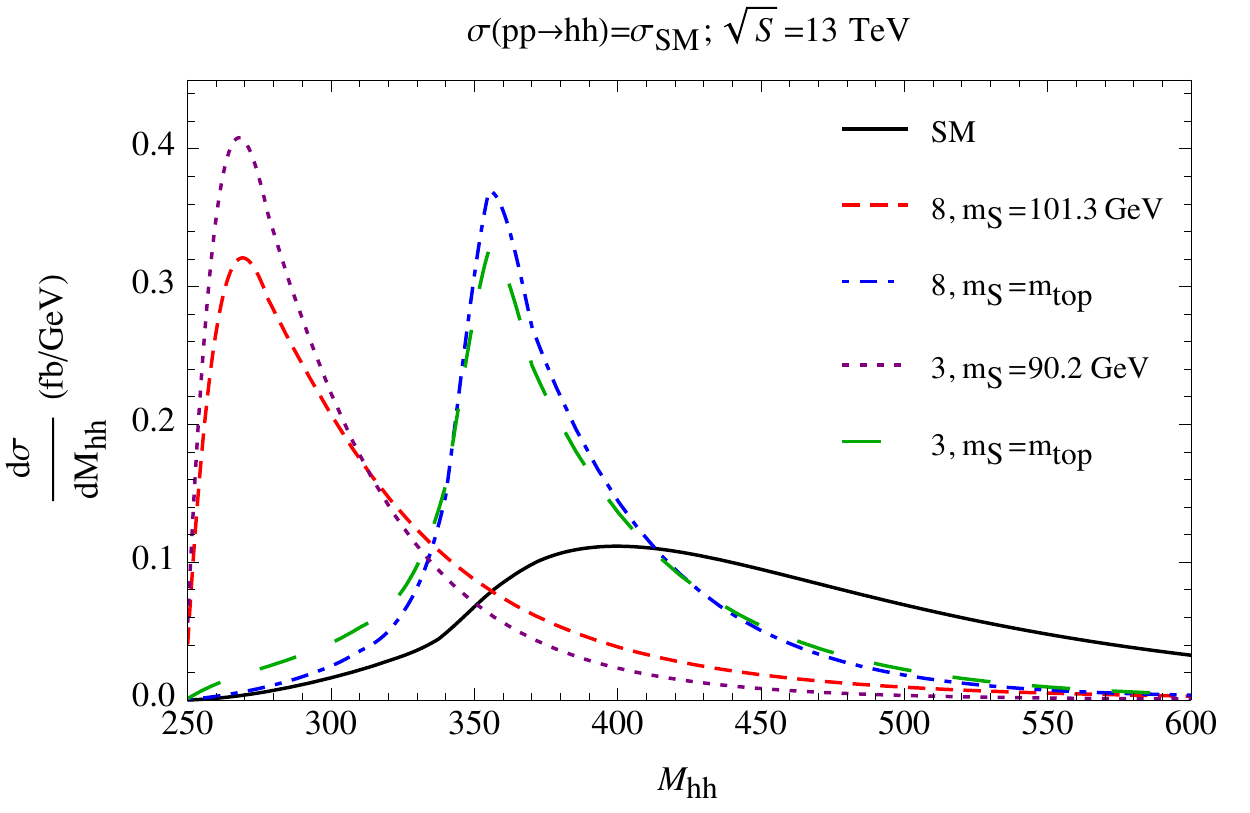}
\end{center}
\vspace{-0.5cm}
\caption{Distributions for $2h$ production when the parameters are tuned to give the SM total cross sections for $1h$ and $2h$ production~\cite{Dawson:2015oha}.
}
  \label{fig:scal3}
\end{figure}

We compare the results  for \hh production when the loop particles are the SM top  with 
those for  a colored scalar  with  $m_s=m_t=173$~GeV.  \reffig{fig:scal1} shows the ratio of the total cross
sections for both $1h$ and $2h$ production, normalized to the lowest order SM predictions in this scenario.  
We note that in order to reproduce the SM rate for $1h$ production using a color triplet scalar (the black dashed line), $\kappa$ needs to be quite large, $\kappa \gtrsim 2$.  If $\kappa$ is tuned to obtain $\sigma/\sigma_{SM}=1$ for $gg\rightarrow h$,
then a color octet intermediate particle  replacing the top quark with positive $\kappa $ (the solid black line) would predict a highly suppressed rate
for $2h$ production (the red dashed line). 
Even when the total rates are identical to the SM predictions, the kinematic 
distributions from color octet and triplet intermediate states are quite
different than those from the SM top, as plotted in \reffig{fig:scal3}.   The scalar needs to be quite light to reproduce the SM rates, and the distribution is
peaked at much lower $m_{hh}$ than the SM prediction.

\begin{figure}[t]
\begin{centering}
\includegraphics[scale=0.6]{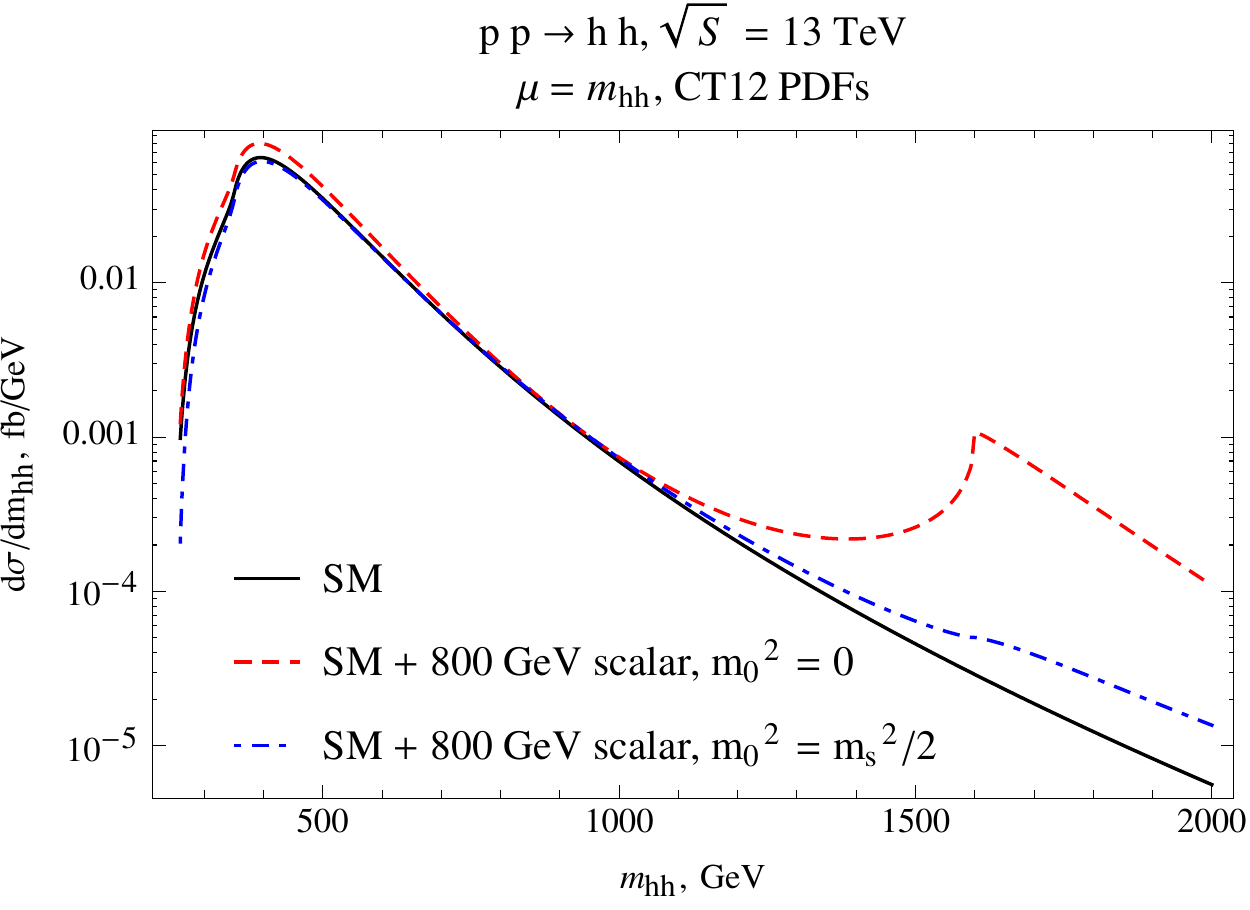}
\par\end{centering}
\caption{Invariant mass distribution in $2h$ production with the SM top quark in addition to an 800 GeV color triplet scalar that gets all (red dashed) or half (blue dot-dashed) of its mass from the Higgs. The SM (black solid) is shown for comparison~\cite{Dawson:2015oha}.}
  \label{fig:topsc_dist}
\end{figure}

 Assuming the top Yukawa is SM-like, adding  an additional scalar receiving all of its mass from electroweak symmetry breaking gives an unacceptably large contribution to the $1h$  production cross section, regardless of the scalar mass and $SU(3)$ representation.  A heavy color triplet scalar with $\kappa = 2 m_s^2 / v^2$, for example, changes the $1h$ production rate by 54\%. Lighter scalars and scalars in other color representations result in even larger deviations. Heavy scalars receiving all their masses from the Higgs have $m_0 = 0$, and are not compatible with LHC  limits on the $1h$ production rate from gluon fusion.
Heavy scalars  decouple quickly in the $1h$ rate and may show up in the high $m_{hh}$ tail of the $2h$ distribution.  The invariant mass distributions for $2h$ production are shown in \reffig{fig:topsc_dist} assuming a SM-like top quark and an additional 800 GeV color triplet scalar. If the scalar receives half of its mass squared from electroweak symmetry breaking, $m_0^2 = m_s^2 / 2$, the $1h$ rate is in roughly $2 \sigma$ tension with the current measurement, and the $2h$ distribution deviates from the SM expectation starting at $2m_s$, roughly speaking. 

%% file: BSMresonance/BSMMSSM.tex
\subsubsection[MSSM]{MSSM \\
\contrib{P. Huang}
}

Now we discuss the modification to double Higgs production in the presence of light stops ~\cite{Belyaev:1999mx,Batell:2015koa,Huang:2017nnw}. We first write down the stop mass matrix
\begin{align}\label{eqn:stopmassmatrix}
   \mathbf{M_{\tilde{t}}^{2}} =
   \left( \begin{array}{cc}
     m_{Q}^{2}+m_{t}^{2}+D_Q & m_{t}X_{t} \\
     m_{t}X_{t} & m_{U}^{2}+m_{t}^{2}+D_U\\
     \end{array} \right).
   \end{align}
The parameters $m_{Q}$ and $m_U$ are soft SUSY breaking mass terms of the left-handed and right-handed stops respectively, $X_t$ is the stop mixing parameter, $D_U$ and $D_Q$ are the $D$-term contributions. The dimensionful trilinear coupling of the Higgs to the stops has a strong dependence on the Higgs mixing parameter $X_t$, which can be larger than the stop masses.
Given that the LHC has excluded stops that are not significantly heavier than the top quarks, a large $X_t$ is preferred to generate relevant contributions to the double Higgs production cross section. However, a large $X_t$ may affect the Higgs vacuum stability~\cite{Frere:1983ag,Claudson:1983et,Falk:1995cq,Kusenko:1996jn,Chowdhury:2013dka,Blinov:2013fta}, which leads to constraints in the stop sector in addition to constraints from the direct stop searches. $X_t$ also contributes to the gluon fusion of a single Higgs production. Including possible modifications in the Higgs coupling to tops, the modification in $\kappa_g$ is given by~\cite{Djouadi:2005gj,Buckley:2012em,Carmi:2012in,Badziak:2016exn,Badziak:2016tzl}
  
 \begin{eqnarray}\label{kappagmodifiedyt}
   \kappa_{g}= \kappa_t  + \frac{\kappa_t}{4} m_t^2 \left[\frac{1}{m_{\tilde{t}_1}^2}+ \frac{1}{m_{\tilde{t}_2}^2} - \frac{\tilde{X}_t^2}{m_{\tilde{t}_1}^2 m_{\tilde{t}_2}^2}\right],
      \end{eqnarray}
with $\kappa_i$ defined as $g_{hii} / g_{hii}^{SM}$.

\begin{figure}[t!]
  \centering
  \includegraphics[width=0.5\textwidth]{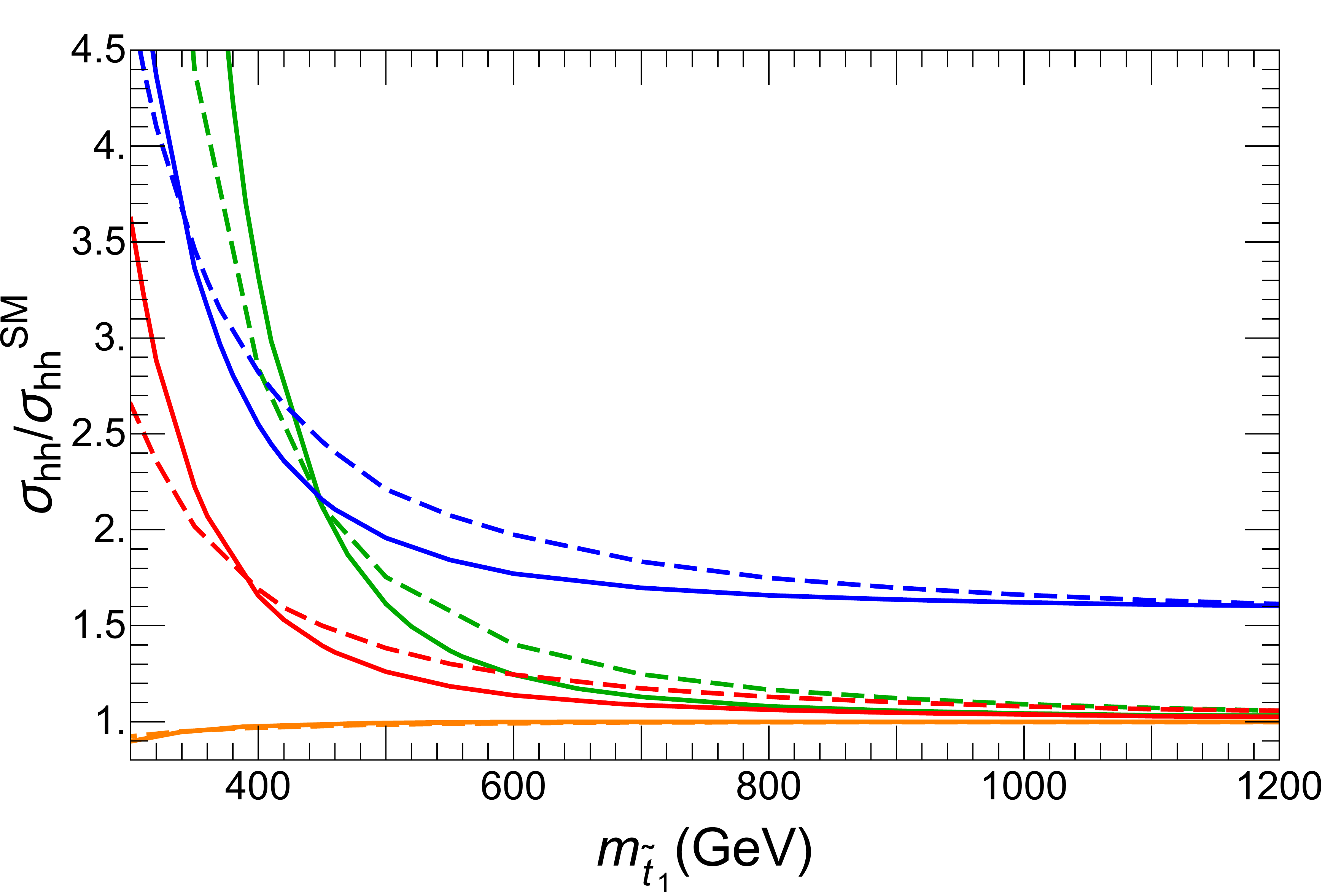} \qquad \includegraphics[width=0.4\textwidth]{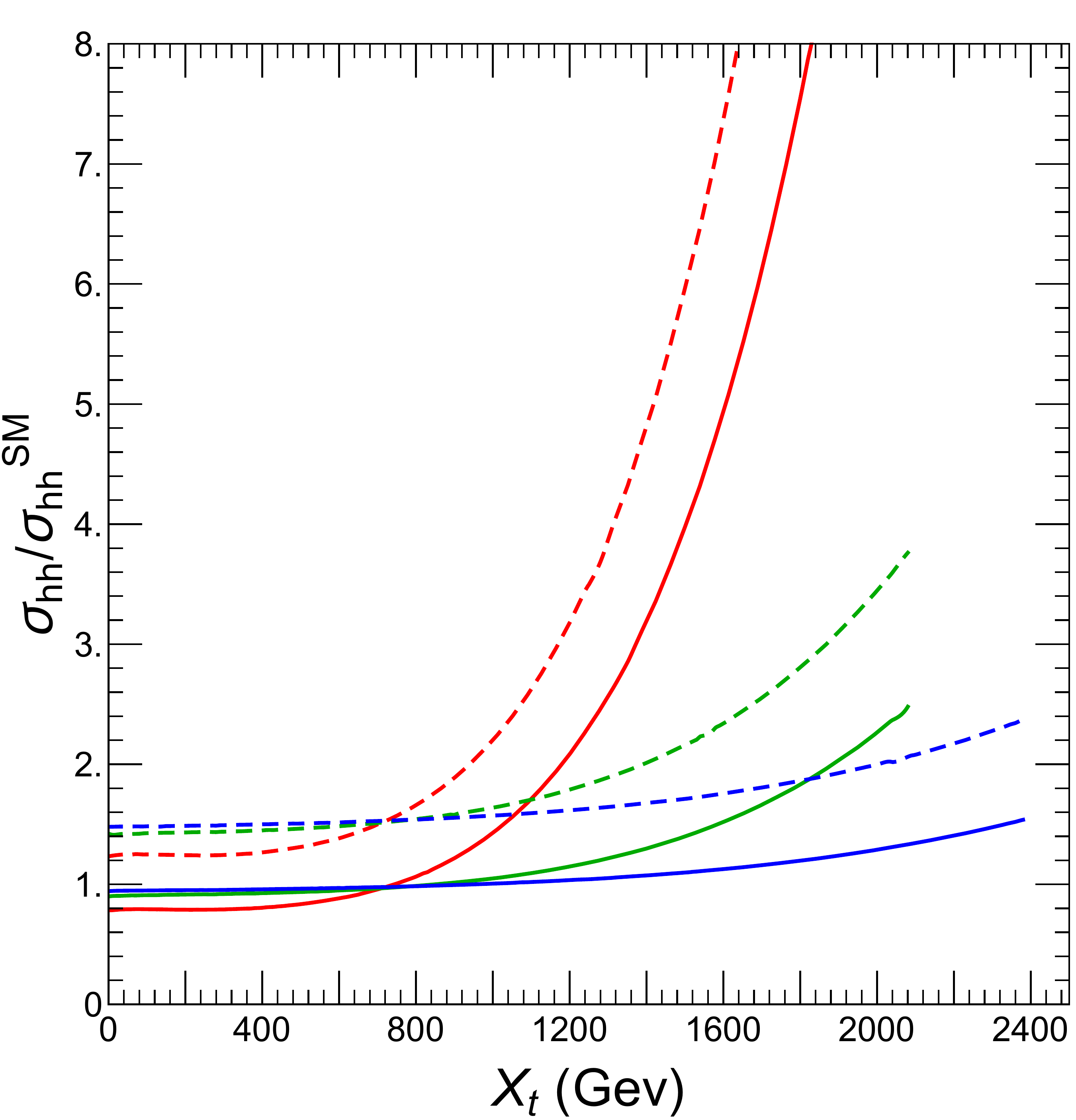}
  \caption{\small Double Higgs production cross section normalized to the SM values as a function of the lightest stop mass(left) and $X_t$(right)~\cite{Huang:2017nnw}.
  The color coding is explained in the main text.
  }
  \label{fig:xsec}
\end{figure}

The contribution of light stops to double Higgs production is summarized in \reffig{fig:xsec} ~\cite{Huang:2017nnw}.
 In the left panel of \reffig{fig:xsec}, we show the double Higgs production cross section normalized to the SM value as a function of the light stop mass using the full one loop calculation (solid lines), and the EFT calculation (dashed lines). $\kappa_t$ is chosen to be 1 for the orange, red and green lines, and 1.1 for the blue lines. For the orange line, $X_t^2$ is chosen to be $m_{\tilde{t}_1}^2 + m_{\tilde{t}_2}^2$. This choice makes the effective Higgs gluon coupling SM like. In the red and blue lines, instead, $X_t^2$ is chosen to saturate the vacuum stability condition,
 \begin{eqnarray}\label{HgsVcmStb}
      A_t^2\le  \left(3.4+0.5\frac{\lvert 1-r\lvert}{1+r}\right)(m_{Q}^2 + m_{U}^2)+60\left( \frac{m_z^2}{2} \cos(2\beta) + m_A^2 \cos^2\beta \right)
      \end{eqnarray}
 in a conservative way by neglecting the $m_A$ and $m_Z$ terms, For the green lines, $X_t^2$ is chosen to saturate the vacuum stability condition with $m_A = 350$~GeV, $\mu = 400$~GeV, and $\tan\beta$ = 1. In the right panel of  \reffig{fig:xsec}, we show the effect of stop mixing parameter  $X_t$ on the double Higgs production cross section for a fixed value of the mass of the lighter stop. Red, green and blue lines represent fixed lighter stop mass of 300, 400 and 500~GeV respectively. Solid lines correspond to $\kappa_t=1$, while dashed lines correspond to $\kappa_t=1.1$. 

The cross section for a given final state depends not only on the double Higgs production cross section, but also on the relevant Higgs decay branching ratios. In the MSSM, some small modifications to the Higgs decay branching ratios are expected. The largest modification is about $\pm 20\%$ for the \bbyyAlt channel~\cite{Huang:2017nnw}. Light stops also lead to modifications to the double Higgs invariant mass distribution. In the presence of a light stop, the amplitudes develop imaginary parts when the invariant mass $m_{hh}$ crosses the 2$m_{\tilde{t}}$ threshold, inducing a second peak in the $m_{hh}$ distribution a little above 2$m_{\tilde{t}}$~\cite{Huang:2017nnw}.

%% file: BSMresonance/BSMGWCosmo.tex
\section[Connection to Cosmology]{Connection to Cosmology \\
\contrib{J. Kozaczuk, A. Long, K. Sinha}
}
\label{sec:cosmology}


Measurements of di-Higgs production test the hypothesis that the Higgs boson couples to new physics with a mass scale $m \sim 100 \, \mathrm{GeV} - 1 \, \mathrm{TeV}$.  
In the hot conditions of the early universe, these particles would have been in thermal equilibrium with the SM plasma.  As the universe expanded and cooled, the presence of this new physics could affect the nature of the electroweak phase transition (EWPT).  
Furthermore the Higgs boson could act as an inflaton field under particular conditions, as described in refs \cite{Bezrukov:2007ep,Masina:2011aa}.

\noindent {\textbf{Electroweak phase transition and electroweak baryogenesis:}} The electroweak phase transition is the dynamical process by which the Higgs field acquired its nonzero vacuum expectation value in the early universe.  
The SM predicts that the phase transition is a smooth, continuous crossover with the Higgs field evolving almost homogeneously from $0$ to $246 \, \mathrm{GeV}$ as the temperature is decreased through the weak scale.  
However, the presence of new physics can easily and dramatically change the predicted nature of the phase transition, even leading to a first order phase transition.  
Unlike the gentle continuous crossover, a first order phase transition is a violent event during which bubbles nucleate, expand, collide, and eventually merge to overtake the whole system.  
Today, our understanding of Higgs physics is too poor to discriminate between even these two qualitatively different scenarios.  

If the cosmological electroweak phase transition was a first order one, it would have profound implications for cosmology.  
The out-of-equilibrium conditions of a first order phase transition provide the right environment for the generation of cosmological relics.  
In this way, a first order electroweak phase transition could explain why our universe has an excess of matter over antimatter on cosmological scales through the mechanism of electroweak baryogenesis~\cite{Cohen:1990it, Morrissey:2012db}.  A strong first-order electroweak phase transition can have other interesting cosmological consequences, such as the dilution of pre-existing thermal relics through entropy injection~\cite{Wainwright:2009mq} and the generation of a stochastic gravitational wave background (discussed below).  

\begin{figure}[!t]
\begin{center}
\includegraphics[width=0.45\textwidth]{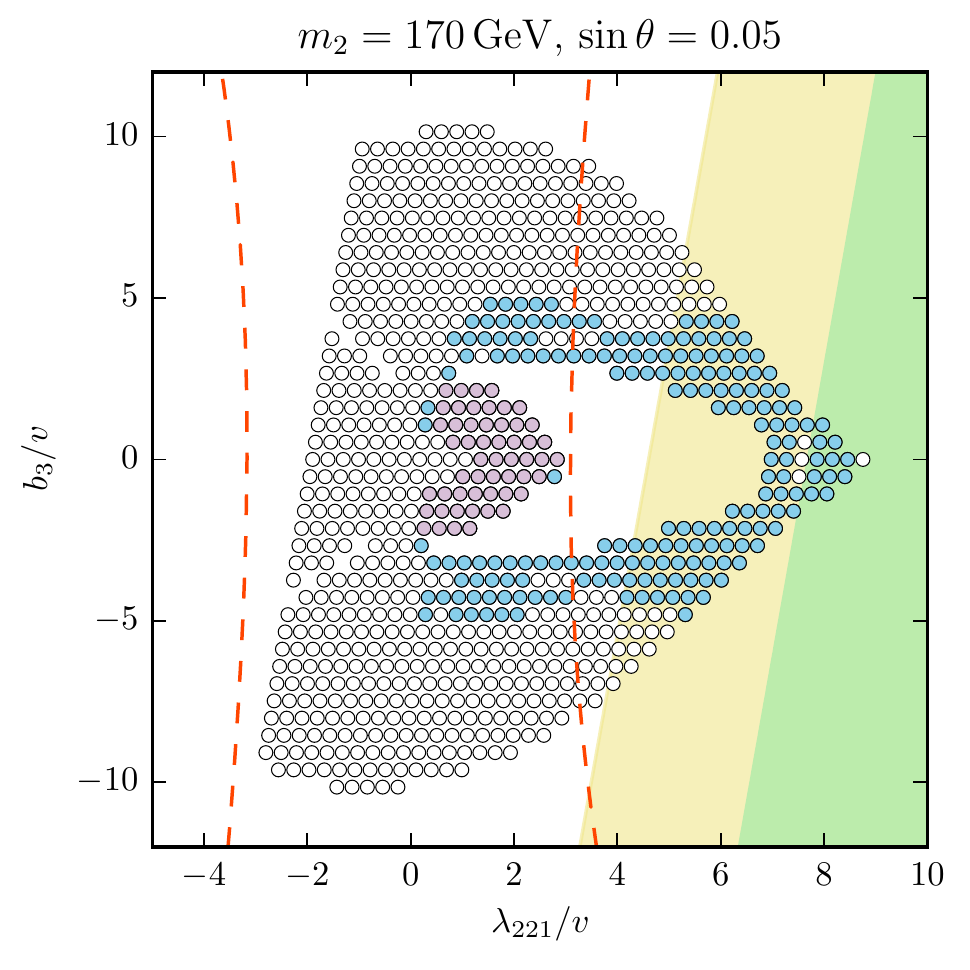} 
\
\includegraphics[width=0.45\textwidth]{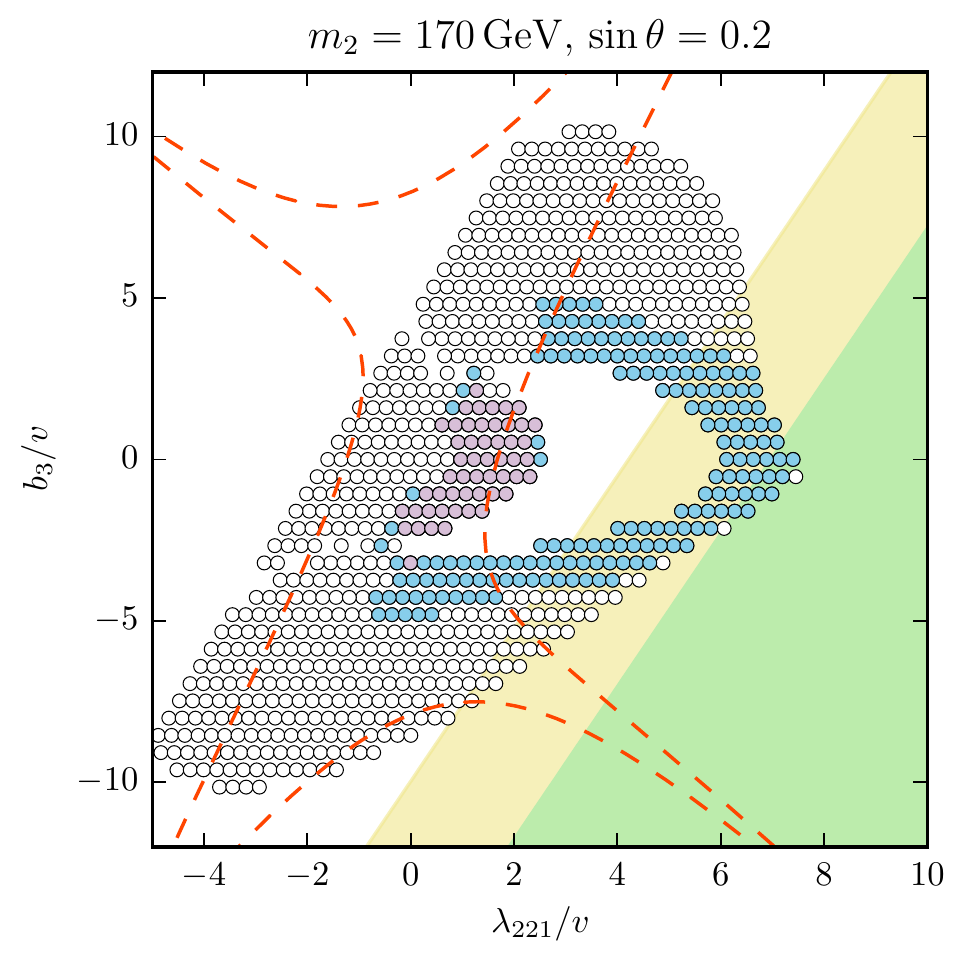}
\vspace*{-0.2cm}
\caption{\label{fig:Kozaczuk}
Figure adapted from Ref.~\cite{Chen:2017qcz} showing slices of the real singlet extension of the SM for a singlet-like scalar mass of 170 GeV and two different mixing angles with the Higgs. Blue and purple shaded points feature a strong first-order electroweak phase transition. Regions outside of the red dashed contours feature deviations in the 125 GeV Higgs self-coupling larger than 30\%. The green (yellow) shaded regions show the discovery (exclusion) reach for pair production of the new scalar at the 14 TeV HL-LHC in a trilepton final state discussed further in Ref.~\cite{Chen:2017qcz}. 
}
\end{center}
\end{figure}

\begin{figure}[!t]
\begin{center}
\includegraphics[width=0.80\textwidth]{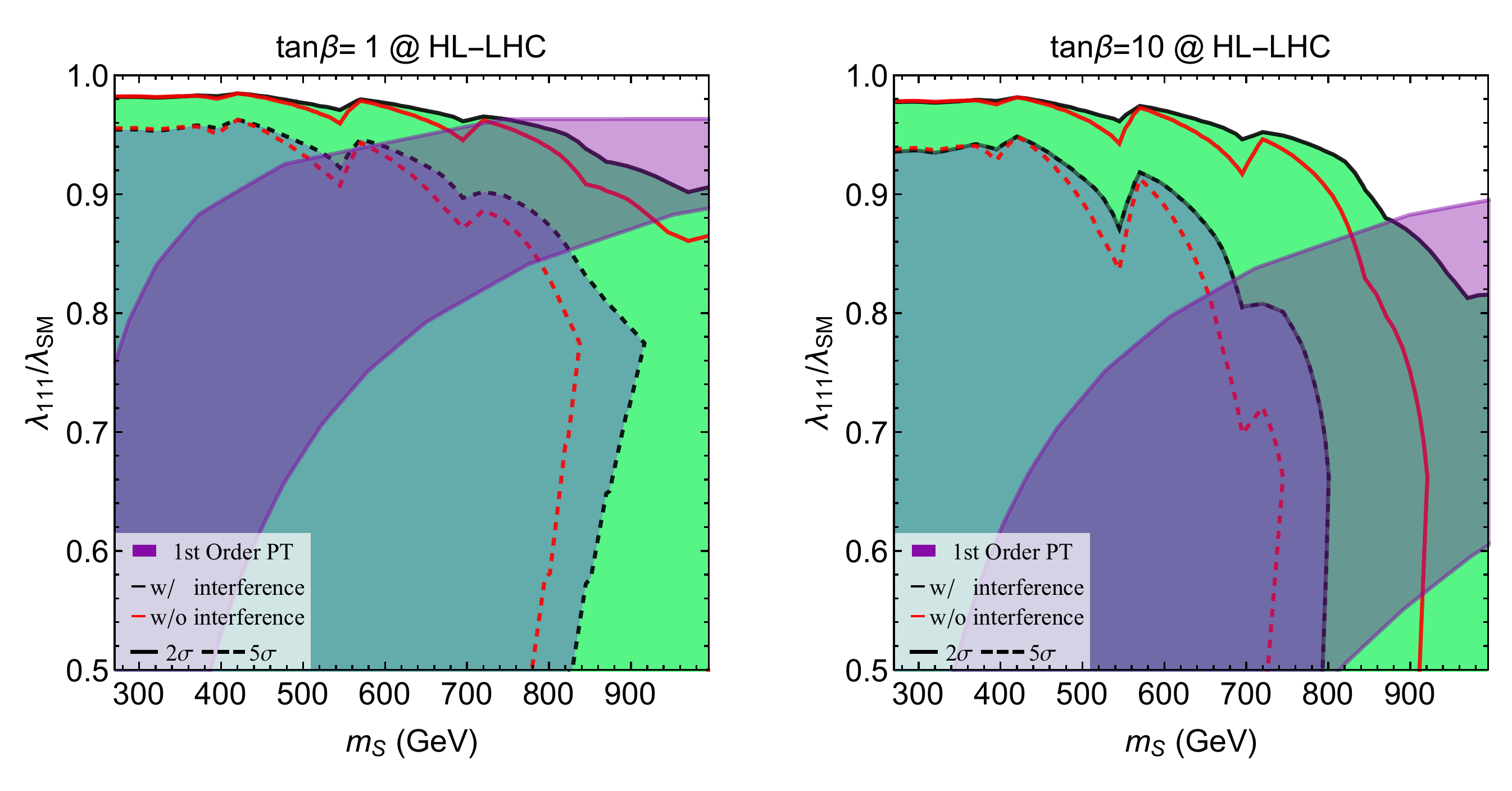}
\vspace*{-0.2cm}
\caption{\label{fig:CLR}
Sensitivity of resonant di-Higgs production (black and red contours) to regions of the singlet model parameter space with a strong first-order electroweak phase transition (purple). Details can be found in Ref.~\cite{Carena:2018vpt} from which this figure was adopted.  
}
\end{center}
\end{figure}

Double Higgs production at colliders allows for a direct probe of the couplings in the Higgs potential responsible for strengthening the electroweak phase transition. A typical example of this in the real singlet extension of the SM is illustrated in \reffig{fig:Kozaczuk} (adapted from Ref.~\cite{Chen:2017qcz}), which shows slices of the parameter space consistent with a strong first-order electroweak phase transition (blue and purple points). Outside of the red dashed contours the deviations in the Higgs self coupling are larger than 30\%. Precise measurements of the double Higgs production rate can thus provide a powerful probe of the electroweak phase transition in this scenario (see also Ref.~\cite{Profumo:2007wc}). Similar conclusions hold in other extensions of the SM as well~\cite{Noble:2007kk, Huang:2016cjm}. For scenarios in which a new scalar heavier than 250 GeV coupled to the Higgs generates a strong first-order EWPT, resonant double Higgs production mediated by the new scalar provides a powerful handle on the nature of electroweak symmetry breaking~\cite{No:2013wsa}. The prospects for such a search at the high luminosity LHC are shown in \reffig{fig:CLR} (adapted from Ref.~\cite{Carena:2018vpt}), again for slices of the singlet model parameter space. Strong first order phase transitions generated by new scalars with masses up to the TeV scale can be probed by resonant di-Higgs production (see also Refs.~\cite{Dolan:2012ac, No:2013wsa, Huang:2017jws}). In models with additional scalars, pair production of the other scalar states can provide a complementary probe of the electroweak phase transition~\cite{Curtin:2014jma, Chen:2017qcz}. The shaded regions of \reffig{fig:Kozaczuk} correspond to the projected HL-LHC sensitivity to pair production of singlet-like scalars in a particular trilepton channel detailed in Ref.~\cite{Chen:2017qcz}. The sensitivity shown is likely conservative, and searches for double scalar pair production involving states other than the 125 GeV Higgs can be a promising avenue for probing electroweak symmetry breaking in the early Universe at the LHC and beyond.

\noindent {\textbf{Complementarity with gravitational wave observations:}} \,\,\, The inhomogeneous nature of a first order phase transition provides the requisite quadrupole moment to source gravitational waves~\cite{Kamionkowski:1993fg}.  
Since gravitational waves are very weakly interacting, they propagate freely until reaching us at Earth today.  
If we can observe this primordial stochastic gravitational wave background, it could provide direct evidence for a first order electroweak phase transition and thereby indicate the presence of new physics coupled to the Higgs.  

Our ability to measure gravitational waves (GWs) has recently been demonstrated in spectacular fashion by the LIGO and VIRGO collaborations~\cite{Abbott:2016blz}.  
Moreover, efforts are underway to build and launch a gravitational wave interferometer in space. The Laser Interferometer Space Antenna (LISA)~\cite{Audley:2017drz} collaboration has recently celebrated a successful pathfinder mission and is expected to be launched in the early 2030s.  
With an interferometer that is no longer tethered to the Earth, the length of its arms can be increased to millions of kilometers, which gives it sensitivity to the $\sim \mathrm{mHz}$ gravitational waves that are expected to arise from a first order electroweak phase transition~\cite{Caprini:2015zlo}.

It is important to understand how future collider measurements, such as Higgs pair production, and observations of a stochastic GW background can complement each other in exploring new physics yielding a strong first-order electroweak phase transition.
\begin{figure}[t]
\begin{center}
\includegraphics[width=0.99\textwidth]{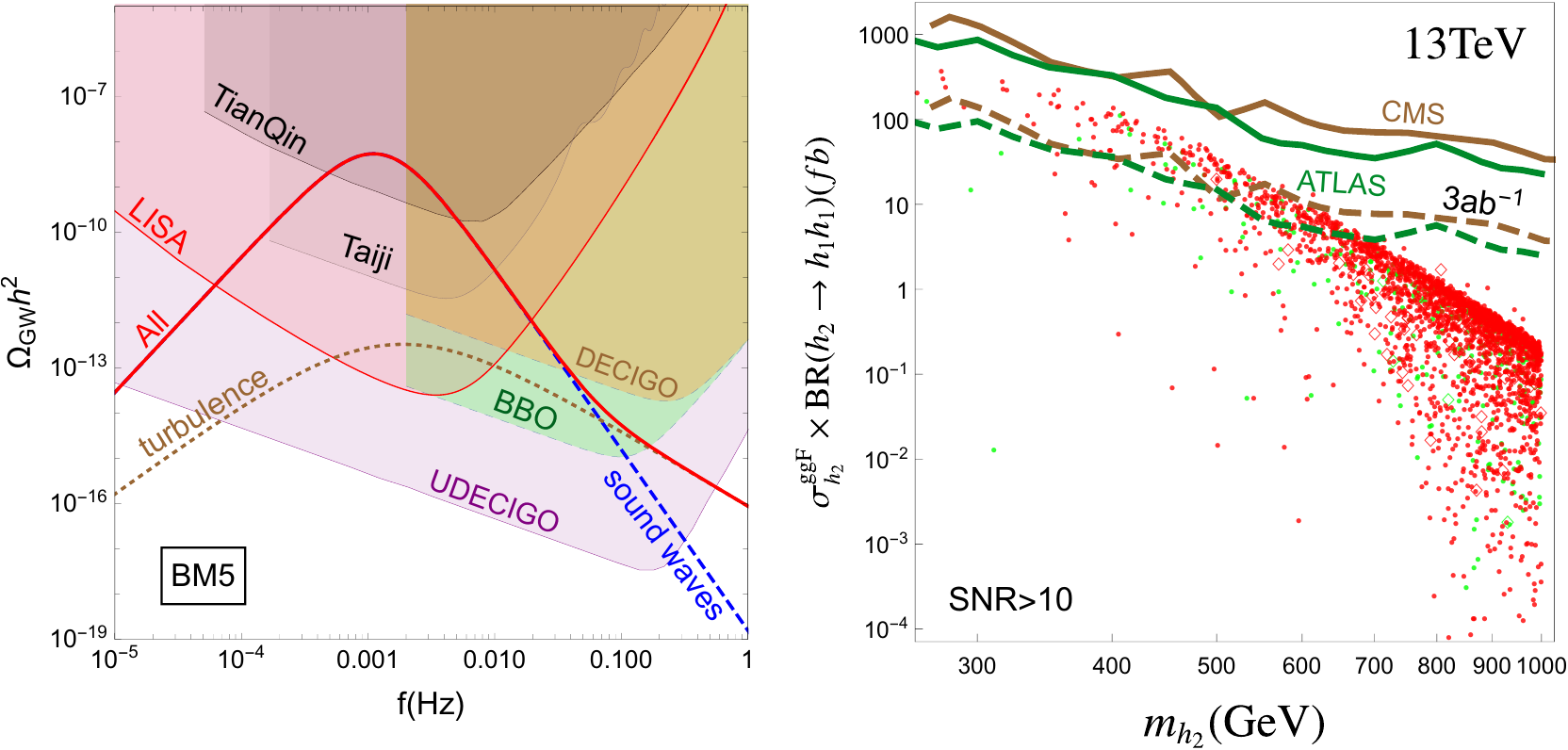}
\end{center}
\vspace*{-0.5cm}
\caption{
\label{gwdihiggscomp}
Left panel: GW spectrum obtained at a benchmark point in the singlet-extended SM. Figure adapted from Ref.~\cite{Alves:2018oct}. Right panel: Current and future ATLAS/CMS di-Higgs sensitivity to points predicting a signal-to-noise ratio larger than 10 at LISA. Figure adapted from Ref.~\cite{Alves:2018jsw}. Details in text.
}
\end{figure}
The simplest template where these questions can be studied is an extension of the SM by a singlet scalar discussed above.
The complementarity between GW and collider measurements has recently been explored in this model by the authors of
Refs.~\cite{Alves:2018oct} and \cite{Alves:2018jsw}, and we summarise the main results here. The left panel of \reffig{gwdihiggscomp} displays the GW spectrum obtained at a benchmark point in this model which is compatible with electroweak precision measurements and all other phenomenological constraints. The mass of the extra singlet is 455 GeV. The total GW signal is shown in red, while the different contributions from sound waves (turbulence) are shown in blue (brown). The color-shaded regions are the experimentally sensitive regions for various GW detectors. The right panel of \reffig{gwdihiggscomp} shows ATLAS (solid green lines) \cite{Aad:2019uzh} and CMS (solid brown lines) \cite{Sirunyan:2018two} limits on resonant di-Higgs production for 36.1 fb$^{-1}$ and 35.9 fb$^{-1}$ of data, respectively, combining several final states. A simple rescaling of the current limits to 3000 fb$^{-1}$ at the HL-LHC (13 TeV) is performed to obtain the corresponding dashed line future projections. For the points on the parameter space giving detectable GWs with a signal-to-noise ratio at LISA larger than 10, the resonant cross sections from gluon fusion at NNLO+NNLL are computed using the  results in \cite{deFlorian:2016spz}. It is clear that resonant di-Higgs studies at the HL-LHC and GW signals from LISA can play complementary roles in exploring this model in the future.

%% file: BSMresonance/BSMDarkMatter.tex
\section[HH and Dark Matter (Missing Energy)]{HH and Dark Matter (Missing Energy) \\
\contrib{M. Blanke, S. Kast, J. Thompson, S. Westhoff, J. Zurita}
} \label{sec:DM}


\newcommand{\mev}{\,\, \mathrm{MeV}}
\newcommand{\met}{\slashed{E}_T}

The final state of two Higgs bosons plus missing {transverse} energy was originally studied in the context of Goldstino dark matter~\cite{Matchev:1999ft}, which is currently the new-physics model that the existing LHC searches~\cite{Aaboud:2018htj,Sirunyan:2017obz} target. More recently it was realised that \emph{di-Higgs plus $\met$} is a signature that also occurs in a plethora of other BSM scenarios including a dark sector~\cite{Kang:2015nga,No:2015xqa,Kang:2015uoc,Brivio:2017ije,Arganda:2017wjh,Chen:2018dyq,Bernreuther:2018nat,Titterton:2018pba}.

In Ref.~\cite{Blanke:2019hpe}, a detailed analysis of the final state with four $b$-quarks and large missing energy in the High Luminosity phase of the LHC was carried out. Here we provide a summary of the most salient findings and refer the reader to Ref.~\cite{Blanke:2019hpe} for further details. The large backgrounds ($V$ + jets, $t\bar{t}$, etc.) and the complex kinematics of the final require a multivariate analysis (MVA), which we summarise here. 

\begin{figure}
\centering{
\begin{minipage}{.3\textwidth}
\includegraphics[width=\textwidth]{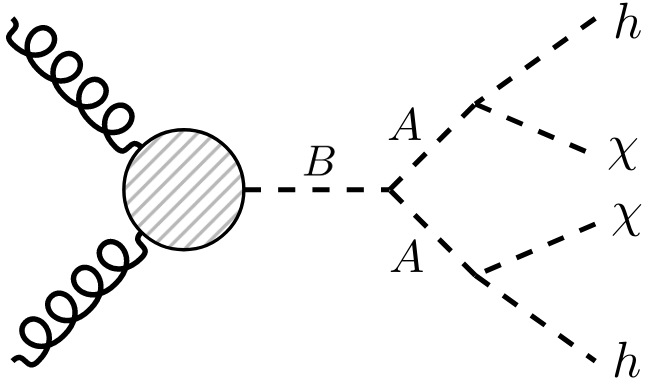}
\begin{center}
symmetric topology
\end{center}
\end{minipage}
\hspace*{1cm}
\begin{minipage}{.3\textwidth}
\includegraphics[width=\textwidth]{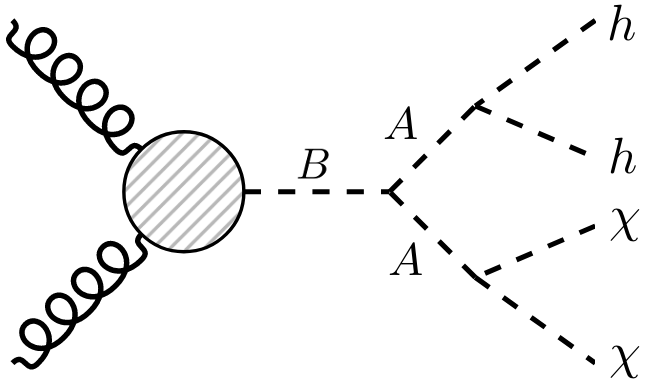}
\begin{center}
resonant topology
\end{center}
\end{minipage}
\caption{\label{fig:topAB} Topologies for a scalar resonance $B$ decaying into $hh+\met$. }}
\end{figure}

Since many new physics models can give rise to the same final state, it is important to define physics scenarios that do not depend {(crucially)} upon the detailed field content, but rather on the masses and couplings that characterise the signature. To this end we have introduced two simplified models targeting two different final state topologies. Both models feature three new scalar particles, dubbed $B$, $A$ and $\chi$ (invisible), with the production mechanism $gg \to B \to AA$, through a dimension five operator $B G_{\mu \nu}^a G^{\mu \nu \, a}$. We then have two options for $A$ to decay: either $A \to h \chi$ for both $A$ bosons (symmetric topology) or  $A \to hh$ and $A \to \chi \chi$ simultaneously (resonant topology). We restrict ourselves to mass spectra where all these states are produced on-shell. Moreover, we assume all new fields to be SM singlets, and we impose a discrete $\mathbb{Z}_2$ parity under which all SM fields and $B$ are even, $\chi$ is always odd and $A$ is even (odd) in the resonant (symmetric) topology. The Feynman diagrams for the $hh + \met$ final state are shown in \reffig{fig:topAB}.

In our study we use the scikit-learn~\cite{Pedregosa:2012toh} implementation of AdaBoost~\cite{Freund:1997xna}, employing the SAMME.R algorithm to perform a Boosted Decision Tree (BDT) classification (70 trees, maximal depth of 3, learning rate of 0.5, minimum node size of 0.025 of the total weights). We apply a $\met > 200$ GeV cut, as we found the inclusive $\met$ trigger better suited for our purposes than triggering on the jets.
 We employ a modified version of the BDRS algorithm \cite{Butterworth:2008iy}, and cluster large-radius jets with $R = 1.2$ $(R = 0.6)$ for the symmetric (resonant) topology (see Sec.~\ref{sec:jetReco} for more details), demanding to have at least one \btagged subjet within each jet, and veto events with leptons. The input variables include the $\pT, \eta, \phi, m$ of the large-radius jets and subjets, global variables such as $\met$, $H_T$, the number of large-radius jets and subjets, and finally variables for the {di-jet and $\met$-jet systems}, e.\,g.\ $\Delta \Phi(J, \met)$, $\Delta R (J_1,J_2)$. We define {the} significance as
\begin{equation}
\Sigma = \frac{S}{\sqrt{\alpha S+B+\beta^2 B^2}} \, ,
\end{equation}
where $S$ and $B$ are the number of signal and background events, $\alpha = 0\ (1)$ for exclusion (discovery) and $\beta$ is the systematic uncertainty, which we fix here to 5 \%. To be conservative we add an additional layer of cautiousness and {define \emph{exclusion} as} $\Sigma(\alpha=0)=3$ (instead of the usual 2) and \emph{discovery} as $\Sigma(\alpha=1)=7$ (instead of the usual value of 5). Moreover, in order not to be pushed to very sparsely populated regions of phase space, we also request $S \geq 20$. 

\begin{figure}[t]
\centering
\includegraphics[width=0.47\textwidth]{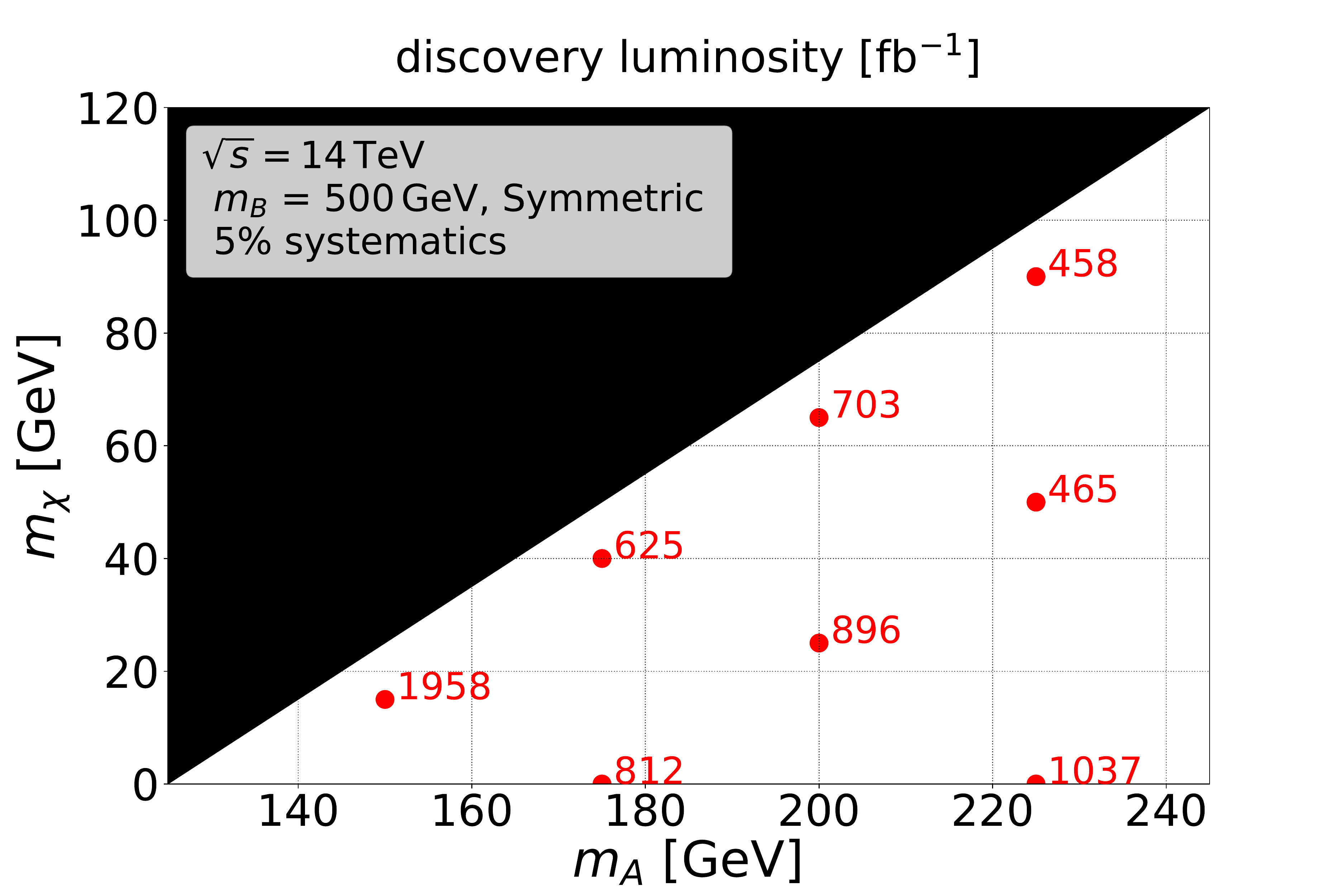} \hspace*{0.4cm}
\includegraphics[width=0.47\textwidth]{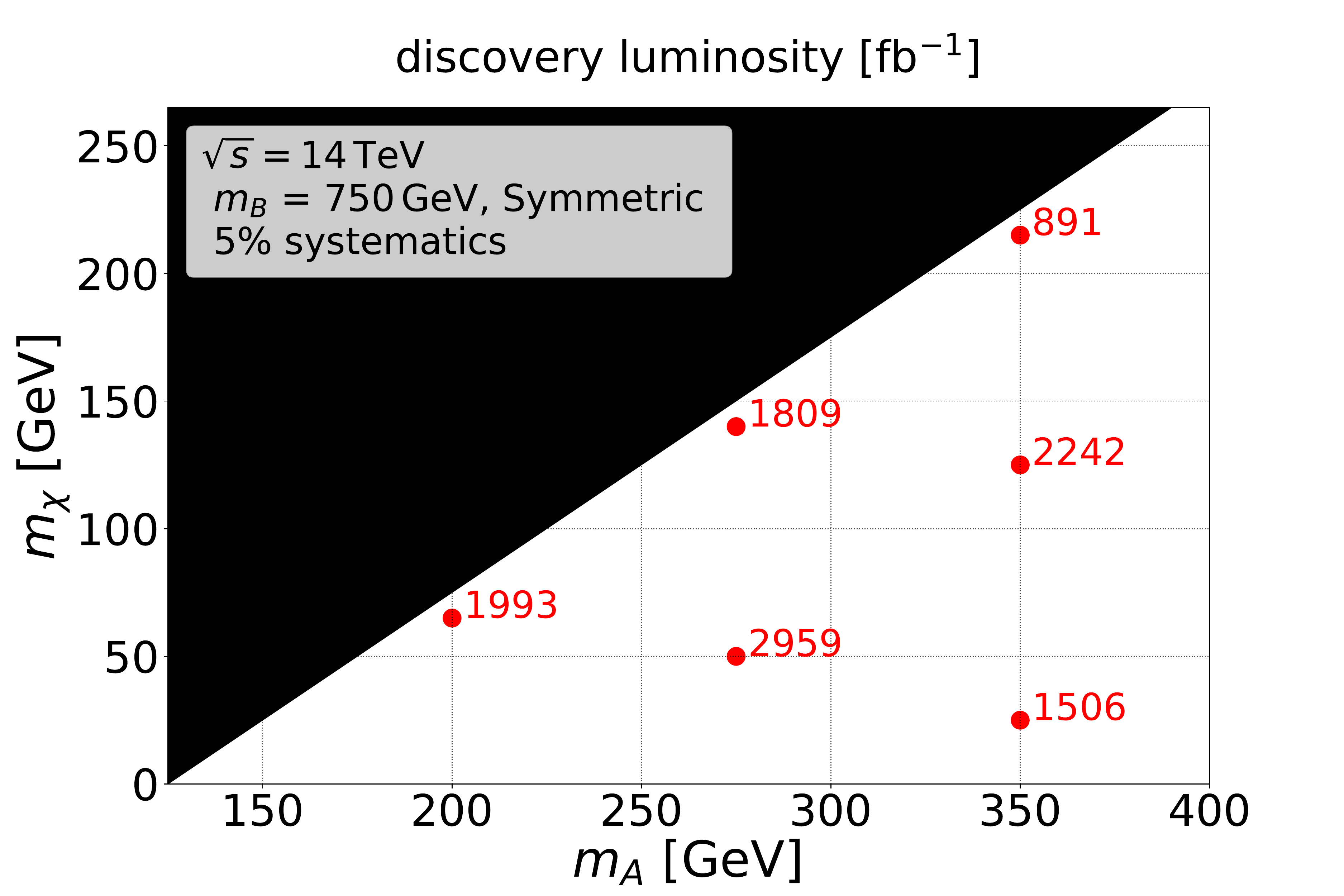} 
\caption{\label{fig:discolumiS}Luminosity required for a discovery (in fb$^{-1}$) at the HL-LHC in the $m_A - m_\chi$ plane for the symmetric model, with $m_B = 500\gev$ (left panel) and $m_B = 750\gev$ (right panel)~\cite{Blanke:2019hpe}.}
\end{figure}

\begin{figure}[t]
\centering
\includegraphics[width=0.47\textwidth]{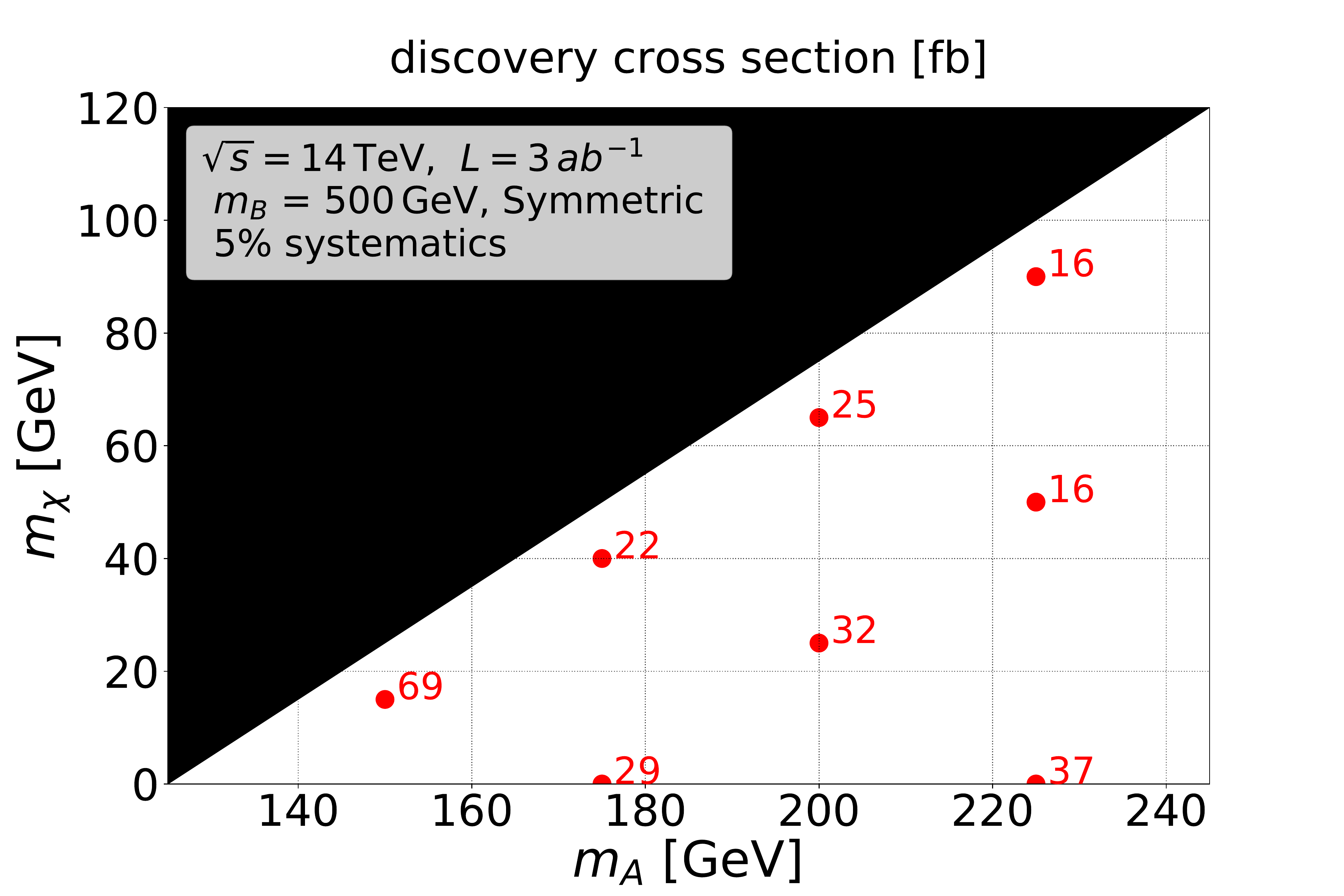} \hspace*{0.4cm}
\includegraphics[width=0.47\textwidth]{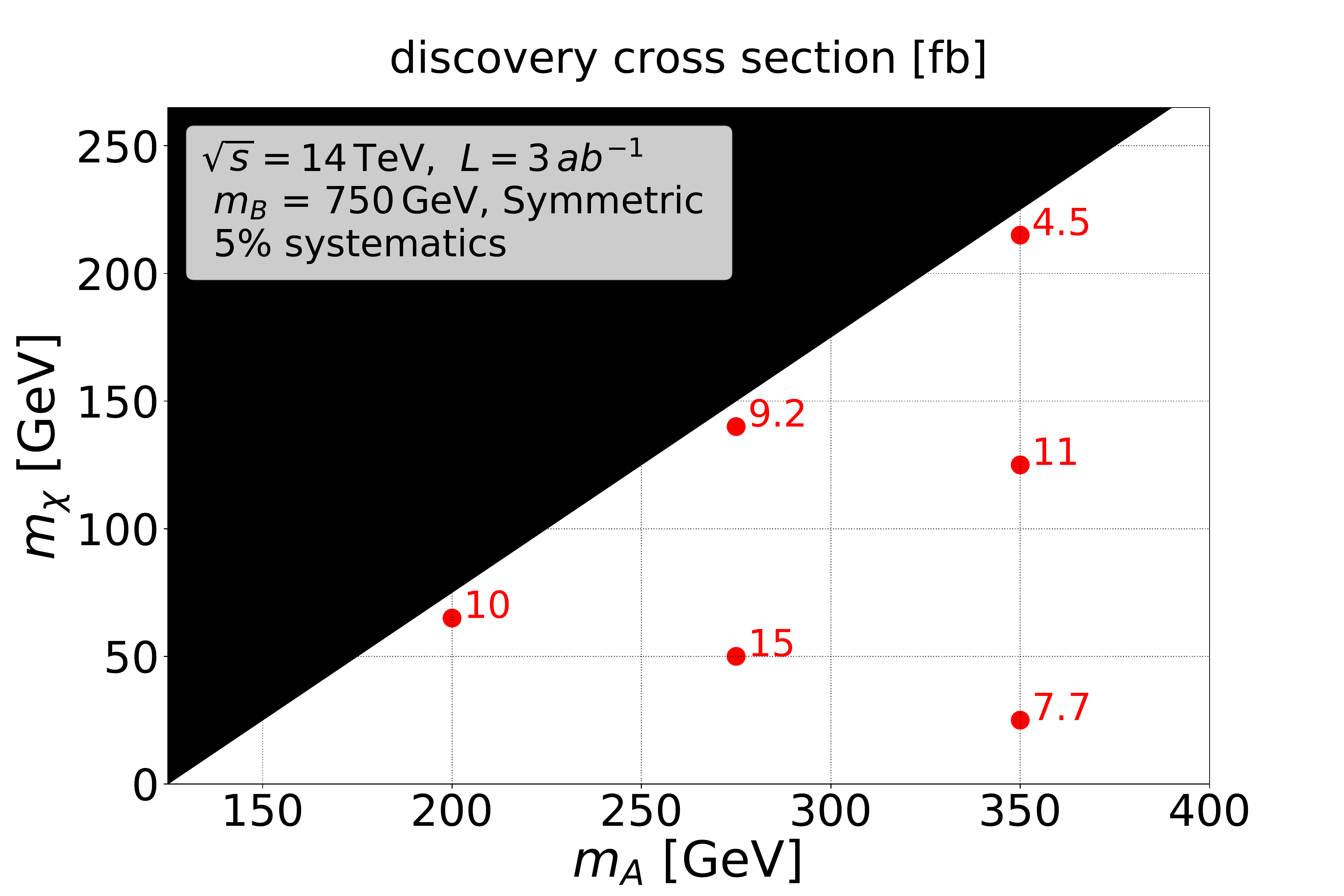} 
\caption{\label{fig:S750xs}Cross sections (in fb) for the $4b 2\chi$ final state required for a discovery at the HL-LHC in the $m_A - m_\chi$ plane for the symmetric model, with $m_B = 500\gev$ (left panel) and $m_B = 750\gev$ (right panel)~\cite{Blanke:2019hpe}.}
\end{figure}

We present our results in terms of the scalar masses for a total integrated luminosity of $3$~ab$^{-1}$. 
In \reffig{fig:discolumiS}, we display the luminosity required to discover the symmetric scenario at the HL-LHC, fixing  $m_B=500\gev$ (left panel) and $m_B = 750\gev$ (right panel).  
Except for one point, all scenarios are well within the reach of the HL-LHC, and thus we also present, in \reffig{fig:S750xs}, the results in terms of the minimal cross section that can be discovered with a luminosity of 3 ab$^{-1}$. The latter has the advantage of providing results that can be readily applied to a larger class of models.

\begin{figure}[t]
\centering
\includegraphics[width=0.47\textwidth]{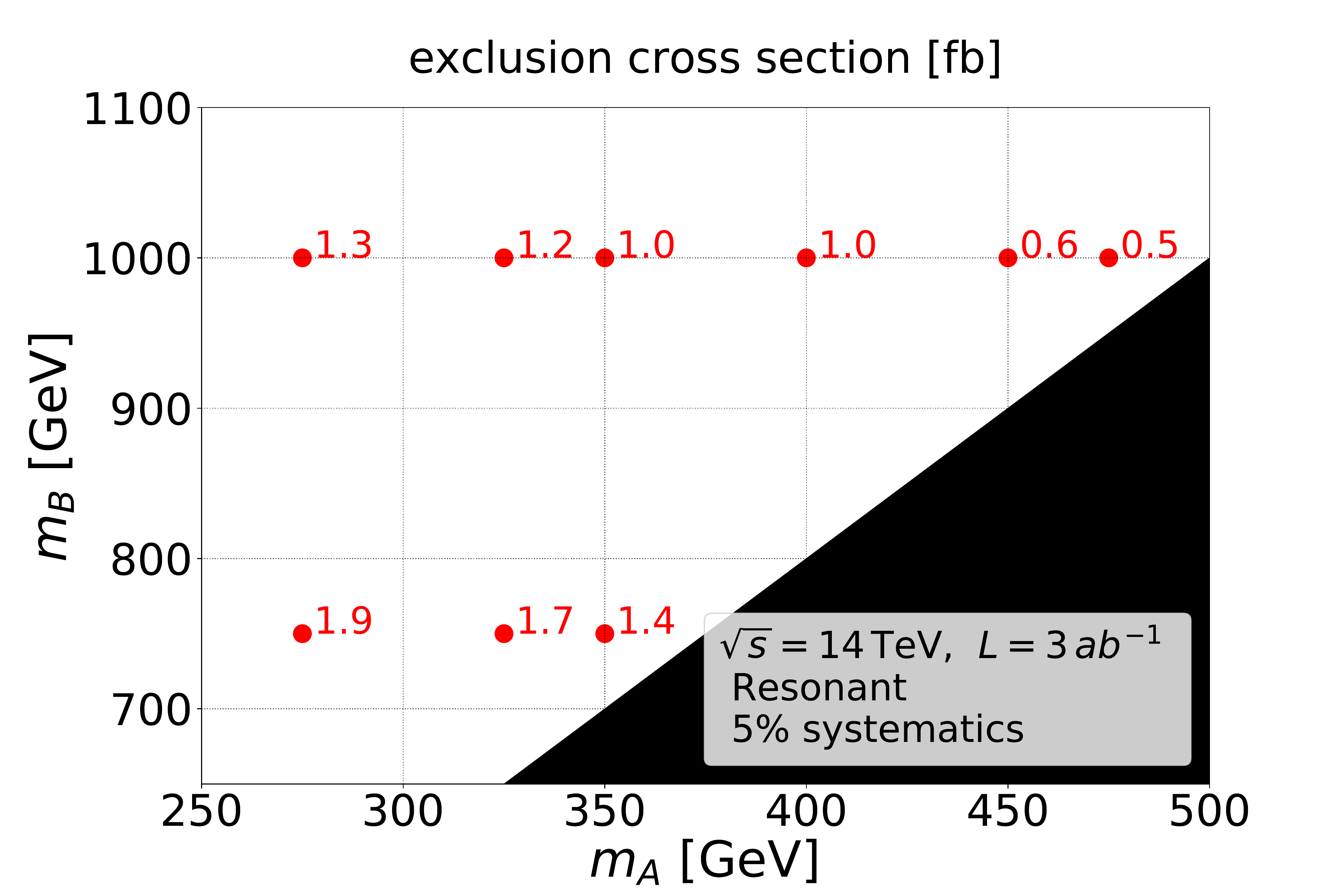} \hspace*{0.4cm}
\includegraphics[width=0.47\textwidth]{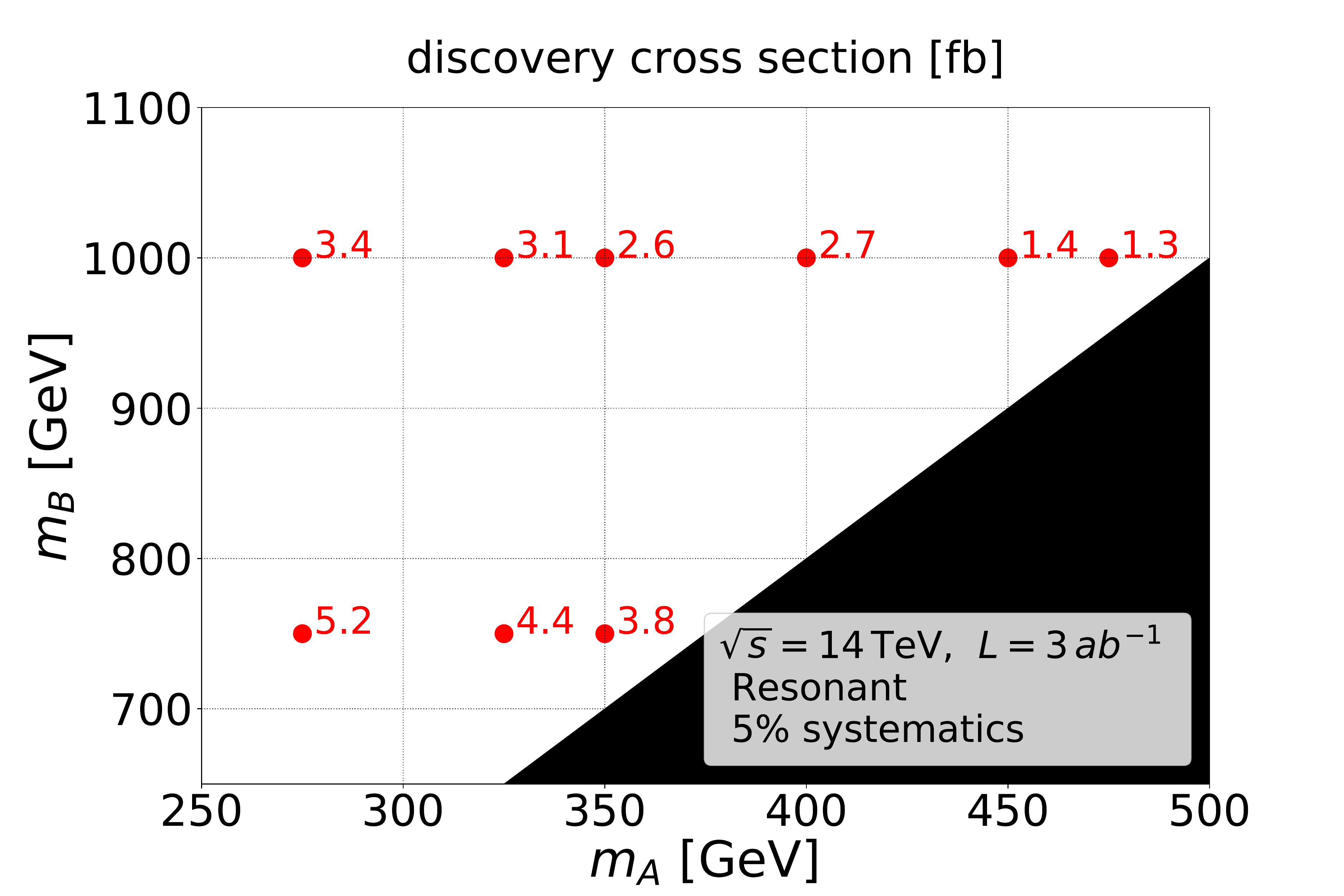} 
\caption{\label{fig:Rxs}Cross sections (in fb) for the $4b 2\chi$ final state required for exclusion (left) and discovery (right) at the HL-LHC in the $m_A - m_\chi$ plane for the resonant model, with $3\,\text{ab}^{-1}$. Here we have fixed $m_\chi =25\gev$, but this parameter is not relevant for the sensitivity provided that $2m_\chi < m_A$~\cite{Blanke:2019hpe}.}
\end{figure}

In the resonant model, the planned HL-LHC luminosity is not sufficient to exclude (left) or discover (right) any of the benchmark scenarios, due to the lower production rates.\footnote{This is due to the additional final states, e.\,g.\ $WW,ZZ+\met$, arising in this model, that can be targeted by complementary searches.} We therefore confine ourselves to presenting the cross sections required to discover a particular resonant benchmark in \reffig{fig:Rxs}. The mass of the lightest scalar, $m_\chi$, does not affect the sensitivity, since the boost of $A$ does not depend on $m_\chi$ or $m_h$. We therefore present our results in terms of the heavier scalar masses $m_A$ and $m_B$.
 From the figure, we see that we can test cross sections in the fb and sub-fb regime. As in the symmetric model, the significance increases when the spectrum is compressed.  

We note that in a complete model involving additional couplings, $\chi$ could either be the dark matter or a long-lived neutral particle that decays outside the LHC detectors. Dark matter direct detection experiments on the one hand and searches for long-lived particles (see e.\,g.\ \cite{Feng:2017uoz,Ariga:2018uku,Evans:2017lvd,Curtin:2018mvb,Alpigiani:2018fgd}) on the other hand thus serve as complementary probes of the nature of $\chi$.

As a final remark, we would like to stress that we have verified, using \texttt{CheckMATE2}~\cite{Dercks:2016npn}, that even with the largest possible cross section displayed here, the search for di-Higgs plus $\met$ is still the most sensitive channel for both symmetric and resonant topologies. We thus encourage the ATLAS and CMS collaborations to expand their di-Higgs portfolio of searches by including a topology-based study of the Higgs pair plus missing transverse energy final state.

%% file: BSMresonance/BSMSummary.tex
\section[Summary: precision goals]{Summary: precision goals for 
the measurement of the Higgs pair production process in the light of new physics}
\label{sec:BSMsummary}

We can usefully summarize this chapter as a set of goals for the measurement of the Higgs pair production process.  While one of the major objectives is certainly the determination of the trilinear Higgs self-coupling,  many models have new particles that can be directly searched for in observables related to the Higgs boson.  One of the most spectacular signatures of new physics in the Higgs sector is resonant double Higgs production (see Secs.~\ref{sec:BSMspin0},~\ref{sec:spin2} and \ref{sec:BSMmodels}).  In models with $s$-channel resonances, the di-Higgs invariant mass spectrum can be distorted compared to the SM distribution, with an additional peak appearing at the mass of the new resonance, as we see, for example, in Fig.~\ref{fig:phenoshape}.  Care must be taken to incorporate interference effects to correctly interpret results (see Sec.~\ref{sec:inteference}).

We have furthermore discussed how large the trilinear Higgs self-coupling can realistically be. In Sec.~\ref{sec:theoretical_constraints_EFT} we have seen how large the modifications to the di-Higgs production cross section can be in concrete models. We are hence now in the position to give a catalogue of precision goals for the measurement of the di-Higgs production process that might be obtained in future experiments, and the implications of each level for the discovery of effects due to  new physics models.

Beyond di-Higgs production, many of the models in Sec.~\ref{sec:BSMmodels} have additional scalar particles.  These new scalars can also be produced in pairs or in association with the observed Higgs boson, expanding di-Higgs production to di-scalar production.  These new modes provide a new, robust phenomenology for colliders.  Searches for these new modes are important for fully mapping out the Higgs potential and, in addition to a modified trilinear Higgs self-coupling, even helping to determine if electroweak baryogenesis is the source of the baryon asymmetry of the Universe (see Sec.~\ref{sec:cosmology}).

The above discussion is focused on probing modifications to the Higgs sector and searching for trilinear scalar couplings.  However, we have seen in Chapter~\ref{chap:EFT} that other effective operators can substantially modify the di-Higgs production process.
In particular, loops of new strongly interacting particles can affect the di-Higgs production cross section, and it is hence important to properly include their effects (see Sec.~\ref{sec:particleinloop}).

We now provide an indicative summary of what experimental precision on the di-Higgs measurement is needed to probe the different BSM phenomena we have surveyed in this chapter:

\begin{itemize}

\item {\bf Bronze:   Precision of 100\%: }   
Measurements at this level are sensitive to models with the largest 
new physics effects, in  which new particles of few hundred GeV mass
 appear in tree diagrams or as $s$-channel resonances. 
 Depending on the model, the heavy new resonance often has 
 sizeable branching ratios also to $VV$ final
 states. We have discussed in Sec.~\ref{sec:BSMspin0} models with singlets
 which allow for sizeable branching ratios of a heavy Higgs boson to light Higgs bosons,
 with values of maximally $BR(H\to hh)=0.4$ for singlet models with $Z_2$ symmetry, while
 larger $BR(H\to hh)$ are possible without $Z_2$ symmetry. 

\item {\bf Silver:   Precision of 25--50\%: }   Measurements at this
 level are sensitive to mixing of the Higgs boson with a heavy scalar
  with a mass of order 1~TeV.  Models of electroweak baryogenesis typically 
  predict this level of deviation in the trilinear Higgs self-coupling. At this level of 
  precision we are able to exclude a physical hypothesis with realistic deviations
  in the Higgs self-coupling, rather
   than just eliminating parts of parameter space.

\item {\bf Gold:   Precision of 5--10\%: }   Measurements at this level 
are sensitive to a broad class of loop diagram effects that might be created
 by light top squarks and or other new particles with strong coupling to the Higgs sector.
Measurements at this level could possibly complement measurements
 on new particles that could be discovered at the HL-LHC.  

\item {\bf Platinum:   Precision of 1\%: }   Measurements at percent
 level are sensitive to typical quantum corrections to the Higgs self-coupling generated by loop diagrams. 

\end{itemize}

In the remainder of this report, we will see how the 
capabilities of the LHC and of future experiments on the
measurement of the di-Higgs production process and the extraction of the Higgs self-coupling align with these goals.

%% file: experimental_status.tex

\input{HH_overview/Experimentaloverview.tex}

\chapter[Detector objects, triggers and analysis techniques]{Detector objects, triggers and analysis techniques}
\textbf{Editors: M.~A.~Kagan, L.~Mastrolorenzo}\\
\label{sec_exp_2dot2}

\input{AnalysisTechniques/overview_AnaTech.tex}
\label{sec:overviewAT}


\section[Jet reconstruction]{Jet reconstruction \\ \contrib{M.~Swiatlowski}}
\label{sec:jetReco}
\input{AnalysisTechniques/jet_reco_AnaTech.tex}

\section[Identification of \bjets]{Identification of \bjets \\ \contrib{M.~A.~Kagan, L.~Mastrolorenzo, C.~Vernieri}}\label{sec:bTagging}
\input{AnalysisTechniques/btagging_AnaTech.tex}

\section[Specific Corrections for \bjet transverse momentum]{Specific Corrections for \bjet transverse momentum \\
\contrib{N.~Chernyavskaya, F.~Micheli, L.~Mastrolorenzo, C.~Vernieri}}\label{sec:bjetreg}
\input{AnalysisTechniques/jet_regression_AnaTech.tex}

\section[Hadronic $\tau$ object identification]{Hadronic $\tau$ object identification \\ \contrib{A.~Bethani, K.~Leney, L.~Mastrolorenzo}}\label{sec:tau}
\input{AnalysisTechniques/tau_reco_AnaTech.tex}

\section[Photon reconstruction]{Photon reconstruction \\ \contrib{E.~Brost, R.~Teixeira de Lima}}
\input{AnalysisTechniques/photon_reco_AnaTech.tex}

\section[Trigger strategies]{Trigger strategies \\ \contrib{J.~Alison}}
\label{sec:bTrigger}
\input{AnalysisTechniques/trigger_AnaTech.tex}

\section[$HH$ specific analysis techniques]{$HH$ specific analysis techniques}
\input{AnalysisTechniques/kin_fit_AnaTech.tex}
\chapter[Overview of HH searches at the LHC] {Overview of HH searches at the LHC }  \label{chap:LHC}
\textbf{Editors: J.~Alison, B.~Di~Micco, A.~Ferrari, C.~Vernieri}\\
\input{HH_overview/IntroChapterHHAnalysis.tex}
\section[\hhbbbb: status and perspectives]{\hhbbbb: status and perspectives \\
\contrib{P.~Bryant, M.~Osherson}}
\label{sec:HH4b}
\input{HH4b/HH4b.tex}
\section[\hhbbyy: status and perspectives]{\hhbbyy: status and perspectives \\
\contrib{E.~Brost, R.~Teixeira de Lima, M.~Gouzevitch}}
\label{sec_exp_2dot4}
\input{yybb/yybb.tex}

\section[\hhbbtt: status and perspectives]{\hhbbtt: status and perspectives \\
\contrib{K.~Leney}}

\label{sec_exp_2dot5}
\input{bbtt/bbtautau.tex}

\section[\hhbbvv: status and perspectives]{\hhbbvv: status and perspectives \\
\contrib{J.~H.~Kim, S.~Shrestha}}
\label{sec_exp_2dot6}
\input{bbVV/bbVV.tex}
\input{bbVV/bbww.tex}

\section[\hh, other signatures: status and perspectives]{\hh, other signatures: status and perspectives \\
\contrib{C.~Veelken}}
\label{sec_exp_2dot7}
\input{HH_other_signatures/HH_other_signatures_main.tex}

\section[\hh production in the VBF mode]{\hh production in the VBF mode \\
\contrib{T.~J.~Burch}}
\label{sec_exp_2dot9}
\input{VBF/vbf.tex}

\chapter{Results presentation}\label{chap:res_pres}
\textbf{Editors: M.~S.~Neubauer, M.~Swiatlowski}\\
\import{./}{HH_overview/results_presentation.tex}

\chapter[LHC results]{LHC results}\label{chap:res_lhc}
\textbf{Editors: B.~Di Micco, J.~Schaarschmidt}\\

\section[\hh results and combination: status and perspectives]{\hh results and combination: status and perspectives}
\label{sec_exp_combination}
\input{HH_overview/Exp-combination.tex}
\section[Constraints on $\klambda$ from single Higgs boson measurements]{Constraints on $\klambda$ from single Higgs boson measurements \\\contrib{B.~Di Micco, S.~Manzoni, C.~Vernieri}}
\label{singleH_exp}

\input{HH_overview/singleH-exp.tex}

%% file: HH_overview/Experimentaloverview.tex
The Higgs self-coupling can be probed at the LHC through Higgs boson pair production and by exploring the radiative corrections to single Higgs measurements.

Both ATLAS and CMS experiments have developed a wide set of searches to test the SM prediction for the Higgs boson self-coupling. In this Part, we aim to provide an exhaustive overview of the current experimental effort to test the Higgs boson self-coupling through both the double and single Higgs boson productions processes. Direct searches for new resonant states decaying to \hh pairs will also be summarised.
We hope that the reader will find this overview useful for experimental studies at the LHC and future colliders.

Chapter~\ref{sec_exp_2dot2} provides an overview of the online selections (trigger), the detector object reconstruction techniques and the calibration strategies specific for \hh final states are reviewed.

Depending on the decay mode of the Higgs boson, a rich variety of signatures is available to probe the production of \hh pairs. Chapter~\ref{chap:LHC} presents an overview of the results of the searches for both non-resonant and resonant \hh production through gluon-gluon fusion from the ATLAS~\cite{Aad:2008zzm} and CMS~\cite{Chatrchyan:2008aa} experiments, based on the data recorded between 2015 and 2018, corresponding to an integrated luminosity of up to about 126~\ifb.



%

A discussion on how ATLAS and CMS collaborations could generalize the presentation of the results and possibly allow their re-interpretations under specific models is given in Chapter~\ref{chap:res_pres}.

The results of all the searches for \hh production at the LHC are presented in Chapter~\ref{chap:res_lhc} under different beyond the SM hypotheses. 
A first attempt to combine statistically the ATLAS and CMS results is also discussed. When possible, recommendations are provided on how to improve the current measurements and expand the interpretation of the experimental results.
The first experimental results from the indirect determination of \klambda via precision measurements of single Higgs processes, as described in Chapter~\ref{chap:EFT} are also presented and discussed.



%% file: AnalysisTechniques/overview_AnaTech.tex



In both ATLAS and CMS, the particles used to reconstruct the Higgs decays are identified by combining the information of several sub-detectors based on different technologies.
A detailed description of the ATLAS and CMS detectors, together with a definition of the coordinate system and the relevant kinematic variables, can be found in Ref.~\cite{Aad:2008zzm,Chatrchyan:2008aa}.
ATLAS and CMS have different concept and detector specifics but similar capabilities.

ATLAS has optimised the detector design to have good stand-alone measurements from each subsystem. Indeed it employees a toroidal magnet field for a stand-alone muon momentum measurement, in addition to the solenoid used for the momentum measurements in the inner detector. CMS instead has put major emphasis on the tracker system, consisting of all silicon detectors, and relies on a very strong solenoid magnet field to achieve excellent transverse momentum resolution.
Both the ATLAS and CMS calorimeter systems have two separate  sub-detectors for the reconstruction of electromagnetic and hadronic showers. 
Both the electromagnetic and hadronic calorimeters for ATLAS have longitudinal and transverse segmentation, while CMS exploits longitudinal segmentation only in the hadronic calorimeter. 
As a result, while the ATLAS calorimeter allows a good calibration of the energy of hadronic objects and stand-alone reconstruction of the jet direction, CMS has to combine with information from the tracking system in order to achieve similar performance for the jet measurements and pile-up subtraction.
Dedicated algorithms are needed in order to identify different particles. Most of them relies on machine learning techniques, such as Neural Networks (NN) or Boosted Decision Tree (BDT), which combine multiple observables at once to achieve good performance~\cite{Hocker:2007ht}. The algorithms are calibrated using data from well understood SM processes and the correction derived are then applied to the simulation to match the observed detector response.
In the following we will describe how jets are reconstructed and identified as initiated from $b$-quarks, as well as the $\tau$ and photon reconstruction algorithms.
The case when two objects are merged due to the high momentum of the parent particle (boosted objects) is also discussed.
In addition, dedicated strategies have been developed to identify final state particles in a very short time to make real time decision whether to save a collision event to disk or discard it. In this Chapter, the online selections (applied at trigger level) and the calibration strategies specific for \hh final states (such as kinematic fit) are reviewed.
In particular differences between ATLAS and CMS strategies are reported when relevant. Possible improvements and limitations of the current algorithms are also discussed. 

%% file: AnalysisTechniques/jet_reco_AnaTech.tex
Quarks and gluons fragment (hadronisation) into a large number of stable particles, which result in narrow cones of hadrons, called ``jets". Jets are reconstructed with the anti-k$_{\mathrm {T}}$~\cite{Cacciari:2008gp} clustering algorithm with a distance parameter $R=0.4$.
Due to the differences between the calorimeter and tracking systems, ATLAS and CMS employ different sets of inputs to the jet clustering algorithm. 
ATLAS uses topological clusters composed of calorimeter cells and applies corrections to the energy measurement based on the longitudinal profile of the energy deposits in the calorimeters, as well as the shape and number of associated inner-detector tracks. 
CMS, on the other hand, uses ``particle flow'' objects~\cite{Sirunyan:2017ulk} which exploits the information from all sub-detectors and aims at reconstructing and identifying all stable particles in the event ($\mu$, e, $\gamma$, $\pi$ etc...).
The particle flow algorithm matches inner-detector tracks to calorimeter energy depositions and perform a combined energy measurement, weighted by the expected resolutions of each detector. 
Thus, it compensates the calorimeter energy resolution with the tracker information at low \pT. Moreover the particle flow approach allows for the subtraction of energy deposits originating from pileup improving the jet energy resolution, especially at low \pT. \\
ATLAS is also considering moving to a particle flow approach for the end of Run 2 analyses~\cite{Aaboud:2017aca}. In the meanwhile, in order to reduce the impact of pileup, ATLAS requires a significant fraction of the tracks associated with each jet below a certain \pT threshold to have an origin compatible with the primary vertex, as defined by the jet-vertex-tagger (JVT) algorithm~\cite{jvt}. 

Techniques are in place to reduce the impact of pileup on the mis-reconstruction of the jet properties, also known as pileup mitigation techniques. Several approaches have been exploited so far. The jet-area method~\cite{Cacciari:2007fd} evaluates the average neutral energy density from pileup interactions and subtracts it from the reconstructed jets. The pileup per particle identification (PUPPI)
algorithm~\cite{Bertolini:2014bba} in CMS assigns a weight to each particle prior to jet clustering based on the likelihood of the particle originating from the hard scattering vertex.
Pileup mitigation will present a significant challenge as the colliders move to higher instantaneous luminosity values. 
Improving the jet resolution by more accurately removing pileup contamination will lead to narrower signal distributions, allowing for increased sensitivity of nearly the entire \hh search program, especially as the pileup is expected to grow to more than three times the current levels at the HL-LHC.

The jets reconstructed in the detectors are calibrated to the particle level (excluding neutrinos) using a multi-stage calibration procedure~\cite{Aaboud:2017jcu, Khachatryan:2016kdb}. It includes MC corrections taking advantage of MC truth and data-driven approaches used to uniform the detectors response and to provide scale factors to correct for simplified detector simulation.

Searches for new resonances decaying to \hh with mass above 1 TeV (high mass), the resulting Higgs bosons have a momentum considerably higher than their mass. Thus, each \hbb decay is reconstructed more efficiently as one hadronic jet with a larger anti-k$_{\mathrm {T}}$ distance parameter: $R=1.0$ (ATLAS) or $R=0.8$ (CMS).
As with $R=0.4$ jets, CMS uses particle flow objects as inputs, whereas ATLAS uses clusters of calorimeter cells only.

The larger jet size leads to increased susceptibility to contamination from underlying events and pileup. Specific algorithms, grooming, are employed to reduce contributions from soft and wide angle radiation by re-clustering the jet constituents, and as a result they help to mitigate pileup effects.


Different ``grooming'' algorithms are applied to large-radius jets to allow a better identification of the substructure of the $W/Z/H$-initiated jets, resulting from the decay to two partons at LO in QCD (2-prong decay). The resolution of the $H$-jet mass is improved by grooming since it filters out contributions from soft and wide angle radiation that affects significantly the estimation of the original invariant mass of the two-prong system.

The optimal choice of the grooming algorithm depends significantly on the specifics of the detector. Few options have been proposed such as trimming~\cite{Krohn:2009th}, employed by ATLAS, pruning~\cite{Ellis:2009me} and soft-drop~\cite{Dasgupta:2013ihk,Larkoski:2014wba}, used by CMS. A review of the performance of these algorithms can be found in Ref.~\cite{Asquith:2018igt}. 

ATLAS has recently validated large-radius jets that incorporate combined inner-detector and calorimeter information~\cite{Aaboud:2018kfi}. These ``track-calorimeter-cluster'' jets use the improved angular reconstruction of the inner detector and the improved energy measurements of the calorimeters in order to significantly improve the resolution at high \pT.

%% file: AnalysisTechniques/btagging_AnaTech.tex

The ability to correctly identify \bjets initiated by \hbb decays is crucial to reduce the otherwise overwhelming background from processes involving jets initiated from gluons ($g$) and light-flavour quarks ($u, d, s$), and
from $c$-quark fragmentation.

The $b$-quark fragmentation process is very peculiar and its properties are fully exploited to achieve good tagging efficiencies.
 The $b$-quarks hadronise in $B$-hadrons and several hadron particles, mostly pions. In particular, the large lifetime ($c\tau\sim 500~\mu$m) and the relative large mass of $B$-hadrons make \bjets unique. A $B$-hadron with $\pT \approx 50$ GeV will fly about half a centimetre in the transverse plane before decaying. Thus, daughter particles are expected to have a sizeable impact parameter with respect to the $B$-hadron point of origin, the primary collision vertex. In addition, $B$-hadrons are much more massive than anything they decay into, thus the decay products have a momentum of few GeV in the $B$ rest of frame. 
They can be identified by looking for (i) a decay vertex displaced from the  primary collision vertex, (ii) tracks with large impact parameter, or distance of closest approach, with respect to the primary collision vertex, (iii) non-isolated leptons from the semi-leptonic decays of $B$-hadrons (soft-leptons). 
The presence of a lepton in a jet is indeed a good signature of the presence of a $B$-hadron given the high rate of semi-leptonic decays ($\sim35\%$).  Moreover, since the $B$-hadrons retain about 70\% of the original $b$-quark momentum~\cite{Abe:1999ki}, usually these leptons have a high \pT relative to the jet \pT which make them easier to identify with respect to other sources of leptons in jets.

The primary source of \bjet mis-identification include jets initiated by a charm quark, and light-flavour quark- or gluon-initiated jets that have displaced vertices or large impact-parameter tracks.
Jets initiated by $c$-quarks , $c$-jets, are misidentified as $b$-jet due to the relatively large mass and lifetime of charm hadrons, and the presence of charm hadrons in the $B$-hadron decay chain. 
Light-flavour quark- or gluon-identified jets can be misidentified as $B$-hadrons due to detector resolution effects in the reconstruction of the secondary vertices and the impact parameters, due to  the hadron interactions with the material or long-lived particle decays (such as kaon and $\Lambda$ particles).

Both ATLAS and CMS have dedicated $b$-tagging algorithms which exploit in turn the secondary vertex, impact-parameter, and soft-lepton information~\cite{Aad:2015ydr,Sirunyan:2017ezt}. As these algorithms are largely complementary, multi-variate techniques based on neural networks are used to combine these different and complementary sets of information in order to yield the highest performance for a \textit{high-level} tagger. 
For jets in a \ttbar simulated sample with $\pT > 20$ GeV, with a selection on high-level taggers that results in a $70\%$ $b$-tagging efficiency, ATLAS achieves mis-tag rates of approximately $0.3\%$ for light flavoured jets and approximately $11\%$ for charm-initiated jets~\cite{ATL-PHYS-PUB-2017-013}. 
The highest performance $b$-tagging algorithm in CMS relies on a deep NN~\cite{guest}, with more hidden
layers and more nodes per layer, capable of combining vertexing information, track related variables and the kinematics of jets reconstructed with the particle-flow algorithm, exploiting the correlations between these variables~\cite{Sirunyan:2017ezt,cmsdp2018033}. 
The performance of the CMS $b$-tagging algorithms has been evaluated for jets with $\pT > 20$~GeV in a simulated sample of \ttbar events. For a $b$-tagging efficiency of $68\%$, the tagger $c$-quark mis-tag rate is $12\%$ 
and the light-quark mis-tag rate is $1.1\%$.
ATLAS outperforms CMS and this is also related to the different tracking performance.


\begin{figure}[!t]
\begin{center}
\includegraphics[width=0.4\textwidth]{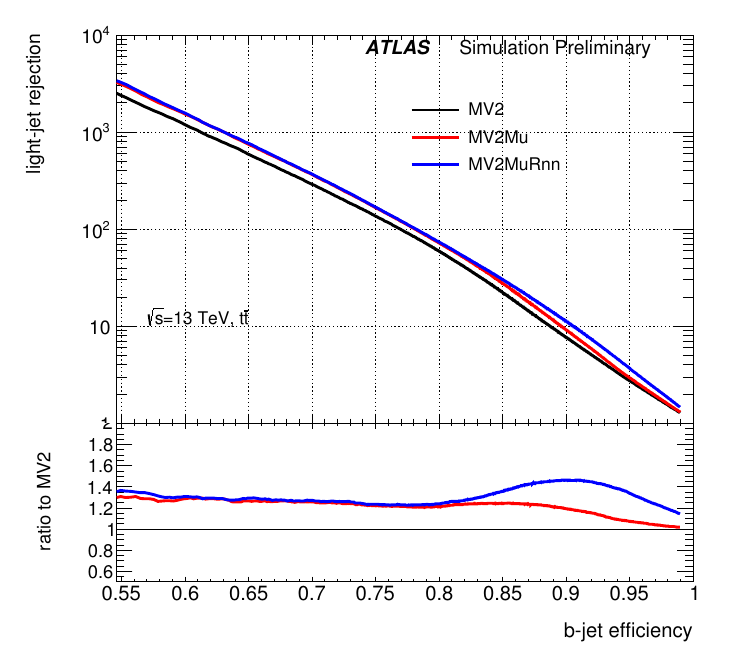}
\includegraphics[width=0.5\textwidth]{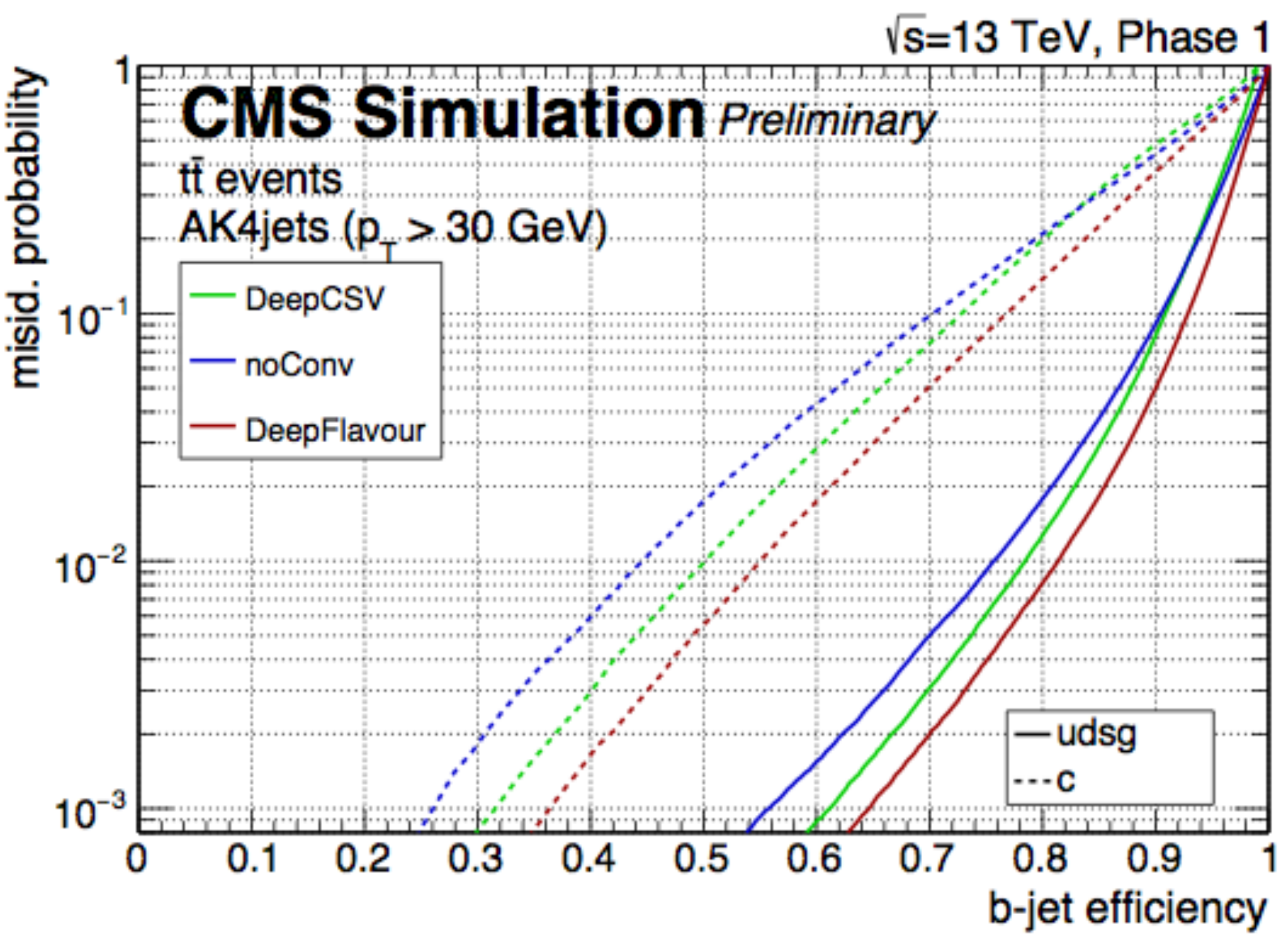}
\caption{\label{fig:DeepAlgoCMSPerf}
Light-flavour rejection as a function of \bjet efficiency for DNN-based algorithm developed by ATLAS (left)~\cite{ATL-PHYS-PUB-2017-013}. Comparison between the DeepCSV and DeepFlavour algorithms developed by CMS (right). The plot shows the $b$-tagging efficiency versus the mis-tag rate from light-jets (continuous line) and $c$-jets (dashed line)~\cite{cmsdp2018033}.
}
\end{center}
\end{figure}

In \refta{tab:btagwp} the commissioned working point for some of the CMS and ATLAS taggers most used in the double Higgs boson searches are listed.

\begin{table}[htbp]
\begin{center}
\begin{tabular}{l|c|c|c|c|c|c} \hline
Working point & \multicolumn{3}{c|}{ATLAS Tagger (MV2)}     & \multicolumn{3}{|c}{CMS Tagger (Deep CSV)}    \\ 
                             & $\epsilon_{b}(\%)$ & $\epsilon_{c}(\%)$ & $\epsilon_{l}(\%)$ &      $\epsilon_{b}(\%)$ & $\epsilon_{c}(\%)$ & $\epsilon_{l}(\%)$ \\      \hline
Very loose                  & 85            & 32& 3&--&--&-- \\
Loose                        & 77                 &    20              &  1                   & 84                 &    40              &  10   \\
Medium                       & 70                 &    11              &  0.3                  & 68                 &    12              &  1.1  \\
Tight                        & 65                 &    6             &  0.2                 & 50                 &    2             &  0.1  \\ \hline
\end{tabular}
\end{center}
\vspace*{-0.3cm}
\caption{Calibrated operating points with relative efficiencies for ATLAS and CMS $b$-tagging algorithms during Run 2, evaluated for jets with $\pT >20$ GeV from $\ttbar$ simulated events~\cite{Sirunyan:2017ezt,ATL-PHYS-PUB-2017-013}.}
\label{tab:btagwp}
\end{table}


The calibration of the tagging efficiency and mis-tag rates of these algorithms is performed by identifying pure samples of $b$-, $c$-, and light-flavour jets and by measuring the tagging efficiency in \ttbar or multi-jet events in data. By comparing to the results from simulation, scale factors can be derived to provide tagging efficiency corrections to the simulation. 
In ATLAS, scale factors for the \bjet tagging efficiency deviate from unity (i.e. no correction) by typically $2-4\%$, with uncertainties of $3-5\%$ except at low- and high-\pT~\cite{Aaboud2018}. 
ATLAS $c$- and light-flavour-quark mis-tag rate scale factors deviate from unity by typically $2-10\%$ and $5-30\%$, with uncertainties of $5-15\%$ and $30-50\%$, respectively~\cite{ATLAS-PLOT-FTAG-2019-003,ATLAS-CONF-2018-001,ATLAS-CONF-2018-006}. 
In CMS, the measured data-to-simulation scale factors for the tight working point of the DeepCSV algorithm range from 0.9 to 1.0. The relative precision on the scale factors is $1\%$ to $1.5\%$ using jets with $70 < \pT < 100$~GeV and rises to $3\%$ to $5\%$ at the highest considered jet \pT. The relative precision on the light-flavour mis-tag scale factors is 5–10\% for the loose working point and  20–30\% for the tight working point. For the $c$-quark mis-identification, the relative precision on the scale factor is 3–5\% for the loose working point, and 10–38\% for the tight working point.
Depending on the usage of the $b$-tagging algorithm in physics searches, reshaping scale factors may be needed to correct for the tagger discriminant distribution. For such scale factors related to $c$-jets in resolved topologies, the total uncertainty is $5\%$ to $10\%$, and the statistical uncertainty in the tagging efficiency dominates over the full jet \pT range.

There are several long-term challenges for $b$-tagging at the HL-LHC. Increased levels of pileup may lead to degradation in tracking performance, both from tracking algorithmic failures producing poor quality or completely fake tracks and from the radiation damage which degrades the pixel detector hit efficiency and resolution. 
In addition, the larger density of tracks in HL-LHC events can lead to $b$-tagging algorithmic challenges, as identifying the tracks from the $B$-hadron decay and rejecting other tracks becomes increasingly challenging. 

\subsection{Boosted \hbb taggers}\label{sec:hbbbosted}
For transverse momenta of the Higgs boson significantly higher than its mass ($\approx 250 $ GeV), the resulting Lorentz boost reduces the angular separation between its decay products. In the case of the decay to $b$-quarks, the Higgs boson is reconstructed as a single large-radius jet (``$H$-jet") and not as two separate jets. 
Then, the composite nature of such a jet is revealed by analysing its substructure. Several phenomenological studies have explored \hbb tagging algorithms (or ``$H$-tagging”) using jet substructure~\cite{Butterworth:2008iy}, though ultimately the optimal performance comes from using both the substructure information of the fat jet and the track and vertex information related to the $B$-hadrons lifetime~\cite{Asquith:2018igt}.


For boosted Higgs-jet identification, ATLAS uses large $R=1.0$ anti-$\mathrm{k_T}$ jets built from calorimeter topological clusters to identify Higgs boson candidates, measure their energy and direction, and estimate a variety of substructure inspired features,  for discriminating the \hbb signal from backgrounds~\cite{ATLAS-CONF-2016-039}. 
For identifying $b$-quark candidates, small $R=0.2$ anti-$\mathrm{k_T}$ jets are built from charged particle tracks only, then the aforementioned suite of $b$-tagging algorithms are applied to these jets~\cite{ATLAS-CONF-2016-039,ATL-PHYS-PUB-2014-013}. 
Such small radius jets can perform $b$-tagging even in dense environments and at small opening angles between $b$-initiated sub-jets as would be expected in boosted jets. 
Utilising tracks to build such jets benefits from the better resolution of tracking detectors over calorimeters. 
New approaches, aimed at providing $b$-tagging for jets where the $b-$quark pair $\Delta R$ is smaller than the track jets radius, have also been developed through the use of variable-radius track jets~\cite{ATL-PHYS-PUB-2017-010}.

In CMS different approaches to identify boosted \hbb candidates have been developed: the subjet $b$-tagging and the double-b tagger~\cite{Sirunyan:2017ezt}. 
In the first approach the subjets are first defined, using the anti-$\mathrm{k_{T}}$ algorithm with 0.4 distance parameter, and then the standard $b$-tagging is applied to each of the subjets. At high \pT the two subjets start to overlap causing the standard $b$-tagging techniques to break down due to double-counting of tracks and secondary vertices when evaluating the $b$-tag discriminants. 
The double-b tagger is a dedicated multivariate (BDT) tagging algorithm which does not define subjets. It fully exploits not only the presence of two $B$-hadrons inside the AK8 jet, but also the correlation between the directions of the momenta of the two $B$-hadrons. There are 27 inputs in total which rely on reconstructed tracks, secondary vertices (SV) as well as the two-SV system.
The performance achieved in terms of background rejection for a given boosted \bb tagging algorithm outperforms those reached by reconstructing and tagging individually the two jets for di-jet transverse momentum $\pT>350-400$ GeV.  

The algorithm has been updated to use a DNN based architecture, known as "Deep-Double-b" \cite{CMS-DP-2018-046}, and more observables exploiting the kinematics of the charged particle flow candidates and secondary vertex information.
For a given \bb tagging efficiency of $70\%$, the inclusive mis-tag rate is reduced more than a factor 2 with respect to the BDT based double-b tagger (from 4\% down to 1.2\%-1.5\%). 

Due to the small cross section of producing events with boosted \hbb or Z$\rightarrow \bb$ jets, the efficiencies of these algorithms are measured using QCD multi-jet events enriched in jets from gluon splitting to \bb (g$\to \bb$) with topology similar to that of boosted \hbb jets~\cite{ATLAS-CONF-2016-039,Sirunyan:2017ezt,CMS-PAS-BTV-15-002}. 

Recently, CMS and ATLAS have reported the first observation of Z to \bb in the single jet topology~\cite{HIG-17-010,ATLAS-CONF-2018-052}, consistent with the SM expectation, in the context of a search of inclusive Higgs boson production at high \pT decaying to \bb.

%% file: AnalysisTechniques/jet_regression_AnaTech.tex
The most sensitive searches for \hh production involve at least one Higgs boson decay to bottom quark-antiquark (\hbb). Improving the invariant mass resolution of the \bjet pair plays a critical role for these searches. 
The jet energy calibration is done as a function of the jet \pT and $\eta$, and taking into account the pileup activity of the event, as the \pT density ($\rho$), which is the corresponding amount of transverse momentum per unit area~\cite{Cacciari:2007fd}.
The jet energy is calibrated in data using QCD di-jet events, which are mostly gluon-initiated jets, without taking into account the additional details of the jet reconstruction. Typical values for the jet energy resolution are 15--20\% at 30 GeV~\cite{Aaboud:2017jcu,Khachatryan:2016kdb}.

Jets initiated from b-quarks contain $B$-hadrons, which have a relatively high probability (35\%) to decay to leptons and neutrinos.
The presence of neutrinos, which escapes detection, in the B hadron semi-leptonic  decay chain results in an even lower response with respect to the light quark/gluon induced jets used in the standard calibration~\cite{Aaboud:2017jcu, Khachatryan:2016kdb}.
In addition, due to the soft particles from the decay of heavy hadrons, \bjets deposit over a wider cone than light jets. Therefore, the standard jet energy calibration does not correct for the energy loss caused by escaping neutrinos or out-of-cone energy leakage, and a dedicated energy correction is needed for \bjets. 

Both ATLAS and CMS have developed specific strategy to correct the \bjet energy and improve the invariant mass resolution
of the reconstructed Higgs boson~\cite{Sirunyan:2018kst,HIGG-2018-04}. 

Both ATLAS and CMS attempt to correct the \bjet energy, by applying a multi-variate technique similar to that used in CDF~\cite{Aaltonen:2011bp}, 

The CMS method~\cite{Sirunyan:2017elk,Sirunyan:2018zkk} uses the regression method to combine various jet and event properties, to get an additional correction beyond the standard CMS jet energy corrections. The regression is essentially a multi-dimensional calibration at the particle level - including neutrinos - which exploits the main \bjet properties. The regression target is the jet \pT at generator level, including the contribution of neutrinos. A specialised BDT is trained on a jet-by-jet basis using a large dataset of simulated \bjets from decay of \ttbar pairs. It provides a correction factor that improves both the \bjet energy scale and its resolution.

Inputs are chosen among variables that are correlated with the b-quark energy and well measured.
They include detailed jet structure information about tracks and jet constituents which differs from light flavour quarks/gluons jets. Information from $B$-hadron decays on the reconstructed secondary vertices are used as well as soft lepton from semi-leptonic decay when available, providing an independent estimate of the b-quark \pT. 
This multi-dimensional regression then combines information about the secondary vertex and tracks associated to a \bjet, jet kinematics, jet composition and individual energy deposits reconstructed by the different CMS sub-detectors, as well as pileup information. 

In Run 1~\cite{PhysRevD.89.012003} the information carried by the variables related to the missing transverse energy ($\mathrm{E_{T}^{miss}}$) has been also exploited as input to the regression. In the absence of real $\mathrm{E_{T}^{miss}}$ in the event, it acts as a kinematic constraint for the momentum balance in the transverse plane. 

In general, the most discriminating variables are those related to the jet kinematic, due to the fact that most of the power of the regression is derived from the neutrinos involved in the semi-leptonic $B$-hadron decays.

The average improvement on the mass resolution, measured on simulated signal samples, when the corrected jet energies are used is about 15-25\%, resulting in an increase in the analysis sensitivity of 10–20\%, depending on the \pT of the reconstructed
Higgs boson and on the analysis strategy.

The validation of the regression technique in data has been performed in Z$(\ell\ell)+\bb$ events, by comparing the $Z$ \pT with the \pT of the $\bb$ system, and in an \ttbar-enriched sample targeting the lepton plus jets final state, by looking at the reconstructed top-quark mass distribution~\cite{Sirunyan:2017elk}.

Very recently, the CMS experiment has improved the regression through the use of a DNN~\cite{Sirunyan:2018kst, Sirunyan:2019wwa}. A dedicated loss function is introduced, allowing simultaneous training of an energy correction and a per-jet resolution estimator\footnote{The loss function combines a Huber function for the energy correction estimation with two quantile estimators~\cite{huber1964}.}. 
%
The DNN regression improves the \bjet resolution by a about
13\%. Therefore this improvement generalises well to \bjets originating from physics processes different from \ttbar production. A larger improvement of roughly 20\% is observed for the di-jet invariant mass resolution. The resolution estimator predicted by the DNN and based on quantile estimators is shown to predict the intrinsic jet resolution with an accuracy of better than 20\% over a \pT range spanning over one order of magnitude. 
The DNN-based regression has been validated in data using events arising from dilepton decay of a Z boson in association with \bjets. It was confirmed that the improvement coming from \bjet energy regression is observed in both data and simulation. 

ATLAS proceeds following the same logic to improve the \bjet energy resolution. Since muons are not included in the standard ATLAS jet calibration, but are present in roughly 15\% of the $B$-hadron decays, \bjets receive an additional $\mu$-in-jet correction. If a muon is found within a jet cone of $\Delta R = \sqrt{\Delta\eta^2 + \Delta\phi^2} = 0.4$, the four-momentum of the muon closest to the jet axis is added to the four-momentum of the jet. 
Additional residual jet \pT corrections are applied to account for escaping neutrinos and equalise the response to jets containing semi-leptonic and hadronic decays of $B$-hadrons. ATLAS has also developed a more sophisticated method using a boosted decision tree (BDT) algorithm with a set of inputs similar to those used by CMS~\cite{ATL-CONF-2019-047}. The improvement of the di-jet invariant mass resolution coming from simple average correction and BDT regression is very similar, and it is of the order of 18\% with respect to the calibration without including muons~\cite{Schopf:2637125}. 

\begin{figure}[!t]
\begin{center}
\includegraphics[width=0.47\textwidth]{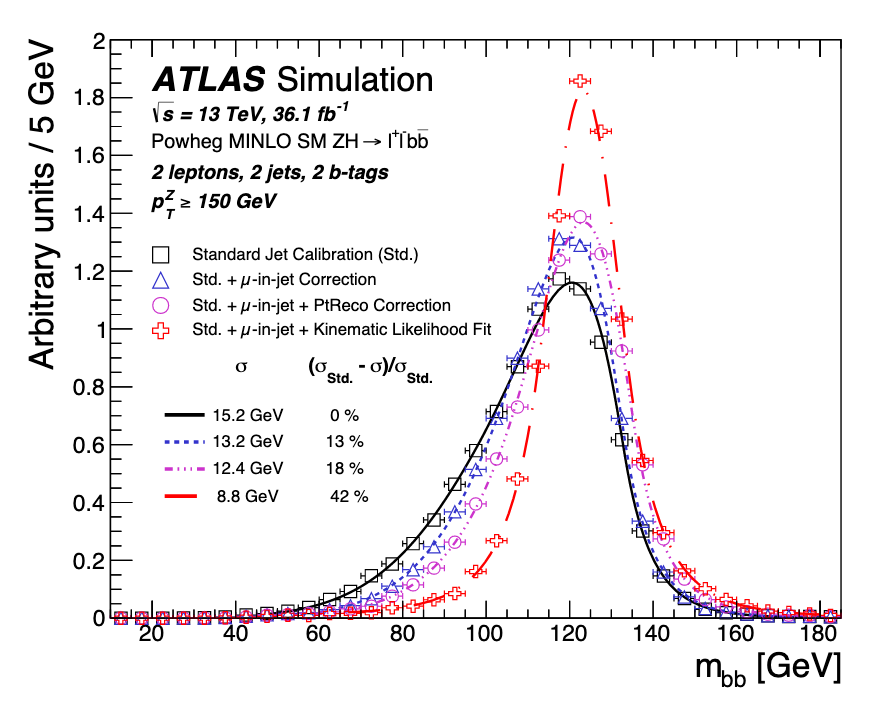}
\includegraphics[width=0.45\textwidth]{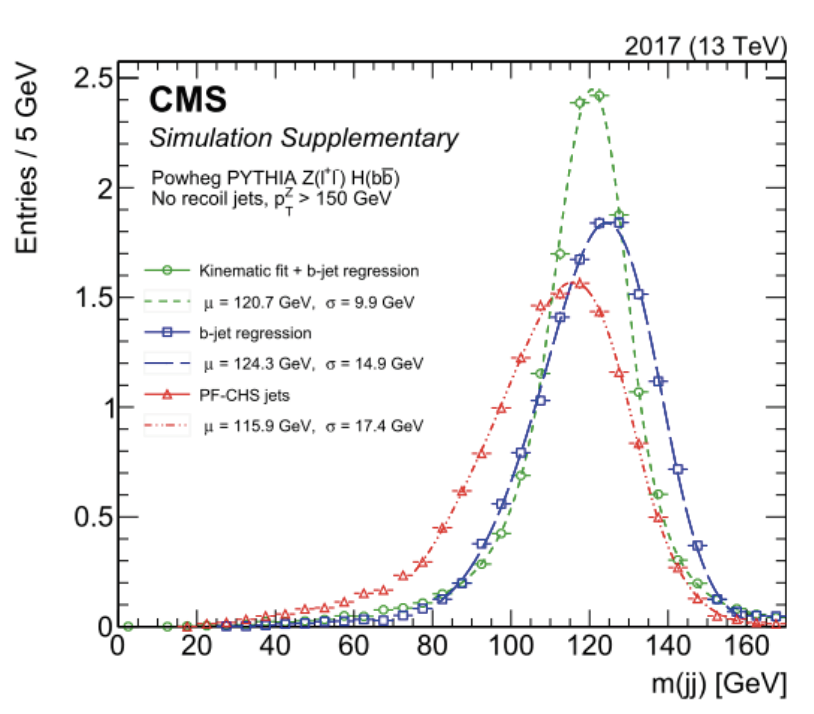}
\caption{\label{fig:kinematicFitPerf}
Di-jet invariant mass distributions for simulated samples of Z$(\ell\ell)\hbb$ events, before and after the $\mu$-in-jet energy correction (left) and the regression procedure
(right) is applied for ATLAS and CMS respectively~\cite{Sirunyan:2018kst,HIGG-2018-04}.}
\end{center}
\end{figure}

In Fig.~\ref{fig:kinematicFitPerf} the impact of the $\mu$-in-jet and the \bjet regression based corrections is shown for simulated \hbb events in ATLAS and CMS.

Both ATLAS and CMS have implemented dedicated algorithms to improve the \bjet energy resolution using machine-learning techniques. These algorithms will further be developed to meet the conditions of the HL-LHC. 
For the HL-LHC phase, a major upgrade of the ATLAS and CMS detectors is planned, with new detectors having a higher granularity, extended coverage, and additional timing layers. 
The high granularity of the detectors and the additional timing information will improve the reconstruction of the \bjets by removing spurious tracks from pileup and by improving the identification of secondary vertices. The additional information coming from detectors can be used for the training of more sophisticated neural network architectures.

%% file: AnalysisTechniques/tau_reco_AnaTech.tex
\label{sec:tauhad}
The $\tau$ is the heaviest lepton with a mass of 1.776 GeV and a lifetime $c\tau\sim\, 87\mu m$. Because of its large mass $\tau$ is the only lepton that can decay hadronically. More precisely, in 65\% of the cases the $\tau$ decays hadronically, typically into either one or three charged mesons (mainly pions) in presence of up to two neutral pions, subsequently decaying into a pair of photons.
While leptonic $\tau$-decays are reconstructed as prompt electrons or muons, hadronic decays of a $\tau$-lepton ($\tau_{\mathrm{h}}$) are reconstructed by combining detailed information of the visible decay products, such as tracks and their impact parameters, and energy clusters corresponding to the $\tau$ candidate.
In both ATLAS and CMS, hadronic $\tau$ detector objects are seeded by jets reconstructed with the anti-$\mathrm{k_{T}}$ algorithm, with a distance parameter $R=0.4$. Further reconstruction in CMS is performed by using the Hadron-Plus-Strips (HPS)~\cite{Sirunyan:2018pgf} method, described further in this section. 

The $\tau$ reconstruction method used by ATLAS~\cite{ATLAS-PUB-2015-045,ATLAS-CONF-2017-029} is based on a set of selection criteria, applied to jets to reject those initiated by quarks and gluons. Information from the inner tracker is used to identify the $\tau$ vertex and the resulting tracks using the number of hits and the distance of closest approach to the $\tau$ vertex, that are compatible with the hadronic $\tau$ decays. The energy of the $\tau$ object is then obtained using dedicate calibration algorithms.

The $\tau$ reconstruction in CMS is based on the HPS algorithm, that combines information from the energy deposited in the calorimeters and from the reconstructed charged tracks. It starts searching for $\tau$ lepton decay products within a particle-flow jet, identifying the most abundant $\tau$ hadronic decays, classified accordingly to the number of reconstructed charged hadrons and electromagnetic energy deposits: 1-prong (1 charged track assumed to originate from a pion),
3-prongs (3 charged tracks with invariant mass compatible with the mass of the intermediate resonances $a_{1}$ or $\rho$) and 1-prong + $\pi^0$ (1 charged track originated from a pion plus electromagnetic deposition).

Both experiments use multi-variate discriminators to reduce mis-identification of quark- and gluon-initiated jets as $\tau$ objects. During the 2015-2017 data taking period, a BDT was used by both experiments and in 2018, ATLAS introduced a $\tau$ identification algorithm based on a recurrent NN (RNN)\cite{Graves}. 
In physics analyses, different selections on the discriminant output score can be applied according to the desired efficiency. 
In ATLAS, three working points referred to as loose, medium and tight are provided, corresponding to different cuts on the BDT output score and hence different efficiencies, as listed in \refta{atlas-tau}. The rejection factors for quark- and gluon-initiated jets is ${\cal O}(10^2)$ to ${\cal O}(10^3)$ depending on the working point as well as \pT and the number of tracks.
In CMS, several multi-variate techniques were probed, with three to six working points. The corresponding efficiencies and background rejection factors are listed in \refta{atlas-tau}. 

\begin{table}[htbp]
\begin{center}
\scalebox{0.85}{
\begin{tabular}{l|c|c|c|c|c|c} \hline
\multirow{2}{*}{Working point} & \multicolumn{2}{c|}{ATLAS} &\multicolumn{4}{|c}{CMS}      \\ 
                            & $\tau$ eff./ jet $\rightarrow$ $\tau$   (1 track) &  $\tau$ eff. / jet $\rightarrow$ $\tau$  (3 tracks)  & $\tau$ eff. & jet $\rightarrow$ $\tau$   & e $\rightarrow$ $\tau$  & $\mu$ $\rightarrow$ $\tau$\\ \hline 
Loose                       &   60\%  / 2\%              &    50\%  / 1\%   & 60\%              & 0.8\%  & 1\% & 0.1-0.5\%    \\
Medium                      &   55\% / 0.8\%             &    40\%  / 0.8\%   & 55\%              & 0.4\%  & 0.2\% &  \\
Tight                       &   45\% / 0.6\%              &    30\% / 0.6\%     & 45\%              & 0.2\%  & 0.1\% & 0.03-0.4\% \\ \hline
\end{tabular}
}
\caption{\label{atlas-tau} Efficiencies for the different $\tau$ hadronic identification working points used by the ATLAS~\cite{ATLAS-CONF-2017-029,TAU-18-001} and CMS~\cite{Sirunyan:2018pgf} collaborations.
 Identification efficiency are evaluated using simulated \htautau or a 2~TeV BSM resonance decay to $\tau^+\tau^-$ events. Jet $\rightarrow$ $\tau$ mis-identification probabilities for $\tau$ objects are evaluated using simulated multi-jet events and reported for inclusive \pT for ATLAS while for 30-60 GeV in the CMS case. e $\rightarrow$ $\tau$ and  $\mu\rightarrow$ $\tau$ mis-identification probabilities are evaluated using $Z/\gamma \rightarrow ee/\mu\mu$ events.}
\end{center}
\end{table}



The efficiency of the hadronic $\tau$ identification criteria has increased enormously since the first data-taking periods in 2010, when the fake rates were more than an order of magnitude larger for similar efficiencies, despite the increase of instantaneous luminosity and concurrent pileup events. Electrons and muons can also be mis-identified as $\tau$ objects, and these backgrounds are suppressed using algorithms that combine information from the inner tracker, calorimeters, and muon detectors. The $e\rightarrow\tau$ and $\mu\rightarrow\tau$ mis-identification probabilities are significantly smaller than for jet$\rightarrow\tau$.

The uncertainties on $\tau$ identification efficiency
correction factor measurement are approximately 5-6\% for the sum of
 the transverse momenta of all charged pions and photons from $\pi^0$ (visible transverse momentum) in the 20-60 GeV range~\cite{ATLAS-CONF-2017-029,Sirunyan:2018pgf}. 
The reconstructed $\tau$ energy scale correction factor is measured with a precision of approximately 1.2–2\% for ATLAS~\cite{ATLAS-CONF-2017-029} and less than 1.2\% for CMS~\cite{Sirunyan:2018pgf}.

\subsection{Boosted \htautau}
Reconstructing the di-$\tau$ system is an integral part of \hh searches that include a \htautau decay. In the case of searches for heavy resonances, it is likely that the di-$\tau$ system is produced with very high transverse momentum and the $\tau$ decay products are more collimated. The higher the energy of the original state the smaller the angular distance between the two $\tau$ objects. 
The $\tau$ reconstruction efficiency drops dramatically as the di-$\tau$ \pT increases. The $\tau$ reconstruction is seeded by anti-$k_t$ jets with a distance parameter $R = 0.4$, corresponding to the maximal distance between jet axes. Therefore $\tau$ pairs within a cone of $\Delta R < $0.4 are merged into the same jet and can not be reconstructed separately, as for the boosted \hbb reconstruction.

To reconstruct highly boosted $\tau$ pairs, a new approach is necessary since they cannot be handled with existing methods by construction. A simple solution would be to reduce the anti-$\mathrm{k_{T}}$ distance parameter, until both $\tau$ objects can be reconstructed separately again. This approach does improve the efficiency of a single $\tau$ reconstruction in a high-momentum regime. 
However, in the case of boosted $\tau$ pairs it is very likely one $\tau$ to have a significantly higher \pT than the other. In the case the $\tau$ pair originates from a scalar particle, as the Higgs boson, then one $\tau$-lepton is likely to give most of its energy to the $\tau$ neutrino because of spin conservation  and the V-A structure of the electroweak interaction. Therefore a better solution is to reconstruct the boosted $\tau$ pair as one object. 
Both ATLAS and CMS have developed specific tools for the identification of boosted di-$\tau$ systems with similar approach.
The sub-structure of wide jets is exploited to look for the presence of two $\tau$ decays. The large-radius jets are used as seeds for the reconstruction and sub-jets are subsequently identified.
ATLAS has employed a multi-variate method~\cite{Kirchmeier:2105592} to discriminate the boosted di-$\tau$ system from other boosted hadronic objects. The observable used as input to the algorithm are similar to those used for standard $\tau$ identification, including calorimeter information on the clusters and energy deposits, as well as tracking inputs related to primary and secondary vertices to estimate the $\tau$ decay length.  Given the distinctive di-$\tau$ decay signature, a higher jet background rejection is likely to be achievable for boosted $\tau$-pairs as compared to boosted \hbb. 

So far the focus of both experiments has been to identify fully hadronic di-$\tau$ system, however everything already stated before applies in the case of the semi-leptonic di-$\tau$ decays as well. In this case the wider jet is investigated to look for the presence of a lepton.  The identification of boosted semi-leptonic di-$\tau$ decays is significantly easier for CMS due to the use of particle flow in the event reconstruction which combines tracker and calorimeter information for particle identification.

CMS has developed a dedicated reconstruction algorithm to reconstruct the Higgs boson decaying to $\tau$ leptons. Higgs decaying to $\tau$ leptons are clustered as large radius jet (distance parameter of 0.8) and if two subjets with \pT $>$ 10 GeV that satisfy the mass drop condition are found, then the two subjets are used as seeds in the standard $\tau$ reconstruction and the HPS algorithm is applied on them to identify hadronic taus. The $\tau$ leptons selected by the HPS algorithm are then required to have $\pT >$ 20 GeV and satisfy a selection on the (MVA) $\tau$-ID isolation. A medium working point is used for the leading $\tau$ and a very loose one is used for the second leading $\tau$ in fully hadronic events.

For boosted Higgs boson decays into $\mu\tau_{\mathrm{h}}$ and $\tau_{\mathrm{h}}\tau_{\mathrm{h}}$, the corresponding efficiencies are about 80\% and 60\% and  the mis-identification rate 10$^{-3}$ and 10$^{-4}$ respectively, depending on the Higgs boson \pT~\cite{CMS-DP-2016-038}.

%% file: AnalysisTechniques/photon_reco_AnaTech.tex
The photon reconstruction algorithm in the ATLAS detector starts by dividing the electromagnetic (EM) calorimeter into angular regions of $\Delta\eta \times \Delta\phi = 0.025\times 0.025$, a tower, with energy given by the sum of all longitudinal cells within that area. 
Towers with an energy larger than 2.5 GeV are used as seeds for the cluster reconstruction with the sliding window algorithm~\cite{Lampl:1099735}, which clusters calorimeter cells within fixed-size rectangle,  with a window size of $3\times5$ towers. 
Clusters are classified as electron, unconverted photons or converted photons according to the presence of matched well reconstructed tracks that are consistent or not with primary vertices in the event. 
These EM clusters are calibrated to account for different effects, in data and simulation, using the transverse and longitudinal segmentation of the EM calorimeter. 
The calibration uses $\mathrm{Z} \rightarrow e^{+}e^{-}$, $\mathrm{Z} \rightarrow (e^{+}e^{-}, \mu^{+}\mu^{-}) \gamma$ and $J/\Psi \rightarrow e^{+}e^{-}$ as standard candles.
This achieves a relative energy resolution of $\sigma/E = 2.5\%\ (1\%)$ for central unconverted photons with $\ET =  20\, (200)$~GeV~\cite{Aad:2014nim}.

In the CMS detector, photon and electron clusters are reconstructed based on energy deposits in the CMS electromagnetic calorimeter (ECAL) crystals. Crystals with an energy deposit above that of their immediate neighbours and above the noise threshold are used as seeds for the clustering algorithm, which works with flexible-sized windows, depending on the energy distribution around the seed crystals~\cite{Khachatryan:2015hwa}. 
Before the clustering, the ECAL crystal responses are corrected for ageing effects, and inter-calibrated by measuring quantities such as the $\pi^{0}/\eta\rightarrow\gamma\gamma$ process and the ratio between the energy of electrons with respect to their momenta measured by the CMS trackers. 
The absolute scale calibration is then performed with $\mathrm{Z} \rightarrow e^{+}e^{-}$ decays as a function of $\eta$. 
In addition, pileup effects are mitigated with a multi-variate regression technique. 
The energy resolution is measured in $\mathrm{Z} \rightarrow e^{+}e^{-}$ events with electrons reconstructed as photons, and achieves a relative resolution of $\sigma_E/E = 1.5\%$ for central electrons with low bremsstrahlung emission~\cite{Khachatryan:2015iwa}.

ATLAS identifies photons with respect to jets with high EM activity (such as $\pi^{0}\rightarrow\gamma\gamma$ from hadronic showers) with rectangular cuts on expected prompt-photon shower shapes~\cite{Aaboud:2018yqu}. 
Due to the longitudinal segmentation of its EM calorimeter, ATLAS uses information based on the energy distribution in the different EM layers for a purer selection. 
This purer version of the algorithm achieves an efficiency of $85-90\%$ $(85-95\%)$ for unconverted (converted) photons in the range of 30~GeV $<\ET<$ 100~GeV~\cite{Aaboud:2018ugz}. 
Additionally, isolation criteria based on vetoing hadronic (track- and calorimeter-based) activity around the cone defined by the photon axis is used to reject $\pi^{0}\rightarrow\gamma\gamma$ from nearby jets that have been reconstructed as a single photon. 
The isolation requirement has a signal efficiency of approximately $98\%$ for SM \hhbbyy events.

Photon identification in CMS is performed both with rectangular cuts and with a multi-variate approach based on a BDT~\cite{Khachatryan:2015iwa}. 
Quantities such as the width of the photon shower in the $\eta$ direction are used to mitigate hadronic background. 
The CMS photon identification BDT includes variables related to extra activity in the detector in the vicinity of the reconstructed photon, and, in the endcaps, extra information obtained from the CMS preshower detector~\footnote{
The preshower detector is located in front of the ECAL in the endcap regions and has a much finer granularity.}, particularly to identify $\pi^0\rightarrow\gamma\gamma$ decays. Since in the endcap regions, the angle between the two emerging photons from the decay of a neutral pion is on average smaller. 
The algorithm was developed during Run 1, focusing on the performance of the CMS \hyy analysis. 
For the Run 2 version of the algorithm, efficiencies on data with $\mathrm{Z} \rightarrow e^{+}e^{-}$ events, where one of the electrons is reconstructed as a photon, are found to be between 75\% and 95\% for photons with $\ET > 20$~GeV \cite{DP17004}.

%% file: AnalysisTechniques/trigger_AnaTech.tex
Identifying \bjets and $\tau$ leptons efficiently at trigger level~\cite{Aaboud:2016leb,Khachatryan:2016bia} is critical for two of the most sensitive \hh final states, \hhbbbb and \hhbbtt. 
The challenges of properly identifying these objects online are outlined in the following sections. 

The ATLAS and CMS triggers consists of two different systems: the L1 Trigger (L1) and the High Level Trigger (HLT).
The first uses custom-built programmable hardware to make an
accept-reject decision in approximately 2.5-3$\mu$s, while the second relies on a shelf processor farm employing the same reconstruction software framework used for the offline reconstruction.

\subsection[\bjet trigger]{\bjet trigger \\
\contrib{J. Alison}}
\begin{figure}
  \begin{center}
    \includegraphics[width=0.65\linewidth]{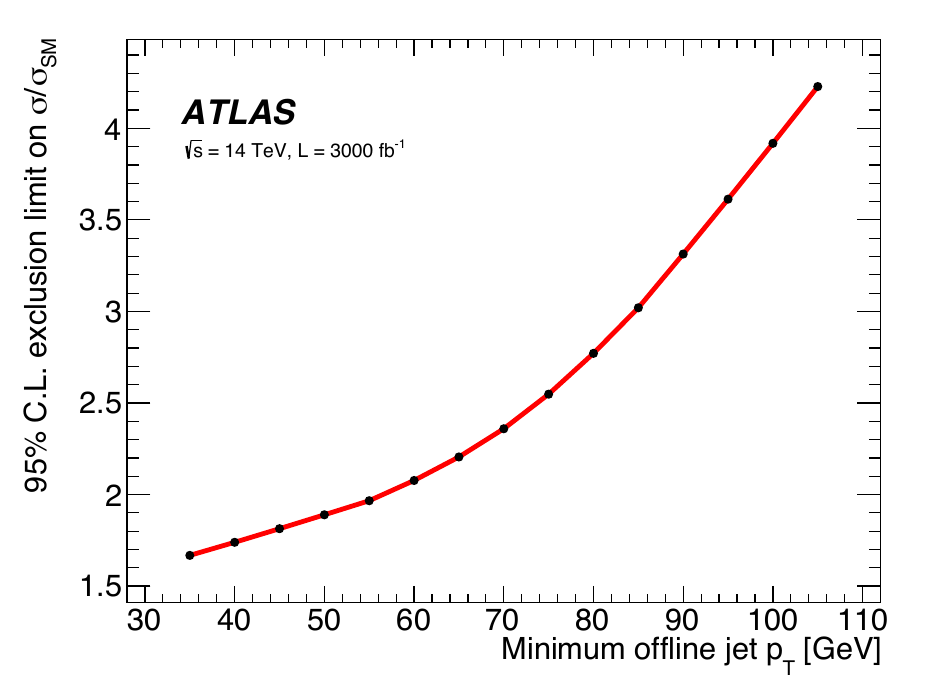}
    \caption{Expected upper limit on the \hhbbbb\ cross section as a function of the minimum jet \pT threshold~\cite{ATLAS-TDR-029}.}
    \label{fig:hhVsJetPt}
  \end{center}
\end{figure}

The all-hadronic \hhbbbb final state poses a great challenge to the online selection criteria due to the overwhelming rate of QCD multi-jet events. 
The \bjet properties are exploited at trigger level to keep an acceptable rate without increasing the jet transverse momentum thresholds, reducing the \hhbbbb\ signal acceptance.
The sensitivity of the projected \hhbbbb\ analysis as a function of the jet threshold is shown in Fig.~\ref{fig:hhVsJetPt}.
An increase of the jet threshold from 60 to 100 GeV reduces the \hhbbbb\ sensitivity by a factor of two.
This loss is greatest for events with relatively low \mhh, the region most sensitive to $\lambda$.
It is therefore crucial to identify \bjets at the trigger level with the highest possible efficiency. 
This is a challenging task for the LHC experiments. 

The L1 triggers do not use inner detector tracks and thus provide no separation between \bjets and jets from light-flavour quarks or gluons.  
As a result, \bjets can only be efficiently collected using relatively inclusive, and consequently high-rate, hadronic L1 triggers.
The output rate of these L1 seeds is a major, and often the most severe, constraint for the \bjet triggers.
The \bjet identification is carried out in the high-level trigger (HLT), when track information becomes available.
Another major limitation to \bjet triggers is the available CPU in the HLT farms to perform track reconstruction. 
While tracking information is necessary for \btagging, it is computationally expensive. 
The large CPU cost, coupled with a high input rate from the inclusive L1 triggers, results in the \bjet triggers demanding a significant fraction of the available HLT CPU. 
The \bjet triggers have to trade performance for speed in order to fit into the allocated resources. 
It is also important for the online \btagging to maintain as much consistency with the offline \btagging algorithms as possible. 
Most of the searches for \hh require \bjets to be identified both online and offline; any inconsistency in the two identification algorithms leads to a reduced overall efficiency.
Maintaining online/offline consistency is particularly challenging as the offline algorithms are constantly evolving -- even after the trigger decisions have been made -- and are not subject to the CPU constraints in the HLT.
The remainder of this section discusses the various inputs to HLT \btagging and summarises the overall performance. 
Differences between ATLAS and CMS are highlighted.

Primary vertex (PV) finding is crucial to \btagging as it defines the reference from which track displacements are measured and it is used to suppress tracks coming from pile-up. 
Both transverse and longitudinal positions of the PV are needed. 
The transverse position is determined from the beam spot position. 
The position of the beam spot is monitored in real-time during data-taking with dedicated HLT triggers. 
The transverse beam spot width is comparable to the accuracy with which the transverse PV position can be measured, $\mathcal{O}(10\mu m)$.
As a result, the PV transverse position is approximated with the beam spot position. 
The longitudinal PV position, however, must be reconstructed on an event-by-event basis. It is used both to reduce the phase-space where tracking is performed, as well as to define the inputs to the \btagging. 
Only tracks pointing to the PV are important for \btagging, so imposing a maximal longitudinal distance between the track and the PV significantly reduces the relevant hit combinations and thus the CPU cost associated to tracking algorithms.

ATLAS and CMS have quite different approaches to reconstruct the online PV. CMS uses an iterative approach that starts with \textit{track-less} vertex finding using jet directions and pixel clusters. 
Pixel clusters matched to the four leading jets in the event are projected to the beam line. 
The position of the PV along the beam line is then determined as a maximum in the projected hits positions. 
This technique is extremely fast and locates the PV along the beam line with an accuracy of about a centimetre. 
This preliminary estimate of the PV position is then used to seed pixel-only track finding. 
The PV finding algorithm is then performed again including the pixel tracks, improving the resolution in $z$ to $\sim100 \mu m$. 
Tracks reconstructed with a combination of hits from both the pixel and strip detectors, consistent with this PV position are then reconstructed and used to further refine the PV position determination, resulting in a final resolution of around 25$\mu m$, comparable to the offline PV  resolution.

The PV finding in ATLAS is done in one step. 
Track reconstruction is performed using inner detector hits matched to $\pT > 30$ GeV jets found at L1. 
The hits are required to fall within $\Delta R < 0.2$ from the jet direction. 
Tracks with $\pT > 1 $ GeV are reconstructed using a configuration of the track finding algorithm optimised for speed~\cite{Aaboud:2016leb}. 
In 2017, to reduce the CPU cost, the threshold on the track \pT was raised to 5 GeV. 
These tracks are then used to reconstruct the PV with an accuracy of $\sim 60 \mu m$ along the beam line.
The PV position is then used as seed to the track finding algorithm in a wider $\Delta R < 0.4$ area around the jet direction. 
A more precise and CPU expensive configuration of the track reconstruction algorithm is used at this final stage.

The HLT track efficiency is one place where there is a significant difference in performance between ATLAS and CMS.
The online track reconstruction in CMS, as evaluated in simulated \ttbar events, has an efficiency that is 10\% lower than that of the corresponding offline reconstruction for tracks with \pT of 1-10 GeV.
In ATLAS for a similar kinematic phase space, the efficiency of the online track reconstruction relative to the offline is better than 98\%.
This difference in track reconstruction performance translates into a difference in online \btagging performance between the two experiments.


The \bjet trigger decisions are ultimately made based on the multiplicities of jets passing various \pT and \btagging thresholds.
Jet reconstruction, discussed in Sec.~\ref{sec:jetReco}, is thus also critical to the \bjet trigger.
The online jet reconstruction follows the procedure used offline as closely as possible. 
Residual differences in the online/offline performance arise mainly from the different track reconstruction used online and from the jet thresholds applied at L1.
Currently neither experiment implements dedicated \bjet\ \pT corrections in the trigger, an obvious potential area for future improvement.

The \btagging algorithm is the final ingredient for the \bjet\ triggers.
The \btagging algorithms used by ATLAS and CMS are described in Sec.~\ref{sec:bTagging}.
The online algorithms follow those used offline as closely as possible. 
The primary differences between the online and offline \btagging arise from differences in the input tracks and from improvements to the offline algorithms that come after the software used in the trigger is frozen.  
In Run 2, ATLAS used the MV2c algorithm both offline and in the trigger~\cite{Gupta:2271945}. 
In the start of Run 2, CMS deployed a version of the CSVv2~\cite{Sirunyan:2017ezt} algorithm at trigger level. 
However, during 2018 the CMS trigger moved to the DeepCSV discussed in Sec.~\ref{sec:bTagging}.

\begin{figure}
  \begin{center}
    \includegraphics[width=0.45\linewidth]{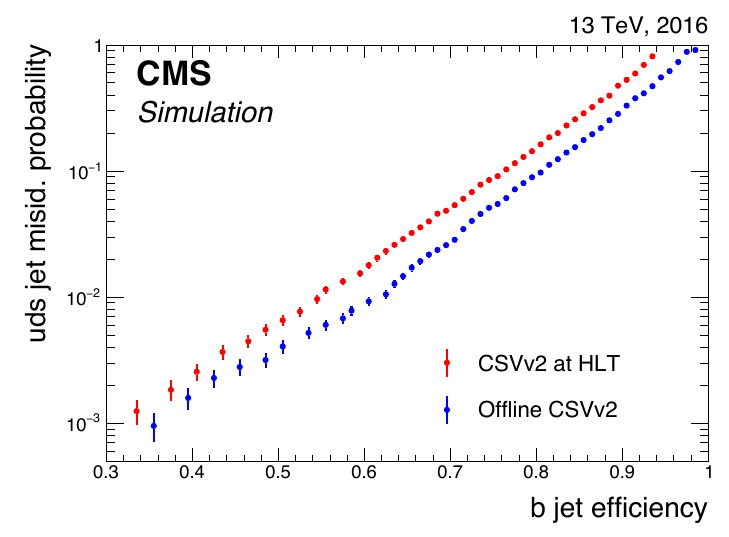}
    \includegraphics[width=0.5\linewidth]{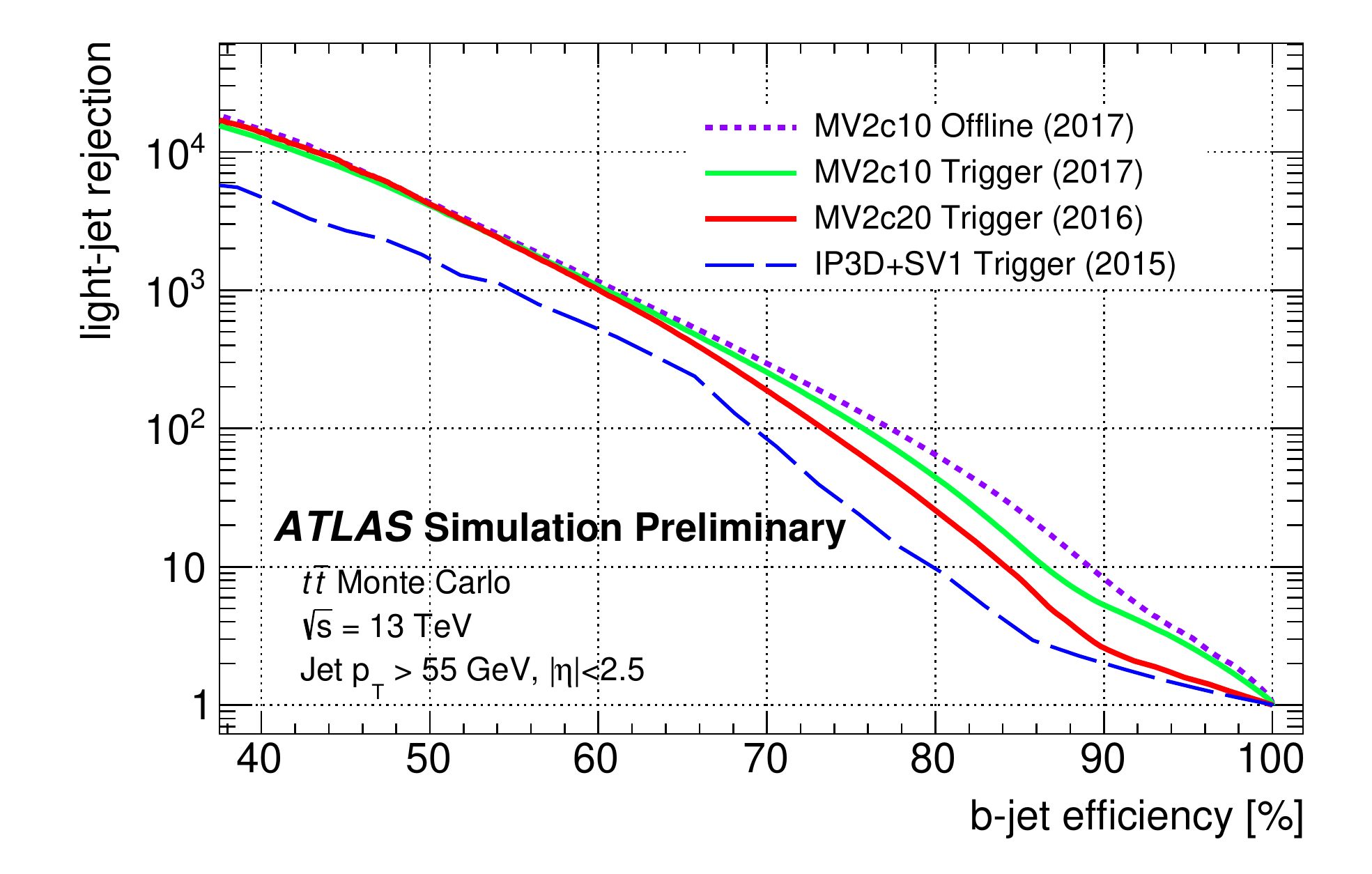}
    \caption{Comparison of online and offline \btagging performance for CMS~\cite{Sirunyan:2017ezt}~(left) and ATLAS~\cite{Gupta:2271945}~(right).}
    \label{fig:OnlineBTaggingPerf}
  \end{center}
\end{figure}

The relative performance of the online and offline \btagging algorithms are shown for both ATLAS and CMS in Fig.~\ref{fig:OnlineBTaggingPerf}.
For a background rejection of 100, the difference in online and offline signal efficiency for ATLAS is $\sim$2\%; in CMS, the corresponding difference is $\sim6$\%.
The worse relative online performance for CMS is likely a result of the lower HLT tracking efficiency because the limited availability of pixel tracks. 



\subsection[$\tau$ trigger]{$\tau$ trigger \\
\contrib{A.~Ferrari, L.~Mastrolorenzo}}
During the Run 2 taking, CMS has developed a new $\tau$ trigger algorithm~\cite{Zabi_2016} for the L1. The recent micro-TCA ($\mu$TCA) technology~\cite{ATCA}, together with more powerful and dedicated Field Programmable Gate Arrays (FPGA), had being deployed at the L1 trigger during the Phase-I upgrade allowing enhanced calorimeter granularity to be used by the online algorithms\footnote{the granularity corresponds to the single calorimeter trigger tower: 5x5 crystals in ECAL in addition to the corresponding projection in HCAL}. The L1-tau algorithm is based on an innovative dynamic clustering technique (used also to trigger electron and photon at L1~\cite{Dev_2017}) capable to combine the information coming from the calorimeters to perform a first online identification of the main $\tau$ hadronic decay mode (1-prong, 1-prong, 1-prong+$\pi^{0}$, and 3-prongs). Together with a cluster-dedicated calibration and an innovative isolation technique to perform an online PU mitigation, the performances obtained allow to effectively use the L1-tau trigger to seed the acquisition of events with hadronic $\tau$ lepton in their final state requiring unprecedentedly low online thresholds. For the different hadronic $\tau$ decay modes considered, the trigger efficiency is found to be close to 100\% for $\tau$ reconstructed online with a $\pT$ fairly above the trigger threshold (to avoid energy resolution effect).

\begin{figure}[!ht]
\begin{center}
\includegraphics[width=0.45\textwidth]{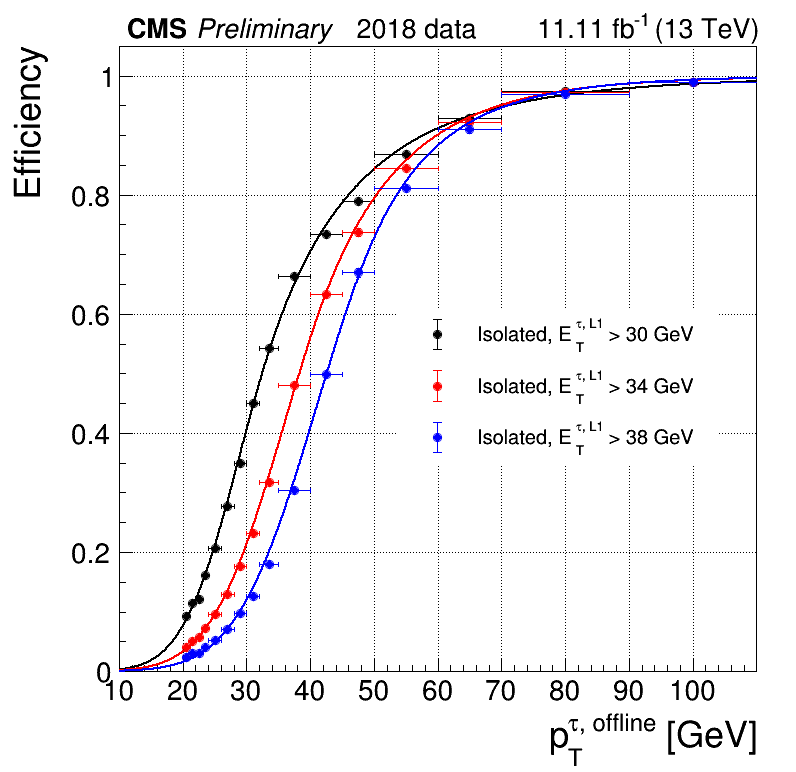}
\caption{\label{fig:tauTrigCMS}
 Level-1 trigger efficiency of isolated $\tau$-seeds (i.e. requiring the L1 $\tau$ candidate to pass a cut on its isolation transverse energy) as a function of the offline $\tau$ \pT~\cite{Zabi_2016}.}
\end{center}
\end{figure}

In ATLAS, the trigger-level identification of $\tau$ objects relies at L1 on the calorimeter information, with a granularity of $0.1 \times 0.1$ in $\eta$ and $\phi$~\cite{ATLAS-CONF-2017-029}. A core region consists of $2 \times 2$ trigger towers and requirements are placed on the transverse energy of the two most energetic adjacent towers as well as an isolation region around the core. At the HLT, topological clusters of calorimeters cells are considered within a cone of radius 0.2 around the L1 $\tau$ object and they are calibrated using the same method as offline $\tau$ objects. Following the use of trigger specific pattern recognition algorithms, requirements of at most 4 and 2 are made on the number of tracks in, respectively, the core ($\Delta R < 0.2$) and isolation ($0.2<\Delta R<0.4$) regions. The hits and tracks identified in this fast-tracking procedure then serve as seeds at the HLT, similarly to the offline reconstruction. Finally, the tracking and calorimeter information is used in a BDT algorithm, similarly to that used for offline identification, with minor differences arising from the fact that vertexing information is not available at trigger level. Three working points (loose, medium and tight) are defined for the online $\tau$ identification. 
These working points were tuned to provide target efficiencies of approximately 0.95 (0.70) after their offline counterpart identification is applied for $\tau$ leptons with one (three) associated tracks

%% file: AnalysisTechniques/kin_fit_AnaTech.tex
The event topology resulting from the Higgs pair production is very peculiar and it could be further exploited to improve the signal reconstruction, as described in the next two sections.

\subsection[Kinematic fit procedure]{Kinematic fit procedure
\\
\contrib{M.~Gouzevitch, C~.Vernieri}}
\label{sec_exp_kinfit}
It has been shown (see \cite{ALEPH:2005aa, Aubert:2003zw, Aubert:2004tea, Aubert:2004aw} and references therein) that the resolution of the measured objects in the final state of $p-p$ collisions can be improved by forcing well-defined kinematic hypotheses through an event-by-event least square fitting technique. The resulting chi-square of the fit can be interpreted as the probability of the proposed kinematic hypotheses to be true for the observed event. 

In the searches for resonances decaying into \hh, the Higgs boson mass,
as measured by both ATLAS and CMS experiments~\cite{Aad:2015zhl}, could be used as a kinematic constraint in the event reconstruction. The kinematic fit procedure is extremely effective for improving the four body invariant mass. Such kinematic constraints are widely used for measurements where a decay proceeds through some known intermediate state.
For example, in the case of \hhbbbb the kinematic fit technique aims to fit the measured quantities, i.e. the four \bjet four vectors, to certain hypotheses within their uncertainty, as described in~\cite{DHondt:2006iej}. On an event-by-event basis, it builds a $\chi^{2}$ function using the four-vectors of the final state objects and their resolutions. The $\chi^{2}$ is minimised by correcting the measured quantities within their resolutions, fulfilling the kinematic constraints by using Lagrangian multipliers. In this case, the number of degrees of freedom allowed in the fit is ten, as there are four jets (three degrees of freedom for each jet) and the two constraints from each di-jet invariant mass. The outcome of the kinematic fit is a set of corrections for each of the measured quantities, which translate into in an improved four-body invariant mass resolution. The information provided by the minimised $\chi^{2}$ is the measure of the probability for the observed event to be compatible with the proposed kinematic. The correction factors are then applied to each jet to improve the four-body invariant mass reconstruction.
This procedure uses $\eta$, $\phi$ and \pT information for each jet and their related uncertainty.
As the jet angles are measured with a better relative resolution than the jet-\pT, the corrections mainly affect the jet transverse momentum. 

The improvement in resolution for the reconstructed signal resonance ranges from 20 to 40\% depending on the mass hypothesis for the CMS \hhbbbb resonant search~\cite{Sirunyan:2018zkk}, resulting in an improvement of the sensitivity of 10–20\%. Similar improvements are also observed in CMS \hhbbtt searches at 13 TeV \cite{Sirunyan:2017djm} or \hhbbyy at 8 TeV \cite{Khachatryan:2016sey}.
The asymmetry of the corrections, due to the jet momentum resolution across the \pT range considered, results in a linear mass shift as function of the resonant mass.
The relative improvement is large for the lowest mass resonant hypotheses, since by construction once the two Higgs boson masses are constrained to the nominal value of the Higgs boson mass, the resolution
of the four-body invariant mass $\sim 2\mH +\Delta( E_{H1}, p_{H1}, E_{H2}, p_{H2})$ is dominated by the precision of the 2\mH $\sim$ 250 GeV term~\cite{Vernieri:1966046}.

The application of the kinematic fit could potentially be extended to other final states involving \bjets, to further improve the resolution of the \mhh invariant mass on top of the dedicated \bjet specific corrections, as the two methods exploit orthogonal information. Indeed the sensitivity of the CMS search for \hhbbtt is enhanced by the use of the kinematic fit, which exploits the four-momenta of both the $\tau$ and \bjets and  the $\pT^{\mathrm{miss}}$ vector in the
event, and is performed under the hypothesis of two 125 GeV Higgs bosons decaying into a
bottom quark pair and a $\tau$ lepton pair. The use of the kinematic fit improves the resolution on \mhh by about a factor of two compared to the four-body invariant mass of the reconstructed leptons and jets~\cite{Sirunyan:2017djm}. The decay products of the $\tau$ leptons are assumed to be collinear in the fit, since they
are highly boosted as they originate from an object that is heavy when compared to their own mass. In the decay of the two $\tau$ leptons, at least two neutrinos are involved and there is no precise measurement of their original  energies. For this reason, the $\tau$ lepton energies are constrained from the balance of the fitted H boson transverse momentum
and the reconstructed transverse recoil, $\pT^{\mathrm{miss}}$, as detailed in Ref.~\cite{Khachatryan:2015tha}.

\begin{figure}[ht]
\centering
\includegraphics[width=0.35\textwidth]{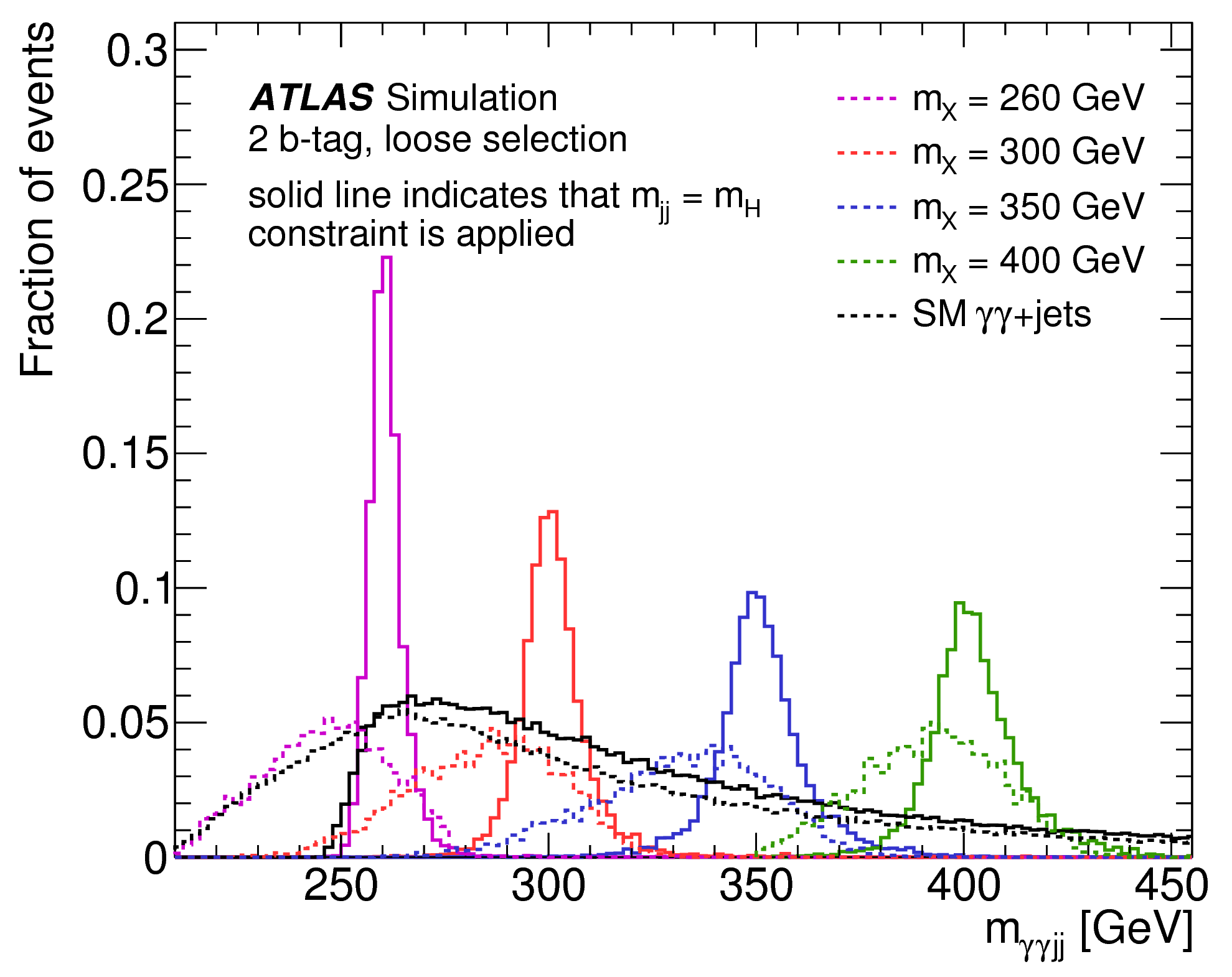}
\includegraphics[width=0.57\textwidth]{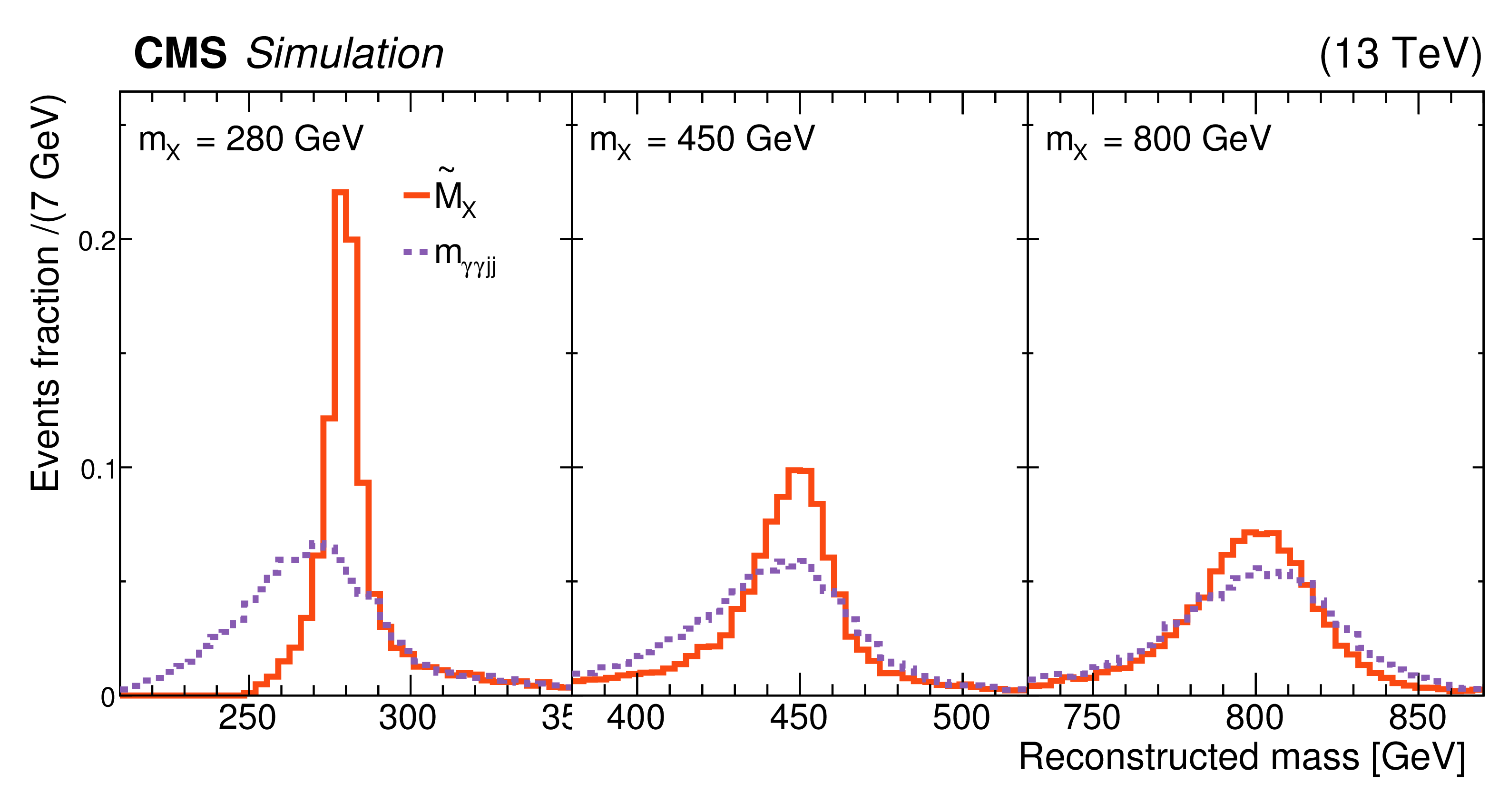}
\caption{Reconstructed \mhh with (solid lines) and without (dashed lines) the dijet mass constraint, for a subset of
the mass points used for the resonant \hhbbyy searches of ATLAS~\cite{Aaboud:2018ftw} (left) and CMS (right)~\cite{Sirunyan:2018iwt}.}
\label{fig:hhbbyy_myybb}
\end{figure}

A simplified version of the kinematic fit is used by the ATLAS \hhbbyy and \hhbbbb searches~\cite{Aaboud:2018ftw, Aaboud:2018knk}, where \mbb is constrained by a simple multiplicative factor $\mh=125/\mbb$ before reconstructing \mhh. This improves the \mhh resolution, on average, by 30--60\% across the resonance mass range of interest as shown in Figure~\ref{fig:hhbbyy_myybb} and sculpts the non-resonant background in the low \mhh range. 

The CMS \hhbbyy~\cite{Sirunyan:2018iwt} search applies two different scaling factors for the \myy and \mbb, and approximates the kinematic fit procedure by defining a modified \mhh estimator. The so called ``reduced'' \mhh mass~\cite{Kumar:2014bca} is shown in the following equation:


\begin{equation}
    \Mtilde = \mjjyy - (\mbb - \mh) - (\myy - \mh).
    \label{eq:reducedmass}
\end{equation}

This estimator subtracts the out-of-cone and resolution effects that impact the \mbb mass more than the jet \pT. While the kinematic fit scales the jet momentum, this method attempts to directly correct the \mbb mass. The \mjjyy is also corrected for the reconstructed \myy value, even if its resolution is much better compared to \mbb . The use of \Mtilde instead \mjjyy improves the \mhh reconstruction by 25 to 30 GeV in absolute, that have the most visible effect at mass resonant hypotheses, as shown in Fig.~\ref{fig:hhbbyy_myybb}. For resonant mass of 300 GeV the resolution reduces from roughly 50 to 20 GeV. CMS also uses an \Mtilde estimator for the boosted \hhbbbb searches~\cite{Sirunyan:2017isc,Sirunyan:2018qca}, reporting an improvement of about 10\% for the dijet mass resolution.

\subsection[$HH$ vertex reconstruction with \hyy decay]{$HH$ vertex reconstruction with \hyy decay\\
\contrib{V. M. M. Cairo, M. Gouzevitch}}
\label{sec_vertex}

For \hyy decays, the Higgs boson mass is computed from the measured photon energies and from their directions relative to the Higgs production vertex.

In general, the hard scatter interaction is identified as the vertex that has the highest total transverse momentum (sumPT) of outgoing charged particles produced in the same $p-p$ collision that generated the Higgs boson. 

For single Higgs boson production through gluon-gluon fusion, there are only two photons coming from the primary vertex at LO. Therefore in absence of additional jets, it is hard to identify the vertex because tracks from the primary $p-p$ collision are due only to the underlying events and are soft. In addition to the primary vertex there are many other vertices due to pileup that could spread out in a region of 10 cm along \textit{z}-axis, therefore it is not possible to identify the right vertex by simply looking at the charged particles. 

The di-photon production vertex is then chosen among all reconstructed primary vertex candidates using multivariate techniques based on track and primary vertex information, as well as the directions of the two photons measured in the calorimeter and inner detector (in the case of photon conversion). In this way, the Higgs boson production vertex is correctly identified with an efficiency of about 80\% \cite{HIGG-2016-21,Sirunyan:2018ouh} for the ggF production mechanism. 

This was optimised in a way that the Higgs boson mass resolution is affected from the wrong vertex identification less than from the photon energy resolution.

The same algorithm used for the identification of the \hyy primary vertex is then used in the case of \hhbbyy searches, but the presence of \hbb allows to exploit the particles produced in the \hbb hadronization which makes it possible to reconstruct and select the correct event of interest with even higher efficiency than that of the \hyy case. In fact, similar performance are achieved by both ATLAS and CMS, which are able to identify the primary vertex correctly in up to 99.9\% of the simulated signal events~\cite{Sirunyan:2018iwt,Aaboud:2018ftw}.

Likewise also in the case of searches for $\hh \to \PGg\PGg\PW\PW^{*}$, the presence of high \pT leptons or jets from the $W$ boson decay could contribute to correctly identify the primary vertex together with the constraints derived from \hyy.

%% file: HH_overview/IntroChapterHHAnalysis.tex

\begin{figure}[htb] 
\begin{center} 
\includegraphics[height=9.0cm]{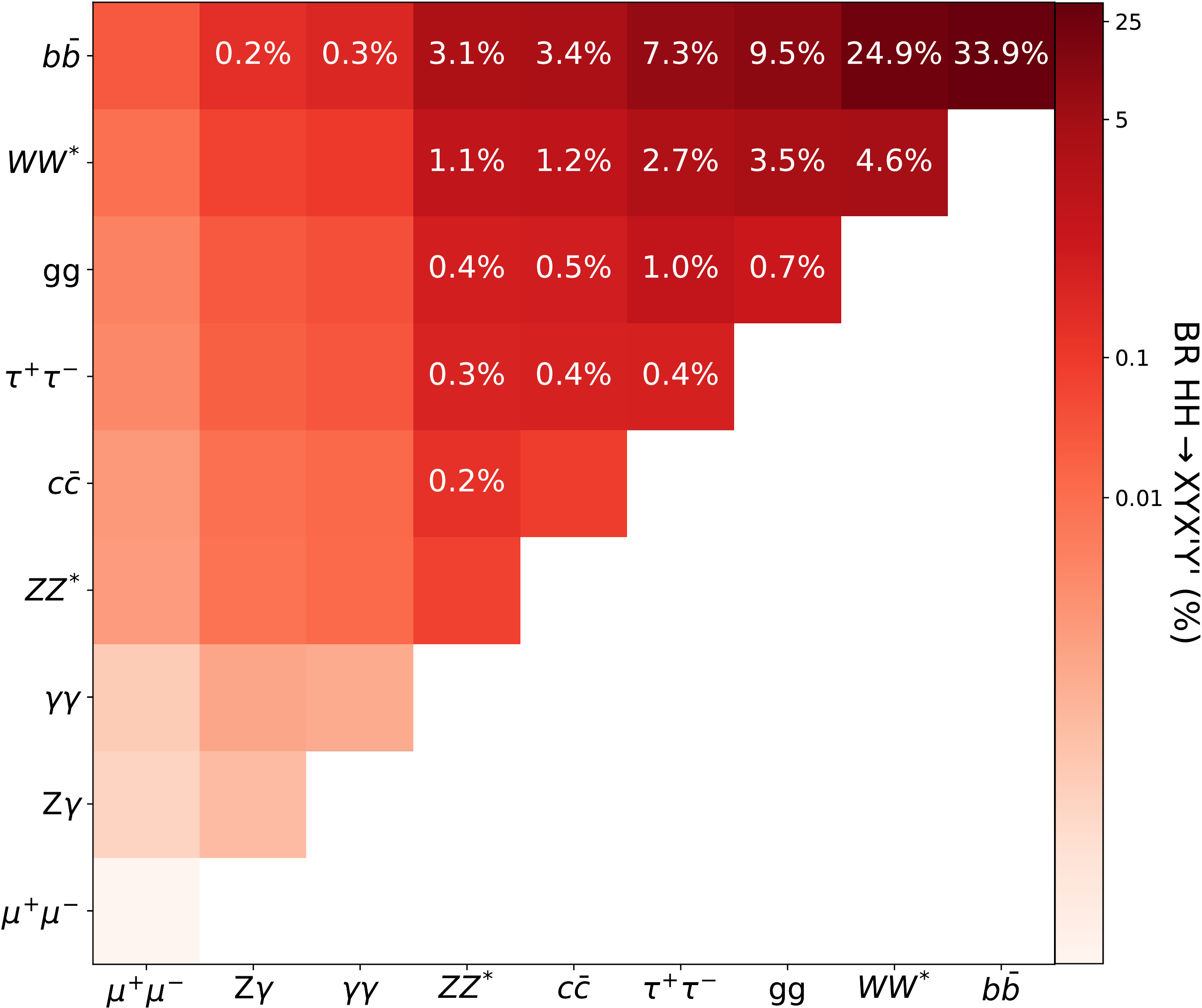}
 \caption{Branching fractions of the decay of an \hh pair to a selected group of final states. The decay modes are shown on each axis by increasing probability. The numerical values are only shown if larger than 0.1\%. 
 The branching fractions of the Higgs boson are evaluated for \mh = 125.0~GeV~\cite{deFlorian:2016spz}.}
 \label{BR-diHiggs} 
 \end{center} 
 \end{figure}

The ATLAS and CMS collaborations have exploited a rich variety of signatures to search for \hh pair production, exploiting the several Higgs boson decay modes shown in Fig.~\ref{BR-diHiggs}.

%


The feasibility of many of them has been considered in several phenomenological studies. The interested reader can consult Refs.~\cite{Dolan:2012rv,Papaefstathiou:2012qe,Baglio:2012np,Goertz:2013kp,Barr:2013tda,Dolan:2013rja,deLima:2014dta,Englert:2014uqa,Dolan:2015zja,Behr:2015oqq,Kling:2016lay,Banerjee:2016nzb,Bishara:2016kjn,Adhikary:2017jtu,Kim:2018cxf,Lu:2015jza,Adhikary:2018ise,Arganda:2018ftn} and references therein.
Sec.~\ref{sec:HH4b}-\ref{sec_exp_2dot7} present an overview of the results of the searches for both non-resonant and resonant \hh production through gluon-gluon fusion from the ATLAS~\cite{Aad:2008zzm} and CMS~\cite{Chatrchyan:2008aa} experiments, based on the data recorded in 2015 and 2016, corresponding to an integrated luminosity of up to about 36~\ifb.

Table~\ref{exp-summary-table1} lists the relevant searches performed by ATLAS and CMS experiments and the corresponding main features. 
\begin{table}[htb]
\begin{center}
{
\begin{tabular}{ccccc}
\hline
\multicolumn{2}{c}{Search channel} & References & Luminosity & Discriminant\\
\hline
\multirow{2}{*}{\bbbb} & 
ATLAS & \cite{Aaboud:2018knk} & 
27.5--36.1 & \mhh \\
                        & 
CMS & \cite{Sirunyan:2018tki} & 
35.9 & BDT \\
\hline
\multirow{2}{*}{\bbyy} & 
ATLAS & \cite{Aaboud:2018ftw} & 
36.1 & \myy/\mhh \\
                                  & 
CMS & \cite{Sirunyan:2018iwt} & 
35.9 & \mbb,\myy (2D) \\
\hline
\multirow{2}{*}{\bbtautau} & 
ATLAS & \cite{Aaboud:2018sfw} & 
36.1 & BDT \\
                              & 
CMS & \cite{Sirunyan:2017djm} & 
35.9 & BDT/$\mathrm{m_{T2}}$ \\
\hline
\multirow{2}{*}{\bbvv} & 
ATLAS & \cite{Aaboud:2018zhh} & 
36.1 & e.c. \\
                              & 
CMS & \cite{Sirunyan:2017guj} & 
35.9 & DNN \\
\hline
\multirow{2}{*}{\wwyy} & 
ATLAS & \cite{Aaboud:2018ewm} & 
36.1 & \myy \\
                              & 
CMS & -- & -- & -- \\
\hline
\multirow{2}{*}{\wwww} & 
ATLAS & \cite{Aaboud:2018ksn} & 
36.1 & e.c. \\
                              & 
CMS & -- & -- & -- \\
\hline
\end{tabular}
}
\end{center}
\vspace*{-0.3cm}
\caption{\label{exp-summary-table1}
Summary of \hh search channels with their
corresponding references, the integrated luminosity of
the dataset used in the analysis and the distribution used to extract the signal (discriminant) - note that e.c. stands for event counting. This table is based on the \hh non-resonant and resonant searches performed with the 2015 and 2016 datasets collected by ATLAS and CMS at 13 TeV.}
\end{table}
The \hhbbbb final state exploits the leading BR for a SM Higgs boson but it suffers from a large multi-jet background. The experimental challenges related to this signature, the current results and potential improvements are discussed in Sec.~\ref{sec:HH4b}. 
Despite the low branching fraction, the \hhbbyy final state has a very good sensitivity to the SM \hh production, thanks to an excellent trigger and reconstruction efficiency of photons, and the excellent invariant mass resolution for the Higgs boson decay to photons, see Sec.~\ref{sec_exp_2dot4}. 
The \hhbbtt final state represents a compromise between the rate and the background contamination. Thanks to the use of multi-variate
analysis techniques, the search performed by the ATLAS collaboration yields to the most stringent limit on \hh production from an individual channel, as discussed in Sec.~\ref{sec_exp_2dot5}. 

The above three final states drive the sensitivity to the SM Higgs boson pair production. However, experiments have also exploited other rare and challenging final states such as $\hhbbvv$, 
where $V = W,\,Z$ (Sec. \ref{sec_exp_2dot6}), $\hh \rightarrow \wwyy$, $\hh \rightarrow \wwww$ and $\hh \rightarrow \tautautautau$ (Sec. \ref{sec_exp_2dot7}).
The current outlook for the non-resonant \hhbbvv channel is challenging and provides ample opportunity for improvement.
Searches for \hh production in final states without \bjets have in general smaller signal yields, but are also typically less affected by backgrounds processes.
As their sensitivity is mainly limited by statistical uncertainties, their sensitivity is expected to scale better with the integrated luminosity, 
as more refined and sophisticated analysis techniques could be employed.






%% file: HH4b/HH4b.tex
Nearly one third of \hh events decay via the \bbbb channel, resulting in the experimental signature of four energetic jets which originate from $b$-quark hadronisation. 
The main challenge for this signature is the large background from multi-jet final states produced by quantum chromodynamics (QCD) processes, which collectively yield rates exceeding that of
the signal by several orders of magnitude.
Other non-resonant processes can contribute to the signal signature, such as the production of top quark pairs, and $W$ or $Z$ bosons in association with \bjets.

\begin{figure}[hbt!]
  \begin{center}
    \subfloat[2015 analysis~\cite{EXOT-2015-11}
      \newline $\approx 15$ events per \ifb
      \newline $\approx 30\%$ statistical uncertainty at the \mhh peak]{
      \includegraphics[height=5.6cm]{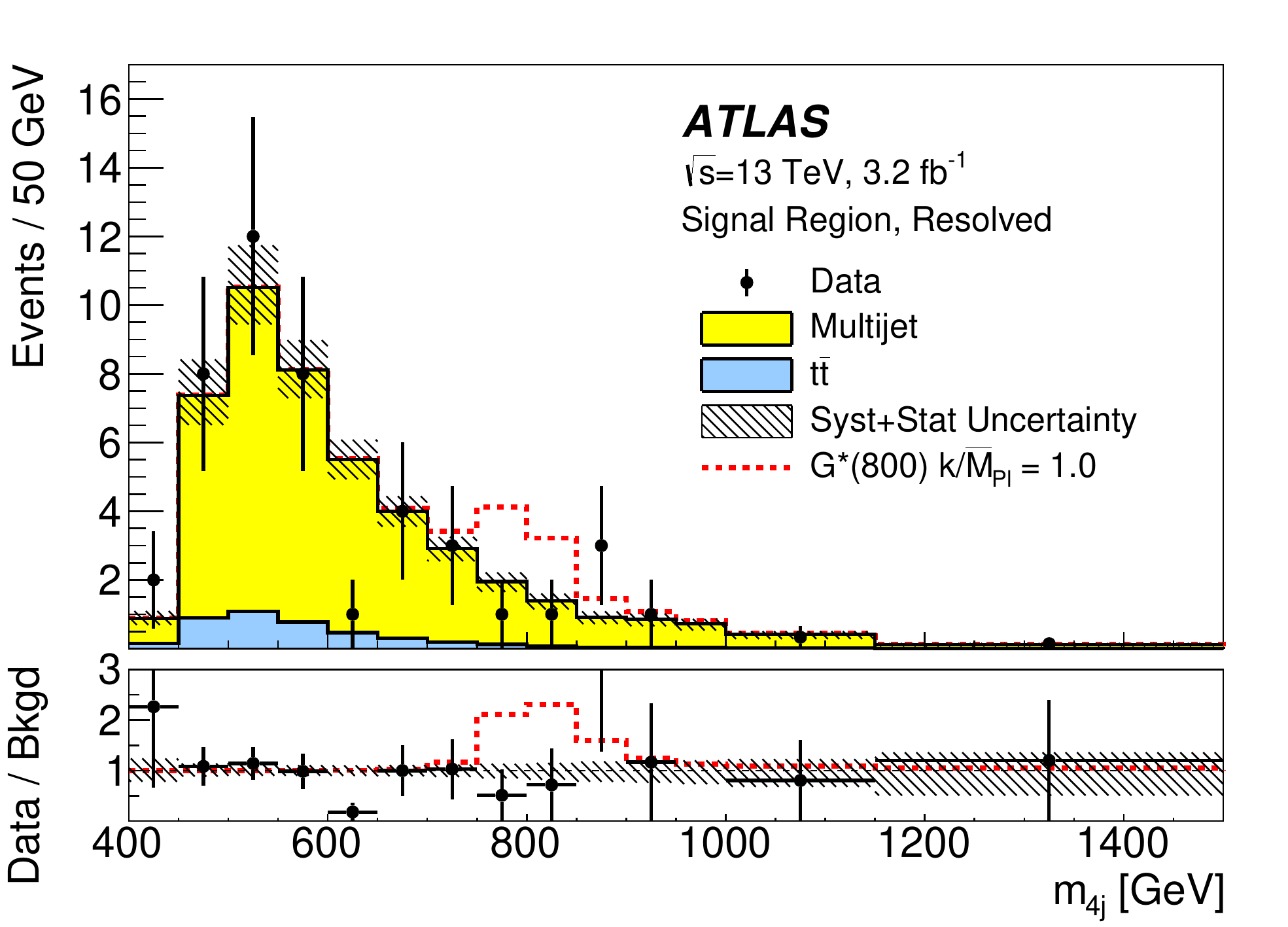}}\hspace{1cm}
    \subfloat[2018 analysis~\cite{Aaboud:2018knk}
      \newline $\approx 300$ events per \ifb
      \newline $\approx  3\%$ statistical uncertainty at the \mhh peak]{
      \includegraphics[height=5.4cm]{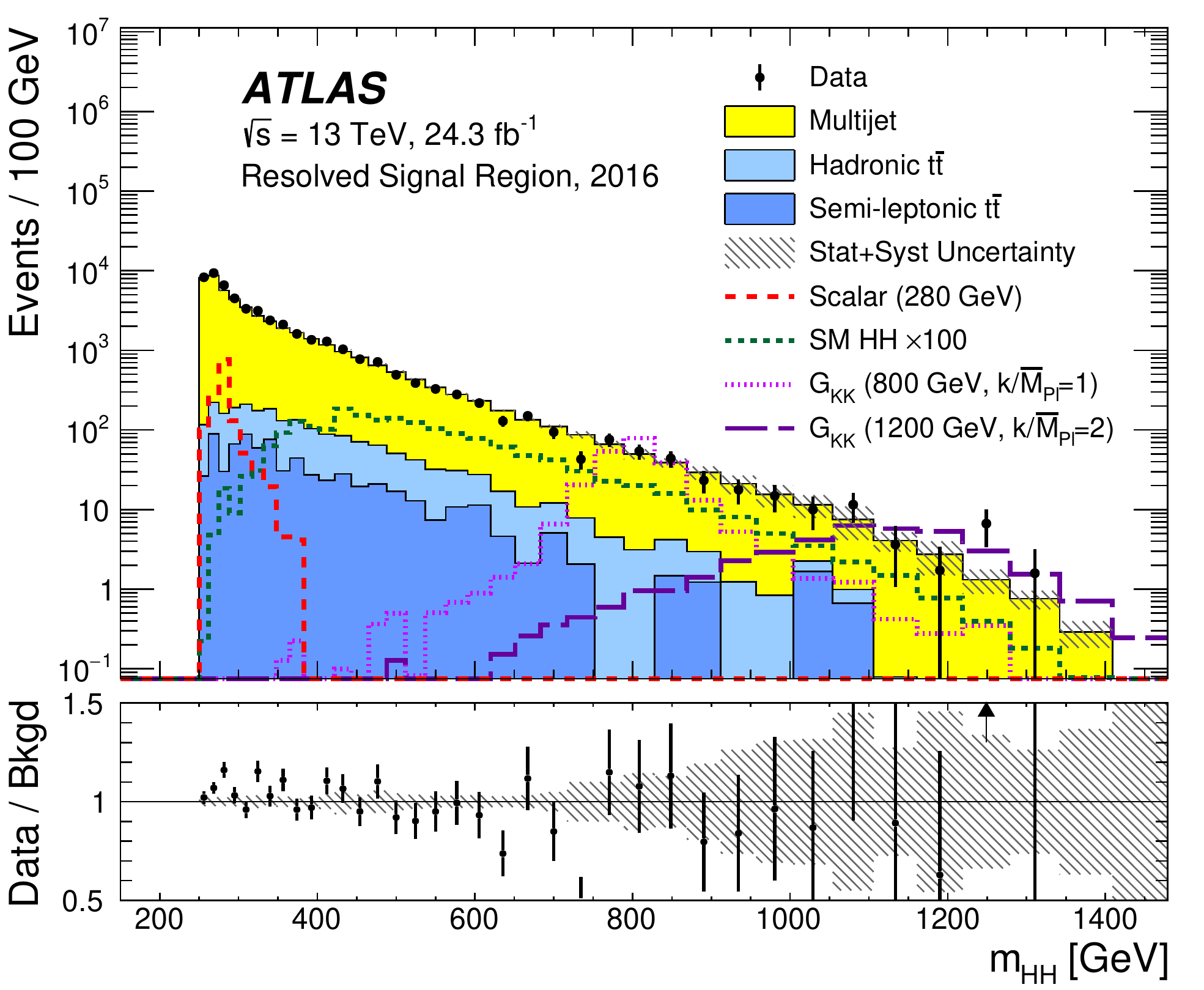}}
    \caption{
    Distributions of \mhh in the signal region of the ATLAS resolved search for 2015~\cite{EXOT-2015-11} (left) and 2016 ~\cite{Aaboud:2018knk} (right) data,
compared to the predicted backgrounds. The hatched bands represent the statistical uncertainties. 
    Note the change in $y$-axis range in the ratio plots.}
  \label{fig:2015vs20184b}
  \end{center}
\end{figure}

As discussed in Sec.~\ref{sec:heft}, most of the impact of modifications of the Higgs boson self-coupling to the \mhh distribution is near the $2\mh$ threshold, where the irreducible multi-\bjet background has a significant contribution. 
Since the start of Run 2, much of the experimental effort has been focusing on extending these searches in the low \mhh range, by employing dedicated trigger strategies, consequently loosening the event selection criteria and modelling the substantially increased background acceptance as illustrated in Fig.~\ref{fig:2015vs20184b}. In the most recent ATLAS search, the loosened kinematic selection requirements have increased the background acceptance by a factor of 20, relative to the restricted phase space probed in the first Run 2 result (Fig.~\ref{fig:2015vs20184b}, left). Combined with the integrated luminosity increase, the statistical uncertainty at the peak of the \mhh distribution has dropped by an order of magnitude to the percent level in the latest Run 2 result (Fig.~\ref{fig:2015vs20184b}, right). By the end of HL-LHC data-taking, we will require a sub-percent level background model -- a daunting task that will require novel data-driven modelling techniques. 

In addition to non-resonant \hh production via gluon-gluon fusion, ATLAS and CMS searches~\cite{Aaboud:2018knk,Sirunyan:2018tki} provide also results
for resonant \hh production in the range $260 < \mhh < 3000$ GeV. The momenta and angles between the decay products of such a resonance vary significantly over this range. In order to increase the sensitivity of this search, different event selection criteria are used for the two main kinematic regions: (i) ``resolved'' with four individually reconstructed \bjets which tests resonance mass hypotheses from $2\times\mh$ up to $1500$ GeV; (ii) ``boosted'' which exploits large-radius jets and substructure techniques (see Sec.~\ref{sec:jetReco} and Sec.\ref{sec:hbbbosted}) to probe resonance mass hypotheses up to 3 TeV. The resolved regime dominates the sensitivity to SM non-resonant \hh production. In addition, the strategy adopted by CMS makes use of a third category, the "semi-resolved". This case, first proposed in Ref.~\cite{Gouzevitch:2013qca}, aims to recover potential events which did not enter the other two categories by considering events where one Higgs candidate merges into a single large-radius jet but the other is reconstructed as two individual \bjets. This analysis moderately improves the sensitivity for \mhh between 750 and 2000 GeV~\cite{Sirunyan:2018qca}. 
For the non-resonant \hhbbbb searches, events are selected online by combining two different trigger selections, both using the \btagging algorithms to identify \bjets.
Events are requested offline to contain four \btagged jets with \pT$>$ 30/40 GeV (CMS/ATLAS). The \btagging efficiency for jets with \pT in the 60–150 GeV range is approximately 70\% (68\%) and gradually decreases for lower and higher jet \pT. This corresponds to a light jet mis-tag efficiency of 0.3\% (1\%) for ATLAS~\cite{ATL-PHYS-PUB-2016-012} (CMS~\cite{Sirunyan:2017ezt}), see Sec.~\ref{sec:bTagging} for more details. After these selection criteria are applied, the dominant background processes are multi-\bjet production (85--90\%) and top-quark pair production (10--15\%). The Z + jets background is estimated to contribute no more than 0.2--0.5\% to the total background, and therefore is neglected.

The main challenge for the signal extraction in the \bbbb final state, is to build a precise model of the multi-jet background without a reliable simulation. The simulation of these final states, due to their large cross section, requires the simulation of a large number of events, which is challenging for the available computing resources. 


In the following the analysis strategies are presented, Sec.~\ref{sec:current4b}, together with their limitations, Sec.~\ref{sec:limitations4b}.
Finally  possible paths forward, where there is clear room for improvement and opportunities for innovation, are discussed, Sec.~\ref{sec:improvements4b}.

\subsection{Analysis strategies}
\label{sec:current4b}

\begin{figure}
  \begin{center}
    \includegraphics[width=0.52\linewidth]{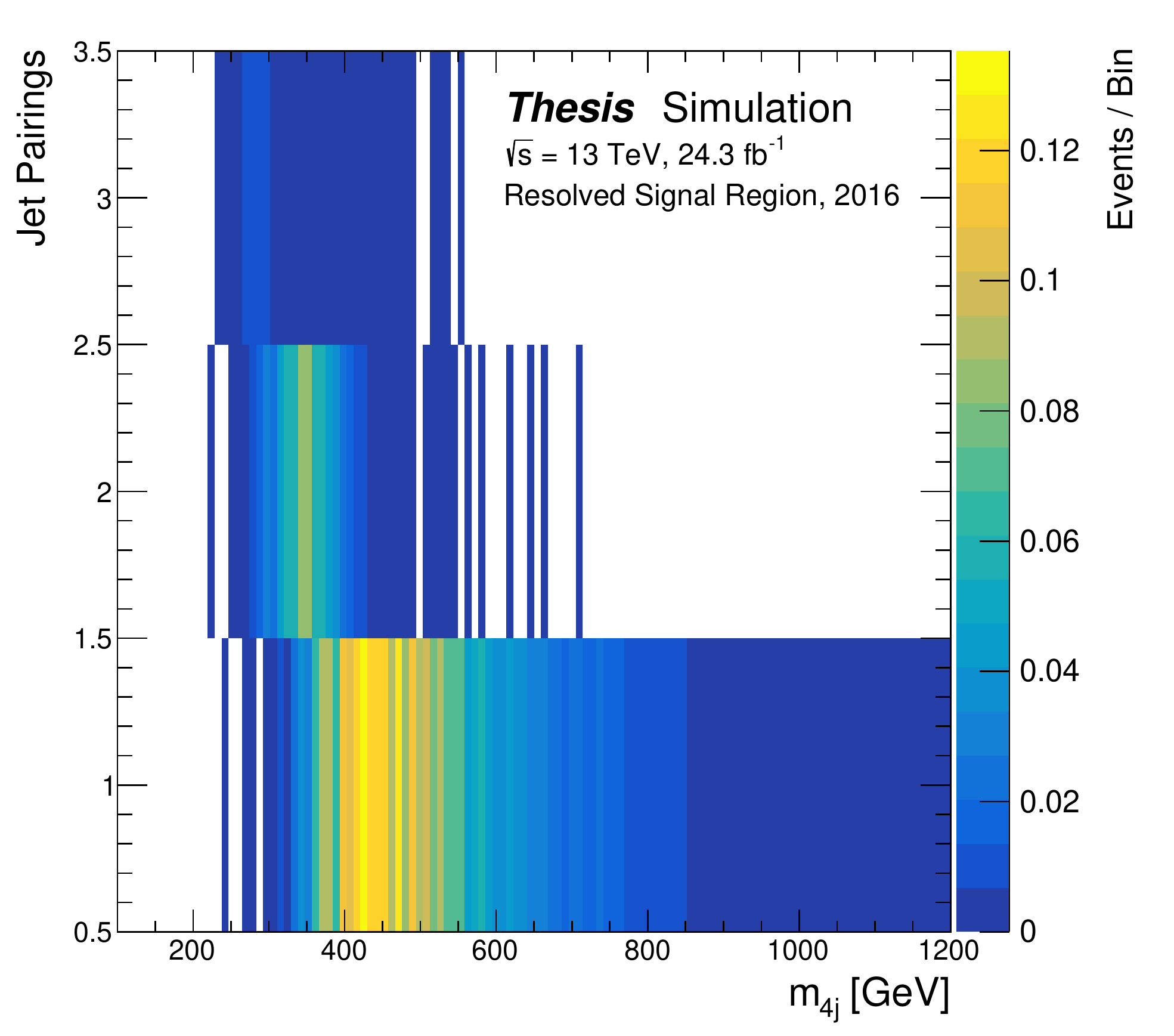}\hfill
    \includegraphics[width=0.44\linewidth]{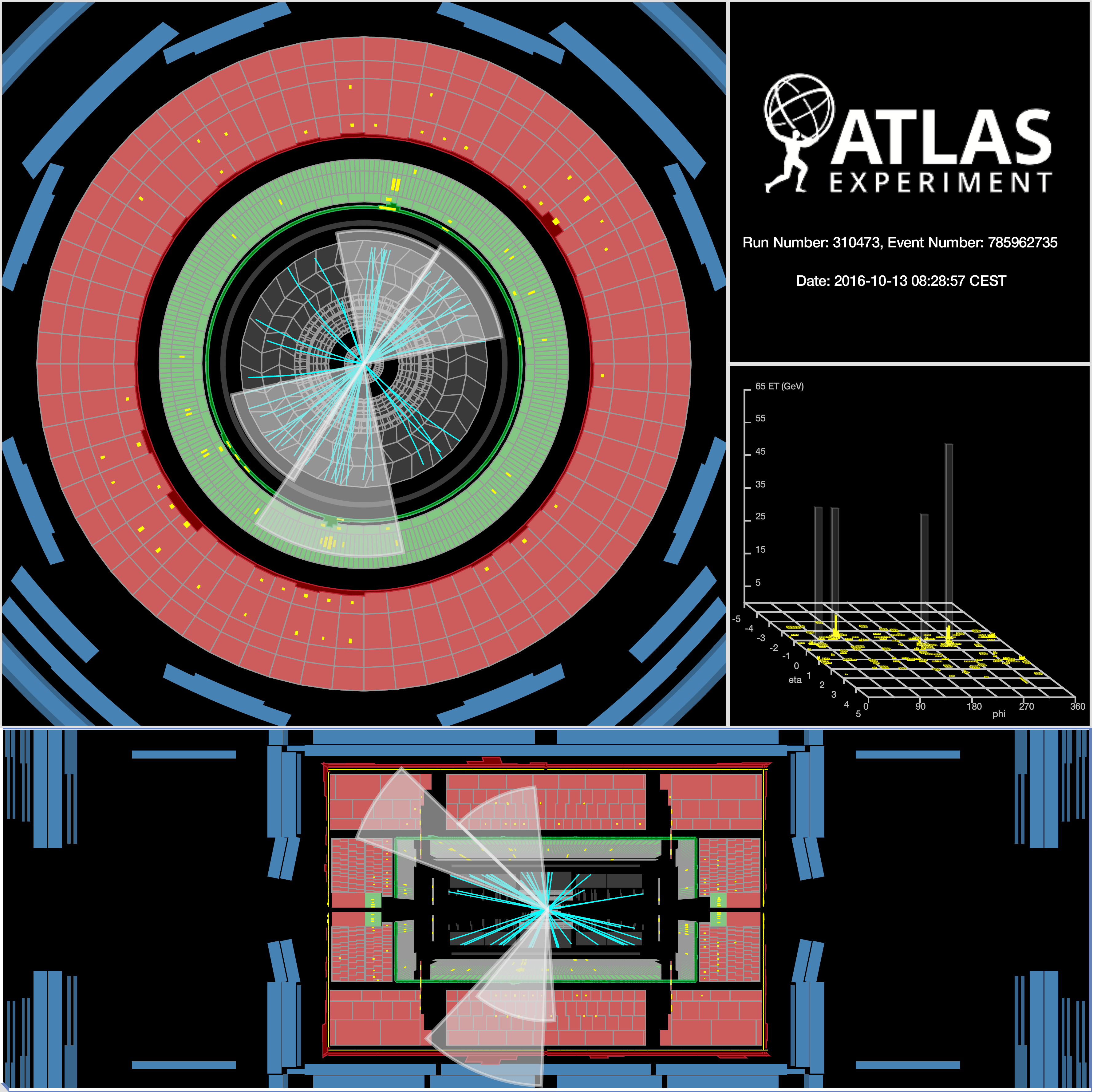}
    \caption{Left: Distribution of the number of jet pairings which pass the $\Delta$R(j,j) selection as a function of the reconstructed \mhh for simulated SM \hh events~\cite{Bryant:2644551}. Right: An event collected by ATLAS during 2016 data taking with 
    \mhh=272 GeV which passes the signal region selection from~\cite{Aaboud:2018knk}. Of the three possible pairings, the two with 
    large di-jet opening angles pass the $\Delta$R(j,j) sliding requirement, while the third pairing with both opening angles approximately equal to twice the jet radius fails. This third pairing is consistent with the topology of the dominant two to two gluon scattering 
    background where the two outgoing gluons split to \bb pairs and results in a low mass large radius jets.}
    \label{fig:eventDisplay4b}
  \end{center}
\end{figure}

Both ATLAS and CMS analysis strategies rely on multi-jet triggers at L1 with online \btagging selections applied at the HLT to a subset of the online jets, as described in Sec.~\ref{sec:bTrigger}.
At L1 the multi-jet trigger selections is required to have a maximum rate of approximately 3 kHz, which demands significant HLT resources for the online \btagging to reduce trigger rates to roughly 40 Hz. 
At HLT, CMS requires four jets with $\pT > $ 30 GeV and two above 90 GeV, and three online \btags corresponding to an offline \btagging efficiency of less than 60\%. ATLAS trigger selection requires four jets with $\pT>35\,$GeV, where at least two are \btagged online with the 60\% working point. 
The CMS trigger efficiency, as evaluated for a resonant signal benchmark as function of \mhh, ranges from 10\% at the $2\mh$ threshold to 60\% above 800 GeV and it is 34\% for the non-resonant hypothesis. 
The ATLAS trigger efficiency is evaluated instead, with respect to the offline requirements and ranges from 65\% at the $2\mh$ threshold to $\gtrapprox99\%$ above 600 GeV. 


The four jets with the highest \btagging score are paired to reconstruct the two Higgs boson candidates. Given these four jets, there are three possible di-jet pair constructions. Both ATLAS and CMS chose the pairing which minimizes the difference between the di-jet masses. CMS performs this minimization over all three pairings for the non-resonant signal and exploits the smaller angular separation of the two \bjets for resonance mass values above 500 GeV. ATLAS reduces the number of considered pairings by applying a sliding selection on the di-jet opening angle as a function of the reconstructed four body mass. The impact of the sliding selection requirements on the signal is shown in Fig.~\ref{fig:eventDisplay4b}. The ATLAS (CMS) approach selects the correct pairing at least 90\% (70\%) of the time for the non-resonant \hh signal hypothesis and across the full range of resonance mass hypotheses (70--95\%). 
A multi-variate classifier able to use all of the di-jet correlation information for all possible pairings would perform better than the $\Delta$R(j,j) sliding requirement, by classifying such events as more background-like than those where none of the pairings are such clear examples of the dominant background.


A requirement on the masses of the Higgs boson candidates is used to define the signal region for the ATLAS search which takes into account of the \mbb resolution:
\begin{equation}
 \sqrt{ \left(\frac{\mH{}_{1}-120~\mbox{GeV}}{10\%\mH{}_{1}~\mbox{GeV}}\right)^2 +\left(\frac{\mH{}_{2}-110~\mbox{GeV}}{(10\%\mH{}_{2}~\mbox{GeV}}\right)^2} < 1.6
\end{equation}
Similarly CMS, for the resonant \hhbbbb search, defines a circular signal region in the two-dimensional space defined by the reconstructed masses of the two Higgs boson candidates, after the regression based corrections, described in Sec.~\ref{sec:bjetreg}, are applied to each \bjet.
The \mhh resolution is further improved by correcting the momenta of the reconstructed $b$-quarks imposing the kinematic constraint of the invariant mass of the Higgs boson candidates to be 125 GeV, as described in Sec.~\ref{sec_exp_kinfit}. The improvement in resolution for the reconstructed signal resonance ranges from 20 to 40\% depending on the resonant mass hypothesis, which results in an improvement of the sensitivity by 10--20\%. 

\subsubsection{Background modelling}
The ATLAS analysis strategy derives the model for high \bjet multiplicity events from the low \bjet multiplicity events, with at least two \bjets.
This procedure relies on the assumption that the ratio of multi-jet production matrix elements with different \bjet multiplicities does not change sharply in the phase space with di-jets near the Higgs boson mass.
This ratio takes into account of the kinematic dependence of the \btagging efficiency and fake rate as well as the different relative contributions of the underlying matrix elements. 
The ratio is then used as a weighting factor to correct low \bjet multiplicity events to match high \bjet multiplicity events and should apply equally well across a broad range of phase space with different di-jet masses, in particular they should apply for events with two di-jet pairs near the Higgs boson mass. 
This assumption is validated in a control region in data, orthogonal to the signal regions used to extract the signal. 
Shape uncertainties in the multi-jet background are estimated to affect the \mhh distribution by 5--30\%, by deriving an alternative background model using the same procedure as in the nominal case, but using data from the control region.

\begin{figure}[hbt!]
  \begin{center}
    \includegraphics[width=\linewidth]{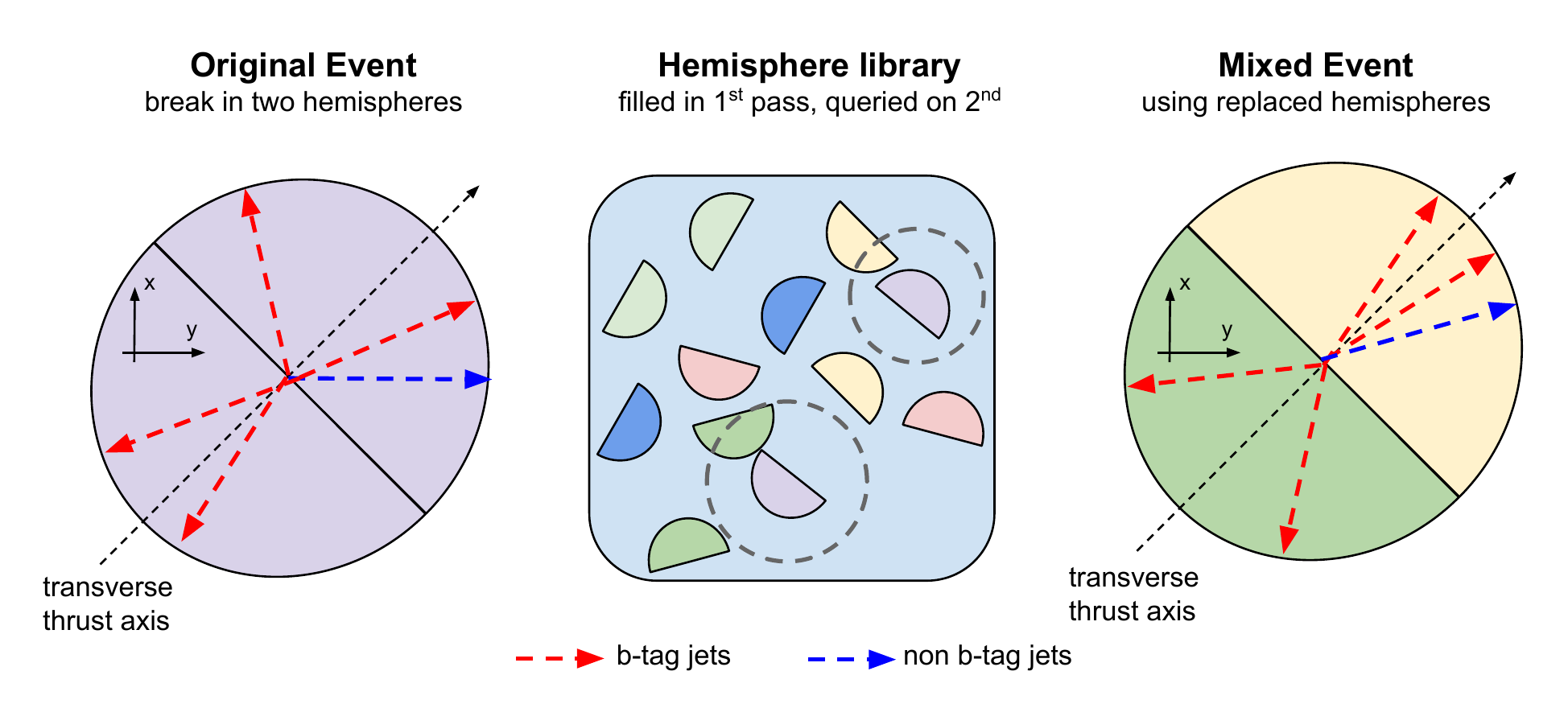}
    \caption{Illustration of the hemisphere mixing procedure from \cite{Sirunyan:2018tki}. }
    \label{fig:hemisphereMixing}
  \end{center}
\end{figure}

For ATLAS, the systematic uncertainty associated to the background model estimate limits the current result and it will become more important with the full Run 2 integrated luminosity.



The ATLAS background estimation method profits of a dedicated trigger selection with only two \bjet, not available in CMS due to different optimisation choice of the \bjet trigger, that favoured the use of lower \pT threshold and a \bjet multiplicity of at least three (see Sec. \ref{sec:bTrigger} for more details).
The CMS collaboration, instead, has developed different background estimation strategies for the resonant and non-resonant \hh searches.
For the non-resonant signal extraction, the so called ``hemisphere mixing'' technique is used, where fake events are generated by mixing and matching di-jet systems from separate events~\cite{Sirunyan:2018tki} as illustrated in Fig.~\ref{fig:hemisphereMixing}. 
This background estimate method does not require the presence of signal depleted control region in data, but it aims at creating an artificial background data set using the whole
original data set as input. Thus, rather than a model of a single distribution, a full model of the original data is produced. 

The \emph{transverse thrust axis} is defined as the axis that maximises the sum of the absolute values of the projections of jets transverse momenta along the axis itself. The event space is divided into hemispheres by cutting along the axis perpendicular to the transverse thrust axis. Artificial events are then built, by picking hemispheres from different events that are similar to the two hemispheres that made up the original event.
The matching algorithm is designed to create fake events with the same kinematic structure as the background process while washing out the correlated structure of the signal process. Because of this, the resulting artificial data sets are unaffected by the presence of a small signal contamination in the original data. This has been verified with signal injection tests.
A BDT classifier, using the \textsc{xgboost} library~\cite{Chen:2016:XST:2939672.2939785}, is employed to separate signal (including other BSM non-resonant hypotheses) from background processes. 
The resulting artificial samples are used to provide a background model in the training of a BDT classifier (training sample), an independent set for its validation and optimisation (validation sample), and a third set used to extract the predicted shape of the optimised BDT (application sample). 
The BDT exploits the \btagging scores, kinematic information of both the \hh system and Higgs candidates, as well as the angles between the \hh system and the leading Higgs boson, for a total of 25 inputs. The BDT distribution for data and the artificial model are compared in control regions and a systematic bias is detected. Thus, the background template is corrected for the bias evaluated from this comparison. 

A search for SM \hhbbbb signal is then performed for an excess in the tail of the BDT output distribution. 

Minor background contamination arising from $\ttbar H$, $ZH$, $\bb H$ do not show a signal-like BDT distribution and their effect is found to be negligible in the selected data at the current level of the search sensitivity.
The systematic uncertainty associated to the shape and normalisation of the background model affects the final result by about 9\% and 30\% respectively.

For the resonant signal extraction, a simultaneous fit to the \mhh spectrum in the signal region is used. The background model is validated in data in dedicated control regions with reduced \btag multiplicity~\cite{Sirunyan:2018zkk}. 
Since the \ttbar contribution
to the background exhibits a shape very similar to that for the multi-jet process, it is implicitly included in the data driven estimate. The systematic uncertainty associated with the choice of the parametric background model is evaluated with pseudo-datasets, generated from an alternative function and fitted with the nominal function to evaluate the bias in the reconstructed signal strength. The measured bias impacts the expected limit by 0.3–1.5~\%.

Data-driven methods to estimate the backgrounds (dominantly multi-jet) are also used in the boosted and semi-resolved regimes. The ATLAS result are obtained with the same approach exploited for the resolved analysis. The CMS results instead, rely on the smooth dependence, in background jets, of the specialised double$-b$ tagging efficiency on the jet mass, in\-tro\-du\-ced in Sec.~\ref{sec:hbbbosted}. This rate can be derived in sidebands of the Higgs boson mass and interpreted as a ratio of events passing to events failing the requirement, so that it can be applied to events with the correct mass, but failing the double-b tagging requirement. The dominant uncertainty in these searches is the uncertainty associated to the substructure requirements for large-radius jet algorithms, which can be as large as 20\%. 

\subsection{Limitations of current analysis strategies}
\label{sec:limitations4b}
\begin{figure}[hbt!]
  \begin{center}
    \includegraphics[width=0.45\linewidth]{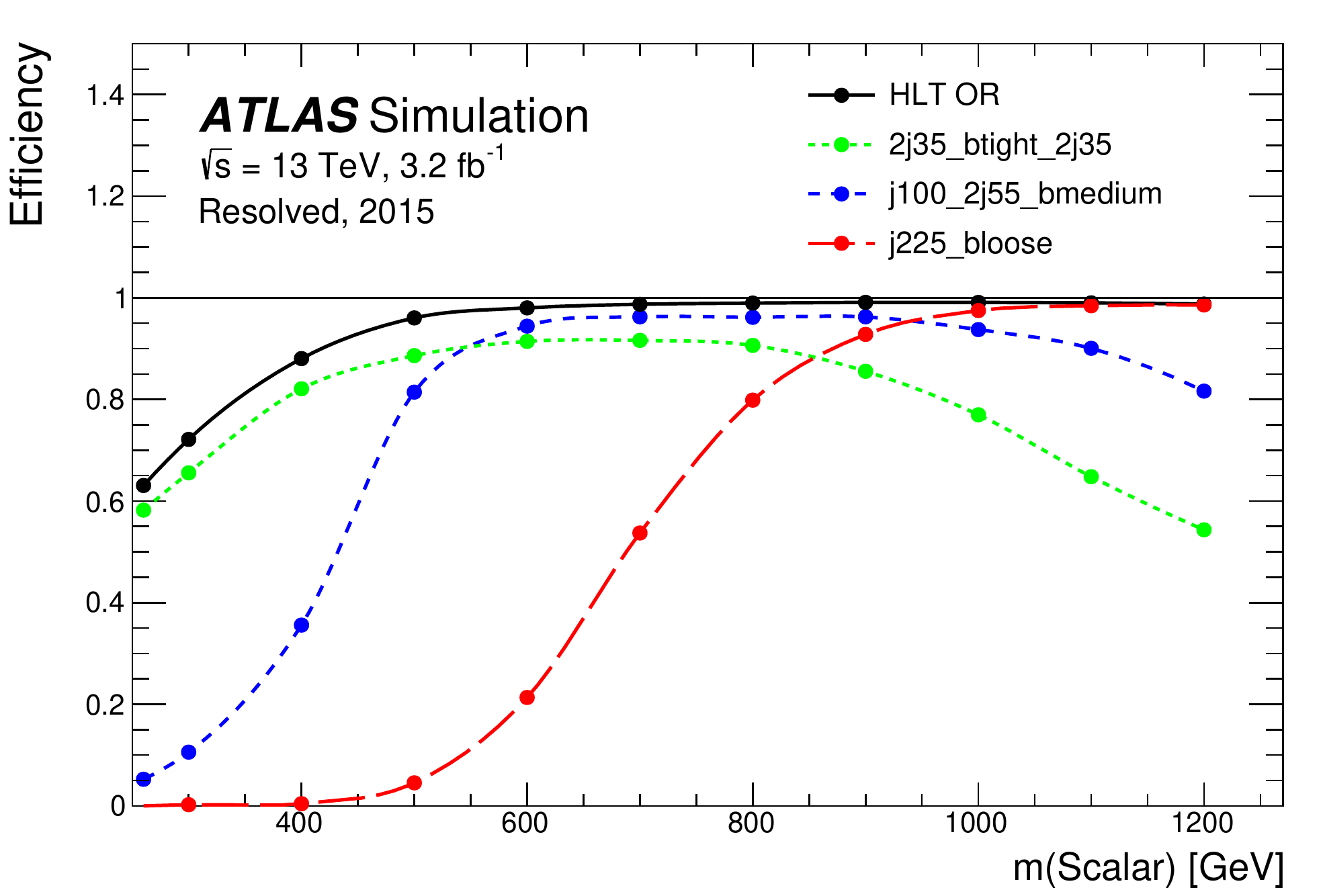}
    \includegraphics[width=0.45\linewidth]{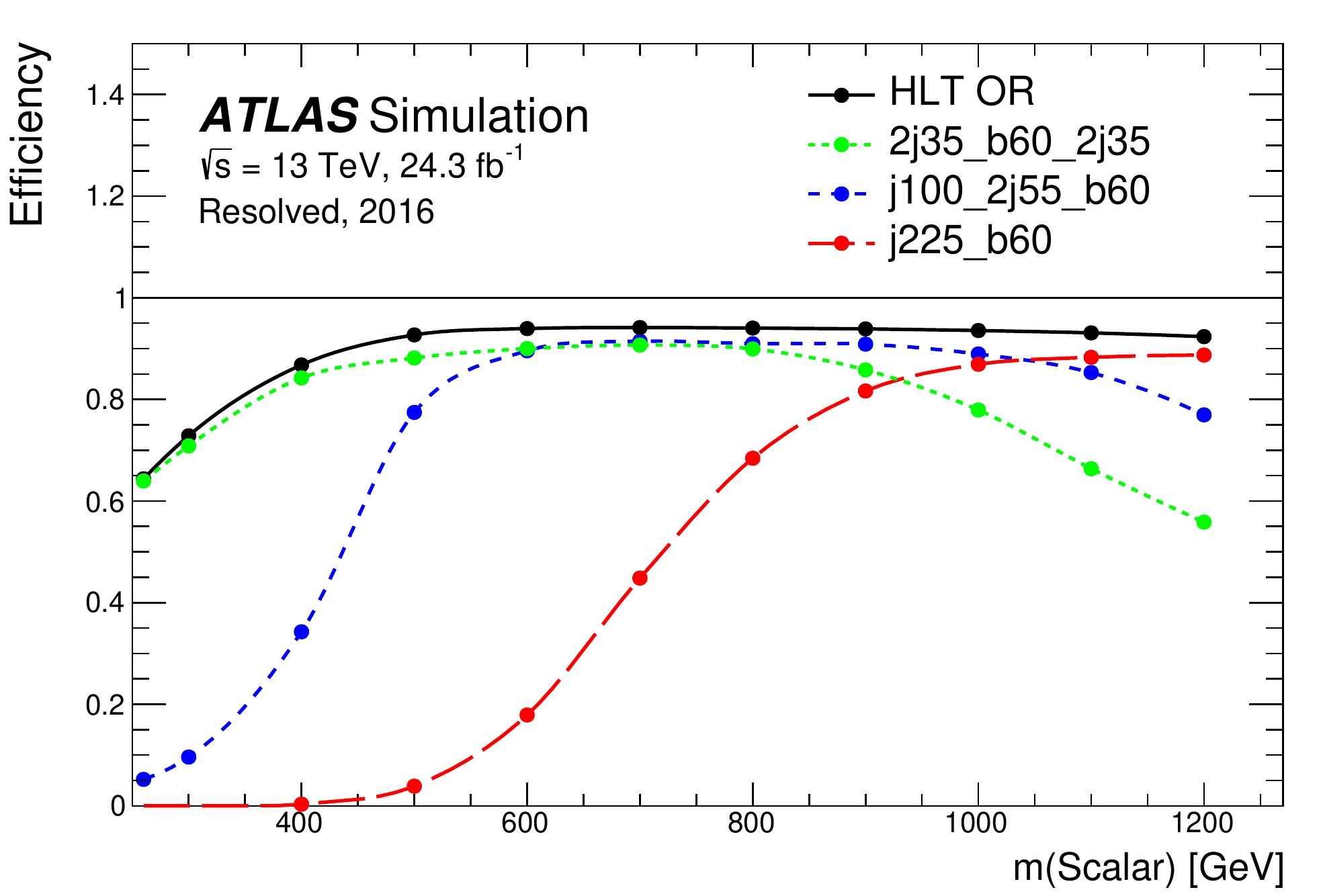}
    \caption{The ATLAS event-level trigger efficiencies for the various signal \hhbbbb hypotheses~\cite{Aaboud:2018knk}. }
    \label{figL1HLT}
  \end{center}
\end{figure}

Both the ATLAS and CMS experiments face significant challenges related to the hardware and software triggers.
The current trigger efficiencies are limited at L1 for events with $\mhh\lessapprox 500$ GeV as illustrated in Fig.~\ref{figL1HLT}.

At the HLT trigger level, maintaining good tracking performance to efficiently reconstruct secondary vertices without using excessive CPU resources is extremely challenging, as discussed in Sec.~\ref{sec:bTrigger}. The required CPU time to perform online \btagging grows non-linearly with pileup.
In fact, in the year 2017 and 2018 the trigger thresholds were increased and tracking algorithms optimised to cope with high instantaneous luminosities, but new techniques will be required to accommodate for the luminosity targets of Run 3.
While ATLAS focused on providing a unified analysis strategy for resonant and non-resonant $\hhbbbb$ searches,
CMS has developed independent strategies and optimised the signal extraction for low-, intermediate- and high-mass resonances.

Both the ATLAS and CMS approaches suffer from statistically limited control regions in data.
The assumptions that go into generating a background model from data do not necessarily hold to a higher degree of precision than can be tested outside of the signal region.
Such uncertainties are difficult to be quantitatively assessed, particularly when they have non-trivial effects on distributions beyond their normalisation.

The recent ATLAS result~\cite{Aaboud:2018knk} attempted to address this by deriving the background model twice, using orthogonal kinematic selections, and using the resulting variation of the background prediction in the signal region to derive systematic uncertainties.
In principle this method accounts for biases in the model due to the extrapolation into the signal region by making one model derivation region kinematically ``closer'' to the signal region.
It also naturally provides a full spectrum (and in principle, high dimensional) uncertainty in the final discriminant distribution with the proper bin-to-bin correlations.
Ideally one would chop the phase space into many orthogonal regions, each progressively closer to the signal region, such that trends in the extrapolation of the models across phase space could be extracted.
Unfortunately, these attempts quickly become limited by the need to validate each model at the statistical precision anticipated in the signal region.
If this requirement is not kept, large systematic uncertainties are required to cover the lack of precision in the model validation. These issues are compounded when trying to model higher dimensional target spaces to improve the sensitivity and model independence of searches.

One of the primary limitations of the current ATLAS background model is the algorithm used to derive the correction factors from low to high \bjet multiplicity.
The method iteratively weights multiple one dimensional distributions, which are selected to encapsulate the primary differences in the scattering processes with as few variables as possible.
This avoids the statistical limitations of high dimensional histograms but may not correctly account for (anti)correlations between the reweighted distributions.
With the integrated luminosity of 27.5\,\ifb used in~\cite{Aaboud:2018knk} one could argue hints of such effects are becoming visible and a new strategy will almost certainly be required for analyses of the full Run 2 data set. 


\begin{figure}[hbt!]
  \begin{center}
    \includegraphics[width=0.5\linewidth]{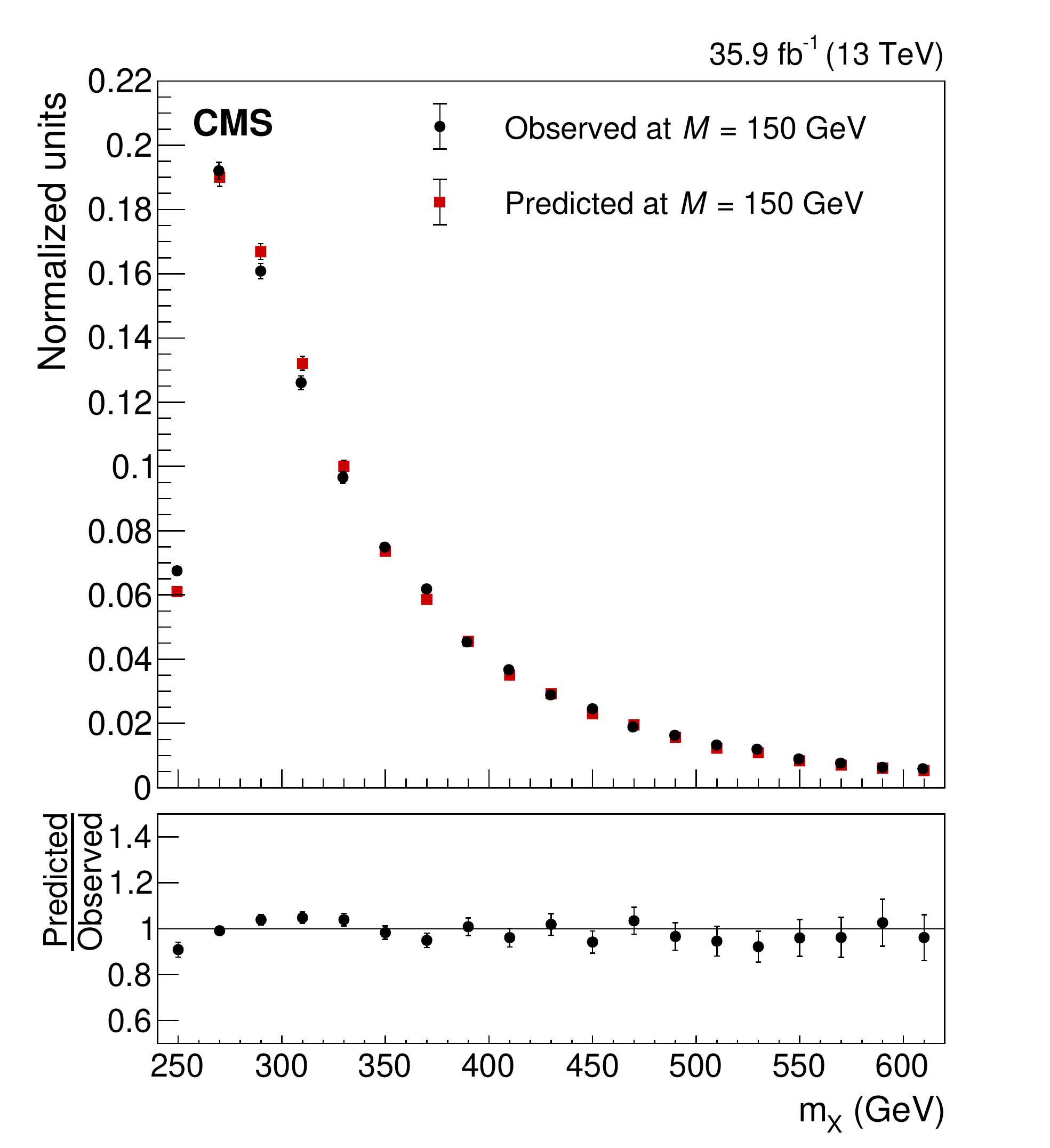}
    \includegraphics[width=0.4\linewidth]{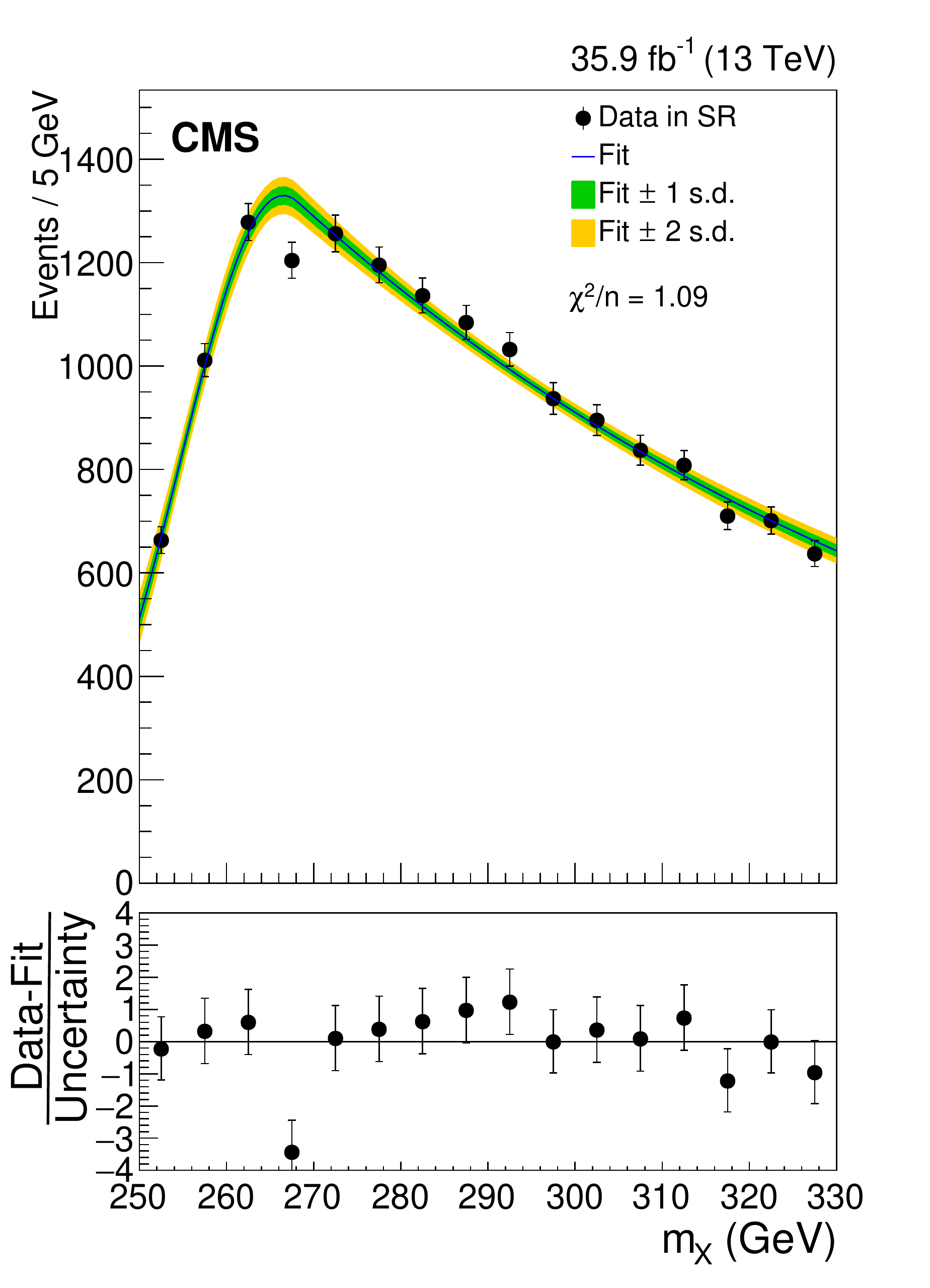}
    \caption{These plots illustrate the validation and result of the CMS background modelling of the \hh mass spectrum~\cite{Sirunyan:2018zkk}.
    Left: observed and predicted \hh mass spectrum in a validation region centred around $\mbb= 150$~GeV.
    Right: A fit to the background-only hypothesis of the \mhh distribution in the signal region in data.}
  \label{fig:fitLMR}
  \end{center}
\end{figure}

In the low mass phase space near the kinematic threshold $\mhh\gtrapprox 250\,$GeV, the CMS and ATLAS searches suffer from the reliability of any potential excess on top of a sharply peaking background, as shown in Fig.~\ref{fig:fitLMR}.

The CMS background model prediction is validated by comparing the prediction for the signal region and the actual signal region in a kinematic sideband defined by moving the Higgs boson mass window from 120 to 150 GeV. ATLAS used a similar background validation method, looking at signal-region-like select \mbb Higgs shifted both below and above the actual Higgs boson mass. 
The background shape has a strong dependence on the di-jet mass selection, as it is shown in Fig.~3 of~\cite{Khachatryan:2015yea} for CMS, but it is properly modelled.

The ATLAS background strategy and the functional fits used by CMS can easily accommodate sub-dominant background sources like \ttbar, $H/Z$+jets and diboson processes using simulated samples.
In the ATLAS approach the simulated backgrounds processes are used in a two step process. First, they are run through the data driven background modelling procedure so that they can then be subtracted from the background model procedure as applied to data. This gives a multi-jet background estimate where the other processes have been removed. Next, the simulated backgrounds are added back into the background model to give the total background. 

The hemisphere mixing method has been successfully used in the search based on the 2016 dataset. With the increasing statistics, the need for an accurate modelling of the \ttbar and electroweak processes will become more relevant. It is not clear that the hemisphere mixing approach used for the CMS non-resonant result can appropriately model the event level correlations of these processes.
Indeed the hemisphere mixing technique relies on its ability to remove the event level correlations of the \hh signal process to avoid signal contamination in the background model. 
This same dilution of event level correlation could subtly impact the \ttbar and electroweak backgrounds such that their contamination in the high signal purity bins of the BDT output is underestimated. 

Furthermore, to avoid tricky statistical issues, the hemisphere mixing, Sec.~\ref{sec:current4b}, can only use each source event once, limiting the statistical precision of the background model to that of the true background. 
If the statistical uncertainty of the published ATLAS background model is set to $\sqrt{N}$~in each bin of the final discriminant, the sensitivity to SM \hh production is reduced by 33\%. This is unsurprising because the ATLAS result~\cite{Aaboud:2018knk} is statistically limited: the sensitivity of a measurement where the background and data have the same statistical uncertainty scales with $1/\sqrt{\mathcal{L}}$.

\subsection{Potential improvements}
\label{sec:improvements4b}

All final states with \bjets are likely to gain from dedicated \bjet energy regressions and calibrations.
In the \bbbb case improvements in the \bjet energy scale reduce the mass resolution for both Higgs bosons allowing for tighter signal region definitions with the same signal efficiency.
CMS has demonstrated this in their most recent resonant search \cite{Sirunyan:2018zkk} where the Higgs boson mass resolution for different resonance mass hypotheses improved by 6-12\%.
The tighter optimal signal region definitions then improved the search sensitivity by 5-20\%.

One promising approach to construct a multi-jet model from lower \bjet multiplicity data is to reweight using a single multivariate classifier output distribution rather than several one dimensional kinematic distributions or sparse high dimensional histograms.
The classifier would be trained to separate low \bjet multiplicity data from high \bjet multiplicity data without any flavour tagging information.
In principle this should appropriately account for (anti)correlations between the classifier input variables and provide a better high dimensional model of the four \bjet data.
ATLAS has already released a search for Higgsino pair production~\cite{Aaboud:2018htj} using a BDT based reweighting scheme using the same event selection as Ref.~\cite{Aaboud:2018knk}.

Multivariate reweighting provides a possible solution to the curse of dimensionality in extrapolating a multi-jet model across \bjet multiplicity but does not address the assumption that the reweighting can be extrapolated across the kinematic phase space.
This assumption could instead be independently verified in a tri-jet sample, if triggers exist to collect such a sample.
The extrapolation across phase space could also be tested in a synthetic sample like that generated by the CMS hemisphere mixing procedure.
Furthermore, at low values of \mhh the signal contamination in events with exactly three \bjets would be negligible and could be used to validate the background procedure with substantially higher statistics than the four \bjet sample.
One would have to use caution with data containing three \bjets, with \mhh$\gtrapprox 500\,$GeV data, as it could offer significant sensitivity to new physics and should be explored as an additional signal selection.

The most obvious approach to improve any search is to perform combined fits with more regions, more dimensions or on especially trained multivariate classifiers. All of these approaches require well understood high dimensional background models.
The following variables, in addition to the Higgs boson candidate masses and \mhh, should be investigated:
\begin{itemize}
\item Angular correlations like $\Delta\eta(H_1,H_2)$, the pseudorapidity separation between Higgs boson candidates, provide discrimination between scalar and tensor resonances and low-mass Higgsino pair production.
\item The ($b$-)jet multiplicity to target VBF \hh production (VBF jet $\eta$ difference and di-jet mass are also relevant in this case).
\item Correlations in other di-jet constructions (in contrast to the Higgs boson candidate construction) may provide a handle in separating the signal from the dominant two-to-two gluon scattering background for $\mhh\lessapprox 400\,$GeV. This combinatoric background can be seen in Fig.~\ref{fig:eventDisplay4b} where, depending on the chosen jet pairing, the displayed event can look like a \hh event with back-to-back \bjets from each Higgs boson or a di-gluon event where each gluon splits to a low-mass collimated \bb pair. 
\end{itemize}

With the full Run 2 dataset of about 300~\ifb, obtained by combining the results from both experiments, it may already be feasible to perform dedicated measurements of SM $ZZ$ and $ZH$ production in the \bbbb final state.
\begin{equation*}
  \begin{split}
    \frac{\sigma(pp\rightarrow ZZ\rightarrow\bbbb)  }{\sigma(pp\rightarrow \hh\rightarrow\bbbb)  } &\approx \frac{15\,\rm{pb}\times0.15^2}{33\,\rm{fb}\times0.58^2} \approx 31 \\[10pt]
    \frac{\sigma(pp\rightarrow ZH\rightarrow\bbbb)  }{\sigma(pp\rightarrow \hh\rightarrow\bbbb)  } &\approx \frac{880\,\rm{fb}\times0.15}{33\,\rm{fb}\times0.58} \approx 7
  \end{split}
\end{equation*}
Measurements of these processes would serve to validate the background model and reduce the impact of the systematic uncertainties.
The same measurements in the \bbtautau final state would benefit from even larger ratios. 
The techniques used to generalise the \hh search to these measurements will also be useful in developing generalised \bbbb searches for additional exotic particles in processes like $Y \rightarrow XH \rightarrow\bbbb$.

%% file: yybb/yybb.tex
\subsection{Overview}

The \hhbbyy final state has the lowest branching fraction among the most sensitive channels, just 0.3\%, but it provides a high signal-to-background ratio by reducing multi-jet events with the identification of two high quality photons. The analysis strategies developed for \hhbbyy, closely follow those for the SM \hyy analyses. \\
Two isolated photons with $\pT> 25$ GeV provide an excellent handle for the triggers. This provides a clear advantage in this final state for $\mhh<400$ GeV, compared to the ones with higher branching fractions, but with trigger strategies requiring higher momenta particles, such as \hhbbbb.\\
Furthermore, the presence of the \hyy~photons provides a clear strategy for event selection and signal extraction. 
As a consequence, the ATLAS and CMS analyses of Run 2 dataset~\cite{Aaboud:2018ftw, Sirunyan:2018iwt} ($\approx$36\ifb) have many similarities. \\
ATLAS excludes SM \hh production at 95\% confidence level with cross sections higher than 0.73 pb (expected 0.93 pb) while CMS 0.79 pb (expected 0.63 pb).  Limits are also set on the modifier of the Higgs self-coupling, with ATLAS constraining $-8.2 < \klambda < 13.2$ and CMS $-11 < \klambda < 17$. 
However, small changes in strategy can lead to significant improvements to sensitivity to SM and BSM \hh production. 
Therefore, it is important to understand the details of each analysis strategy.


\subsection{Signal modelling}

\label{sec:hhbbyy_signal_modeling}

The \hhbbyy~final state benefits from having a fully reconstructable final state.
In contrast with other final states, such as \hhbbbb and other fully hadronic channels, there are no combinatoric issues in the identification of the Higgs boson candidates. 
Therefore, one expects to see clear peaks consistent with the Higgs boson mass in both the di-jet and di-photon invariant mass spectra. 
Due to the good energy resolution and low reconstruction uncertainties for photons at the LHC experiments, the di-photon mass resolution is small relative to the di-jet mass, with $\sigma_{\gamma\gamma}/\textrm{M}_{\gamma\gamma} = 1.3\% (1.5)$ for the most sensitive signal region in the CMS (ATLAS) search. 
CMS quotes $\textrm{M}_{\textrm{jj}}/\sigma_{\textrm{jj}} = 15\%$ for that same category after applying the \bjet energy regression, derived using \hhbbbb~signal events as described in Sec.~\ref{sec:bjetreg}. 

In order to avoid issues with the statistical precision of the simulated samples, both the ATLAS and CMS searches model the peaks from Higgs boson decays with the double-sided Crystal-Ball (DSCB) function for \myy~(ATLAS and CMS) and \mjj~(CMS). 
The DSCB function is chosen for its Gaussian core and power-law asymmetric tails:

\begin{equation} \label{eq:bbyy_dscb}
f(x;\mu, \sigma, \alpha_{L}, p_{L}, \alpha_{R}, p_{R}) = N \cdot 
\begin{cases} 
A_{L} \cdot \left( B_{L} - \frac{x - \mu}{\sigma} \right)^{-p_{L}} & \mbox{for } \frac{x - \mu}{\sigma} > - \alpha_{L}, \\
A_{R} \cdot \left( B_{R} + \frac{x - \mu}{\sigma} \right)^{-p_{R}} & \mbox{for } \frac{x - \mu}{\sigma} > \alpha_{R}, \\
e^{  \frac{\left( x - \mu \right)^{2}}{\sigma^{2}} } & {}^{\mbox{for } \frac{x - \mu}{\sigma} < - \alpha_{L}}_{\mbox{and } \frac{x - \mu}{\sigma} > \alpha_{R}}
 \end{cases},
 \end{equation}
 where $A_{L}, A_{R}, B_{L}, B_{R}$ are normalization constants defined by:
 \begin{eqnarray}
 A_{k} &=& \left( \frac{p_{k}}{\left| \alpha_{k} \right|} \right)^{p_{k}} \cdot e^{-\frac{\alpha^{2}}{2}}, \\
 B_{k} &=& \frac{p_{k}}{\left| \alpha_{k} \right|} - \left| \alpha_{k} \right|.
 \end{eqnarray}
 
 One of the main benefits of using the DSCB function to model the Higgs boson is the ability to describe the effects of systematic uncertainties in its shape with extra parameters in Eq.(~\ref{eq:bbyy_dscb}). 
 Therefore, scale and resolution effects can be mapped into variations of the mean and width of the DSCB Gaussian core, respectively, keeping the tail parameters fixed.  
 This description simplifies the final steps of the searches, which involve unbinned parametric fits to describe the continuous background and extract the signal.
 
 The DSCB parameters are determined from a fit to the simulated \hhbbyy~signal. 
 In ATLAS, the fit is performed in the \myy~distribution, while in CMS, the fit is performed simultaneously in the \myy and \mjj distributions\footnote{It has been checked using  simulations that the correlations between the two distributions are negligible.}, with $f(\myy,\mjj) = g(\myy)\times h(\mjj)$ where $g(x)$ and $h(x)$ are DSCB functions. 
 The modelling of the \myy~and \mjj~distributions is shown in Fig.~\ref{fig:hhbbyy_signal_modeling}.
 
 \begin{figure}[ht]
\centering
\includegraphics[width=0.45\textwidth]{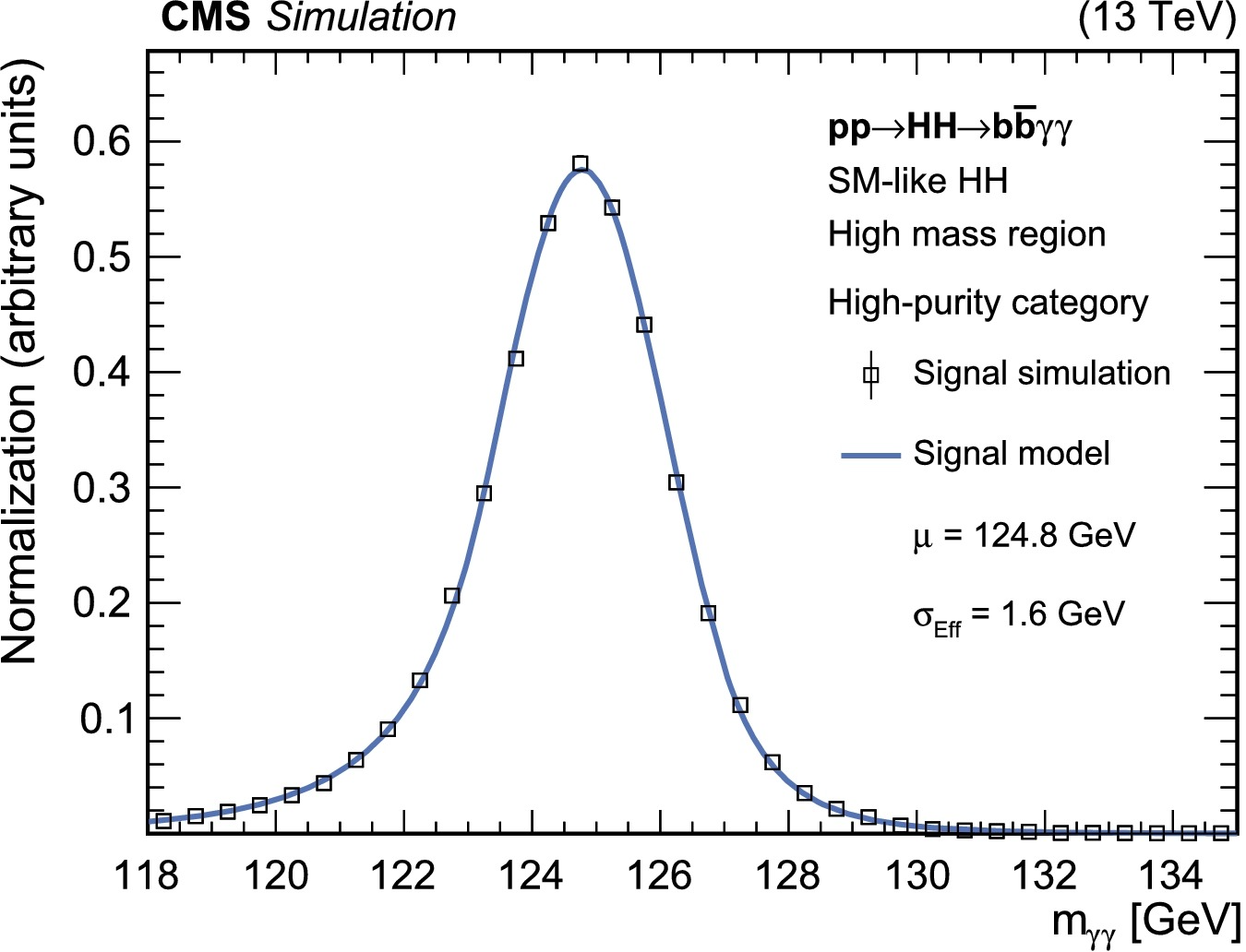}
\includegraphics[width=0.45\textwidth]{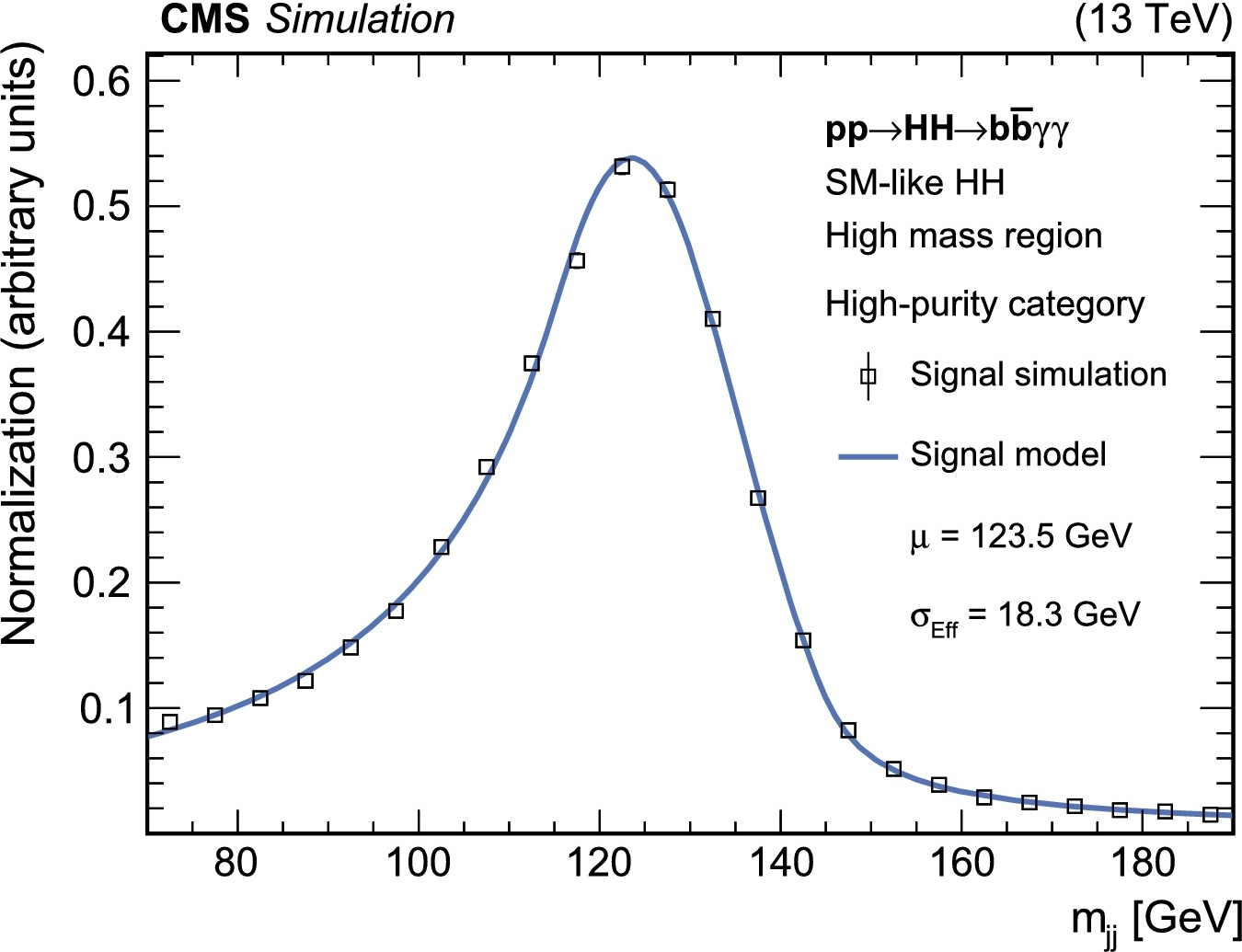}
\caption{\myy~(left) and \mjj~(right) signal modelling in the CMS \hhbbyy~analysis. The blue lines represent the double-sided Crystal Ball parametric fit to the SM \hh signal simulation (squares)~\cite{Sirunyan:2018iwt}.}
\label{fig:hhbbyy_signal_modeling}
\end{figure}
 
 
 The ATLAS and CMS searches use slightly different strategies to simulate the \hhbbyy signal. The ATLAS non-resonant signal is modelled at approximate NLO, using \MGAMCNLO \cite{Alwall:2014hca} with CT10 NLO pdf set \cite{Lai:2010vv}, 
 reweighted in \mhh to take into account the full top quark mass dependence, and parton showering uses \textsc{Herwig++}~\cite{Bellm:2017bvx} using parameter values from the UEEE-5-CTEQ6L1 tune \cite{Seymour:2013qka}. CMS models the non-resonant \hhbbyy signal at LO in \MGAMCNLO\ using the PDF4LHC15\_NLO\_MC pdf set \cite{Carrazza:2015hva, Dulat:2015mca, Harland-Lang:2014zoa, Ball:2014uwa}, and parton showering uses \pythia~\cite{Sjostrand:2014zea} with the CUET8PM1 \cite{Khachatryan:2015pea} tune. A comparison of non-resonant \hhbbyy production at LO and NLO is shown in Fig.~\ref{fig:hhbbyy_LO_NLO}. The transverse momenta of the jets is harder in the LO simulation, and therefore the signal acceptance is higher at LO.
 
\begin{figure}[ht]
\centering
\includegraphics[width=0.45\textwidth]{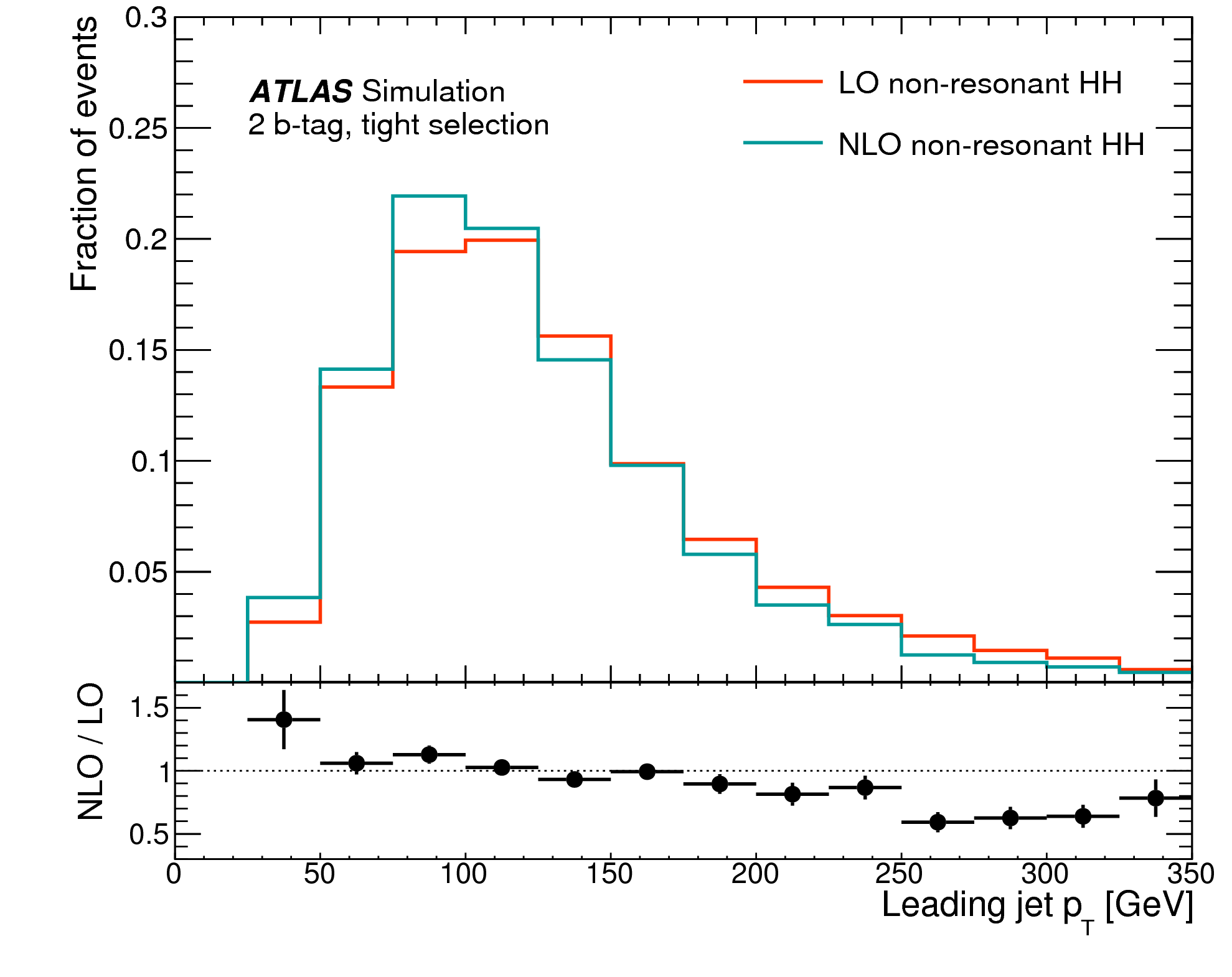}
\includegraphics[width=0.45\textwidth]{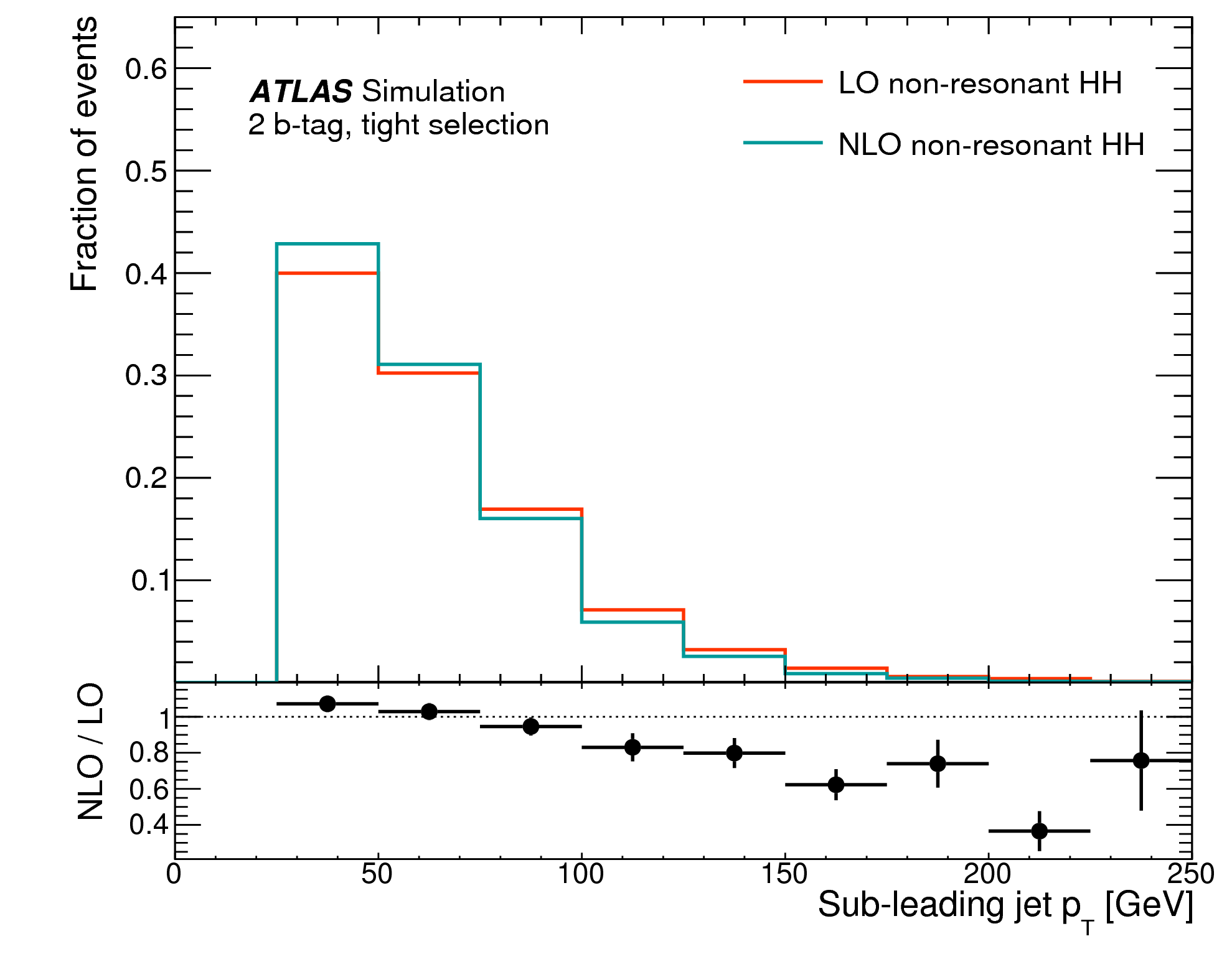}
\caption{Comparison of ATLAS \hhbbyy signal at LO (red) and NLO (green) for the transverse momenta of the leading (left) and sub-leading jets (right)~\cite{Aaboud:2018ftw}.}
\label{fig:hhbbyy_LO_NLO}
\end{figure}

\subsection{Event selection and reconstruction}


The online selection strategy for the \hhbbyy~analyses follows closely the approaches from the \hyy~analyses, utilising the \hyy~targeted di-photon triggers. 
These triggers offer lower online thresholds on the photon \pT than the jet corresponding triggers thanks to the good quality of the trigger-level reconstructed photons. 

In ATLAS, the di-photon trigger requires two clusters of energy deposits in the electromagnetic calorimeter with transverse energy above 35~GeV and 25~GeV for the leading and sub-leading cluster, respectively. 
These clusters are required to have shapes that are consistent with photon-initiated electromagnetic showers and that are isolated from other electromagnetic activity~\cite{ATLAS-CONF-2018-028}. 
The CMS di-photon trigger requires the leading (sub-leading) transverse isolated energy deposit to be above 30 (18)~GeV, and that the invariant mass of the di-cluster system be above 90~GeV~\cite{Sirunyan:2018kta}. 
Both ATLAS and CMS triggers are nearly fully efficient for the \hyy~and \hhbbyy~photons that pass the kinematic requirements. 
The offline object selection of the \hhbbyy~analyses is seeded by finding good quality photons and jets, which must be consistent with the hadronisation of the $b$-quarks. 
The overall strategies of the ATLAS and CMS analyses are similar, and begin by selecting photons close to their trigger thresholds, with extra criteria inspired by \hyy~analyses - such as $\ETy/\myy$ requirements, which preserve the shape of the \myy~distribution. 

One important distinction between the ATLAS and CMS \hhbbyy analyses is their use of the $b$-tagging information (see Sec.~\ref{sec:bTagging}) to classify event categories.
%
The ATLAS \hhbbyy search categorises events according to the number of \bjets: 
\begin{enumerate}[label=(\roman*)]
\item two $b$-tags, defined by selecting events with exactly two jets which pass the 70$\%$ efficient $b$-tagging working point;
\item one $b$-tag, exactly one jet passes the 60$\%$ efficient $b$-tagging working point. 
\end{enumerate}
Events with more than two $b$-tagged jets are vetoed in order to be orthogonal with the \hhbbbb~analysis, and those with no $b$-tagged jets are not considered as signal events. 
The \hbb~candidate is then reconstructed with the two $b$-tagged jets, in the two $b$-tag region, and with the $b$-tagged plus an extra jet, in the one $b$-tag region. This extra jet is selected with a BDT trained with kinematic information of each possible \hbb~candidate reconstructed with the $b$-tagged jet and the non-$b$-tagged jets in the event. 
More details of this approach will be discussed in Sec.~\ref{sec:hhbbyy_ML}.
After the \hbb~candidate is defined, loose and tight jet selections are defined, depending on the \pT of the jets: loose if the leading jet $\pT > 40$~GeV; tight if the leading jet has $\pT > 100$~GeV and the sub-leading $\pT > 30$~GeV.


CMS exploits the full distribution of the probability that the jets are $b$-tagged (the $b$-tagging score). 
First, the \hbb~candidate is reconstructed using the jets with the highest $b$-tagging score and their scores are then used as inputs for the multivariate event categorisation, described in Sec.~\ref{sec:hhbbyy_ML}. The angular correlations between the four objects used to reconstruct the \hyy~and \hbb~candidates, the helicity angles, are also exploited by the event categorisation algorithm to classify signal versus $\gamma\gamma$+jets background events.

Helicity angles have been historically used in analyses such as the SM $H\rightarrow ZZ^{*}\rightarrow l^{+}l^{-}l^{+}l^{-}$ searches and subsequent measurements, as they have been shown to distinguish between different spin and parity hypotheses for the Higgs boson~\cite{Bolognesi:2012mm}. 
Some of these angles are also sensitive to the tensor structure of a resonance production mechanism~\cite{Chizhov:2011wt}.
Similar to the four-lepton final state, the \hhbbyy~analysis also profits from having four final state objects that can be used to measure such angles. 

Three helicity angles have been found to bring the most sensitivity in the CMS \hhbbyy search, as shown in Fig.~\ref{fig:hhbbyy_cms_helicity}. They are defined in the Collins-Soper (CS) references frame~\cite{PhysRevD.16.2219}.
The CS frame boosts to the rest frame of the Higgs bosons and defines fixed axes such that measured variables are sensitive to spin
and CP properties of the Higgs boson. It minimises the dependence of the angles on the transverse momentum of the \hh system, as follows:
\begin{itemize}[noitemsep]
    \item $|\cos(\theta^{\textrm{CS}}_{\hh})|$: $\theta_{\hh}$ is the angle between the momentum of the \hyy~candidate and the line that bisects the acute angle between the colliding protons.
    \item $|\cos(\theta^{\textrm{CS}}_{\gamma\gamma})|$, $|\cos(\theta^{\textrm{CS}}_{\textrm{jj}})|$: $\theta^{\textrm{CS}}_{\gamma\gamma}$ and $\theta^{\textrm{CS}}_{\textrm{jj}}$  are the angles between the Higgs bosons and their decay products in the CS reference frame. The two photons or jets used to define the angle are chosen randomly.
    \end{itemize}


\begin{figure}[ht]
\centering
\includegraphics[width=0.3\textwidth]{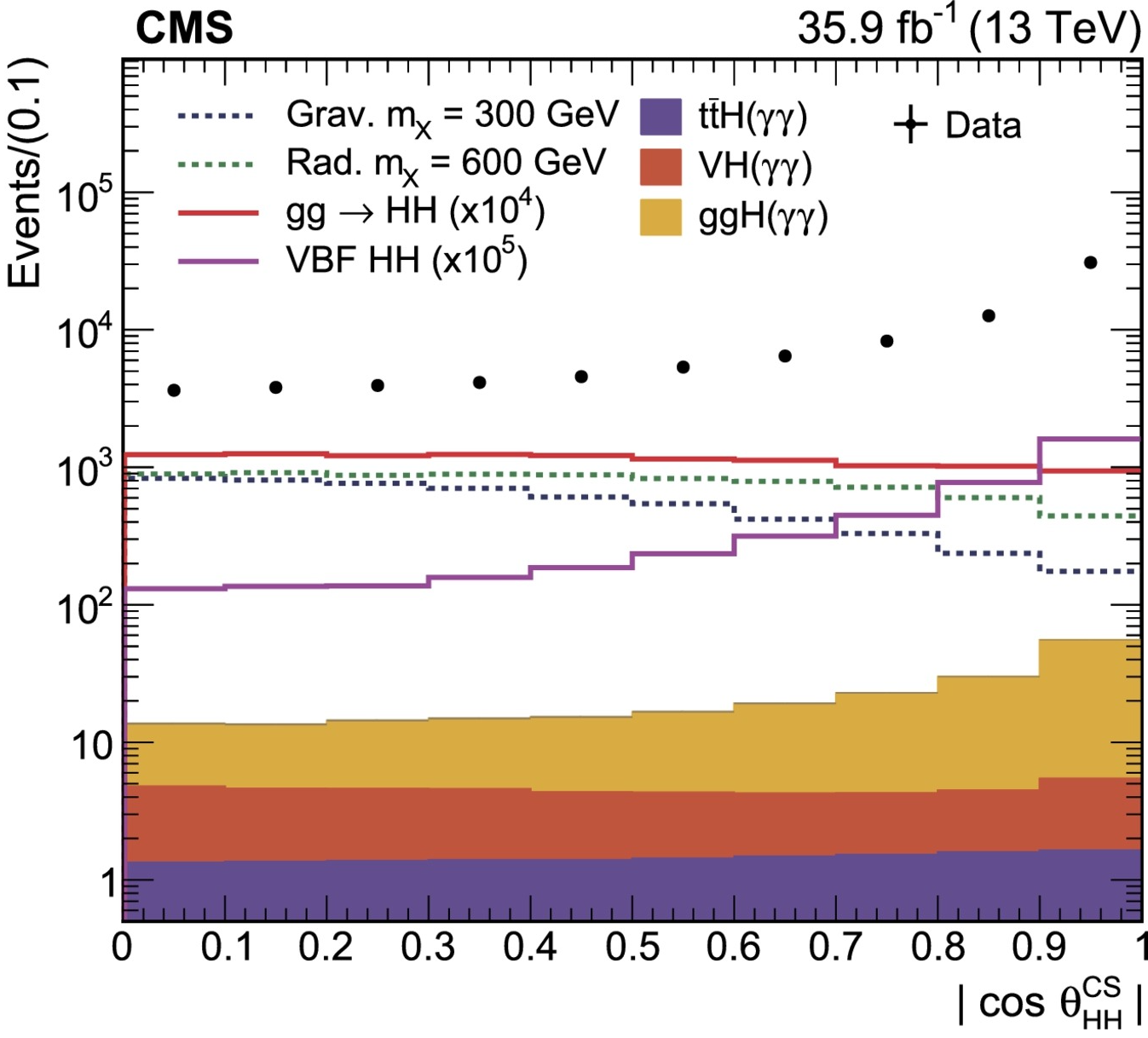}
\includegraphics[width=0.3\textwidth]{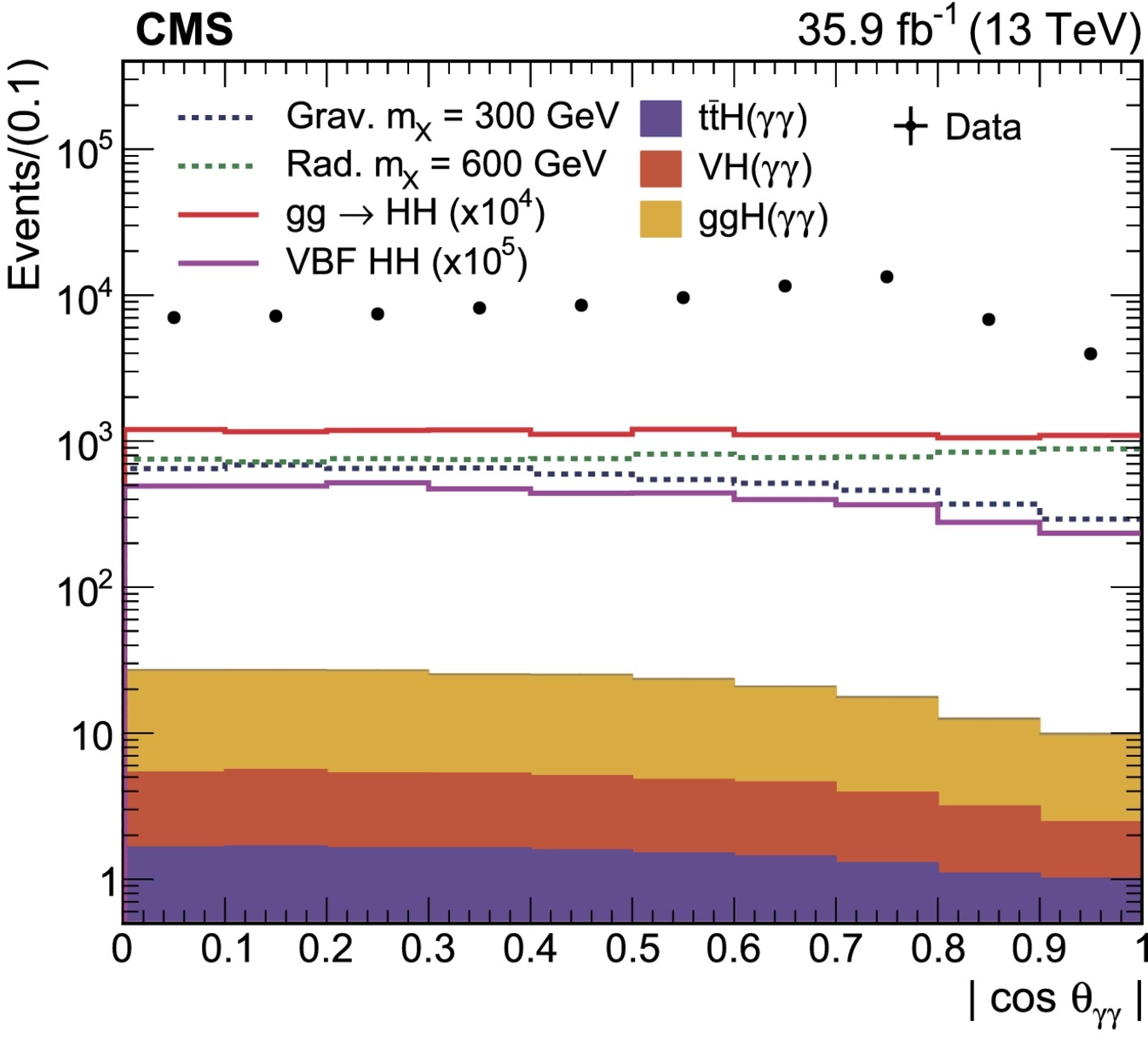}
\includegraphics[width=0.3\textwidth]{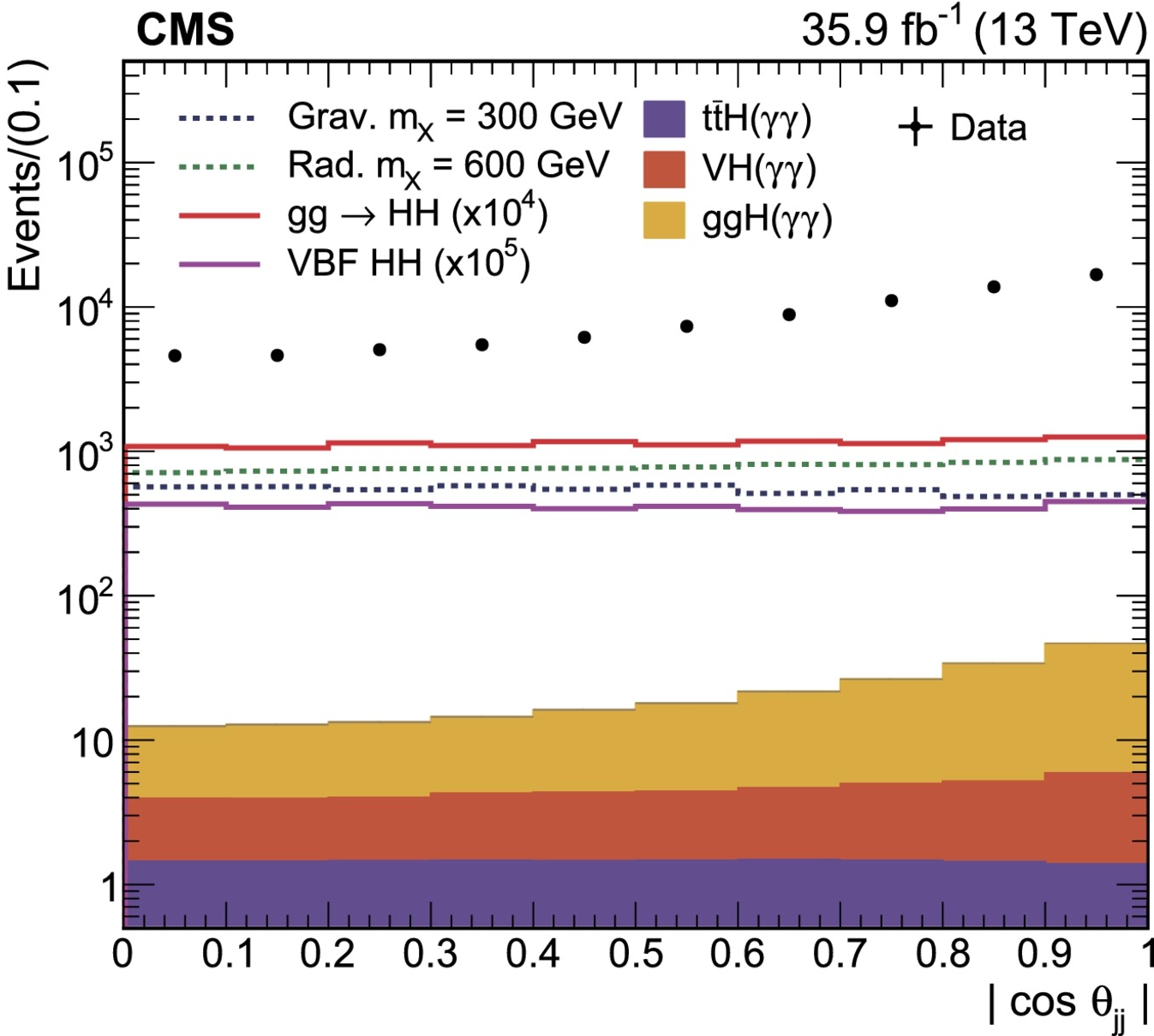}
\caption{Distributions of the three helicity angles for data (dots), $\gamma$+jets background, different signal hypotheses and three single Higgs boson samples ($\ttbar H$, $VH$, and ggF) after the selections on photons and jets~\cite{Sirunyan:2018iwt} have been applied.}
\label{fig:hhbbyy_cms_helicity}
\end{figure}

The output of this algorithm classifies events as more signal- or continuum-background-like, separating events into high, medium and low purity categories. Only the two highest purity categories are used for the signal extraction. 

The \hhbbyy final state can be fully reconstructed and the \mhh spectrum is a particularly important observable for the resonant \hhbbyy searches. The \mhh estimator described in Sec.~\ref{sec_exp_kinfit} is actually used for the signal extraction.

\subsection{Background modelling}

Searches for \hhbbyy are affected by both backgrounds from single Higgs boson production and by non-resonant backgrounds with continuum \myy spectra.

The dominant backgrounds to the \bbyy final state are those in which two objects identified as photons (either prompt photons or jets misidentified as photons) are produced in association
with jets (referred to as $\gamma$+jets). The simulation of these final states poses a major challenge because of large effects from higher orders in QCD. Furthermore, the knowledge of the fragmentation effects for a jet misidentified as a photon is quite limited. For these reasons, these contributions are modelled entirely from data in both ATLAS and CMS \hhbbyy searches with maximum-likelihood fits to parametric shapes. 

However, the choice of a specific function, or families of functions, for the background modelling leads to extra systematic uncertainties related to possible biases in the signal estimate. 
This uncertainty is derived by generating pseudo-data from a certain function choice (truth function) and performing the signal extraction with another function of choice (fit function). 
There are then two alternative approaches: either by testing different truth functions against the fit function, from which an uncertainty due to this choice can be extracted; or the number of degrees of freedom can be increased in the fit function to reduce the bias of fitting different truth function to a negligible level (defined formally as a maximum of 14\% of the statistical uncertainty\footnote{The bias is estimated by how much the definition of standard deviation around the unbiased expected signal strength ($\mu$) has to be inflated to cover 68.3\% of the bias expected $\mu$, in alternative of adding a bias term that corrects the bias $\mu$. A 14\% bias with respect to the unbiased standard deviation, requires to inflate the definition of standard deviation by 1\%, which is much smaller than the systematic uncertainties in the analysis.}).

With the large amount of data to be analysed in the next iterations of these searches, this uncertainty might become the dominant one, justifying the pursuit of alternative background estimation methods, such as Gaussian Processes (GPs)~\cite{Frate:2017mai} and envelope~\cite{Dauncey_2015} methods.

In the GPs approach, instead of defining the parametric description $f(x)$, where $x$ is the fitted observable (in this case \myy~or \mjjyy), $f(x)$ is modelled as a Gaussian and the correlation between two points $x$ and $x'$ is given by a {\it covariance kernel} $\Sigma(x,x')$. 
The choice of the fit function becomes the physics inspired definition of a covariance kernel, which could encode detector specific information, such as energy resolution and scale uncertainties when fitting the invariant mass distribution. 
Moreover, the GPs fit is enough flexible to allow for any function that respects the covariance relation defined by the kernel.

The envelope method includes the bias uncertainties in the fitting procedure. All possible pa\-ra\-me\-tri\-sa\-tions of the background are considered while performing the maximum likelihood fit, with a penalty proportional to their number of degrees of freedom.

Single Higgs boson processes, with two additional jets and with a subsequent decay of the Higgs boson to two photons, are 5--14\% of the total background. Additional jets can be effectively initiated by $b$-quarks, or by lighter quarks and misidentified as a \bjet. The SM single Higgs boson background contribution is estimated using a parametric model fitted to simulated samples.
The SM single Higgs boson background is particularly challenging for the \bbyy searches, as the \myy~peak, which is the most important handle for signal discrimination, appears as a background feature. 
However, other event characteristics that are dependent on the Higgs boson production mechanism can be exploited. 
Some of these features might also be helpful to reduce the continuous background, therefore a combined background mitigation procedure can be devised.
The CMS parametric fit of \myy~and \mjj distributions, described in Sec.~\ref{subsubsec:bbaa_extr}, mitigates the impact of the single Higgs boson background. Alternatively a machine learning based multi-classification algorithm, exploiting kinematic properties of the four-body system and other event observables, such as the number of jets and \bjets, and missing transverse momentum, could be investigated.

New physics may enhance single Higgs production too, 
both \hh and $\ttbar H$ can be enhanced by modifications to the Higgs-top Yukawa coupling. Performing a simultaneous signal extraction of the \hhbbyy and $\ttbar H\to\gamma\gamma$ signals could account for this possible scenario. The ability to constrain the SM single \hyy~backgrounds will play an important role in the future.


%

\subsection{Signal extraction}
\label{subsubsec:bbaa_extr}

The signal extraction in the \hhbbyy~searches is similar to the procedure used in the SM \hyy~measurements, to take advantage of the excellent mass resolution of the \hyy channel, by fitting the resonant \hyy peak on top of the continuous and monotonically falling background. 
For \hhbbyy searches, the presence of the \hbb resonant peak becomes an extra handle to constrain or reduce background processes. 
The ATLAS and CMS analyses signal extraction strategies differ particularly in the usage of the \hbb~mass spectrum.

The ATLAS approach is an unbinned, maximum-likelihood fit to the \myy~distribution 
and the \hbb~resonance is used to reduce the background by requiring the compatibility of \mjj~peak with the Higgs hypothesis ($80/90 < \mjj < 140$~GeV for the loose/tight selection). 
The relative contribution of $\gamma\gamma$, $\gamma\text{j}$, $\text{j}\gamma$ and jj produced in associations with jets, to the continuum background is determined from data by varying the photon identification and isolation criteria. The functional form used to model the background is then chosen using events simulated with \textsc{Sherpa}~\cite{Gleisberg:2008ta}. The accuracy of the simulation is tested in events passing all the event selection requirements but failing the \btagging and a correction factor, up to 5\%, is derived as function of \myy and applied in the one and two $b$-tag categories. The $\gamma\gamma\bb$ is the dominant contribution to the continuum background in the two $b$-tag category ($\approx$80\%), while $\gamma\gamma\text{bj}$ ($\approx$60\%) dominates in the one $b$-tag category, as shown in Fig.~\ref{fig:hhbbyy_atlas_bkgmodel}.

\begin{figure}[ht]
\centering
\includegraphics[width=0.45\textwidth]{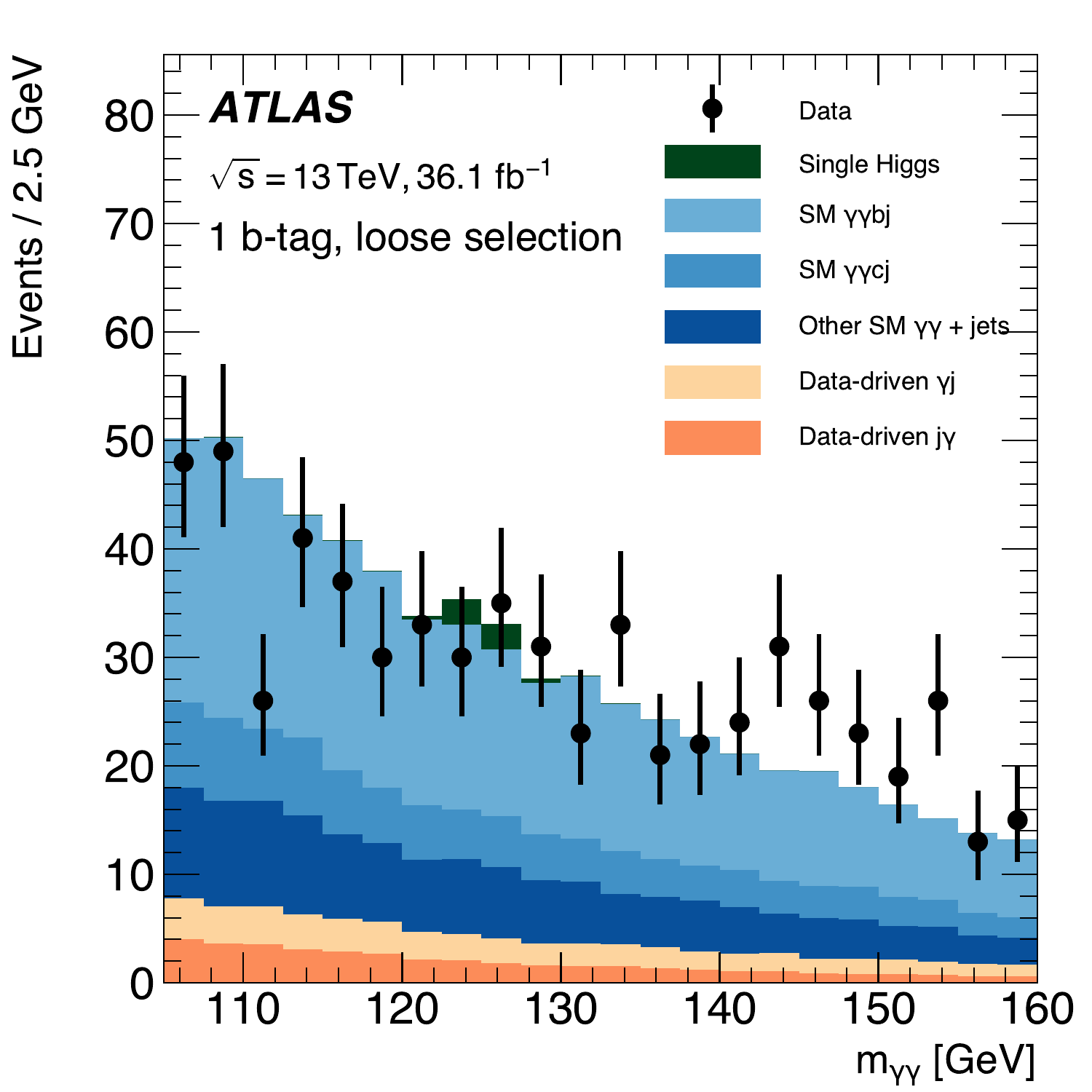}
\includegraphics[width=0.45\textwidth]{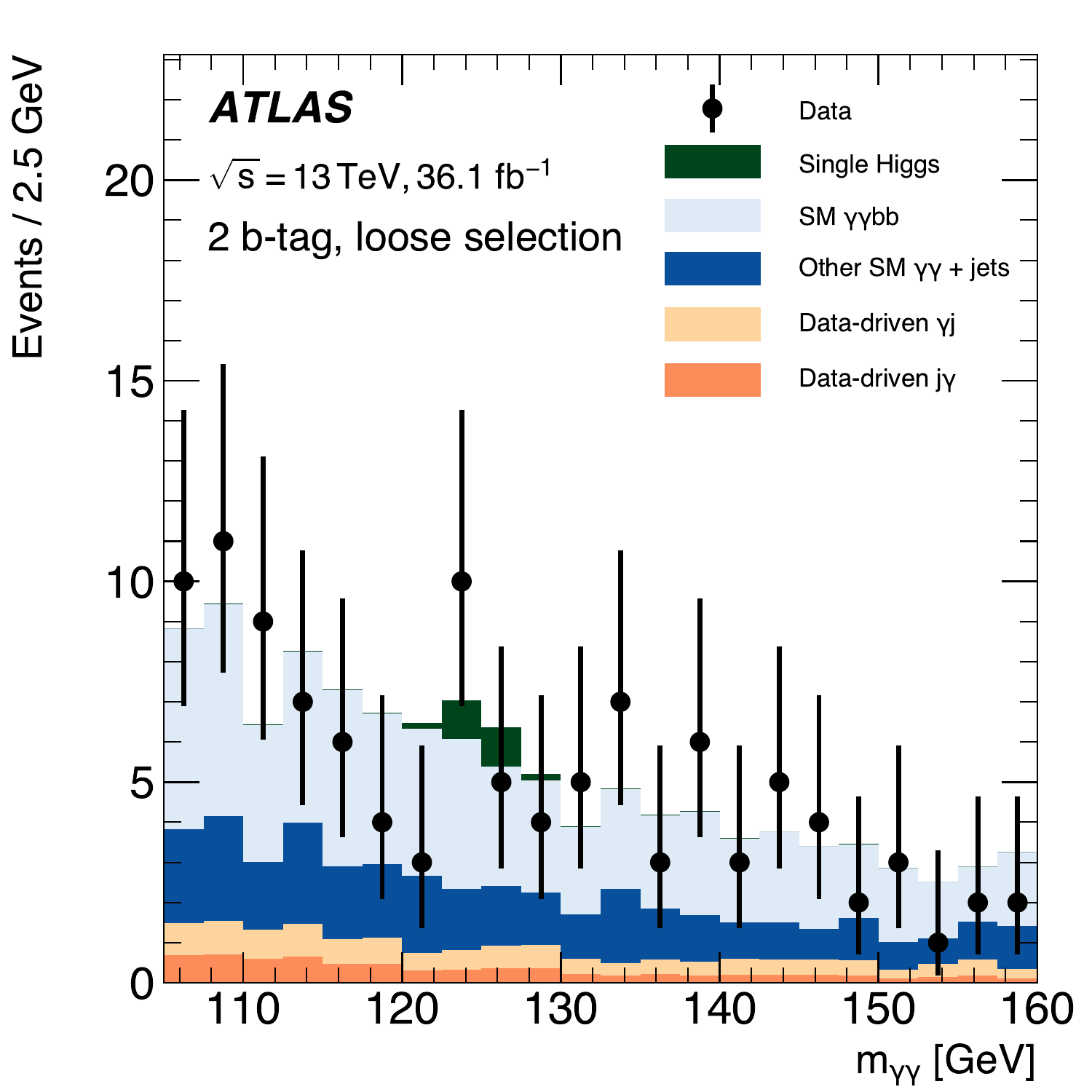}
\caption{The expected number of background events for the continuum $\gamma\gamma$+jets production, other continuum $\gamma$+j production (orange) and single Higgs boson production (green) is compared to the observed data (black points) for the \myy distribution in the one (left) and two $b$-tag (right) categories~\cite{Aaboud:2018ftw}.}
\label{fig:hhbbyy_atlas_bkgmodel}
\end{figure}

The full background fit is shown in Fig.~\ref{fig:hhbbyy_atlas_fit}, for the loose and tight selections in the two $b$-tag category.

\begin{figure}[ht]
\centering
\includegraphics[width=0.45\textwidth]{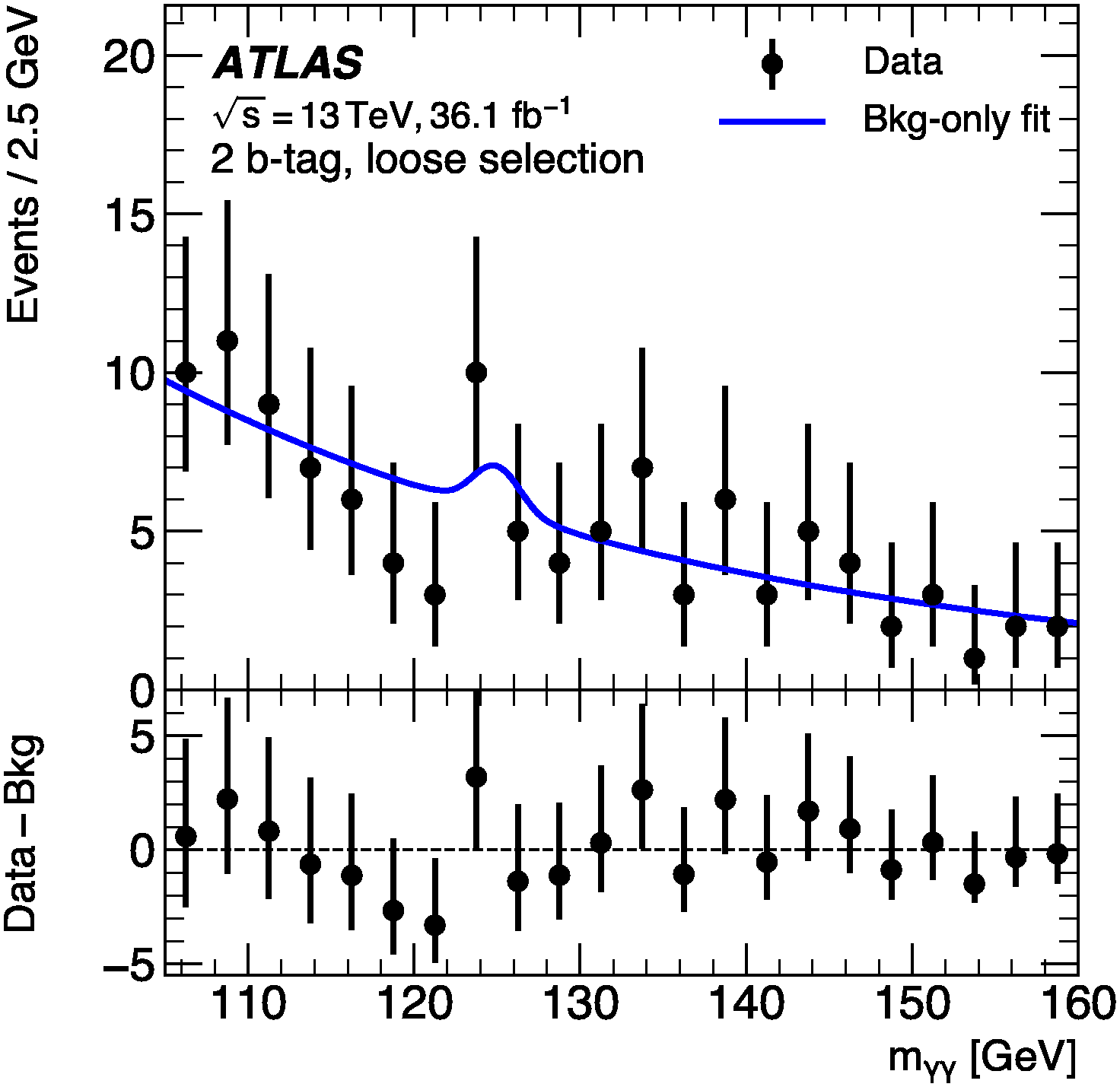}
\includegraphics[width=0.45\textwidth]{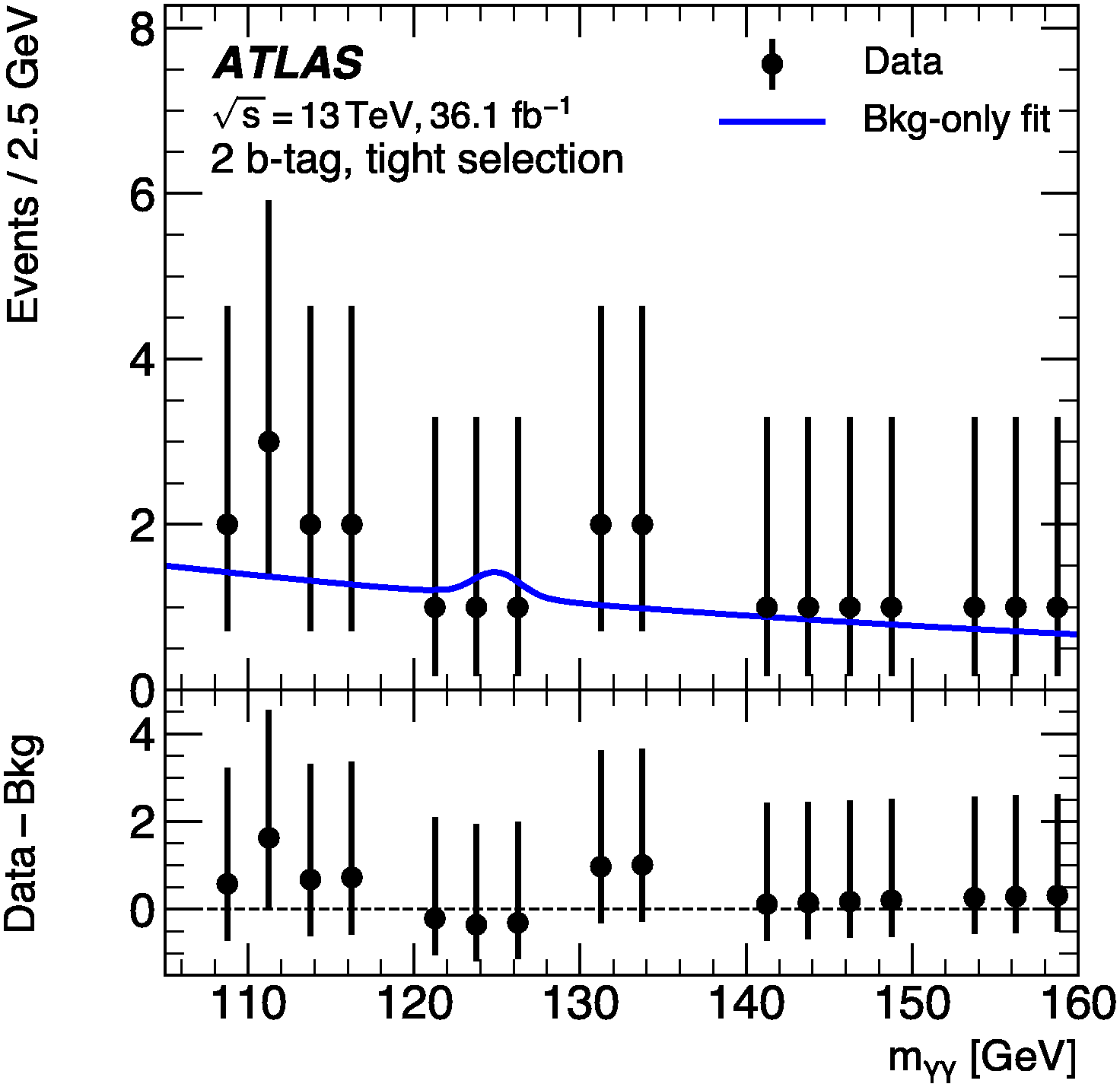}
\caption{\myy~distributions for the two $b$-tag category after the loose (left) and tight (right) event selection requirements are applied~\cite{Aaboud:2018ftw}.}
\label{fig:hhbbyy_atlas_fit}
\end{figure}

In the CMS analysis, the signal extraction is performed simultaneously in the \myy and \mjj distributions~(2D fit). 
It assumes that the background can be described by a two dimensional parametric function and that can be factorised, similarly to the parametric signal model described in Sec.~\ref{sec:hhbbyy_signal_modeling}. 
This hypothesis is tested by checking if possible correlations between \myy~and \mjj~would be statistically significant with the typical expected number of background events in the analysis signal regions. 
The validity of this assumption is therefore dependent on the size of the dataset, and has to be checked again with the increase of the integrated luminosity. 
The projections in \myy~and \mjj distributions for the most sensitive categories to the SM \hh production are shown in Fig.~\ref{fig:hhbbyy_cms_fit}.

\begin{figure}[ht]
\centering
\includegraphics[width=0.45\textwidth]{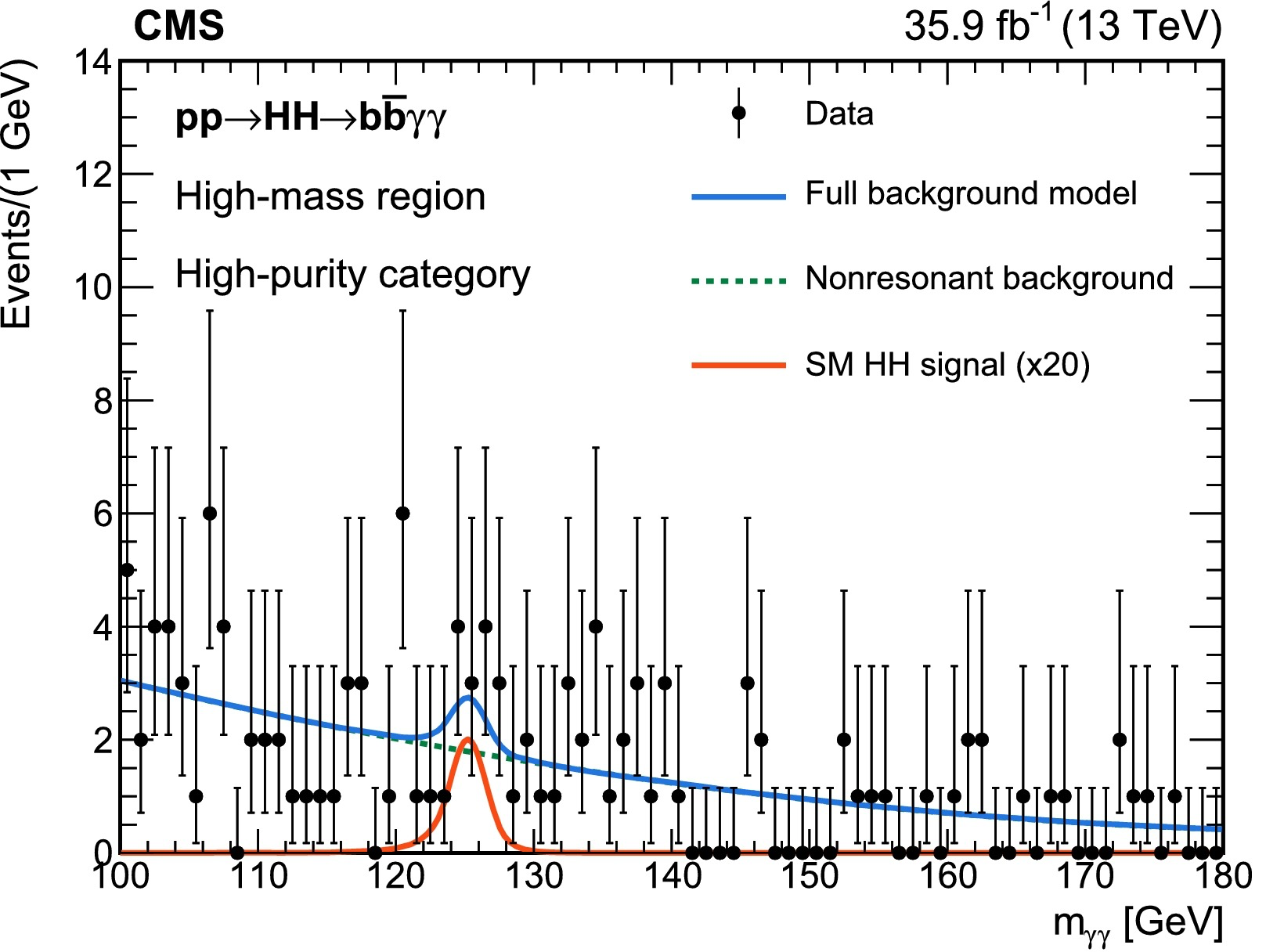}
\includegraphics[width=0.45\textwidth]{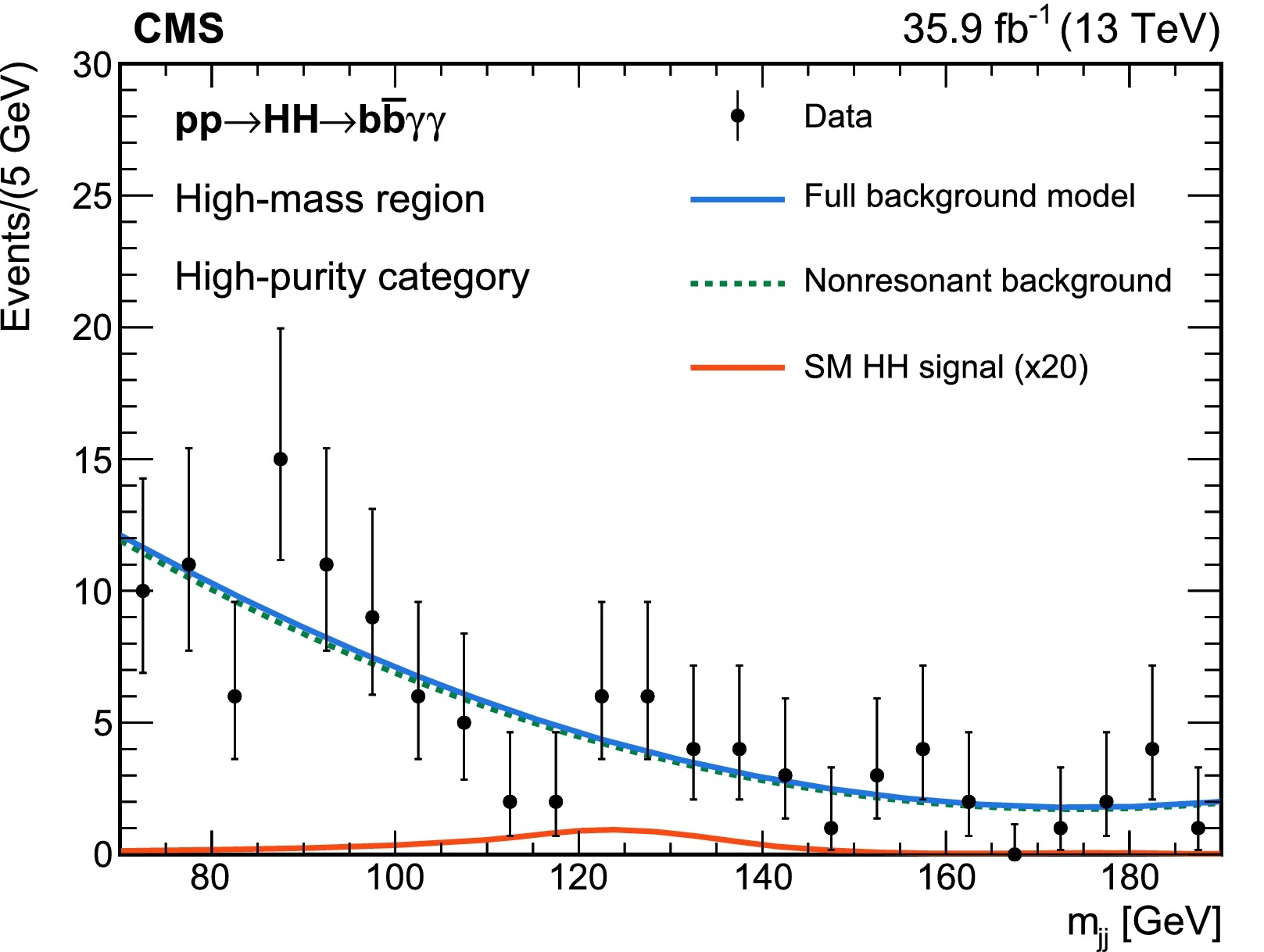}
\caption{\myy~and \mjj~projections of the 2D maximum-likelihood fit for the signal extraction, in the most sensitive category to the SM \hh production~\cite{Sirunyan:2018iwt}.}
\label{fig:hhbbyy_cms_fit}
\end{figure}

The ATLAS approach simplifies the continuum background description, as it does not depend on the accuracy of the \myy--\mjj correlations modelling, for both the signal and background hypotheses. 
On the other hand, the CMS 2D fit approach constrains better the non-resonant background exploiting fully the \mjj distribution and it improves the search sensitivity by $\approx 10\%$. 

\subsection{Systematic uncertainties}

The \hhbbyy searches are currently limited by the statistics of the Run 2 dataset. 
Theoretical uncertainties on the PDF and scale variations are applied to the non-resonant signal model and they amount to 3--6\%. Uncertainties on the normalisation of single Higgs boson background processes are also taken into account, corresponding to 1--20\%.

Photon trigger efficiency, as well as photon energy scale and resolution uncertainties impact the signal model and acceptance by 1--5\%. The largest experimental uncertainties come from the jet energy scale and resolution (1--5\%), and from flavour-tagging uncertainty (10--20\%).

ATLAS additionally applies an uncertainty due to the continuum background fitting process, and a 100$\%$ uncertainty on the ggF and WH single Higgs boson production modes in events with extra heavy-flavour particles.

\subsection{Machine learning in \hhbbyy: use and challenges} \label{sec:hhbbyy_ML}

Both ATLAS and CMS analyses use machine learning methods for event categorisation and signal classification. 
ATLAS uses a BDT method for the selection of the \hbb~candidate, when only one jet is $b$-tagged. 
The $b$-tagged plus non-$b$-tagged jet pairings are built for the signal (only one pairing is correct) and the continuum background (no pairing is correct) using simulated events. 
The BDT is trained to classify correct and incorrect pairings based on the kinematic information of the paired jets: paired jets \pTj, di-jet \pT and $b$-tagging information, \mjj, paired jets $\eta^{\textrm{j}}$, di-jet $\eta^{\textrm{jj}}$, $\Delta\eta$ between the paired jets. 
The ranking of the jets according to the closest match between the di-jet mass and the Higgs boson mass, highest jet \pTj and highest di-jet \pT is exploited as well.
Each di-jet pairing in an event is given a BDT score, the di-jet with the highest score is then selected to reconstruct the \hbb~candidate.


In the CMS analysis, a BDT is trained for signal classification against the continuum background. 
The training variables are: the $b$-tagging scores of the jets that form the \hbb~candidate, the three helicity angles and the \hh transverse balance variables \pTjj/\mjjyy~and \pTyy/\mjjyy. The training is performed with the ensemble of all non-resonant \hh production hypotheses (SM plus the shape benchmark used for BSM reinterpretations as explained in Sec.~\ref{sec:shape_bench}) as signal. This choice allows for the final classifier performance to be generalised to various BSM \hh production hypotheses. 
Events that pass all the analysis selection criteria except the identification and isolation requirements for one photon candidate, are used as background events for the training.
This choice is validated by comparing the input distributions in the training dataset with the signal selection events that fall outside of a mass window of 30~GeV around the Higgs boson mass in \myy. 



A common issue with classifiers trained with specific target signals is how their performance can be generalised to other signal hypotheses, for which the kinematic properties might change substantially. 
Both ATLAS and CMS searches deal with this challenge by defining different kinematic regimes with the four-body invariant mass, populated by different signal hypotheses and background compositions, in which dedicated training or cut based analyses can be performed. 
In addition, CMS chooses to use an ensemble of different signal simulated samples as the signal hypothesis for the BDT training to guarantee a uniformity of the sensitivity to different final states.

A different approach, already used by other \hh analyses such as the CMS \hhbbVV search, is to train a discriminant based on a parameterised neural network (NN).
The NN training is performed as a function of a certain model parameter, such as the X resonance mass when looking for resonant $\textrm{X} \rightarrow \hhbbyy$ signals, or the Higgs boson self-coupling modifier \klambda. 
The performance for each individual model parameter is similar to the performance of a network trained using that single hypothesis as the target signal.
Therefore, the parameterised NN effectively trains different NNs for each model parameter in a single training procedure.
Additionally, this NN is also able to interpolate between the model parameters used for training.




%% file: bbtt/bbtautau.tex
The \hhbbtt final state has a  branching fraction of 7.3\% for a SM Higgs boson with mass of 125~GeV and a relatively small background contribution from other SM processes, compared to the \hhbbbb search. Three final states of the $\tau$-lepton pair are combined for the \hhbbtt searches: $\tau\tau\rightarrow e\tau_{\mathrm{h}}$, $\tau\tau\rightarrow \mu\tau_{\mathrm{h}}$, and $\tau\tau\rightarrow \tau_{\mathrm{h}}\tau_{\mathrm{h}}$.  These three final states all together account for 88\% of $\tau\tau$ decays.  The case where both $\tau$-leptons decay to lighter charged leptons ($\ell = e/ \mu$) and their associated neutrinos account for the remaining 12\% of $\tau\tau$ decays, but this category of events have not been considered by either the ATLAS or CMS experiments yet~\cite{Aaboud:2018sfw,Sirunyan:2017djm}.

The reconstruction of \hhbbtt events poses several challenges, including the reconstruction of hadronic objects: \bjets and hadronically decaying $\tau$-leptons, $\tau_{\mathrm{h}}$ (as described in sections \ref{sec:bTagging} and \ref{sec:tauhad}), the rejection of objects that mimic these, and the reduction of backgrounds. The most irreducible backgrounds are $\mathrm{Z} \rightarrow\tau\tau$ produced in association with heavy-flavour jets, \ttbar pairs and multi-jet processes, in which quark- and gluon-initiated jets are misidentified as $\tau_{\mathrm{h}}$. Single Higgs boson production, particularly in association with a Z boson or a top pair, is becoming an important background contribution as the size of the available dataset increases.  


\subsection[Analysis strategies]{Analysis strategies \\ \contrib{K. Androsov, A. Bethani, A. Betti, H. Fox, M. Gallinaro, K. Leney}}
\label{subsubsec:bbtautau_trigger}

At the trigger level, both the ATLAS and CMS experiments require the presence of an isolated lepton or hadronic $\tau$ object, depending on the final state.

For the fully hadronic channel ($\tau_{\mathrm{h}}\tau_{\mathrm{h}}$), CMS requires di-$\tau$ triggers, while ATLAS uses both single- and di-$\tau$ triggers, described in Sec.~\ref{sec:bTrigger}. For the semi-leptonic channels ($\ell\tau_{\mathrm{h}}$), ATLAS uses single lepton and lepton-plus-$\tau$ on line selections, while CMS uses only single lepton triggers.  This use of lepton-plus-$\tau$ triggers by ATLAS allows the use of lower \pT thresholds for the analysis object selection, which result in a 3\% gain on the final sensitivity for the semi leptonic channels. 

Similar trigger strategies are planned for the future, however the increased instantaneous luminosity will force the lepton and jet \pT thresholds to be raised unless new techniques, exploiting track information for instance, are used.

In order to reconstruct a \hhbbtt candidate event, it is necessary to identify any electron or muon from a leptonic $\tau$ decay, one or two hadronically decaying $\tau$ leptons ($\tau_{\mathrm{h}}$), the jets originating from the two $b$-quarks, and the missing transverse momentum of the event. The latter arises predominantly from the neutrinos accompanying the $\tau$-lepton decays, although neutrinos in semi-leptonic $B$-hadron decays may also contribute.
Both collaborations use a medium operating point for hadronic $\tau$ identification, as described in Sec.~\ref{sec_exp_2dot2}.

In addition to the hadronic $\tau$ objects, electrons or muons, the presence of two jets within the tracker acceptance is required. Jets may be required to be tagged as originating from the hadronisation of $b$-quarks.  The operating point used in the ATLAS and CMS analyses provides approximately 70\% $b$-tagging efficiency with a mis-identification rate of approximately 0.3\% and 1\%, respectively, for light-flavoured jets. The optimisation of the \bjet tagging efficiency and the requirement on the number of \bjets, depends on the  background suppression. In the ATLAS search, both jets are required to be $b$-tagged, while in the CMS analysis events are split into three exclusive categories depending on the number of $b$-tagged jets (0, 1, or 2). 

In the CMS analysis, further classification into ``resolved'' and ``boosted'' categories is used in the case of resonant \hh production, for invariant mass values (\mhh) above 700 GeV~\cite{Sirunyan:2017djm}, where high \pT \hbb candidates are reconstructed more efficiently as a large-radius jet, as described in section~\ref{sec:hbbbosted}.
The event is classified as boosted if it contains at least one AK8 jet of invariant mass larger than 30~GeV and $\pT>170$~GeV that is composed of two sub-jets.  Otherwise, the event is classified as resolved.
In order to improve the resolution and to enhance the sensitivity of the resonant analysis, the invariant mass is reconstructed using a kinematic fit, as detailed in Sec.~\ref{sec_exp_kinfit}.

Different observables related to the event kinematic are used to discriminate between signal and background, with various differences depending on the signal model and the considered \htautau decay mode. 

In both the resonant and non-resonant production modes, \mhh is one of the most discriminating variables for background rejection.  An essential part of reconstructing the \hh mass is to first reconstruct the mass of the two sub-systems, \mtt and \mbb. The \mbb is improved by applying dedicated \bjet specific energy corrections as discussed in Sec.~\ref{sec_exp_2dot2}. 
Accurately reconstructing the mass of a resonance decaying to a pair of $\tau$-leptons is challenging because of the presence of multiple neutrinos from $\tau$-lepton decays, which lead to a kinematic description of the system that is under-constrained.
The ``collinear approximation" is a simple but frequently used technique to address this problem. It is based on the observation that the neutrinos are produced nearly collinear with the corresponding visible $\tau$-lepton decay and that all the missing transverse energy in the event comes from the neutrinos of the $\tau$-lepton decays. Then, \mtt is directly calculated from the masses and momenta of the visible products of the $\tau$-lepton. 
This technique gives a reasonable mass resolution, when the two neutrino momenta are not back-to-back or when the di-tau system transverse momentum is large enough to compensate the resolution effects on the reconstructed missing transverse energy.
In order to better reconstruct \mtt, both the ATLAS and CMS collaborations have developed algorithms including dynamic likelihood techniques~\cite{Kondo:1988yd,Kondo:1991dw} to account for the invisible part of the four-momentum due to the neutrinos. The Missing Mass Calculator (MMC)~\cite{MMC} is used by ATLAS, whereas the Secondary Vertex Fit (SVfit)~\cite{Bianchini:2014vza,Bianchini:2016yrt} is used by CMS.  Both algorithms calculate the best estimate of the $\tau\tau$ invariant mas on an event by event basis, using constraints from the measurements of the visible decay products and the missing transverse energy.

In the case of the MMC algorithm, the estimate exploits the fact that the solutions of the under constrained kinematic system are not all equally probable. Then, additional constraints from the $\tau$ kinematics are applied. In this case, the distance $\Delta$R between the neutrino(s) and the visible decay products is parametrised and provides a probability density function that is then incorporated in a global event likelihood. The most probable value provides the final estimation of \mtt.

In a similar way, the SVfit \mtt values are reconstructed by combining the measured observables, the $x$ and $y$ components of the missing transverse energy, with a probability model, that includes terms for the $\tau$ decay kinematics. The model makes a prediction for the probability to observe the missing transverse energy values measured in the event, given a parameterisation of the kinematics of the $\tau$ pair decay and it provides a probability density function as a function of the unknown parameters. The best estimate for the \mtt is the value that maximises this probability.

After selecting events compatible with a di-$\tau$ plus \bjets final state, the ATLAS search requires that the MMC-based \mtt be above 60~GeV, while CMS uses an elliptical selection in the \mtt-\mbb plane around the SM Higgs boson mass:
\begin{equation}
  \frac{(\mtt-116~\mbox{GeV})^2}{(35~\mbox{GeV})^2}+\frac{(\mbb-111~\mbox{GeV})^2}{(45~\mbox{GeV})^2}{} < 1
\end{equation}

ATLAS uses three categories of signal events, based on the trigger selection, while CMS defines nine categories in total depending on the $\tau$ final state and number of \bjets: ($e\tau_{\mathrm{h}}, \mu\tau_{\mathrm{h}}, \tau_{\mathrm{h}}\tau_{\mathrm{h}}$) $\times$ (one \bjet, two \bjets, boosted).  For both experiments, the most sensitive category is the $\tau_{\mathrm{h}}\tau_{\mathrm{h}}$ where both jets pass the $b$-tagging requirements.

Both experiments use BDTs trained on different kinematic variables to improve the analysis sensitivity. 
ATLAS uses BDTs to separate the signal from multi-jet, \ttbar and Z+\bjet backgrounds. The input variables include angular information, the full di-$\tau$ mass including neutrinos, \mbb corrected for neutrinos in semi-leptonic $B$-decays (see Sec.~\ref{sec:bjetreg}), \mhh and in the leptonic category also thee transverse mass. 

ATLAS uses BDTs for all categories of signal events, while CMS uses a BDT only in the semi leptonic resolved categories.  Furthermore, while ATLAS uses the output BDT scores directly to extract the signal, CMS applies a cut on the BDT output score and then uses the \mhh distribution as the final discriminant for the resonance search, and the ``stransverse mass" ($m_{\mathrm{T}2}$) for the non-resonant analysis.  The $m_{\mathrm{T}2}$ variable exploits the fact that the stransverse mass of the $t \to Wb$ system is constrained by the top quark mass, and therefore $m_{\mathrm{T}2}$ is bounded for the \ttbar background (without resolution effects), while this is not the case for the \hhbbtt signal~\cite{Barr:2013tda}. 

\begin{figure}[ht]
\centering
\includegraphics[width=0.45\textwidth]{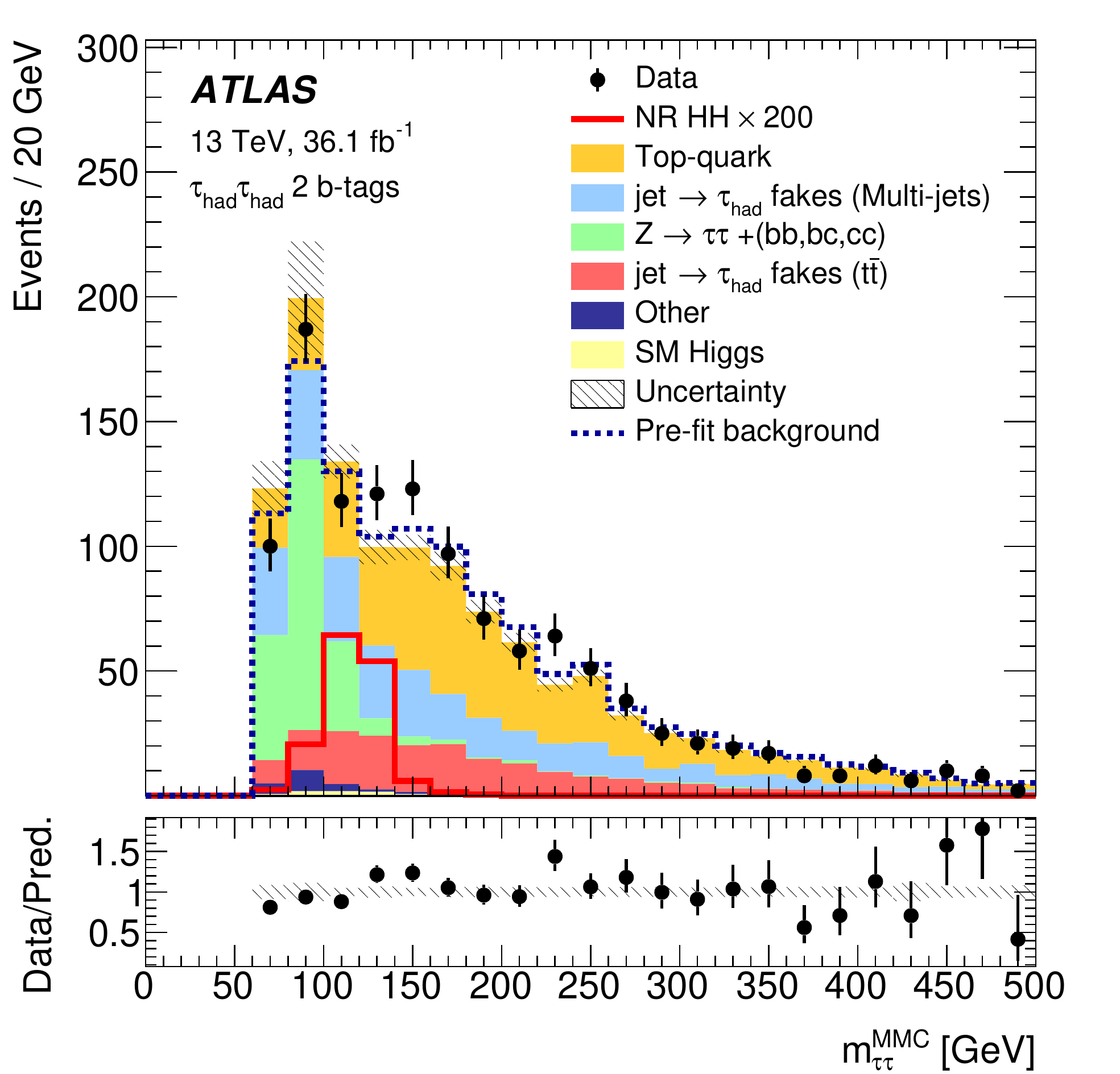}
\includegraphics[width=0.45\textwidth]{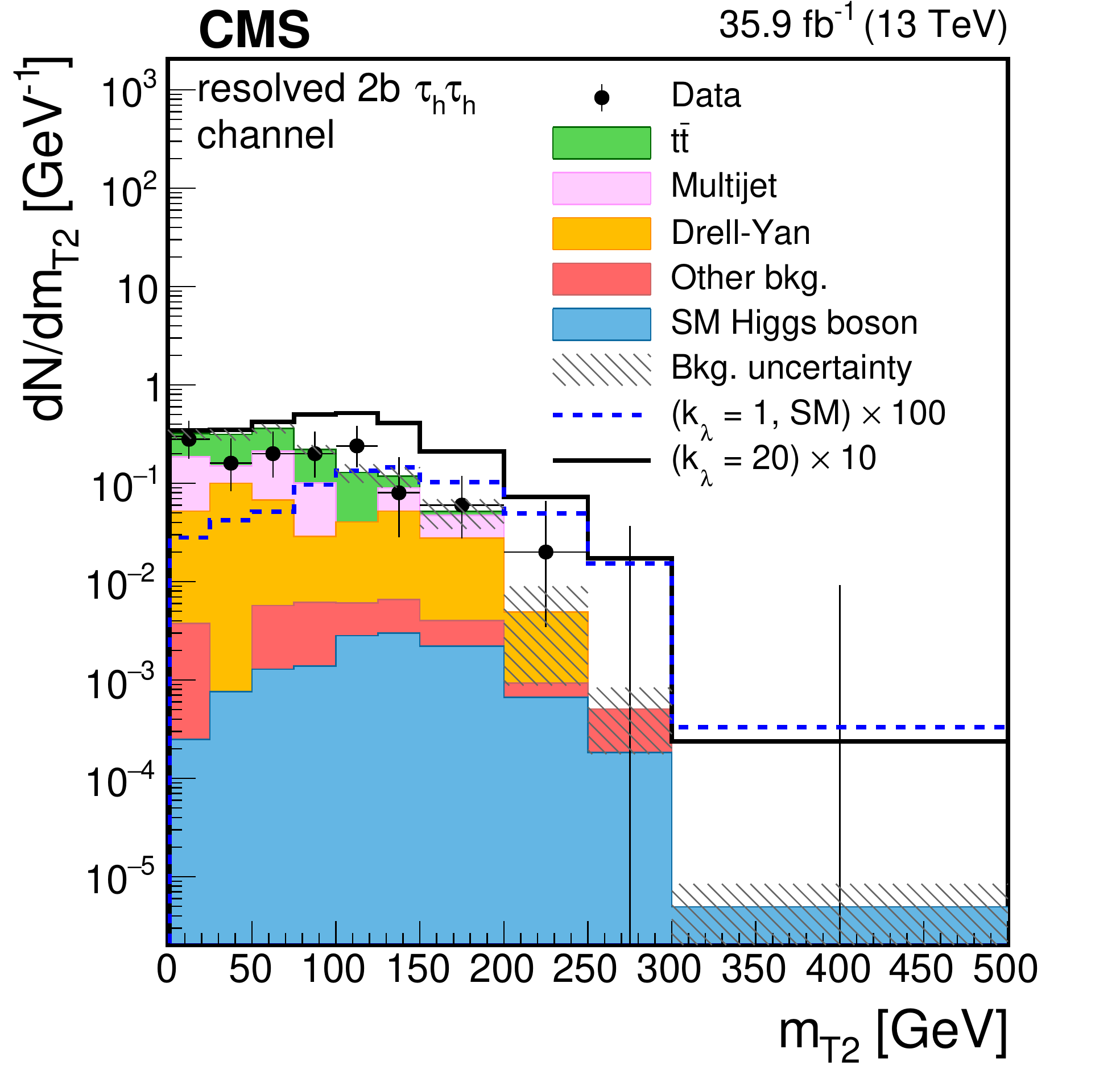}
\caption{The distribution of the MMC-based \mtt (left) ~\cite{Aaboud:2018sfw} and $m_{\mathrm{T}2}$ (right) variables in the $\tau_{\mathrm{h}}\tau_{\mathrm{h}}$ and two \bjets category~\cite{Sirunyan:2017djm}.}
\label{fig:hhbbtt_var}
\end{figure}

The MMC-based \mtt and $m_{\mathrm{T}2}$ distributions are shown for the $\tau_{\mathrm{h}}\tau_{\mathrm{h}}$ and two \bjets category in Fig.~\ref{fig:hhbbtt_var} for simulated signal and background events.

\subsection[Modelling of background contributions]{Modelling of background contributions \\ \contrib{K. Androsov, A. Bethani, A. Betti, H. Fox, M. Gallinaro, K. Leney}}
\label{subsubsec:bbtautau_ttbar}

One of the main backgrounds in the \hhbbtt search comes from \ttbar events with $\ell+\tau_{\mathrm{h}}$\ final states $\left(\ttbar\rightarrow W(\rightarrow\ell\nu)b\, W(\rightarrow\tau_{\mathrm{h}}\nu\nu)\bar{b}\right)$, or lepton and jet final states ($\ttbar\rightarrow W(\rightarrow\ell\nu)\bar{b}\, W(\rightarrow q \bar{q})\bar{b}$) where the jet is incorrectly reconstructed as a $\tau_{\mathrm{h}}$ object. Due to the relatively large top quark pair production cross section ($\approx 832$~pb at $\sqrt{s}=13$~TeV~\cite{Czakon:2011xx,Czakon:2013tha}) and final states similar to the signal process, this is the dominant source of background.
In both experiments the \ttbar model relies on simulation, where ATLAS uses the \textsc{POWHEG-BOX} generator~\cite{Alioli:2010xd} with a NNLO+NNLL precision for the cross section, whereas CMS uses the \textsc{POWHEG~2.0} generator ~\cite{Campbell:2014kua} with a NLO precision for the cross section.  
In ATLAS, the component of the \ttbar background in which the reconstructed $\tau_{\mathrm{h}}$ objects are matched to a hadronically decaying $\tau$-lepton at truth level is estimated from simulation. Its normalisation is further constrained in data using the low BDT output score region of the $\tau_{\ell}\tau_{\mathrm{h}}$ channel. In the $\tau_{\ell}\tau_{\mathrm{h}}$ channel the component of \ttbar in which the reconstructed $\tau_{\mathrm{h}}$ object is mis-identified is estimated in an entirely data-driven way.  In the $\tau_{\mathrm{h}}\tau_{\mathrm{h}}$ channel, the simulation is corrected using a $\tau$ fake-rate derived from data.



Events with a boosted \hbb candidate are assigned to a dedicated boosted category in the CMS analysis.  A BDT discriminant based on the kinematic differences between the \hh and \ttbar processes is used in the lepton+jet final states in order to reduce the large amount of \ttbar background.

The production of Z bosons in association with heavy-flavour jets provides a significant background to the \hhbbtt signal process.  Both experiments take the shape of $\mathrm{Z}/\gamma^{*} \to \tau^+\tau^-$+jets from simulation and normalise it using control regions defined in the data. ATLAS uses the \textsc{SHERPA} generator while CMS uses \MGAMCNLO~\cite{Alwall:2014hca,Hirschi:2015iia}.

The modelling of this background is limited by the current understanding of the hadronisation of jets initiated from $b$ or $c$ quarks. Cross section predictions for this background do not match the observations in data, and large correction factors need to be applied. In both ATLAS and CMS experiments these are derived from control regions dominated by $\mathrm{Z} \rightarrow \mu\mu$+jets events. In both cases, the selection is similar to that of the signal region, and an additional cut on $\mathrm{m}_{\mu\mu}$ is applied.  ATLAS applies an additional selection on the $b$-tagged jet pair invariant mass \mbb in order to reduce the SM ZH process and provides a single normalisation factor that is used to correct the ${Z}+\bb/{bc}/{c\bar{c}}$ processes. CMS performs a simultaneous fit in the three categories and provides three normalisation factors, depending on the number of \bjets from the hard process: Z + 0, 1 or at least 2 \bjets at the generator level.

The scale factors and their uncertainties are applied to the $\mathrm{Z}/\gamma^* \rightarrow \ell^+\ell^-$ simulated processes and are propagated to the background estimation to correctly account for higher-order effects. The use of finer granularity event categories may further constrain these sources of background and reduce their associated uncertainties.  


In the ATLAS searches, fake factors are derived in control regions with inverted isolation requirements on the light lepton  ($\tau_{\ell}\tau_{\mathrm{h}}$ channel) or events where the two $\tau$ objects have the same-sign charge ($\tau_{\mathrm{h}}\tau_{\mathrm{h}}$ channel).  The fake factors are then applied to a template region where reconstructed $\tau_{\mathrm{h}}$ objects fail the nominal ID requirements, but still pass a very loose requirement on the $\tau$ ID BDT score (in order to maintain a selection of jets that have $\tau_{\mathrm{h}}$-like properties).  The fake factors are binned in \pT and number of associated tracks.

For all channels in the CMS search, control regions are constructed by inverting the requirements on the sign of the $\tau_{\mathrm{h}}$-pair charge product, and the $\tau_{\mathrm{h}}$ isolation.  The three control regions are therefore defined as: same sign (SS) isolated, opposite sign (OS) anti-isolated, SS anti-isolated.
The shape of the multi-jet template is estimated from the SS isolated region, while the normalisation is estimated as the ratio of the yields of OS anti-isolated and SS anti-isolated regions multiplied by the yield in the SS isolated region.

\subsection[Limitations of the current result and perspectives]{Limitations of the current result and perspectives \\ \contrib{M. Gallinaro, T. Vickey}}
\label{subsubsec:bbtautau_limitations}

For the non-resonant \hhbbtt production, observed limits of 12.7 and 31.4 times the rate predicted by the SM have been set by the ATLAS and CMS experiments respectively.
A BDT was not used in the CMS $\tau_{\mathrm{h}}\tau_{\mathrm{h}}$ or boosted channels, since following the semi-leptonic resolved analysis strategy of cutting on the BDT output score and using the \mhh variable as the final discriminant was not feasible in these channels, due to limited statistics in the final selection.  This approach of cutting on the BDT output score leads to a larger statistical uncertainty on the final result, and additionally makes it harder to constrain the nuisance parameters associated to background processes.

Another source of the difference in the results obtained by the two experiments is that the CMS selection has significantly lower efficiency for the signal in all categories due to less efficient $b$-tagging.  This effect is amplified in the two \bjet categories.  For example, in the most sensitive $\tau_{\mathrm{h}}\tau_{\mathrm{h}}$ two \bjet category the expected yield of non-resonant SM \hh events is $0.75 \pm 0.14$ events ($0.55 \pm 0.10$ in the last two bins of the BDT) in ATLAS, while CMS expects $0.21$ events in their signal region.  These limitations have a direct impact on the signal extraction strategy chosen by CMS and on the sensitivity of the final limits.

As the result of (i) the limited statistics in the final selection for CMS, (ii) the absence of a multivariate analysis for the most sensitive category and (iii) the choice of selecting events based on the BDT score instead of the extracting the signal from its distribution, the final signal sensitivity obtained by CMS is considerably weaker, by a factor 1.7, than the result obtained by ATLAS. 

An independent study using CMS data~\cite{Giraldi:2623854} has shown that results comparable to those published by ATLAS can be obtained if multivariate techniques are used for all three channels and if the signal is extracted using the continuous BDT score.  This confirms that by improving the analysis strategy CMS results could reach similar sensitivity to those reported by the ATLAS experiment.
Although the analyses are currently dominated by statistical uncertainties on the data, the impact of systematic uncertainties will become increasingly important as the size of the available dataset increases.  
The dominant source of systematic uncertainties are the multi-jet and \ttbar background normalisation, which are 5--30\% depending on the final state and category and 10--17\% respectively; the knowledge of the $\tau_{\mathrm{h}}$ and $b$-tagging efficiency, which impact the overall signal normalisation up to 10--16\% and 6--8\% respectively the ATLAS and CMS non-resonant searches. 

Reducing these uncertainties is therefore an important way to improve the sensitivity of the \hhbbtt searches in the future.

%% file: bbVV/bbVV.tex
\newcommand{\de}{\partial}
\newcommand{\Mb}{\bar{M}}
\newcommand{\Mpl}{M_{\textrm{Pl}}}
\newcommand{\Mp}{M_{\textrm{Pl}}}
\newcommand{\reef}[1]{(\ref{#1})}

\newcommand{\cL}{\mathcal{L}}
\newcommand{\cR}{\mathcal{R}}
\newcommand{\BR}{{\rm BR}}
\newcommand{\blu}{\color{blue}}
\newcommand{\ros}{\color{red}}
\newcommand{\eq}[1]{Eq.~(\ref{#1})}
\newcommand{\lag}{\mathcal{L}}
\newcommand{\op}{\mathcal{O}}
\newcommand{\lp}{\left(}
\newcommand{\rp}{\right)}
\newcommand{\nn}{\nonumber}
\newcommand{\vev}[1]{\langle {#1} \rangle}
\newcommand{\pslash}{\cancel{p}}
\newcommand{\be}{\begin{equation}}
\newcommand{\ee}{\end{equation}}
\newcommand{\bea}{\begin{eqnarray}}
\newcommand{\eea}{\end{eqnarray}}
\newcommand{\bc}{\begin{center}}
\newcommand{\ec}{\end{center}}
\newcommand{\ba}{\begin{array}}
\newcommand{\ea}{\end{array}}
\newcommand{\noindentFR}[1]{\vspace{5mm}\noindent {\bf #1}\\}

\newcommand{\deltapt}{p_{T,VV}}
\newcommand{\ptvmin}{p_{T,V}^{(min)}}
\newcommand{\thetastar}{\theta^{*}}
\newcommand{\aq}{a_{q}^{(3)}}
\newcommand{\ptv}{p_{T,V}}
\newcommand{\rf}[1]{{\color{purple} #1}} 
\newcommand{\com}[1]{{\color{red} \sffamily (#1)}} 
\newcommand{\TeV}{\,\mathrm{TeV}}
\newcommand{\GeV}{\,\mathrm{GeV}}
\newcommand{\MeV}{\,\mathrm{MeV}}
\newcommand{\keV}{\,\mathrm{keV}}
\newcommand*{\ipb}{\mbox{pb$^{-1}$}}
\newcommand{\pb}{\,{\rm pb}^{-1}}
\newcommand{\ab}{\,{\rm ab}^{-1}}
\newcommand{\Br}{{\mathrm{Br}}}

\newcommand\TwoFigBottom{-2}
\newcommand{\cg}{c_{ggh}}
\newcommand{\cgg}{c_{gghh}}
\newcommand{\ctt}{c_{tt}}
\newcommand{\ct}{c_{t}}
\newcommand{\cb}{c_{b}}

\newcommand{\mpt}{{{\cancel p}_T}}
\newcommand{\mptvec}{{\cancel{\vec p}}_T}


The $\hh \to \bbww$ final state has the second largest branching fraction, providing desirable statistics and leaving much flexibility to consider all its different sub-channels, depending on the $W$ decay mode: fully hadronic, semi-leptonic and di-lepton final states. In addition, in the fully hadronic and di-lepton state, this channel has the same final state objects as \bbzz, which could provide additional sensitivity.
Given the large statistics, the \hhbbvv channel, where $V$ is either $W$ or $Z$, is quite important, necessitating a careful study. However, it has been relatively overlooked, mostly due to the large \ttbar background. The current outlook for the non-resonant \hhbbvv channel is challenging and provides ample opportunity for improvement. In this section, we summarise the current experimental status and explore potential solutions to improve sensitivity in this channel.

Double Higgs production could also be used as a probe of a new scalar particle $S$, ubiquitous in many well-motivated extensions of the SM \cite{Haber:1984rc,Branco:2011iw}. The new scalar can mix with the Higgs boson acquiring couplings with the SM particles. If the $S$ mass is larger than twice the Higgs mass, $S$  can decay into two Higgs bosons, and it manifests as a resonance in the $HH$ invariant mass. On the other hand, if $S$ is lighter than twice the Higgs mass, the resonant double-Higgs production is forbidden. In this particular scenario, the mixed non-resonant $H S$ production \cite{Chen:2017qcz} provides an alternative window to search for an evidence of new physics. The $S$ boson will dominantly decay into two on-shell $W$ or $Z$ bosons. Therefore, for both mass regimes, the \hhbbvv channel is ranked high in terms of branching fractions, with a higher priority of the \hhbbww decay chain. 


ATLAS has reported results of a search for Higgs boson pair production where one Higgs boson decays via \hbb, and the other decays via \hww with subsequent decays of the $W$ bosons into $\ell\nu q\bar{q}$, where $\ell$ is either an electron or a muon~\cite{Aaboud:2018zhh}. One of the $W$ bosons is off-shell. The small contamination from leptonic $\tau$ decays is not explicitly vetoed in the analysis. CMS has reported results in the final state with two leptons such that it is sensitive to both \bbww and \bbzz channels, again with one of the gauge bosons being off-shell~\cite{Sirunyan:2017guj}.

\subsection{\hhbbww$(\ell\nu q\bar{q}$)}
\label{sec:bbww_strategies}
For the analysis of this channel data were collected using a set of single lepton triggers (triggers requiring the presence of at least one high \pT electron or muon) with increasing lepton \pT thresholds through the data taking as function of the instantaneous luminosity, in order to keep the total event rate below the requirements of the data acquisition system. Events are required to contain at least one reconstructed electron or muon matching a trigger-lepton candidate. In order to ensure that the leptons originate from the interaction point, requirements on the transverse and longitudinal impact parameters of the leptons relative to the primary vertex are imposed. 


Four different event selections have been optimised for: non-resonant \hh, resonant \hh production for resonance masses below 600~GeV, in the 600-1500~GeV mass range, and above 1500~GeV. For the latter category the \hbb candidates are reconstructed as large-radius jets and identified with boosted reconstruction technique described in Sec.~\ref{sec:hbbbosted}. 
The \bjets are identified using operating points such that the $b$-tagging efficiency is 85\% and 77\% in the resolved and boosted case respectively. 

The dominant background process for both boosted and resolved searches is the top-quark background, which ranges from more than 50\% to 90\% depending on the kinematic regime.
All the background processes are estimated from simulation, except the normalisation of the \ttbar process and the multi-jet background which are derived from data. 
Exploiting kinematic constraints, in particular the masses of the $W$ and Higgs bosons, each event can be fully reconstructed despite the presence of one neutrino in the final state.  

\begin{figure}
\begin{center}
\includegraphics[width=0.49\textwidth]{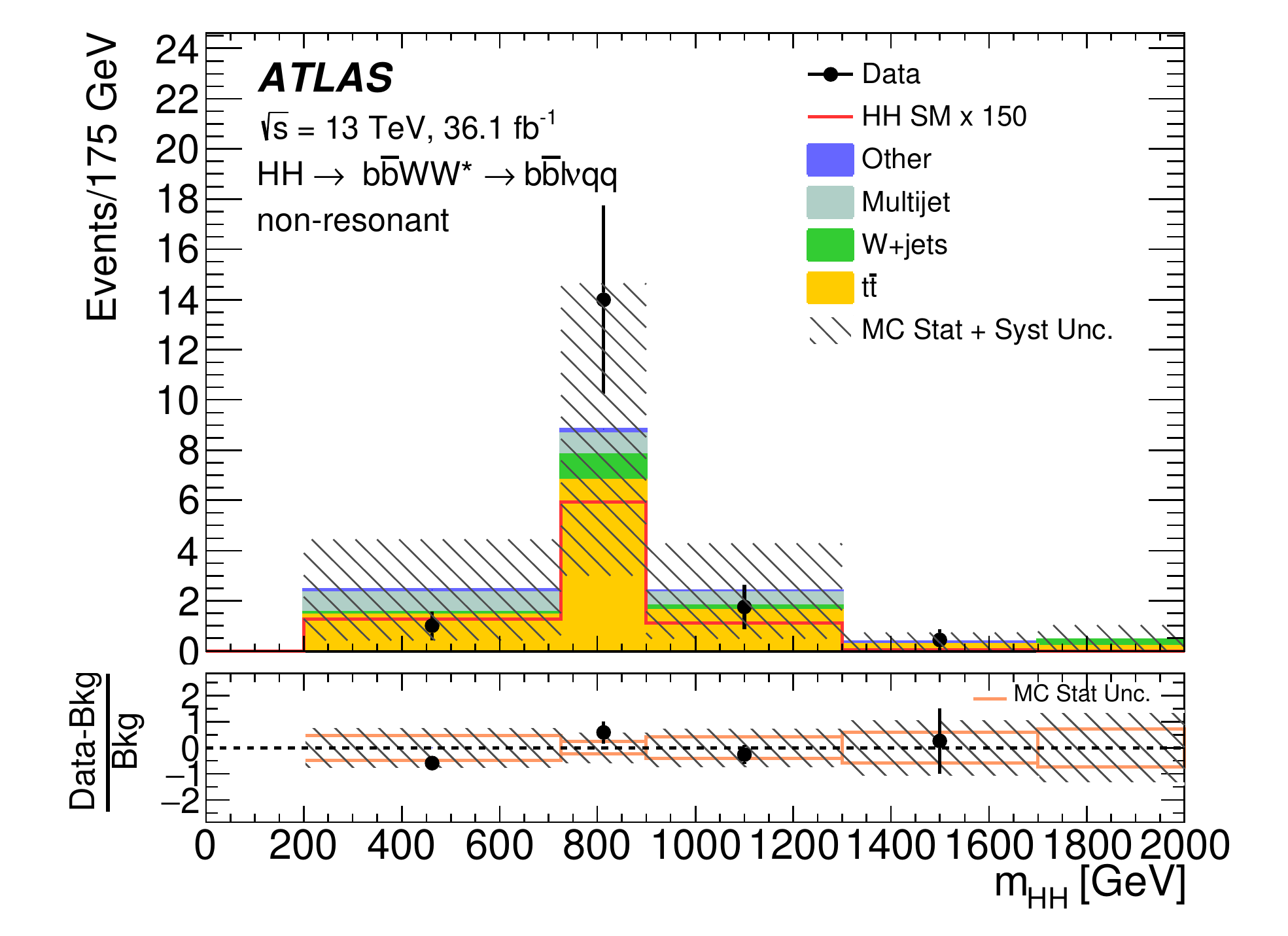}
\includegraphics[width=0.49\textwidth]{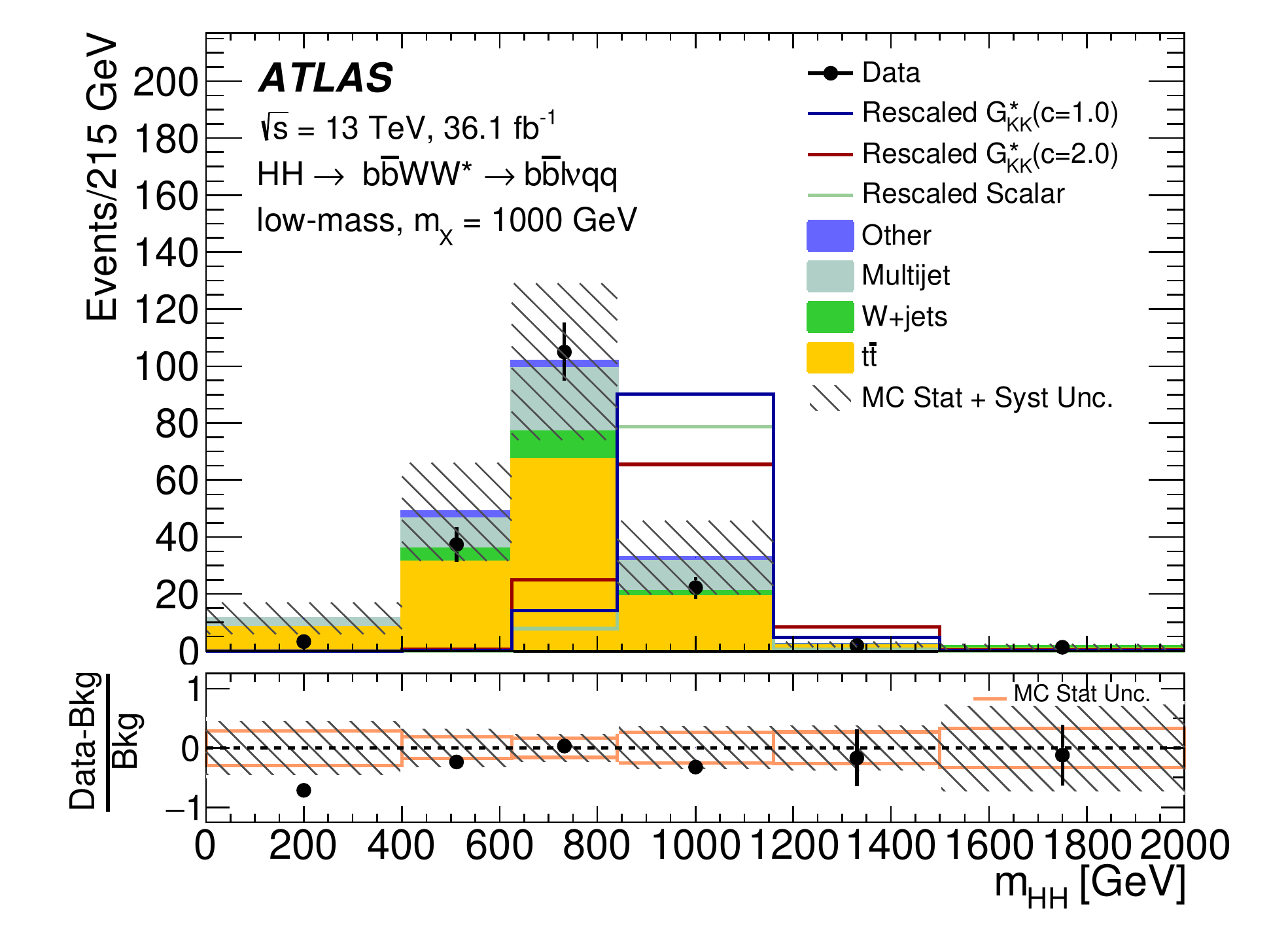}
\end{center}
\vspace*{-0.5cm}
\caption{Distributions of \mhh for the non-resonant \hh search (left) and for the search of a resonance (right) using the selections of the resolved analysis. The lower panel shows the fractional difference between data and the total expected background with the corresponding statistical and total uncertainty. The signal distributions are scaled arbitrary for presentation~\cite{Aaboud:2018zhh}.}
\label{fig:mhh_1}
\end{figure}

The invariant mass of the \hh system (\mhh) after applying all selection requirements for the resolved analysis is shown in Fig.~\ref{fig:mhh_1}.
Data are generally found to be in good agreement with the expected background predictions within the total uncertainty. 
The dominant systematic uncertainties for the resolved regime are \ttbar modelling (18\%), flavour tagging (30\%), JES/JER (20\%) and data samples in control regions (60\%).
In the resolved analysis, a counting experiment is performed after applying all selection requirements, which include a requirement on \mhh in the searches for resonant \hh production. In the boosted analysis, the fully reconstructed \mhh shape is used to extract the signal. 

The resolved and boosted analyses have non trivial overlap of events. In fact, a set of energy deposits in the calorimeter can be reconstructed both as two separate jets and one large-radius jet. 
The expected limit in the boosted analysis is higher than that from the resolved analysis for masses greater than 1300~GeV in the case of the scalar interpretation, and for masses greater than 800~GeV in the spin-2 hypothesis. 

For the non-resonant signal hypothesis the observed (expected) upper limit on the $\sigma(pp \to \hh) \times {\mathcal{B}}(\hh \to \bbww)$ at 95\% CL is:
\[
\sigma(pp \to \hh) \times {\mathcal{B}}(\hh \to \bbww) < 2.5 \, \left
  (2.5^{+1.0}_{-0.7} \right )  \,
{\mathrm{pb}},
\]
which corresponds to 300~(300$^{+100}_{-80}$) times the cross section predicted by the SM.

These results from ATLAS are dominated by large backgrounds and associated systematic uncertainties. Additionally, further optimization of the trigger and the computation of the neutrino longitudinal momentum are needed. It will also be interesting to see how the sensitivity improves when adding fully-hadronic and di-lepton channels. Finally, techniques discussed in Sec.~\ref{sec:bbww_prospective} and multivariate analysis also appear promising and should definitely be explored in the next iteration of this search. 

\subsection[\hhbbvv($\ell\nu\ell\nu$)]{\hhbbvv($\ell\nu\ell\nu$) \\
\contrib{B.~Di~Micco, S.~Shrestha}} 
For the di-lepton analysis, data were collected in Run 2 with a set of di-lepton triggers\footnote{Di-lepton triggers require the presence of two leptons at level 1 and at HLT.} with asymmetric \pT thresholds. Events with two oppositely charged leptons are selected using asymmetric \pT requirements, chosen to be above the corresponding trigger thresholds, for leading and subleading leptons of 25~GeV and 15~GeV for $ee$ and events with one electron and one muon where the muon has a higher \pT than the electron ($\mu e$), 20~GeV and 10~GeV for $\mu \mu$ events, and 25~GeV and 10~GeV for events with one electron and one muon where the electron has a higher \pT than the muon ($e\mu$). Electrons in the pseudo-rapidity range $|\eta| < 2.5$ and muons in the range $|\eta| < 2.4$ are considered.

Jets are required to be separated from a selected lepton by a distance of $\Delta R > 0.3$ and are considered to be $b$-tagged if they pass the working point of the algorithm at which the efficiency is 70\%, see Sec.~\ref{sec:bTagging}.

The top-quark background is the single most-dominant background, which accounts for almost 85--90\% of the total background and is estimated from simulation. The Drell-Yan production in the same flavour channels amounts to 7--10\% of the total and is estimated with data-driven techniques. Other backgrounds have almost negligible contribution. 

Deep neural network (DNN) discriminators are used to improve the signal to background separation. As the dominant background process (\ttbar production) is irreducible, the DNNs rely on information related to event kinematics. The variables provided as input to the DNNs exploit the presence in the signal of two Higgs bosons decaying into two \bjets on one side, and two leptons and two neutrinos on the other, which results in different kinematics for the di-lepton and di-jet systems between signal and background processes. 
Two parameterised DNNs are trained: one for the resonant search and one for the non-resonant search. In order to extract the best fit signal cross sections, a binned maximum likelihood fit is performed using templates built from the DNN output distributions
in the three \mjj regions and in the three channels ($e^{+}e^{-}$, $\mu^{+}\mu^{-}$, and $e^{\pm}\mu^{\pm}$).


The major uncertainty source is the top-background modelling (5--13\%), followed by simulated sample size (up to 20\%).
The results obtained by CMS are in agreement, within uncertainties, with the SM predictions. 
For the SM \hh hypothesis, the data exclude a product of the cross section and branching ratio of 72~fb, corresponding to 79 times the SM prediction. The expected exclusion is $81^{+42}\rm_{-25}$~fb, corresponding to $89^{+47}\rm_{-28}$ times the SM prediction. 

ATLAS has also presented preliminary results in this channel \cite{ATLAS_EPS} using an integrated luminosity of 139 fb$^{-1}$, only for the non-resonant signal model.
The analysis follows a similar approach of the CMS analysis using a DNN to separate signal from \ttbar, $Z\to e^+e^-, \mu^+ \mu^-$ ($Z\to ll$) and $Z\to \tau^+ \tau^-$. The DNN produces four outputs: $p_{HH}, p_{\ttbar}, p_{Z\to ll}$ and $p_{Z\to \tau^+ \tau^-}$.
The four DNN outputs are combined in a single variable using the relation:
\[
d_{HH} = ln\left(\frac{p_{HH}}{p_{Z\to ll} + p_{Z \to \tau^+ \tau^-} + p_{\ttbar}}\right)
\]


The observed (expected) results are upper limits at 95\% CL equal to  40 (29) times the SM cross section that, when the different integrated luminosity is taken into account, are slightly better than the CMS results.

The results from CMS are dominated by large \ttbar background and associated systematic uncertainties, that already exceed the statistical precision. It will be interesting to see how sensitivity improves when adding fully hadronic and single lepton \hww final states. Finally, techniques discussed in Section~\ref{sec:bbww_prospective}, in addition to the already used multivariate analysis, also appear to be  promising and should be explored in the next iteration of the analysis. 

%% file: bbVV/bbww.tex
\subsection[New kinematic observables for \hhbbww]{New kinematic observables for \hhbbww
\\ \contrib{V.~D'Amico,~B.~Di~Micco,~J.H.~Kim,~K.~Kong,~K.~T.~Matchev,~M.~Park}}
\label{sec:bbww_prospective}
The sensitivity to double Higgs boson production in the \hhbbww final state, where both \W bosons decay leptonically, could be improved by the use of two novel kinematic observables, {\it Topness} and {\it Higgsness}~\cite{Kim:2018cxf,Kim:2019wns,Graesser:2012qy}. These functions, which could be generalised to other final states as well, capture features of the dominant \ttbar background and the \hh signal events, respectively, and result to be effective in separating these two different event topologies. For the \hhbbww$(\ell \nu \ell \nu)$ final state other two observables are combined, the subsystem $M_{T2}$ (or subsystem $M_2$)~\cite{Lester:1999tx,Burns:2008va,Barr:2011xt} for \ttbar production and the subsystem $\sqrt{\hat {s}}_{\text{min}}$ (or subsystem $M_1$)~\cite{Konar:2008ei,Konar:2010ma,Barr:2011xt} for \hh production. 
The $M_{T2}$ variable is defined as:
\begin{equation}
\label{eq:mt2}
    M^2_{T2} \equiv M^2_2 = \min\limits_{\cancel{p}_{T1}+\cancel{p}_{T2}=\cancel{p}_{T}}
\left[ \max \{ m^{2}_{T}\left(p_{Tl^{-}},\cancel{p}_{1} \right), m^{2}_{T}\left(p_{Tl^{+}},\cancel{p}_{2} \right) \} \right]
\end{equation}
where $\cancel{p}_{T1}$ and $\cancel{p}_{T2}$ are the neutrino transverse momenta, $\cancel{p}_T$ is the the measured missing transverse momentum, $\cancel{p}_1$ and $\cancel{p}_2$ are the neutrino four-momenta. The minimisation is performed on the eight components of the two neutrino four momenta with the constraint that the sum of their transverse momenta is equal to the measured missing transverse momentum.
The $M_1$ variable is defined as:
\begin{equation}
\label{eq:m1}
    M_1 =  \sqrt{M^2_{\rm vis} + |\vec{p}_T|^2} + |{\cancel{\vec p}_T}| \quad M^2_{\rm vis} = E^2_{\rm vis} - |\vec{p}_{T\, \mathrm{vis}}|^2 - p_{z\,\mathrm{vis}}^2  
\end{equation}
where $E_{\rm vis}$ and $\vec{p}_{T, \mathrm{vis}}$ are the sum of the energy and of the transverse momenta of all visible particles respectively.

\begin{figure}
    \centering
   \includegraphics[width=0.6\textwidth]{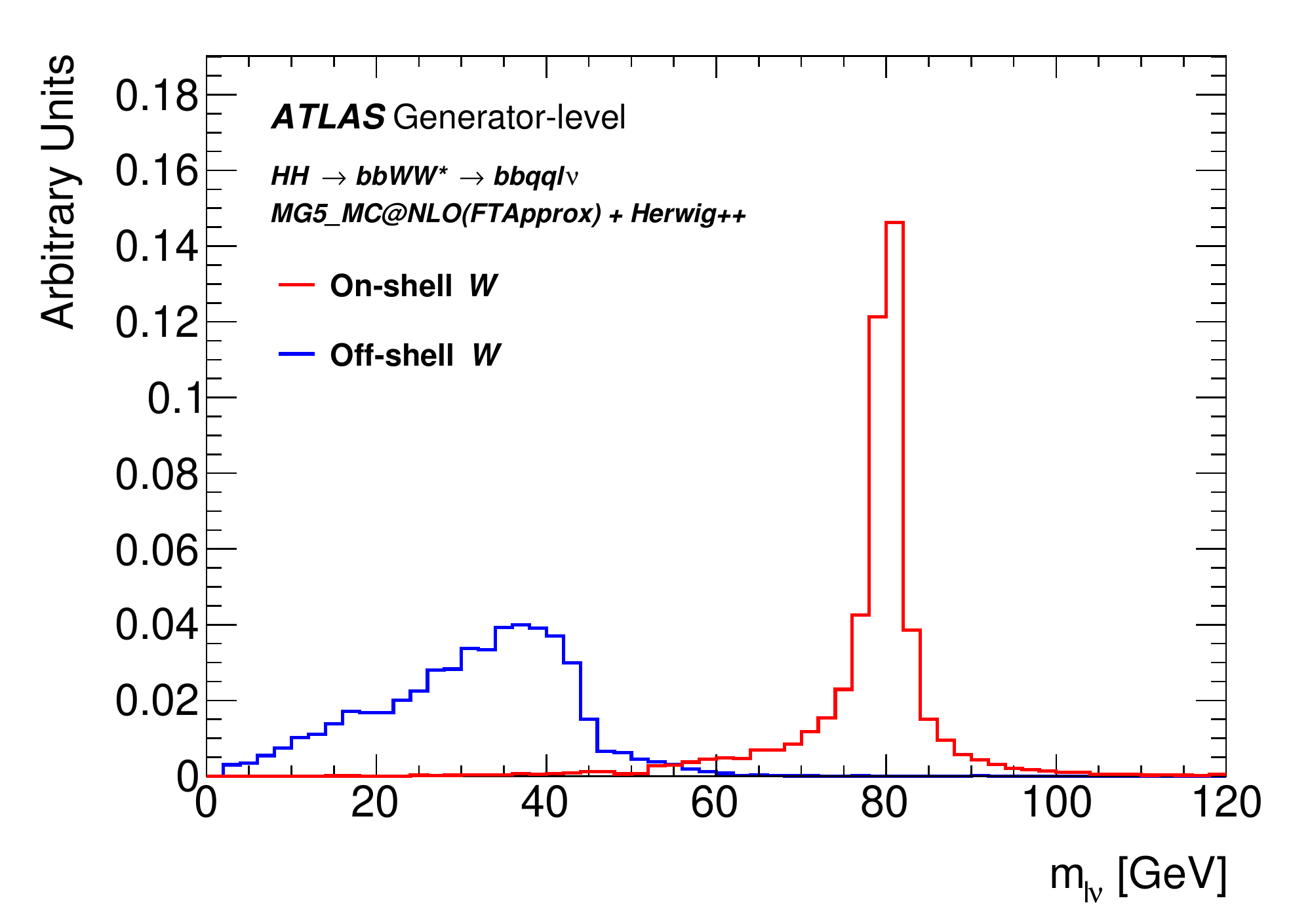}
    \caption{
    Distribution of the invariant mass of the lepton-neutrino system for simulated signal $\hh \to \bbWW$ events without selection requirements, separated for the on-shell and the off-shell W boson. Signal sample generated with {\sc MG5\_aMC@NLO} using the FTApprox approximation and with a {\sc Herwig++}  parton-shower simulation. The distributions are normalised to unit area~\cite{ATL-PHYS-PUB-2019-040}.}
    \label{fig:mwwst}
\end{figure}

The {\it Topness} variable quantifies the degree of consistency of the event kinematic with the di-lepton \ttbar production, with six unknowns (the three-momenta of the two neutrinos, $\vec p_{\nu}$ and $\vec p_{\bar\nu}$) and four on-shell constraints, $m_t$, $m_{\bar t}$, $m_{W^+}$ and $m_{W^-}$. An estimate of the neutrino momenta can be obtained by minimising the following quantity:
\begin{equation}\label{eq:tt} 
\chi^2_{ij} \equiv \min_{\tiny \mptvec = \vec p_{\nu T} + \vec p_{ \bar\nu T}}  \Bigg[ 
\frac{\left ( m^2_{b_i \ell^+ \nu} - m^2_t \right )^2}{\sigma_t^4}    +
\frac{\left ( m^2_{\ell^+ \nu} - m^2_W \right )^2}{\sigma_W^4}    
+ \frac{\left ( m^2_{b_j \ell^- \bar \nu} - m^2_t \right )^2}{\sigma_t^4}  + \frac{\left ( m^2_{\ell^- \bar\nu} - m^2_W \right )^2}{\sigma_W^4}   \Bigg]  
\end{equation}
subjected to the missing transverse momentum constraint, $\mptvec = \vec p_{\nu T} + \vec p_{ \bar\nu T}$. 
Since there is a two-fold ambiguity in the pairing of a $b$-quark and a lepton, {\it Topness} is defined as the smaller of the two $\chi^2$:
\begin{eqnarray}
\label{eq:T}
T &\equiv&  { \min} \left ( \chi^2_{12} \, , \, \chi^2_{21} \right ) \, .
\end{eqnarray}
In double Higgs boson production, a selection on the invariant mass \mbb is used to identify \hbb candidates and to reduce the SM backgrounds. {\it Higgsness} characterises the decay of the other Higgs boson, $\hww \to \ell^+ \ell^- \nu \bar\nu$. It is defined as follows: 
\begin{eqnarray}
H &\equiv&    {\rm min} \left [
 \frac{\left ( m^2_{\ell^+\ell^-\nu \bar\nu} - \mh^2 \right )^2}{\sigma_{h_\ell}^4}    \right. 
 + \frac{ \left ( m_{\nu  \bar\nu}^2 -  m_{\nu\bar\nu, \text{peak}}^2 \right )^2}{ \sigma^4_{\nu}}
  +  {\rm min} \left ( 
\frac{\left ( m^2_{\ell^+ \nu } - m^2_W \right )^2}{\sigma_W^4} + \right. \nonumber \\ 
& + & \left. \frac{\left ( m^2_{\ell^- \bar \nu} - m^2_{W^*, \text{peak}} \right )^2}{\sigma_{W_*}^4}  \, , \right. 
 \left.  \left .
\frac{\left ( m^2_{\ell^- \bar \nu} - m^2_W \right )^2}{\sigma_W^4} + 
\frac{\left ( m^2_{\ell^+ \nu} - m^2_{W^*, \text{peak}} \right )^2}{\sigma_{W_*}^4}  
\right )  \right ]   \, ,  \label{eq:hww} 
\end{eqnarray}
where $m_{W^*}$ is the invariant mass of the lepton-neutrino pair coming from the off-shell $W$. The $m_{W^*}$ distribution has an end-point at around $ \mh - m_\mathrm{W}$  (see Fig.~\ref{fig:mwwst} for the $\ell \nu qq$ events which yield to results similar to $\ell \nu \ell \nu$ events), and its peak is located at 
\begin{eqnarray}
m_{W^*}^{\text{peak}} = \frac{1}{\sqrt{3}} \sqrt{ 2 \left ( \mh^2 + m_W^2 \right ) - \sqrt{\mh^4 + 14 \mh^2 m_W^2 + m_W^4}} \;.
\end{eqnarray}
\begin{figure}[t]
\centering
\includegraphics[width=6.cm]{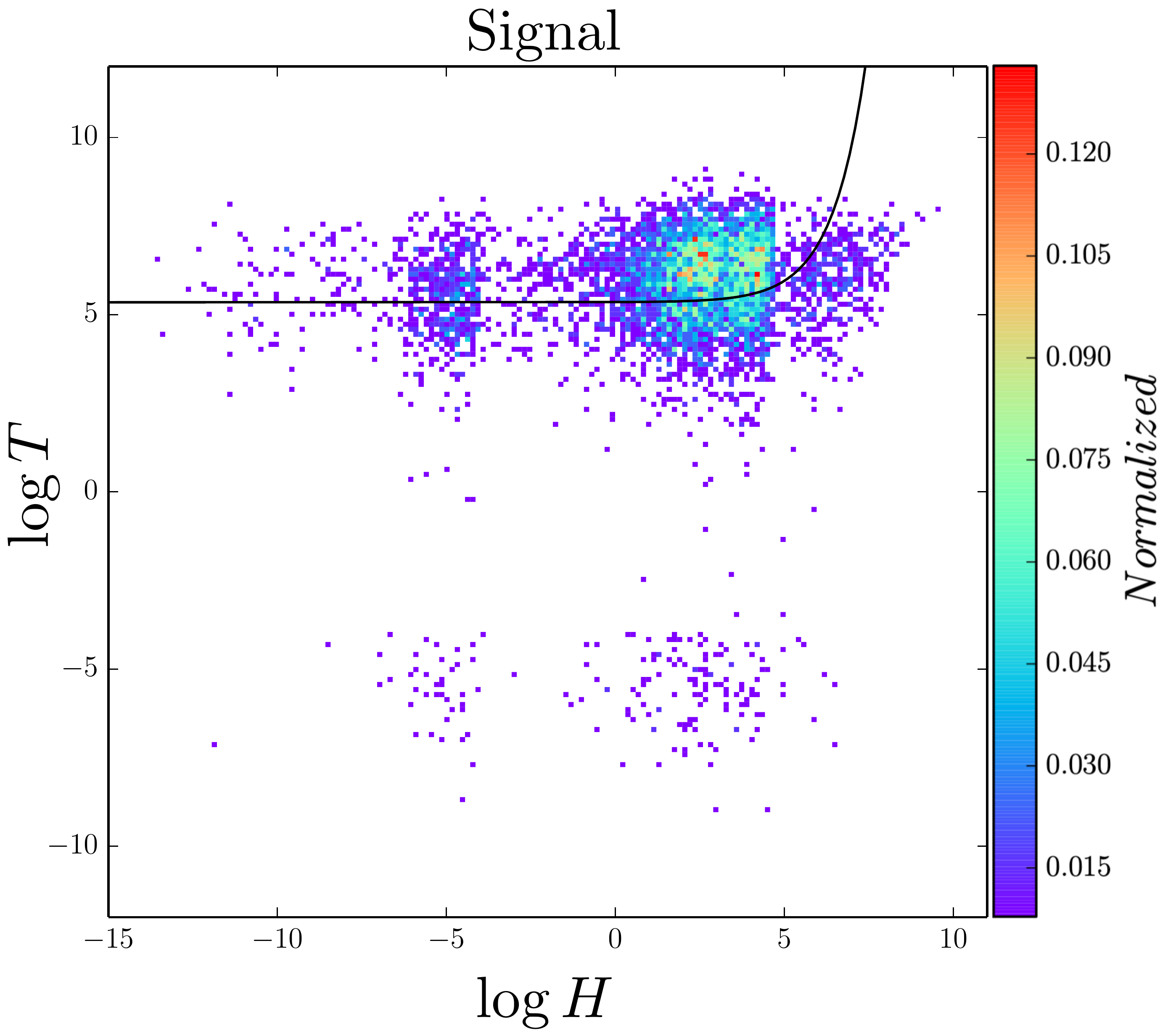}   \hspace*{-0.1cm}
\includegraphics[width=6.cm]{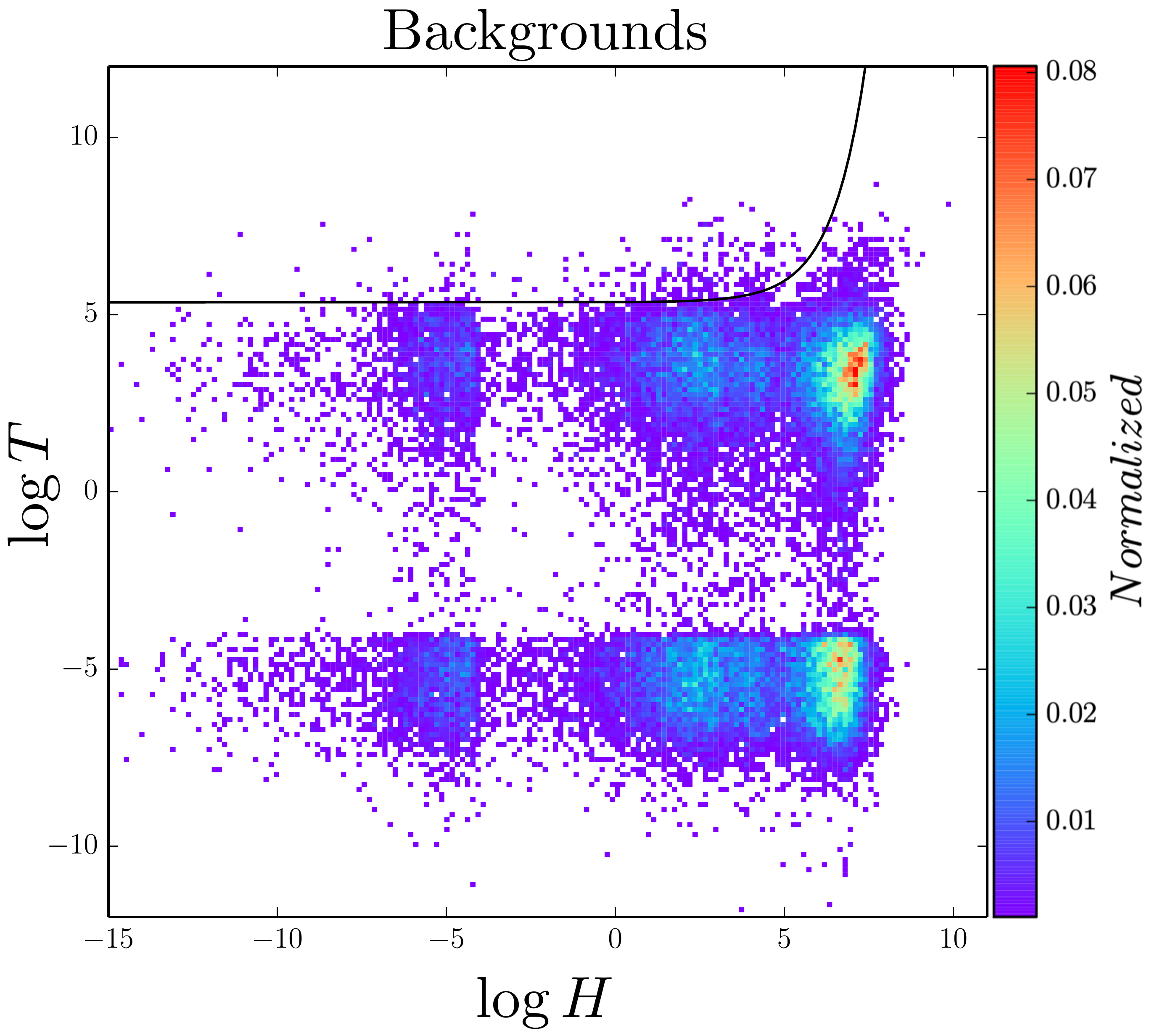} 
\caption{\label{fig:scatter} 
 Distribution of ($\log H$, $\log T$) for simulated signal (\hh) and backgrounds (\ttbar, $\ttbar{H}$,$\ttbar{V}$, $\ell\ell b j$, \bbtautau and others) events after loose selection requirements as defined in Ref.~\cite{Kim:2018cxf}.}
\end{figure}
Note also that $m_{\nu\bar\nu}^{\text{peak}} = m_{\ell\ell}^{\text{peak}} \approx 30$~GeV is the location of the peak in the $\textrm{d}\sigma/\textrm{d} m_{\nu\bar\nu}$ or $\textrm{d}\sigma/\textrm{d} m_{\ell\ell}$ distribution~\cite{Kim:2018cxf,Cho:2012er}. The $\sigma$ parameters in Eq.~(\ref{eq:tt}) and (\ref{eq:hww}) stand for the experimental uncertainties and intrinsic particle widths. In principle, they can be treated as free parameters, and tuned by a  neutral network or a boosted decision tree. For the studies shown in the following the values $\sigma_t=5$~GeV,  $\sigma_W=5$~GeV,  $\sigma_{W^*}=5$~GeV,  $\sigma_{h_\ell}=2$~GeV, and $\sigma_\nu = 10$~GeV have been used. 
%

The {\it Higgsness} and {\it Topness} distributions are shown in Fig.~\ref{fig:scatter} for simulated signal and all backgrounds (\ttbar, $\ttbar{H}$, $\ttbar{V}$, $\ell\ell b j$, \bbtautau and others) events.  
Simulated signal and background events include for parton shower and hadronisation simulation, as well as semi-realistic detector effects, as described in Ref.~\cite{Kim:2018cxf,Kim:2019wns}

The dominant \ttbar events are expected to be on the lower right corner with smaller {\it Topness} and larger {\it Higgsness}. The \hh events are, on the other hand, expected to have smaller {\it Higgsness} and larger {\it Topness}. A selection in the ($\log H$, $\log T$) is then used to separate signal and backgrounds. 

\begin{figure*}[t]
\centering
\includegraphics[width=6.2cm]{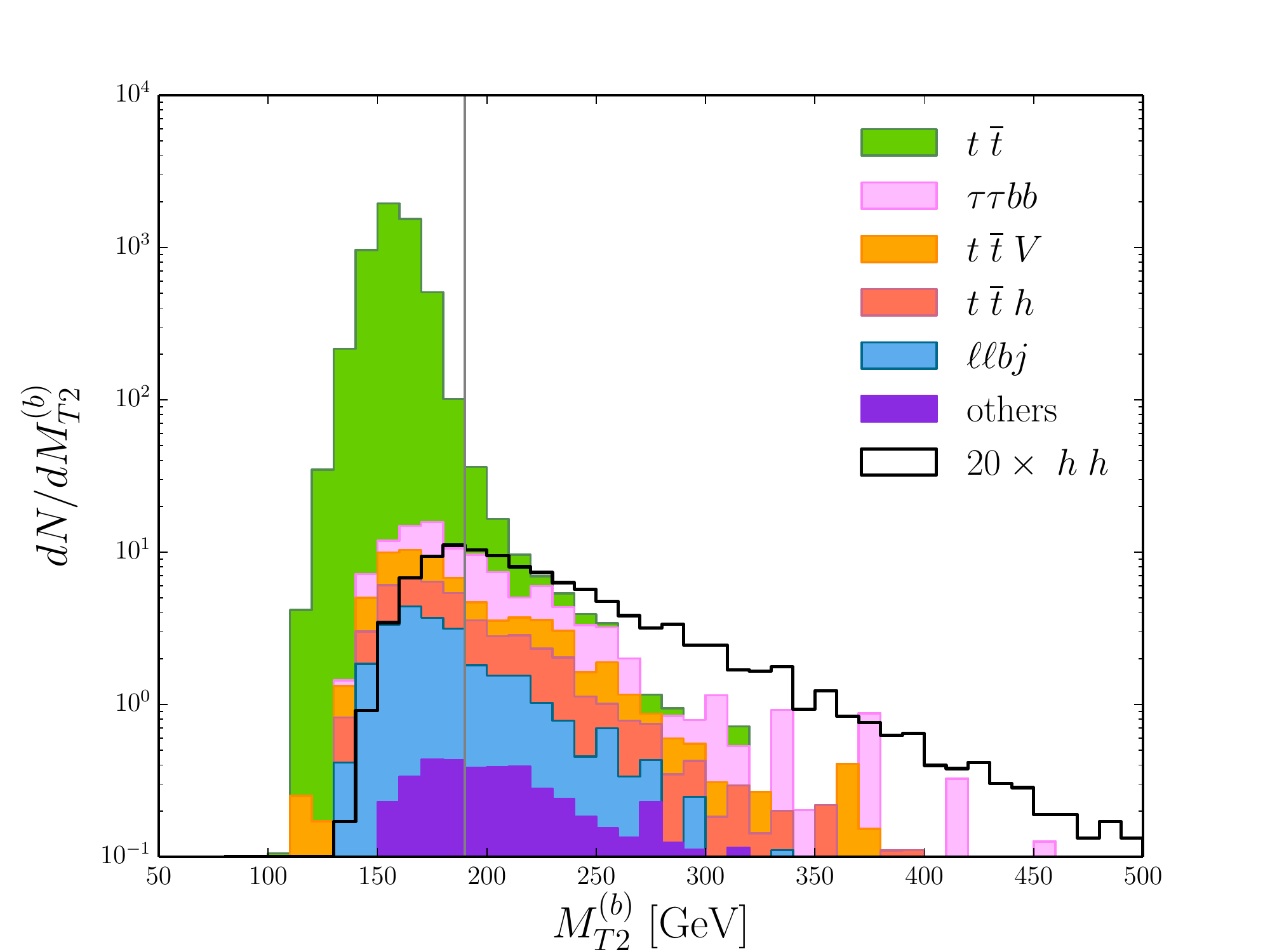}  \hspace*{-0.6cm}
\includegraphics[width=6.2cm]{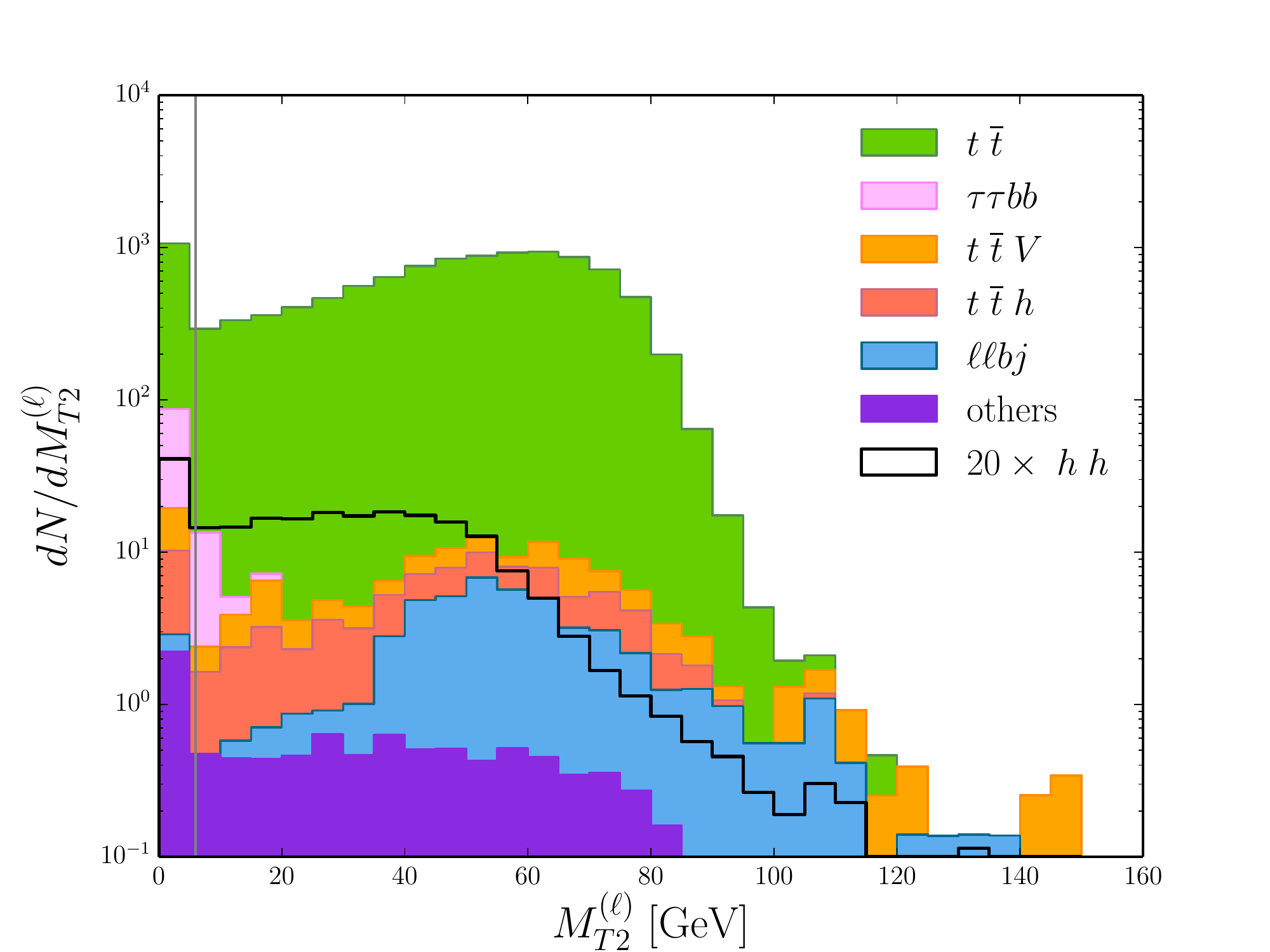} \\ \hspace*{-0.7cm}
\includegraphics[width=6.2cm]{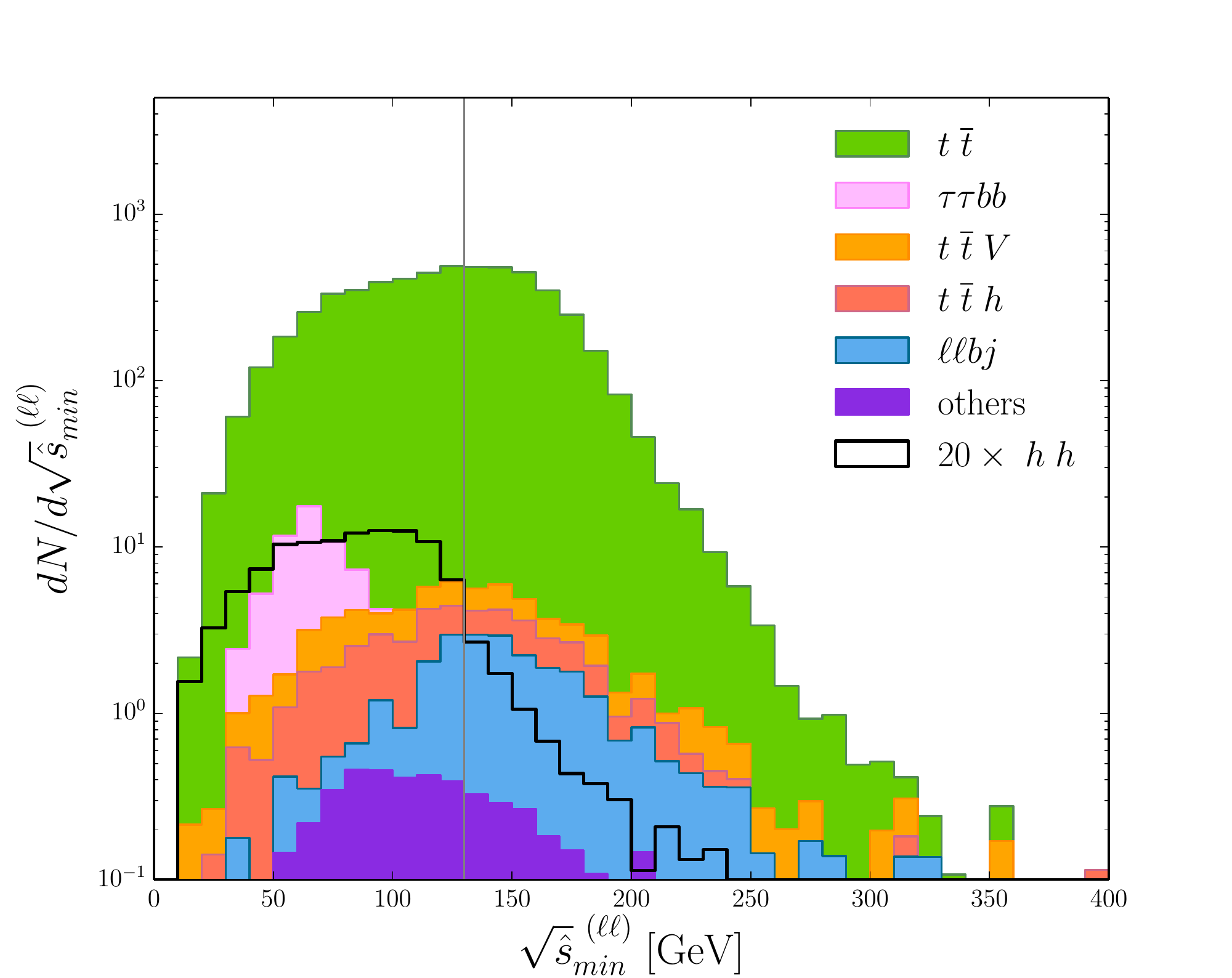}  \hspace*{-0.3cm}
\includegraphics[width=5.3cm]{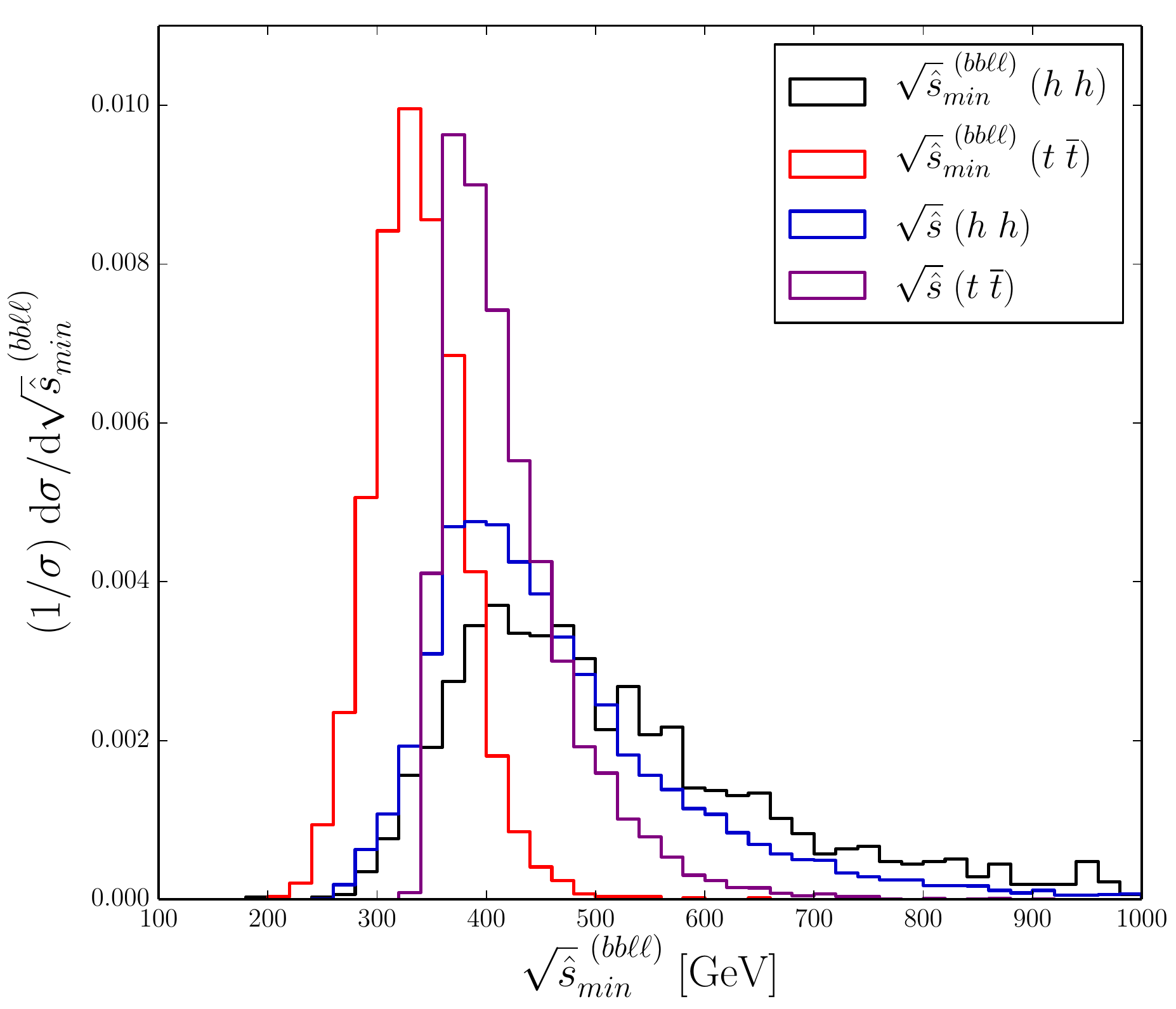}
\caption{\label{fig:newcuts} Distributions of $M_{T2}^{(b)}$, $M_{T2}^{(\ell)}$ and $\sqrt{\hat{s}}_{\text{min}}^{(\ell\ell)}$ for the signal (\hh) and all backgrounds (\ttbar, $\ttbar{H}$, $\ttbar{V}$,  $\ell\ell b j$, \bbtautau and others) events~\cite{Kim:2018cxf,Kim:2019wns}. 
The vertical lines at $M_{T2}^{(b)} = 190$~GeV, $M_{T2}^{(\ell)}= 6$~GeV and $\sqrt{\hat{s}}_{\text{min}}^{(\ell\ell)}=130$~GeV show the optimised cuts. 
The lower-right panel shows the distributions of $\sqrt{\hat{s}}_{\text{min}}^{(bb\ell\ell)}$ as defined in Eq.~(\ref{eq:smin}) as compared to the $\sqrt{\hat{s}}\equiv M_1$ for the \hh and \ttbar events.}
\end{figure*}

Along with {\it Higgsness} and {\it Topness}, the $M_{T2}$ variable, Eq.~(\ref{eq:mt2}) could be exploited for both the \hbb ($M_{T2}^{(b)}$) and leptonic ($M_{T2}^{(\ell)}$)~\cite{Burns:2008va} candidates, as well as $\hat{s}_{\text{min}}^{(\ell\ell)}$ for $\hww \to \ell^+\ell^- \nu \bar \nu$~\cite{Konar:2008ei,Konar:2010ma}. 
%
%
In the case of $M_{T2}^{(b)}$, the two \W bosons play the role of two missing neutrinos.
The $M_{T2}^{(b)}$ and $M_{T2}^{(\ell)}$ distributions are shown in Fig.~\ref{fig:newcuts} (upper panels). 
The vertical lines at $M_{T2}^{(b)} = 190$~GeV and $M_{T2}^{(\ell)}= 6$~GeV represent optimised cuts, suppressing \ttbar and \bbtautau (Drell-Yan) backgrounds, respectively.

The $ \hat{s}_{\text{min}}^{({\rm v})} $ variable~\cite{Konar:2008ei,Konar:2010ma,Barr:2011xt} is defined as:
\begin{eqnarray}\label{eq:smin}
 \hat{s}_{\text{min}}^{({\rm v})} = m_{{\rm v}}^2 + 2 \left ( \sqrt{ |\vec P_{T}^{\rm v} |^2 + m_{\rm v}^2 }  |\mptvec| - \vec P_{T}^{\rm v} \cdot \mptvec \right ) ,
\end{eqnarray}
where the script $({\rm v})$ represents a set of visible particles under consideration. The $m_{\rm v}$ and $\vec P_{T}^{\rm v}$ denote their invariant mass and transverse momentum, respectively.

The $ \hat{s}_{\text{min}}^{({\rm v})} $ variable provides the minimum value of the Mandelstam invariant mass $\hat s$ which is consistent with the observed visible four-momentum vector. Figure~\ref{fig:newcuts} (lower-left panel) demonstrates that the $\sqrt{\hat{s}}_{\text{min}}^{(\ell\ell)}$ distribution has an endpoint at around \mh for \hh events. All other backgrounds, however, extend above this point. This justifies the use of $\sqrt{\hat{s}}_{\text{min}}^{(\ell\ell)} <130$~GeV as a cut to reduce the backgrounds. Figure~\ref{fig:newcuts} (lower-right panel) shows distributions of $\sqrt{\hat{s}}_{\text{min}}^{(\bb\ell\ell)}$ and the true $\sqrt{\hat{s}}$ for \hh and \ttbar events. First, one can observe that $\sqrt{\hat{s}}_{\text{min}}^{(\bb\ell\ell)}(\hh)$ provides a good measure of the true $\sqrt{\hat{s}} (\hh)$, while $\sqrt{\hat{s}}_{\text{min}}^{(\bb\ell\ell)}(\ttbar)$ peaks lower, near the $2 m_t$ threshold. Secondly, both $\sqrt{\hat s}(\hh)$ and $\sqrt{\hat s }(\ttbar)$ peak at $\sim 400$ GeV. This implies that while the two top quarks are produced near threshold ($2~m_t$), the two Higgs bosons are produced well above the corresponding $2\mh$ threshold. Consequently, the two top  quarks are more or less at rest, while the two Higgs bosons are expected to be relatively boosted and their decay products tend to be more collimated. This observation motivates the use of simple kinematic variables such as $\Delta R_{\ell\ell}$, $\Delta R_{\bb}$, $m_{\ell\ell}$ and \mbb to further separate signal and background  events~\cite{Kim:2018cxf,Kim:2019wns}.

\begin{figure}
    \begin{center}
    \includegraphics[width=0.49\textwidth]{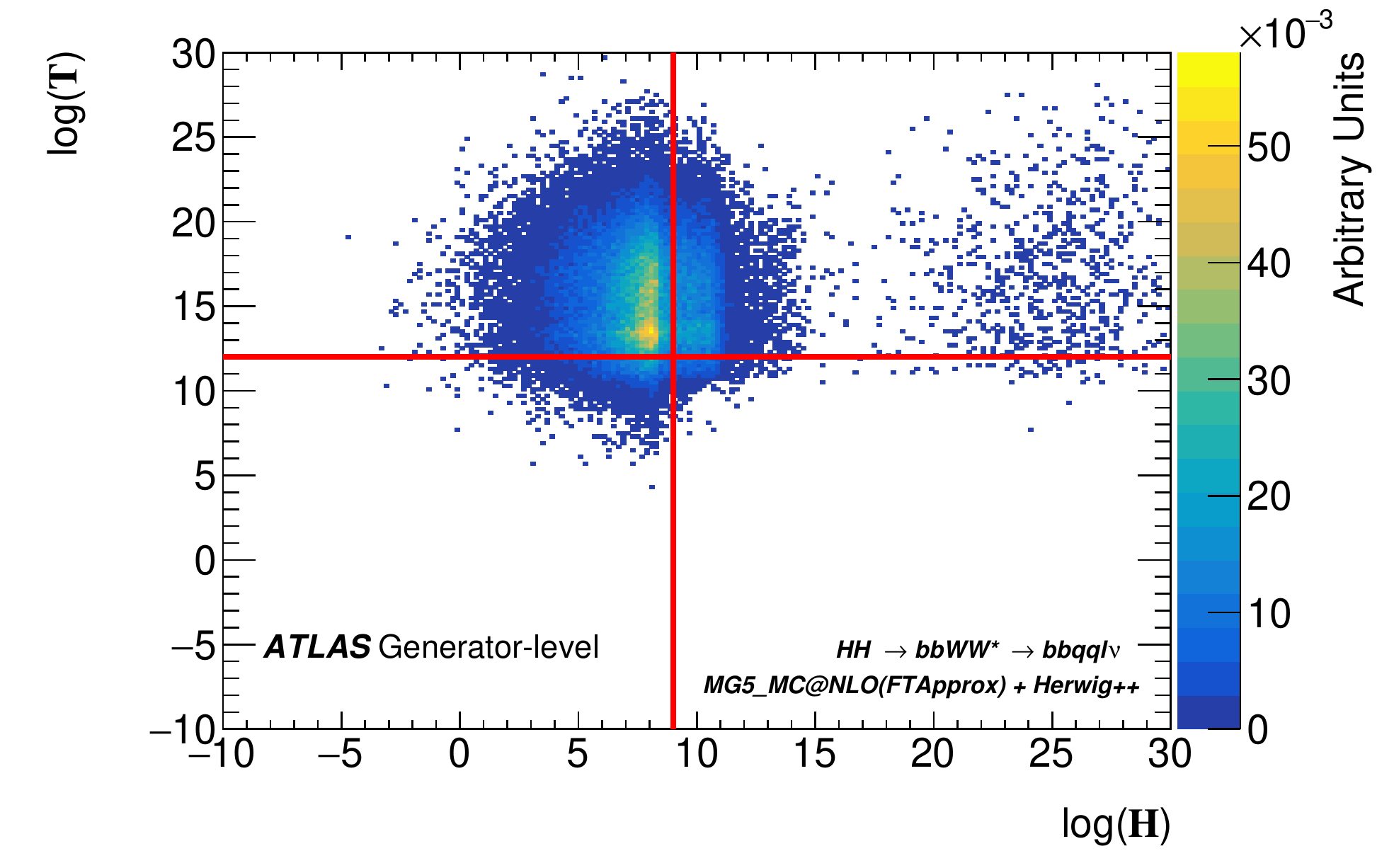}
    \includegraphics[width=0.49\textwidth]{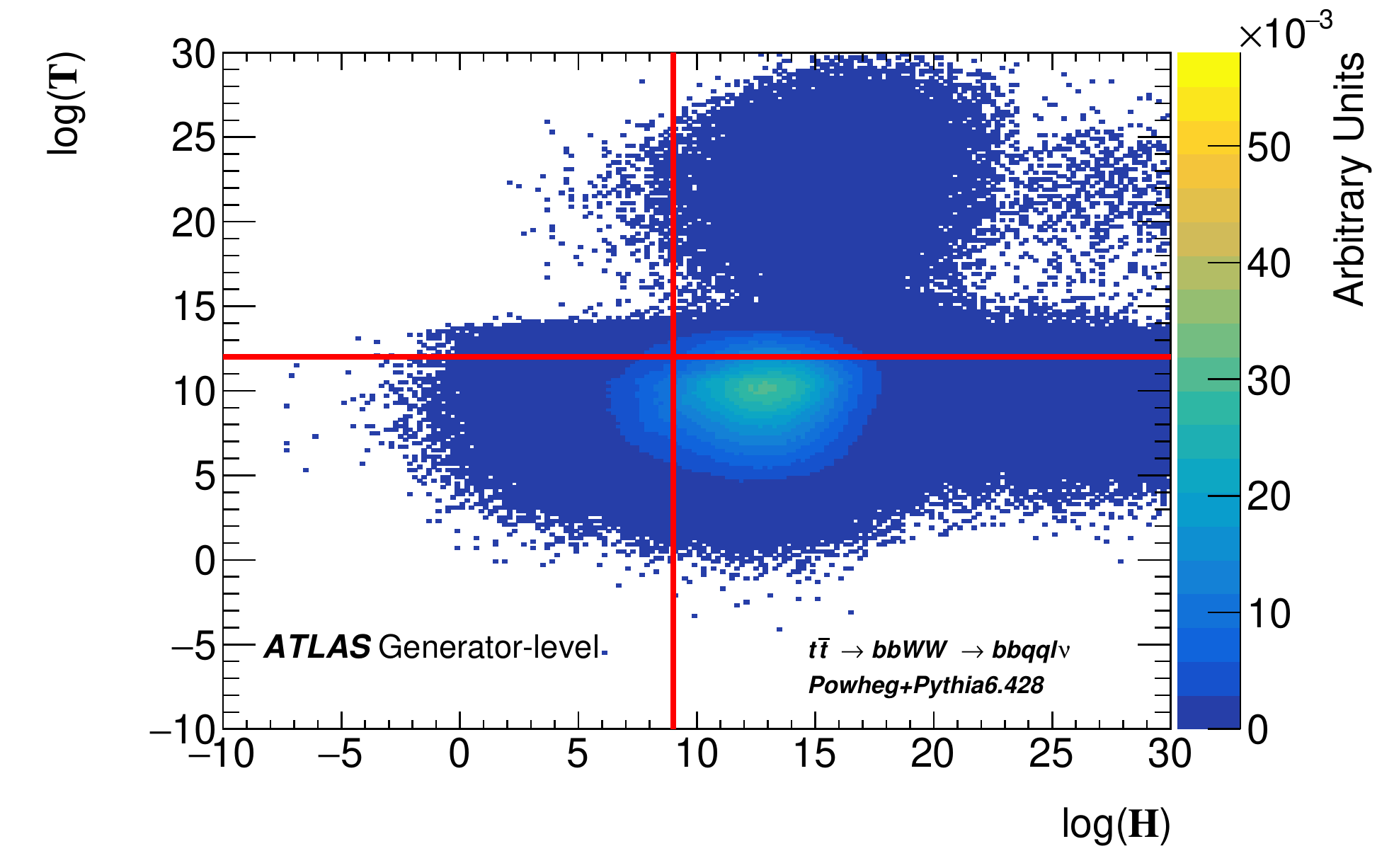}
    \caption{
    Distribution of {\it Higgsness} and {\it Topness} ($log(\textbf{H})$, $log(\textbf{T})$) for simulated signal $\hh \to \bbWW$ and background $\ttbar\to \rightarrow  bbqql\nu$ events without selection requirements. The signal has been simulated with {\sc MG5\_aMC@NLO} using the FTApprox approximation and with a {\sc Herwig++}  parton-shower simulation, while the background sample is generated with {\sc Powheg} and {\sc Pythia 6.428}. The distributions are normalised to unit area. Red lines are drawn to give a visible reference for a possible separation between signal and background~\cite{ATL-PHYS-PUB-2019-040}.} 
   \label{fig:HTJJ}
    \end{center}
\end{figure}

The new observables presented in this section are quite general and can be easily applied to 
different topologies. For the \hhbbww$(\ell \nu jj$) final state, the {\it Topness} variable is defined through Eq.~(\ref{eq:T}) where 
\begin{equation}
\chi^2_{ij} \equiv  \min\limits_{p_{z}^{\nu}}
\Bigg[\frac{\left(m_{b_{i}\ell\nu}^{2}-m_{t}^{2} \right)^2}{\sigma_{t}^{4}}
+\frac{\left(m_{\ell\nu}^{2}-m_{W}^{2} \right)^2}{\sigma_{W}^{4}} 
+\frac{\left(m_{b_{j}{jj}}^{2}-m_{t}^{2} \right)^2}{\sigma_{t}^{4}}
+\frac{\left(m_{ jj}^{2}-m_{W}^{2} \right)^2}{\sigma_{W}^{4}}\Bigg].
\end{equation}
In this expression $p_z^{\nu}$ is the longitudinal neutrino momentum, $b_1$ and $b_2$ are the \bjets in the final state, $jj$ is the di-jet system, $m_{b_i \ell\nu}$ is the invariant mass of the lepton, neutrino, \bjet system and $m_{{b_j}{ jj}}$ that of the \bjet plus di-jet system.
The {\it Higgness} is defined by the identity:
\begin{eqnarray}
    H &\equiv& \min\limits_{p_{z}^{\nu}}
\left[\frac{\left(m_{\ell\nu { jj}}^{2}-m_{h}^{2} \right)^2}{\sigma_{h}^{4}}
+\min \left( \frac{\left(m_{\ell\nu}^{2}-m_{W}^{2} \right)^2}{\sigma_{W}^{4}}  
+\frac{\left(m_{ jj}^{2}-m_{W_{\text{peak}}^{*}}^{2} \right)^2}{\sigma_{W^{*}}^{4}} ,\right.\right.
\\ \nonumber
&&\qquad \qquad \quad 
\frac{\left(m_{ jj}^{2}-m_{W}^{2} \right)^2}{\sigma_{W}^{4}}
\left.\left.+\frac{\left(m_{l\nu}^{2}-m_{W_{\text{peak}}^{*}}^{2} \right)^2}{\sigma_{W^{*}}^{4}}
\right)\right]\\ \nonumber
\end{eqnarray}
The distribution of the {\it Higgness} and {\it Topness} variables are shown in Fig.~\ref{fig:HTJJ} for simulated signal and \ttbar events.

%% file: HH_other_signatures/HH_other_signatures_main.tex
Searches for \hh production in channels without \bjets have in general smaller signal yields, but are typically less contaminated by backgrounds than those with \bjets.
As the sensitivity of searches without \bjets is mainly limited by statistical uncertainties, we expect that their sensitivity will scale better with the integrated luminosity than the \bjets final states.
ATLAS has recently investigated both the $\hh \to \yyww$~\cite{Aad:2015xja, Aaboud:2018ewm} and $\hh \to \wwww$~\cite{Aaboud:2018ksn} final states,
while CMS has studied for the first time $\hh \to \tautautautau$~\cite{SlidesHH4tau}.
The branching fractions of these channels for SM \hh bosons are $9.85 \cdot 10^{-4}$, $4.67 \cdot 10^{-2}$, and $4.00 \cdot 10^{-3}$, respectively.
Phenomenological studies of the $\hh \to \yyww$ and $\hh \to \wwww$ channels
have been published in Refs.~\cite{Adhikary:2017jtu} and~\cite{Baur:2002rb, Baur:2002qd, Li:2015yia, Ren:2017jbg, Adhikary:2017jtu}, respectively.


\subsection{$\hh\rightarrow \gamma\gamma W W^{*}$}

\begin{figure*}[ht!]
\centering
\includegraphics[height=0.46\textwidth,angle=270]{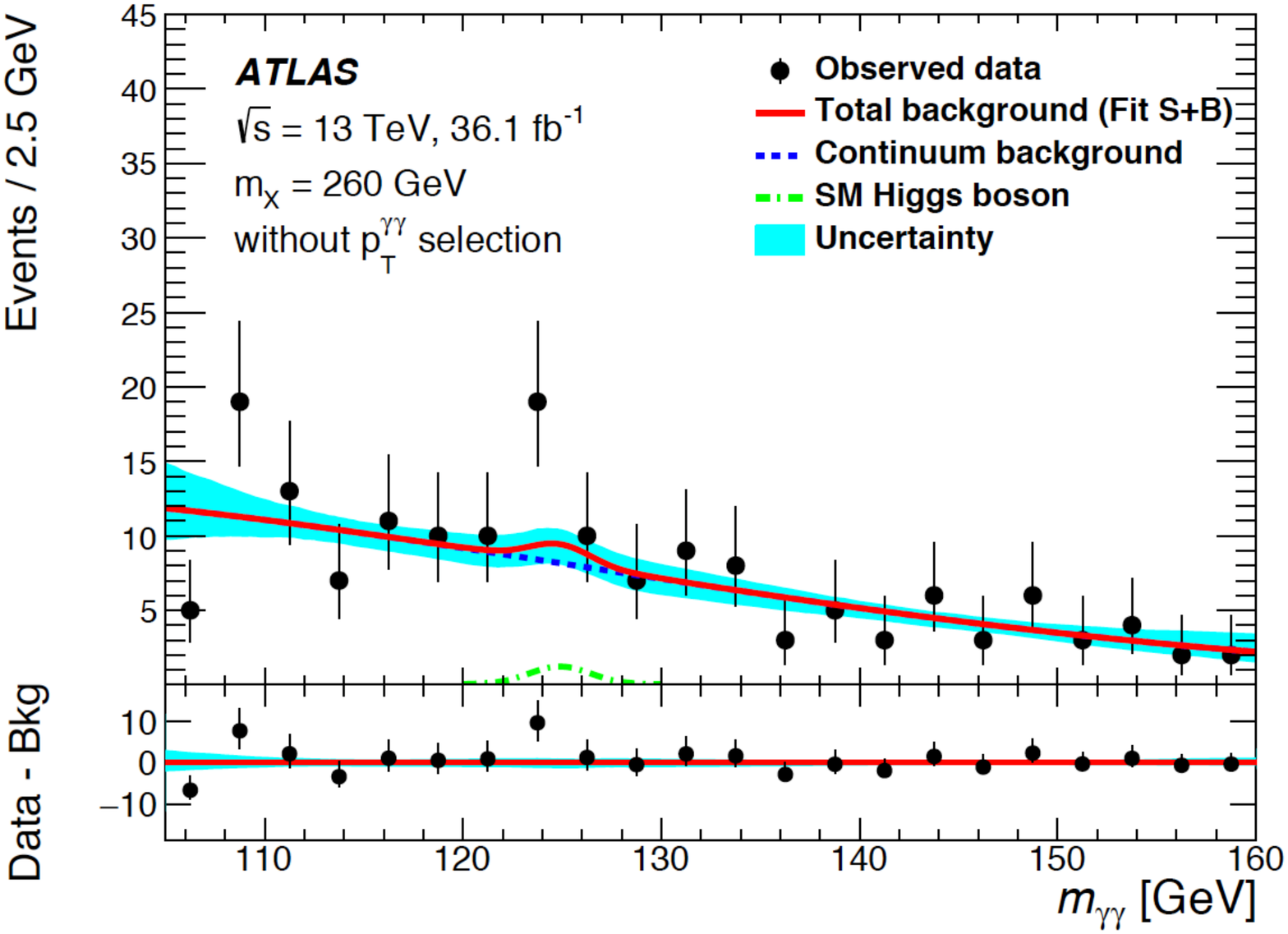} \hfil
\includegraphics[height=0.46\textwidth,angle=270]{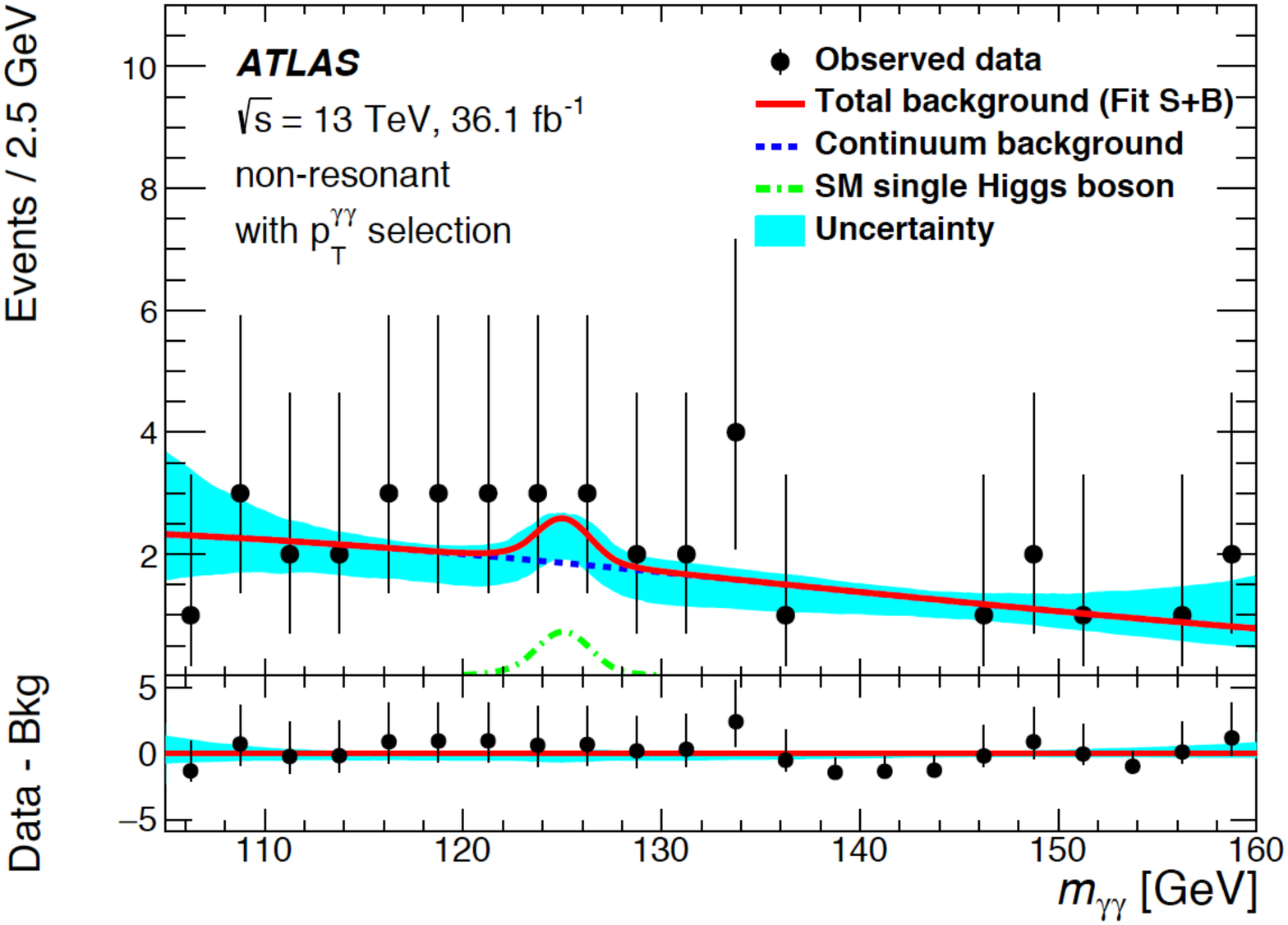}
\caption{
Distribution in \myy observed in the ATLAS analysis of $\hh \to  \gamma\gamma W W^{*}$, compared to the expected contribution from SM single Higgs boson plus SM non-resonant \hh production (dash-dotted line) and other backgrounds (dashed line), when no cut on the \pT of the di-photon system is applied (left) and with a cut of $\pT^{\gamma\gamma} > 100$~GeV applied (right)~\cite{ Aaboud:2018ewm}.
}
\label{fig:mGG_HH_ggWW_ATLAS}
\end{figure*}

Events in the $\hh \to \gamma\gamma W W^{*}$ channel are selected in the final state $\gamma\gamma \ell \nu jj$,
covering $34.3\%$ of the total $\hh \to \gamma\gamma W W^{*}$ signal.
The search looks for both SM non-resonant and resonant \hh production in the mass range between $260$ and $500$~GeV~\cite{Aad:2015xja, Aaboud:2018ewm}.
The signal is extracted by means of a maximum-likelihood fit to the distribution in mass of the photon pair, \myy.
In the non-resonant analysis and in the search for resonances of mass $400$~GeV and higher, the \pT of the di-photon system, $\pT^{\gamma\gamma}$, is required to exceed $100$~GeV, in order to reduce backgrounds.
Within a mass window centred on $\mH = 125.09$~GeV and of size equal to 2 times the experimental resolution on \myy, 7 events are observed in the data, in agreement with an expected background of $6.1 \pm 2.3$ events.
In the search for resonances of mass below $400$~GeV, where no $\pT^{\gamma\gamma} > 100$~GeV cut is applied, $33$ events are observed in the data, while $24 \pm 5.0$ events are expected from background processes.
The expected signal contribution from SM non-resonant \hh production amounts to $3.8 \cdot 10^{-2}$ ($4.6 \cdot 10^{-2}$)
in case the requirement on $\pT^{\gamma\gamma}$ is applied (not applied).
The distributions in \myy, obtained when no cut on $\pT^{\gamma\gamma}$ is applied and with the $\pT^{\gamma\gamma} > 100$~GeV cut applied,
are shown in Figure~\ref{fig:mGG_HH_ggWW_ATLAS}.
The event yields, as well as the distributions in \myy, observed in the data agree with the SM expectation in both cases.
As no evidence for a \hh signal is observed, the analysis proceeds by setting an upper limit on the \hh signal cross section.
The observed (expected) limit on the cross section for non-resonant \hh production with SM kinematics amounts to 230 (160) times the SM prediction.
In the corresponding Run 1 analysis, 4 events were observed in the signal mass window of the \myy distribution,
compared to $1.65 \pm 0.47$ events expected from background processes and $7.2 \cdot 10^{-3}$ signal events expected from SM non-resonant \hh production, and an observed (expected) upper limit of 1150 (680) times the SM cross section was set.

\subsection{$\hh\rightarrow W W^{*} W W^{*}$}

\begin{figure*}[ht!]
\centering
\includegraphics[width=80mm,angle=270]{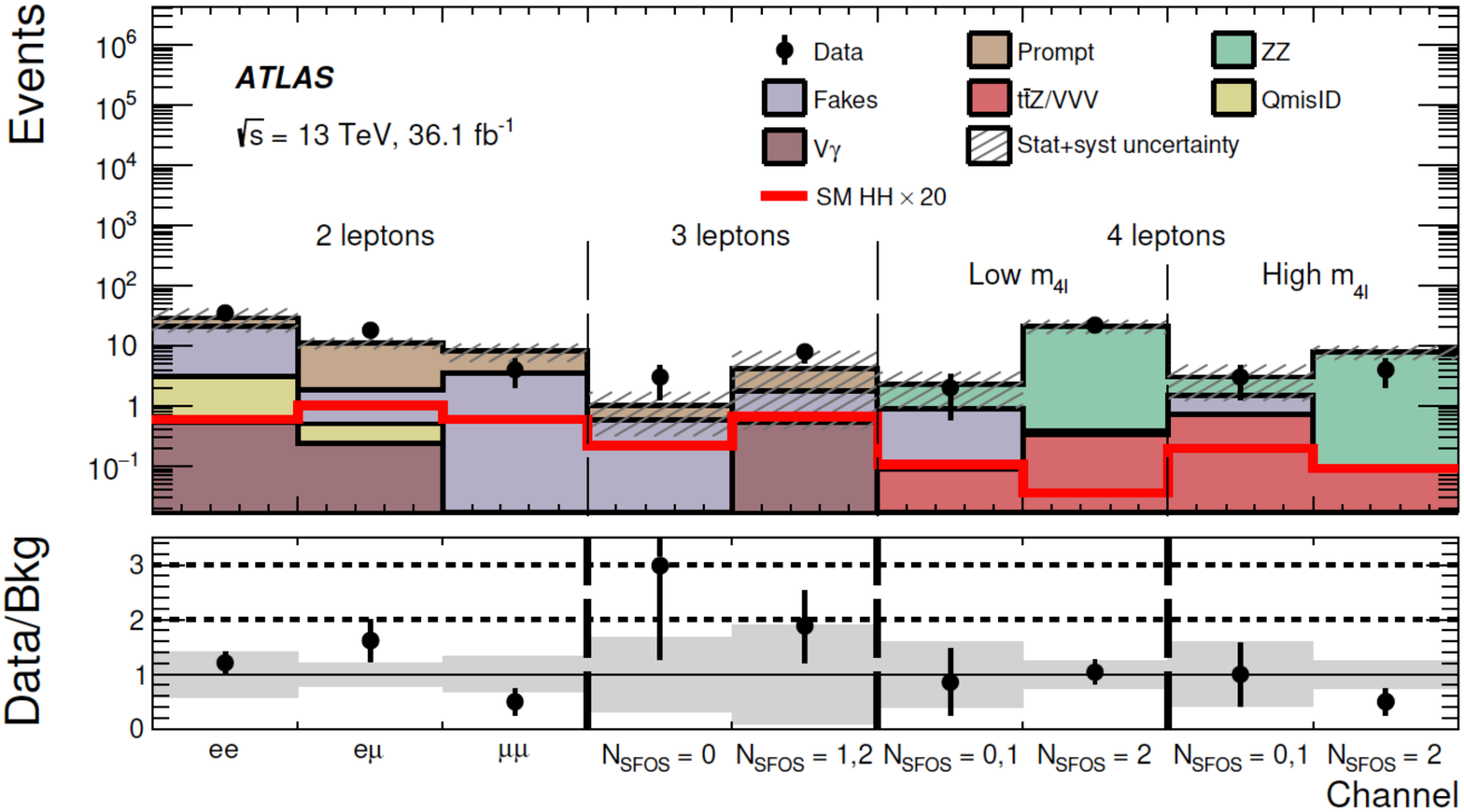}
\vspace*{-1.0cm}
\caption{
Event yields observed in the ATLAS analysis of $\hh \to W W^{*} W W^{*}$~\cite{Aaboud:2018ksn}, compared to the expected contribution of background processes and to a non-resonant \hh signal of SM kinematics and production rate amounting to 20 times the SM value. The symbol $N_{\textrm{SFOS}}$ denotes the number of lepton pairs of same-flavour and opposite-charge, while the low and high $m_{4\ell}$ categories refer to events in which the mass of the 4 lepton system is below and above 180~GeV, respectively.
}
\label{fig:evtYield_HH_WWWW_ATLAS}
\end{figure*}

The ATLAS analysis of $\hh \to W W^{*} W W^{*}$~\cite{Aaboud:2018ksn} selects events in a combination of final states with 2, 3, and 4
leptons.
In the di-lepton channel, the contamination from background processes is reduced by requiring the two leptons to be of the same charge.
The combination of the 2, 3, and 4 lepton final states covers $10.7\%$ of the total $\hh \to W W^{*} W W^{*}$ signal.
Similar to the $\hh \to \gamma\gamma W W^{*}$ analysis, the analysis of the $\hh \to W W^{*} W W^{*}$ final states exploits both SM non-resonant and BSM resonant production in the mass range $260$ to $500$~GeV.
In addition, the presence of heavy scalars $\textrm{S}$ of mass $135 < m_{\textrm{S}} < 165$~GeV originating from the decay of resonances $\textrm{X}$ of mass $280 < m_{\textrm{X}} < 340$~GeV, $\textrm{X} \to \textrm{SS}$ is probed.
An automatic optimisation of event selection criteria ("rectangular cuts"), implemented in the package TMVA~\cite{Hocker:2007ht}, is employed in order to enhance the ratio of signal over background events, before the signal gets extracted by means of a maximum likelihood fit to the event yields in nine event categories.
Events selected in the di-lepton channel are analysed in three event categories, containing events with either two electrons ($e e$), two muons ($\mu\mu$), or one electron plus one muon ($e \mu$), respectively.
In the 3 lepton channel, events containing zero and events containing one or more pairs of leptons of the same flavour and opposite charge are analysed separately.
Events selected in the 4 lepton channel are analysed in four event categories, based on the multiplicity of same flavour and opposite charge lepton pairs and the mass of the 4 lepton system.
The event yields observed in the data is compared to the SM expectation for the \hh signal and for background processes in Figure~\ref{fig:evtYield_HH_WWWW_ATLAS}.
The data is in agreement with the SM expectation.
The analysis proceeds by setting upper limits on the \hh signal cross section.
The combined fit of the nine event categories yields an observed (expected) limit on the cross section for non-resonant \hh production with SM kinematics of 160 (230) times the SM prediction.

\subsection{$\hh \rightarrow \tau^{+}\tau^{-}\tau^{+}\tau^{-}$}

The CMS search for $\hh \to \tautautautau$~\cite{SlidesHH4tau} is performed in the final state with 2 leptons and 2 $\PGt_{\textrm{h}}$, corresponding to 31.2\% of the total $\hh \to \tautautautau$ signal.
The analysis is performed in six event categories, based on the flavour of the leptons ($ee$, $\mu\mu$, $e\mu$) and on their charge (same-sign, opposite-sign). 
Events containing pairs of leptons of the same flavour, opposite charge, and mass within the range 70 to 110~GeV are rejected, in order to remove background arising from $Z/\gamma^{*} \to \ell^{+}\ell^{-}$ Drell-Yan production.
The multi-jet background is estimated from data, while the contribution of other backgrounds is modelled using the MC simulation.
The signal extraction is based on a maximum-likelihood fit to the distribution in mass of the 2 leptons plus 2 $\PGt_{\textrm{h}}$ system in case of the three event categories containing opposite-sign lepton pairs.
In the event categories with same-sign lepton pairs, the small number of background events precludes the usage of a shape
analysis and the event yields are instead used as input to the maximum-likelihood fit ("cut and count" analysis).
The analysis is still blinded.
Based on the expected signal acceptance and efficiency and on the expected background contamination, the analysis is expected to be sensitive to resonant \hh signals produced with a cross section of order 10~pb.

\subsection{Potential improvements}

A common feature of the three channels $\hh \to \yyww$, $\wwww$, and $\tautautautau$
is that their sensitivity is limited by small signal yields and sizeable statistical uncertainties with the present data. 
Significant gains in analysis sensitivity have been achieved in the "established" channels $\hh \to \bbbb$, $\bbyy$, and $\bbtt$
during LHC Runs 1 and 2, thanks to improvements in the analysis methods (up to a factor 2-3 improvement in sensitivity for the same luminosity).
Significant potential exists to likewise improve the sensitivity of the "new" channels $\hh \to \yyww$, $\wwww$, and $\tautautautau$.

In the $\hh \to \yyww$ channel, potential improvements include the use of multivariate methods to
enhance the separation of the \hh signal from backgrounds, the reconstruction of the mass of the \hh system by means of an algorithm similar to the ``High Mass Estimator" (HME) algorithm developed for the analysis of resonant \hh production in
$\hh \to \bbww$, described in Ref.~\cite{Huang:2017jws}, the replacement of the $\pT^{\gamma\gamma} > 100$~GeV cut by event categories based on $\pT^{\gamma\gamma}$, and the extension of the analysis to the $\gamma\gamma\ell\nu\ell\nu$ and  $\gamma\gamma jjjj$ final states.

Potential improvements to the sensitivity of the $\hh \to \wwww$ channel comprise the substitution of the
"rectangular cuts" that are employed for separating the \hh signal from backgrounds by more modern multivariate methods such as BDTs or NNs, and by upgrading the analysis from a "cut and count" approach to a shape analysis, based on the output of a BDT or NN. Besides improving the separation of the \hh signal from the background, we expect that a shape analysis based on the output of the BDT or NN will have the further benefit of providing useful constraints to the systematic uncertainties, compared to the simple "cut and count" approach.
Non-prompt and fake leptons constitute a sizeable source of background in particular in the final state with 2 leptons of the same charge, where it amounts to 30-40\% of all backgrounds. We expect significant reductions of this background may be achievable thanks to anticipated improvements in the identification of leptons with multivariate methods in the future.

Potential improvements to the sensitivity of the $\hh \to \tautautautau$ channel are expected from extending the analysis to cover further final states (4 leptons, 3 leptons plus 1 $\PGt_{\textrm{h}}$, 1 lepton plus 3 $\PGt_{\textrm{h}}$, 4 $\PGt_{\textrm{h}}$) and to determine reducible backgrounds other than multi-jet production from data instead of from the MC simulation. 
The latter is expected to not only reduce the systematic uncertainties, but also the statistical uncertainties on the background expectation, as samples of backgrounds with large cross sections, for example Drell-Yan production, have a higher event statistics already in the LHC Run 2 data, compared to the event statistics presently available by MC simulation.
Moreover, the current CMS analysis of \hh production in the final state with 2 leptons and 2 $\PGt_{\textrm{h}}$ neglects the signal contribution arising from the decays $\hh \to \tautauww$ and $\wwww$. We expect
these decays to provide a significant contribution to the overall \hh signal yield. 

\begin{figure*}[ht!]
\centering
\includegraphics[width=0.46\textwidth]{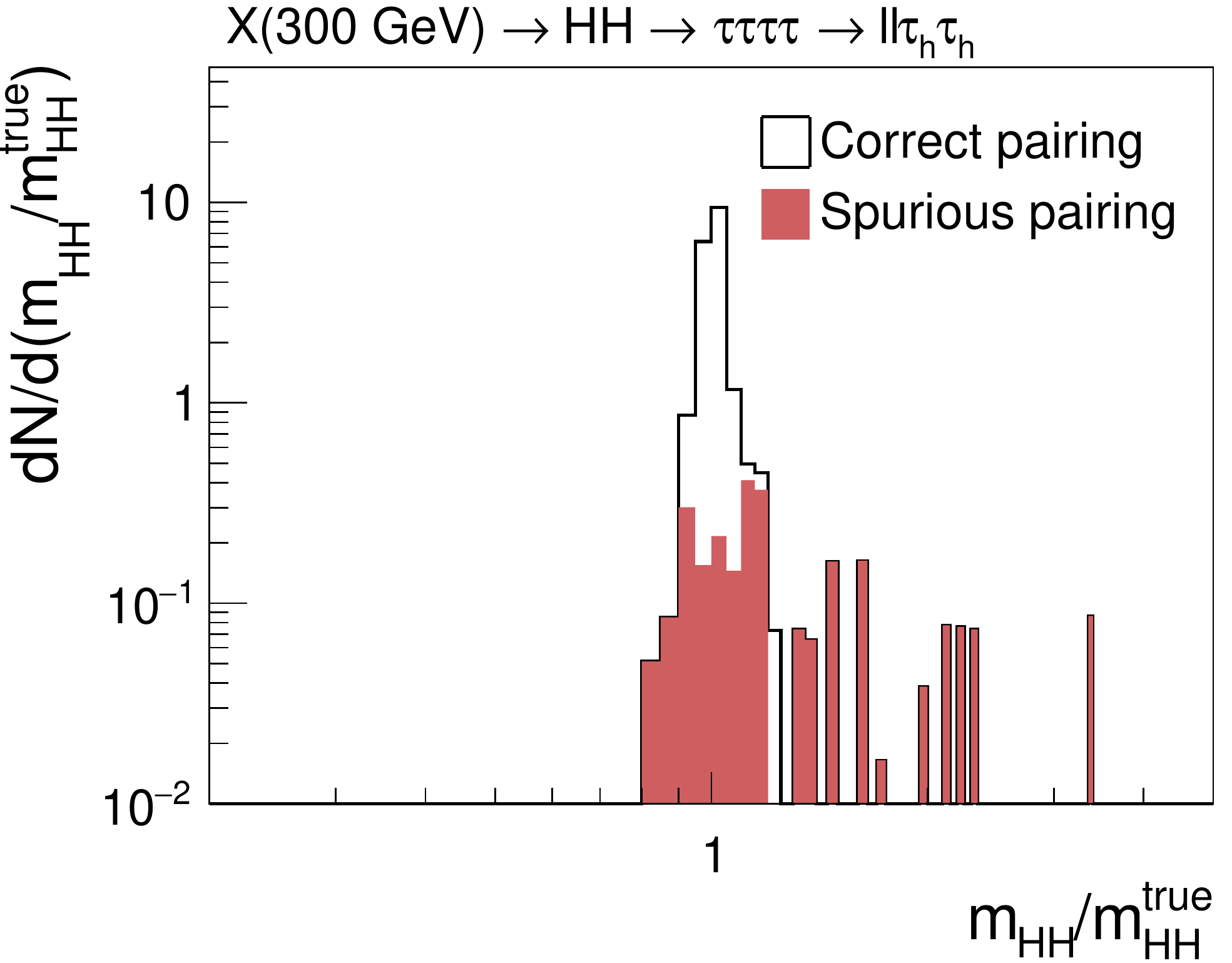} \hfil
\includegraphics[width=0.46\textwidth]{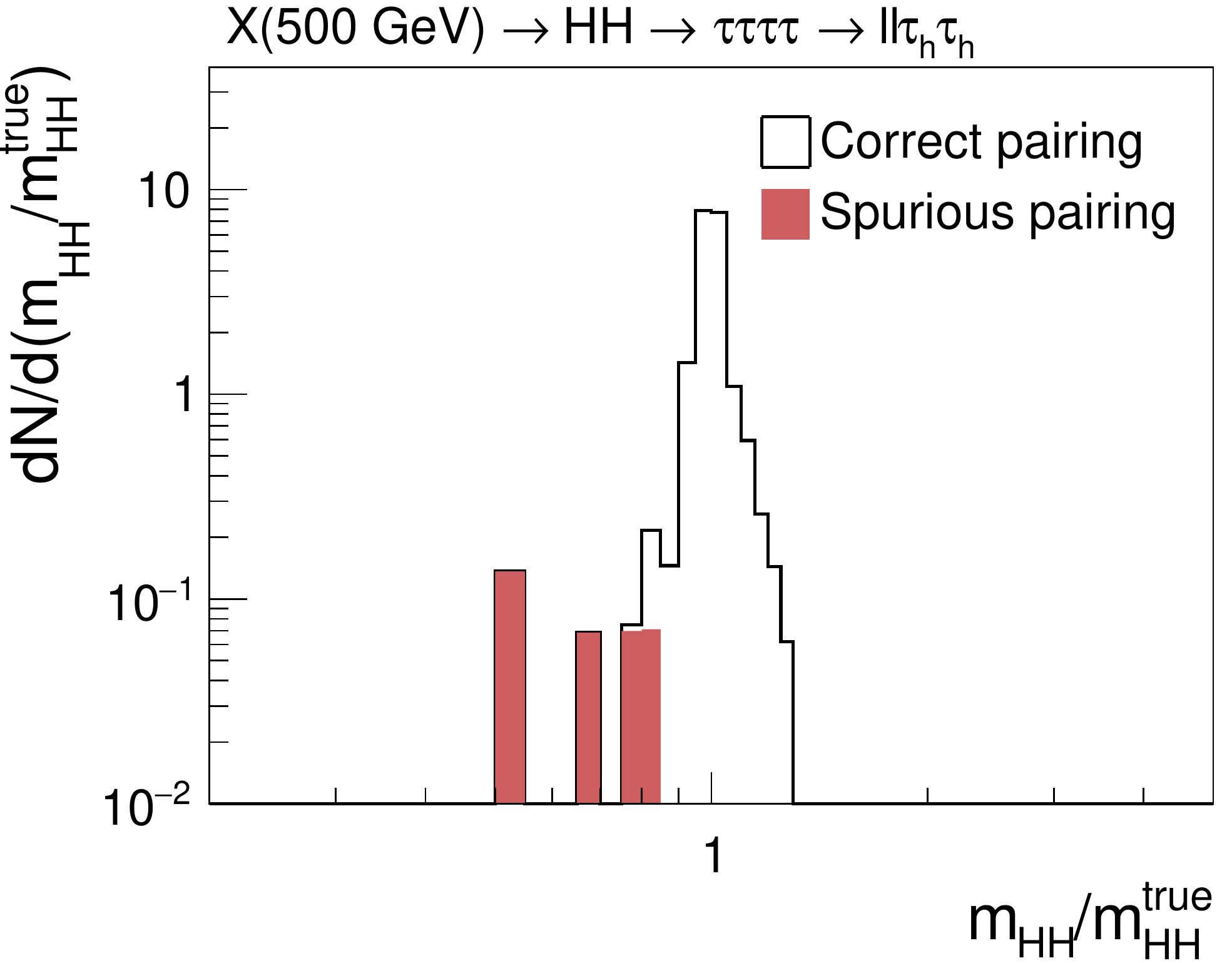}
\caption{
Distributions in the ratio of reconstructed to true mass of the $\hh$ system, $\mhh/\mhh^{true}$, in simulated $\hh \to \tautautautau$ signal events of true $\hh$ mass $300$~GeV (left) and $500$~GeV (right)~\cite{Ehataht:2018nql}.
The x-axis ranges from $0.2$ to $5$.
}
\label{fig:SVfit4tauResolution}
\end{figure*}

A further improvement in the sensitivity of the $\hh \to \tautautautau$ channel may be achieved by using an algorithm for reconstructing the mass of the $\hh$ system, presented at the workshop.
The algorithm is based on a dynamical likelihood approach~\cite{Kondo:1988yd,Kondo:1991dw} and represents an extension of the SVfit algorithm~\cite{Bianchini:2014vza,Bianchini:2016yrt} that is used in the CMS \hhbbtt analysis presented in section~\ref{sec_exp_2dot5}.
Measurements of the energies and momenta of the visible $\tau$ decay products and of the missing transverse energy are combined in a probability model for the $\hh \to \tautautautau$ decay with constraints on the mass of each $\tau^+\tau^-$ pair to equal $\mH = 125.09$~GeV. 
Details of the algorithm are given in Ref.~\cite{Ehataht:2018nql}.
The algorithm achieves a resolution on \mhh, the mass of the $\hh$ system, of $22\%$ ($7\%$) in simulated $\hh \to \tautautautau$ signal events in which the Higgs boson pair originates from the decay of a narrow resonance $X$ of mass $m_{X} = 300$ ($500$)~GeV and produces a final state with 2 leptons and 2 $\PGt_{\textrm{h}}$. 
The quoted resolutions include the effect that the algorithm chooses an incorrect assignment of the 2 leptons and 2 $\PGt_{\textrm{h}}$ to the first and second $H$ boson in $13\%$ ($2\%$) of simulated signal events at $m_{X} = 300$ ($500$)~GeV, which causes the Higgs mass constraint to be applied to the wrong combinations of leptons and $\PGt_{\textrm{h}}$, thereby degrading the resolution on $\mhh$.
In case the algorithm could be improved to always choose the correct assignment, the resolution on \mhh would improve to $4\%$ ($6\%$) for signal events of $m_{X} = 300$ ($500$)~GeV.
Distributions in the ratio of reconstructed to true mass of the $\hh$ system are shown in Figure~\ref{fig:SVfit4tauResolution}, separately for simulated $\hh \to \tautautautau$ signal events in which the correct assignment ("correct pairing") is chosen and events in which the incorrect assignment ("spurious pairing") is chosen by the algorithm.

In summary, we expect that the sensitivity of channels without \bjets will increase faster compared to the sensitivity of
channels with \bjets as more LHC data becomes available in the future and more refined and sophisticated analysis techniques get utilised in the new channels. 
In our view, it is a worthwhile effort to study the feasibility of these new channels in preparation for the upcoming HL-LHC data-taking period.

%% file: VBF/vbf.tex
At the current LHC centre of mass energy of 13 TeV, 
the VBF \hh production cross section is an 
order of magnitude smaller than the gluon-gluon fusion (ggF) process, which is 
the predominant mode studied so far in this review (see the detailed discussion in Chapter~\ref{chap:th_status} and in particular \refta{table:xsec2}). 
Being initiated by quarks rather than gluons, the 
signature of the VBF production mode differs 
greatly from ggF. Its distinctive topology is characterized by the presence of two separated quarks in the final state, which are reconstructed as high energetic jets in the forward region of the detector with large separation across the beam direction. Both the large invariant mass (\mjj) and rapidity separation of the outgoing VBF jets are particularly effective in isolating this peculiar signature.

\begin{figure}[htp]
\begin{center}
\includegraphics*[width=0.95\textwidth]{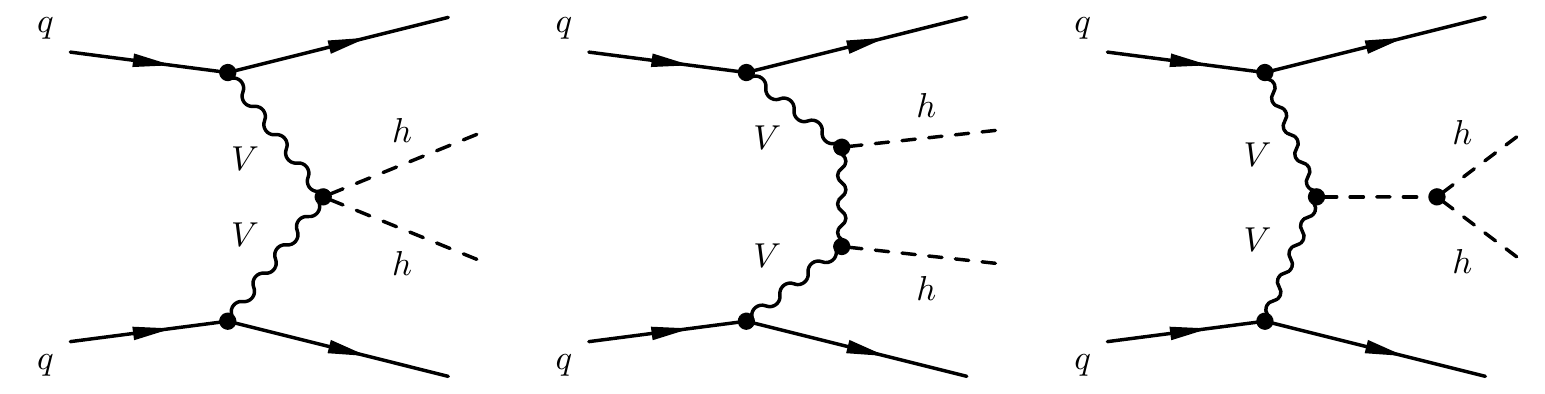}
\end{center}
    \caption{Leading order Feynman diagrams for Higgs pair production via vector boson fusion. The HHVV vertex (left), corresponding to the $c_{2V}$ coupling, is not probed by single Higgs boson processes. The HVV vertex (center) corresponds to the $c_{V}$ coupling, is constrained by single Higgs boson measurements. The HHH vertex (right) involves the Higgs boson self-coupling.}
    \label{fig:vbf_feyn}
\end{figure}

The VBF \hh production 
proceeds at tree level through the three Feynman 
diagrams shown in Fig.~\ref{fig:vbf_feyn}. The left, middle, and right diagrams scale with $c_{2V}$ , ${c_{V}}^2$ and ${c_V}{\klambda}$, respectively, where $c_{2V}$ and ${c_{V}}$ are the coefficients of the $HHVV$ and $HVV$ couplings, normalized to their SM values.

A study of this process at the LHC and future colliders with $\sqrt{s}=100$~TeV, to explore the sensitivity to higher-dimension operators, is reported in Ref.~\cite{Bishara:2016kjn}. Here the emphasis was on the large \mhh domain, where the behaviour of the longitudinal-longitudinal component of the amplitude is characterized by the destructive interference between the first two diagrams:
\begin{equation}
A({ V_L V_L\to \hh}) \sim \frac{\hat{s}}{v^2}(\delta_c) + {\cal O}(m_W^2/\hat{s})
\label{eq:amp_vbfhh}
\end{equation}
where:
\begin{equation}
\delta_c = c_{2V}-c_{V}^2 
\label{eq:delta_c}
\end{equation}
The quantity $\delta_c$ vanishes in the SM as well as in BSM extensions where the Higgs boson belongs to an SU(2) doublet, and the growth of the amplitude with energy is suppressed. The study of the high \mhh behaviour is therefore a powerful probe of $\delta_c$ and of the gauge structure of the Higgs sector.

\begin{figure}[htp]
    \centering
    \includegraphics[width=0.75\textwidth]{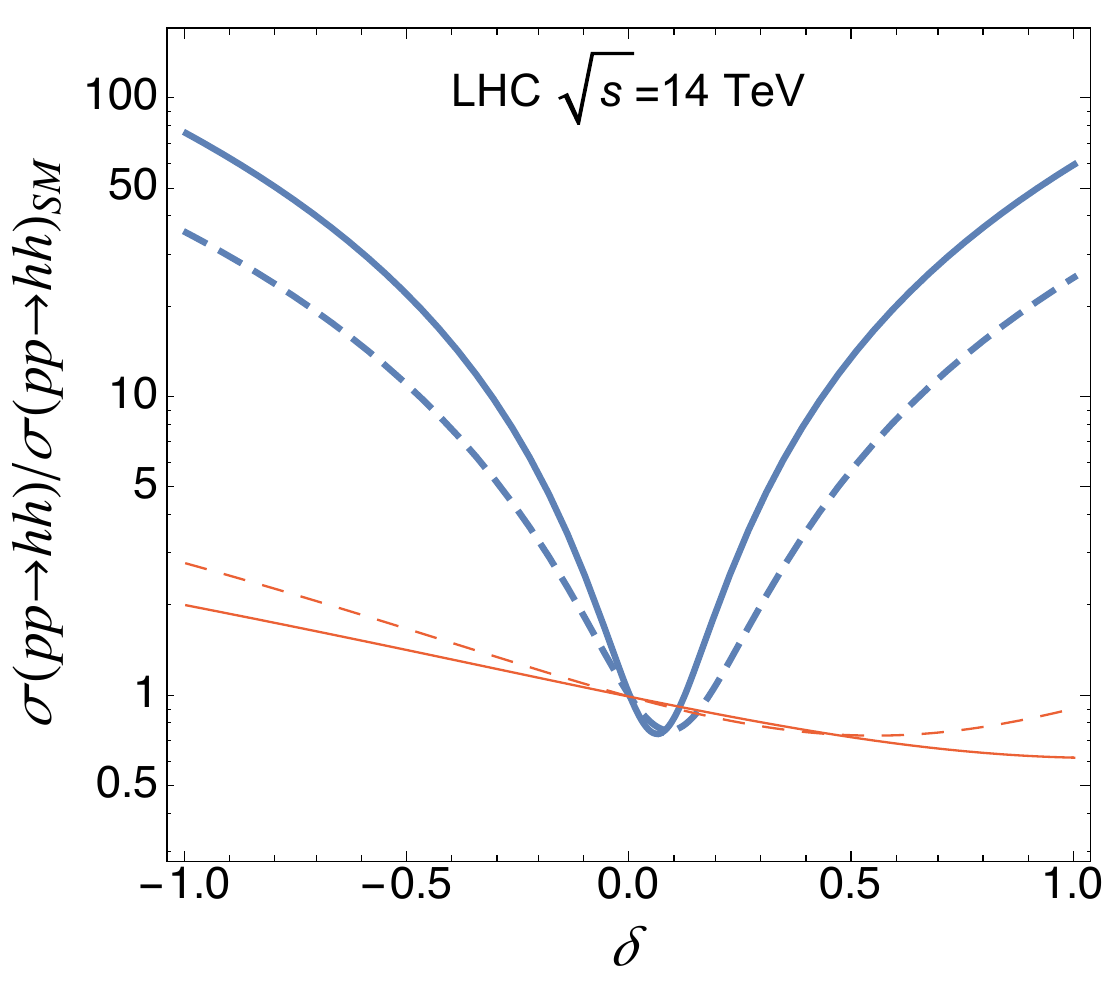}
    \caption{VBF \hh production cross section as a function of the coupling deviation from the SM value for the $HHVV$ ($HHH$) vertex in blue (red). The solid line is after acceptance cuts, the dashed line is after analysis cuts applied on the rapidity difference and \mjj ~\cite{Bishara:2016kjn}.}
    \label{fig:vbf_lhc}
\end{figure}

While the constraints 
on $c_{V}$ are currently derived by searches for
single Higgs boson VBF production at 
$1.21^{+0.22}_{-0.21}$ times the value predicted 
by the SM~\cite{ATLAS:2018doi}, the other 
two couplings are far less constrained. The HHVV 
vertex is unconstrained from current data, hence searches 
for VBF \hh production provide the only direct 
probe to the associated parameter. Furthermore, 
enhancements of this coupling with respect to its 
SM prediction may yield to a significant increase of the 
VBF \hh cross section, by as much as two orders of 
magnitude at twice the value predicted by the SM, as 
shown in Fig.~\ref{fig:vbf_lhc}~\cite{Bishara:2016kjn}. 
Such an enhancement would be noticeable with the full 
Run 2 data. Furthermore the VBF \hh production probes the Higgs self-coupling as well, resulting in an additional constraint.

A phenomenological study that exploits the \bbbb final state is reported in Ref.~\cite{Bishara:2016kjn}, applying boosted jet tagging techniques -- justified by the high \pT of the Higgs bosons in the relevant kinematic region -- to minimise the dominant background processes. An example of the impact of $\delta_c\ne 0$ is shown in Fig.~\ref{fig:vbfHH2} of Sec.~\ref{sec:other_probes}, for VBF \hh production at future colliders. In that figure the di-Higgs mass spectrum, the rapidity separation and invariant mass of the VBF jets, in the SM and in a $c_V=1$, $c_{2V}=0.8$ scenario are compared to the expected backgrounds (in the parton-level simulation). These observables have long been used in searches for single Higgs boson through VBF production at the LHC. 
After the detector simulation of fully showered events, Ref.~\cite{Bishara:2016kjn} carried out a detailed study of the shape of the mass distribution, reporting that at the LHC with an integrated luminosity of 300\ifb the $c_{2V}$ coupling can be measured with about 40\% precision at the 68\% CL. This results in a strong motivation to extend the current searches to the \hh VBF production mode during Run 2 and 3, although this analysis of the \bbbb final state shows clearly that the VBF channel is less sensitive than ggF to the Higgs boson self-coupling\footnote{The sensitivity to \klambda arises from the threshold region $\mhh\sim2\mH$ where multi-jet background buries the signal even for large modifications of the Higgs couplings with respect to their SM values.}.  

\begin{figure}[htp]
    \centering
    \includegraphics[width=0.33\textwidth]{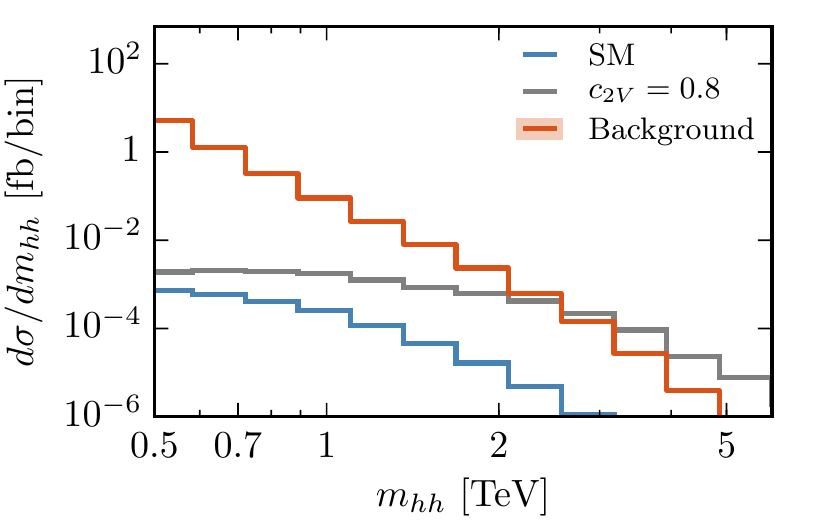}
    \includegraphics[width=0.33\textwidth]{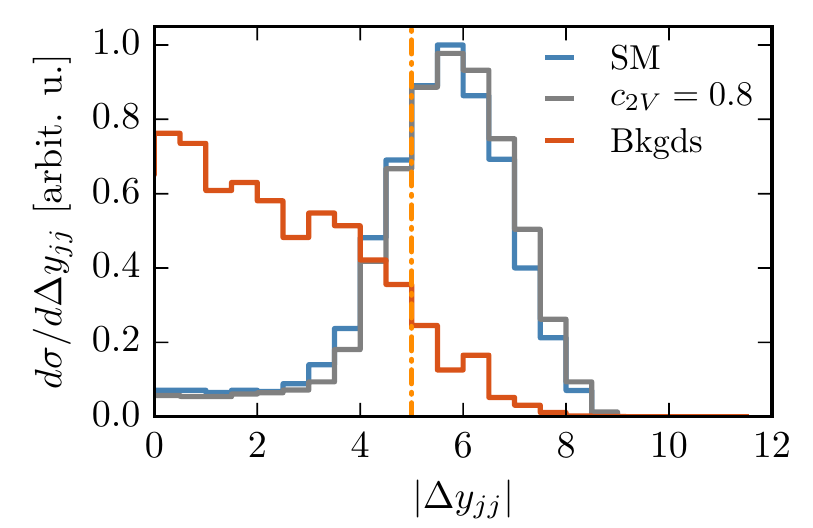}
    \includegraphics[width=0.33\textwidth]{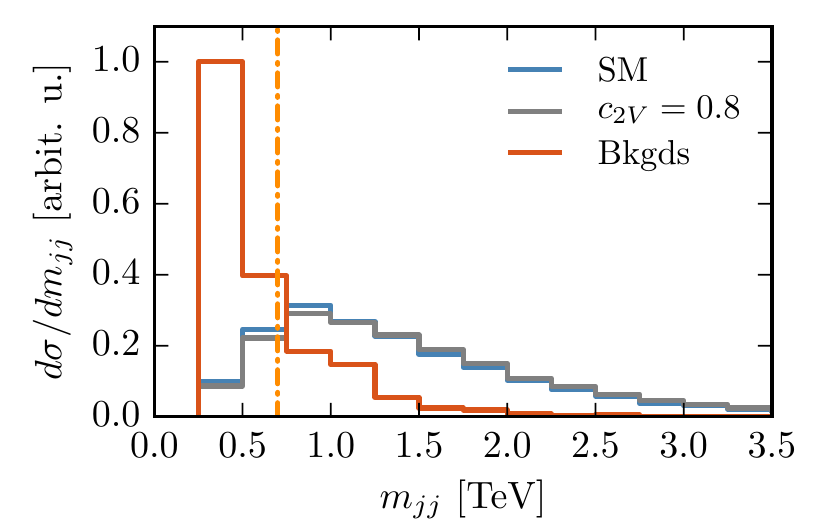}
    \caption{Distribution of the \mhh, the difference in rapidity $|\Delta y_{jj}|$ (left) and invariant mass \mjj (right) of the 
    VBF jets associated to a \hh pair in a phenomenological study~\cite{Bishara:2016kjn} at 14 TeV. Distributions are also shown for a $c_{2V}$ value at 0.8 times the SM prediction. The background includes multi-jet events, \ttbar and Higgs boson production via ggF where additional radiation can mimic VBF jets.}
    \label{fig:mjj_eta}
\end{figure}


In the CMS \hhbbyy search~\cite{Sirunyan:2018iwt}, a VBF signal model has been 
considered for the first time experimentally. However, signal events 
in this analysis were chosen via a BDT trained on ggF 
\hh events, thereby limiting the potential for 
sensitivity improvement. The efficiency times 
acceptance for SM-like VBF \hh events using this 
model is 13\%, with 10\% in the high mass region 
(greater than 350~GeV) and 3\% in the low mass 
region (below 350~GeV). Ultimately, considering 
this VBF \hh signal in the analysis designed to 
target ggF \hhbbyy  
improves the sensitivity by 1.3\% (while the VBF \hh cross section represents 5\% of the total one). 
Figure~\ref{vbf-cms-bbyy} illustrates the small 
impact of including a VBF signal on the overall 
sensitivity of the search for ggF 
\hhbbyy in CMS. 
A dedicated category with event selections 
designed to specifically target VBF production 
will lead to a much better improvement of the 
sensitivity when combined with an analysis 
targeting ggF \hh production, due to the 
aforementioned signal purity obtainable 
in such a category. In addition to taking 
advantage of VBF specific \mjj and 
$\Delta \eta$ distributions, different 
\mhh regimes can be used in order 
to isolate VBF production from the dominant 
ggF production mode: at large values of \mhh, 
VBF production is enhanced relative to the ggF 
production of \hh pairs~\cite{Dolan:2015zja}. Given 
the difficulty of measuring SM \hh production 
due to its small cross section, any such gain is 
invaluable in improving an analysis. Furthermore, 
should \hh production be observed due to a BSM 
enhancement, such a model would be vital in 
understanding the source and nature of that 
said enhancement.

\begin{figure}
    \includegraphics[width=.32\textwidth]{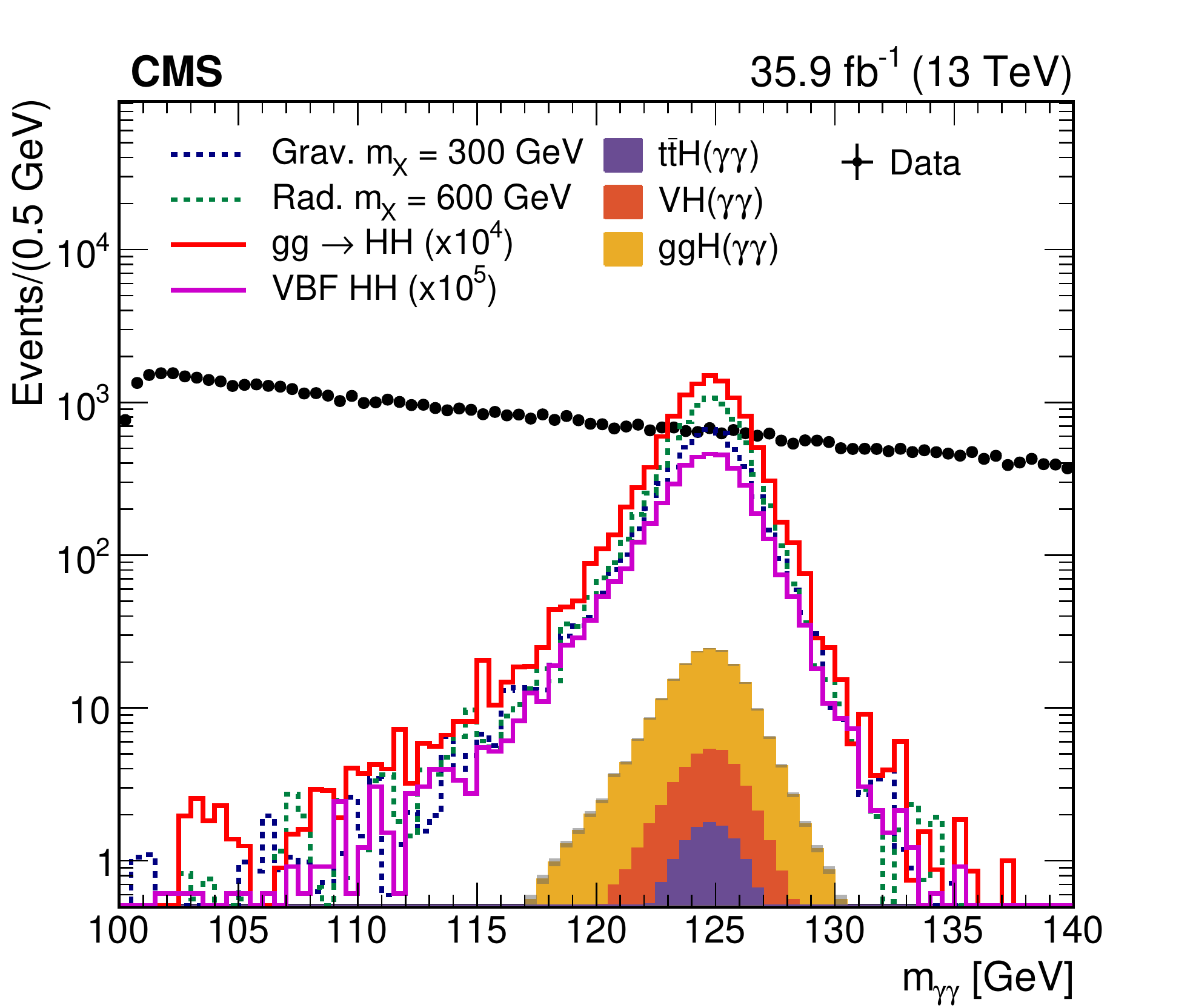}
    \includegraphics[width=.32\textwidth]{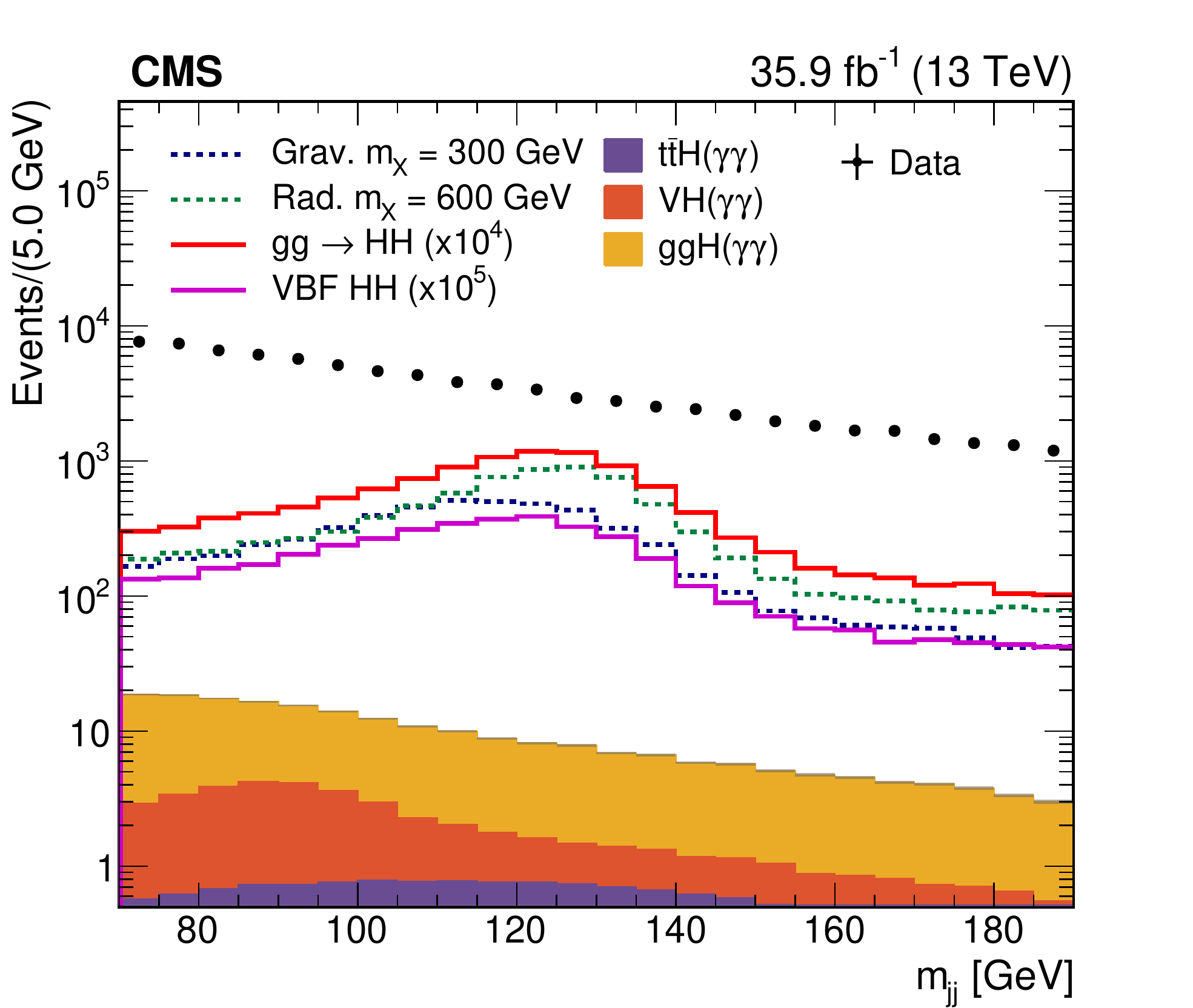}
    \includegraphics[width=.32\textwidth]{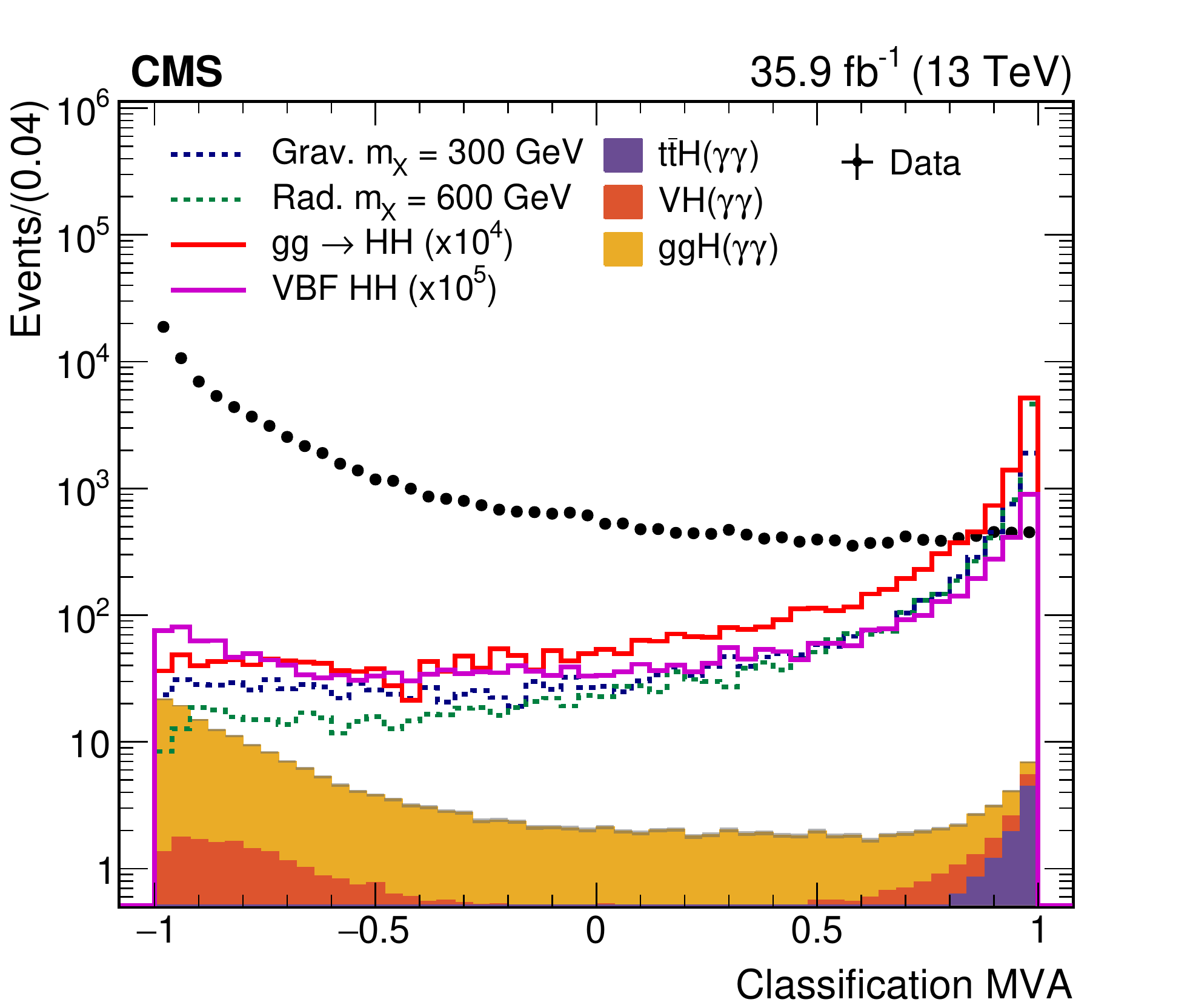}
    \caption{Higgs candidate \myy (left) and \mjj (middle) distributions, as well as the BDT classifier score (right) in the CMS \hhbbyy analysis after kinematic selection criteria are applied. The contribution of the VBF \hh process is shown in pink, normalized to $10^5$ times its cross section~\cite{Sirunyan:2018iwt}.}
\label{vbf-cms-bbyy}    
\end{figure}

The very first experimental search targeting VBF \hh production has been presented by ATLAS in the \bbbb final state~\cite{ATL-CONF-2019-047}, using the data collected during 2016--2018 and corresponding to an integrated luminosity of 126\,\ifb. 
The analysis strategy follows very closely the analogous search in the same final state for the ggF initiated process and described in Sec.~\ref{sec:HH4b}.
The main differences are the use of a multivariate jet energy regression, described in Sec.~\ref{sec:bjetreg}, to correct the energy of \bjets which improves by about 10\% the jet energy resolution; and of course of VBF specific selection requirements. The event selection requires at least four central ($|\eta| < 2$) \btagged jets with \pT $>$ 40 GeV and at least two forward ($|\eta| > 2$) jets with \pT $>$ 30 GeV to ensure compatibility with the VBF \hh production mode. The two forward jets with highest \pT have to satisfy requirements on both their angular separation, $|\Delta\eta| > 5$, and their invariant mass, $\mjj > 1$ TeV.  In addition to the VBF \hh production, also the $VHH$ process, resulting in the $HHjj$ topology has been taken into account, although it is found to have a negligible contribution after the VBF specific event selection requirements. 

This search sets 95\% CL upper limit on the non-resonant VBF \hh production cross section of 1600 fb, where the expected value is 1000 fb.
The results are also interpreted as a function of $c_{2V}$, while $c_V$ and \klambda are set to their SM values. The observed (expected) excluded range is $c_{2V} <$ -1.00  and $c_{2V} >$ 2.67 ($c_{2V} <$-1.07 and $c_{2V} >$ 2.78), which is not sensitive yet to the SM prediction ($c_{2V}=1$). This search tests large deviations of $c_{2V}$ from their SM predictions, which result in a harder \mhh spectrum and higher momentum for the \bjets from the Higgs boson decay.


Similarly to the searches for resonant ggF Higgs boson pair production, VBF \hh production involving an intermediate resonance may be considered.
This kind of search would be complementary to the ggF searches, as in this case the vector bosons are the ones coupling to the new resonance, which then decays to a pair of Higgs bosons.
However, this production mode is not particularly well studied (see Ref.~\cite{Brooijmans:2014eja} for an analysis in the context of a model with warped extra dimensions), since its very small cross section poses a question on the ability of the LHC to impose significant constraints through this type of search.
Nevertheless,
the VBF \hh search reported in Ref.~\cite{ATL-CONF-2019-047} has been interpreted also in the context of resonant production and results are reported in the resonance mass range of 260–1000 GeV, where two classes of signals have been tested to perform a rather inclusive search under both the narrow and broad hypotheses for the resonance width.


%% file: HH_overview/results_presentation.tex
%

Experimental collaborations typically publish the results of their BSM searches in the context of particular models. The CLs method~\cite{Read2002} is used to derive exclusion limits on some model parameter space, by comparing the compatibility of the data to background-only and signal plus background hypotheses. Then, a test statistics from the likelihood ratio is used to discriminate between the two hypotheses. Reasonably well motivated, but somehow arbitrary benchmark models are often used for the signal hypothesis. Examples include generic resonances decaying to \hh of spin-0~\cite{Haber:1984rc,Branco:2011iw} or spin-2 with particular choices of the natural width~\cite{Randall:1999ee}, as well as the SM \hh process with the production cross section re-scaled by an arbitrary factor. The observed (and expected) upper cross section limits in the context of these models are often the final result of the publication, and the specific values obtained are often published as HEPData~\cite{Maguire:2017ypu}.

Often the theoretical community devises new models to which existing searches might be sensitive. Given the lengths of time between experimental publications (and the difficulty from experimental collaborations to interpret their data in all available models)
theorists might want to assess the sensitivity of existing published results to their model. If the hypothesised signal is sufficiently similar to those already presented in benchmark models used by the experiments (in width, resulting kinematic, etc.), the results on the benchmarks can be directly applied. On the other hand, if the properties of the signal differ significantly, the current presentation of results for \hh searches via specific benchmarks leave few options for re-interpretation.

There are several methods to explore to fully exploit the scientific potential of the experimental results. 
This section describes some possibilities discussed by the community and their relative strength. LHC experiments are encouraged to provide more information in their publications to allow possible re-interpretations of their results within the particle physics community. The impact of the available experimental results would consequently increase by allowing a more rapid testing of exciting new theoretical predictions.

\section{Examples from other BSM searches}
The difficulty of the re-interpretation of the Higgs pair production searches lies in their sophisticated profile likelihood fits. As these analyses fit a full distribution (\mhh or an MVA score, typically), it is difficult to provide an upper limit on the cross section without an assumption on the shape of the signal distributions.

A possible alternative is to use a simpler approach, such as event counting after the application of analysis cut, ``cut and count'', that is less dependent from the signal shape, since this information is not directly used in the likelihood fit. Instead, a simple implementation of the analysis selections, along with parameterised efficiency provided by the experimental collaborations, is sufficient to calculate a predicted signal yield, which can be compared to the model-independent cross section upper limits. The physics groups of both ATLAS and CMS experiments searching for SUSY and exotic new particles, provide information that is interpretable in this manner for many of their searches~\cite{Aaboud:2017vwy,Sirunyan:2017cwe}. A number of frameworks to facilitate combination of such results and to allow for the fast testing of models against different experimental results are available, such as \textsc{CHECKMATE}~\cite{Dercks:2016npn} and \textsc{GAMBIT}~\cite{Athron:2017qdc}. In particular, GAMBIT provides a framework for quickly simulating a provided signal model, testing its yields in a variety of encoded signal regions from a variety of analyses, and for testing the compatibility of a signal to the data by calculating the combined likelihood over these analyses (assuming complete orthogonality between different analyses) by summing over the various individual log likelihoods. For example, GAMBIT has recently re-interpreted and combined several ATLAS and CMS SUSY searches to set new constraints on chargino and neutralino production at the LHC, showing a small excess that individual analyses published by the experimental collaborations were not sensitive to~\cite{Athron:2018vxy}. This result enables the experimental groups to focus on a potential hint of new physics revealed by the existing data.

While some searches are amenable to simple, re-interpretable ``cut and count'' approaches, some sensitivity would be lost by the \hh searches if this approach were to be taken. This is because the shape of the signal distributions provides important information for the discrimination against background. Adopting this approach for the presentation of di-Higgs results is therefore disfavoured.

\begin{figure}
    \centering
    \includegraphics[width=0.75\textwidth]{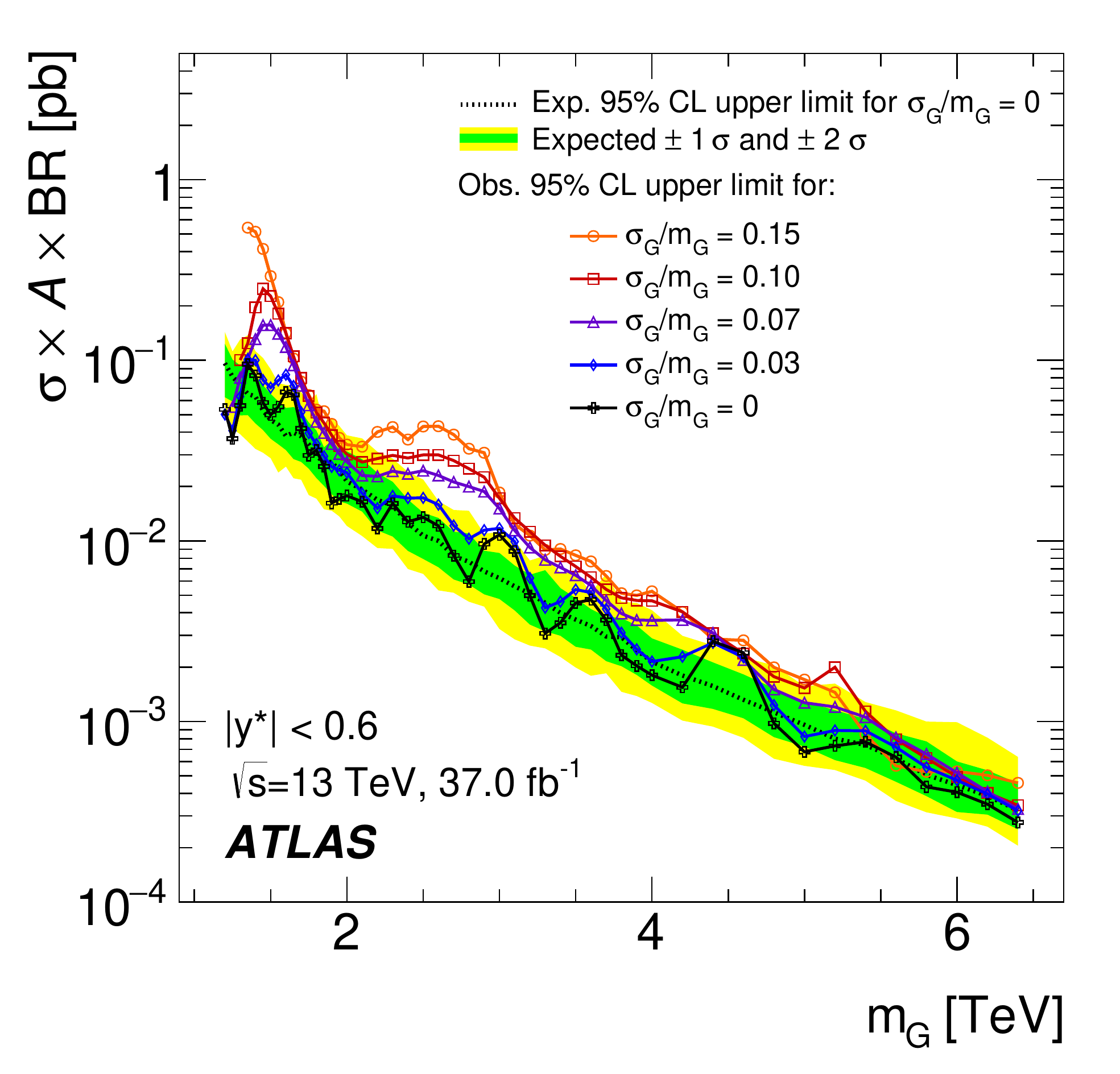}
    \vspace*{-0.3cm}
    \caption{Upper cross section limits on generic gaussian-shaped resonances decaying to di-jets as a function of the mean mass, for various widths hypotheses~\cite{Aaboud:2017yvp}.}
    \label{fig:dijets}
\end{figure}

\begin{figure}
    \centering
    \includegraphics[width=0.75\textwidth]{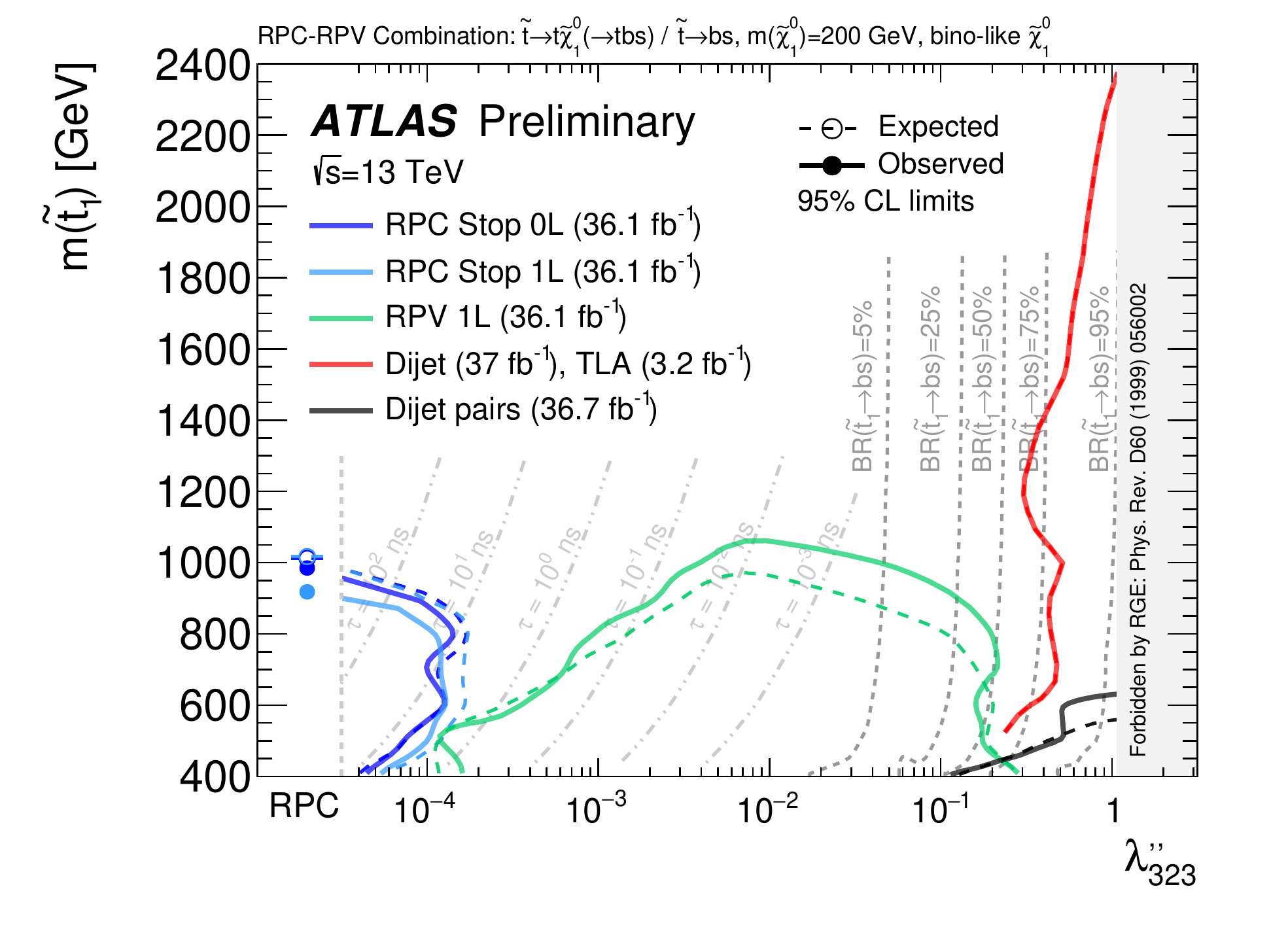}
    \vspace*{-0.3cm}
    \caption{ATLAS SUSY results, re-interpreted in the context of long-lived and prompt $R$-parity violating models~\cite{ATLAS-CONF-2018-003}. The excluded stop mass is shown as a function of the $\lambda_{323}''$ RPV coupling parameter. The red line is the limit from the di-jet resonance research, re-interpreted using the Gaussian limit strategy.}
    \label{fig:dijets_limits}
\end{figure}

Another commonly used method for generic result presentation is utilised by several ATLAS and CMS searches for di-jet resonances~\cite{Aad:2014aqa}, where the signal is extracted by fitting the di-jet mass distribution, similarly to \hh searches. Results are provided in the context of specific models and, additionally, experimental efficiencies and upper cross section limits on generic gaussian-shaped signals of various widths are also provided in HEPData. These allow to map the experimental results to various classes of models, as most resonant signals will look similar to a Gaussian with some width which depends on the details of the model.
An example from ATLAS's 13 TeV di-jet resonance search is shown in Fig.\ref{fig:dijets}~\cite{Aaboud:2017yvp}. This was used to re-interpret the results in the context of stop production in $R$-parity violating models~\cite{ATLAS-CONF-2018-003}, allowing for the strongest observed limits on stop particles at the LHC, as shown by the red lines in Fig.~\ref{fig:dijets_limits}. This approach is particularly promising for the presentation of \hh results, as many interesting BSM models predict resonant signals whose shapes are approximately Gaussian. However, for non-resonant signals, where the shapes are not as easy to parameterise this method would not work. Moreover, MVA techniques would limit the use of such simple signal parameterisation.

\begin{figure}
    \centering
    \includegraphics[width=0.55\textwidth]{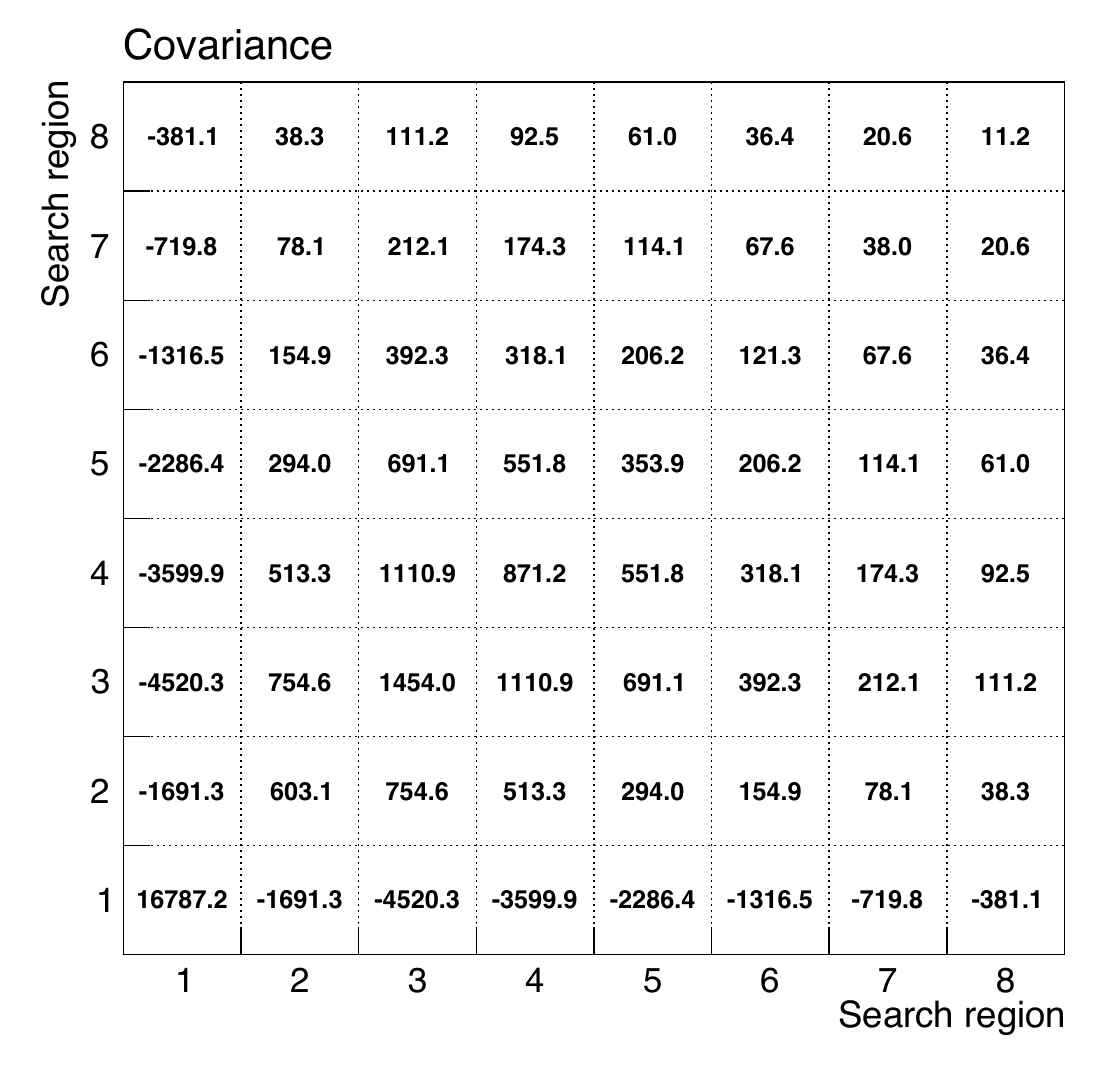}
    \vspace*{-0.2cm}
    \caption{The covariance of backgrounds in various search regions, which can be used to calculate a simplified likelihood function~\cite{SimpLH}.}
    \label{fig:cms_covariance}
\end{figure}

\begin{figure}
    \centering
    \includegraphics[width=0.55\textwidth]{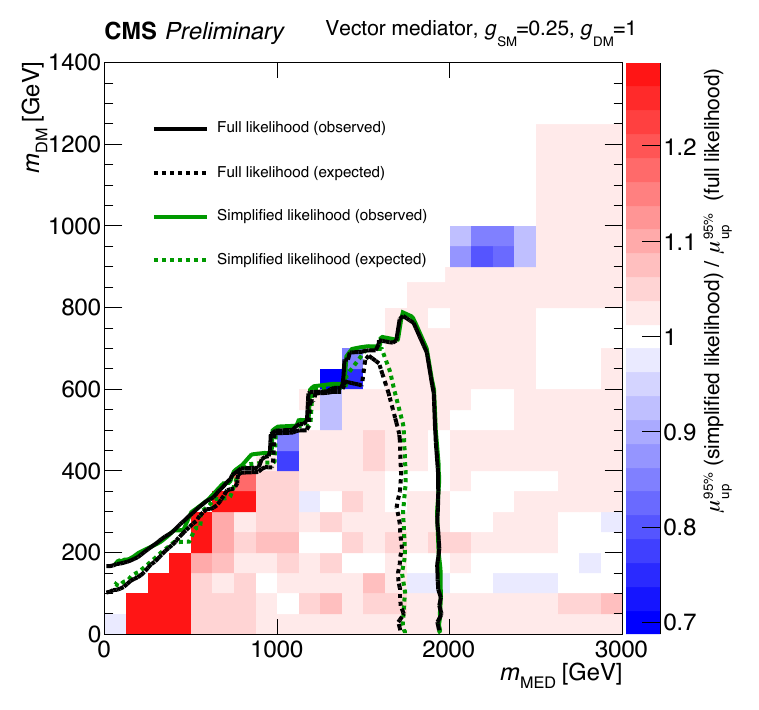}
    \caption{The expected and observed limits, for an example dark matter model, where the fit uses either the full, experimental likelihood or the simplified likelihood~\cite{SimpLH}.}
    \label{fig:cms_simp_example}
\end{figure}

The CMS Collaboration, in the context of SUSY searches, has adopted a new way of share the results with the HEP community by publishing ``simplified'' likelihoods~\cite{SimpLH}\footnote{See also~\cite{Cranmer:2013hia} for important considerations regarding systematic uncertainties.}. 
The covariance matrices for the various elements of the uncertainties on the background model are published along with the recipes for reconstructing the likelihood. An example is shown in Fig.~\ref{fig:cms_covariance}, and an example of the results one can obtain using the simplified likelihood compared to the full likelihood for a dark matter search are shown in Fig.~\ref{fig:cms_simp_example}. The likelihood can therefore cover arbitrarily complicated functions and numbers of signal regions. This method is promising for the presentation of di-Higgs results, as a full description of the \mhh distribution and the relationship of the uncertainties among bins can be succinctly encapsulated in the covariance matrix. 

\section{Options for the future}
One possibility to improve the re-interpretability of the published results is to simply make available the statistical objects and the code used to develop the profile likelihood for each specific search. These fits are usually interpreted in the \textsc{RooFit} framework~\cite{Verkerke:2003ir}. The full information on the signal and background shapes and a complete list of systematic uncertainties affecting both is contained in the so called ``workspace'' within \textsc{ROOT} files. Although the binary format of these containers utilised by \textsc{RooFit} makes the replacement of the signal model with an arbitrary shape difficult.

\textsc{PyHF}~\cite{pyhf} is a ROOT-free implementation of the underlying \textsc{HistFactory}~\cite{Cranmer:2012sba} probability distribution functions that addresses this issue by describing the workspace in human-readable \textsc{JSON}~\cite{json} format. 

\begin{figure}
    \centering
    \includegraphics[width=0.65\textwidth]{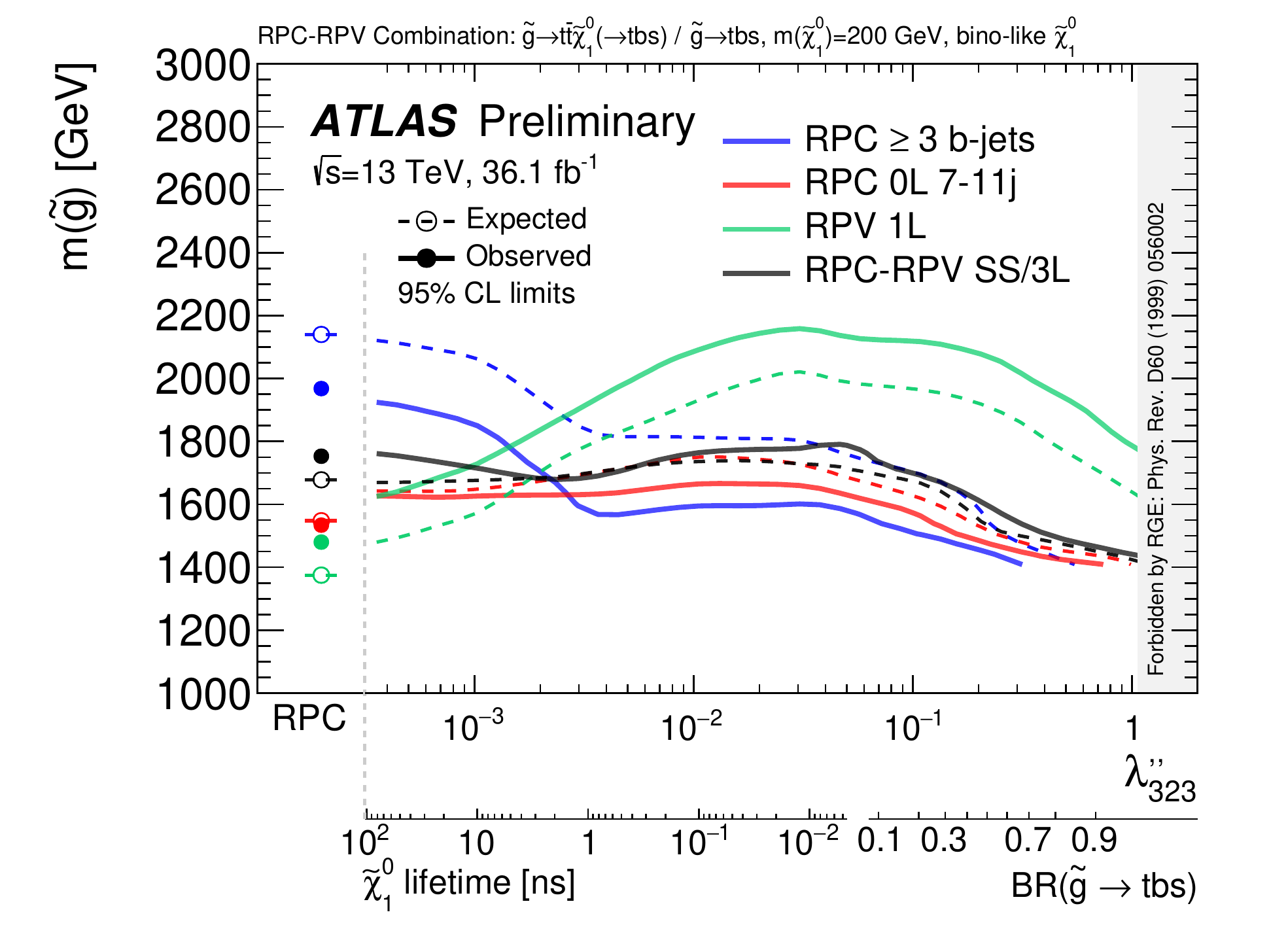}
    \caption{ATLAS SUSY results, re-interpreted in the context of long-lived and prompt $R$-parity violating models~\cite{ATLAS-CONF-2018-003}. The excluded gluino mass is shown as a function of the $\lambda_{323}''$ RPV coupling parameter. The blue lines, corresponding to a 36.1\ifb search for gluinos in final states with many $b$-jets~\cite{Aaboud:2017hrg}, were obtained entirely using the RECAST framework.}
    \label{fig:recast_example}
\end{figure}

Another similar and promising avenue is \textsc{RECAST}~\cite{Cranmer:2010hk}, currently used by the ATLAS Collaboration. It provides a container-based archiving system for the full implementation (selection and statistical analysis) of a search. For an analysis preserved in the RECAST framework, a set of \textsc{LHE} files describing a specific BSM signal can be provided and the full simulation and statistical interpretation can be run automatically, producing a final CLs value for a particular BSM signal. This is very useful for collaborations to quickly re-spin analyses using their full simulation and analyses methods. Figure~\ref{fig:recast_example} shows an example from the ATLAS SUSY re-interpretation effort in RPV scenarios~\cite{ATLAS-CONF-2018-003}, where the blue lines were obtained entirely using the RECAST framework implementation of an existing Run 2 search for gluinos in final states with many \bjets~\cite{Aaboud:2017hrg}.

The use of \textsc{RECAST} is currently limited internally to the collaborations. Any additional interpretation of the published results would need to go through the normal approval and publication process within the ATLAS collaboration. Some effort is being invested to streamline these processes for simple interpretations, but this remains a difficult prospect. So while \textsc{RECAST} would provide the most accurate re-interpretations possible (by using the full, official detector simulations and no compromises in the statistical analysis), the suitability as a tool for commonplace re-interpretation is limited, at least in its current form. 

\section{Machine-learning vs interpretability}

A potential challenge, especially as analyses techniques become more sophisticated, is how to re-interpret results obtained with multi-variate analyses such as BDTs or NNs. While truth-level distributions produced without full detector simulation (or, smeared distributions produced via a partial simulation such as \textsc{DELPHES}~\cite{deFavereau:2013fsa}) may be sufficient to re-produce the characteristics of simple analyses, the more complicated MVA approach may encode aspects of the detector that are more difficult to reproduce. 

While this is a valid concern, the level of agreement between the truth-level and fully simulated samples can be assessed directly by the experimental collaborations. In some cases, such as the ATLAS SUSY stop 1-lepton search~\cite{Aaboud:2017aeu}, this has already been done, and the agreement between the fully-simulated and truth-level inputs, run through the same Boosted Decision Tree, agree to within 10\%. As the accuracy of most re-interpretation approaches is similar, this shows that at least in some cases, using a BDT does not necessarily preclude re-interpretation. This same analysis in fact published the \textsc{XML} configuration files used by the TMVA~\cite{Hocker:2007ht} implementation of the BDT, allowing for others to easily re-run exactly the same selection. Similar possibilities exist for NNs from a \textsc{Keras} model~\cite{chollet2015keras}, for example, where the model can be saved in \textsc{JSON} or \textsc{YAML}~\cite{YAML} formats. Publishing the full configuration of the MVA is an approach that other analyses could follow to ensure the results can continue to live beyond the initial publication. 

\section{Considerations for non-resonant signatures and EFT}
\label{sec:RPNR}

Re-interpretation is important for both resonant and non-resonant \hh signatures. The discussion so far has focused on resonant signatures, but most applies transparently to the non-resonant signatures as well. As long as results are described as cross section upper limits on known models, or likelihoods are published where arbitrary signals can be included directly, limits on new signal types can be calculated or extracted. Of course, depending on the particular model to be tested the assumptions of the analysis may not be optimal, but the sensitivity of a given analysis to a particular non-resonant signal can always be derived.

There are several other approaches to consider as well, especially in the context of EFT, discussed in Chapter~\ref{chap:EFT}, which can generate a large variety of potential signal dynamics. As described in Sec.~\ref{sec:shape_bench}, CMS and the authors of \cite{Carvalho:2015ttv} have proposed the use of shape benchmarks to form a basis of possible signals for an HEFT analysis. CMS interprets the results of the non resonant \hh searches for each of these shape benchmark hypotheses, and theorists can study their own particular model and identify the shape benchmark which best describes the signal, allowing a quick and simple extraction of the upper limit on the cross section. This approach has the advantage of making the cross section limit setting trivial in the case the signal under test is sufficiently similar to a benchmark. On the contrary,  it cannot be used for a particularly unique signal which is not included by any of the benchmarks.

Experiments provide upper limits to the \hh cross section production for different benchmarks. Anyone who has interest to explore a particular portion of the EFT phase space can use those benchmarks to obtain an estimate of how strongly this portion is constrained by \hh data. A map of the EFT phase space to the specific shape benchmarks can be derived with a dedicated tool available at \url{http://rosetta.hepforge.org/} ~\cite{Carvalho:2017vnu}. This map can be then used to estimate the upper limits on each combination of EFT parameters

\begin{figure}[tb!]
\begin{center}
\includegraphics[width=0.5\textwidth]{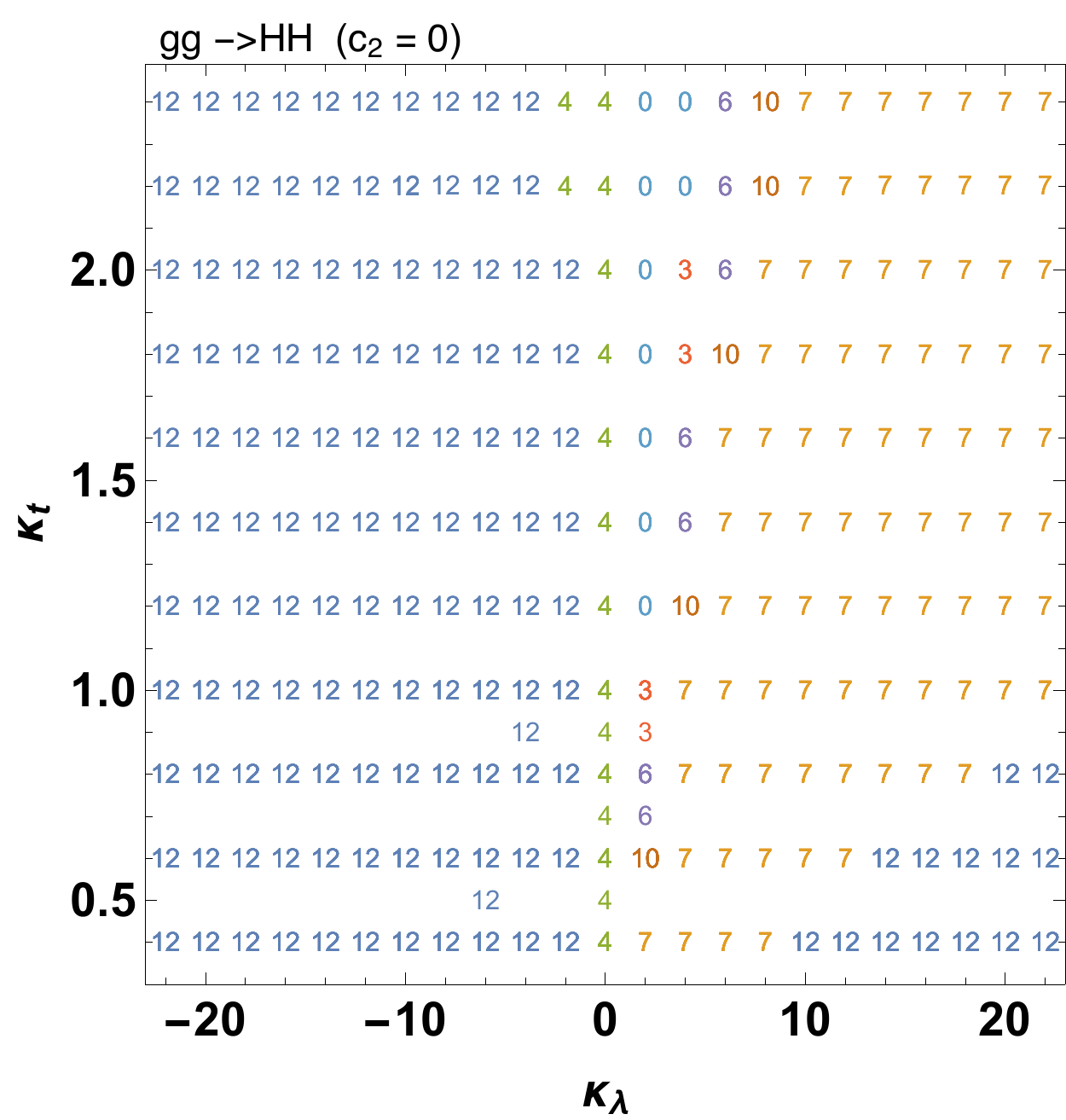}\\
\includegraphics[width=0.45\textwidth]{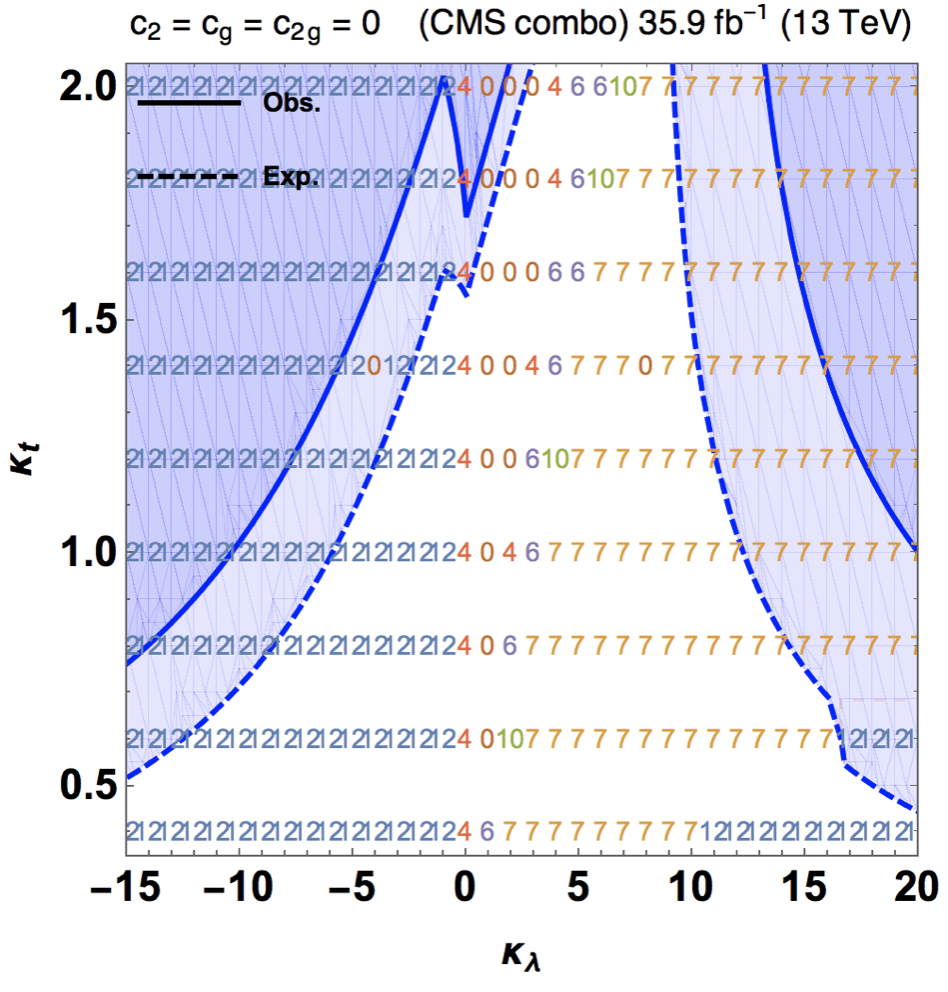}
\includegraphics[width=0.45\textwidth]{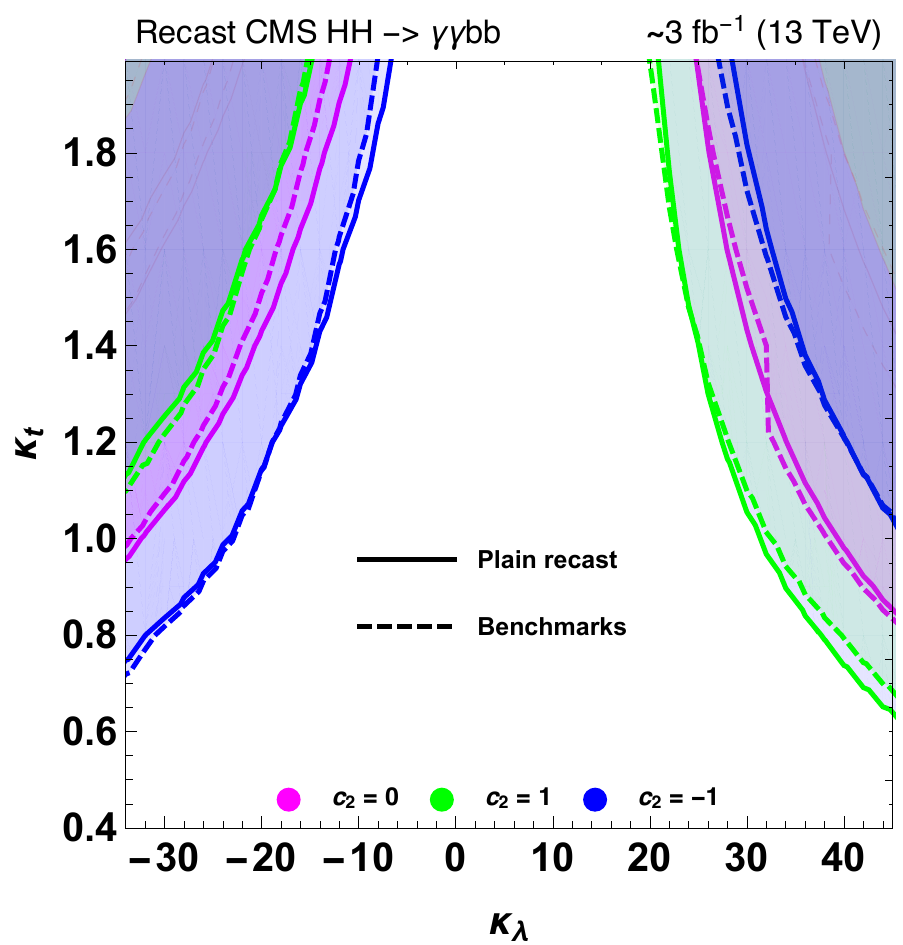}
\end{center}
\caption{Mapping of benchmarks into $\kl \times \kt$ phase space (top) obtained using recast tool \url{http://rosetta.hepforge.org/}~\cite{Carvalho:2017vnu};  benchmark mapping and excluded region of the EFT parameter phase space obtained using CMS combined limits on the benchmarks with a data sample of 35.9\ifb collected at $\sqrt{s}=13\,$TeV \cite{Sirunyan:2018two} (bottom-left); upper limits obtained using benchmark mapping compared to the upper limit obtained using directly EFT shapes if the recast approach from Ref.~\cite{Carvalho:2017vnu} based on 3\ifb of CMS data collected at $\sqrt{s}=13\,$TeV (bottom-right);}
\label{fig:bench}
\end{figure}

An example of this procedure is shown in Fig.~\ref{fig:bench}, where the map between $\kappa_t$, \klambda and the shape benchmarks is provided on the top.
The exclusion limits derived by the CMS collaboration will be discussed in Sec.~\ref{sec:non_res:comb} and in Fig.~\ref{fig:comb:eft} the upper limit for each benchmark by combining different final states is shown. Benchmark limits are then reported on the EFT map and compared to theory predictions in bottom-left of Fig.~\ref{fig:bench}. The jump in limits at $\kt \approx 2$ and $\kl \approx 0$ is the typical feature of the discrete benchmark approach. When the benchmark changes between two neighbouring points there is a discontinuity in the limit. If this discontinuity happens in the vicinity of the 95\% CL exclusion boundary, it propagates to the exclusion limits. Nevertheless it is a rare effect and the benchmark approach allows to have a stable estimate the excluded regions on EFT parameters. 

We also show in Fig.~\ref{fig:bench} bottom-right, the direct comparison of limits obtained with the shape benchmarks and those directly derived from an EFT analysis. The limits were obtained by a simple counting experiment based on a recast of public CMS results ~\cite{Carvalho:2017vnu}. The observed difference was rather small.

If an excess is present in the data related to a non-resonant production it is possible to spot it in one of the benchmarks. 
For example just looking at the upper limits on the SM-like production may hide an excess at very low or very high \mhh values incompatible with the SM. 
A detailed analysis could be then be performed on different EFT points that belong to a given cluster using their real \mhh shapes \cite{Carvalho:2016rys}. 

The Simplified Template Cross Section Method (STXS) described in Sec.~\ref{singleH_exp} takes a similar approach for single Higgs measurements by defining simple fiducial regions for cross section measurements of Higgs properties and kinematics. As single Higgs boson production has been convincingly observed in many channels and phase space regions, these are cross section measurements and not just upper limits. Many of these regions, for example those for $\ttbar H$ production with high \pT Higgs bosons, are potentially sensitive to deviations from the SM values of the self-coupling parameter. Generically, EFTs can cause simultaneous changes in several of these fiducial regions, but because the results are presented as easily interpretable measured cross sections, it is possible to determine the compatibility of a particular EFT model with the data. This can play an important role for understanding which EFT model should be explored by the experimental collaborations, as obviously excluded phase space points can be ruled out before expensive simulation is performed.

\section{Conclusions and recommendations}

We have reviewed several ways of enabling di-Higgs searches published by the experimental collaborations to be re-interpreted after their publication within the HEP community. With the timescale between publications potentially increasing as the LHC datasets grow, the ability to re-interpret existing searches will be critical to allow the LHC data to be used to their full potential in a timely manner in the coming years. Several different options are possible, but experimental collaborations are encouraged to consider the following:

\begin{enumerate}[label=(\roman*)]
    \item if possible, provide the full likelihood developed by the analyses, preferably in an easy to modify format such as the \textsc{PyHF JSON};
    \item if the first item is not possible, consider providing covariance matrices which allow for the reproduction of the likelihoods used by the searches to some degree of accuracy. This requires potentially more work from the analysis teams, with smaller accuracy for re-interpretations, so it is less preferred than the first option;
    \item where possible, provide the full configuration of the machine learning algorithms used by analyses. Ideally these would be accompanied by a detailed comparison between fully simulated samples and truth-level samples, so that re-interpretations can assess the applicability of using the MVA without the full detector simulation.
\end{enumerate}
By following these practices, the impact of the searches for BSM signals published by the LHC experiments will increase beyond their initial publications, and the results will be fully exploited within the HEP community for years to come.

%% file: HH_overview/Exp-combination.tex
The ATLAS and CMS collaborations have searched for Higgs boson pair production at 8 and 13 TeV, as reviewed in detail in Chapter~\ref{chap:LHC}. 
These searches have tested both resonant and non-resonant \hh production for new physics contributions. In particular for the non-resonant case, results have been provided either by assuming the SM prediction for the \hh kinematic and that only the total cross section is affected by BSM contributions, or by assuming BSM effects would impact only the Higgs boson self-coupling while all the other couplings are unaffected and equal to their SM values.

Although observing non-resonant \hh production at the level predicted by the SM is likely not possible until the end of the HL-LHC data taking, it nevertheless remains extremely important to probe this process with current dataset to constrain BSM models allowing for large increase of the \hh production cross section.

The \hh decay final states that have been studied are  described in detail in the previous chapters. Here all results and their combination are summarised and discussed.

The combination is performed by building a single likelihood function using all signal and background normalisation regions, correlating properly theoretical and experimental systematic uncertainties and ensure consistency in the definition of the parameters of interest, for resonant searches, sometime different mass values are probed by different channels, therefore an interpolation procedure needs to be applied in order to properly test each mass point.


\section{Non-resonant production mode}
\label{sec:non_res:comb}

\subsection{SM-like production}

ATLAS has searched for the non-resonant production in the \hhbbyy, \hhbbbb, \hhbbtt, \hhbbww, \hhwwyy and \hhwwww final states. For the \hhbbww final state, only the single lepton channel has been included in the combination. CMS has instead combined searches in the \hhbbyy, \hhbbbb, \hhbbtt and \hhbbVV channels, using only the di-lepton channel for \hhbbVV. 
The 95\% CL expected and observed upper limits on the signal strength $\mu = \sigma_{\hh} /\sigma_{\hh}^{SM}$ are reported in \refta{exp-summary-table} for each individual final state. Their combination has allowed the two experiments to set an observed (expected) upper limit on \hh production at 6.9 (10), 22.2 (13) times the SM, for ATLAS and CMS respectively,  using data collected in 2015 and 2016 at 13 TeV~\cite{Sirunyan:2018two,Aad:2019uzh}.

\begin{table}[ht!]
\begin{center}
{
\begin{tabular}{lc|cc}
\hline
Search channel & Collaboration & \multicolumn{2}{|c}{95\% CL Upper Limit}  \\
&  & observed & expected \\
\hline
\multirow{2}{*}{\bbbb} & 
ATLAS &
13 & 21 \\
                        & 
CMS & 
75 & 37 \\
\hline
\multirow{2}{*}{\bbyy} & 
ATLAS & 
20 & 26 \\
                                  & 
CMS & 
24 & 19 \\
\hline
\multirow{2}{*}{\bbtautau} & 
ATLAS &  
12 & 15 \\
                              & 
CMS &
32 & 25 \\
\hline
\multirow{2}{*}{\bbvv ($\ell \nu \ell \nu$)}* & 
ATLAS & 
40 & 29 \\                            & 
CMS & 
79 & 89 \\
\hline
\multirow{2}{*}{\bbww ($\ell \nu q q$)} & 
ATLAS & 
305 & 305 \\
                              & 
CMS & 
 -- & --  \\
\hline
\multirow{2}{*}{\wwyy} & 
ATLAS & 
230 & 160 \\
                              & 
CMS &  -- & -- \\
\hline
\multirow{2}{*}{\wwww} & 
ATLAS & 
160 & 120 \\
                              & 
CMS &  -- & -- \\
\hline

\multirow{2}{*}{Combined} &
ATLAS &
6.9 & 10 \\
                             &
CMS &  22 & 13 \\
\hline
\end{tabular}
}
\end{center}
\vspace*{-0.4cm}
\caption{List of \hh searches at the LHC based on the $p-p$ data collected by ATLAS and CMS at 13 TeV and corresponding to about 36~\ifb. Observed and expected upper limits on the SM \hh production cross section are normalised to the SM prediction~\cite{deFlorian:2016spz}. The ATLAS search for \bbvv$(\ell \nu \ell \nu)$ is not included in the combination and uses 139\ifb of integrated luminosity.}\label{exp-summary-table}
\end{table}

The best final state for the non-resonant \hh production is \bbtautau in ATLAS, and \bbyy in CMS. For each experiment the combined sensitivity of all channels together is improved by 30\% with respect to the best channel. This can be easily explained by a relatively comparable sensitivity of the \hhbbyy, \hhbbtt and \hhbbbb final states.

The expected upper limit on non-resonant SM \hh production cross section for ATLAS decreases by 71\% with respect to the best single channel (\bbtautau), and for CMS by 68\%, with respect to the best limit provided by the \bbyy channel. 
This shows that the combination significantly outperforms single channel performance, as a result of the comparable sensitivity of the \hhbbyy, \hhbbtt and \hhbbbb final states in particular.

 The differences between the ATLAS and CMS sensitivities in each channel, are also the result of different optimisation of the experimental analysis strategies, besides object reconstruction performance. ATLAS employs BDT discriminators for all the analysis categories in the \hhbbtt search, boosting the sensitivity of this final state with respect to the analogues CMS search. Likewise CMS uses a sophisticated MVA categorisation for the \bbyy search, while the equivalent ATLAS search does not. Future improvements in the analysis techniques would lead to a further increase of the sensitivity, in addition to the larger integrated luminosity that will become available.
Besides \bbtautau and \bbyy, in ATLAS also the \bbbb is one of the main final state contributing to the combined result, thanks to the good \btagging performance and improved \bjet triggers (see Sec.~\ref{sec:bTagging} and ~\ref{sec:bTrigger}). 
Both the ATLAS and CMS experiments have used the value of 33.53~fb as the \hh production cross section predicted by the SM, it has been recently updated from Ref.~\cite{deFlorian:2016spz}. A more recent evaluation recommends a value of  31.05~fb (see \refta{table:xsec2}), this was used to derive the HL-LHC projections~\cite{Grazzini:2018bsd} reported in Chapter~\ref{chap:hl-lhc}. For more details on the theoretical prediction, see Chapter~\ref{chap:HHcxs}. 
The impact of systematic uncertainties is currently not negligible. ATLAS has evaluated the expected sensitivity to the SM non-resonant production in the ideal case where no systematic uncertainties are considered and quotes an improvement of about 13\% on the upper limit. 

\subsection{Higgs self-coupling constraint}

As described in Sec.~\ref{subsec:EFTtheory} it is possible to consider special classes of new physics models, that modify only the Higgs boson self-coupling, $\lambda_{HHH}$, as $\klambda = \lambda_{\rm HHH}/\lambda_{\rm HHH}^{\rm SM}$.

 Different techniques have been developed to test several $\klambda$ values, limiting the number of events to simulate.
The gluon-gluon fusion $\hh$ production process depends on the box and triangle amplitudes as described in Sec.~\ref{sec:production_modes},
and the differential $pp \to \hh$ cross section can be expressed as a second degree polynomial in $\klambda$,

\begin{equation}
\frac{d\sigma}{d\Phi} = A + B\klambda + C \klambda^2
\end{equation}

where $d\Phi$ represents the infinitesimal phase space volume. This expression is valid at all order in QCD, higher order QCD corrections will affect, in fact, the values of the $A,B,C$ coefficients but not the functional dependence from $\klambda$.

In a first approach, this feature can be used to simulate \hh events for any value of $\klambda$ using only three different hypotheses for $\klambda$ by solving the following system of equations:

\begin{eqnarray}
\left (\frac{d\sigma}{d\Phi}\right )_1 = A + B\kappa_{\lambda_1} + C \kappa_{\lambda_1}^2 \\
\left (\frac{d\sigma}{d\Phi}\right )_2 = A + B\kappa_{\lambda_2} + C \kappa_{\lambda_2}^2 \\
\left (\frac{d\sigma}{d\Phi} \right)_3 = A + B\kappa_{\lambda_3} + C \kappa_{\lambda_3}^2 
\end{eqnarray}
and computing the dependence of the coefficient $A,B,C$ from the differential cross section in a given phase space. In practice this is achieved with a linear combination of the three reference hypotheses with coefficients obtained by the inversion of the $3\times3$ coefficient matrix obtained from the equation above. A natural choice for two $\klambda$ values of the reference samples is $\klambda = 0, 1$, corresponding to the box-only and the SM cases. In order to optimise the signal generation, the third value can be chosen close to the expected sensitivity, which corresponds to $\klambda = 10, 20$ depending on the individual final state. 

\begin{figure}[hbt]
    \centering
    \includegraphics[width=0.45\textwidth]{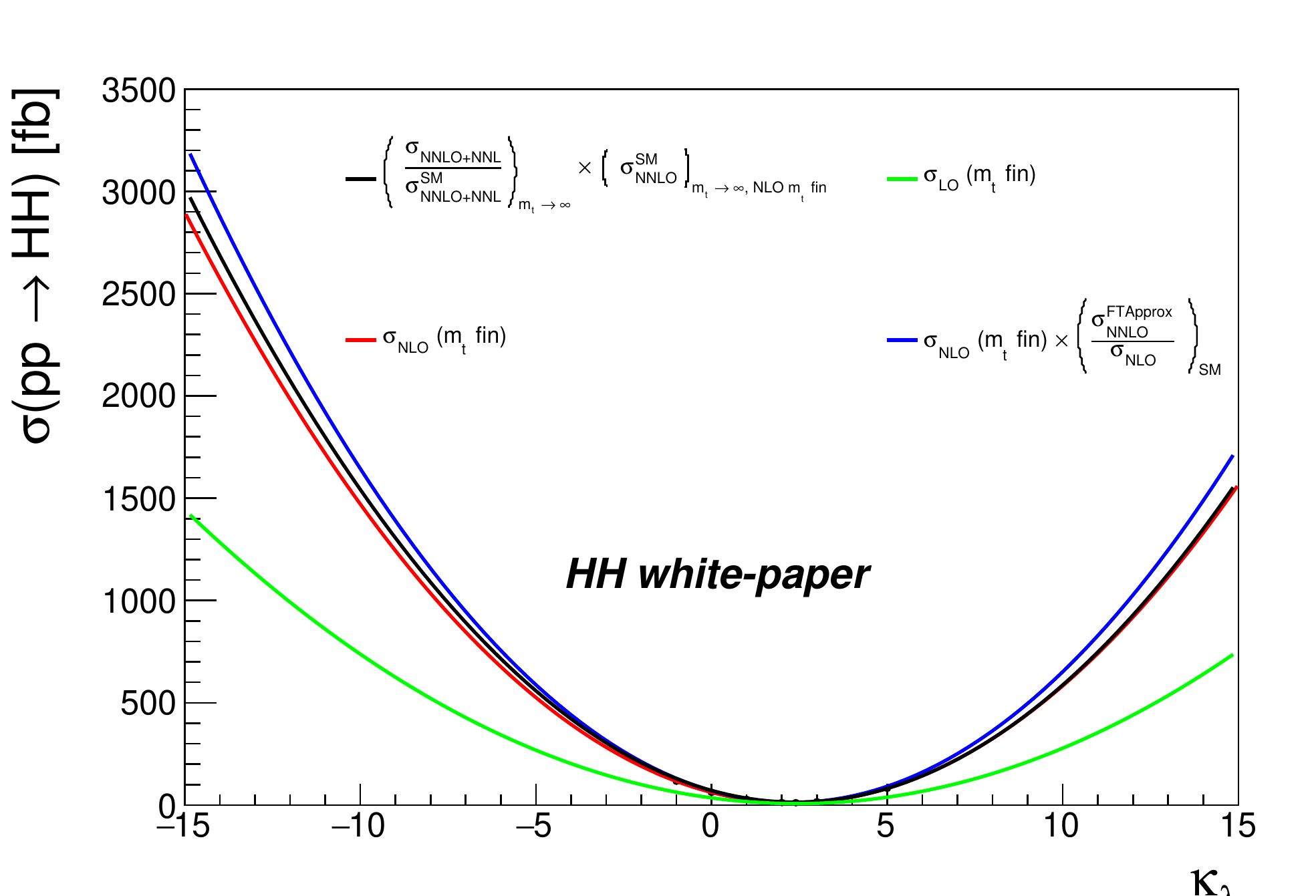}
     \includegraphics[width=0.45\textwidth]{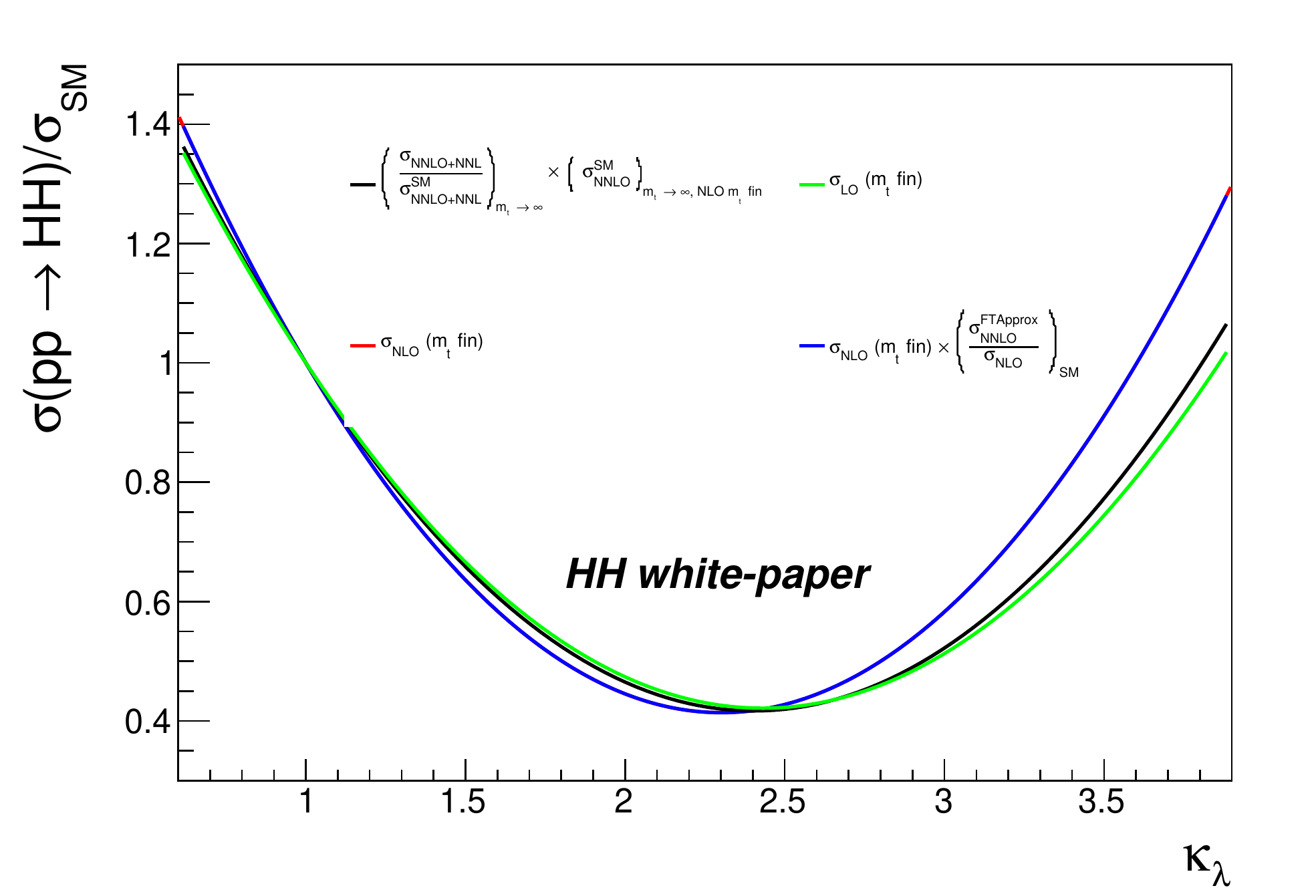}
    \caption{Left: $pp \to \hh$ production cross section as a function of $\klambda$. Right: ratio of the  $pp \to \hh$  to its SM expectation, obtained for $\klambda =1$. Different calculations, as used by the LHC experiments are shown. }
    \label{fig:cross_plots}
\end{figure}

The three samples have to be properly normalised to the best cross section prediction, which can be also parameterised as a second degree polynomial with coefficients $a,b,c$. 
Figure~\ref{fig:cross_plots} shows the comparison of several cross section predictions: LO; NNLO+NNLL in the $m_t \to \infty$ approximation rescaled with the NNLO+NNLL SM cross section obtained including finite $m_t$ NLO contribution and $m_t \to \infty$ NNLO corrections;   finite $m_t$ NLO for all $\klambda$ values rescaled with the NNLO SM cross section obtained with the FTApprox method (partial $m_t$ finite). 

The corresponding second degree polynomial parameters are shown in Table~\ref{tab:cross_parab_summary}. The ratio of the parameters to their LO computation is also shown and it is almost equal for all parameters for the prediction shown in the last row, that is actually used by the LHC experiments, but shows differences up to 15\% with the recent finite $m_t$ NLO computation, second and third row of the table.

\begin{table}[hbt]
    \centering
    \begin{tabular}{lcccccc}
    \hline
    computation & A [fb] & A/A(LO) & B [fb] & B/B(LO)& C [fb] & C/C(LO)\\
    \hline
    LO $m_t$ fin  & 35.0 & & -23.0 & & 4.73 & \\
    NLO $m_t$ fin & 62.6 & 1.79 & -44.4 & 1.93 & 9.64 & 2.04 \\
    NLO $m_t$ fin $\times$ NNLO SM FTApprox & 70.0 & 2.00 & -49.6 & 2.16 & 10.8 & 2.28 \\
   NNLO + NNLL $m_t \to \infty$ $\times$ & &  &  &  &  & \\
    \hspace{1 cm} NNLO+NLL SM (partial $m_t$ fin) & 71.3 & 2.04 & -47.7 & 2.08 & 9.93 & 2.10\\
    \hline
    \end{tabular}
    \caption{Second order polynomial parameters ($A+B\cdot \klambda + C\cdot \klambda^2$) for different computations as used by the LHC experiments and new recommendations. The column X/X(LO) shows the ratio of the parameter with respect to their LO prediction. }
    \label{tab:cross_parab_summary}
\end{table}
\clearpage

A second approach~\cite{Carvalho:2016rys} is derived from the clustering technique, as described in Sec. \ref{sec:shape_bench}, and requires the production of a large number of samples at the LO at generator level (LHE files). Then, the following ratio is built:

\begin{equation}
{\rm R}_{ HH} \equiv \frac{\sigma_{ HH}}{\sigma_{ HH}^{\rm SM}} 
\end{equation}

where the A, B, C coefficients are extracted in slices of $\mhhAlt$ and $\cos\theta^*_{ HH}$ once for all. These weights can then be used to reweight \hh events for any value of \klambda. The corresponding cross section value is then obtained by rescaling the best SM prediction:

\begin{equation}
\sigma_{\hh} = \sigma_{\hh}^{\rm best~precision} \cdot \text{R}_{\hh}\,,
\label{eq:cxNNLO}
\end{equation} 

While the first method properly takes into account the best predictions available up to date and is well suited to test several $\klambda$ values, it is rather complex to extend to a large number of EFT parameters. 
The second method takes advantage that the relative coefficients are rather independent of the QCD order, at which the calculations are performed, as already observed in the first method. It is less precise but one can account once for all in 5D EFT space the 15 parameters that are necessary to describe it. 
In the following ATLAS uses the first method while CMS the second.

\begin{figure}[ht!]
\begin{center}
\includegraphics[width=0.50\textwidth]{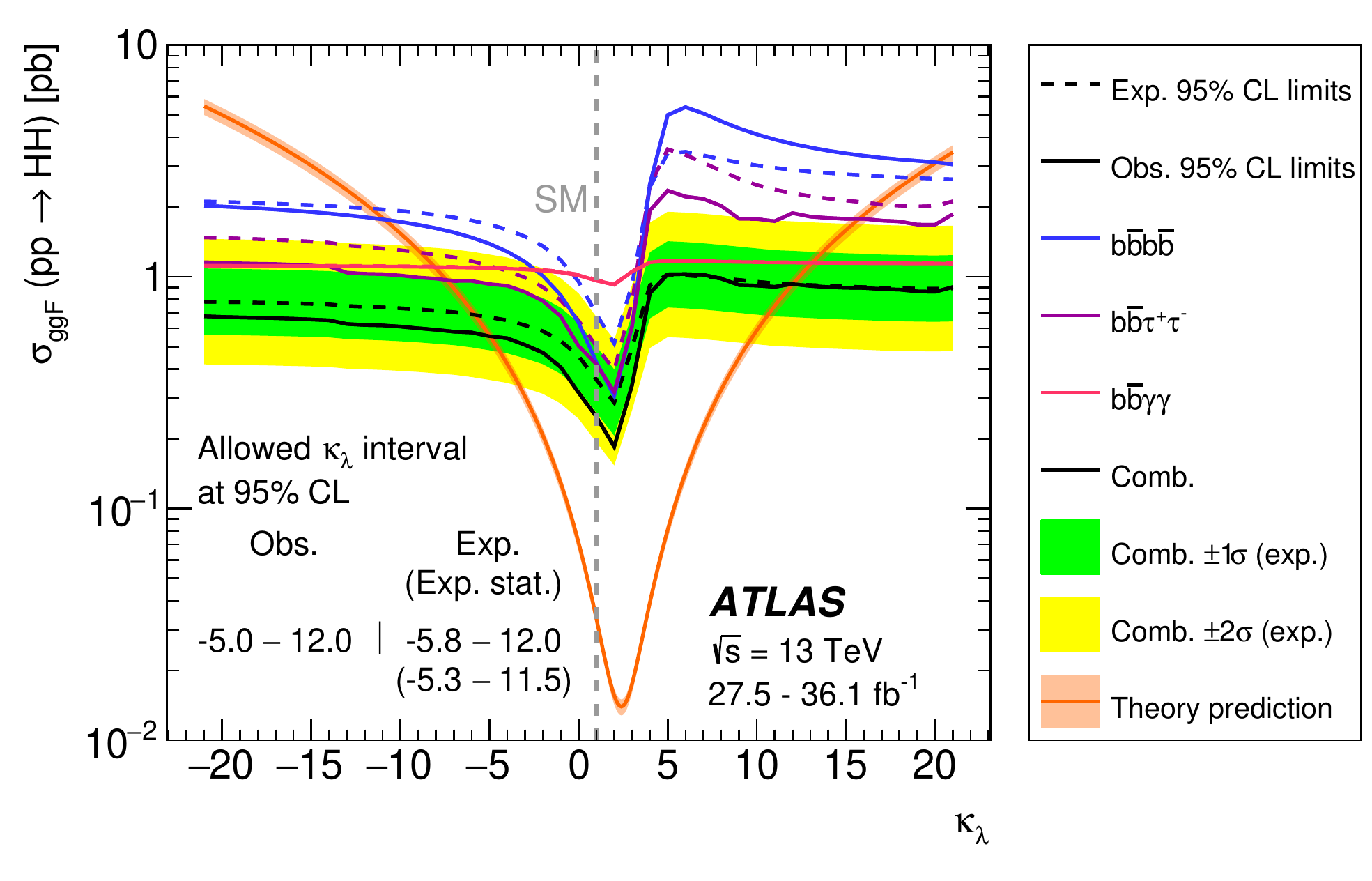}
\includegraphics[width=0.48\textwidth]{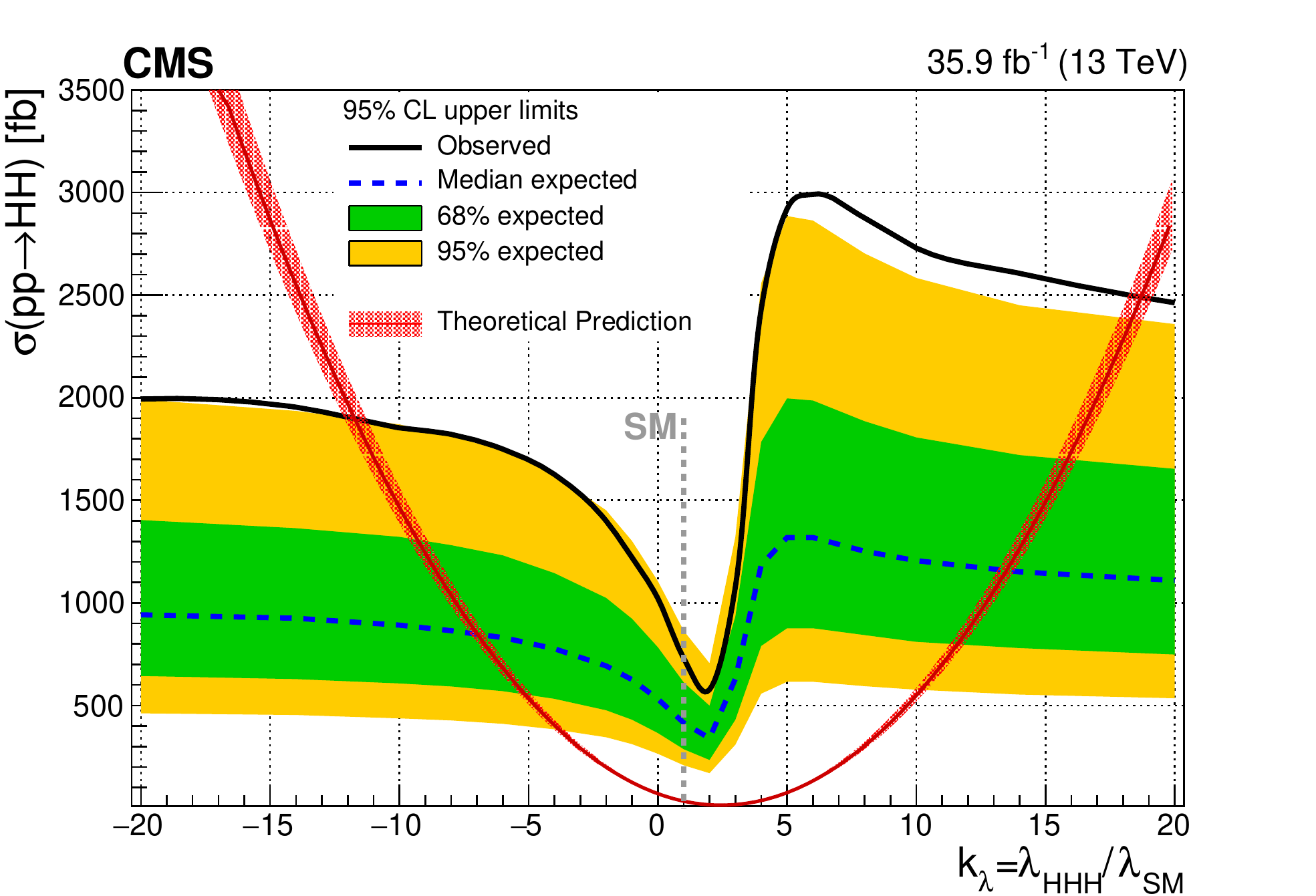}
\caption{Expected and observed 95\% CL upper limits on the \hh production cross section as a function of $\klambda$ for ATLAS (left) and CMS (right)~\cite{Sirunyan:2018two,Aad:2019uzh}. The SM expectation and its uncertainty are also reported. All other Higgs boson couplings are set to their SM values. }
\label{fig:comb:lambda}
\end{center}
\end{figure}

Any modification to the \klambda value would affect both the \hh production cross section and decay kinematics. These effects are fully simulated for each \klambda value considered in the scan performed by the ATLAS and CMS collaborations. Modifications to the Higgs boson decay branching fractions through one loop electroweak corrections (see Sec.~\ref{tril-single}) are not considered in the analyses of the two collaborations, although they can modify the results up to 10\%. Figure~\ref{fig:comb:lambda} shows the upper limit on $\sigma(pp \to \hh)$ for a given value of \klambda published by the ATLAS and CMS collaborations.

\begin{figure}[ht!]
\begin{center}
\includegraphics[width=0.51\textwidth]{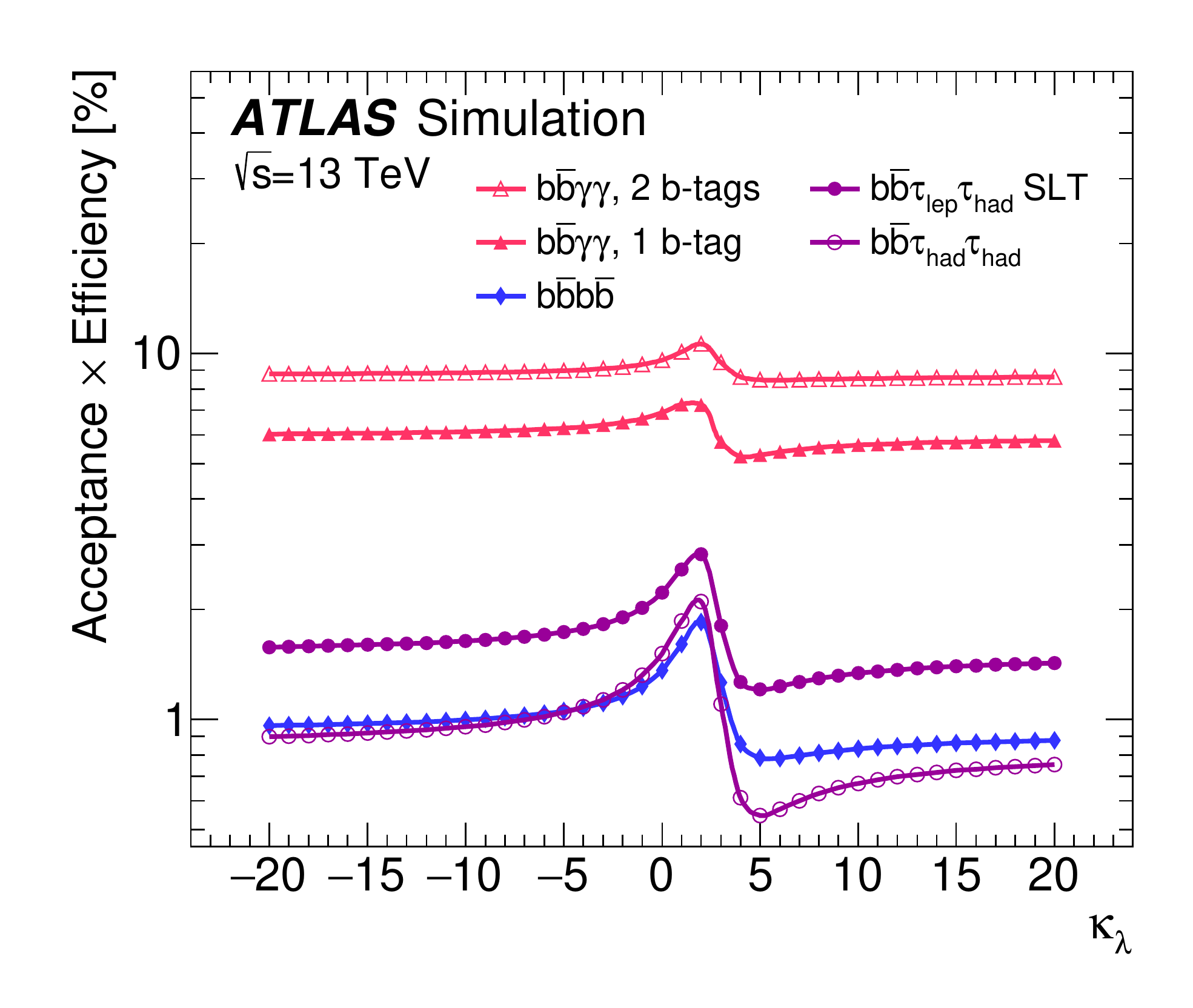}
\end{center}
\vspace*{-0.8cm}
\caption{\hh signal acceptance times efficiency as a function of \klambda for the ATLAS \hhbbbb, \hhbbtt and \hhbbyy searches~\cite{Aad:2019uzh}. }
\label{fig:hhaccept}
\end{figure}

\begin{figure}
     \centering
     \includegraphics[width=0.7\textwidth]{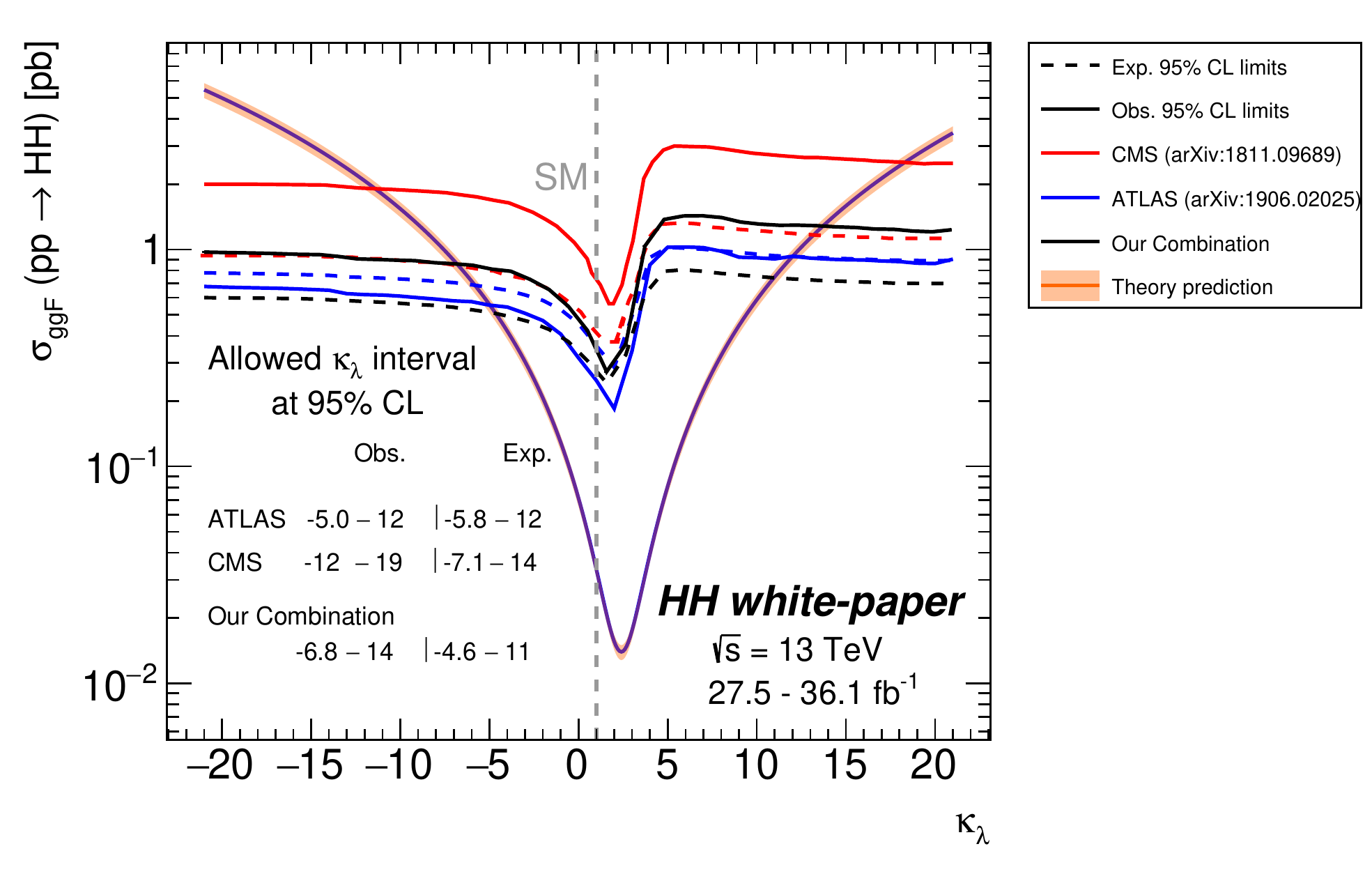}
     \vspace*{-0.5cm}
     \caption{95\% CL upper limits on the $\sigma(pp \to \hh)$ cross section as a function of $\klambda$. The ATLAS and CMS limits are shown together their statistical combination. }
     \label{fig:comb:ATLAS_CMS}
 \end{figure}

The shape of the upper limit curve follows the signal acceptance, shown in Fig.~\ref{fig:hhaccept}, for the \hhbbtt and \hhbbyy case. In the \bbbb search also the invariant mass of the four \bjets, which is used to extract the signal, is affected by $\klambda$, while the BDT score and the \myy distributions used to extract the signal in the \hhbbtt and \hhbbyy analysis respectively, do not show a \klambda dependence as strong as in the \bbbb final state.  
The dependence of the signal acceptance from \klambda is shown in Fig.~\ref{fig:hhaccept}.
The maximum of the acceptance is obtained for $\klambda \sim 2$, where the cross section is minimum as shown in Fig.~\ref{fig:comb:lambda}. This $\klambda$ value corresponds to the maximum destructive interference between the box and the triangle diagrams, resulting in a harder \mhh spectrum (see Fig.~\ref{fig:chhh_3D} and Fig.~\ref{fig:mhh_eft})  that increases the signal acceptance.
For $|\klambda|>10$ the triangle diagram becomes dominant and the upper limit becomes symmetric in $\klambda$.
 The ATLAS and CMS combined upper limits on the \hh cross section as function of $\klambda$ are shown in Fig.~\ref{fig:comb:ATLAS_CMS}.
 The combination of the ATLAS and CMS results is also shown. It has been derived from the published ATLAS and CMS expected and observed upper limits, assuming that the likelihoods have gaussian shape and that, for each of the two experiments, the observed and expected likelihoods differ only by a shift on the mean value, while they have the same width. 
 The combination of ATLAS and CMS results has been performed without including correlation of systematic errors between the two experiments. Due to the $~2\sigma$ excess in CMS results, the combined observed result is slightly worse than the ATLAS one.
 
The corresponding intervals where the \klambda is observed (expected) to be constrained at 95\% CL are listed in ~\refta{tab:comb:summary} for the main channels.

\begin{table}[ht!]
\begin{center}
{
\begin{tabular}{lccc}
\hline
Final state & collaboration & \multicolumn{2}{c}{allowed \klambda interval at 95\% CL} \\
               &             & observed & expected \\
\hline
\multirow{2}{*}{\bbbb} & ATLAS & -11 -- 20 & -12 -- 19 \\
                        & CMS   & -23 -- 30  & -15 -- 23  \\
\hline
\multirow{2}{*}{\bbtautau} & ATLAS & -7.3 -- 16 & -8.8 -- 17 \\
                              & CMS   & -18 -- 26  & -14 -- 22  \\
\hline
\multirow{2}{*}{\bbyy} & ATLAS & -8.1 --13 & -8.2 -- 13 \\
                                  & CMS   & -11 -- 17 & -8.0 -- 14 \\
\hline
\hline
\multirow{2}{*}{Combined} & ATLAS & -5.0 -- 12  & -5.8 -- 12 \\
                          & CMS   & -12 -- 19 & -7.1 -- 14 \\
\hline
Our combination & Both experiments & -6.8 -- 14 & -4.6 -- 11 \\
\hline
\end{tabular}
}
\end{center}
\vspace*{-0.5cm}
\caption{\label{tab:comb:summary} The observed and expected 95\% CL intervals on \klambda for the combination and the in\-di\-vi\-dual final states  analysed for non-resonant \hh production at 13 TeV with about 36~\ifb. All other Higgs boson couplings are set to their SM values~\cite{Sirunyan:2018two,Aad:2019uzh}. 
The \bbbb CMS values are obtained by extrapolating the published CMS values outside the published range [-20,20].}
\end{table}

\subsection{More general EFT scans}

As discussed in Chapter~\ref{chap:EFT}, BSM contributions could be constrained in a model independent approach using the EFT formalism. In the HEFT model (Sec.~\ref{sec:heft}) five anomalous Higgs boson couplings (Eq.~\ref{eq:ewchl}) relevant for \hh production are identified: $c_{hhh} \equiv \klambda$, $c_t \equiv \kappa_{t}$ and three additional interaction vertexes $c_{tt}$, $c_{gghh}$ and $c_{ggh}$.

\begin{figure}
    \centering
    \includegraphics[width=0.65\textwidth]{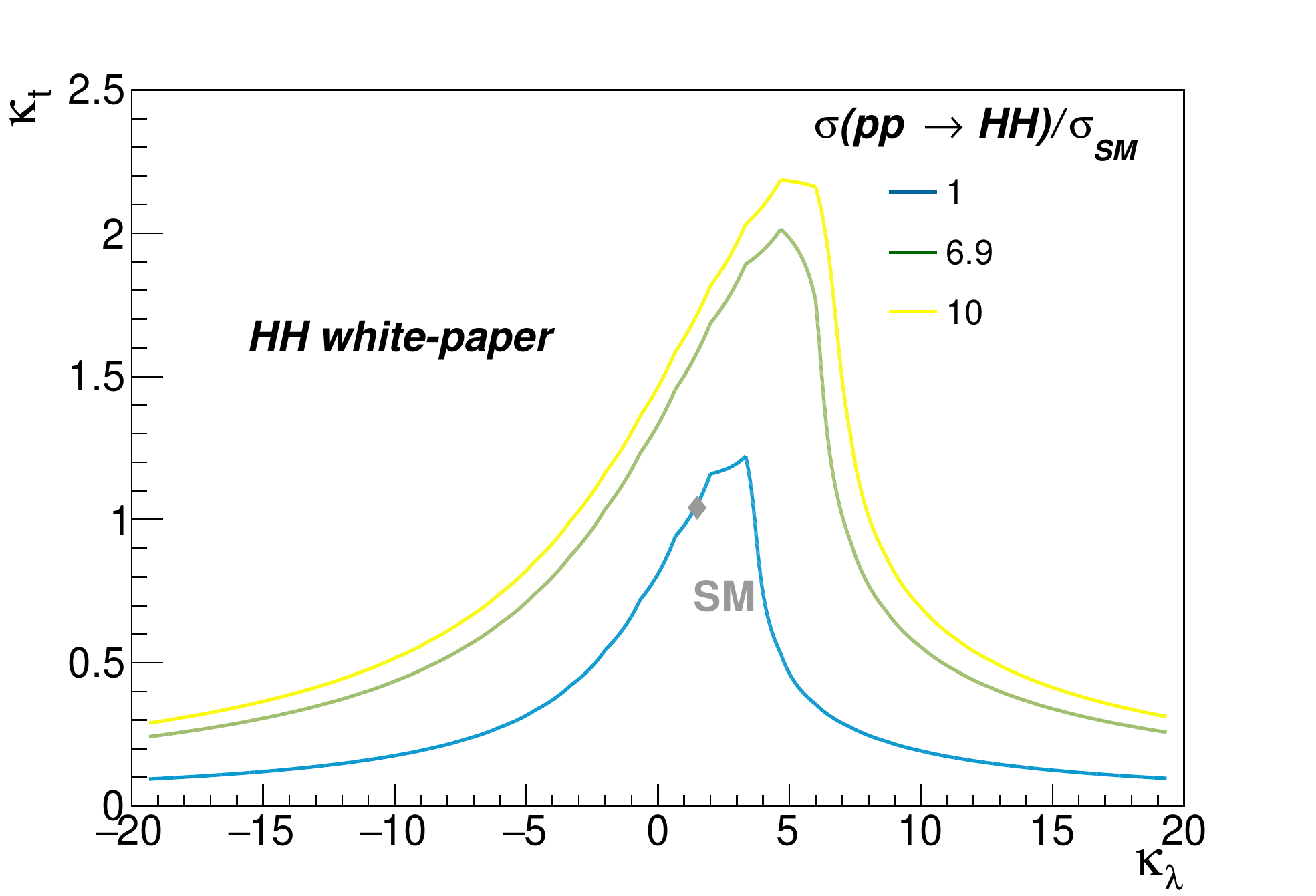}
    \caption{Contour level of $\sigma(pp \to \hh)/\sigma_{\rm SM}$ as function of $\kappa_t$,$\klambda$, under the assumption of no additional Higgs coupling vertices, as derived in Eq.\ref{eq:eqtop}. The diamond indicates the SM predicted value. The reference values of 6.9 and 10 correspond to the best available observed and expected upper limits on the $\sigma(pp \to \hh)$ cross section as measured by the ATLAS experiment.} \label{fig:klkt}
\end{figure}

When imposing no new interactions in the model: $ c_{tt} = c_{gghh} = c_{ggh}=0$, the $pp \to \hh$ cross section depends only from $\kappa_t$ and $\klambda$ through the diagrams in Fig.~\ref{fg:pp2hh} (a). 
The BSM amplitude of the process can then be written as:
\[
\mathcal{A} = \kappa_t^2 \mathcal{A}_1 + \kappa_t\kappa_{\lambda}\mathcal{A}_2
\]
where $\mathcal{A}_1$ and $\mathcal{A}_2$ are given by the SM top-box and triangle diagrams.
The cross section is proportional to $|\mathcal{A}|^2$ therefore the following expression holds:
\begin{equation}
\sigma_{pp \to HH}(\kappa_t,\klambda)
\propto \left (\kappa_t^4\overline{|\mathcal{A}_1|^2} + 2\kappa_t^3\klambda \overline{\Re \mathcal{A}_1 \mathcal{A}^{*}_2}  + \kappa_t^2\klambda^2 \overline{|\mathcal{A}_2|^2} \right)
\end{equation}
where with the overline we indicate the average of the quantity over the phase space of the process, factorising $\kappa_t^4$ we obtain
\begin{equation}
\sigma_{pp \to HH}(\kappa_t,\klambda)
\propto \kappa_t^4\left [\overline{|\mathcal{A}_1|^2} + 2\left (\frac{\klambda}{\kappa_t}\right ) \overline{\Re \mathcal{A}_1 \mathcal{A}^{*}_2}  + \left ( \frac{\klambda}{\kappa_t} \right)^2 \overline{|\mathcal{A}_2|^2} \right]
\label{eq:eqtop}
\end{equation} 
From this expression it is clear that it is impossible to extract $\klambda$ constraints from \hh production without assumptions on $\kappa_t$, this is more evident in the representation in Fig.~\ref{fig:klkt}.
The $\kappa_t$ and $\klambda$ parameters can be constrained also using single Higgs measurements as described in Sec.~\ref{singleH_exp}, these measurements impose a different correlation pattern between $\kappa_t$ and $\kappa_{\lambda}$, therefore a future combination of single $H$ and \hh measurements is expected to provide a more model independent determination of $\klambda$.

\begin{figure}[ht!]
\begin{center}
\includegraphics[width=0.99\textwidth]{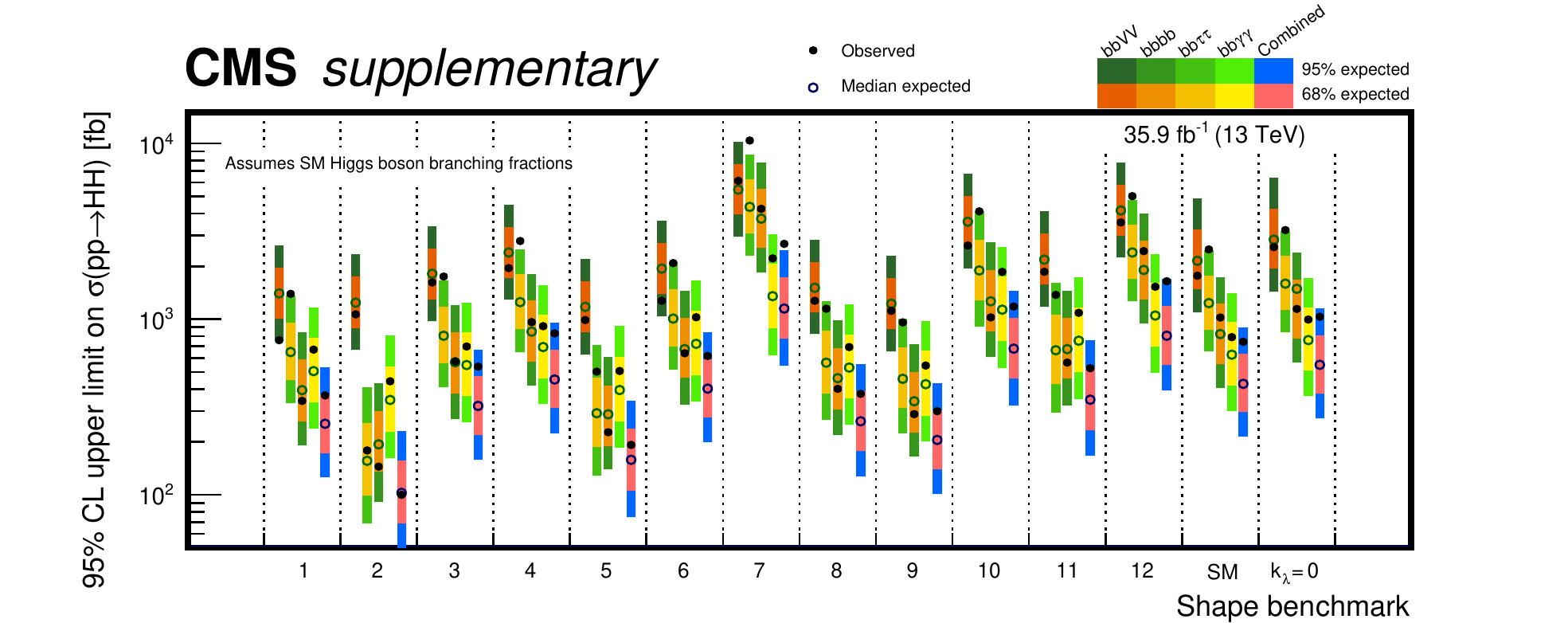}
\caption{The 95\% CL upper limits on the non-resonant \hh cross section for different EFT benchmark topologies (bins 1 to 12). Each benchmark represents a possible modification in both the predicted rate and kinematic distributions. The last two bins show the 95\% CL upper limits for $\klambda=1$ (SM) and 0. Each of the four final states is shown separately together with their combination~\cite{Sirunyan:2018two}.}
\label{fig:comb:eft}
\end{center}
\end{figure}

A five dimensional scan of the HEFT couplings is computationally excessive, therefore a clustering strategy has been developed to group together
possible combinations of coupling values that present similar kinematic properties as discussed in detail in Sec.~\ref{sec:shape_bench}. Twelve
clusters have been identified, in addition to the SM ($\klambda=1$) and the $\klambda=0$ scenarios. Within each
cluster, the representative points in the EFT space shown in \refta{tab:EFTpnt} are identified as benchmarks.
Each benchmark predicts a different \mhh distribution as shown in Fig.~\ref{fig:EFTbenchmhh}, that affects the signal acceptance and the final discriminant of the analyses determining different sensitivities for different benchmark points. The CMS experiment has adopted this approach and provided the observed and expected exclusion limits on the \hh cross section for the different EFT benchmarks, which are shown in Fig.~\ref{fig:comb:eft}. 

\section{Resonant \hh production mode}
In addition to the non-resonant production, searches for resonant \hh are performed in the \mhh range from 250 to 3000 GeV, for spin-0 under the narrow width approximation\footnote{The width of the signal mass distribution is much smaller than the experimental resolution.} and spin-2 resonances (see Sec.~\ref{sec:spin2}).

\begin{figure}[ht!]
\begin{center}

\includegraphics[width=0.63\textwidth]{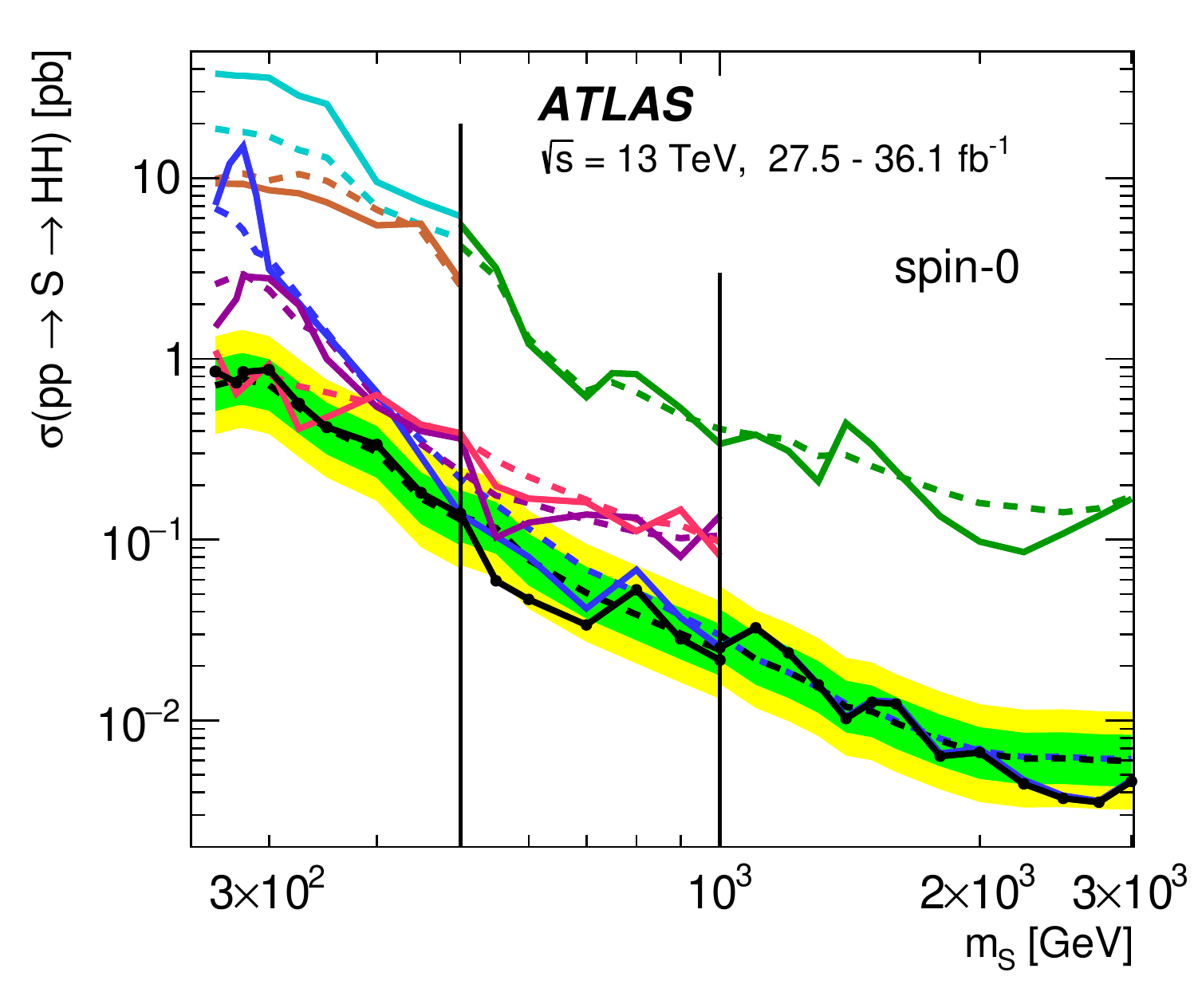}\includegraphics[width=0.22\textwidth]{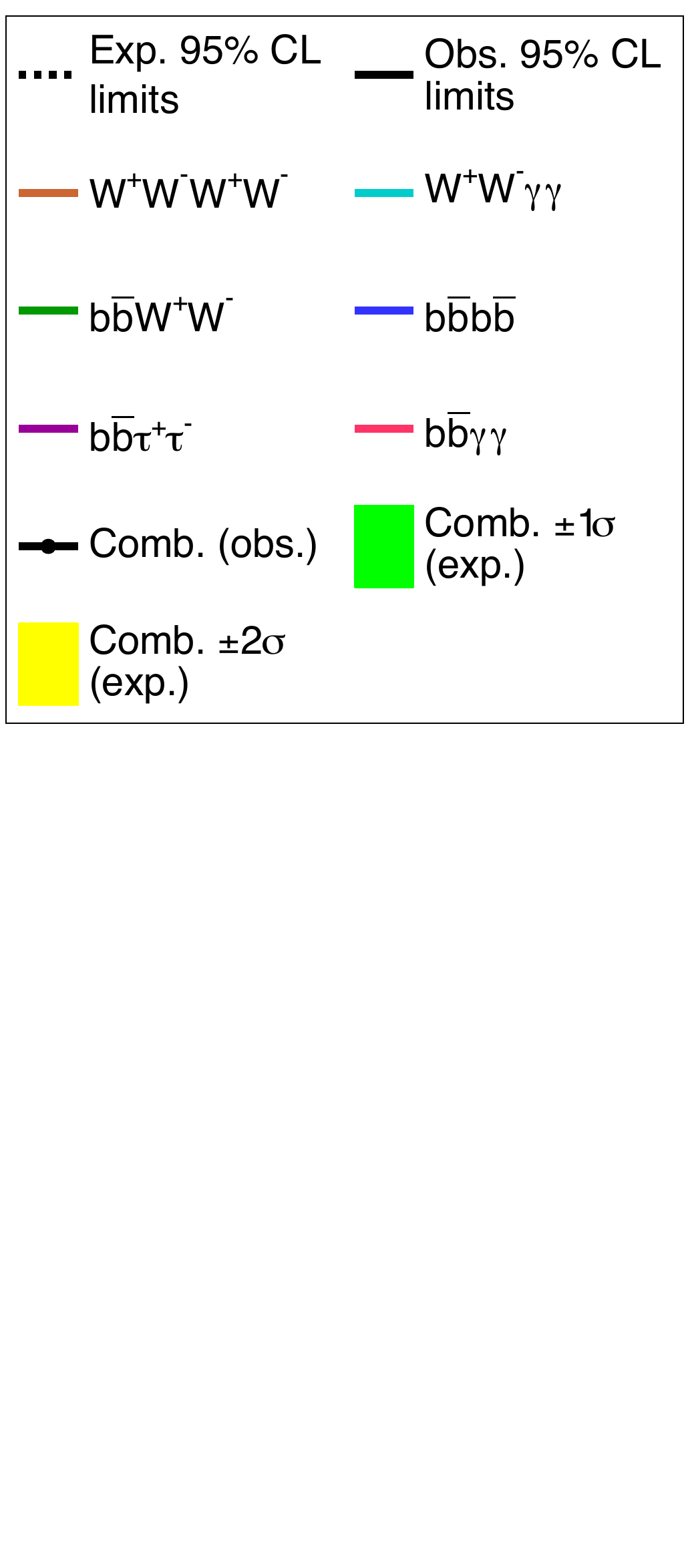} \\

\includegraphics[width=0.72\textwidth]{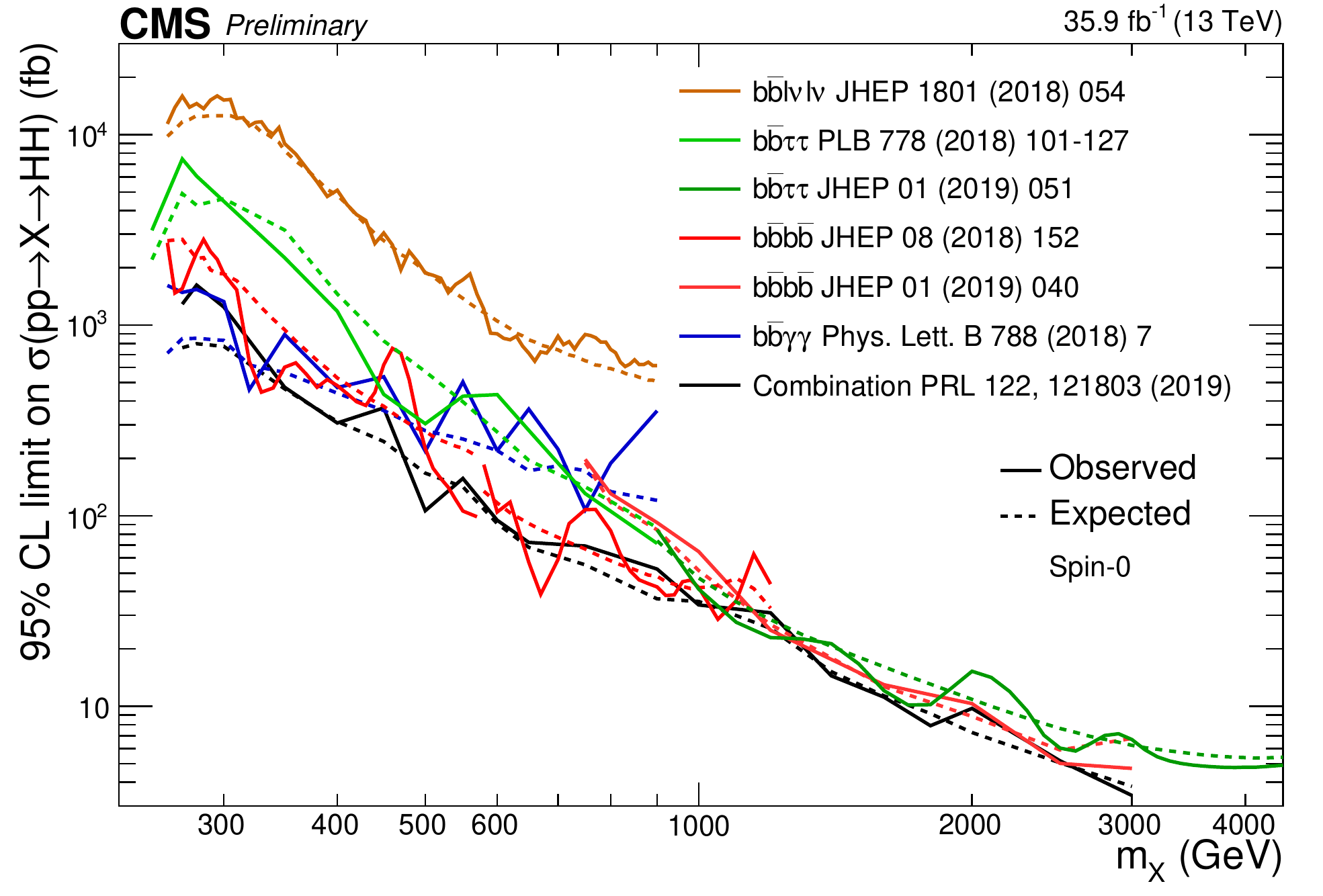}
\end{center}
\caption{
Expected and observed 95\% CL exclusion limits on the production cross section of a narrow, spin zero resonance (S or X) decaying into a pair of Higgs bosons.
Top: ATLAS combination and breakdown by final state for m$_{S}<$3~TeV~\cite{Aad:2019uzh};
Bottom: CMS combination and breakdown by final state for m$_{S}<$3~TeV~\cite{Sirunyan:2018two}.}
\label{fig:comb:scalar}
\end{figure}

\begin{figure}[ht!]
\includegraphics[width=0.5\textwidth]{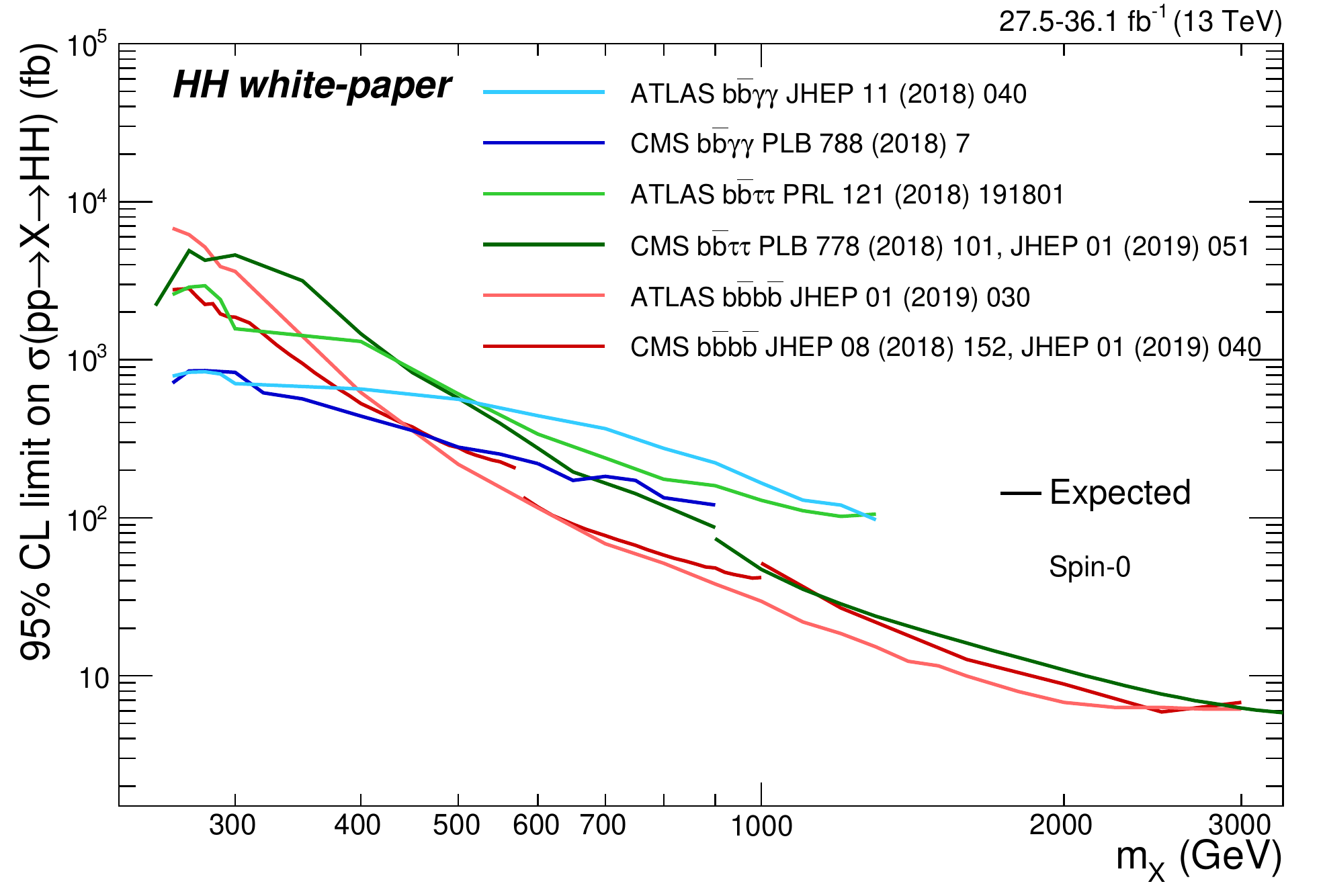} 
\includegraphics[width=0.5\textwidth]{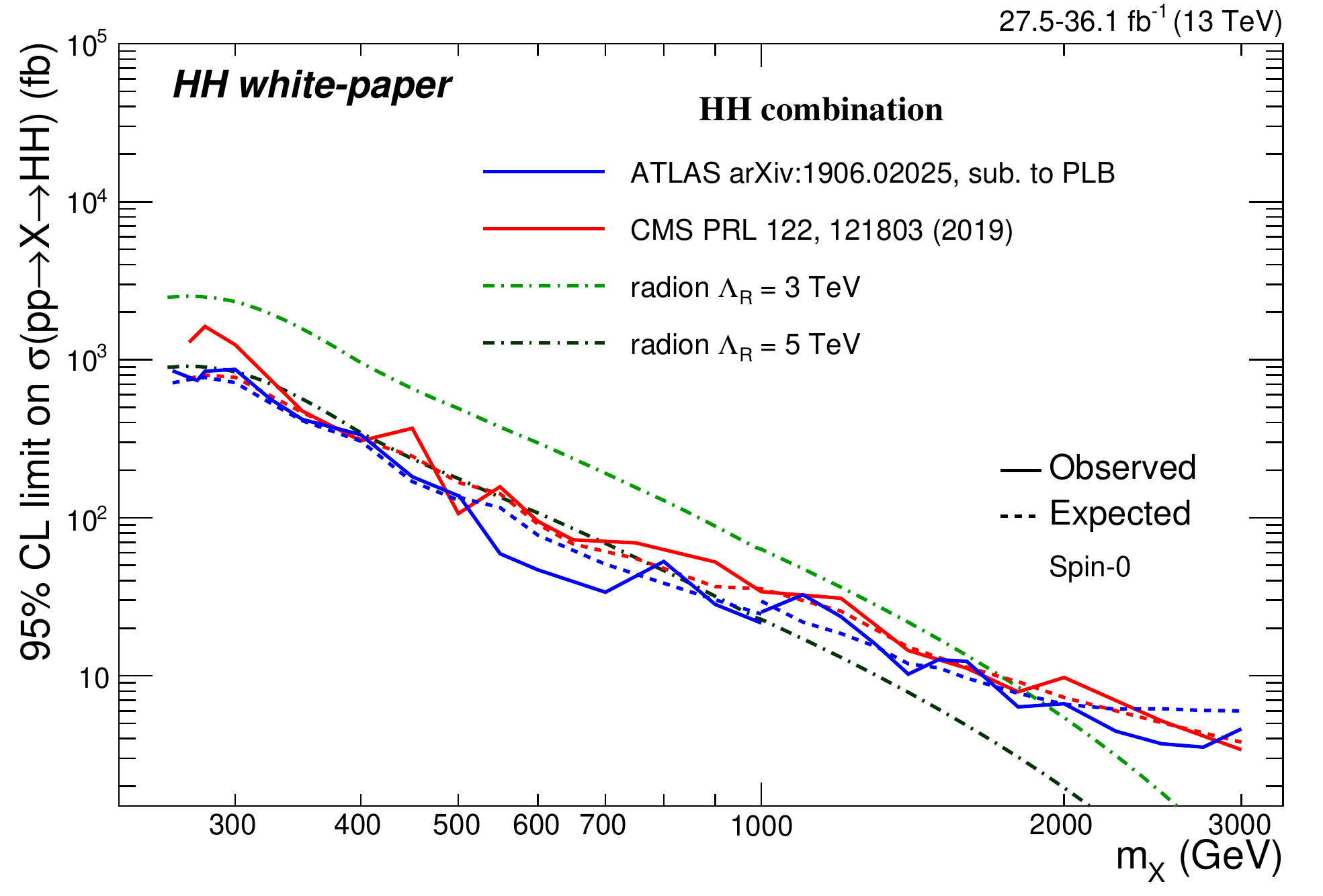}
\caption{
Left: expected 95\% CL upper limits on the resonant production cross section of a narrow, spin-0 resonance (S or X) decaying into a pair of Higgs bosons for the most sensitive channels: \bbbb, \bbtautau and \bbyy.
Right: expected and observed 95\% CL combined exclusion limits on the resonant spin-0 \hh production. A theory prediction from bulk radion model is overlaid \cite{Goldberger:1999uk, Oliveira:2014kla}.}
\label{fig:spin0_ATLAS_CMS_comp}
\end{figure}

For the resonant hypothesis the ATLAS and CMS collaborations have both analysed the \bbbb, \bbww, \bbtautau and \bbyy channels. ATLAS has also included \wwww and \wwyy, while CMS \bbvv has investigated the di-lepton final state. 
No evidence for a signal is observed, and upper limits at 95\% CL have been set on the production cross section for spin-0  resonances and they are shown in Fig.~\ref{fig:comb:scalar}. In the same figure limits obtained by the combination of all channels are also shown.
The most sensitive channels for both the ATLAS and CMS collaborations are \bbbb, \bbtautau and \bbyy. Their sensitivity for the ATLAS and CMS experiments are compared in Fig.~\ref{fig:spin0_ATLAS_CMS_comp}.

\begin{figure}[ht!]\begin{center}
\subfloat[]{\includegraphics[width=0.6\textwidth]{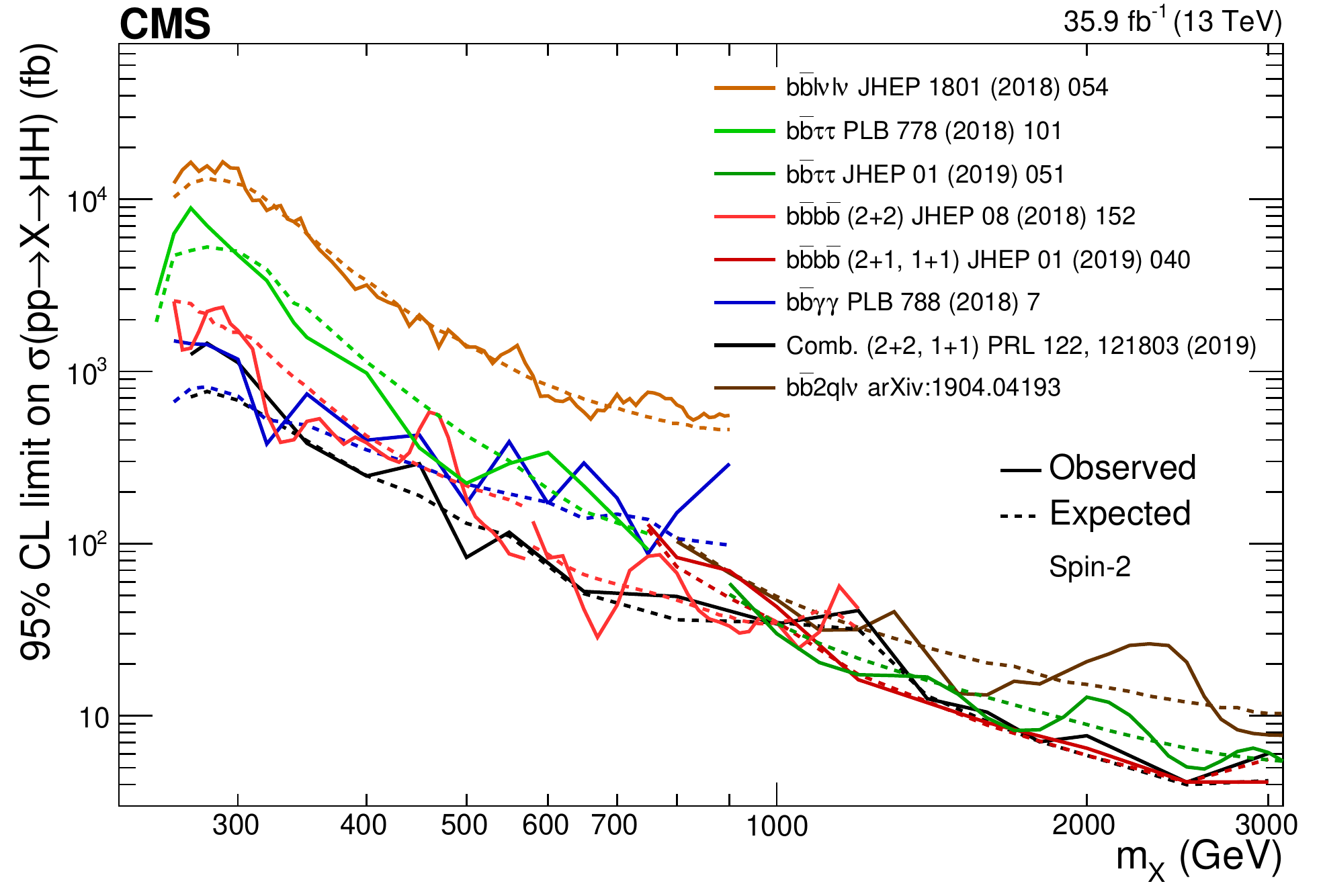}}\\
\subfloat[]{\includegraphics[width=0.42\textwidth]{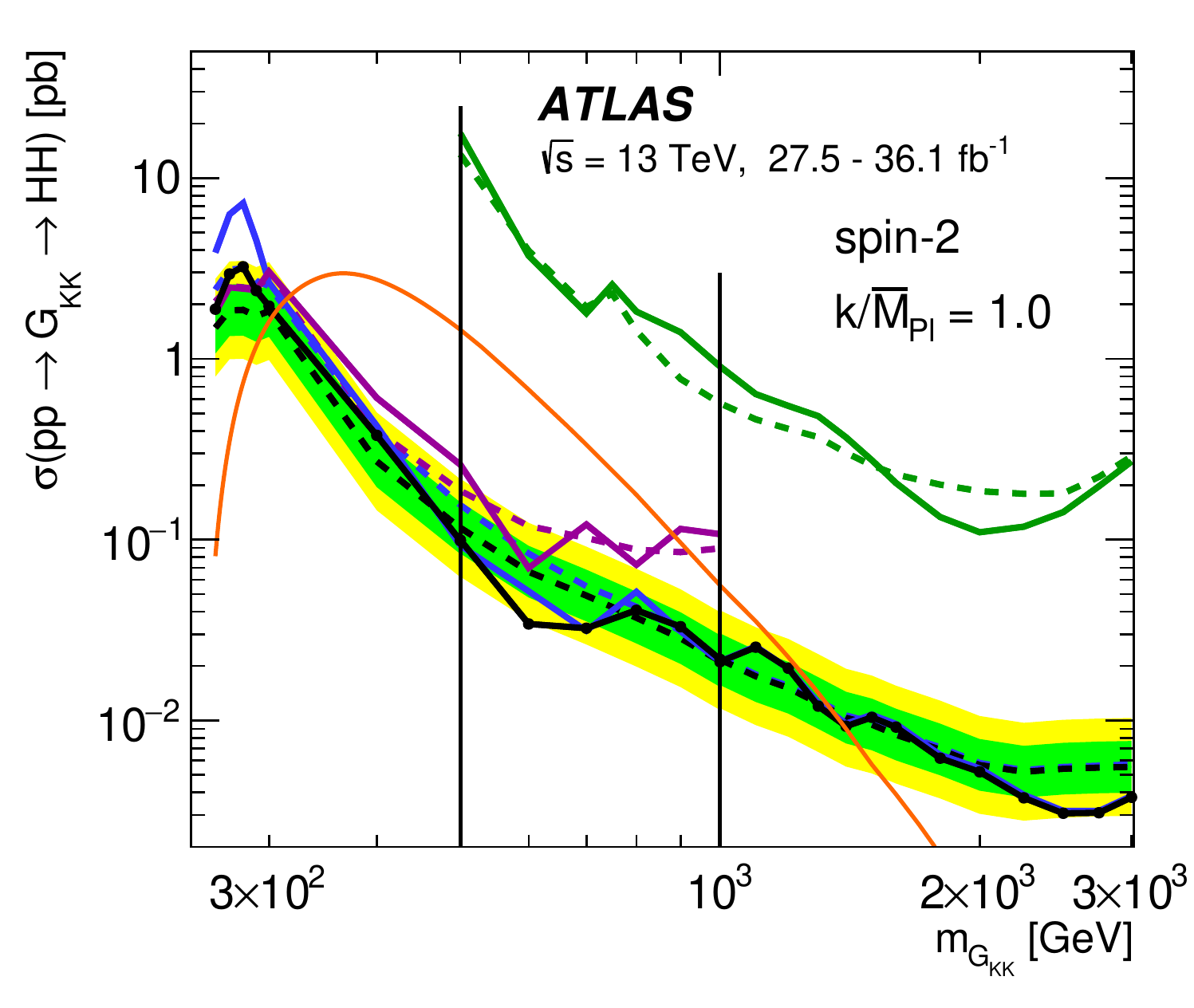}\includegraphics[width=0.42\textwidth]{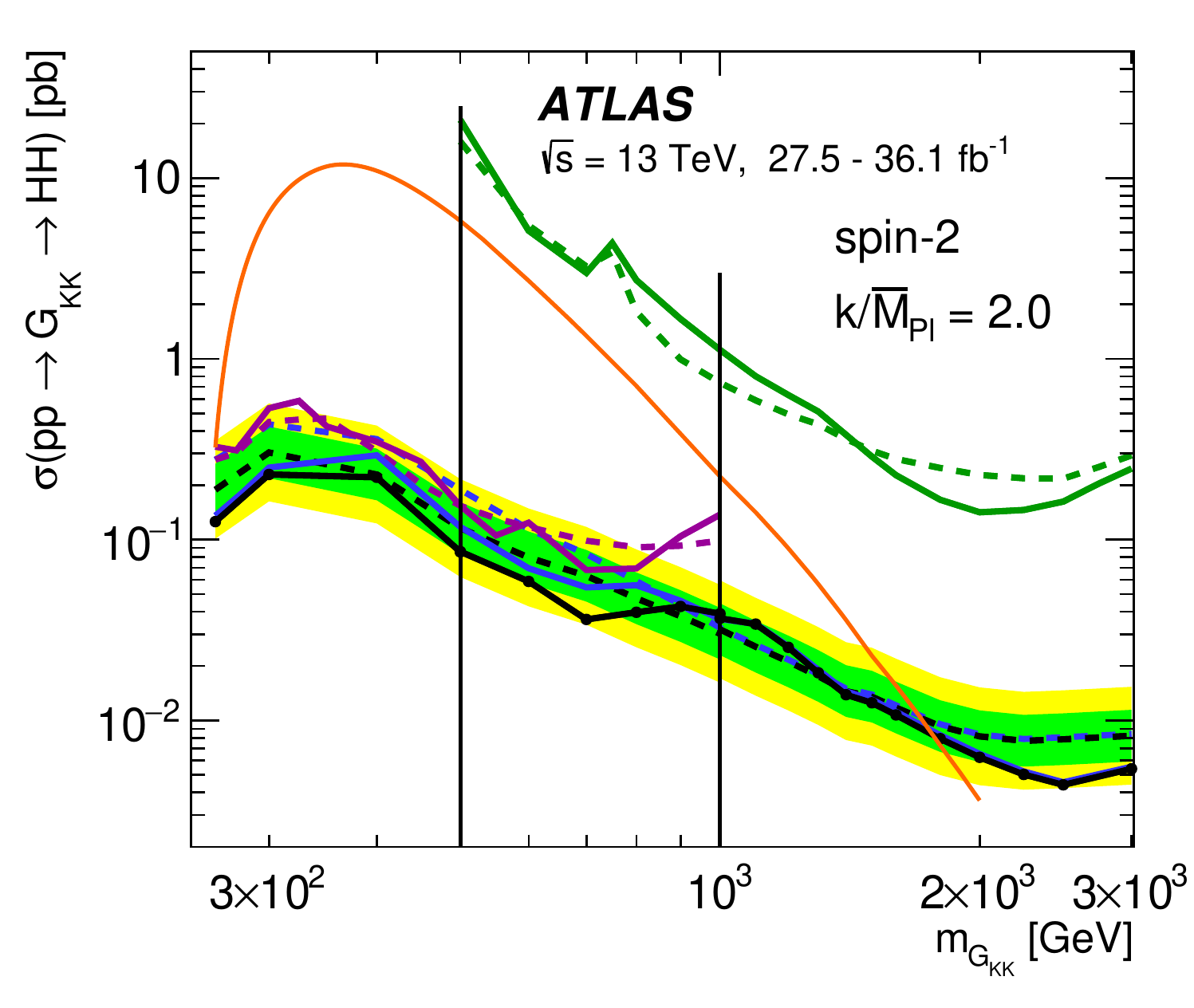}\includegraphics[width=0.23\textwidth]{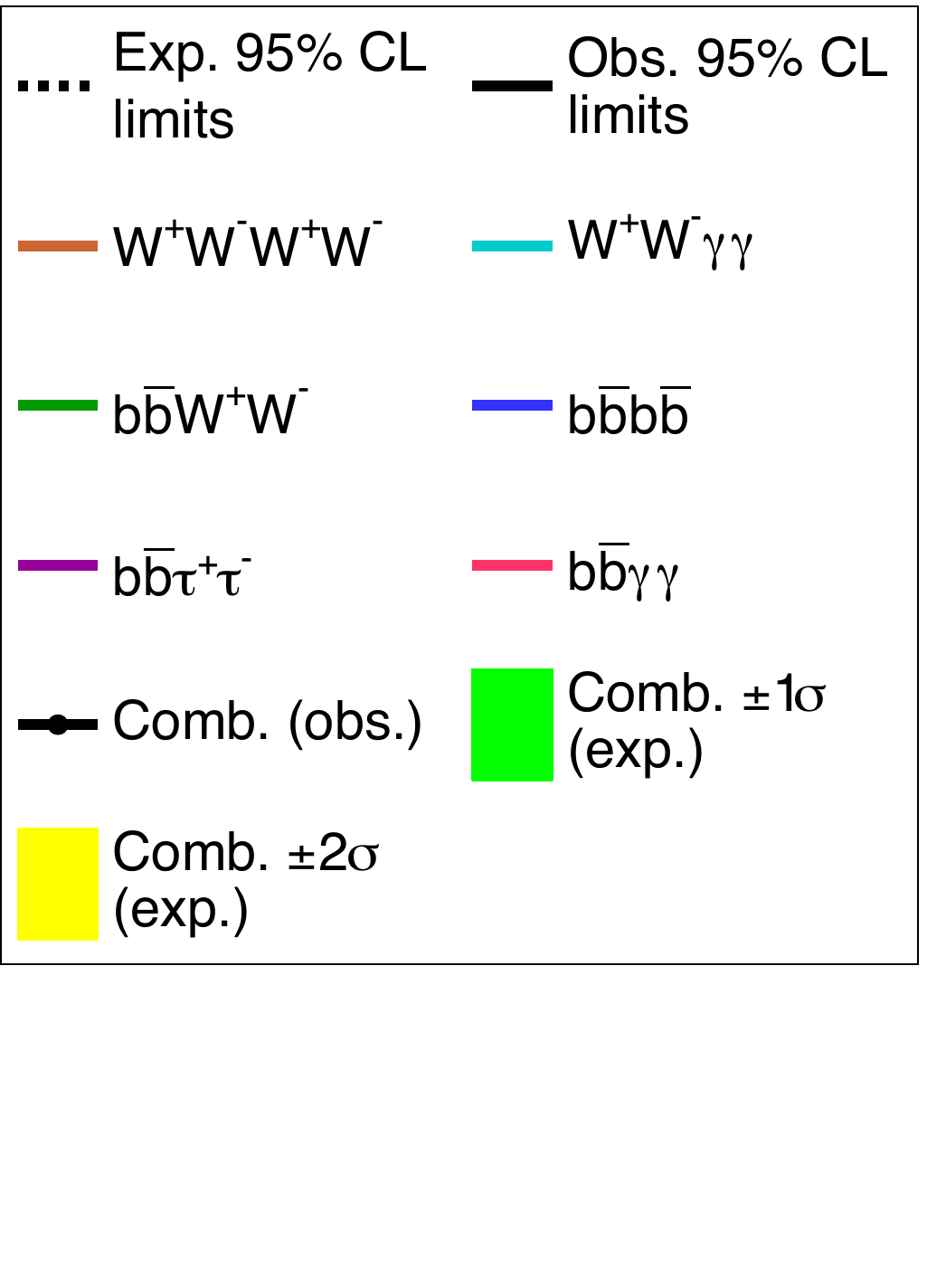}} 
\end{center}
\caption{
Expected and observed 95\% CL exclusion limits on the production cross section of a spin-2 resonance decaying into a pair of Higgs bosons.
Top: CMS combination and breakdown by final state with $k/\bar{M}_{\rm pl}=0.5$~\cite{Sirunyan:2018two};
Bottom left: ATLAS combination and breakdown by final state for $k/\bar{M}_{\rm pl}=1.0$~\cite{Aad:2019uzh};
Bottom right: ATLAS combination and breakdown by final state for $k/\bar{M}_{\rm pl}=2.0$~\cite{Aad:2019uzh}.}
\label{fig:ATLAS_spin2}
\end{figure}

The spin-2 model has been tested for several values of the $k/\bar{M}_{\rm pl}$, namely $< 0.5$ from CMS and 1.0 and 2.0 from ATLAS, as shown in Fig.~\ref{fig:ATLAS_spin2}. As $k/\bar{M}_{\rm pl}$ increases, the resonance width becomes larger and the narrow width approximation, used by the CMS collaboration, is valid only for $k/\bar{M}_{\rm pl}<0.5$. The ATLAS analyses take into account the  natural width of the resonance in the simulation of the signal processes.

At higher resonance masses, the \bbbb channel dominates the sensitivity in both experiments, thanks to the large branching fraction of \hbb, the good signal efficiency and the decreasing background at high \mhh value. Also the boosted \bbtautau final state, which has been investigated by CMS, significantly contributes to the combination for resonance masses above 1~TeV. ATLAS has also investigated \bbww in the single lepton final state~\cite{Aaboud:2018zhh} in a regime where the $\hbb$ system is boosted and the $WW$ decay is resolved, but it is not as competitive as the \bbbb final state at high mass. CMS additionally has explored the \bbww single lepton channel where both Higgs bosons are boosted, and demonstrated good sensitivity for resonances below 1.5 TeV in mass~\cite{Sirunyan:2019quj}.
All this has been possible thanks to the developments of dedicated techniques used to identify boosted \hbb events, as discussed in Sec.~\ref{sec:jetReco} and Sec.~\ref{sec:hbbbosted}.
\clearpage

\section{Interpretation in complete models}
\begin{figure}[ht!]
\centering
\includegraphics[width=0.65\textwidth]{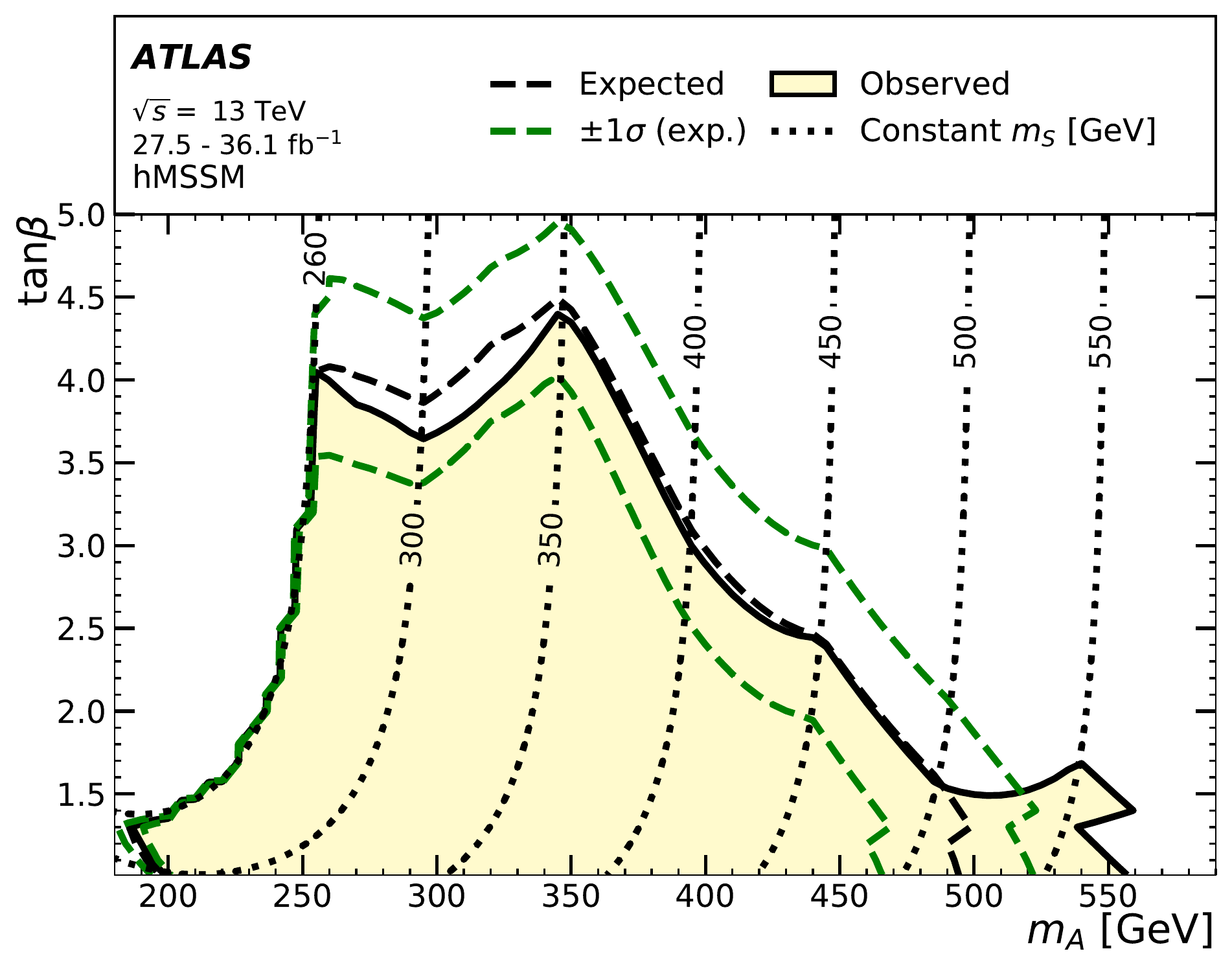}\\
\caption{Expected and observed 95\% CL exclusion limits in the tan$\beta$-$m_A$ parameter space of the hMSSM model \cite{Aad:2019uzh}.}
\label{fig:comb:hMSSM}
\end{figure}

\begin{figure}[ht!]
\centering
\includegraphics[width=0.45\textwidth]{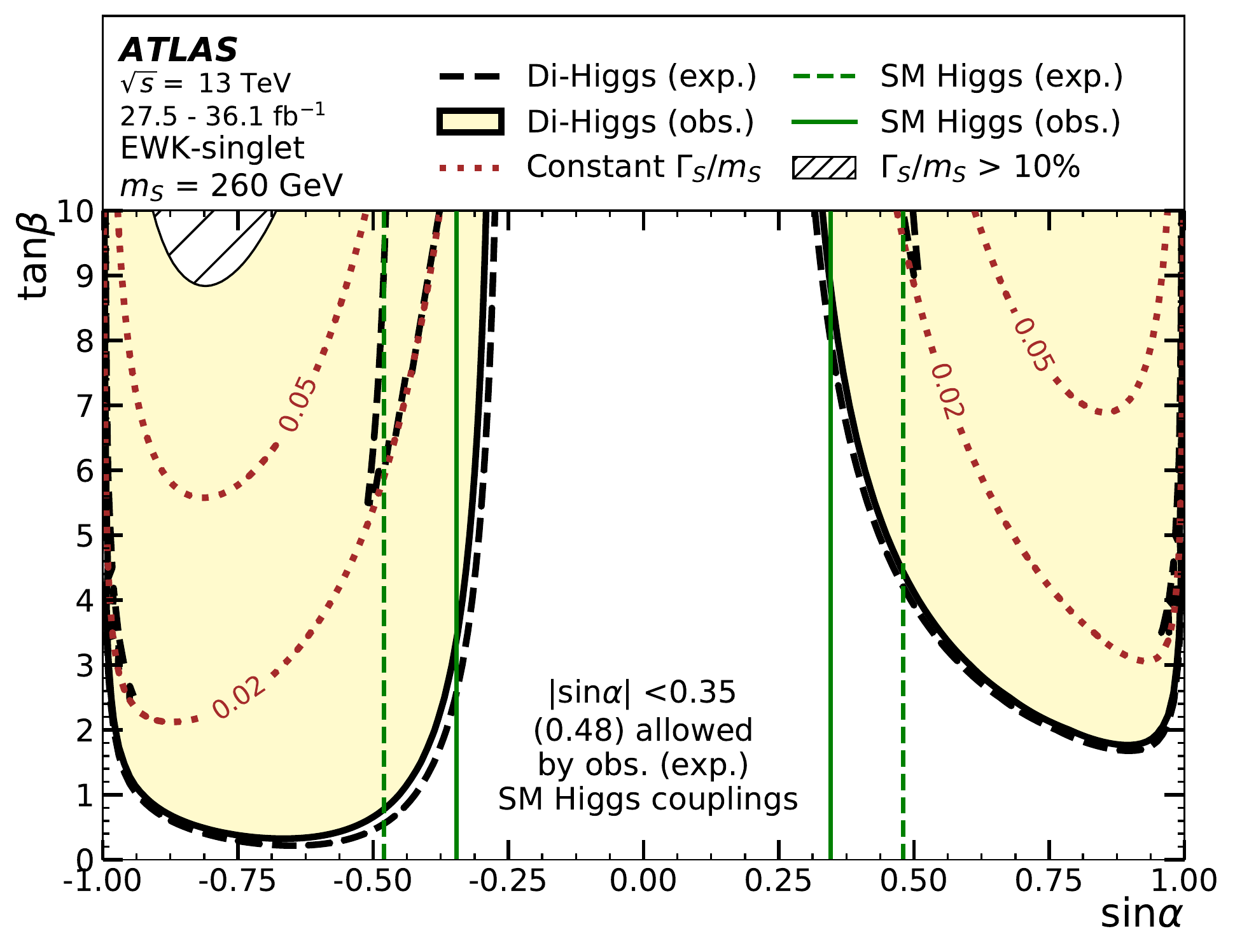}
\includegraphics[width=0.45\textwidth]{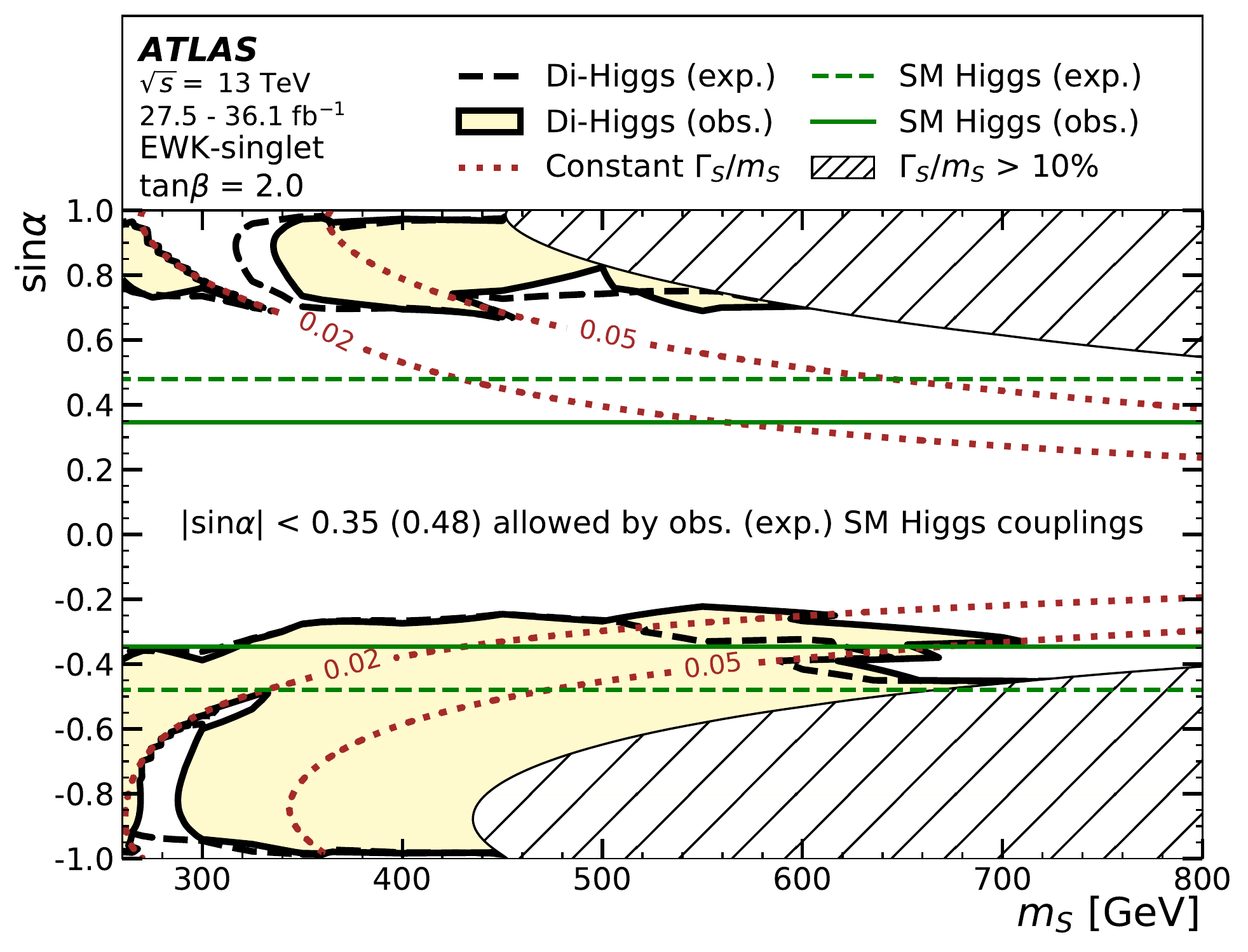}
\\
\caption{Expected and observed 95\% CL exclusion limits in the: sin$\alpha$-tan$\beta$ parameter space for $m_S = 260$ Gev (left), and the sin$\alpha$-$m_{S}$ parameter space for tan$\beta = 2$ of the EWK-singlet model \cite{Aad:2019uzh}.}
\label{fig:comb:EWKsing}
\end{figure}

The ATLAS collaboration has provided interpretations of spin-0 resonance limits in two models, the hMSSM model (Sec.~\ref{sec:hmssm})
and the EWK-singlet model (Sec.~\ref{sec:BSMspin0}, with a $Z_2$ symmetry).
The exclusion limits in the model parameter space are shown in Fig.~\ref{fig:comb:hMSSM} and  Fig.~\ref{fig:comb:EWKsing} respectively.
The interpretation is derived in the narrow width approximation, in this sense results are valid only when the scalar resonance width is much smaller than the detector resolution. This happens when $\Gamma_S/m_S < 2$\% in the \bbyy case, 5\% in the \bbtautau case and 10\% in the \bbbb case. Regions where $\Gamma_S/m_S > 2$\% have been removed from Fig.~\ref{fig:comb:hMSSM}, while in Fig.~\ref{fig:comb:EWKsing} regions where $\Gamma_S/m_S>$10\% were removed and are indicated with a dashed region, while only the \bbbb and \bbtautau channels are combined when $\Gamma_S/m_S>$2\% and only \bbbb results are shown for $\Gamma_S/m_S > 5$\%.
  
\section{Impact of systematic errors and concluding remarks}

The results presented in this chapter, based on 27.5--36.1~\ifb $p-p$ collision data, are limited by the size of the available dataset rather than systematic uncertainties. The overall impact of systematic uncertainties and their leading contributions are shown for the ATLAS analyses in ~\refta{tab:comb:sys}\footnote{The related CMS publications do not provide this information, but we don't expect large differences in systematic uncertainties between the ATLAS and CMS searches.}.

\begin{table}[htbp]
\begin{center}
\begin{tabular}{l|c|c|c|c|c|c|c}
\hline
Upper limit percentage variation& NR & \multicolumn{2}{c|}{Spin-0} & \multicolumn{2}{c|}{Spin-2
                                     $k/\overline{M}_{\text{Pl}} = 1$} &
                                                                    \multicolumn{2}{c}{Spin-2 $k/\overline{M}_{\text{Pl}} = 2$} \\
\cline{3-8} 
&   & 1~TeV & 3~TeV & 1~TeV & 3~TeV & 1~TeV & 3~TeV \\
\hline
Simulation statistics &  3\% &  1\% &  -   &  2\% &  -   &  1\% &  -   \\
Background modelling   &  5\% &  7\% &  9\% & 11\% & 15\% & 16\% & 21\% \\
Signal theory         &  1\% &  -   &  -   &  -   &  1\% &  -   &  -   \\
Tau                   &  2\% &  -   &  -   &  -   &  -   &  1\% &  -   \\
Jet                   &  -   &  1\% &  2\% &  2\% &  3\% &  5\% &  4\% \\
\btagging           &  1\% &  2\% &  -   &  3\% &  -   &  4\% &  -   \\
\hline
All                   & 13\% & 12\% & 11\% & 19\% & 18\% & 29\% & 25\%
  \\
  \hline
\end{tabular}
\end{center}
\vspace*{-0.5cm}
\caption{Percentage variations of the upper limits on the cross section of various signal
  models due to systematic uncertainties for the ATLAS analysis\cite{Aad:2019uzh}\footnotemark. The variations are
  calculated by computing the ratio between the difference of the upper limits
  obtained including all systematic uncertainties with the one
  obtained by removing
  the systematic uncertainty under study, and the nominal upper limit
  including all systematic uncertainties. The variations from the
  six leading systematic uncertainties and from all systematic
  uncertainties (``All'') are listed. The row ``All'' is obtained by removing
  all systematic uncertainties. When the fractional change is less than 1\%, ``-'' is shown. ``NR''
indicates the non-resonant signal model.} \label{tab:comb:sys}
\end{table}

\footnotetext{The table is extracted from the auxiliary material available on the ATLAS webpage \url{https://atlas.web.cern.ch/Atlas/GROUPS/PHYSICS/PAPERS/HDBS-2018-58/}}
Assuming that systematic errors will remain sub-dominant while keeping the current analysis sensitivity,  an expected limit of 5 times the SM prediction is reachable with the analysis of the full Run 2 dataset by the ATLAS and CMS collaborations.

\begin{figure}
    \centering
    \includegraphics[width=0.7\textwidth]{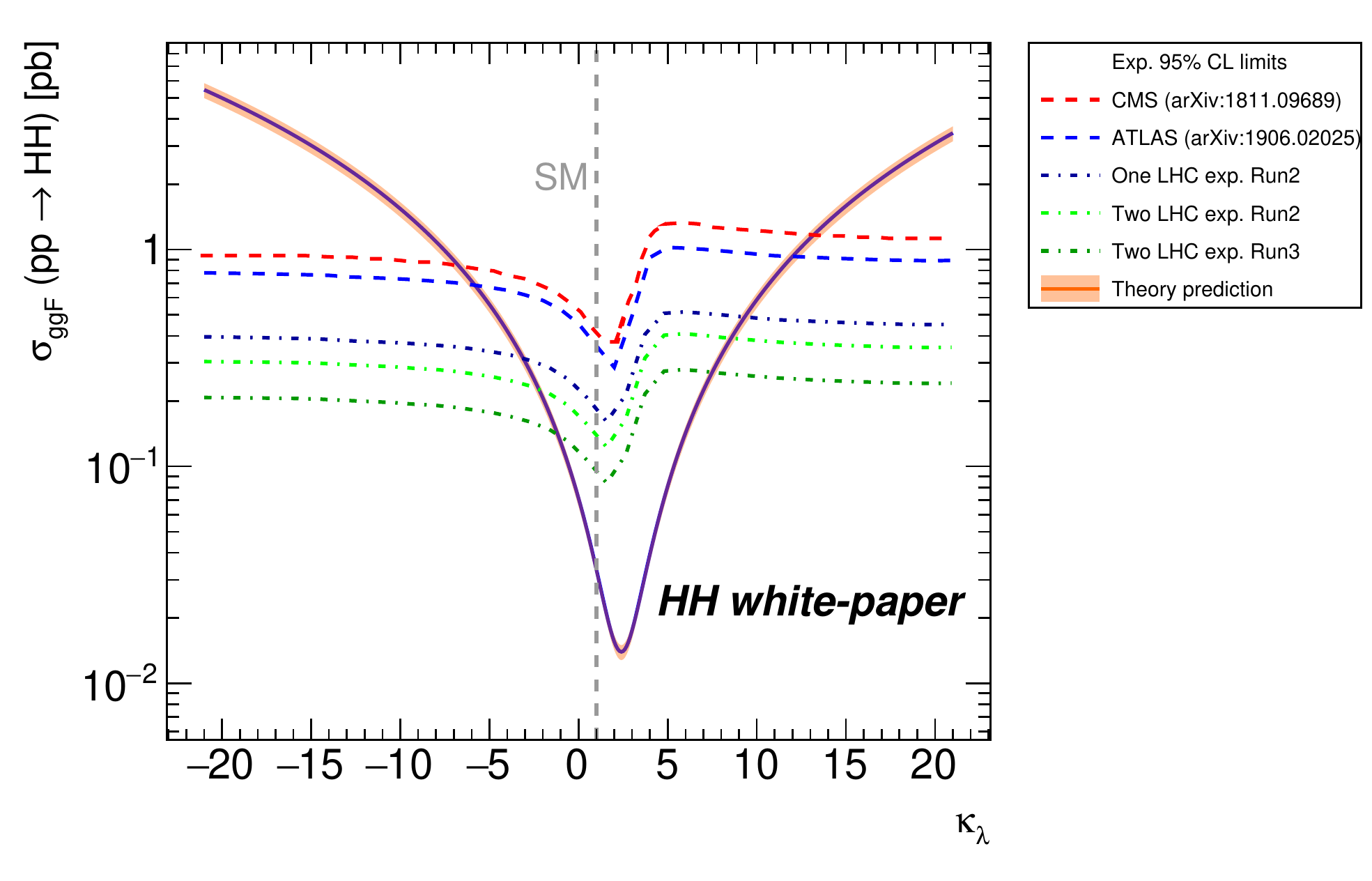}
    \vspace*{-0.3cm}
    \caption{Expected limit on the $pp \to HH$ cross section as a function of $\klambda$ for one of the two LHC experiments and their combination estimated for the Run 2 integrated luminosity (140\ifb) and the Run 2 and 3 (300\ifb).}
    \label{fig:klambda:extrapolation}
\end{figure}

In the non-resonant analysis the dominant sources of systematic uncertainties are associated to the background estimation methods, especially for the modelling of the multi-jet component in the \bbbb and \bbtautau analyses, for which no reliable simulation is yet available. The data-driven approaches used extrapolate the multi-jet parameterisation from signal-free control regions in data.
The precision of these data-driven methods is expected to improve with the increasing size of the available dataset, while the uncertainty associated to the extrapolation technique is an intrinsic limitation of this approach.

Other sources that contribute to systematic uncertainties are related to objects reconstruction and identification. Both ATLAS and CMS experiments are working on more precise evaluations of these object-related uncertainties and improvements on this subject are expected in the future. Further contributions are the uncertainty on the integrated luminosity and the limited MC statistics. For the latter, future progress in fast simulation techniques and truth level filtering techniques will allow to generate larger simulated samples and therefore hopefully mitigate these statistical limitations.


For searches of high mass resonances, where the signal is expected to appear on the tails of steeply falling invariant mass distribution of the non-resonant background,  the analyses are now strongly limited by the number of events available in the current dataset.


In the future, more final states will be combined. The orthogonality between the various searches must be carefully ensured, especially when combining similar final states, such as \bbww, \bbzz and \bbtautau. They can all result in a signature with two \bjets, two leptons and missing transverse energy. 

A potential combination of ATLAS and CMS based on the full Run 2 dataset will be considered, which increases the sensitivity significantly compared to single-experiment results as shown in Fig.~\ref{fig:comb:ATLAS_CMS}. The treatment of systematic uncertainties and correlations between ATLAS and CMS needs to be carefully studied. A useful reference is the ATLAS and CMS Higgs coupling combination~\cite{Khachatryan:2016vau} and the procedure for this combination outlined in~\cite{CMS-NOTE-2011-005} by the LHC Higgs Combination Group. Theoretical uncertainties on the signal process should be correlated, likewise simulation-based background uncertainties may be correlated, while the data-driven background uncertainties should not. A harmonised treatment of the signal process in terms of MC generators, theoretical uncertainties, as well as the same mass points, will facilitate such a combination. 

By combining the ATLAS and CMS data, the expected upper limit on the non-resonant cross section should reach the sensitivity of about 3.5 times the SM prediction at the end of Run 2 (140\ifb), and to 2.4 times the SM at the end of Run 3 (300\ifb). If instead the impact of the systematic uncertainties will not be sub-dominant contributions, the sensitivity at the end of Run 3 would be about 5 times the SM expectation, and it would be completely driven by the systematic uncertainties, assuming no improvements on the analysis strategy.
Concerning $\klambda$, Fig.~\ref{fig:klambda:extrapolation} shows the expected limit on the $pp \to \hh$ cross section as a function of $\klambda$ using Run 2 and Run 3 extrapolations assuming negligible systematic errors.
Without important improvements to the analysis strategies, but assuming it will be possible to reduce the impact of the systematic errors, $\klambda$ is expected to be constrained in the interval $-1.2 < \klambda < 7.5$ at 95\% CL at the end of Run 3.

%% file: HH_overview/singleH-exp.tex
In addition to the direct determination of the Higgs self-coupling through the study of Higgs boson pair production, an indirect measurement is also possible exploring the NLO EW corrections to single Higgs measurements, as discussed in detail in Sec.~\ref{tril-single}. 
The first experimental constraint on \klambda from single Higgs measurements has been determined by the ATLAS experiment~\cite{ATLAS-PUB-19-009}, by fitting data from single Higgs boson analyses taking into account the NLO $\klambda$ dependence of the cross section and the branching fractions of the ggF, VBF, $WH$, $ZH$ and $\ttbar{H}$ production modes and the $\gamma\gamma$, $WW$, $ZZ$, $\tau\tau$ and \bb decay modes, as listed in \refta{tab:list}. Differential information has also been exploited through the use of the Simplified Template Cross Section categories (described in Sec.~\ref{sec:RPNR}).


\begin{table}

\begin{center}
\begin{tabular}{lcc}
\hline
Measurement & Reference & $\mathcal{L}$ \\
\hline 
$\ttbar{H}$ (\hbb and multileptons final states) & \cite{HIGG-2017-02,HIGG-2017-03} & 36.1 . \ifb\\
\hgg  (including $\ttbar{H}$) &\cite{ATLAS-CONF-2018-028,HIGG-2016-21,Aaboud:2018urx}& 79.8\ifb \\
\hzz (including $\ttbar{H}$) &\cite{HIGG-2016-22,ATLAS-CONF-2018-018}   & 79.8 \ifb \\
VH \hbb&\cite{HIGG-2018-04,ATLAS-CONF-2018-053} &79.8 \ifb\\
\hww & \cite{HIGG-2016-07}&36.1 \ifb\\ 
\htautau&\cite{HIGG-2017-07}&36.1 \ifb\\
 \hline
\end{tabular}
\caption{List of the ATLAS measurements of Higgs production and decay modes, combined to derive a constraint on the value of the Higgs boson self-coupling, \klambda. The measurements used in this analysis are based on data collected at 13 TeV corresponding to an integrated luminosity of up to 79.8\ifb.\label{tab:list}}
\end{center}
\end{table}


Each analysis separates the measured events into orthogonal kinematic and topological categories depending on the reconstructed final state. These categories partially account for the kinematic dependence and they have been optimised to maximise the sensitivity to their associated truth-level region. Although the gluon-gluon fusion production mode is subdivided in bins of jet multiplicity and transverse momentum of the Higgs boson, \pTH, differential corrections are not yet available\footnote{They would involve higher order calculations including two loop corrections} and therefore the corresponding STXS bins related share the same parameterisation as for the inclusive ggF production. Nevertheless, such contributions have been evaluated in the Heavy Top Quark expansion\cite{Gorbahn:2019lwq}, that is valid for $\pTH  << m_t$, i.e. $\pTH < 150$ GeV (see Sec.~\ref{tril-single}) and result to be small.
The $\ttbar{H}$ production mode is considered inclusively in one single bin, as no differential measurement is available yet. 
The $gg\rightarrow ZH$ cross section is not parameterised as a function of \klambda, because the theoretical computation is still missing and it should contribute mostly in high \pTH regions where the sensitivity to \klambda is expected to be small.



The values of the kinematic dependent $C_1$ linear coefficients, Eq.(~\ref{eq:muf3}), that parameterise the sensitivity of the measurement to \klambda have been derived for each STXS region defined in the measurement. The values obtained are reported in \refta{tab:c1}.

\begin{table}
\small
{\def\arraystretch{1.3}
\begin{center}
\begin{tabular}{l|l|c|c|c}
\hline 
\multicolumn{2}{c|}{\multirow{2}{*}{STXS region}}&VBF &WH & ZH \\ \cline{3-5}

\multicolumn{2}{c|}{}&\multicolumn{3}{c}{$C_1^i \times 100$}\\
\hline
\multirow{5}{*}{VBF+V(\text{had})H}&VBF-cuts $+\, p_\text{T}^{j1}<200$ GeV, ${}\le 2j$&0.63&0.91&1.07\\
&VBF-cuts $+\, p_\text{T}^{j1}<200$ GeV, ${}\ge 3j$ &0.61&0.85&1.04\\
&VH-cuts $+\, p_\text{T}^{j1}<200$ GeV&0.64&0.89&1.10\\
&no VBF/VH-cuts, $p_\text{T}^{j1}<200$ GeV &0.65&1.13&1.28\\
&$p_\text{T}^{j1} > 200$ GeV  &0.39&0.23&0.28\\
\hline
\multirow{4}{*}{$qq\to H \ell \nu$}&$p_\text{T}^V < 150$ GeV &&1.15&\\
&$150 < p_\text{T}^V < 250$ GeV, 0$j$ &&0.18&\\
&$150 < p_\text{T}^V <250$ GeV, ${}\ge 1j$ &&0.33&\\
&$p_\text{T}^V > 250$ GeV &&0&\\
\hline
\multirow{2}{*}{$qq\to H \ell \ell$}&$p_\text{T}^V < 150$ GeV &&&1.33\\
&$150 < p_\text{T}^V < 250$ GeV, 0$j$ &&&0.20\\
\multirow{2}{*}{$qq\to H \nu \nu$}&$150 < p_\text{T}^V <250$ GeV, ${}\ge 1j$ &&&0.39\\
&$p_\text{T}^V > 250$ GeV &&&0\\
\hline 
\end{tabular}
\end{center}}
\caption{$C_1^i$ coefficients for each region of the STXS
  scheme for the VBF, WH and ZH production modes. The definition of the STXS regions can be found in Ref.~\cite{ATLAS-PUB-19-009} . In the VBF categories, ``VBF-cuts''~\cite{deFlorian:2016spz} indicates selections applied to 
target the VBF di-jet topology, with requirements on the di-jet invariant mass ($m_{jj}$)
and the difference in pseudorapidity between the two jets; the additional $\leq 2j$ and $\geq 3j$ region separation is performed indirectly by requesting $p_\text{T}^{Hjj}\lessgtr 25$~GeV. ``VH-cuts''
select the $W,Z \to jj$ decays, requiring an $m_{jj}$ value close to the
vector boson mass~\cite{deFlorian:2016spz}. The
$C_1^i$ coefficients of the $p_\text{T}^V > 250$ GeV regions are negligible,
O$(10^{-6})$, and are set to $0$.}\label{tab:c1}
\end{table}


The NLO EW $K$-factors, that includes one loop EWK correction not involving $\klambda$, are computed inclusively for each production mode and not in STXS bins, as in the regions of phase space where these corrections are most significant (typically for high Higgs boson transverse momentum), the sensitivity to the Higgs boson trilinear coupling is minimal~\cite{Maltoni:2017ims}. 
The selection efficiencies have been evaluated as function of \klambda and a negligible dependency has been found, thus they are assumed to be constant. The exception is $\ttbar{H}$ production mode, which shows a 10\% increase for $\klambda < - 10$, but in this interval the reduction of the cross section due to the \klambda dependence is about 80\%, therefore largely dominating over the efficiency variation. This assumption will be re-evaluated in the ggF production mode once a complete computation of the differential NLO EW corrections will become available.


 \begin{figure}[htbp] 
\begin{center} 
\includegraphics[width=0.45\textwidth]{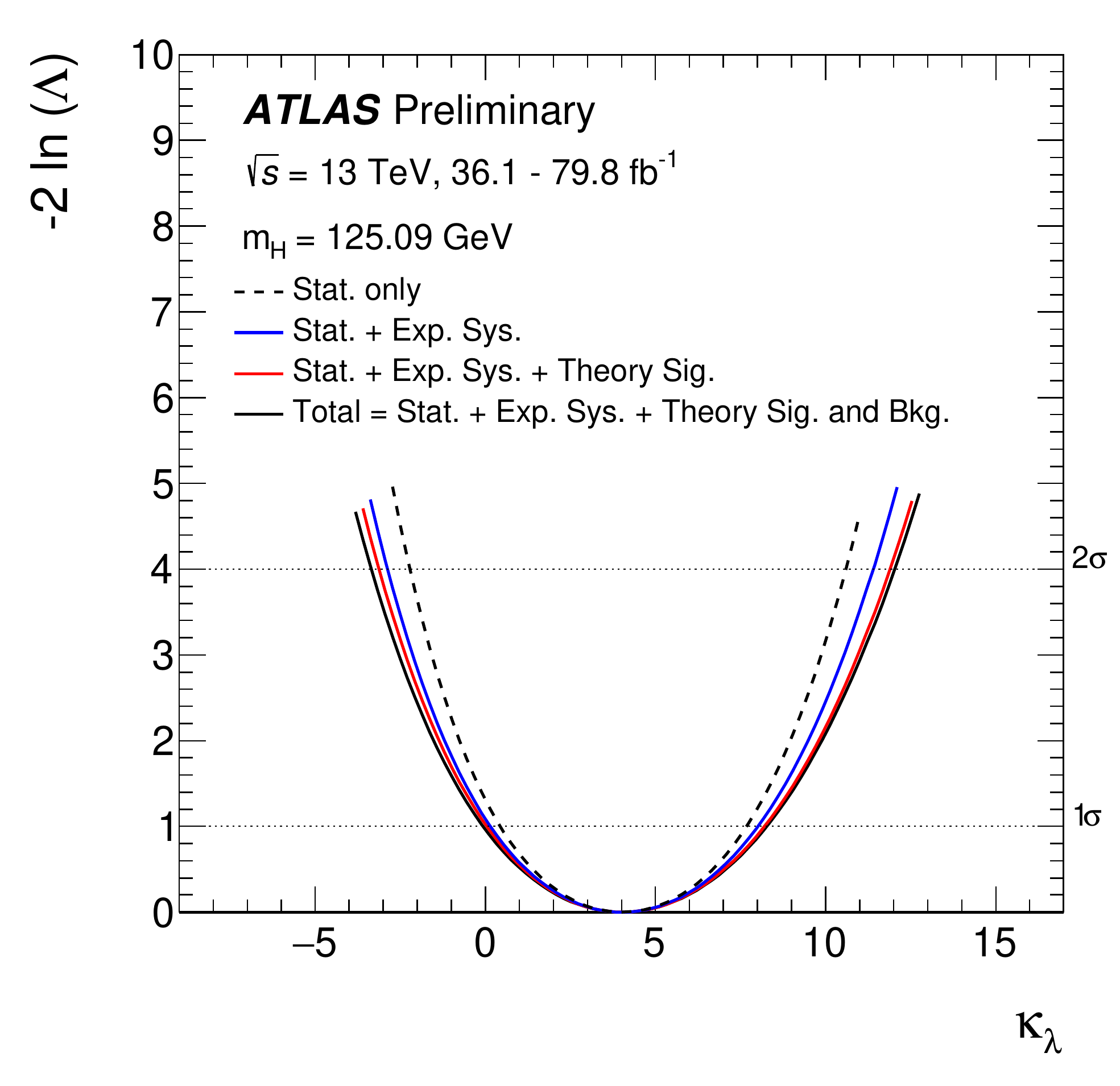}
 \caption{The profile likelihood scan performed as a function of \klambda on data~\cite{ATLAS-PUB-19-009}.}
 \label{fig:fig4a} 
 \end{center} 
 \end{figure}

 A likelihood fit is performed to constrain the value of the Higgs boson self-coupling \klambda, while  all other Higgs boson couplings are set to their SM values ($\kappa_{i,\mathrm{F}} = \kappa_{i,\mathrm{V}} = 1 $ in Eq.~\ref{eq:mui} and~\ref{eq:muf2}). Thus, for a large variety of BSM scenarios, where new physics modifies only the Higgs boson self-coupling, the constraints on \klambda derived through the combination of single Higgs measurements can be directly compared to the constraints set by double Higgs production measurements.
 The profile likelihood scan performed as a function of \klambda is shown in Fig.~\ref{fig:fig4a}. The central value and uncertainty of the modifier of the trilinear Higgs boson self-coupling is determined to be:
 \begin{equation}
 \klambda = 4.0^{+4.3}_{-4.1} =4.0^{+3.7}_{-3.6} \mathrm{(stat.)} ^{+1.6}_{-1.5} \mathrm{(exp.)} ^{+1.3}_{-0.9} \mathrm{(sig. th.)} ^{+0.8}_{-0.9} \mathrm{(bkg. th.)}
 \label{eq:lambdaresult}
 \end{equation}
 where the total uncertainty is decomposed into components for statistical, experimental and theory uncertainties on signal and background modelling. The 95\% CL allowed interval for \klambda is $-3.2 < \klambda < 11.9$ (observed) and $-6.2 < \klambda < 14.4$ (expected). This interval is competitive with the one obtained from the direct \hh searches
 using an integrated luminosity of 36.1~\ifb, which is  $-5.2 < \klambda < 12.1$ (observed) and  $-5.8 < \klambda < 12.0$ (expected).
The dominant contributions to the \klambda sensitivity derive from the di-boson decay channels $\gamma\gamma$, ZZ, WW and from the ggF and $\ttbar{H}$ production modes.
The differential information currently provided by the STXS binning in the VBF, $WH$ and $ZH$ production modes does not improve the sensitivity to \klambda significantly. However, differential information should help most in the $\ttbar{H}$ production mode. 
A dedicated optimisation of the kinematic binning, including  the most sensitive ggF and $\ttbar{H}$ production modes, still needs to be fully theoretically and experimentally explored and might improve the sensitivity in the future.

While the sensitivity on \klambda derived from single Higgs processes in an exclusive fit is comparable to those from \hh direct searches, the constraints become significantly weaker when BSM deformations to the single Higgs couplings are taken properly into account.
Two additional fit configurations with a simultaneous fit to (\klambda, $\kappa_{\mathrm{V}}$) or to (\klambda, $\kappa_{\mathrm{F}}$) have been considered. These fits target BSM scenarios where new
 physics could affect only the Yukawa type terms ($\kappa_{\mathrm{V}}=1$) of the SM or only the couplings to vector bosons ($\kappa_{\mathrm{F}}=1$), in addition to the Higgs boson self-coupling (\klambda)~\cite{Carena:2015moc}. This set of results provides a rough indication of the simultaneous sensitivity to both Higgs boson self-coupling and single Higgs boson couplings with the data statistics currently
available for the input analyses.


Figure~\ref{fig:fig6ab} shows negative log-likelihood contours on (\klambda, $\kappa_{\mathrm{V}}$) and (\klambda, $\kappa_{\mathrm{F}}$). The constraining power of the measurement is reduced by including additional degrees of freedom to the fit. In particular, the sensitivity to \klambda is degraded by 50\% (on the expected lower 95\% C.L. exclusion limit) when determining simultaneously $\kappa_{\mathrm{V}}$ and \klambda.

These observations have been confirmed recently by a preliminary CMS result\cite{CMS:2020gsy}, based on a part of Run 2 dataset. 


Similarly, the sensitivity to $\klambda$ from double Higgs measurements completely vanishes if the coupling to the top quark ($\kappa_t$) is left free to float, due to a $\kappa_t^4$ dependence of the total $pp \to \hh$ cross section (see Sec.~\ref{sec:non_res:comb} and Fig.~\ref{fig:klkt}). Therefore a determination of $\klambda$ which would take into account BSM contributions affecting $\kappa_t$, $\kappa_F$ or $\kappa_V$ would be possible only through a simultaneous analysis of both single and double Higgs measurements. As the experimental sensitivity increases, the addition of more differential information, in particular for $\ttbar{H}$ and ggF, would allow the  inclusion of more relevant EFT operators in the analysis, as $\kappa_t$ and $c_{gg}$ (cf. discussion at the end of Sec.~\ref{tril-single}).

 \begin{figure}[htbp] 
\begin{center} 
\includegraphics[width=0.45\textwidth]{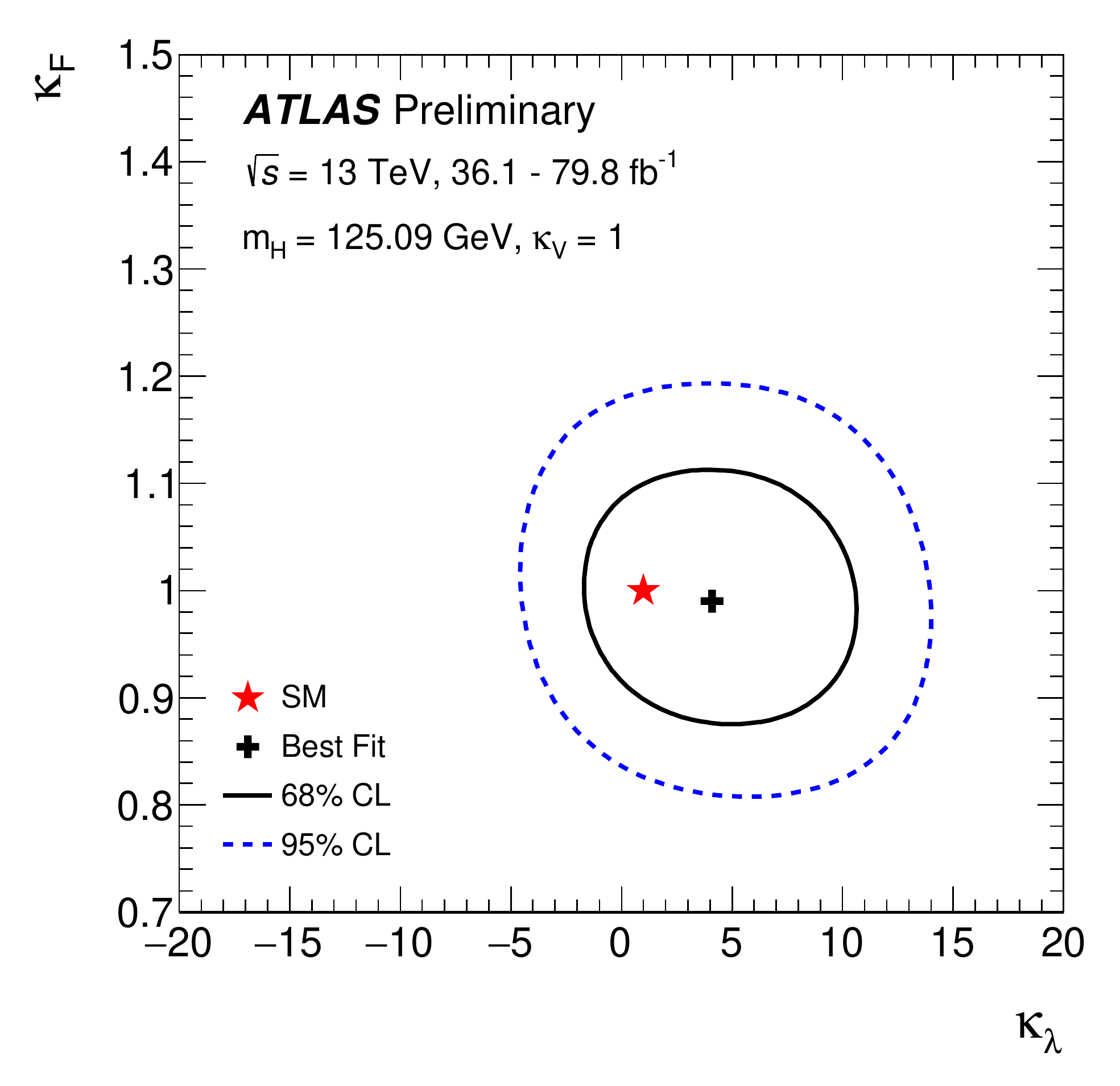}
\includegraphics[width=0.45\textwidth]{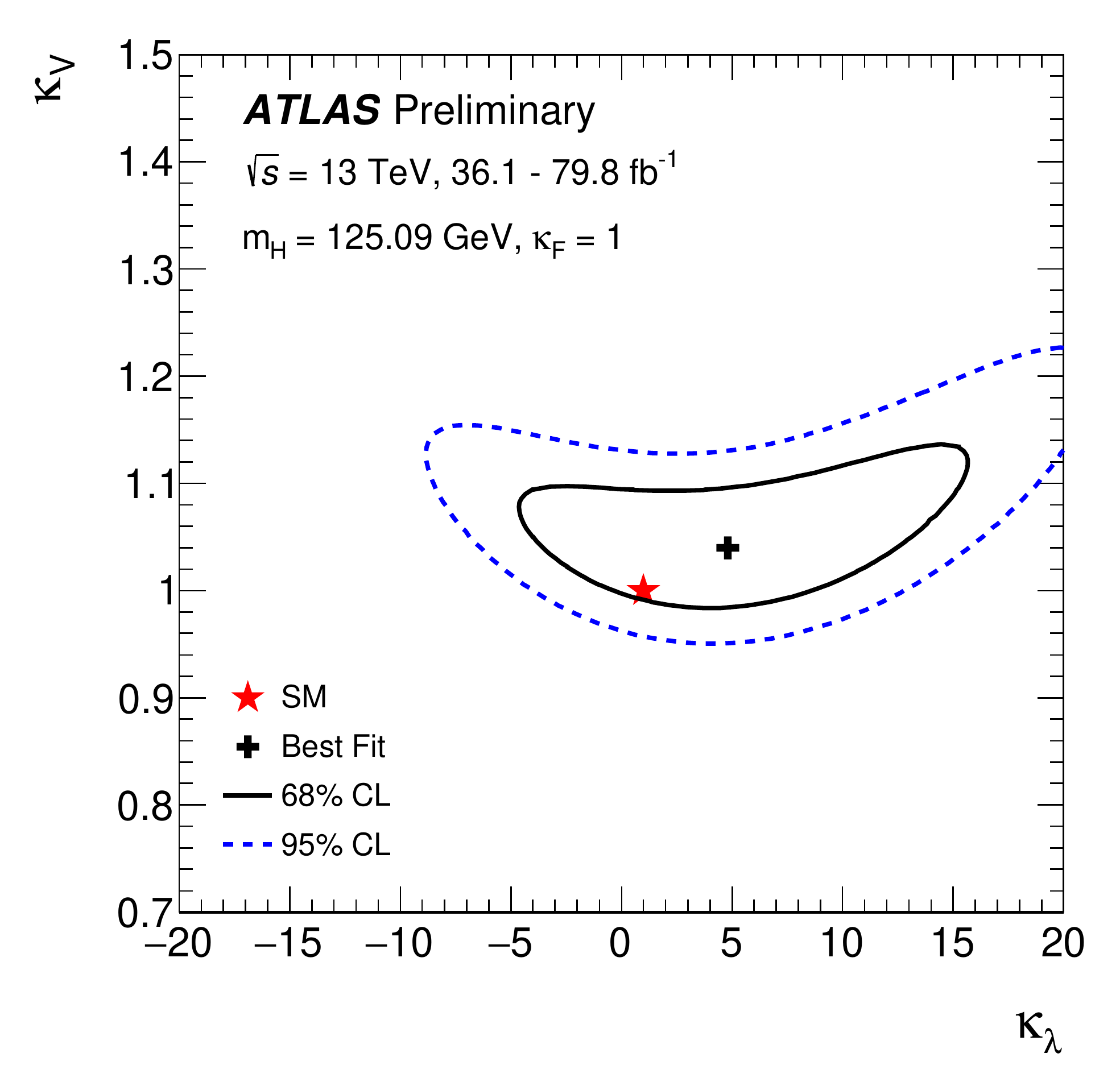}
 \caption{Negative log-likelihood contours at 68\% and 95\% CL as function of (\klambda, $\kappa_{\mathrm{F}}$) under the assumption of $\kappa_{\mathrm{V}}=1$
(left), and as function of (\klambda, $\kappa_{\mathrm{V}}$) under the assumption of $\kappa_{\mathrm{F}}=1$) (right). The best fit value is indicated by a cross while the SM hypothesis is indicated by a star~\cite{ATLAS-PUB-19-009}.}
 \label{fig:fig6ab} 
 \end{center} 
 \end{figure}

%% file: HH_future.tex
\input{HHFuture/introPart3.tex}

\chapter{Higgs self-coupling at HL-LHC}
\textbf{Editors: N. de Filippis, M.~Selvaggi}\\
\label{chap:hl-lhc}
\input{HHFuture/HH_HL-LHC.tex}
The content of this chapter is based on the studies presented in~\cite{Cepeda:2019klc,SystWeb}.
\chapter{Higgs self-coupling at future $e^+e^-$ colliders}
\textbf{Editor: M.~Peskin}\\
\label{chap:epem}

\input{HHFuture/HHineev3.tex}
\chapter{Higgs self-coupling at future hadron colliders}
\textbf{Editor: M.~Selvaggi}\\
\label{chap:HHFutureHadron}
\input{HHFuture/HH_pp.tex}
\chapter*{Higgs self-coupling at future colliders: Summary}
\addcontentsline{toc}{chapter}{Higgs self-coupling at future colliders: Summary}
\input{HHFuture/intro_future.tex}

%% file: HHFuture/introPart3.tex
\contrib{M.~E.~Peskin, M. Selvaggi}

Even after collecting all of its projected integrated luminosity, the bounds that can be obtained on the Higgs boson self-coupling at the LHC will still be quite loose. These results will be sensitive to possible large or resonant enhancements of the self-coupling, discussed in Chapter 3.   However, they will not yet be able to establish that the  self-coupling is non-zero if its true value is close to that predicted by the Standard Model.  In most scenarios, then, colliders beyond the LHC will be needed to  establish the the size of Higgs self-coupling both qualitatively and quantitatively.  The expected capabilities of such future colliders are the topic of the final part of this document.

In Chapter~\ref{chap:hl-lhc} we present the projections for the HL-LHC, a machine that will collide protons at a centre of mass energy of 14~TeV and deliver about 3~$\mathrm{ab}^{-1}$.
The projections for circular and linear lepton colliders are then discussed in Chapter~\ref{chap:ee}.
Finally, in Chapter~\ref{chap:helhc-fcchh} the prospects for future high energy (27 and 100~TeV) hadron colliders are reviewed.
As a summary, the capabilities of the various colliders are compared at the end of this Part.

%% file: HHFuture/HH_HL-LHC.tex
\\[2ex]
\contrib{S.~Gori, C.~Vernieri}
\\
By the end of the LHC Run 3 in 2024, the ATLAS and CMS experiments are each expected to have collected about 300~\ifb of integrated luminosity. 

A long Shutdown 3 (LS3) scheduled between 2024 and the middle of 2027 would close the Phase II of the LHC program~\footnote{based on the information available in Spring 2020.}. It will feature upgrades of the accelerator and the experiments for the High Luminosity phase of the LHC (HL-LHC), when the instantaneous peak luminosity will reach $7.5\times 10^{34}$ \instl, corresponding to about 200 inelastic $p-p$ collisions per beam-crossing on average. The HL-LHC is expected to run at a centre of mass energy of 14 TeV and with a bunch spacing of 25 ns. 
The CMS and ATLAS collaborations are expected to collect an integrated luminosity of about 3000-4000~\ifb in approximately ten years. The ATLAS and CMS experiments will undergo major upgrades to maintain the excellent performance of the event reconstruction, in order to fully profit of the HL-LHC potential, despite the challenging radiation levels and data taking conditions.

The inner detectors are expected to be completely replaced in both experiments with new all-silicon, radiation tolerant tracking systems, extending
their coverage to $|\eta|<$ 4.0~\cite{ATLAS-TDR-030,Collaboration:2017mtb,Collaboration:2272264}, enabling pileup jet rejection in the forward region.
The existing readout electronics of the calorimeters and muon spectrometers will be completely replaced due to both the limited radiation tolerance and the incompatibility with the upgraded trigger systems~\cite{ATLAS-TDR-027,ATLAS-TDR-028, ATLAS-TDR-026,Collaboration:2283189,Collaboration:2293646}. 

The large increase of pileup is one of the main experimental challenges for the HL-LHC physics program. Both ATLAS and CMS are planning for the first time to exploit the time spread of the interactions to distinguish between collisions occurring very close in space but well separated in time, with a timing resolution of 30 ps per track. CMS is developing a timing detector sensitive to minimum ionizing particles (MIPs) between the tracker and the electromagnetic calorimeters, covering the region of $|\eta| < 3$~\cite{Collaboration:2296612}. ATLAS is pursuing a High-Granularity Timing Detector, based on low gain avalanche detector technology, covering the pseudorapidity region between 2.4 and 4.0~\cite{Allaire:2302827}. 

The upgraded detectors will be read out at an unprecedented data rate and both the trigger and the data acquisition systems (DAQ) will undergo a substantial upgrade~\cite{ATLAS-TDR-029,Collaboration:2283193,Collaboration:2283192}.
Following the current design, the ATLAS and CMS trigger systems will continue to feature two levels: a first hardware-based first level (L1) consisting of custom electronic boards and a second software-based level, running on standard processors.
ATLAS proposes a L1 trigger with a maximum rate of 1 MHz and 10 $\mu$s latency, and a hardware-based tracking sub-system as co-processor to achieve further rejection. 
The CMS upgraded L1 trigger will allow a maximum rate of 750 kHz, and a latency of 12.5 $\mu$s and will include, for the first time, tracking information and high-granularity calorimeter information.
Selected events will be stored permanently at a rate of 7.5/10 kHz (CMS/ATLAS) for offline processing and analysis. 



\section{Measurement of the Higgs boson self-coupling at HL-LHC}
The study of the double Higgs boson production is one of the key goal of the HL-LHC physics program. Despite the small production cross section compared to the single Higgs boson production, more than 10$^5$ \hh pairs per experiment are expected to be produced by the HL-LHC. 

An overview of the main \hh production modes at $\sqrts=14\tev$ and the corresponding theoretical predictions is provided in Chapter~\ref{chap:HHcxs} and in particular in \refta{table:xsec2}.


The ATLAS and CMS collaborations have derived their projected sensitivity to the \hh production at the HL-LHC either through extrapolations from existing Run 2 results or using parametric simulations of the expected detector performance, assuming an average pileup of 200 collisions per bunch crossing. 
Only the production of \hh pairs through gluon-gluon fusion is considered, as the other production mechanisms are more than an order magnitude smaller. 
Although analyses targeting the VBF production mode will benefit at the HL-LHC from the extended tracker acceptance and, consequently,  the improved ability to identify forward jets from the hard-scattering interaction, which will yield to an increased background rejection in this channel.


To derive the HL-HLC expected sensitivity, several assumptions have been made to model the systematic uncertainties from theoretical and experimental sources. 

Theoretical uncertainties have been assumed to be reduced by a factor of two with respect to those used in the Run 2 analyses, thanks to the expected developments in both higher-order calculation as well as  in the reduction of PDF uncertainties. Experimental systematic uncertainties are assumed to scale as $\sqrt{\mathcal{L}}$, where $\mathcal{L}$ is the integrated luminosity, until a pre-defined lower limit is reached, depending on the intrinsic detector limitations, according to detailed simulation studies of the upgraded detector.
The common recommendations for the systematic uncertainties for HL-LHC studies are summarised in \refta{tab:systHLLHC}~\cite{SystWeb}. 
All the uncertainties related to the limited number of simulated events are neglected, assuming that large simulation samples will be available. The uncertainty on the luminosity is set to 1\%, which is the goal of ATLAS and CMS to be able to fully exploit the HL-LHC physics potential \footnote{This will demand the design of hardware for luminosity monitoring with performance intrinsically linear with pileup and radiation hardness.}.
Uncertainties are kept at the same value as in the latest public results available. It is assumed that the degradation due to higher pileup conditions will be compensated by improvements in the reconstruction algorithms.     

\begin{table}[htbp]
\begin{center}
\begin{tabular}{cc} \hline
Source  &  Uncertainties\\
\hline
Luminosity & 1-1.5\% \\
Muon efficiency (ID, iso) & 0.1-0.4\% \\ 
Electron Efficiency (ID, iso)& 0.5\% \\ 
Tau efficiency (ID, trigger, iso) & 5\% (if dominant  2.5\%) \\
Photon efficiency (ID, trigger, iso) & 2\% \\
Jet Energy Scale & 1-2.\% \\
Jet Energy Resolution & 1-3\% \\
b-jet tagging efficiency & 1\% \\
c-jet tagging efficiency & 2\% \\
light jet mis-tag rate  &5\% (at 10\% mis-tag rate)\\
 \hline
\end{tabular}
\end{center}
\vspace*{-0.3cm}
\caption{Summary of the systematic uncertainties used to extrapolate the results at the HL-LHC by ATLAS and CMS. These are representative values. The dependence for example of \pT and $\eta$ and the operating points, if applicable, need to be taken into account~\cite{SystWeb}.}
\label{tab:systHLLHC}
\end{table}



\section{Double Higgs boson production measurements}
For the study of the \hh production at HL-LHC, the most promising decay channels from the Run 2 searches were exploited by the ATLAS and CMS collaborations~\cite{ATLAS-CONF-2018-053,CMS-PAS-FTR-18-019,Cepeda:2019klc}: \bbbb, \bbtautau, \bbgg.
In addition CMS has investigated also the potential of \bbww ($WW \to \ell \nu \ell' \nu')$  and \bbzz ($ZZ \to \ell\ell\ell'\ell'$) with $\ell, \ell' = $e, $\mu$ at the HL-LHC.
The ATLAS and CMS studies are described in the following two paragraphs.

\paragraph{ATLAS projections}

Studies by the ATLAS collaboration were made by extrapolating the recent results obtained at $\sqrt{s}=13$~TeV with Run 2 data  with approximately 24.3~\ifb and 36~\ifb of integrated luminosity for the \bbbb and the \bbtautau analyses respectively. 
While the challenging data-taking conditions at HL-LHC could worsen the \btagging efficiency, the new inner tracker detector as well as novel reconstruction techniques could provide a sizeable improvement. It was estimated that the upgrades of the inner tracker~\cite{Collaboration:2017mtb} would lead to an 8\% improvement in \btagging efficiency. 

For the \bbbb decay channel, the dominant systematic uncertainty is associated to the modelling of the multi-jet
background, using control regions in data, which is left unchanged with respect to the published results. The high number of pileup events at the HL-LHC poses challenges in maintaining high acceptance when triggering on multi-jet final states. The sensitivity has been studied as a function of the minimum online jet \pT requirement, and the minimum jet \pT used in the offline analysis is set by the four-jet trigger threshold. An increase of the jet \pT threshold to 75 GeV would degrade the sensitivity by 50\%
relative to the 40 GeV offline threshold of the corresponding Run 2 result used for this extrapolation~\cite{ATLAS-TDR-029}.
The ATLAS results for the \bbtautau (\ensuremath{\mu \tau_{h},  e\tau_{h} } and \ensuremath{\tau_{h} \tau_{h}}, based on the Run 2 data, currently set the world’s strongest limit by a single channel. 
The Run 2 BDT distributions, used to separate the signal from the background processes, are scaled to the integrated luminosity of 3000~\ifb, taking into account the change of cross section with the increased centre of mass energy. In the Run 2 analysis one of the dominant systematic uncertainty is due to the limited statistics of the simulated samples used to estimate background processes and it is neglected in these extrapolations.

The analysis of the \bbgg channel is based on truth level particles convoluted with the detector resolution, efficiencies and fake rates, as derived from fully simulated samples using the upgraded ATLAS detector layout and assuming a pileup of 200 collisions per bunch crossing. The event selection makes use of a multivariate analysis with a BDT exploiting the full kinematic information of the event~\cite{ATLAS-CONF-2018-053}. The systematic uncertainties follow the prescriptions summarised in \refta{tab:systHLLHC}. Their effect is very small since this channel will still be dominated by statistical uncertainties at the end of the HL-LHC operations.

\paragraph{CMS projections}

The CMS estimates of the sensitivity to \hh production were derived using a parametric simulation based on the \textsc{DELPHES}~\cite{deFavereau:2013fsa} software\footnote{The parameterisation is based on the results obtained with a full simulation of the CMS detector and dedicated reconstruction algorithms.}, which provides a model of the CMS detector response in the HL-LHC conditions. 

The \bbgg is the most sensitive decay channel thanks to the excellent \hyy invariant mass resolution and the improved \btagging performance, as expected from the inclusion of the timing detector~\cite{Collaboration:2296612}. The inclusion of the track timing information~\footnote{By using timing information he number of spurious reconstructed secondary
vertices is reduced by 30\%.} provides an increase in the \btagging efficiency of about 4–6\% depending on the pseudorapidity, evaluated for the same mis-tag rate.
A multivariate kinematic discriminant is employed to suppress the background contributions, mostly originated from non-resonant \ensuremath{\gamma \gamma}  production in association with heavy flavour jets.

For the \bbtautau decay channel, (\ensuremath{\mu \tau_{h},  e \tau_{h} } and \ensuremath{\tau_{h} \tau_{h}}), the separation of the \hh signal from the background processes (mostly \ttbar and Drell-Yan production of \ensuremath{\tau^{+} \tau^{-}} pairs) is achieved with a machine learning approach based on a deep neural network (DNN), using a wide set of kinematic variables.

Two complementary strategies were explored to identify the \bbbb signal contribution depending on the event topology. In the case where the four jets from the \bbbb decay could all be reconstructed separately, ``resolved" topology, the use of multivariate methods was explored to efficiently discriminate the \hh signal from the overwhelming multi-jet background. 
Alternatively when the two Higgs bosons are produced with a high Lorentz boost, they are reconstructed as two large radius jets (``boosted” topology), as described in Sec.~\ref{sec:jetReco}. While the large majority of SM \hh events falls in the resolved category,  boosted topologies help to suppress the multi-jet background and provide sensitivity to BSM scenarios where the differential \hh production cross section is enhanced at high \mhh by the presence of $gg\hh$ or $\ttbar\hh$ effective contact interactions.\\

\paragraph{Comparison between ATLAS and CMS}

Different assumptions and optimisation strategies have been adopted by the ATLAS and CMS collaborations which have impacted the reported sensitivity at the HL-LHC. A short summary of the main ones follows:

\begin{itemize}

\item ATLAS assumed a better \btagging performance with respect to CMS (approximately 5\% larger efficiency for the same mis-tag rate);
\item A second local minimum appears for the ATLAS likelihood while it is almost absent for CMS, and it is mostly due to the \bbgg contribution, as shown in Fig.~\ref{fig:comb_HH_experiment}-right. This is a result of the different optimisation strategies of the two experiments. ATLAS \bbgg analysis has been optimised to maximise the sensitivity to the SM signal and events with low \mhh values are not used, while the CMS analysis accounts for a dedicated category of events with low values of \mhh to gain sensitivity to different hypotheses of the \klambda value.
\item  ATLAS employs a dedicated $\gamma \gamma$+HF QCD simulation sample for the \bbgg analysis, while CMS uses an inclusive $\gamma \gamma$+jets sample. The limited statistics of the CMS simulated samples in the signal region reduced the quality of the BDT training and prevented from using the category with the best expected signal over background ratio.
\item the \ensuremath{m_{\gamma \gamma}} resolution for the \bbgg analysis takes into account the variable ageing of the calorimeters with respect to the Run 2 values. The degradation of the CMS \ensuremath{m_{\gamma \gamma}} resolution is due to the expected, slow but steady, ageing of the crystals in the barrel  of CMS ECAL with the accumulated luminosity and radiation dose~\cite{Collaboration:2283187}, while the ATLAS LAr calorimeter is not expected to suffer from ageing~\cite{ATLAS-TDR-027};
\item For the $\tau_{h}$ reconstruction performance, ATLAS has extrapolated from Run 2 performance, while CMS parameterised the efficiency and the fake rate for the HL-LHC scenario, resulting in a worse sensitivity in the fully hadronic decay channel.
\end{itemize}

\subsection{Results}\label{sec:results_HL_LHC}
The sensitivity of all channels studied for HL-LHC is shown in \refta{tab:comb_significance}. From the table is evident that the \bbgg and \bbtautau decay channels provide the best sensitivity, followed by \bbbb. The \bbww and \bbzz analyses although limited by the small branching fraction, provide additional sensitivity when combined with the other channels.
Results from the analyses of different decay channels have been statistically combined within each collaboration. Systematic uncertainties associated to common backgrounds and the \hh signal were taken into account as correlated nuisance parameters across the corresponding decay channels. The uncertainties associated to the same physics objects, such as the to the \btag efficiency, were also correlated. 

Considering only statistical uncertainties, the combined significance of the ATLAS (CMS) analysis was found to be 3.5$\sigma$ (2.8$\sigma$) for the SM \hh production rate. This is reported in \refta{tab:comb_significance}, where the individual values for each channel are also shown. 

\begin{table}[tb!]
\begin{center}
\begin{tabular}{lcccc} \hline
 & \multicolumn{2}{c}{\textbf{Statistical-only}} & \multicolumn{2}{c}{\textbf{Statistical + Systematic}}\\
 \hline
 & ATLAS & CMS & ATLAS & CMS \\
\hline
\hhbbbb & 1.4 & 1.2 & 0.61 & 0.95 \\
\hhbbtt & 2.5 & 1.6 & 2.1 & 1.4 \\
\hhbbyy & 2.1 & 1.8 & 2.0 & 1.8 \\
\hhbbVV & - & 0.59 & - & 0.56 \\
\hhbbZZ & - & 0.37 & - & 0.37 \\
\hline
Combination &  3.5 & 2.8 & 3.0 & 2.6 \\
 &  \multicolumn{2}{c}{4.5} & \multicolumn{2}{c}{4.0}\\
 \hline
\end{tabular}
\end{center}
\vspace*{-0.3cm}
\caption{Significance in standard deviations of the individual channels as well as their combination, under the assumption of a SM rate for \hh production~\cite{Cepeda:2019klc}.}
\label{tab:comb_significance}
\end{table}


The combined sensitivity to the self-coupling modifier parameter \klambda is assessed by generating an Asimov dataset containing the background plus SM signal. The individual contributions to the scan of the likelihood as a function of \klambda, for each decay mode, are shown in Fig.~\ref{fig:comb_HH_experiment}-right.
The structure of the likelihood function, characterised by two local minima, is a result of the quadratic dependence of the total cross section on \klambda, while the relative height of the two minima depends on the analysis acceptance as a function of $\kappa_{\lambda}$ and the relative sensitivity to differential \mhh information. 
Considering only statistical uncertainties, \klambda is constrained at 95\% confidence level (CL) to $-0.4\leq \klambda \leq 7.3$ and $-0.18\leq \klambda \leq 3.6$ for ATLAS and CMS, respectively. 


A simple statistical combination of ATLAS and CMS analyses was also performed, by treating all channels as uncorrelated contributions. This is a reasonable assumption, despite the theory and the luminosity uncertainties being expected to be correlated between the experiments, since their impact is negligible on the individual results.
A combined significance of $4\sigma$ can be achieved, when all systematic uncertainties are included, reaching $4.5\sigma$ neglecting all of them. 

The combined likelihood scan as a function of \klambda is reported in Fig.~\ref{fig:comb_HH_experiment}-left. The 95\% (68\%) CL intervals is $0.1\leq \klambda \leq 2.3$ ($0.5\leq \klambda \leq 1.5$). The hypothesis corresponding to the absence of self-coupling (\klambda=0) would be excluded at the 95\% CL in these projections for HL-LHC. 
The lower limit on \klambda is slightly higher for CMS thanks to the contribution of the \hhbbbb, \hhbbVV and \hhbbZZ, while the upper limit is similar.

\begin{figure}[!htb]
\centering 
\includegraphics[width=0.49\textwidth]{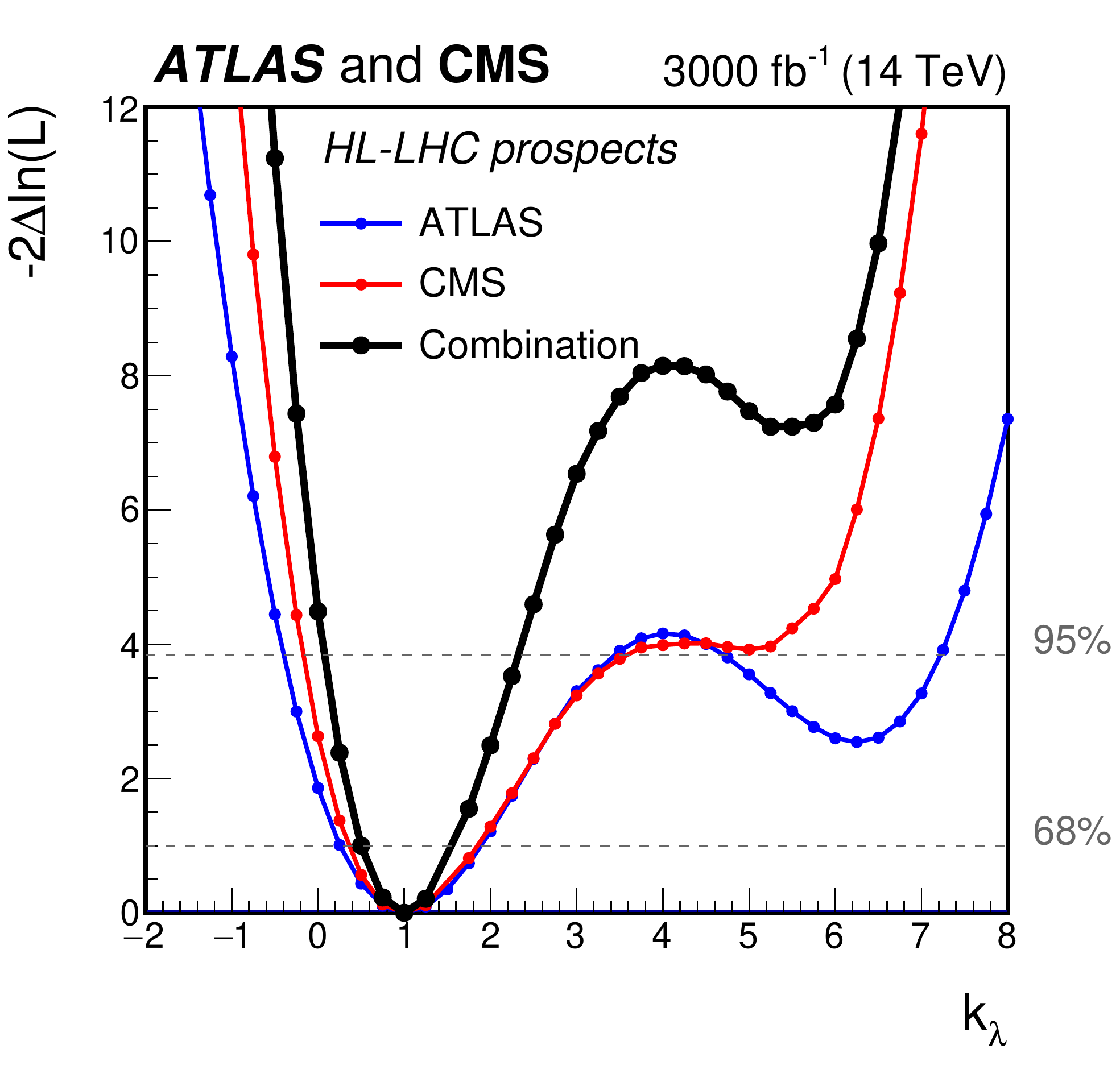} 
\includegraphics[width=0.49\textwidth]{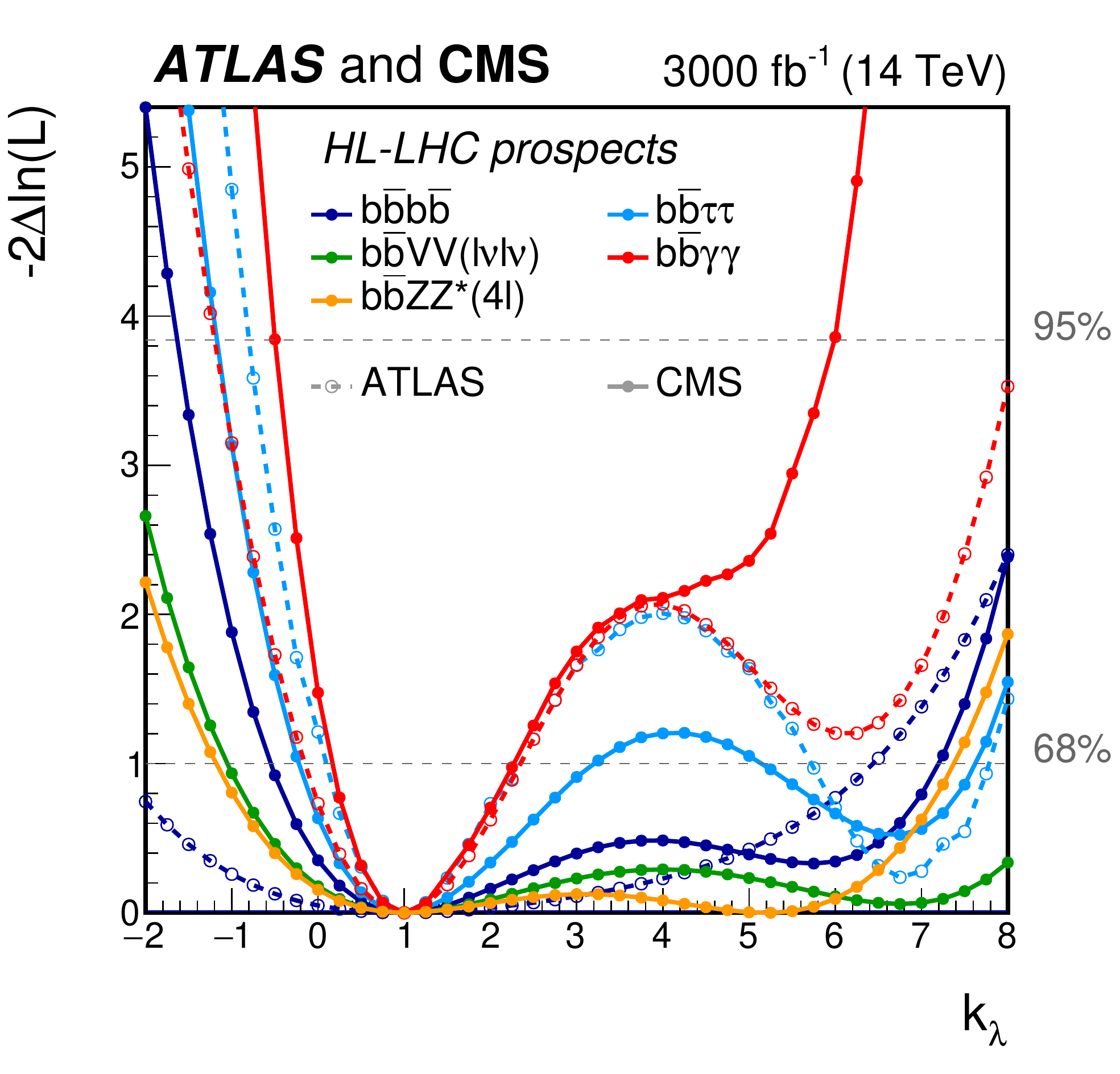}
\vspace*{-0.2cm}
\caption{Minimum negative-log-likelihood as a function of \klambda. Left: The ATLAS, CMS and combined results. Right: Results are shown by decays channels for ATLAS and CMS separately~\cite{Cepeda:2019klc}. Since the \hhbbVV and \hhbbZZ channels are exploited only by the CMS experiment, the likelihoods for those two channels are scaled to 6000\ifb in the combination.} 
\label{fig:comb_HH_experiment} 
\end{figure}

The expected measured values of \klambda for the different channels, as well as the combined measurement, are shown in the first box of Fig.~\ref{fig:comb_HH}.



\section{Single Higgs measurements}
\label{sec:singleH_fut}

As discussed in Secs.~\ref{tril-single} and~\ref{singleH_exp}, a complementary strategy to extract information on the trilinear coupling is through precise measurements of single Higgs production, decays and kinematic distributions.

In particular differential cross section measurements as a function of the Higgs boson transverse momentum, \pTH, are used to extract an indirect constraint on the Higgs boson self-coupling, as they allow to disentangle the effects of modified Higgs boson self-coupling values from other effects such as the presence of anomalous
Higgs couplings to the top quark.
The kinematic dependence of these deviations are determined by reweighting signal events, on an event by event basis, using the tool described in Ref.~\cite{EWreweightingtool}, similarly to the procedure adopted in Sec.~\ref{singleH_exp} for the LHC results.
The CMS experiment has performed the first HL-LHC analysis of this kind, for the $\ttbar H$ production mode followed by the decay \hyy \cite{CMS:2018rig}, exploiting both hadronic and leptonic \ttbar decay modes.

The left panel of Fig.\ref{fig:ttH-Hgg} shows the expected $\ttbar{H}$ and $tH$ differential cross sections times branching ratio, for the fiducial phase space defined in \cite{CMS:2018rig}, in bins of \pTH, which derives \klambda . dependent corrections to the tree level cross sections as a function of the kinematic properties of the event. 
Assuming $3~\mathrm{ab}^{-1}$ of HL-LHC data, uncertainties at the level of 20--40\% in the differential cross sections are expected. 

The profile log-likelihood scan as a function of \klambda is shown in the right panel of Fig.~\ref{fig:ttH-Hgg}. For simplicity, in the scan, all other Higgs boson couplings are assumed to have their SM values. In particular the Higgs coupling to top quarks is set to its SM value, $\kappa_t=1$. A constraint of $-4.1\leq \klambda\leq 14.1$ can be derived at 95\% CL. Slightly more stringent results can be obtained considering only the statistical uncertainty. 

\begin{figure}[!htb]
\centering 
\includegraphics[width=0.49\textwidth]{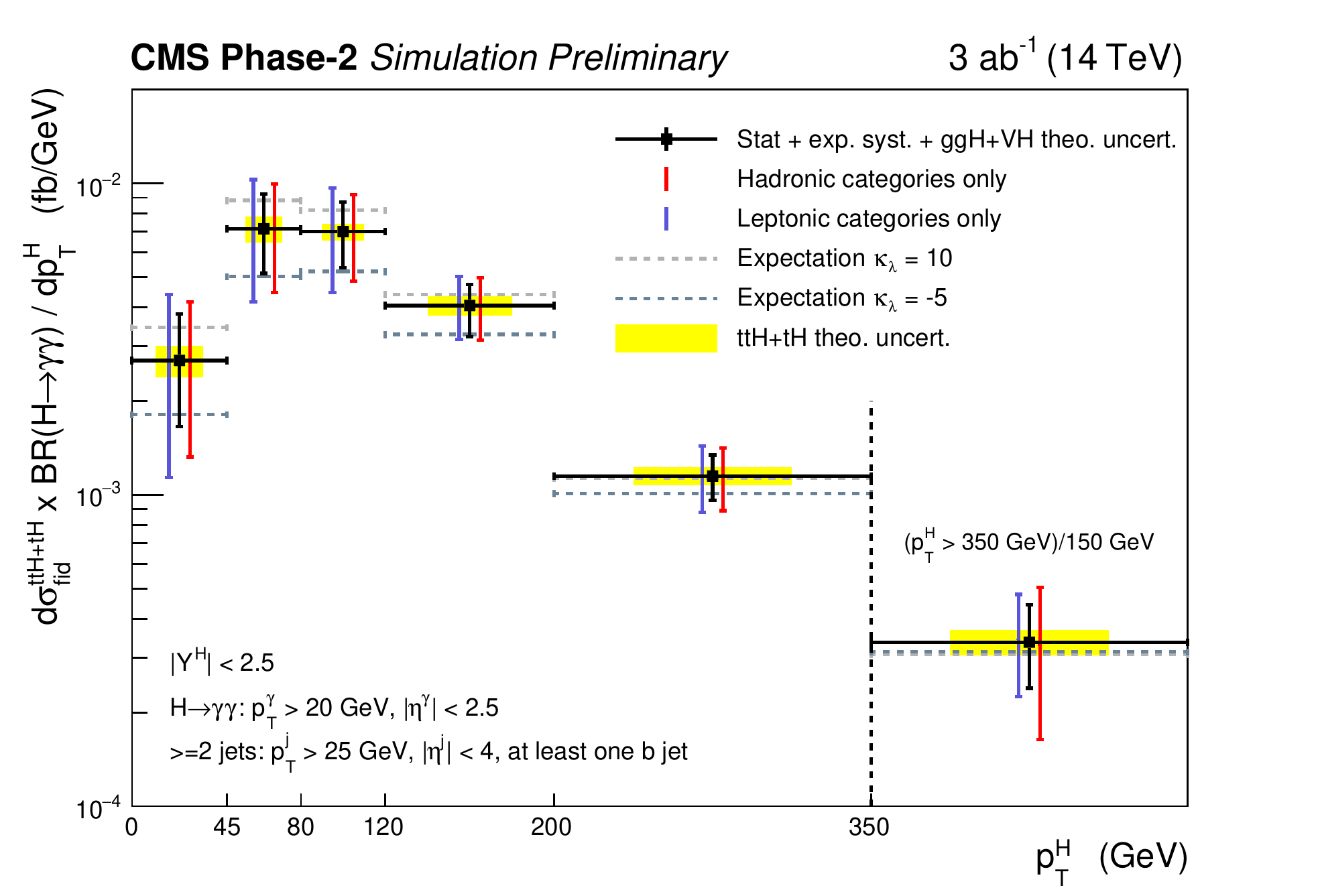} 
\includegraphics[width=0.49\textwidth]{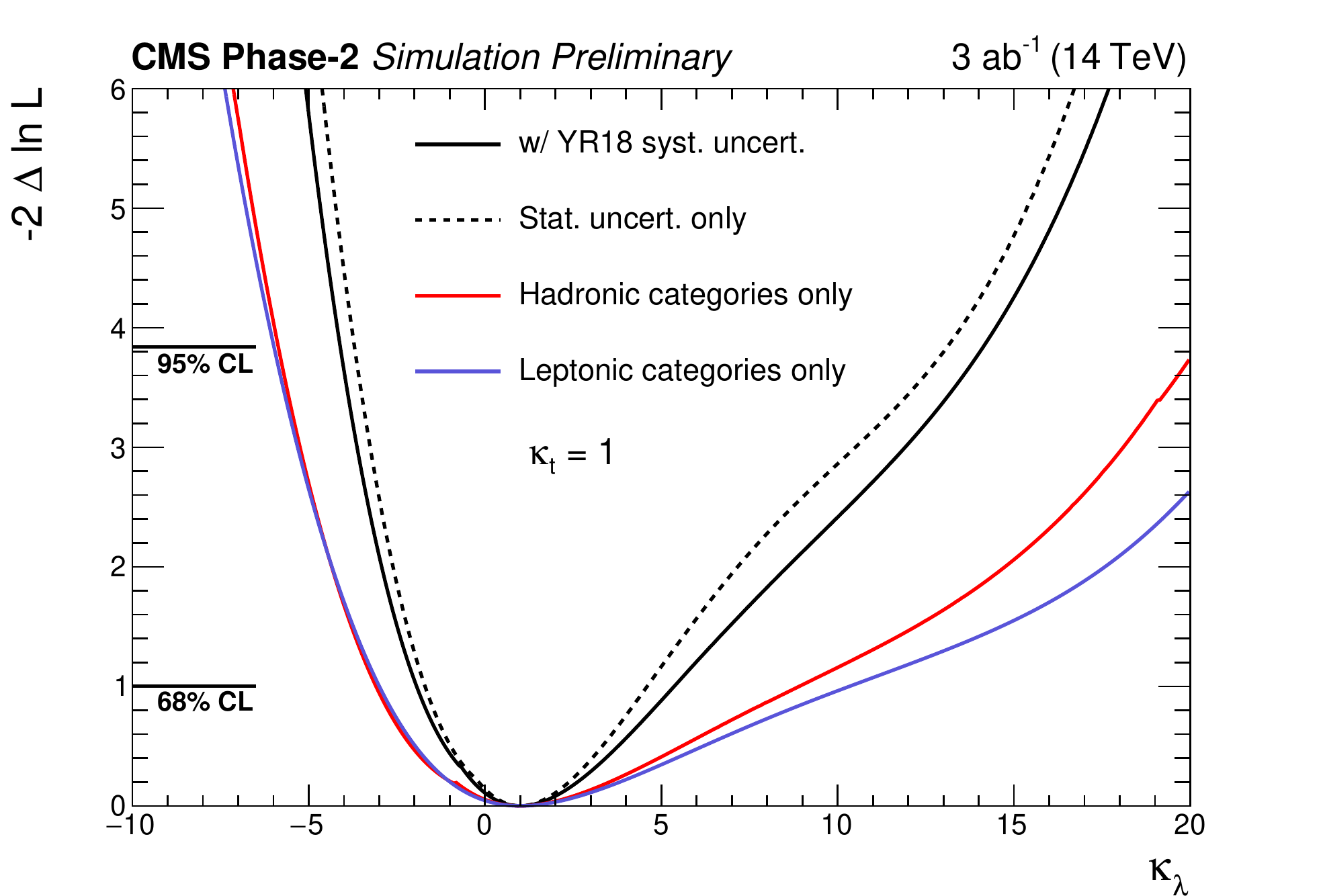}
\caption{Left, The expected differential $\ttbar{H}$ and $tH$ cross sections times branching ratio, along with their respective uncertainties, in bins of \pTH. 
Right, Profile log-likelihood scan as a function of \klambda~\cite{CMS:2018rig}. The simulation assumes 3~ab$^{-1}$ of data.} 
\label{fig:ttH-Hgg} 
\end{figure}

The obtained bounds are much weaker compared to the direct measurement using double Higgs pair production. They make use however of only one production mode and one decay channel, resulting less competitive than the present limit from single Higgs measurement published by the ATLAS collaboration (see Sec.~\ref{singleH_exp}). Combining all other production mechanisms and decays, should therefore lead to more stringent constraints on the self-coupling. This highlights an interesting complementarity on the determination of the Higgs self-interaction strength between double and single Higgs production mechanisms.

Many models that predict sizeable deviations in \klambda, also predict deviations in the other Higgs couplings. Therefore, global fits that include single Higgs differential measurements, as well as direct constraints from double Higgs measurements are needed to constraints all parameters (see Sec.~\ref{singleH_exp}). The SMEFT framework described by 9 free parameters, or the HEFT with 5 parameters can be used to perform these global fits (see Chapter~\ref{chap:EFT}).

As discussed in Sec.~\ref{tril-single} (see, in particular, the right panel of Fig.~\ref{fig:hllhcchi2}), while \hh measurements are expected to drive the bound on \klambda, differential single Higgs data is nonetheless relevant as it can help lifting the degeneracy between minima around $\delta\klambda\sim 5$.

A summary plot for the different expected direct and indirect constraints on the Higgs boson self-coupling at the HL-LHC is provided in Fig~\ref{fig:comb_HH}. In particular the results from a global fit are compared to the constraints derived by assuming BSM effects would impact \klambda only (exclusive fit).


\begin{figure}[!htb]
\centering 
\includegraphics[width=0.8\textwidth]{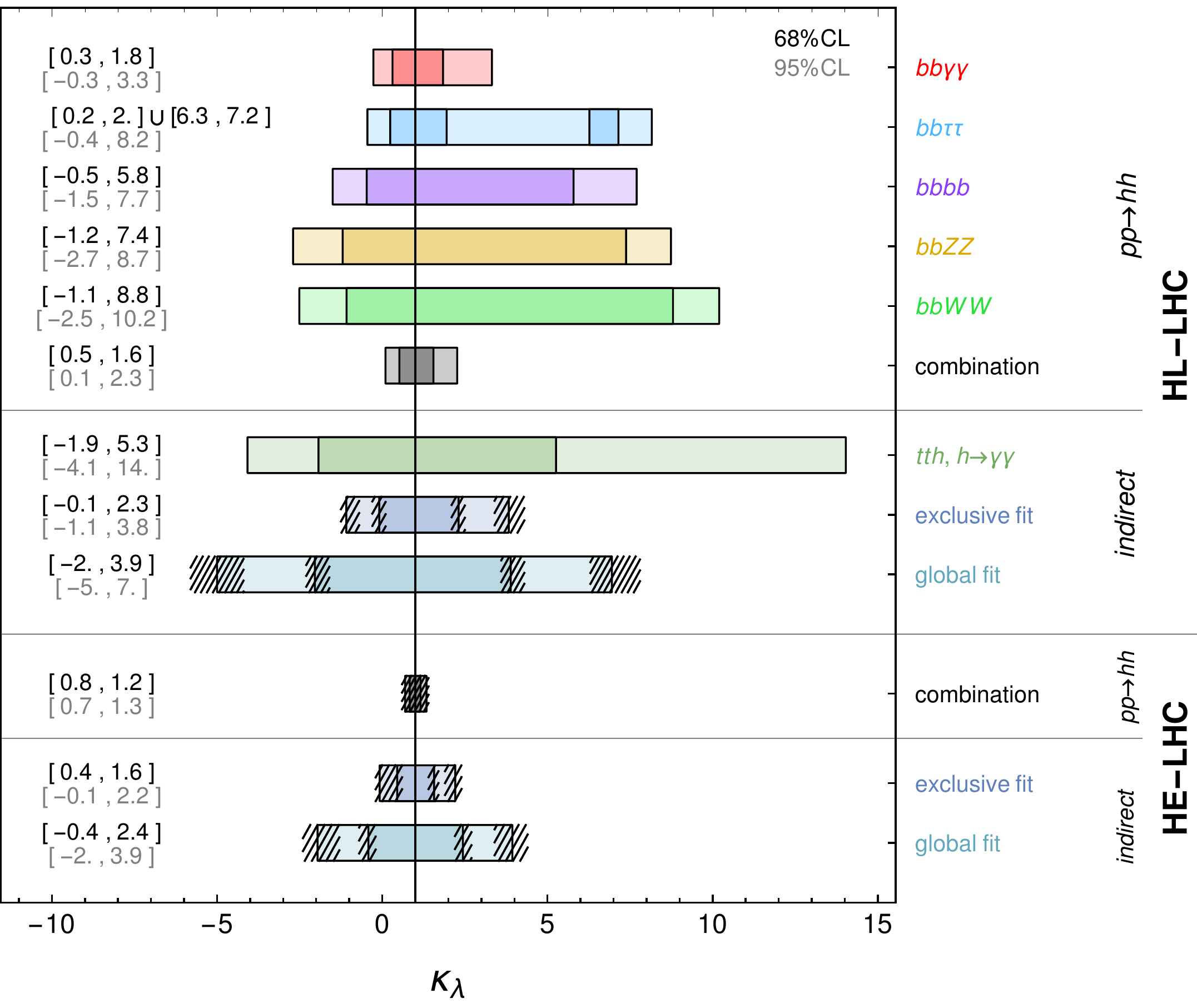}
\caption{Summary of the expected measured values of \klambda from the several channels. The first box corresponds to \hh searches at the HL-LHC; the second box to indirect probes using single Higgs measurements at the HL-LHC; the third box to the HE-LHC \hh searches. The lines with error bars show the total uncertainty on each measurement while the boxes correspond to the statistical uncertainties only. In the cases where the extrapolation is performed only by one experiment, same performance are assumed for the other experiment and this is indicated by a hatched bar~ \cite{Cepeda:2019klc}.} 
\label{fig:comb_HH} 
\end{figure}

%% file: HHFuture/HHineev3.tex
\label{chap:ee}

\def\Dlr{\mathrel{\raise1.5ex\hbox{$\leftrightarrow$\kern-1em\lower1.5ex\hbox{$D$}}}}

\section{Introduction}

The  Higgs self-coupling can be measured at $e^+e^-$  colliders in two
different ways, in parallel to the ways that this coupling is studied
at hadron colliders.  On one hand, the Higgs self-coupling can be measured
using single Higgs production processes.  Here, the use of $e^+e^-$ beams
offers two advantages.   First, $e^+e^-$  colliders promise an intrinsic precision in 
single Higgs measurements that will be higher than at hadron colliders,
reaching or exceeding the 1\% level.  Second, $e^+e^-$ colliders offer
a
larger number of independent single Higgs
observables that can be used to great effect
in the interpretation of the measurements. In particular,
runs of $e^+e^-$ colliders at two different energies
offer a way to lift degeneracies that plague a global fit
of inclusive analyses at hadron colliders and prevent them from separating
the Higgs self-couplings from the  Higgs couplings
to the other SM particles.  On the other hand, the self-coupling 
can also  be measured using the double Higgs 
production processes.  At $e^+e^-$ colliders, this method uses the 
reactions  $e^+e^-\to ZHH$   (double Higgs-strahlung) and
$e^+e^-\to \nu\bar\nu HH$ (vector boson fusion). 

In this chapter, we will discuss all of these aspects in turn.   We
will set the stage in Sec.~\ref{sec:eescope}  by describing the proposed
next-generation $e^+e^-$ Higgs factories in terms of their expected
energies and integrated luminosities. All of these proposed
facilities can carry out the
single Higgs analysis, though only colliders that reach a centre-of-mass (CM)  energy of
at least 500~GeV have access to double Higgs production. 
In Sec.~\ref{sec:oneHv1},
we will describe the capabilities of these facilities for
determining  the Higgs self-coupling through 
single Higgs measurements.   The discussion here and in the next few
sections is given in the simplest model context, in which a
variable $\kappa_\lambda$ is added to the SM and determined in a
1-parameter fit.  This analysis shows a significant
improvement compared to what can be achieved at
HL-LHC with the single Higgs technique.
However, one should  be careful in interpreting
the result obtained in this 1-parameter fit as, in any BSM
scenario, other parameters will affect the single Higgs
measurements and should be properly taken into account in a
global fit.  This point is discussed in some detail later in the chapter.

In Sec.~\ref{sec:genHHee}, we  discuss general
features of the $HH$ reactions available at $e^+e^-$ colliders.
Following this, we review in Secs.~\ref{sec:eedirectILC} and
\ref{sec:eedirectCLIC} the analyses of $HH$ production reactions
 and the projected accuracy of the determination of
$\kappa_\lambda$.

In Secs.~\ref{sec:eedirectEFT} and \ref{sec:eeindirect}, we will
revisit these measurements in the context of the general description
of new physics effects by the SMEFT.
Once we have determined that $\kappa_\lambda$ is
not equal to 1, we are in the domain of physics beyond the SM.   The
same new physics that alters $\kappa_\lambda$  can also, in principle,
alter the other couplings of the Higgs boson and, thus, can create
changes in measured single- and double-Higgs cross-sections that have
nothing to do with the Higgs potential.    It turns out that $e^+e^-$
measurements offer incisive tools for separating the effect of
$\kappa_\lambda \neq 1$ from other new physics effects on the primary
observables.   Thus, in most cases, it is possible to determine the
magnitude of
$\kappa_\lambda$ specifically and to  separate its effects from those of
other, quite general, effects of new physics.
 We present the analysis first, in
Sec.~\ref{sec:eedirectEFT}, for the determination from double-Higgs production
and then, in Sec.~\ref{sec:eeindirect}, for the determination from
single Higgs production.

Sec.~\ref{sec_ee:quartic}  reviews a first attempt to determine
both the Higgs cubic and quartic couplings from $e^+e^-$ observables.
 The analysis here is
given in a
2-parameter model that does not include other possible new physics
effects
expected in BSM scenarios.   We note that, as in the 1-parameter fit of single Higgs
measurements,  inclusion of 
these other BSM parameters can dramatically change the results and 
prevent  the  assignment of  a robust bound on the quartic
self-coupling.   Finally,
Sec.~\ref{sec:eeconclude} gives some conclusions.

\section{Scope of the proposed $e^+e^-$ Higgs factories}
\label{sec:eescope}
\contrib{A. Blondel, P. Janot}

At this time, four proposals for next-generation $e^+e^-$ colliders are
under consideration in different regions of the world.   Two of
these are circular colliders---the Circular Electron-Positron Collider (CEPC) and the Future
Circular (Lepton) Collider  (FCC-ee). The other two---the International
Linear Collider (ILC) and the Compact Linear Collider (CLIC)---are single-pass
linear colliders.   All four facilities aim at providing
 precision measurements of the electroweak and
Higgs interactions at high energy and, in particular, measuring
parameters of the Higgs boson with unprecedented precision.  The
measurements of $\klambda$ and most other Higgs boson parameters
at $e^+e^-$ colliders are expected to be statistics-limited, so it is
important to pay attention to the sizes of the  total data sets
proposed for these machines.   The purpose of this section is to
provide a guide to the data sets currently proposed for the various
stages of each machine.

The two circular colliders, CEPC and FCC-ee, have similar physics programs. The CEPC conceptual design report was presented in 2018~\cite{CEPCStudyGroup:2018rmc,CEPCStudyGroup:2018ghi}.  The FCC-ee conceptual design report was presented in 2019~\cite{Abada:2019}. Both colliders propose a program of data-taking at 240 GeV, plus a program of precision electroweak measurements at the Z pole and WW threshold.  
In each case there are two detectors, taking a total of 5.6 ab$^{-1}$ at 240 GeV in the CEPC proposal and 5 ab$^{-1}$ in the FCC-ee proposal.  The FCC-ee plan also includes a second stage at 350 GeV and 365 GeV to reach the $t\bar t$ threshold, with 0.2 + 1.5 ab$^{-1}$  of data.  For the CEPC, a run at the $t\bar t$ threshold has been studied, but it is not part of the proposed program at this time.  The FCC-ee group has also discussed a scenario with 4 detectors, taking a total of 12 ab$^{-1}$  in the 240 GeV stage and 5.5 ab$^{-1}$  at 365 GeV~\cite{Blondel:2018aan}. We will refer to this scenario below as “FCC-ee (4IP)”. It has been suggested that the addition of Energy Recovery Linacs to the FCC-ee will allow running at 500 GeV; that proposal is still at a preliminary stage~\cite{Litvinenko:2019txu}.

The CLIC conceptual design report was presented in 2012~\cite{Aicheler:2012bya}. An update of the design and run plan was recently presented in \cite{Charles:2018vfv} and the corresponding physics case described in\cite{deBlas:2018mhx}.   CLIC is proposed to be constructed in three stages, the first at 380 GeV with 1 ab$^{-1}$  of integrated luminosity, the second at 1.5 TeV with 2.5 ab$^{-1}$ , and the third at 3 TeV with 5 ab$^{-1}$ .

The FCC-ee proposal with two detectors and the CLIC proposal were included in the report \cite{Bordry:2018gri} that sets out the future collider projects at CERN to be considered in the update of the European Strategy for Particle Physics.

The ILC completed its technical design report in
2013~\cite{Behnke:2013xla}.
A recent review of the ILC
design and physics capabilities can be found in
\cite{Bambade:2019fyw}.   The ILC is proposed to be  constructed in two
stages, the first at 250~GeV  with  2~ab$^{-1}$ of
integrated luminosity, the second at 500~GeV  with 4~ab$^{-1}$, with
an additional 
 short run at 350~GeV with 200~fb$^{-1}$.  The
current proposal for the ILC does not include a run at 1~TeV, but this
is within the capabilities of the ILC technology.  Parameters for 
1~TeV running were already given in the 2013 TDR.  The reports
\cite{Barklow:2015tja} and \cite{Bambade:2019fyw} describe a possible
run at 1~TeV taking 8~ab$^{-1}$ of data.

\section{Determination of the Higgs self-coupling from single Higgs
  reactions ---  CEPC,
  FCC-ee, CLIC, ILC}
 \contrib{A. Blondel, C. Grojean, P. Janot}\label{sec:oneHv1}

We first consider the determination of the trilinear Higgs
self-coupling from single Higgs reactions. The idea of this analysis
is similar to that presented in Sections~\ref{singleH_exp} and~\ref{sec:singleH_fut}.   If
$\klambda \neq 1$, loop diagrams containing the triple-Higgs
vertex will produce radiative corrections to Higgs boson production
cross-sections and decay rates that are proportional to
$\klambda$.  These radiative
corrections are at the percent level.   Since $e^+e^-$ colliders are
designed to measure Higgs boson cross-sections and branching ratios at
or below 
this level, their measurements can provide interesting constraints on
$\klambda$.   The study of these constraints was initiated by
McCullough~\cite{McCullough:2013rea}, who pointed out that these
give a radiative correction that,
for $(\klambda - 1)= 1$,
increases  the cross-section for $e^+e^-\to ZH$  by about 1.5\% at 
$\sqrt{s} \sim$~240--250~GeV.  Some more subtle aspects of
McCullough's analysis will be discussed in Sec.~\ref{sec:eeindirect}.

\begin{figure}
\begin{center}
  \includegraphics[width=0.60\hsize]{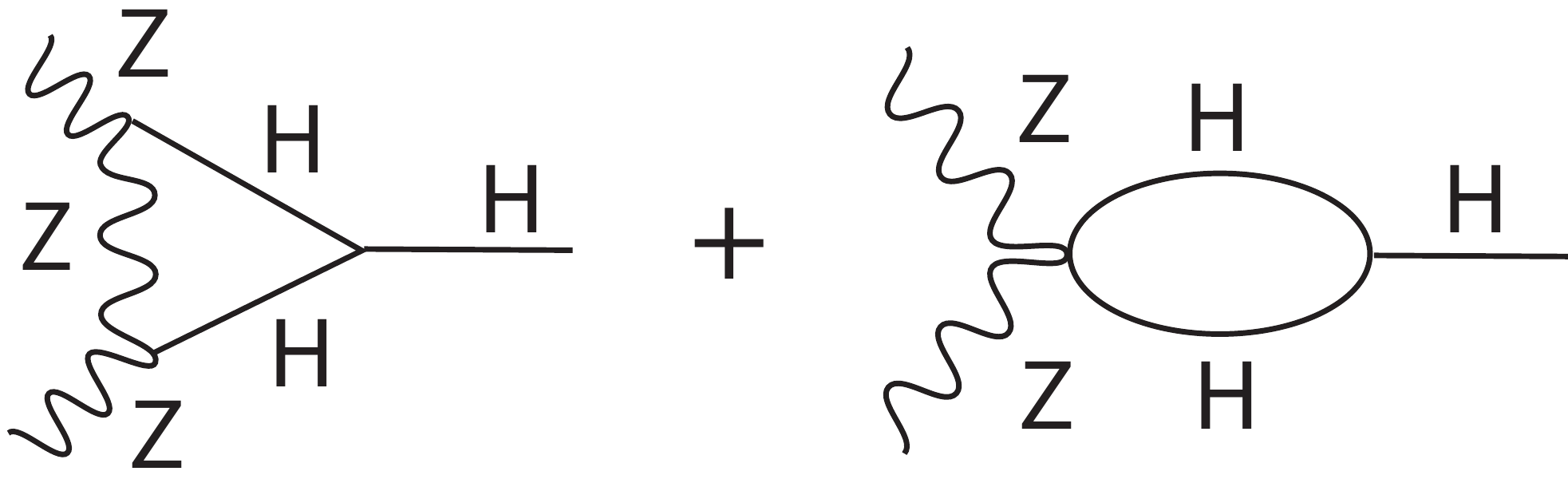}
\end{center}
\vspace*{-0.3cm}
\caption{Feynman diagrams contributing to the shift of the $HZZ$
  vertex due to the 1-loop effect of the Higgs self-coupling~\cite{McCullough:2013rea}.}
\label{fig:ZZH}
\end{figure}

For $e^+e^-$ colliders, the most important such loop diagrams
are those shown in Fig.~\ref{fig:ZZH}.   These diagrams correct the
$HZZ$ vertex that appears in the production reaction $e^+e^-\to ZH$
and the decay $H\to ZZ^*$.   The very similar diagrams with external
$W$ bosons correct the
$HWW$ vertex that appears in the production reaction $e^+e^-\to
\nu\bar\nu H$
and the decay $H\to WW^*$.   The radiative correction to the $Ht\bar
t$ vertex, which contributes to the decay $h\to gg$ in 2 loops, gives
a smaller effect and will not be considered here.

\begin{table}
\begin{center}
 \begin{tabular}{lcc}
collider  &   1-parameter  &  full SMEFT \\ \hline
CEPC 240 &         18\%            &       -         \\  \hline
FCC-ee 240 &       21\%         &        -         \\ 
FCC-ee 240/365 &     21\%         &    44\%       \\ 
FCC-ee (4IP)           &     15\%         &      27\%        \\ \hline
ILC 250          &       36\%               &        -        \\ 
ILC 250/500    &      32\%            &        58\%        \\  
ILC 250/500/1000 &      29\%           &     52\%       \\  \hline
CLIC 380           &          117\%         &         -         \\ 
CLIC 380/1500          &         72\%          &       -      \\ 
CLIC 380/1500/3000         &     49\%              &      -             \\ \hline
\end{tabular}
\end{center}
\vspace*{-0.3cm}
\caption{Uncertainties on the value of $\klambda$ expected from
  precision measurements of single Higgs observables at $e^+e^-$
  colliders, from ~\cite{deBlas:2019rxi}.  The collider scenarios are
  listed by name and CM energy.   More details on each can be found in
  Sec.~\ref{sec:eescope}. Results are given for a 1-parameter fit to the SM plus
  a varying $\klambda$ and for a fit that includes the
  possibility of other new physics effects modelled by the
  SMEFT.   Cases in which the SMEFT analysis does not close  are denoted 
  by ``-''.  The  physics of the SMEFT 
 analysis is described in Sec.~\ref{sec:eeindirect}. In  \cite{deBlas:2019rxi}, the projected
 uncertainties from single Higgs analyses  are  presented combined with
 an assumed independent uncertainty of 50\% from the HL-LHC $HH$
 analysis.  We have removed that combination here to clarify the size
 of the constraint that comes specifically from  $e^+e^-$ colliders.}
 \label{tab:HFCfit}
\end{table}

Recently, the ECFA Higgs@Future Colliders working
group has performed fits to the expected set of single Higgs
measurements to assess the sensitivity to deviations of the Higgs
self-coupling from its SM value~\cite{deBlas:2019rxi}.    These fits use the expected measurement accuracies for the various single Higgs
 observables given in the references cited  in the previous section.   The results are shown in Table~\ref{tab:HFCfit}.   

 The table lists uncertainties from a 1-parameter fit, corresponding to the model in which the SM is modified only by a shift of the parameter $\kappa_\lambda$,  and a fit to a larger model including the complete set of new physics effects that can be parametrized by dimension-6 SMEFT operators.   The ECFA Higgs@Future Colliders group has reported its results as combined with an expected 50\% uncertainty in $\kappa_\lambda$ expected from the HL-LHC.  To clarify the extra information that will come from  $e^+e^-$ measurements,  the values given in the table remove the 
HL-LHC contribution and quote results from $e^+e^-$ measurements alone.   In some cases of the multi-parameter fit, the analysis does not close and the $e^+e^-$ results alone do not give a competitive constraint.   Those cases are indicated in the Table by a ``-''.

In all cases, the 1-parameter analysis seems to indicate a substantial
sensitivity to the Higgs self-coupling.   Including the
possibility of other new physics effects weakens this sensitivity, but, for some scenarios, the constraint is still a powerful one.  We discuss the physics of the multi-parameter fit in Sec.~\ref{sec:eeindirect}.

\section{$HH$ production processes at $e^+e^-$ colliders}
\label{sec:genHHee}

\begin{figure}[t]
\centering
\includegraphics[width=0.7\hsize]{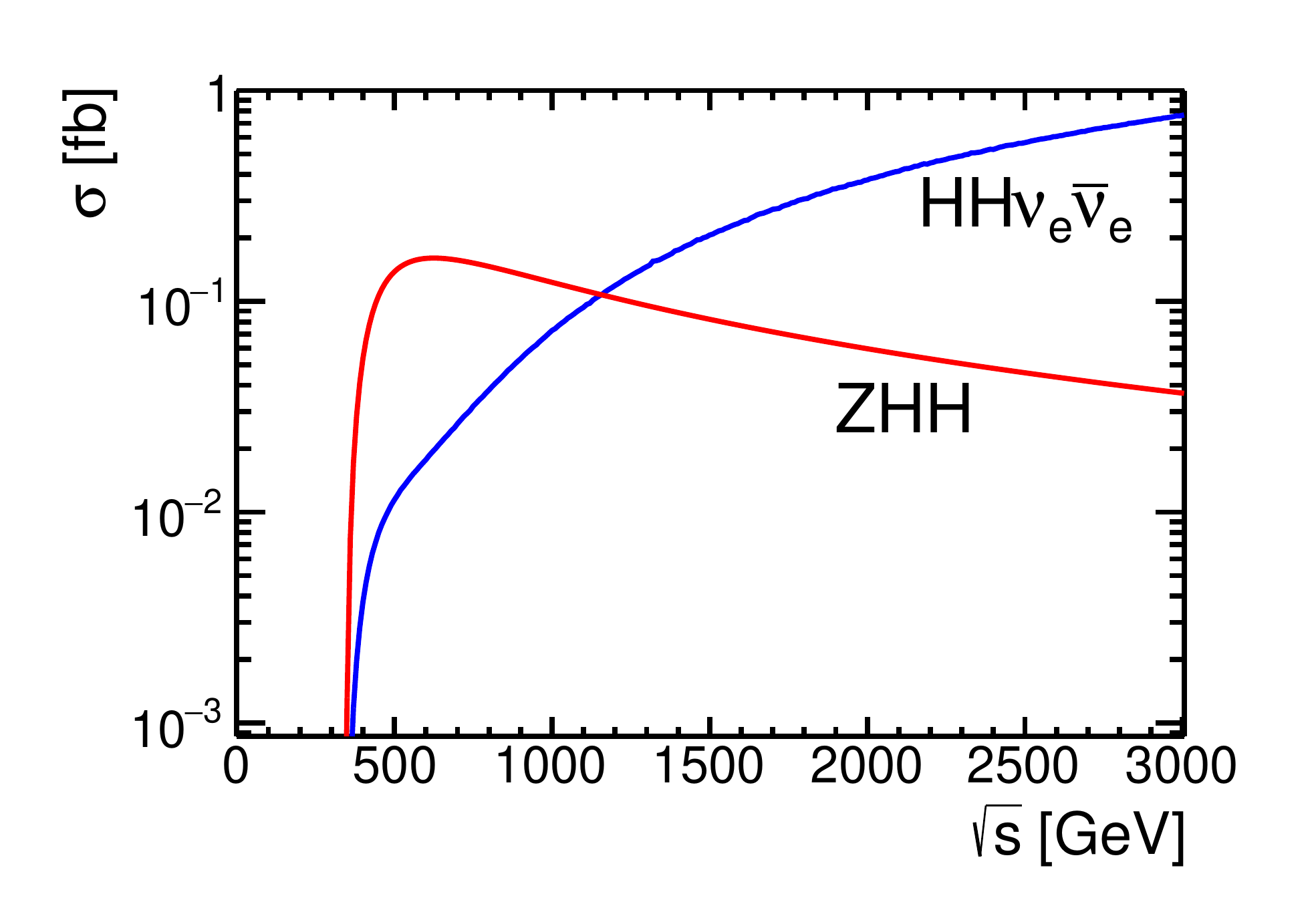}
\vspace*{-0.3cm}
\caption{Cross-sections for the double Higgs production processes
  $e^+e^- \to ZHH$ and $e^+e^- \to \nu\bar{\nu}HH$, as a function of
  $\sqrt{s}$ for $m_H=125\,$GeV.   The cross-sections are shown for
unpolarised beams.  These cross-sections are higher -- in the latter
case, by almost a factor 2 -- when the $e^-$ beam is highly polarised
in the left-handed sense. } 
\label{fig:sigzhh_vvhh}
\end{figure}

The cross-sections for the $ZHH$ and $\nu\bar\nu HH$ production
processes at $e^+e^-$ colliders 
are shown in Fig.~\ref{fig:sigzhh_vvhh}.   The cross-sections are
shown in this figure for unpolarised beams.   Planned analyses at
linear $e^+e^-$ colliders will make use of polarised beams.  Since the
$\nu\bar\nu HH$ process, in particular, requires the initial state
$e^-_Le^+_R$, working with polarised beams can raise the cross-section
significantly, by almost a factor of 2.  Still, these cross-sections
are very small, and the processes are difficult to recognise even in
the relatively clean environment of an $e^+e^-$ collider.

In both cases, the $HH$ production processes are multi-body reactions
whose
cross-sections increase slowly from threshold.   Energies much higher
than the nominal threshold energies of 250~GeV and 341~GeV are needed  to
produce a significant event sample. 
 The $ZHH$ process  is thus not accessible at 350~GeV, but it 
can be studied at an $e^+e^-$ centre-of-mass energy of
500~GeV.   The $\nu\bar\nu HH$ reaction, which 
is a 4-body process, requires  still higher energies, optimally, CM
energies above 1~TeV.   The ILD group has
studied these reactions at CM energies of
500~GeV and 1~TeV~\cite{Tian:2013qmi, KurataHHH,Duerig:2016dvi}.    The
CLICdp group has studied these reactions at CM energies of 1.5~TeV and
3.0~TeV~\cite{Roloff:2019crr}.   We describe the analyses
below.
In both cases, the analyses are done in the framework of full simulation using detailed detector models.   This simulation framework is reviewed in Secs.~6 and 7 of~\cite{Bambade:2019fyw}.

\begin{figure}[t]
\centering
\includegraphics[width=0.95\hsize]{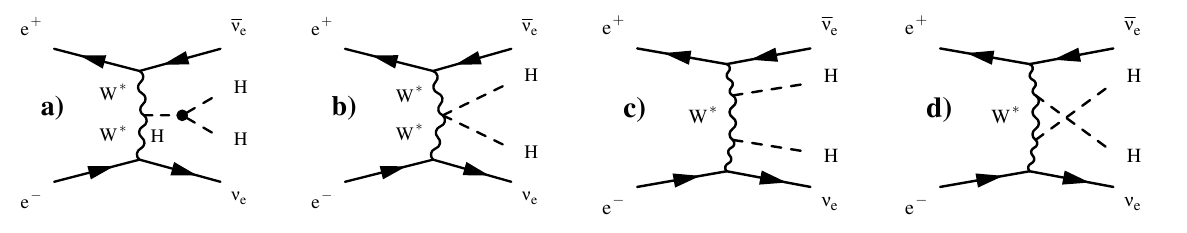}
\caption{Diagrams contributing to $e^+e^- \to \nu\bar{\nu}HH$.} 
\label{fig:hhdiagrams}
\end{figure}
%
The diagrams for both processes  include a diagram with
the Higgs self-coupling in interference with diagrams in which the two
Higgs bosons are radiated separately from $W$ or $Z$ propagators.  The
SM diagrams for $e^+e^-\to \nu\bar\nu HH$ are shown in Fig.~\ref{fig:hhdiagrams}.
Note that both processes appear at the tree level in the SM and, since
we are in the electroweak world, the tree level is a good
approximation to the full result.  

\begin{figure}[tb]
  \centering
    \includegraphics[width=0.7\hsize]{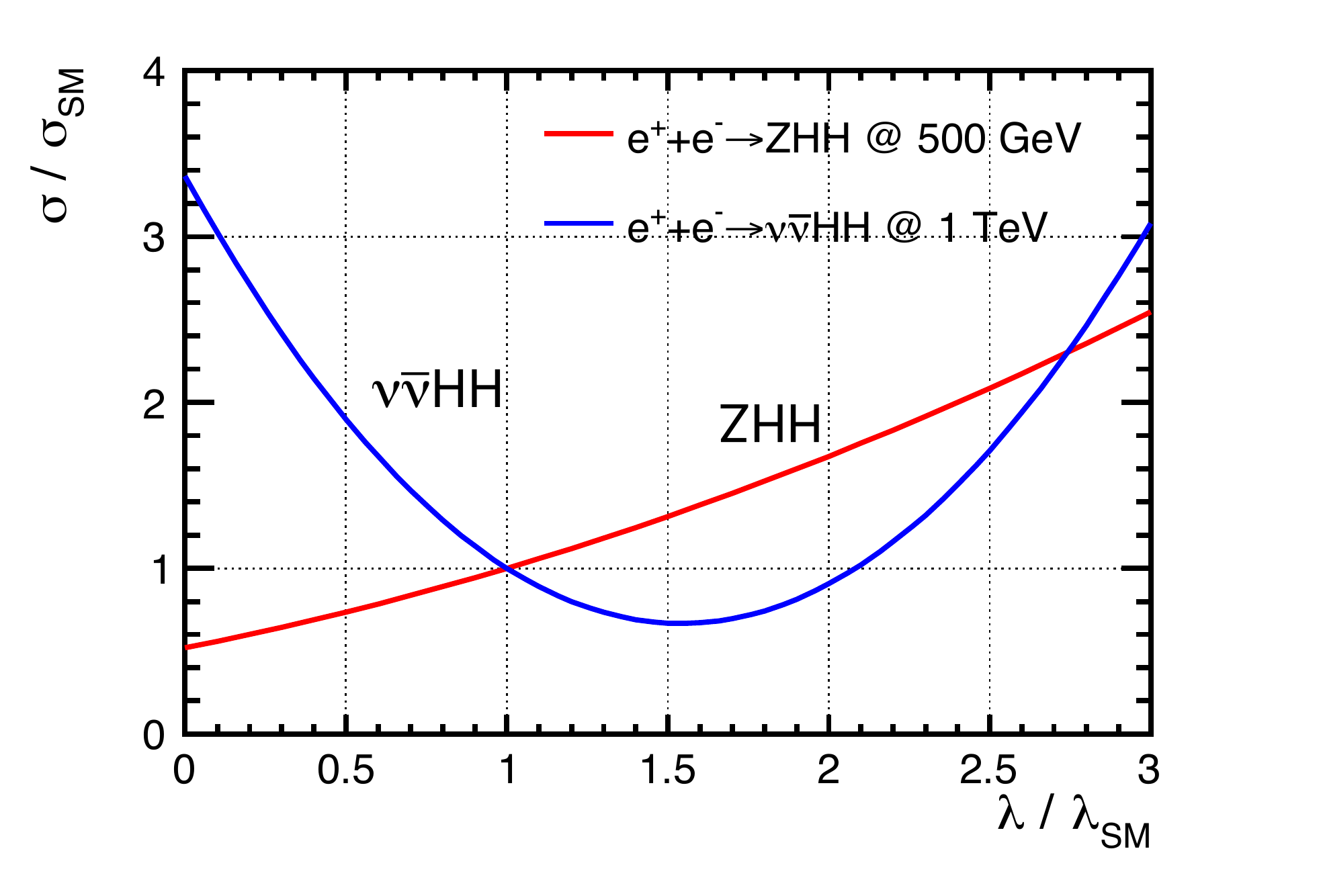} 
    \vspace*{-0.3cm}
    \caption{The dependence of the cross-sections for \hh production as a
      function of the Higgs self-coupling $\lambda$ for
      $e^+e^-\to ZHH$ ~(red line) and for 
      $e^+e^-\to\nu\bar{\nu}HH$ ~(blue line).
      The values of  both $\lambda$ and $\sigma$ are scaled to their 
      SM values. Note that the exact results depend on the assumed beam
    polarisations, here taken in the ILC scheme.}
  \label{fig:HHHBSM}
\end{figure}

A remarkable feature of the $e^+e^-$ reactions is that the two
processes have opposite dependence on $\klambda$.   That is, the 
self-coupling diagram interferes constructively with the other SM
diagrams in the case of $ZHH$ and destructively in the case of
$\nu\bar\nu HH$.   The dependence of the cross-section on the  variation of
the Higgs self-coupling is shown in Fig.~\ref{fig:HHHBSM}.    This
means that, whatever the sign of the deviation of $\klambda$
from 1, one of the two processes will have an increased cross-section
and will thus have increased statistical sensitivity to the actual value of $\klambda$.

\section{Measurement of $HH$ production --- ILC}
 \contrib{J. Tian}
\label{sec:eedirectILC}

The ILC in its second stage will have sufficient CM energy 
to observe the reaction $e^+e^-\to ZHH$. 
Full simulation studies show that discovery of this
double Higgs-strahlung process is possible within the planned program
of the ILC  with    4~ab$^{-1}$ of data at 500~GeV.  The run plan assumes polarisation of 80\% for the
electron beam and 30\% for the positron beam, with the beam
polarisations divided  among LR/RL/LL/RR  polarisations as
40\%/40\%/10\%/10\%.  Using the ILD detector model in full simulation
analyses, the ILD group studied the extraction of the $ZHH$ process
from background in the di-Higgs decay channels  $hh\to
b\bar{b}b\bar{b}$~\cite{Tian:2013qmi,Duerig:2016dvi} and $hh\to
b\bar{b}W^+W^-$~\cite{KurataHHH}.  At this point in the ILC program,
the absolute branching ratios of the Higgs boson into the $b\bar b$ and $WW^*$ channels will already be known with sub-percent
accuracy~\cite{Bambade:2019fyw}.  The measurements of $\sigma\cdot BR$
for the two channels can then be applied directly to  determination of the $HH$ production cross-section.  


\begin{figure}[tb]
  \centering
  \includegraphics[width=0.4\hsize]{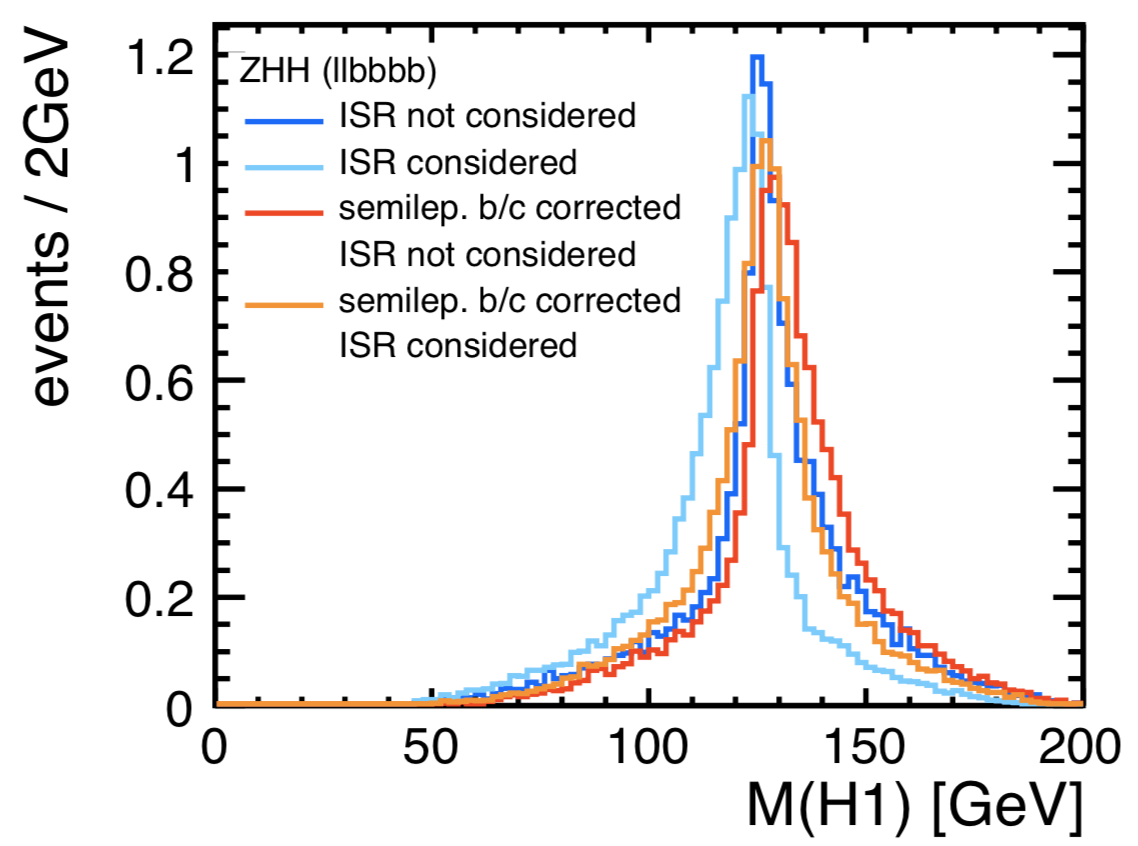}
  \quad
 \includegraphics[width=0.4\hsize]{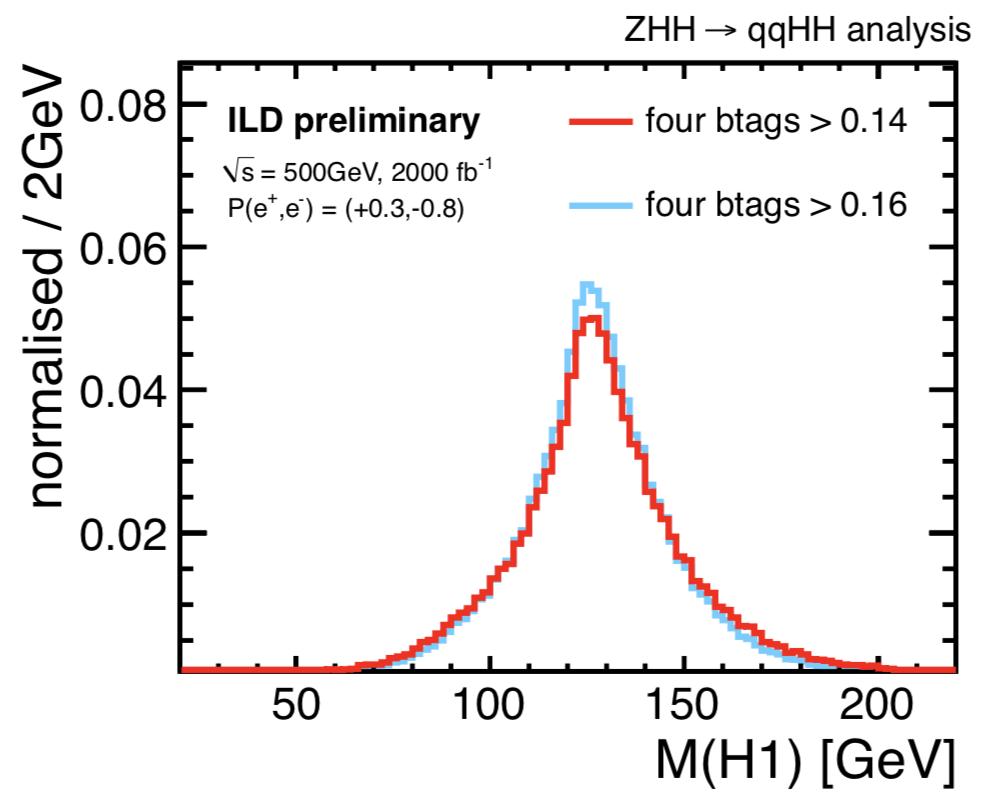}
 \caption{Reconstruction of the Higgs boson mass in the measurements
  of the $e^+e^-\to ZHH$ cross-section described in
\cite{Duerig:2016dvi}: left: highest energy Higgs boson in events
with $Z\to \ell^+\ell^-$; right: highest energy Higgs boson in
 events with  $Z\to q\bar q$. }
  \label{fig:eehmasses}
\end{figure}

The measurement of the $ZHH$ cross-section in the $b\bar{b}b\bar{b}$ mode
is described in detail in the Ph.D. thesis of
D\"urig~\cite{Duerig:2016dvi}.   She presents three  parallel analyses for
the channels of $Z$ decay to charged leptons, to neutrinos, and to
quarks (including $b\bar{b}$).  In each case, a kinematic fit is
carried out on the 4- or 6-jet system, and the output parameters from  this fit are  supplied to multivariate classifiers. The reconstructed Higgs
masses in the cases of charged leptonic and hadronic $Z$ decays are
shown in Fig.~\ref{fig:eehmasses}.   The  most important
backgrounds come from $ZZZ$, $ZZH$, and continuum $b\bar{b} q\bar{q}
q\bar{q}$ production.  The final selections correspond to efficiencies
of 36\%, 19\%, and 19\% for the charged lepton, neutrino, and quark
decay channels, respectively. 

Scaling the results from the $b\bar{b}b\bar{b}$ and $b\bar{b}WW$
analyses to the expected luminosity of 4~ab$^{-1}$, the combination
of the various channels yield a precision of 16.8\% on the $HH$ total cross
section.  Assuming a 1-parameter fit to  the SM with only the Higgs self-coupling as a free parameter,  this corresponds to an uncertainty of 27\% on that coupling $\klambda$.

At still higher energies, vector boson fusion becomes the dominant  \hh
production channel.  In a linear collider, luminosity is expected to
increase linearly with CM energy.   Thus, studies at 1~TeV assume a
data sample of 8~ab$^{-1}$~\cite{Bambade:2019fyw}. It is shown in 
 \cite{Tian:2013qmi,KurataHHH} that the $HH$ production cross-section can readily be observed. In the same context of varying the trilinear Higgs coupling only, $\klambda$ can be determined to a precision of 10\%. 

\begin{figure}[tb]
  \centering
   \includegraphics[width=0.48\hsize]{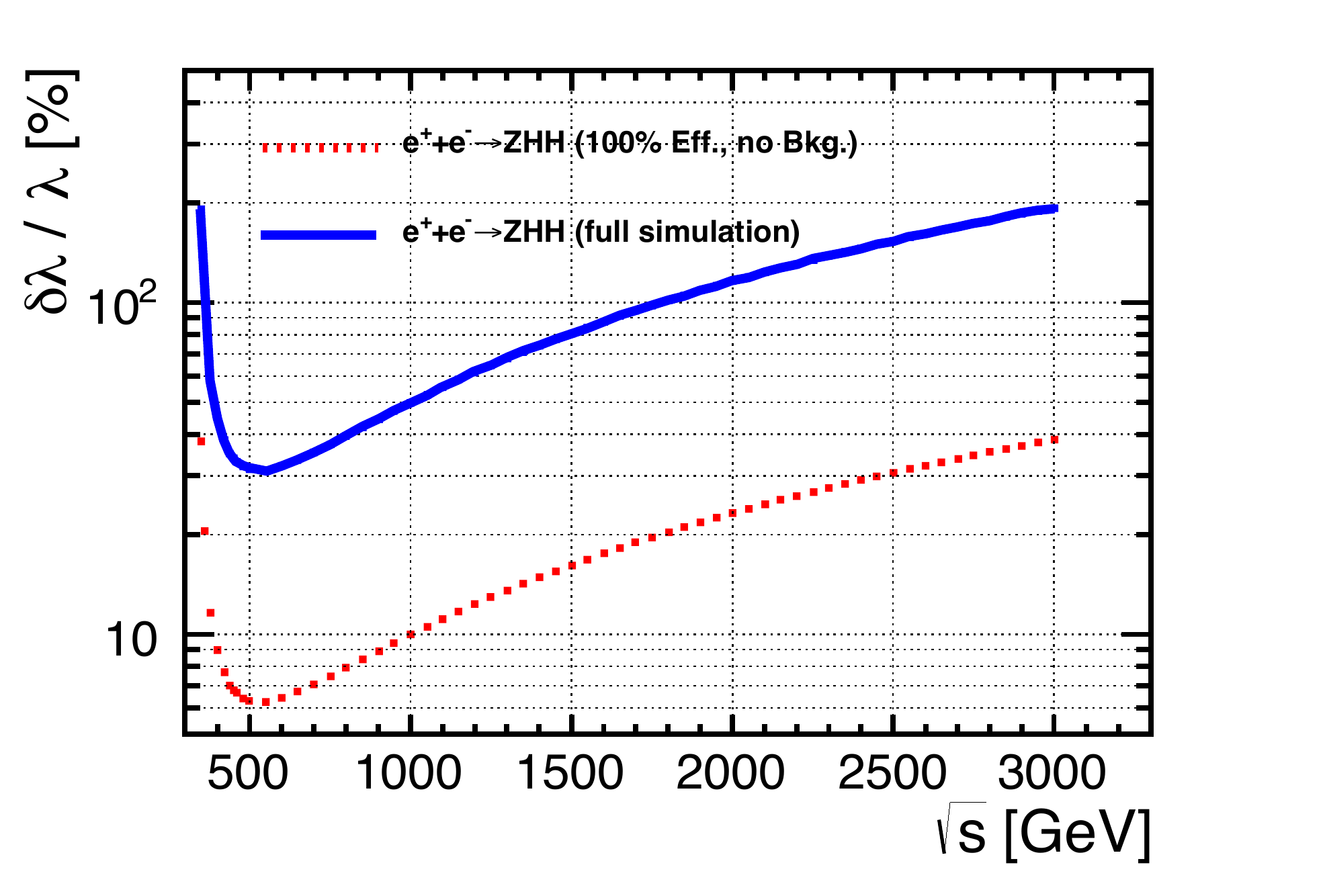}
   \  \ 
    \includegraphics[width=0.48\hsize]{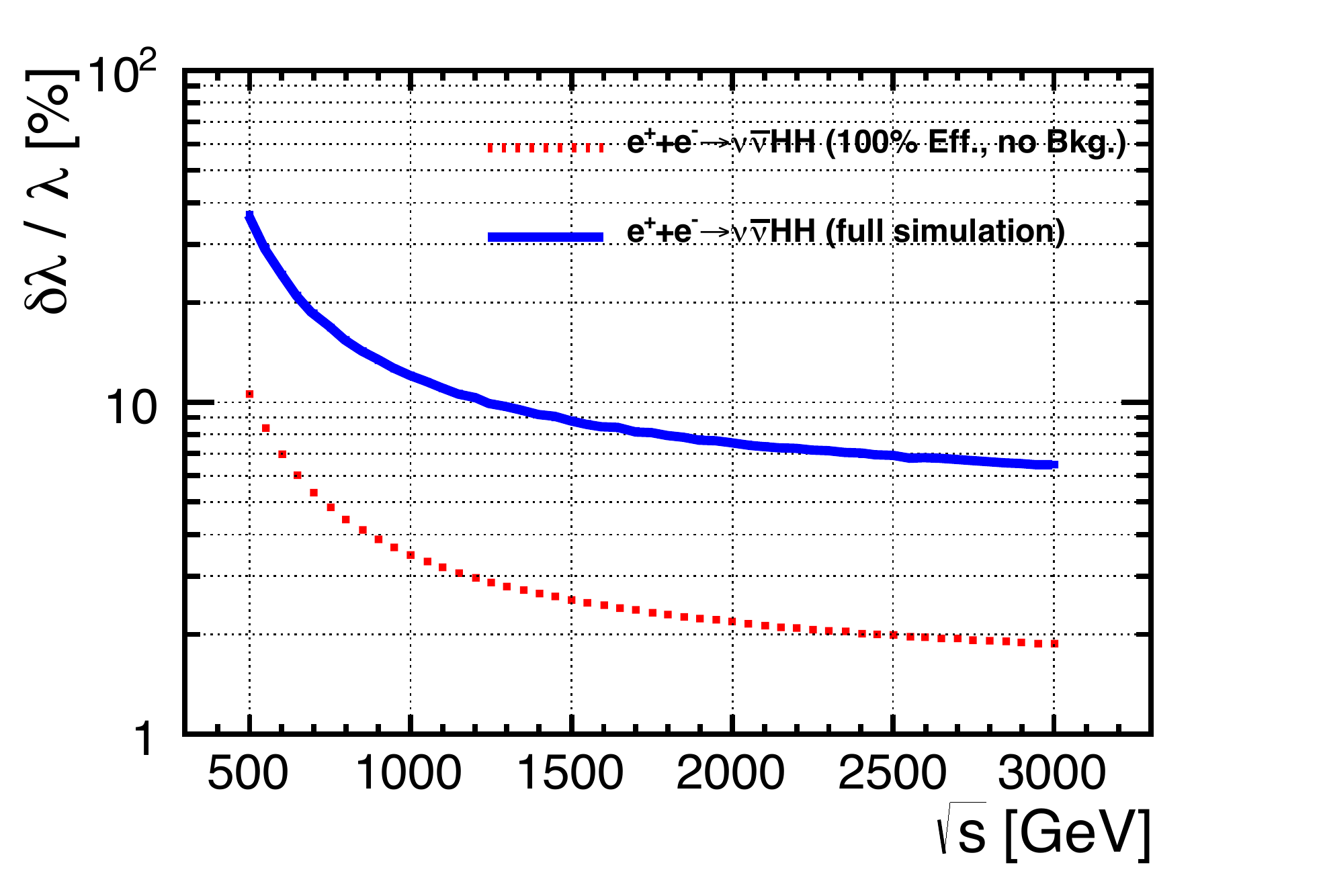}  
  \caption{The expected precision of $\lambda$ as a function of $\sqrt{s}$ for $e^+e^-\to ZHH$ ~(left) and for $e^+e^-\to\nu\bar{\nu}HH$ ~(right). The two lines in each plot correspond to the ideal situation using Monte Carlo truth (red/dotted) and the realistic situation (blue/solid) using current full-simulation analyses.
The same integrated luminosities of 4~ab$^{-1}$ is assumed at all values of $\sqrt{s}$.}
  \label{fig:HHHSensitivity}
\end{figure}

The impact of the centre-of-mass energy on the trilinear Higgs coupling
measurement is studied by extrapolating the full simulation results done 
at 500~GeV and 1~TeV to other energies.  The extrapolation is done in
such as way as to take into account the dependence on $\sqrt{s}$ for both the total cross-sections and the interference contributions~\cite{TianHHH:2015}.
The results are shown in Fig.~\ref{fig:HHHSensitivity} as the blue
lines for the two reactions. 
In addition to the results from realistic full simulations, the expectations for the ideal case,  assuming no background and 100\% signal efficiency, are shown as the red lines in the figure. The differences between the blue and the red lines are large, a factor of 4--5.  This suggests that there is much room for improvement in the clustering algorithm used to identify 2-jet systems with the Higgs boson mass, which would  lead to improvement in the final results.  Improvements could also come from better flavour-tagging algorithms and inclusion of additional  signal channels such as $Z\to\tau^+\tau^-$. The figure does imply that $\sqrt{s}=500$--600 GeV is optimal for $e^+e^-\to ZHH$. On the other hand,
 $\sqrt{s}$ energies of  1~TeV or above would be needed for optimal
 measurement of $e^+e^-\to\nu\bar{\nu}HH$.


\begin{figure}[tb]
  \centering
  \includegraphics[width=0.67\hsize]{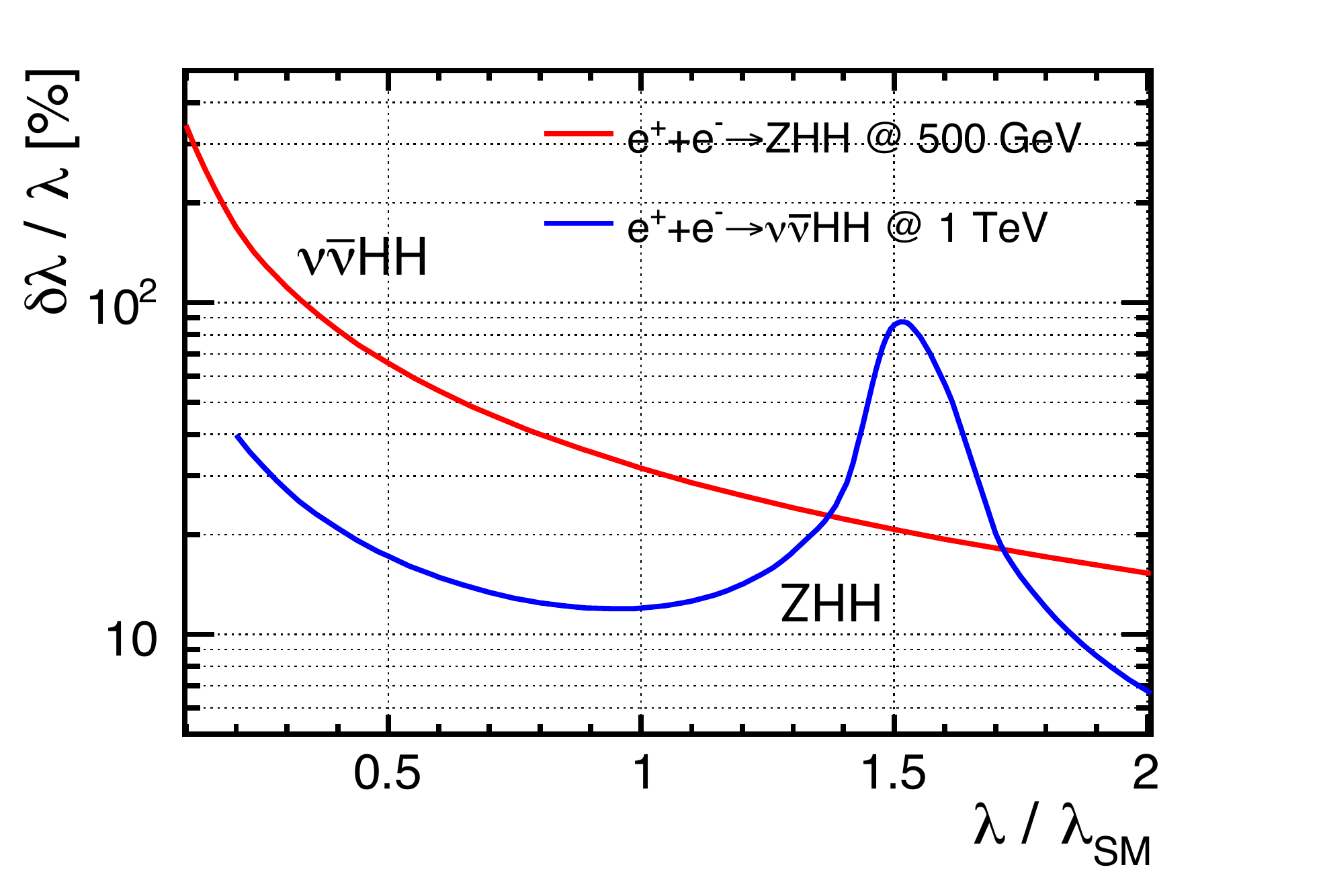}
  \caption{Expected precision of $\lambda$ when $\lambda$ is enhanced
    or suppressed from its SM value.}
  \label{fig:HHHBSMresult}
\end{figure}

Since large deviations of the trilinear Higgs coupling are expected in some new physics models, it is interesting to see how the expected precision would change in that case.
Using the dependence of the cross-section of the two reactions on the
Higgs self-coupling shown in Fig.~\ref{fig:HHHBSM}, we can  convert the
expected precision of the ILC measurements just described to precision on
the Higgs self-coupling at highly enhanced or suppressed values.
The results are shown in Fig.~\ref{fig:HHHBSMresult}.    The
two reactions, useful at 500 GeV and 1 TeV
respectively, are complementary in determining the trilinear Higgs coupling. 
If the trilinear Higgs coupling is indeed a factor of 2 larger than
its SM value, as expected in models of electroweak baryogenesis 
described in Sec.~\ref{sec:cosmology}, the $ZHH$ process at 500~GeV is especially useful and would already provide a measurement with 15\% precision  on the enhanced value of the self-coupling.

\section{Measurement of $HH$ production --- CLIC }
 \contrib{P. Roloff, U. Schnoor}
\label{sec:eedirectCLIC}

In the CLIC program, the Higgs self-coupling would be studied at
the 1.5~TeV and 3~TeV stages.   The planned integrated 
luminosities for these stages are 2.5~ab$^{-1}$ and 5~ab$^{-1}$,
respectively, with 80\% polarisation of the electron beam  and an
unpolarised positron beam.   The luminosity is planned to be divided between the L and R orientations at 80\%/20\%.   As in the ILC program, the CLIC program of single Higgs measurements will produce values
of the Higgs branching fractions to the major decay modes with
sub-percent accuracy~\cite{Abramowicz:2016zbo,Robson:2018zje}.

The study \cite{Roloff:2019crr} describes analyses 
of the $ZHH$ and $\nu\bar\nu HH$ reactions by the CLICdp group.  These
studies are based on full-simulation analyses with the \verb+CLIC_ILD+
detector model. In this  study, the second CLIC stage is taken
to be at 1.4~TeV.  All relevant background processes are included in the 
simulation data.

The $\nu\bar\nu HH$ process is studied in the
$b\bar b b\bar b$ and $b\bar b W W^*$ Higgs decay channels.
The main background contributions originate from diboson production and $ZH$ production. A Boosted Decision Tree (BDT) is used to extract the
$\nu\bar\nu HH$  signal in the dominant decay channel of \bbbb  production. The measurement benefits from the clean collision environment of $e^+e^-$ linear colliders as well as the excellent heavy flavour
tagging capabilities and the accurate jet energy resolution realised in the CLIC detector model.

In the studies at 1.4~TeV, evidence for $\nu\bar\nu HH$ production is
found with a significance of 3.6~$\sigma$, and the $ZHH$ process can be
observed at this stage with a significance of 5.9~$\sigma$.  In the
studies at 3~TeV with $e^-$ beam polarisation,
 the $\nu\bar\nu HH$ reaction is observed already
with 700~fb$^{-1}$.  With the total integrated luminosity of
5~ab$^{-1}$, the  $\nu\bar\nu HH$ cross-section can be measured with
a precision of 7.3\%, assuming that it takes the SM value.

\begin{figure}[tb]
  \centering
    \includegraphics[width=0.6\hsize]{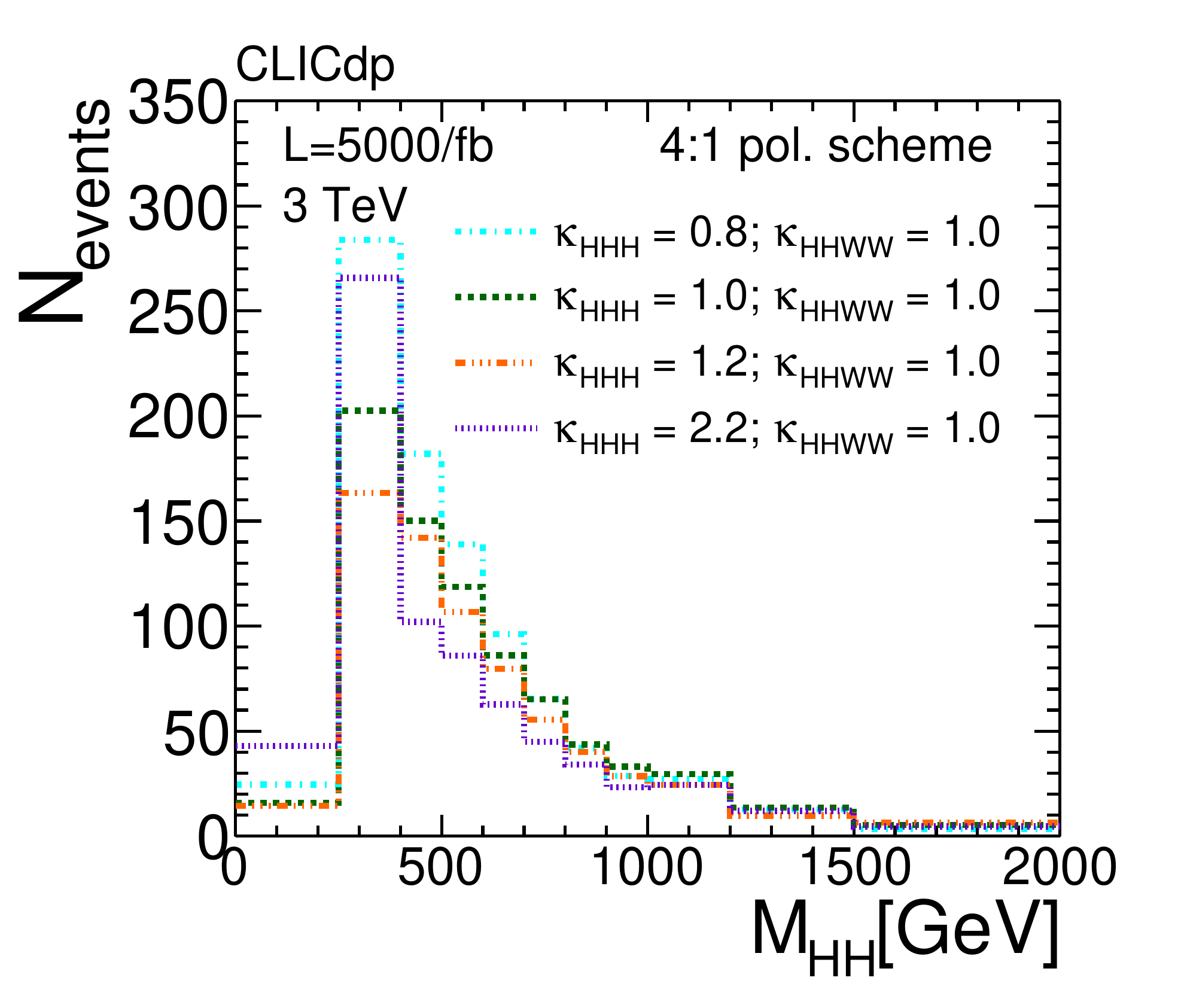} 
    \vspace*{-0.2cm}
    \caption{\hh invariant mass distribution measured in CLIC
      full-simulation studies at 3~TeV for different values of Higgs
      self-coupling modifier $\kappa_{HHH} = \lambda/\lambda_{SM}$.
      The $WWHH$ coupling is held fixed at the SM value.   The
      simulation assumes 5~ab$^{-1}$ of data and the 80\%/20\%
                     division of polarisation described in the
                             text~\cite{Roloff:2019crr}.}
  \label{mHHvary}
\end{figure}

The extraction of the Higgs self-coupling at CLIC is based on the
total cross-section measurements, combined with information from the
differential cross-section to distinguish effects of the self-coupling diagram in Fig.~\ref{fig:hhdiagrams}.   Because of the destructive interference in the  $\nu\bar\nu HH$ process, there is an ambiguity in the
interpretation of the total cross-section result, since the cross
section for  $\lambda/\lambda_{SM} = 2.2$ is the same as that
predicted in the SM; see Fig.~\ref{fig:HHHBSM}. This ambiguity can be
resolved using the $ZHH$ measurement, but it is also resolved by
measuring the \hh invariant mass  distribution in the $\nu\bar\nu HH$ reaction.
Figure~\ref{mHHvary} shows how the measured shape of this distribution changes as $\lambda/\lambda_{SM}$ is varied.   As $\lambda/\lambda_{SM}$
increases, this mass distribution decreases noticeably in the region
$ 500 < m(HH) < 1000$~GeV while the peak of the distribution at
$m(HH)\sim 400$~GeV rises dramatically. 

  The value of $\lambda/\lambda_{SM}$ can then be 
  extracted from a template fit to the 
  binned distribution of the invariant mass of the
 reconstructed Higgs boson pair in bins of the BDT response.
This value can be combined with the result of the
$ZHH$ cross-section measurement at 1.4~TeV  to
extract the value of the trilinear Higgs self-coupling and its uncertainties.
 The resulting constraint that will be provided by the CLIC
 measurements is found to be
 \beq
 0.93  \leq \lambda/\lambda_{SM} \leq   1.11 \
 \eeq{CLICvalue}
 at the  68 \% C.L.  
 Section 2.2.1 of \cite{deBlas:2018mhx} describes
 a fit in which this result is combined with
 a global fit to single Higgs observables using the SMEFT framework.
The final constraint on $\klambda$ is essentially unchanged from Eq.~\leqn{CLICvalue}.

\section{SMEFT interpretation of $HH$ measurements}
 \contrib{J. Tian}
\label{sec:eedirectEFT}

Up to this point, we have considered $HH$ production only using the
model in which 
$\klambda = \lambda/\lambda_{SM}$ is free to vary
while the other possible new physics effects have been ignored. This is
probably too stringent an assumption. A modification of the Higgs
sector that can give rise to a large change in the Higgs self-coupling
will probably also affect other SM couplings~\cite{Huang:2016cjm}.
These changes will independently lead to changes in the prediction for
the $HH$ production cross-section.   A robust search for a
deviation in the Higgs self-coupling should take this into account.

\begin{figure}
\begin{center}
\includegraphics[width=0.8\hsize]{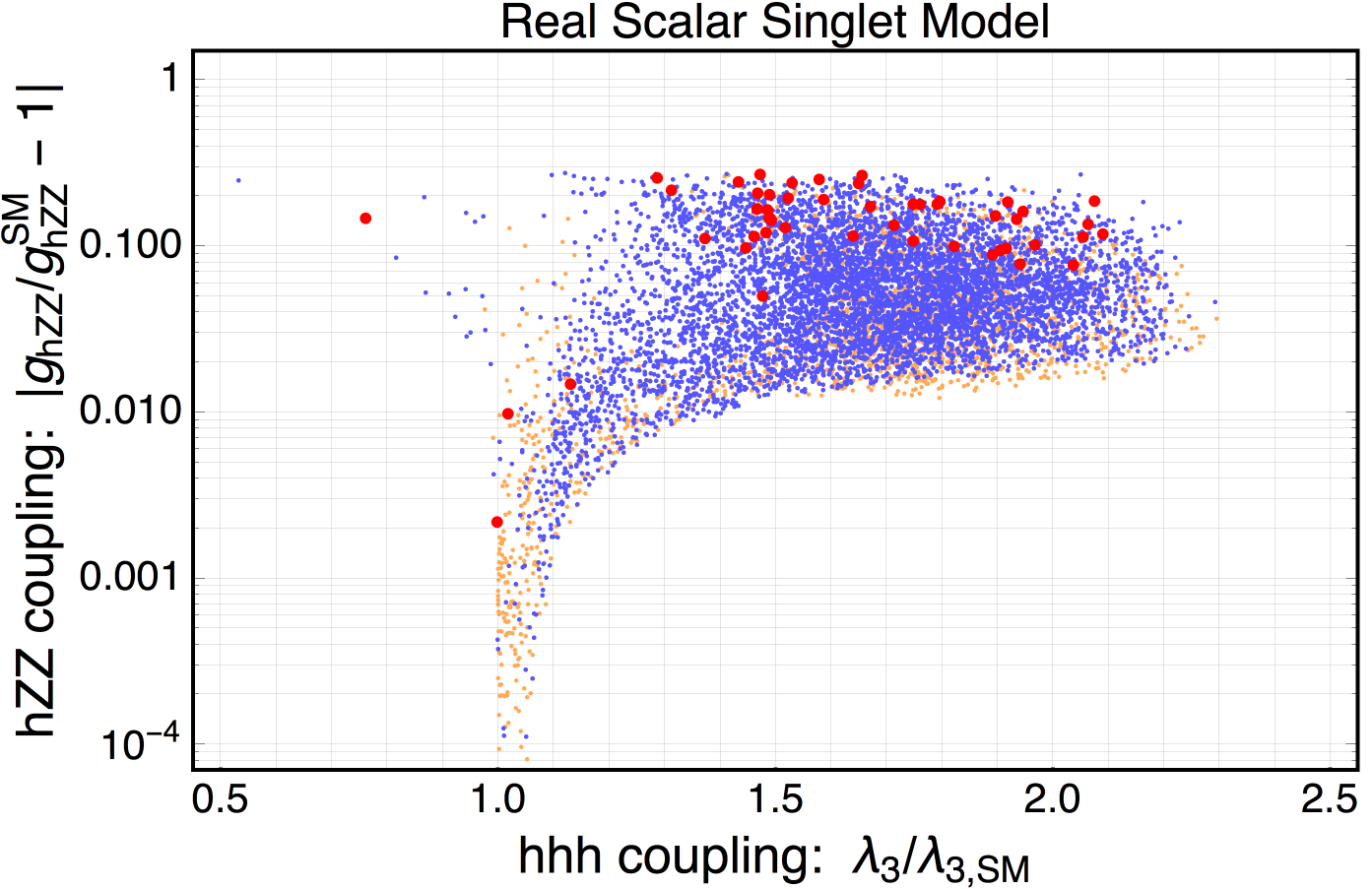}
\end{center}
\vspace*{-0.3cm}
\caption{Parameter scan of the models considered in
  \cite{Huang:2016cjm}, showing the predicted shifts of values of the $HHH$ and $HZZ$  couplings from the SM predictions.   The colour of each point indicates the strength of the  electroweak phase  transition in the
  corresponding model, with blue and red indicating strong
  first-order transitions.}
 \label{fig:eeHuang}
\end{figure}

An example of this effect of new physics is shown in
Fig.~\ref{fig:eeHuang}.  This figure refers to a class of models in
which the Higgs self-coupling is modified through mixing with a SM
singlet scalar field.  This is a subset of the models considered in
Sec.~\ref{sec:BSMspin0}, without separately observable scalar resonances.
These models can still  generate a large shift in
the Higgs self-coupling to produce a first-order electroweak phase transition as required for successful electroweak
baryogenesis.  This mechanism also generates smaller tree-level shifts in
the $HWW$ and $HZZ$ couplings.   As the figure shows, some models in
this class generate shifts of the single Higgs couplings visible at
the HL-LHC, while other models produce smaller effects, requiring the higher precision available
at  $e^+e^-$ colliders for their observation.  These modifications of the
Higgs boson couplings to vector bosons will modify the SM predictions
for the $HH$ production cross-sections in a manner that is independent
of the modification generated by a shift in $\kappa_\lambda$.  
To claim a measurement of $\klambda$, the  influence of the
altered $HVV$ couplings must be  separated out.
  
In Secs.~\ref{subsec:EFTtheory} and~\ref{sec:eft_fit_GPZ}, we have 
discussed such a more general analysis for the process
 $gg\to \hh$ at hadron
colliders.  We described using the SMEFT to take into account possible
new physics effects on the $HH$ production cross-section that are
independent of changes in the Higgs self-coupling. We have shown that the
SMEFT can be used as a tool to quantify the influence of these
orthogonal effects of new physics
and that, in principle at least, these effects can be controlled by
making a global fit to the SMEFT parameters. 

A similar analysis has been carried out  for the reaction $e^+e^-\to ZHH$ at 500~GeV~\cite{Barklow:2017awn}.  Because the $e^+e^-$ cross-section
depends on fewer operators than the cross-section at hadron colliders,
it is  possible to include the effects of all relevant dimension-6 operators that appear in the SMEFT.  When this is done,  it is seen that 
new physics effects different from the shift in the Higgs
self-coupling can potentially have a major influence on the $HH$  cross
section, easily swamping the variation due to $\kappa_\lambda$.
Fortunately, the high precision expected for  single Higgs and other measurements at an  $e^+e^-$ collider will allow these effects to be controlled.

\begin{figure}
\begin{center}
\includegraphics[width=0.70\hsize]{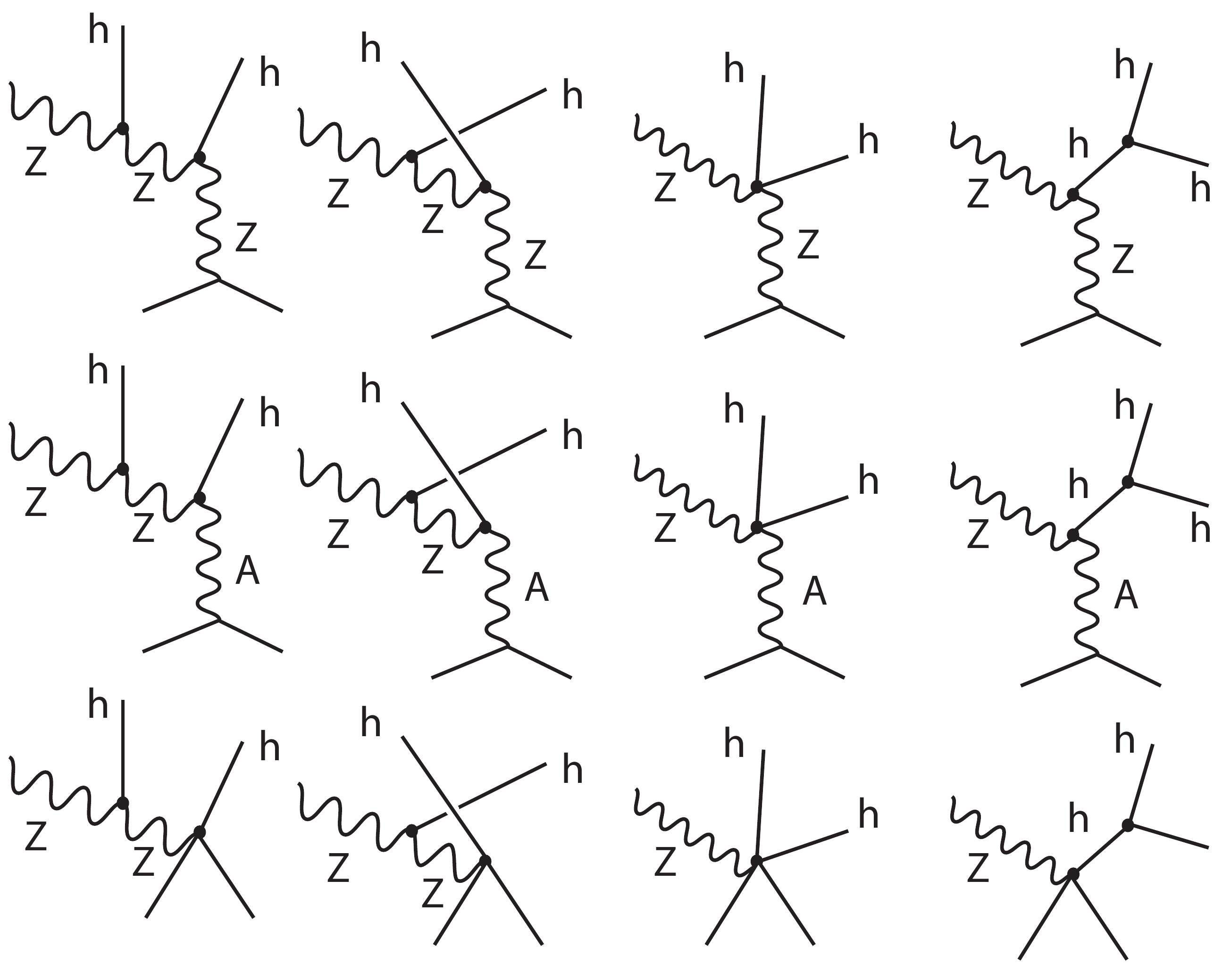}
\end{center}
\vspace*{-0.3cm}
\caption{Feynman diagrams contributing to $e^+e^-\to ZHH$ in the SMEFT
  with dimension-6 operators included.  The vertices shown are also
  typically modified from their SM values by dimension-6 perturbations~\cite{Barklow:2017awn}.}
\label{fig:Zhhdiagrams}
\end{figure}

The full set of diagrams contributing to $e^+e^-\to ZHH$ in the SMEFT at
tree level, including SM vertices and all contributing dimension-6 operators, is shown in Fig.~\ref{fig:Zhhdiagrams}.  One should note that, in general, the vertices in these diagrams are not equal to the SM vertices but rather include extra pieces due to the dimension-6 perturbations.

The complete variation of the tree level cross-section with the SMEFT coefficients is given in \cite{Barklow:2017awn}.  Some of the smaller terms in the complete expression are difficult to explain without reference to the renormalization scheme used there. Here we will write  a simplified
formula that gives the dependencies on the most important SMEFT
coefficients.  We assume here the case of unpolarised beams. 
In the SMEFT, $\klambda$ receives two different contributions
from  coefficients of dimension-6 operators.  In particular, we saw in
Eq.~\leqn{eq:klrel} that
 \beq
 \kappa_\lambda = 1 +  c_6  - \frac{3}{2} c_H \ , 
 \eeq{recallklrel}
where the parameter $c_6$ is the coefficient of an operator that
modifies the Higgs potential and $c_H$ is a universal rescaling of all Higgs couplings that originates from a modification of the Higgs field
 kinetic term.   When we speak of a new physics modification of the
 Higgs potential within the SMEFT, we are speaking specifically about
 the generation of a nonzero value for $c_6$.

In terms of these two parameters and other coefficients of electroweak
dimension-6 operators, the ratio of the unpolarised total cross-section  for $e^+e^-\to ZHH$ at 500~GeV to its SM value is given at the tree level  by the expression
\beq
\sigma/\sigma^{SM}(ZHH) = 1 + 0.56  c_6 - 4.15  c_H + 15.1 ( c_{WW}) 
+  62.1 (c_{HL} + c^\prime_{HL}) - 53.5 c_{HE} + \cdots \ ,
\eeq{UZhh}
where the omitted terms are less important.
The coefficients listed here come from the SMEFT Lagrangian terms
\beqa
\Delta {\cal L} &=& \frac{c_H}{2 v^2} \partial^\mu(\Phi^\dagger \Phi)
\partial_\mu(\Phi^\dagger \Phi) 
- \frac{ c_6\lambda}{v^2} (\Phi^\dagger \Phi)^3 
+ \frac{c_{WW}}{2v^2} \Phi^\dagger \Phi
W^a_{\mu\nu} W^{a\mu\nu} \CR 
& & + i \frac{c_{HL}}{v^2}  J_H^{\mu}  (\bar L
\gamma_\mu L)  + 4 i \frac{c'_{HL} }{ v^2}  J_H^{a \mu } (\bar L
\gamma_\mu t^a L) +  i \frac{c_{HE}}{ v^2} J_H^{\mu} (\bar e
\gamma_\mu e) \ ,
\eeqa{firstL}
where $W_{\mu\nu}^a$ is the SU(2) field strength, $L$ is the
left-handed lepton doublet $(\nu_e,e)_L$, and $J_H^\mu$, $J_H^{a\mu}$
are the Higgs currents $(\Phi^\dagger \Dlr{}^\mu \Phi)$ and
$(\Phi^\dagger t^a \Dlr{}^\mu \Phi)$, respectively.
In the full expression, the cross-section depends on a total of 17
SMEFT coefficients.  Some of the numerical factors in Eq.~\leqn{UZhh} 
are uncomfortably large.  And, unfortunately, because the phase
space for $e^+e^-\to ZHH$ is very restricted at 500~GeV and the cross
section is quite small, there is no useful additional information from
the differential distributions to separate the various dependencies.

However, it turns out that the accuracy of precision measurements at
$e^+e^-$ colliders is enough to solve the problem. The parameters
$c_{HL}$, $c'_{HL}$, $c_{HE}$ are tightly constrained by precision
electroweak measurements, even at the current LEP level of precision.
At a linear $e^+e^-$ collider, the parameters $c_H$ and $c_{WW}$ are
 constrained by the measurement of the total  cross
section for  $e^+e^-\to ZH$, the polarisation asymmetry in this total
cross-section, and the Higgs branching ratio to $WW^*$.  The analysis
\cite{Barklow:2017suo} describes a global fit to the data set that
will be acquired in the ILC program.  The  parameters in this fit include the full set of dimension-6 operators that contribute to the measured cross
sections at the tree level.   From the results of this analysis, it
is found that, apart from the $c_6$ term, 
 the expression in Eq.~\leqn{UZhh} can be evaluated with an
uncertainty of 2.8\% from the $c_H$ and $c_{WW}$ terms and
0.9\% from the $c_{HL}$, $c'_{HL}$, and $c_{HE}$ terms.  These
errors are slightly correlated, so the total uncertainty from
SMEFT coefficients other than $c_6$  (including those not
mentioned here)
is 2.4\%.   This gives a systematic error on the extraction of $c_6$ of 5\%, to be added in quadrature to the larger statistical error estimated in
Sec.~\ref{sec:eedirectILC}~\cite{Barklow:2017awn}.  In many of the
models shown in Fig.~\ref{fig:eeHuang}, the value of $c_{WW}$ is 
large enough that it makes a significant correction to the value of
the $HH$ production cross-section.  Nevertheless, this correction will
be known from the   single Higgs data and can be subtracted without
loss of accuracy in the determination of $c_6$.

With the contributions from these additional dimension-6 operator coefficients under control, the measurement of the total cross-section for the reaction $e^+e^-\to ZHH$ can be interpreted as a model-independent 
measurement of $c_6$ within the broad class of models describable by
the SMEFT.

It would be interesting to perform a similar analysis  for the $\nu\bar\nu HH $ process. To do this, we would need to have 
the analogue of  Eq.~\leqn{UZhh} for this reaction. We expect a result of a similar form.  One difficulty to be
aware of it that the factors in front of the coefficients  $c_{HL}$,
$c'_{HL}$, and $c_{HE}$
grow as  $s/m_Z^2$.   However, these parameters can be controlled to
an even greater degree than was taken into account in
\cite{Barklow:2017awn} by the improvements in our knowledge of precision
electroweak observables that are expected from
 these $e^+e^-$ colliders~\cite{Abada:2019,Fujii:2019zll}.

\section{SMEFT interpretation of single Higgs reaction 
measurements}
 \contrib{C. Grojean}
\label{sec:eeindirect}

Just as for the $HH$ determination of the Higgs self-coupling, it is
important to ask whether the determination of the self-coupling from
single Higgs measurements can be affected by other new physics
contributions.  Here again we can use the SMEFT to quantify these
effects and eventually to separate them from the effects of the
self-coupling. 

However, there is an important difference between the situation for
the single Higgs determination and that described in the previous
section.    In Eq.~\leqn{UZhh}, the contributions from all of the
SMEFT parameters appeared with numerical coefficients of order 1.
However, for the corresponding expressions in the 
single Higgs case, while  most of the SMEFT parameters
enter with order-1 coefficients, the contribution from $c_6$ which we
are most interested in has a coefficient of order 1\%.   This is
expected, since most of the relevant SMEFT operators enter these
formulae at the tree-level, while $c_6$ enters only at the 1-loop
level.  For example, for the ratio of the unpolarised cross-sections 
for $e^+e^-\to ZH$ at 250~GeV, the formula corresponding to Eq.~\leqn{UZhh}
is
\beq
\sigma/\sigma^{SM}(ZH) = 1 + 0.015  c_6 -  c_H + 4.7 ( c_{WW})  
+  13.9 (c_{HL} + c^\prime_{HL}) - 12.1 c_{HE} + \cdots \ .
\eeq{UZoneh}
Thus, very strong constraints are needed on all of the additional
variables in Eq.~\leqn{UZoneh} to extract any information about
$c_6$.   The extraction of $c_6$ is also more subtle than
in the case of $HH$ production, since the same
data that supplies these constraints is also used to determine the
value of $c_6$.

\begin{figure}
\begin{center}
\includegraphics[width=0.70\hsize]{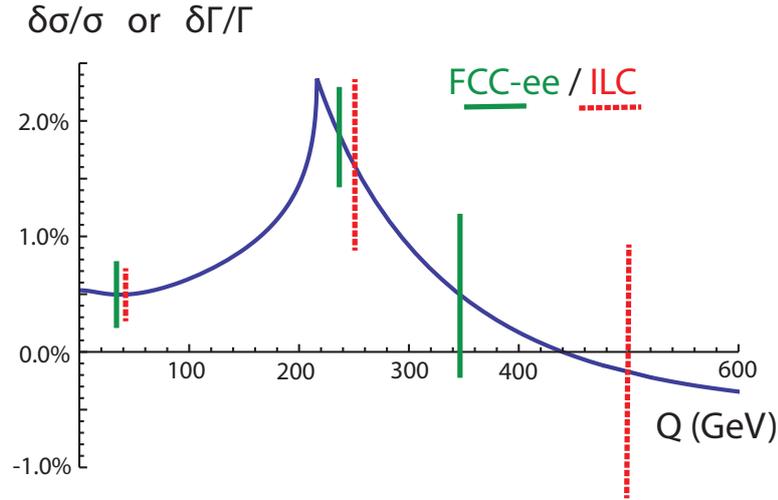}
\end{center}
\vspace*{-0.3cm}
\caption{Relative enhancement of the  $e^+e^-\to ZH$ cross-section
  and the $h\to W^+W^-$ partial width, in \%,  for $\klambda =
  1$, due to the 1-loop diagrams shown
  in Fig.~\ref{fig:ZZH}.   One $Z$ or $W$ leg is off-shell at the
  invariant $Q^2$ while the other $Z$ or $W$ and the Higgs boson are
  kept on-shell.  The vertical lines show the uncertainties expected
  from proposed $e^+e^-$ colliders from single measurements of the
  relevant
quantities,  green/solid for FCC-ee, red/dashed for ILC.   For $Q> 200$~GeV, these
are 1~$\sigma$
error bars  for  measurements of
  $\sigma(e^+e^-\to ZH)$.  For $Q\sim 40$~GeV,  these are 
1~$\sigma$ errors on  $\Gamma(h\to WW^*)$ from the SMEFT fits to the
full collider programs for FCC-ee and ILC reported
in \cite{deBlas:2019rxi}.}
\label{fig:McC}
\end{figure}

In \cite{DiVita:2017vrr}, Di Vita and collaborators explained that the expected accuracy of single Higgs measurements at $e^+e^-$ colliders will be
such that it is feasible to extract a value of $c_6$ despite this
difficulty.   They noted, in particular, that the enhancement of the
$HZZ$ and $HWW$ couplings highlighted by McCullough and discussed in 
Sec.~\ref{sec:oneHv1} has a special feature that aids this process.   
This radiative correction arises from the Feynman diagrams shown in
Fig.~\ref{fig:ZZH}.
It is useful to consider these diagrams as being evaluated with the
Higgs boson and one $Z$ or $W$ boson on mass shell while the other
vector boson is off-shell at a variable momentum invariant $Q = \sqrt{Q^2}$.
The value of the sum of diagrams has a characteristic dependence on
$Q$ that cannot be reproduced as a sum of effects of point like
dimension-6 SMEFT operators.   This is shown in Fig.~\ref{fig:McC}.
The diagrams give an enhancement that is not monotonic as a function of
$Q$ but rather has a sharp cusp at the $ZH$ threshold  ($Q = m_H
+ m_Z$).   Measurements at $e^+e^-$ Higgs factories will measure
this function at several different values of $Q$:  at values of $Q$
equal to the CM energies at the various collider
stages in the cross-section $\sigma(e^+e^-\to ZH)$, at $Q \sim 40$~GeV
in 
the partial width $\Gamma(H\to WW^*)$,  at
 $Q \sim 30$~GeV in the partial width $\Gamma(H\to ZZ^*)$, and at $Q^2 \lesssim
0$
in the vector boson fusion cross-section  $\sigma(e^+e^- \to \nu\bar\nu H)$.
Fig.~\ref{fig:McC} shows the expected accuracy of the three most
important of these measurements in the FCC-ee and ILC programs and
indicates how the set of three measurements can provide independent
values for the SMEFT parameters $c_6$, $c_H$, and $c_{WW}$.

Some caution should be used in interpreting this plot directly.  The errors
shown for $\Gamma(H\to WW^*)$ are those from the SMEFT fits done in 
\cite{deBlas:2019rxi} using the expected results from the full
FCC-ee and ILC programs.  Thus, they use the values of the indicated
cross-section plus other data.   A full SMEFT analysis would include
many other measurements than the three indicated here, including
other measurements that put powerful constraints on $c_{WW}$.  On the
other hand, such an analysis would be based on  17 SMEFT parameters,
not just the few indicated in Eq.~\leqn{UZoneh}. 

The analysis that we have described does not include possible loop
corrections to the other Higgs couplings, for example, the influence
of the 
loop corrections to the $Hbb$ vertex or the $Htt$ vertex on
$\Gamma(H\to gg)$. 
In these cases, however, the vertex is measured only at one value of
$Q$, so the effect of $c_6$ is indistinguishable from a simple
$Q$-independent shift of the coupling strength, which is controlled by
a separate SMEFT parameter.  Because of this,
only the corrections to
the $HZZ$ and $HWW$ couplings give sensitivity to $c_6$. 

\begin{figure}[t]
\begin{center}
\includegraphics[width=0.9\hsize]{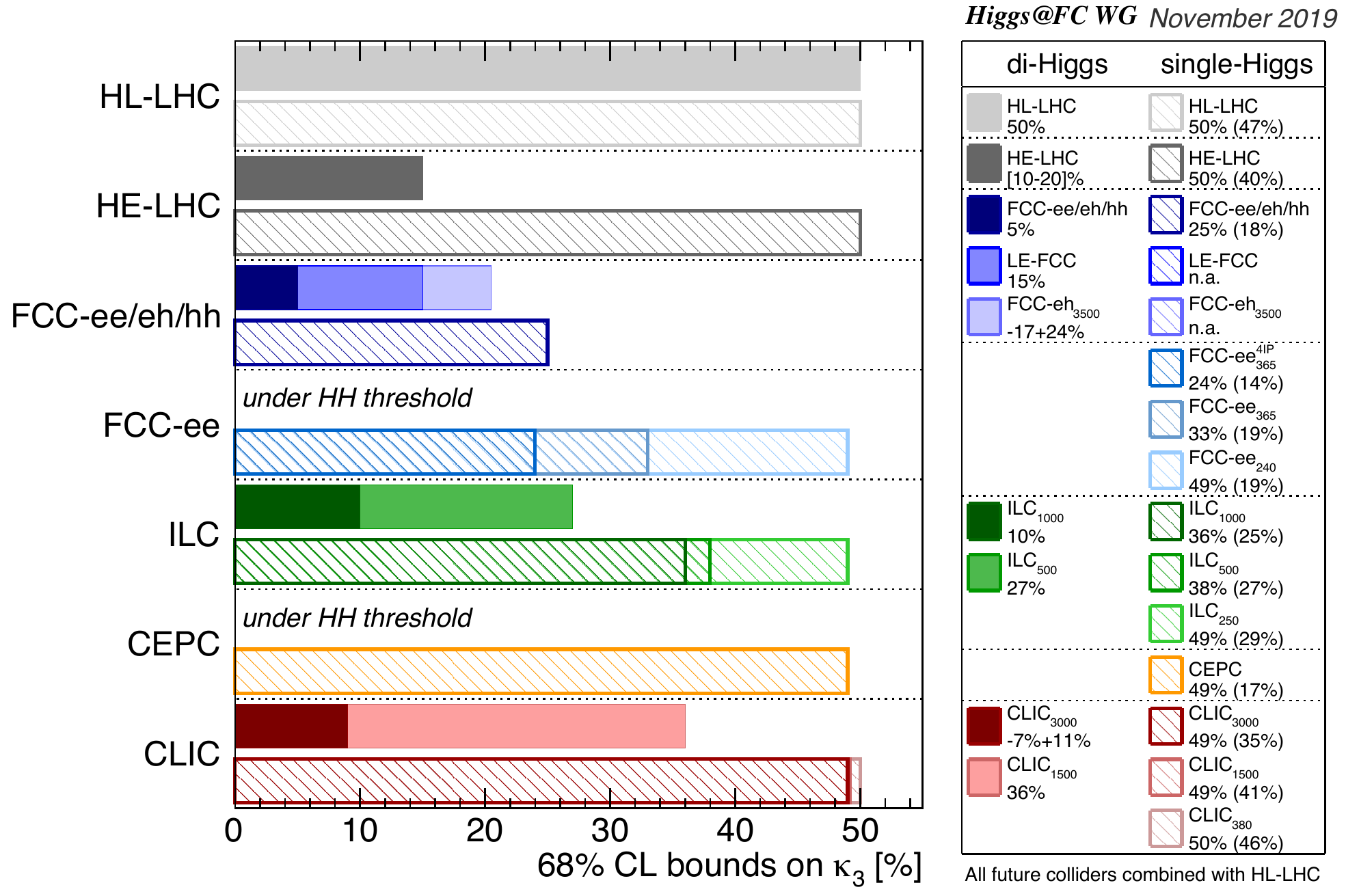}
\end{center}
\vspace*{-0.3cm}
\caption{Uncertainties on the Higgs self-coupling  projected for the
   High-Luminosity LHC and for other future colliders, at various
   stages,
   by the ECFA Higgs@Future Colliders working group~\cite{deBlas:2019rxi}.
   The results  are presented as uncertainties on  $\kappa_3 =
   \kappa_\lambda$. In the bar graphs, results from the direct method
   are shown with solid bars and result from the indirect method with
   hatched bars. The estimates for the indirect determination of
   the self-coupling are based on a multi-parameter SMEFT analysis
   which also takes into account projected results from the LHC.  Estimates in parentheses correspond to a 1-parameter fit without other new physics effects.  The results for all $e^+e^-$ colliders include the projected  single-$H$ and $HH$ results from the HL-LHC, approximated by a 50\% uncertainty in
   $\klambda$. }
\label{fig:HFC}
\end{figure}

A complete fit of the SMEFT parameters to the expected single Higgs
data from the proposed Higgs factories has recently been carried out
by the ECFA Higgs@Future Colliders working
group~\cite{deBlas:2019rxi}.   The results of this analysis for the
projected uncertainty in $c_6$ are shown in Fig.~\ref{fig:HFC}.  The
results are those shown in the second column of
Table~\ref{tab:HFCfit},
except that the numbers in the figure are combined with an expected 50\%
uncertainty on $c_6$ from the measurement  of the $HH$
production cross-section at the HL-LHC.

The results in Fig.~\ref{fig:HFC} and Table~\ref{tab:HFCfit} show that
it is very important for the closure of the 17-parameter fit to have
data on $e^+e^-\to ZH$ at two different CM energies.  The cross
section for this reaction falls off rapidly at energies above 250~GeV,
so the plan of the FCC-ee to take data at 250~GeV and 365 GeV is more
optimal from this perspective.  For ILC, there is some compensation in
that the function shown in Fig.~\ref{fig:McC} has a larger variation
with $Q$ from 250~GeV to 500~GeV.   For CLIC, though running above
1~TeV allows the excellent measurements of $HH$ production described
above, the $ZH$ process is well measured only at the 380~GeV stage and
so $c_6$ is poorly constrained by the single Higgs analysis. 

Finally, though, the FCC-ee and ILC programs would be expected to
yield a measurement of $c_6$ from single Higgs processes with an
uncertainty of  40--60\%, independently of any results from $HH$
production.  This is
comparable to the expected precision from the HL-LHC.
This indirect determination
would then provide a welcome independent measurement
of the self-coupling. This measurement is expected to be statistics-limited and so would benefit from an increase in the running time or, in the case of circular machines, doubling the number of detectors.  We emphasise again that this measurement is  essentially free of model dependent assumptions within the broad class of models that can
be described by the SMEFT.

\section{The quartic Higgs self-coupling}\label{sec_ee:quartic}
 \contrib{F. Maltoni, D. Pagani}

Up to this point in our discussion of $e^+e^-$ probes of the Higgs
potential, we have only considered dimension-6 operators in the
SMEFT.  For operators that specifically modify the Higgs potential, 
we have considered only one higher-dimension operator, the operator
with coefficient $c_6$ in Eq.~\leqn{lsmeft} whose main role is to
shift the coupling $\lambda_3$.   More general modifications of the
Higgs potential are available from operators of higher dimension.  It
is relevant to ask whether inclusion of this possibility affects the
determination of $c_6$ or $\lambda_3$.

This question was studied for the first time
in Ref.~\cite{Maltoni:2018ttu}.   This paper considered the
two-parameter Higgs potential modification
\beq
   \Delta\mathcal{L} =  - \frac{\bar c_6}{v^2}\left(\Phi^\dagger\Phi - \frac{v^2}{ 2} \right)^3
 - \frac{\bar c_8}{v^4}\left(\Phi^\dagger \Phi- \frac{v^2}{ 2}\right)^4 \ ,
\eeq{sixeightL}
as in eq. \leqn{VNP},  including the dimension-6 and also a dimension-8 perturbation.  From
this effective Lagrangian one can define
\beqa
  c_6 &\equiv& (2 v^2/m_h^2) \bar c_6 \CR
  c_8 &\equiv& 4 (2 v^2/m_h^2) \bar c_8 \ .
  \eeqa{sixeightdefs}
Then, similarly to eq.~\leqn{eq:klrel},
\beqa
    \kappa_\lambda &\equiv& \frac{\lambda_3}{ \lambda_3^{SM}  } = 1 +
    c_6 \CR
   \kappa_4 &\equiv& \frac{\lambda_4 }{ \lambda_4^{SM}  } = 1 +
   6  c_6  +  c_8  \ .
\eeqa{kappasixeight}
In this context, it is possible to analyse processes with single Higgs
production, $HH$ production, and $HHH$ production to deduce
constraints on $c_6$ and $c_8$.  It is important to note that the
restriction of the fit to two possible operators is a simplification
with respect to the analyses described in the previous sections.   First,
$c_H$ and other possible dimension-6 operator coefficients
are  not
included in this analysis.  We have already seen in
Sec.~\ref{sec:eedirectEFT}, though, that 
the relevant operator coefficients would already be strongly constrained by 
expected measurements of single Higgs observables at $e^+e^-$
colliders.
But, further,  the effects of many more new operators appearing at dimension 8
are not included.   This  analysis assumes that  their coefficients are
suppressed
with respect to $c_8$.  Nevertheless, this analysis represents a first
step toward answering this question.

The study \cite{Maltoni:2018ttu} derives constraints on these
two parameters by considering the deviations in the total cross
section for Higgs production processes at $e^+e^-$ colliders of
increasing energy, treating both $s$-channel $Z$ and $W$ fusion
reactions, and both  tree-level and one-loop effects of
the higher-dimension operators.    The specific processes considered,
and the loop order at which the two couplings first appear, are:
\beq
\begin{tabular}{ccc}
Process &   $ \lambda_3$  & $\lambda_4$  \\ \hline
$ZH\ , \ \nu_e\bar\nu_e H$  &   one-loop  &  two-loop \\
$ZHH\ ,\ \nu_e\bar\nu_e HH$  &   tree  &  one-loop \\
$ZHHH\ ,\  \nu_e\bar\nu_e HHH$  &   tree  &   tree  \\
\end{tabular}
\eeq{eeHHprocesses}
In Ref.~\cite{Maltoni:2018ttu}, both tree and one-loop effects are
included for each process.   Two-loop corrections  are not included.
It is important to note that, while the one-loop effects on single Higgs
cross-sections can be computed without reference to an underlying EFT
framework, the one-loop corrections to the $HH$ and $HHH$ cross-sections are
UV-sensitive and require renormalization.  Using the  SMEFT formalism and taking advantage of the 
parameterisation in Eq.~\leqn{sixeightL}, the authors of 
Ref.~\cite{Maltoni:2018ttu} worked out these UV-finite expressions.

By changing the values of $c_6$ and $c_8$, one can
independently change the values of $\lambda_3$ and $\lambda_4$.   As
already noted, UV-finite radiative corrections due to $c_6$
and $c_8$  are taken into account.  
The analogous study for 
future hadron colliders, was first presented in
Ref.~\cite{Borowka:2018pxx}.   This is technically more challenging
because of the need to consider two-loop calculations in the $gg$
fusion channel.   That analysis is described in Sec.~\ref{sec:other_probes}.

\begin{figure}[t]
\begin{center}
\includegraphics[width=0.44\hsize]{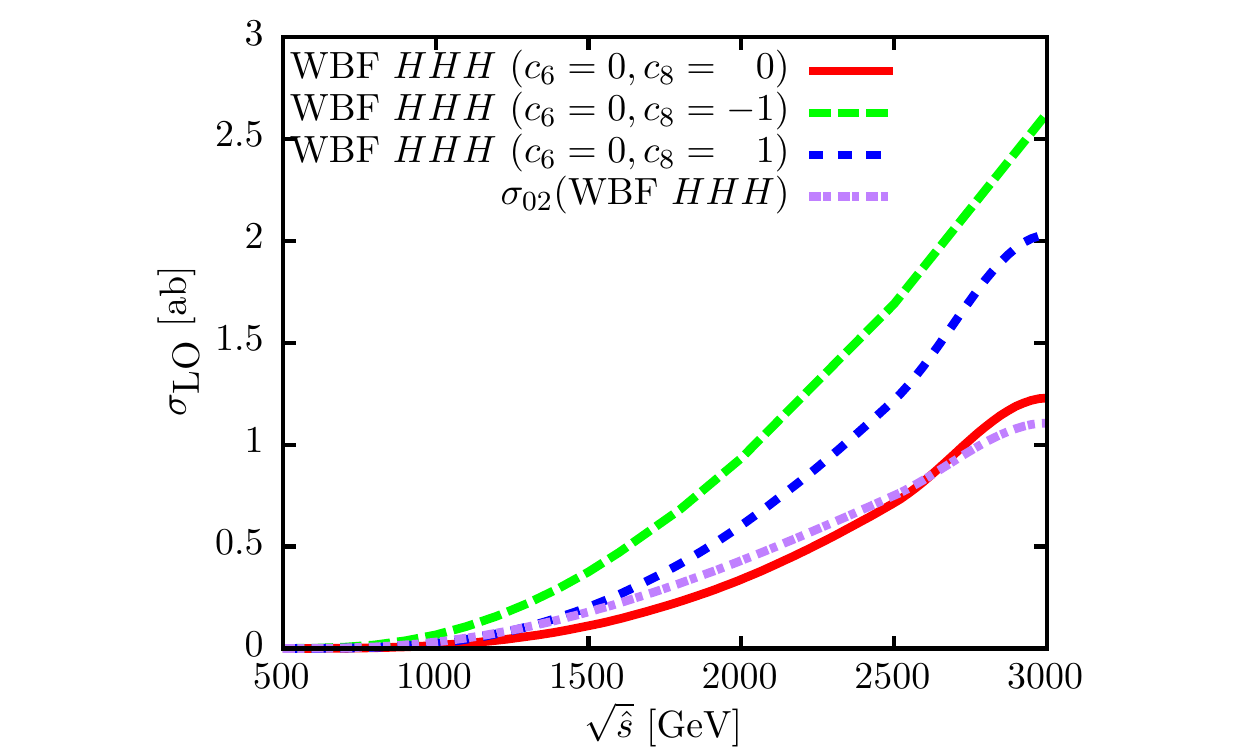}\ \ 
\includegraphics[width=0.44\hsize]{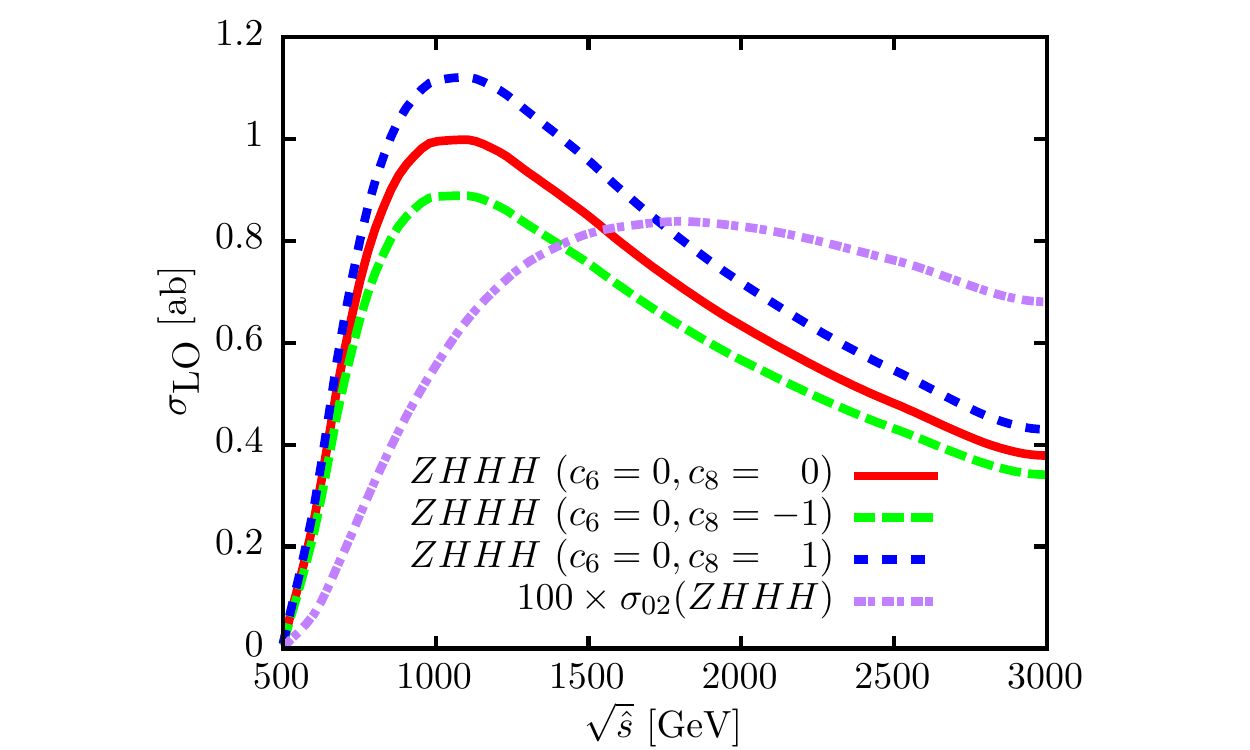}\ \ 
\end{center}
\vspace*{-0.4cm}
\caption{Leading order total cross-sections for $ZHHH$ (left) and $W$ fusion
  $HHH$ (right)  in $e^+e^-$ collisions, for representative
  values of $c_6$ and $c_8$.  The red solid curves
  are the SM values.  The results refer to
  the ideal beam polarisation choice  $P_{e^-} = -1.0$, $P_{e^+} = +1.0$.}
\label{fig:HHHcs}
\end{figure}

With this formalism in hand, Ref.~\cite{Maltoni:2018ttu} presented
constraints on the $ c_6$ and $c_8$ couplings that
would be obtained from the run programs of the proposed $e^+e^-$
colliders
CEPC, FCC-ee, ILC, and CLIC.   The strongest constraints apply to
ILC and CLIC at energies of 500~GeV and above, where \hh and
eventually $HHH$ production is observable.   In this case, we can
consider
the possibility that $c_8$ can be large and quote joint
constraints
on the two variables.   In particular, we can compare the allowed
range of $c_8$ to the theoretical limit 
$c_8\lesssim
31$ quoted in Eq.~\leqn{EFT:c8bound}.  A limitation of this strategy
is that the $HHH$ cross-sections at $e^+e^-$ colliders are quite
small. The SM cross-sections are shown in Fig.~\ref{fig:HHHcs}.

\begin{figure}[ht]
\begin{center}
\includegraphics[width=0.47\hsize]{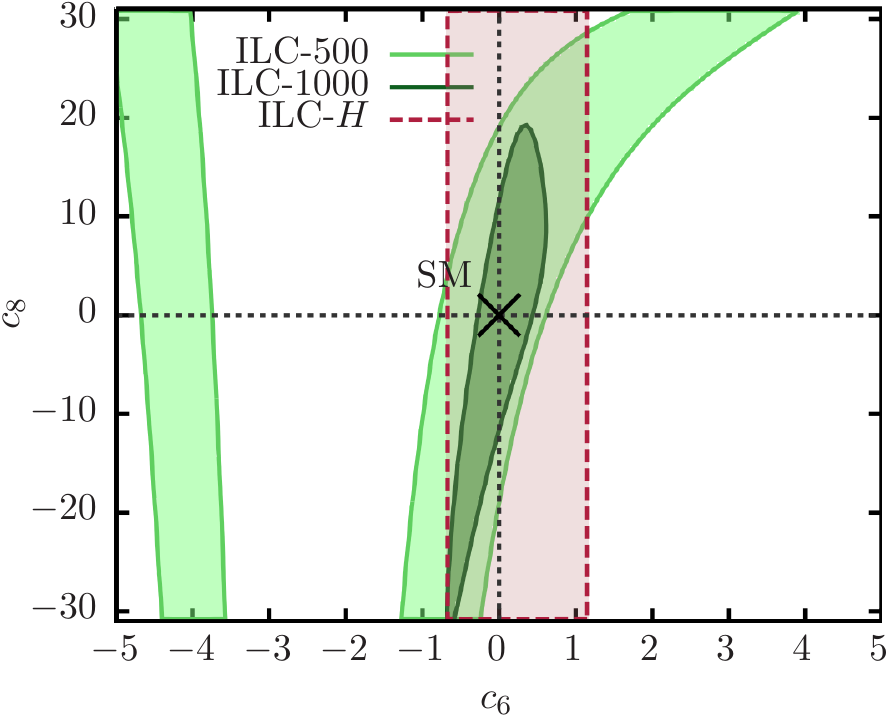}\ 
\includegraphics[width=0.47\hsize]{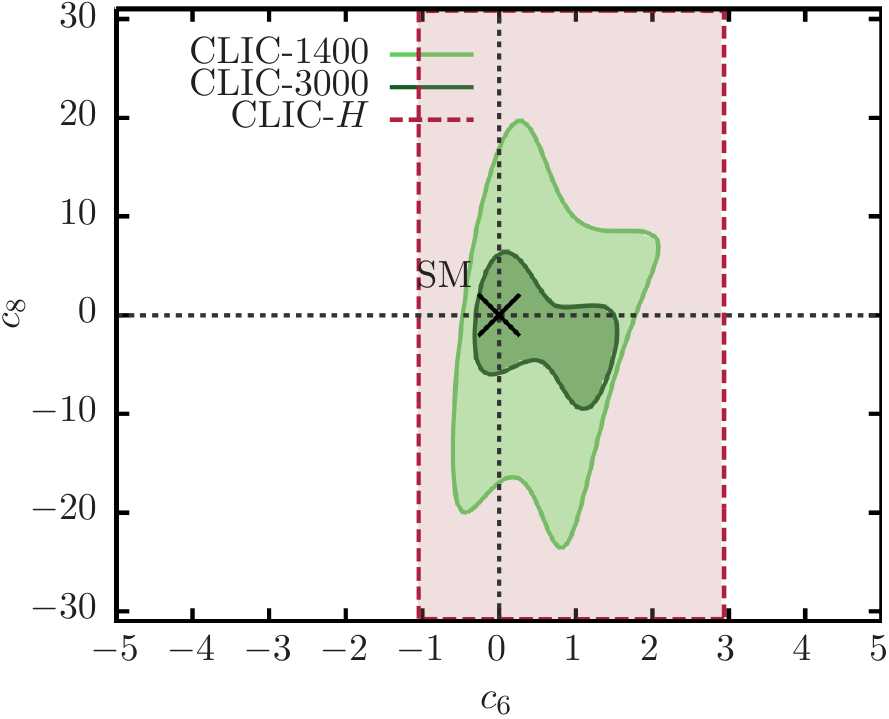}\\ 
\includegraphics[width=0.47\hsize]{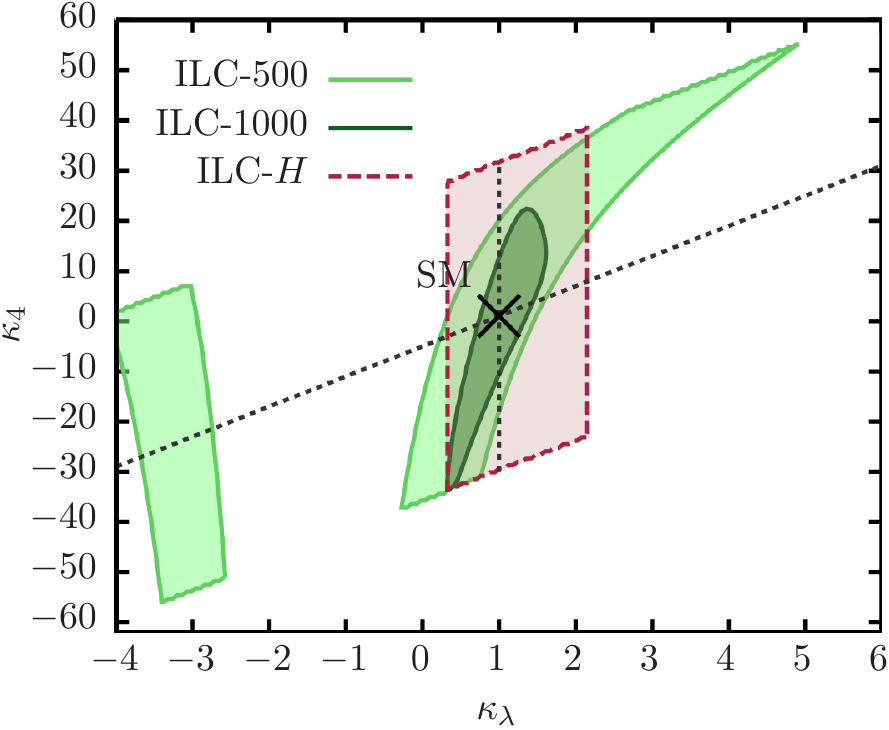}\ 
\includegraphics[width=0.47\hsize]{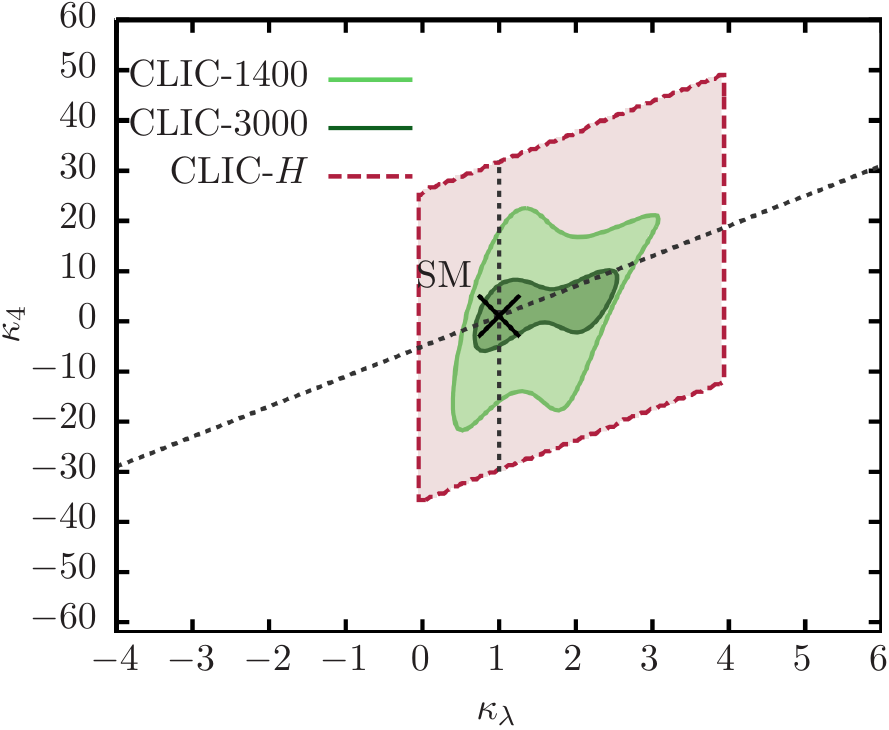}
\end{center}
\vspace*{-0.5cm}
\caption{Combined 90\% CL constraints on the cubic and quadratic Higgs
  self-couplings from the $e^+e^-$ colliders ILC (left) and CLIC
  (right). 
 The upper plots show the 
constraints in the $(c_6, c_8)$ plane; the bottom
plots show the constraints in the $(\kappa_\lambda,\kappa_4)$ plane.
The red regions marked ILC-H and CLIC-H refer to a combination of all
single
 Higgs measurements at all energy stages for each collider under
 study.
 In all cases only the perturbative region described in
 Sec.~\ref{sec:theoretical_constraints_EFT}, $ |c_6| < 5$, $
 |c_8| < 31$,  has been considered. }
\label{fig:Pagani}
\end{figure}

In Fig.~\ref{fig:Pagani},  we show as light and dark green regions the
90\% CL regions in this parameter space that would be obtained from
successive stages of running at ILC and CLIC from the measurements of 
$HH$ and $HHH$ production cross-sections, assuming that the true
answer is the SM cross-section.   The top plots show these 
constraints in the  $(c_6, c_8)$ plane; the bottom
plots show the constraints in the $(\klambda,\kappa_4)$ plane.
For ILC, the analysis considers results from $ZHH$ production at
500~GeV  and its combination with results from $ZHHH$ and $W$ fusion
$HH(H)$ at 1~TeV.  (The relevant cross-sections for $HHH$ production
are too small to be measured at 500~GeV.) For CLIC, the analysis
considers $ZHH(H)$ and $W$ fusion $HH(H)$ production at 1.4~TeV and 
3~TeV stages.   The luminosities and polarisations assumed are given
in Table~3 of \cite{Maltoni:2018ttu}; these differ in some details
from the current proposals in Refs.~\cite{Bambade:2019fyw} and \cite{Robson:2018enq}.
We also show as red bands the limits on $c_6$ from single
Higgs measurements at ILC and CLIC, using the indirect method
described in the Sec.~\ref{sec:eeindirect}, and assuming that the
two-loop dependence
on $c_8$  is negligible.   The 
CLIC constraints are stronger than ILC ones, thanks to the higher
energies and the greater accessibility of $HHH$ processes. On the
other hand, the $c_6$ constraints from single Higgs
production are stronger 
at the ILC.   Both this constraint and the measurement of $W$ fusion
at higher energies can be used to remove the two-fold ambiguity seen
in the plots for ILC500. Scenarios in which  the measured cross
sections differ from those of the SM have been investigated, assuming
that  they correspond to the more general case $|c_6| < 5$
and and $c_8 = 0$. For positive and large values of $c_6$, the 
constraints on both $c_6$ and $c_8$ 
become stronger than in the SM case, analogously to what we have already 
pointed out in the discussion of Fig.~\ref{fig:HHHBSMresult} for the 
case of $c_6$ only.

The conclusion of this study is that the first coarse bounds on the
value of 
$c_8$,   and in turn on $\lambda_4$, can be set at future
$e^+e^-$
 colliders. The SM rate for triple Higgs
production is not measurable at these $e^+e^-$ colliders, even at the
highest energies considered.  But the cross-section strongly depends
on $\lambda_4$, and so it is possible to obtain significant constraints.  
The combination of results from double and triple Higgs production
at high energies improves the constraints.  The $W$ fusion channel
will give  the strongest bounds. For this reason, by increasing the
energy, the precision of the constraints on $\lambda_3$ and
$\lambda_4$ will improve, regardless of the true value of
$\lambda_3$.   
The constraints that can obtained at CLIC at 3~TeV via $W$ boson
fusion $HHH$ production are similar to those that would be obtained
 at a future 100~TeV hadron collider.

 \section{Conclusions}
 \label{sec:eeconclude}

The conclusion of this section have now become clear:
\begin{itemize}
    \item An $e^+e^-$ collider with significant integrated luminosity at
      two different CM energies (e.g., 250 and 350~GeV or 250 and
      500~GeV)  will be able to  determined the Higgs self-coupling
      from measurements of single Higgs reactions. This determination
      would be robust and 
      model-independent, in the sense that it would be  insensitive to
      the presence of other new physics effects as parameterised by the
      SMEFT.    For currently proposed $e^+e^-$ colliders, the precision on the
      Higgs self-coupling would be 44\% for FCC-ee and 58\% for ILC,  comparable to the
      precision expected from HL-LHC.  With four interaction points or
      double the running time, the FCC-ee precision would improve to
      27\%. 
  \item The ILC at 500~GeV would also be able to determine the Higgs
self-coupling from  the measurement of the $ZHH$ production cross
section. 
    This measurement would also  be  robust and 
      model-independent, in the same sense as above.   The expected
      precision on the self-coupling is 27\%. This could be
      combined with the single Higgs determination to reach a
      precision of  24\%.
    \item The ILC at 1~TeV or CLIC in its proposed program at 1.5 and
      3~TeV would be expected to determine the Higgs self-coupling to
     a precision of  10\% by 
      the measurement of $HH$  production  using the $ZHH$ and
      $\nu\bar\nu HH$ channels.   Though it is likely that this
      measurement would be model-independent in the sense above, that
      issue needs further
      study.
    \item The FCC-ee(4IP), ILC and CLIC programs would all be able to provide very strong evidence ($> 4 \sigma$) for or against an increase of the Higgs self-coupling by a factor 2, as actually expected in models of electroweak baryogenesis.
 \end{itemize}

    There are reasons to guess that our knowledge of the Higgs
    self-coupling might be even better than that described here.
    First, all estimates from $e^+e^-$ colliders are based on current
    full-simulation analyses.   We have shown  in
    Fig.~\ref{fig:HHHSensitivity} 
     large gaps
    between the results of these analyses and those of ideal analyses with
    perfect signal/background discrimination.   There is considerable
    room to be more clever when we are directly working with data.
    
    Second, these results from $e^+e^-$ colliders will be combined with
    results from $pp$ colliders operating in the same time frame.
    In particular, all
     four proposed Higgs factories are expected to constrain the
     parameter $c_{\Phi G}$ in Eq.~\leqn{lsmeft}, which contributes
     at the tree level in a SMEFT analysis of $gg\to HH$, by measuring
     the $Hgg$ coupling to 1\% precision~\cite{Bambade:2019fyw}.
    Complementarity  between $e^+e^-$ and $pp$ measurements will eventually
      lead us to the most precise understanding of the Higgs self-coupling.
 

%% file: HHFuture/HH_pp.tex
\newcommand*{\flow}[1]{\ensuremath{Flow}_{n,5}}
\newcommand*{\tauone}{\tau_{\ensuremath{1}}}
\newcommand*{\tautwo}{\tau_{\ensuremath{2}}}
\newcommand*{\tauthree}{\tau_{\ensuremath{3}}}
\newcommand*{\tautwoone}{\tau_{\ensuremath{21}}}
\newcommand*{\tauthreeone}{\tau_{\ensuremath{31}}}
\newcommand*{\tauthreetwo}{\tau_{\ensuremath{32}}}
\newcommand*{\Zp}{\ensuremath{Z^{\prime}}}
\newcommand*{\ZpSSM}{\ensuremath{Z^{\prime}_{\mathrm{SSM}}}}
\newcommand*{\hht}{\ensuremath{H_{\ensuremath{T}}}}
\newcommand*{\ptSub}[1]{\ensuremath{p_{\text{T} #1}}}
\newcommand*{\ptSup}[1]{\ensuremath{p_{\text{T}}^{\ensuremath{#1}}}}
\newcommand*{\ptSupPar}[2]{\ensuremath{p_{\text{T}}^\ensuremath{\text{#1}}(\ensuremath{{#2}})}}
\newcommand*{\etaSup}[1]{\ensuremath{\eta^{#1}}}
\newcommand*{\etaSub}[1]{\ensuremath{\eta_{#1}}}
\newcommand*{\etaSubPar}[2]{\ensuremath{\eta^\ensuremath{\text{#1}}(\ensuremath{{#2}})}}
\newcommand*{\etaPar}[1]{\ensuremath{\eta(\ensuremath{{#1}})}}
\newcommand*{\ptEl}{\ensuremath{p_{\text{T}}^{e}}}
\newcommand*{\ptMu}{\ensuremath{p_{\text{T}}^{\mu{}}}}
\newcommand*{\ptZp}{\ensuremath{p_{\text{T}}^{\ensuremath{Z^{\prime}}}}}
\newcommand*{\mZp}{\ensuremath{M_{\ensuremath{Z^{\prime}}}}}
\newcommand*{\mll}{\ensuremath{m_{\ensuremath{\ell \ell}}}}
\renewcommand*{\kl}{\ensuremath{\kappa_{\lambda}}}
\newcommand*{\effg}{\ensuremath{\epsilon_{\gamma}}}
\newcommand*{\misg}{\ensuremath{\epsilon_{j \rightarrow \gamma}}}
\newcommand*{\effb}{\ensuremath{\epsilon_{b}}}
\newcommand*{\misl}{\ensuremath{\epsilon_{l \rightarrow b}}}
\newcommand*{\misc}{\ensuremath{\epsilon_{c \rightarrow b}}}
\newcommand*{\mislc}{\ensuremath{\epsilon_{l(c) \rightarrow b}}}
\newcommand*{\maa}{\ensuremath{m_{\gamma\gamma}}}
\newcommand*{\mfourl}{\ensuremath{m_{4\ell}}}
\newcommand*{\mmumu}{\ensuremath{m_{\mu\mu}}}
\newcommand*{\mlla}{\ensuremath{m_{\ell\ell\gamma}}}
\newcommand*{\ninoone}{\ensuremath{\tilde{\chi}_{1}^{0}}}
\newcommand*{\ninotwo}{\ensuremath{\tilde{\chi}_{2}^{0}}}
\newcommand*{\chinoonepm}{\ensuremath{\tilde{\chi}_{1}^{\pm}}}
\newcommand*{\chinoonemp}{\ensuremath{\tilde{\chi}_{1}^{\mp}}}
\newcommand*{\nhits}{\ensuremath{N_{\mathrm{layer}}^{\mathrm{hit}}}}
\renewcommand*{\mjj}{\ensuremath{M_{\ensuremath{jj}}}}
\newcommand*{\stopq}{\tilde{t}_{1}}
\newcommand*{\chilsp}{\tilde{\chi}_0^{1}}
\newcommand*{\topq}{t}
\newcommand*{\nb}{N_{b}}
\newcommand*{\nt}{N_{t}}
\newcommand*{\metvec}{\vec{\pt}^{\textrm{miss}}}
\newcommand*{\metmag}{\pt^{\textrm{miss}}}

\newcommand*{\vh}{\ensuremath{\mathrm{VH}}}
\newcommand*{\vbf}{\ensuremath{\mathrm{VBF}}}

\newcommand*{\hzzfourl}{\ensuremath{H \rightarrow ZZ^* \rightarrow 4\ell}\xspace}
\newcommand*{\hfourl}{\ensuremath{H \rightarrow 4\ell}}
\newcommand*{\hfourmu}{\ensuremath{H \rightarrow 4\mu}}
\newcommand*{\heemumu}{\ensuremath{H \rightarrow 2e2\mu}}
\newcommand*{\hmumua}{\ensuremath{H \rightarrow \mu\mu\gamma}}
\newcommand*{\hx}{\ensuremath{H \rightarrow X}}
\newcommand*{\hy}{\ensuremath{H \rightarrow Y}}

\newcommand*{\hmumu}{\ensuremath{H \rightarrow \mu^{+}\mu^{-}}}
\newcommand*{\hrhoa}{\ensuremath{H \rightarrow \rho\gamma}}
\newcommand*{\hja}{\ensuremath{H \rightarrow J/\psi\gamma}}
\newcommand*{\haa}{\ensuremath{H \rightarrow \gamma \gamma}}
\newcommand*{\hza}{\ensuremath{H \rightarrow Z \gamma}}
\newcommand*{\hlla}{\ensuremath{H \rightarrow \ell^{+}\ell^{-} \gamma}}
\newcommand*{\zmumua}{\ensuremath{Z \rightarrow \mu^{+}\mu^{-} \gamma}}
\newcommand*{\hzalla}{\ensuremath{H \rightarrow Z \gamma \rightarrow \ell^{+}\ell^{-} \gamma}}

\newcommand*{\qqaa}{\ensuremath{q q \rightarrow \gamma \gamma}}
\newcommand*{\fourl}{\ensuremath{4\ell}}
\newcommand*{\mumu}{\ensuremath{\mu^{+}\mu^{-}}}
\newcommand*{\lla}{\ensuremath{\ell^{+}\ell^{-} \gamma}}
\renewcommand*{\mh}{\ensuremath{m_{H}}}
\renewcommand*{\br}[1]{\ensuremath{\text{BR({#1})}}}
\newcommand*{\riso}[1]{\ensuremath{\text{rel.Iso(\ensuremath{{#1}})}}}
\newcommand*{\rbr}[2]{\ensuremath{\text{BR({#1})/BR({#2})}}}
\newcommand*{\sci}[2]{\ensuremath{{#1}\times 10^{#2}}}
\newcommand*{\sqrtsfcc}{\ensuremath{\sqrt{s}=\text{100 TeV}}}
\newcommand*{\sqrtslhc}{\ensuremath{\sqrt{s}=\text{14 TeV}}}
\newcommand*{\sqrtshelhc}{\ensuremath{\sqrt{s}=\text{27 TeV}}}
\newcommand*{\invab}{\ensuremath{\mathrm{ab}^{-1}}}
\newcommand*{\invpb}{\ensuremath{\mathrm{pb}^{-1}}}
\newcommand*{\piz}{\ensuremath{\mathrm{\pi}^{0}}}
\newcommand*{\sob}{\ensuremath{\mathrm{S/B}}}
\newcommand*{\sigprod}{\ensuremath{\sigma_{\mathrm{prod}}}}
\newcommand*{\lumi}{\ensuremath{\mathcal{L}}}
\newcommand*{\intlumifcc}{\ensuremath{\mathcal{L}=30\text{ ab}^{-1}}}
\newcommand*{\intlumihelhc}{\ensuremath{\mathcal{L}=15\text{ ab}^{-1}}}
\newcommand*{\alphas}{\ensuremath{\alpha_S}}
\newcommand*{\deff}{\ensuremath{\delta_{\epsilon}}}
\newcommand*{\muss}{\ensuremath{\sigma_{\text{obs}}/ \sigma_{\text{SM}}}}
\newcommand*{\dmuOvermu}{\ensuremath{\delta\mu / \mu}}
\newcommand{\comphep}   {\sc{c}\rm{omp}\sc{hep}}
\newcommand{\zjets}{\ensuremath{\PZ/\PGg^*\text{+jets}}\xspace}
\newcommand{\wjets}{\ensuremath{\PW\text{+jets}}\xspace}
\newcommand{\VPtmiss}{\ensuremath{\vec {P}_{\mathrm{T}}^{\text{miss}}}\xspace}
\newcommand{\MtW}{\ensuremath{m_{\mathrm T}(\PW)}\xspace}
\newcommand{\tGu}{\ensuremath{\rm t\rightarrow \rm ug}\xspace}
\newcommand{\tGc}{\ensuremath{\rm t\rightarrow \rm cg}\xspace}
\newcommand{\kLu}{\ensuremath{\lvert\kappa_{\rm tug}\rvert/\Lambda}\xspace}
\newcommand{\kLc}{\ensuremath{\lvert\kappa_{\rm tcg}\rvert/\Lambda}\xspace}
\renewcommand{\GeV}{\ensuremath{\,\text{Ge\hspace{-.08em}V}}\xspace}
\renewcommand{\TeV}{\ensuremath{\,\text{Te\hspace{-.08em}V}}\xspace}
\newcommand{\MET}{\ensuremath{E_{\mathrm{T}}^{\text{miss}}}\xspace}
\newcommand{\ptvecmiss}{\ensuremath{{\vec p}_{\mathrm{T}}^{\kern1pt\text{miss}}}\xspace}

\newcommand{\cred}{\color{red}}
\newcommand{\cgreen}{\color{X575}}
\renewcommand{\nn}{\nonumber}
\newcommand{\lt}{\lambda_3}
\newcommand{\lf}{\lambda_4}
\newcommand{\kf}{\kappa_4}
\newcommand{\klt}{\kappa_\lambda_3}
\newcommand{\klf}{\kappa_\lambda_4}
\newcommand{\cbs}{\bar{c}_6}
\newcommand{\cbe}{\bar{c}_8}
\newcommand{\cbstr}{\bar{c}_6^{\rm true}}
\newcommand{\cbetr}{\bar{c}_8^{\rm true}}
\newcommand{\MSbar}{{\rm \overline{MS}}}

\newcommand{\amc}{{\sc MadGraph5\textunderscore}a{\sc MC@NLO}}

\newcommand{\py}{{\sc Pythia8}}
\newcommand{\delphes}{{\sc Delphes}}

\label{chap:helhc-fcchh}

\begin{figure}
  \centering
  \includegraphics[width=0.48\columnwidth]{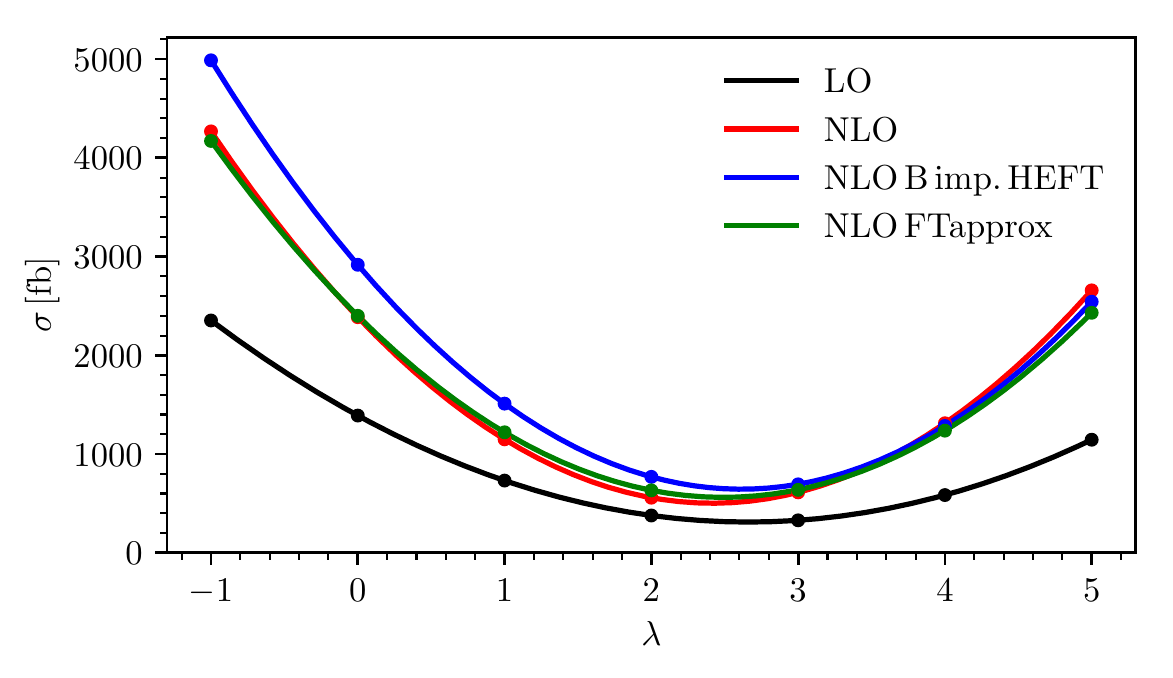}
  \includegraphics[width=0.46\columnwidth]{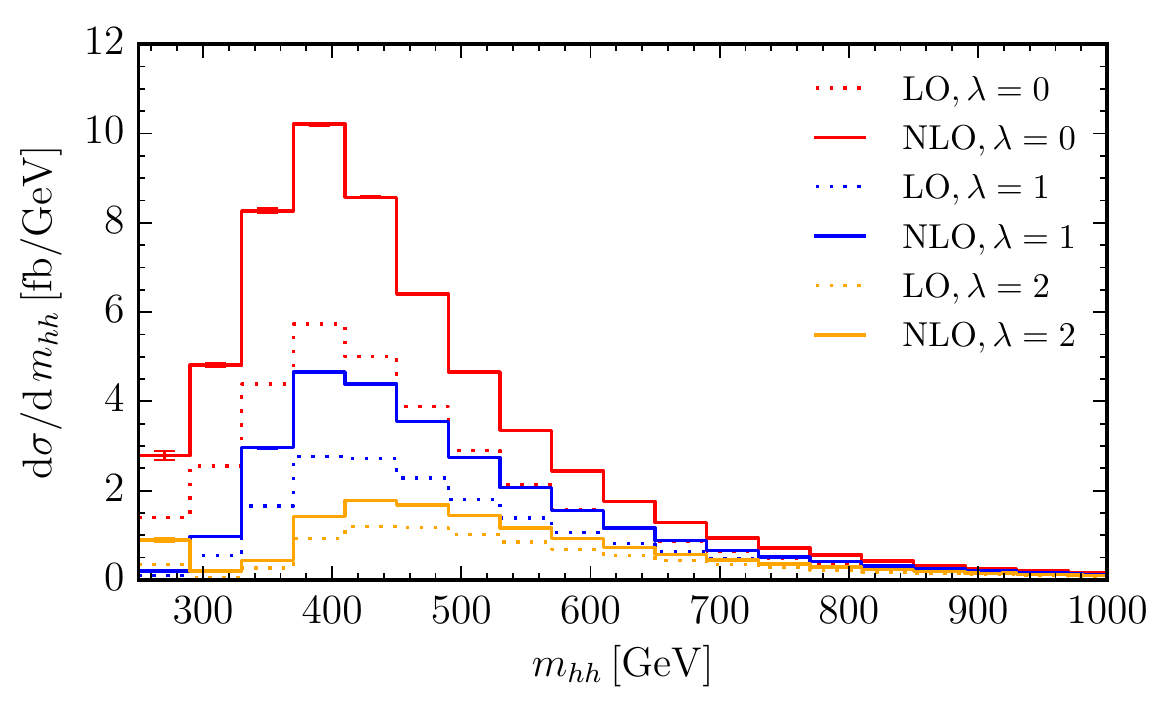}
  \vspace*{-0.2cm}
  \caption{Left: Total \hh cross section at \sqrtsfcc\ as a function of the self-coupling modifier $\kl$. Right: Differential (N)LO \mhh distribution at \sqrtsfcc\ for different values of $\kl$. From Ref.~\cite{Borowka:2016ypz}}
  \label{hh_vs_kl}
\end{figure}

At future hadron colliders, the Higgs self-coupling can be probed predominantly via Higgs pair production. Additional indirect constraints can also be obtained from single Higgs production.  However, as will be discussed later they are not competitive with the direct method when 5-10\% (\emph{silver} level) precision is within reach. 

The cross sections for several production channels are given in \refta{table:xsec2-future}, where the quoted systematic uncertainties reflect today's state of the art.
The most studied channel, in view of its large rate, is gluon-gluon fusion. In the SM, a large destructive interference between the leading diagram with a top-quark loop and that with the self-coupling occurs. While this interference suppresses the SM rate, it makes it more sensitive to possible deviations from the SM couplings, the sensitivity being enhanced after NLO corrections are included, as shown in the case of gg$\to$\hh in Ref.~\cite{Borowka:2016ypz}, where the first NLO calculation of $\sigma$(gg$\to$\hh) inclusive of top-mass effects was performed. As shown in Fig.~\ref{hh_vs_kl}(left), for values of $\klambda$ close to 1, $1/\sigma_{\hh} d\sigma_{\hh}/d\klambda\sim -1$ , and a measurement of $\klambda$ at the few percent level therefore requires the measurement and theoretical interpretation of the Higgs pair rate at a similar level of precision. Table~\ref{table:xsec2-future}, already discussed in Chapter~\ref{chap:HHcxs}, shows that the current theoretical systematic uncertainty on the signal is at the 5\% level (for a complete discussion see Chapter~\ref{chap:HHcxs}), which is already competitive with the statistical and experimental systematic uncertainties that are achievable at a \sqrtsfcc\ collider. It is furthermore reasonable to expect a further reduction of such uncertainties to the percent level. In Fig.~\ref{hh_vs_kl}(right) the differential \mhh distribution is shown for different values of the Higgs self-coupling. At high \mhh the s-channel ``triangle'' contribution is suppressed and the box diagram dominates. Conversely at low \mhh the triangle diagram, that contains information on the Higgs self-coupling is enhanced. Information on the Higgs self-coupling can thus be extracted from the differential \mhh distribution.
\begin{table}[t!]
\renewcommand{\arraystretch}{2.0}
\begin{center}
\begin{tabular}{l|c|c|c}
\hline
$\sqrt{s}$ & 14 TeV & 27 TeV & 100 TeV \\ \hline
ggF \hh &  $36.69^{+2.1\%}_{-4.9\%}\pm 3.0\%$ & $139.9^{+1.3\%}_{-3.9\%}\pm 2.5\%$ & $1224^{+0.9\%}_{-3.2\%}\pm 2.4\%$ \\ 
VBF \hh &  $2.05^{+0.03\%}_{-0.04\%}\pm 2.1\%$ & $8.40^{+0.11\%}_{-0.04\%} \pm
2.1\%$ & $82.8^{+0.13\%}_{-0.04\%}\pm 2.1\%$ \\ 
$Z\hh$ & $0.415^{+3.5\%}_{-2.7\%} \pm 1.8\%$ & $1.23^{+4.1\%}_{-3.3\%} \pm 1.5\%$ & $8.23^{+5.9\%}_{-4.6\%} \pm 1.7\%$ \\ 
$W^+ \hh$ & $0.369^{+0.33\%}_{-0.39\%}\pm 2.1\%$ & $0.941^{+0.52\%}_{-0.53\%}\pm 1.8\%$ & $4.70^{+0.90\%}_{-0.96\%}\pm
1.8\%$ \\ 
$W^- \hh$ & $0.198^{+1.2\%}_{-1.3\%}\pm 2.7\%$ & $0.568^{+1.9\%}_{-2.0\%}\pm 2.1\%$ & $3.30^{+3.5\%}_{-4.3\%}\pm 1.9\%$ \\ 
$\ttbar H$ & $0.949^{+1.7\%}_{-4.5\%}\pm 3.1\%$ & $5.24^{+2.9\%}_{-6.4\%}\pm 2.5\%$ & $82.1^{+7.9\%}_{-7.4\%}\pm 1.6\%$ \\ 
$tj\hh$ & $0.0367^{+4.2\%}_{-1.8\%}\pm 4.6\%$ & $0.254^{+3.8\%}_{-2.8\%}\pm 3.6\%$ & $4.44^{+2.2\%}_{-2.8\%}\pm 2.4\%$ \\ \hline
\end{tabular} 
\caption{Signal cross sections (in fb) for various \hh production mechanisms (from Chapter~\ref{chap:HHcxs}).
\label{table:xsec2-future}}  
\end{center} 
\end{table}

The Higgs self-coupling can be probed via a number of different Higgs boson decay channels. Given the small cross section, typically at least one of the Higgs bosons is required to decay to a pair of $b$-quarks. The \bbyy decay mode has been singled out as the \emph{golden} channel despite the small branching ratio (BR $=$ 0.25\%). The second most sensitive decay mode is \bbtt with a large branching fraction (BR $\approx$ 6\%). Other sensitive final states include \bbbb and \bbzz($4\ell$).

The results are presented in terms of the achievable precision on the self-coupling modifier $\klambda$. The results of 27 TeV (HE-LHC) and 100 TeV (FCC-hh) studies are summarised in \refta{tab:kappalambda_coll}. Projections at 27 TeV colliders studies are extracted from Ref.~\cite{Goncalves:2018yva} and the recently published in Ref.~\cite{Cepeda:2019klc}. Most of the material presented in the following section summarises the results obtained as part of the 100 TeV collider Conceptual Design Report studies (CDR) FCC-hh detector performance studies~\cite{Mangano:2018mur} and the Ref.~\cite{Contino:2016spe}.

\begin{table}[th]
\renewcommand{\arraystretch}{1.5}
 \begin{center}
\begin{tabular}{l|c|c}
  \hline
 & HE-LHC (27 TeV, \intlumihelhc)
 & FCC-hh (100 TeV, \intlumifcc)   \\ \hline
 $\delta\klambda$
 & 10-20\%
 & 5-7\%
 \\
\hline
\end{tabular}
\caption{\label{tab:kappalambda_coll}
Expected precision on the direct Higgs self-coupling measurement at future 27 and 100 TeV $p-p$ colliders. }
\end{center}
\end{table}

\section[The Higgs-self coupling at \sqrtshelhc]{The Higgs-self coupling at \sqrtshelhc \\ \contrib{F.~Maltoni, D.~Pagani, M.~Selvaggi,  A.~Shivaji, X.~Zhao}}\label{sec:self_coupling_27}

The results presented in Chapter~\ref{chap:hl-lhc} performed in the context of HL-LHC have been extended to provide estimates of the prospects at the HE-LHC, assuming a centre of mass collision energy of 27 TeV and \intlumihelhc\ of data~\cite{Cepeda:2019klc}.

The detector performance is assumed to be that of the HL-LHC ATLAS detector. Comparisons between simulation at centre of mass energy of 14 and 27 TeV have been performed and have shown that the kinematics of the Higgs boson decay products, as well as the \hh invariant mass distribution, are similar. However the Higgs particles produced in double Higgs production tend to point more frequently in the forward region at 27 TeV, which slightly decreases the acceptance (by around 10\%). This effect has not been taken into account and the impact is expected to be small. The event yields for the various background processes have been scaled by the luminosity increase and the cross section ratio between the two centre of mass energies.

\begin{figure}[tb]
\centering
\includegraphics[width=0.7\textwidth]{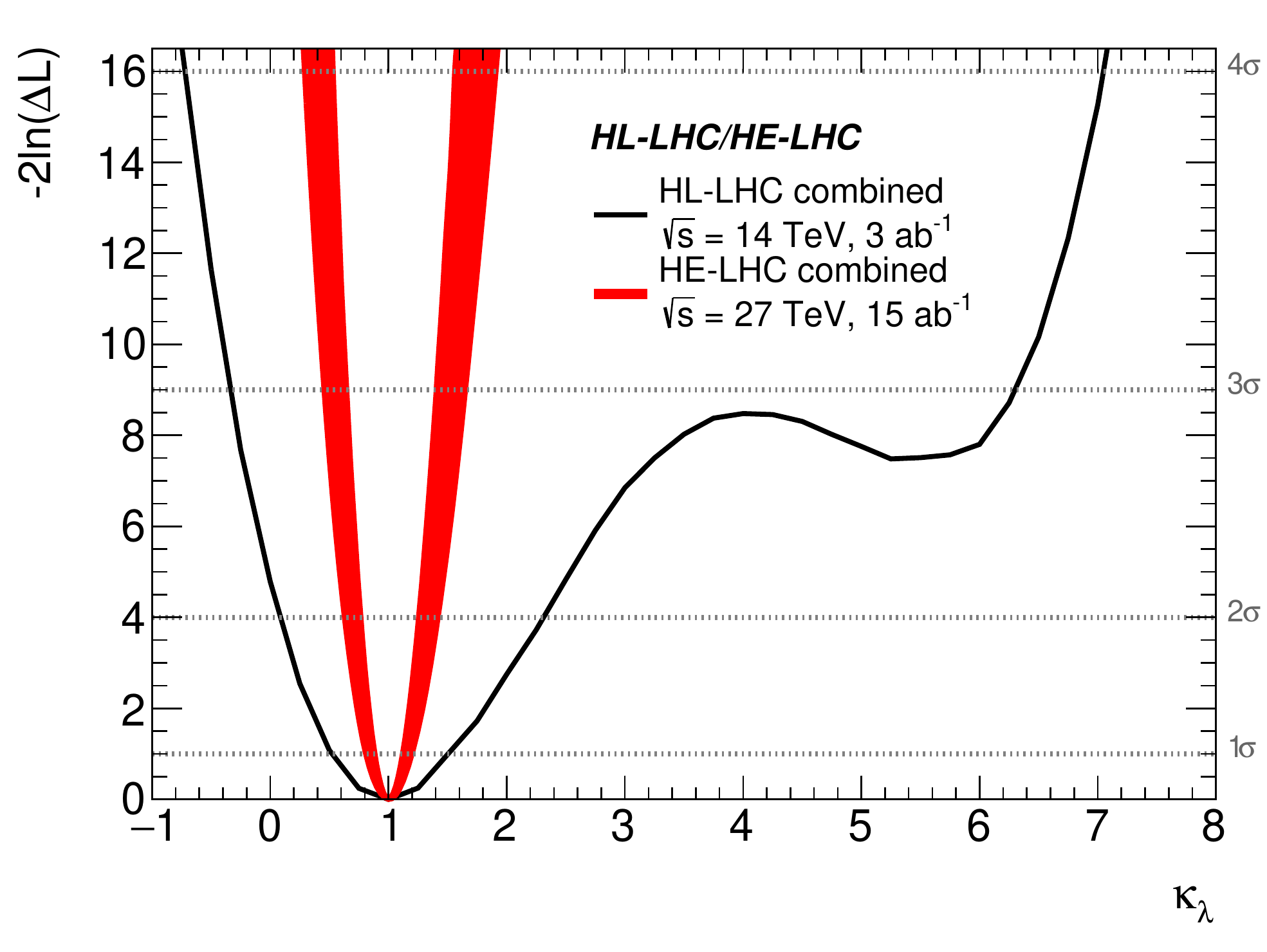}
\vspace*{-0.2cm}
\caption{Expected sensitivity for the measurement of the Higgs self coupling through the measurement of direct \hh production at HE-LHC. The black line corresponds to the combination of ATLAS and CMS measurements with HL-LHC data presented in Chapter~\ref{chap:hl-lhc}, with systematic uncertainties considered. The red band corresponds to an estimate of the sensitivity using a combination of the \bbyy and \bbtt channels, without systematic uncertainties considered.}
\label{fig:HH_HELHC_comb}
\end{figure}

Without including systematic uncertainties a significance of 7.1 and 10.7 standard deviations has been found for the \bbyy and \bbtt channels respectively. The hypothesis of no Higgs self-coupling is expected to be excluded by these channels with a significance of 2.3 and 5.8 standard deviations respectively. Finally the $\klambda$ parameter is expected to be measured with a 68\% CL precision of $\delta \kl \approx 40\%$ and $\delta \kl \approx 20\%$ for the two channels respectively. Fig.~\ref{fig:HH_HELHC_comb} shows that the $\klambda$ parameter could be measured with a precision of 10 to 20\% under these assumptions. It should be emphasised that these results rely on assumptions of experimental performance in very high pileup environment O(800-1000) that would require further validation with more detailed studies. We also stress that no systematic uncertainties have been considered.

\begin{figure}
        \centering
        \includegraphics[width=0.45\linewidth]{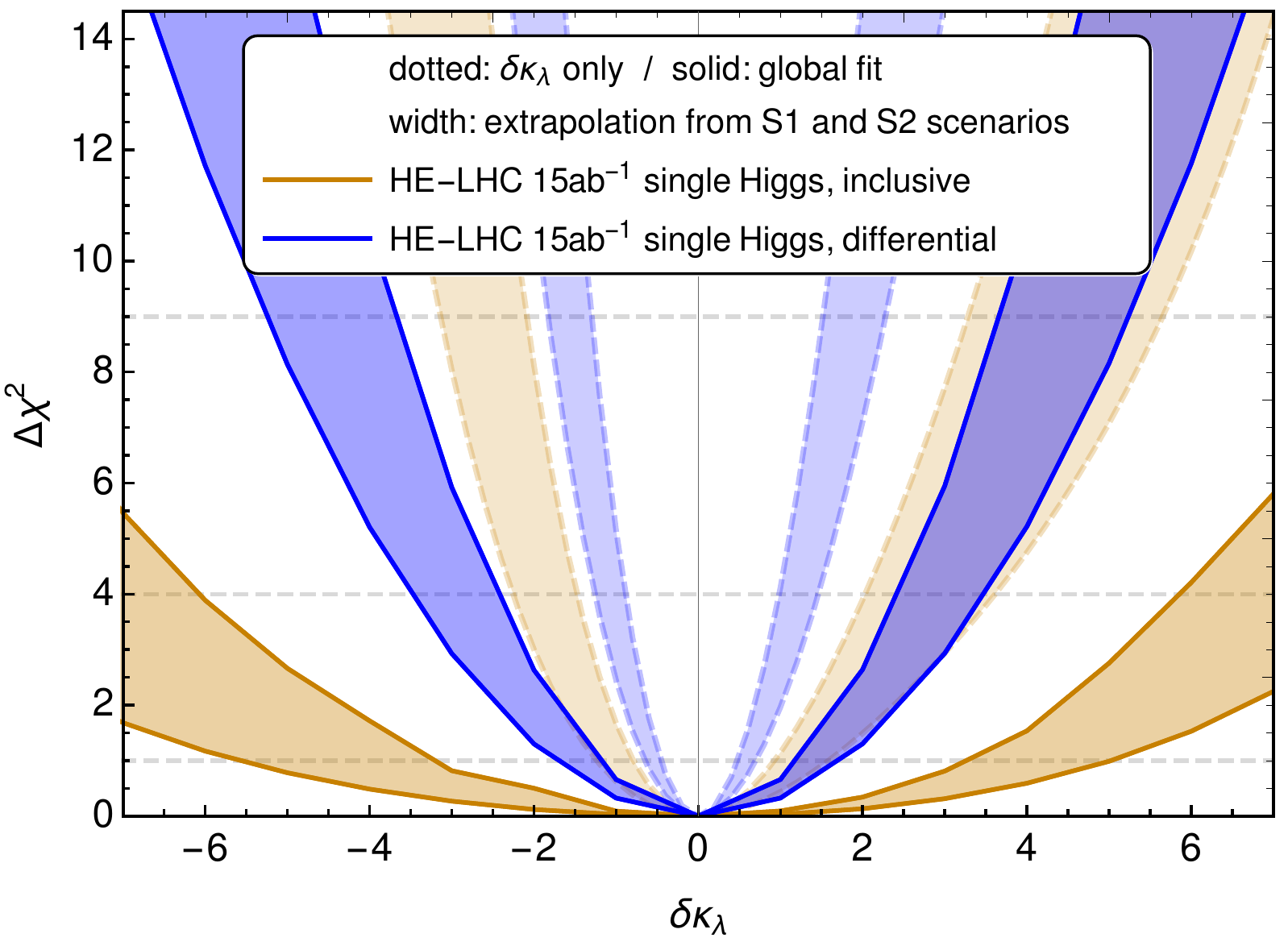}\hfill
        \includegraphics[width=0.45\linewidth]{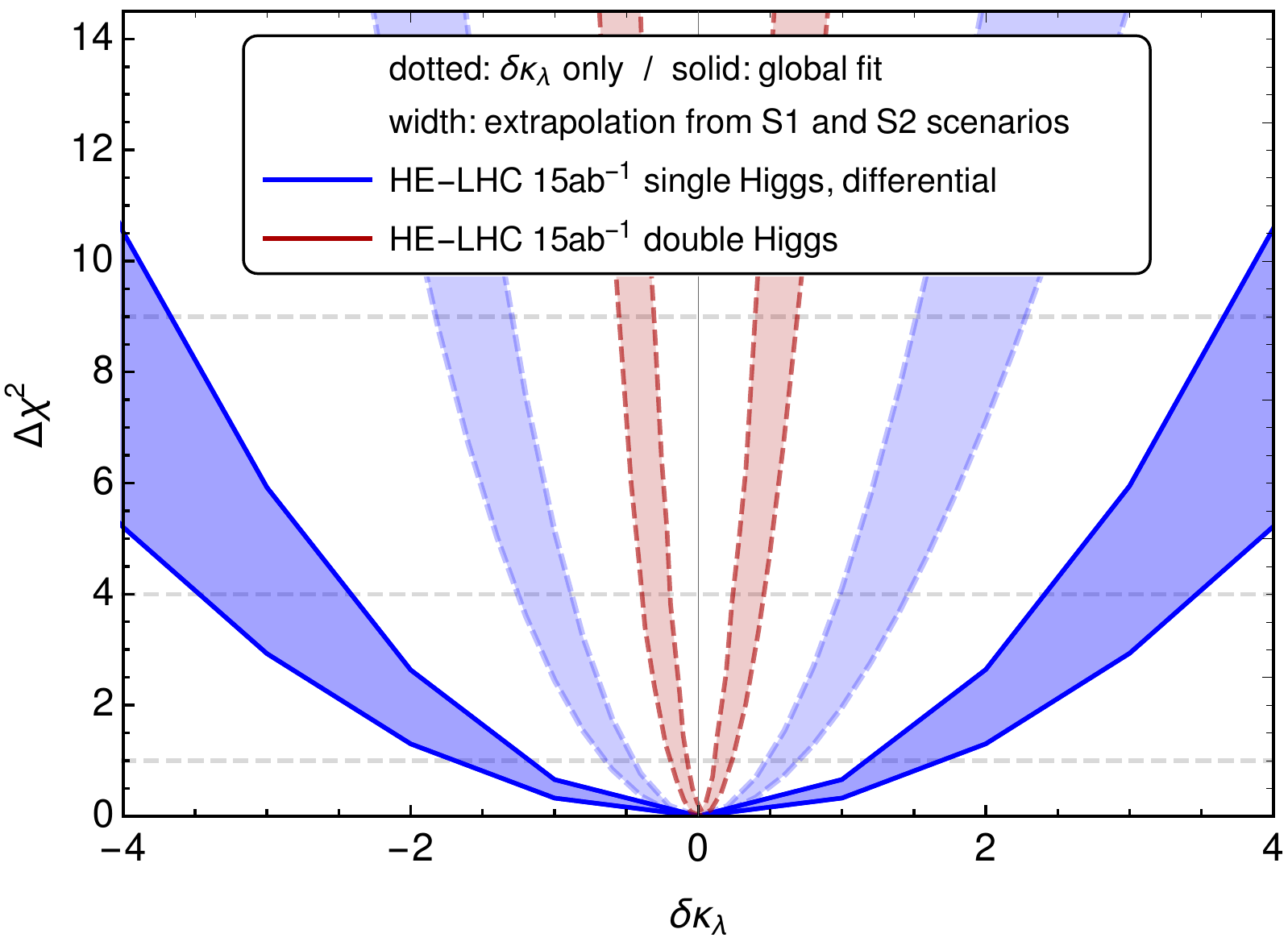}
        \caption{Expected precision on the Higgs self-coupling using single and double Higgs processes at the HE-LHC. The widths of the bands correspond to the differences between a conservative and an aggressive scenario of assumed systematic uncertainties.}
        \label{fig:helhcchi2}
\end{figure}

In contrast, phenomenological studies~\cite{Goncalves:2018yva} focusing on the \bbyy channel alone find $\delta \kl \approx 15\%$. 
Projections for the HE-LHC  assuming the same uncertainties as for the HL-LHC (see Chapter~\ref{chap:hl-lhc}), have been discussed in Ref.~\cite{Cepeda:2019klc} and are shown in Figures~\ref{fig:helhcchi2}. Inclusive and differential single Higgs measurements are shown in the left plot. A global fit that includes the effect of all possible deviations from the SM, gives at best a precision of 200\% on the self-coupling. If only the self-coupling modifier is allowed to vary, one finds 50\% in the best case scenario using differential single Higgs information. The right plot shows a comparison of the achievable precision using single and double Higgs measurements. Double Higgs production dominates the sensitivity with a 10-20\% precision, depending on the assumed scenario of systematic uncertainties. 

\section{The Higgs self-coupling at \sqrtsfcc}

All studies have been performed using simulation assuming an integrated luminosity \intlumifcc\ at \sqrtsfcc\ assuming the reference Future Circular Collider hh (FCC-hh) detector~\cite{Benedikt:2018csr}. The detector simulation has been performed with the fast simulation tool \delphes~\cite{deFavereau:2013fsa} using the reference FCC-hh detector parameterisation. The signal samples have been generated at leading order with \amc and \py\ accounting for the full top mass dependence, for several values of the self-coupling modifier, \kl. A self-coupling dependent K-factor to match NNLL+NNLO accuracy was been derived from~\cite{Borowka:2016ypz} and applied to the signal samples.

\subsection[\hhbbyy]{\hhbbyy \\ \contrib{G.~Ortona, M.~Selvaggi}}

The \bbyy decay mode provides a very clear signature of two photons and two \bjets in the final state allow the large background rates to be controlled. The main backgrounds are $\gamma\gamma$+jets, $\gamma$+ jets (with at least one jet being mis-identified as a photon) as well as single Higgs production. The $\ttbar H$ sample was also generated at LO with up to one extra jet merged with the parton shower. The latest NLO cross section $\sigma_{\ttbar H} = 34$ pb was used for this sample~\cite{Contino:2016spe}. The gluon-gluon fusion single Higgs contribution was generated at LO in the infinite top mass approximation with an extra \bb pair.
The \vh\ sample has been produced at LO with up to two extra jet merged. The \vbf\ contribution was found to be negligible. The QCD backgrounds $\gamma\gamma$+jets, and $\gamma$+ jets are simply generated at LO. All samples have been generated using the 5 flavour scheme (5F) with a vanishing $b$-quark mass.

\subparagraph{Detector and performance assumptions}

The FCC-hh detector is assumed to have a similar performance to the HL-LHC detectors. The photon identification efficiency is assumed to be $\effg=$~95\% for $|\eta| < 2.5$ and $\effg=$~90\% for $2.5 < |\eta| < 4.0$ regardless of the photon \pT. The light jet to photon mis-identification probability (fake-rate) is parameterized by the function $\misg = 0.002 \exp(-\pT[\mathrm{GeV}]/30)$. We assume a resolution on the di-photon pair invariant mass $\delta \maa =$~1.3 GeV. The \btagging efficiency $\effb$ and the light (charm) mis-tag rates $\mislc$ are assumed to be $\effb=$~85\% and $\mislc=$~1 (5)\%. These numbers are similar to what has been assumed for the HL-LHC detectors. 

\begin{figure}
  \centering
  \includegraphics[width=0.32\columnwidth]{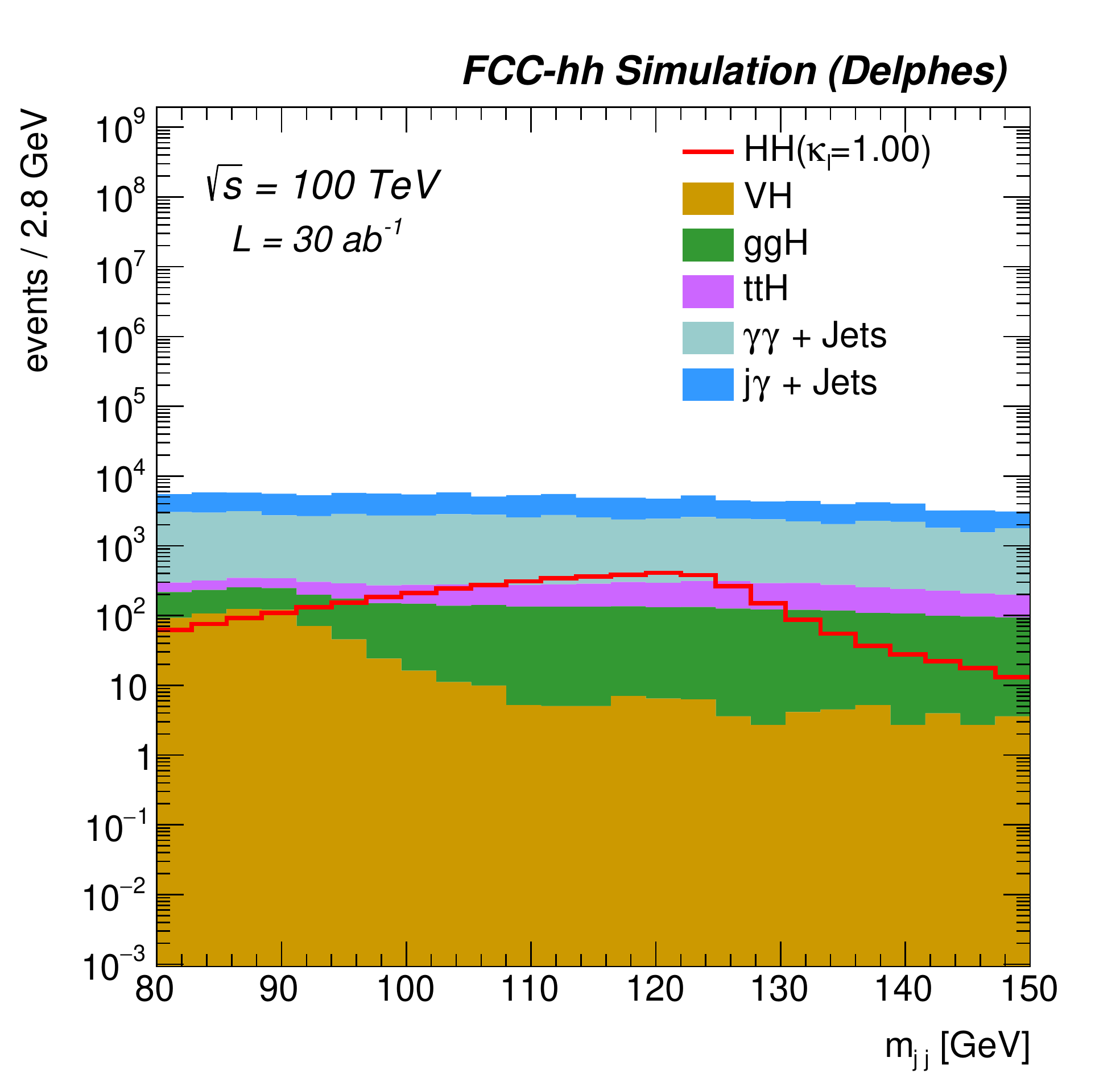}
  \includegraphics[width=0.32\columnwidth]{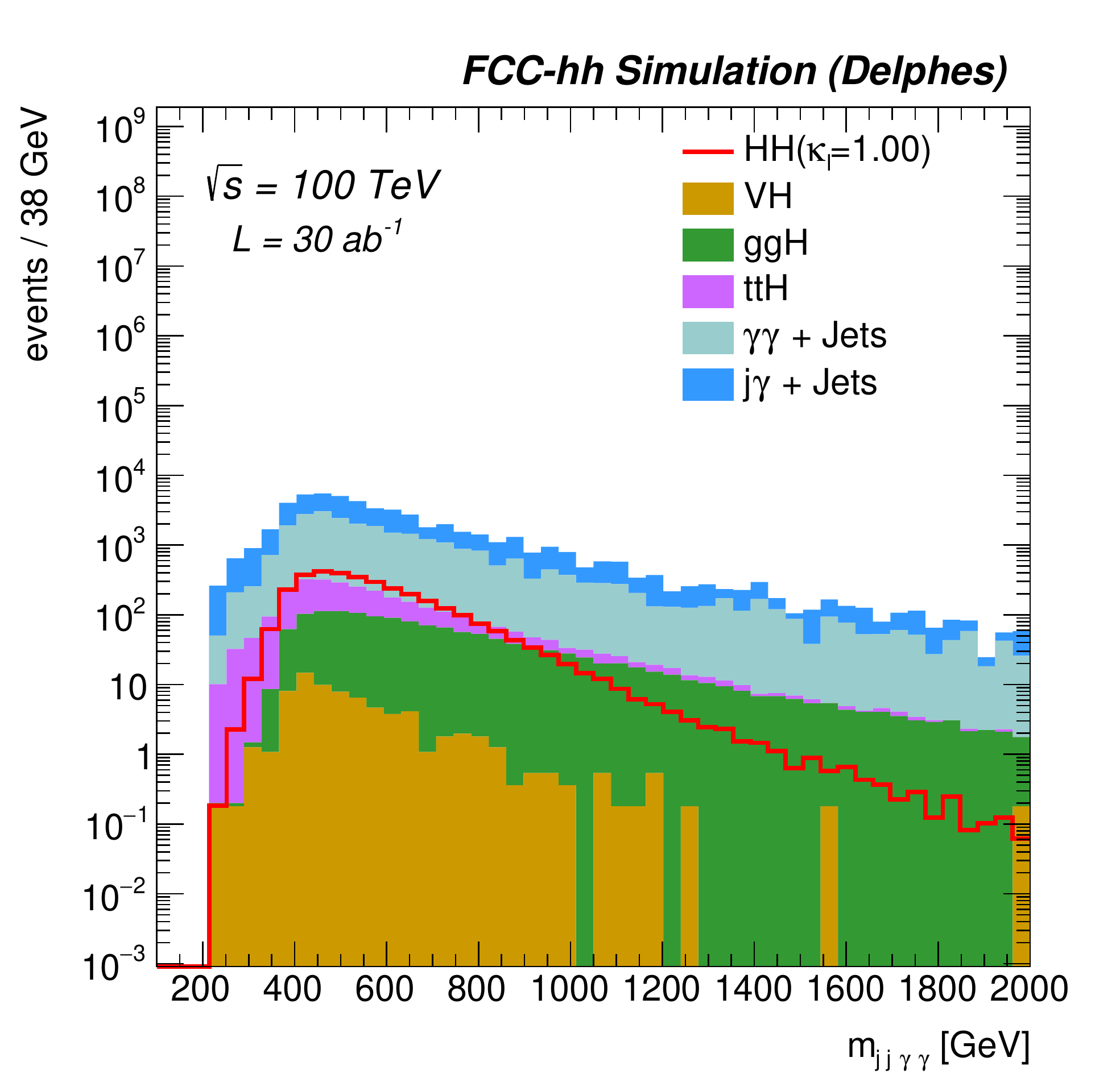}
  \includegraphics[width=0.32\columnwidth]{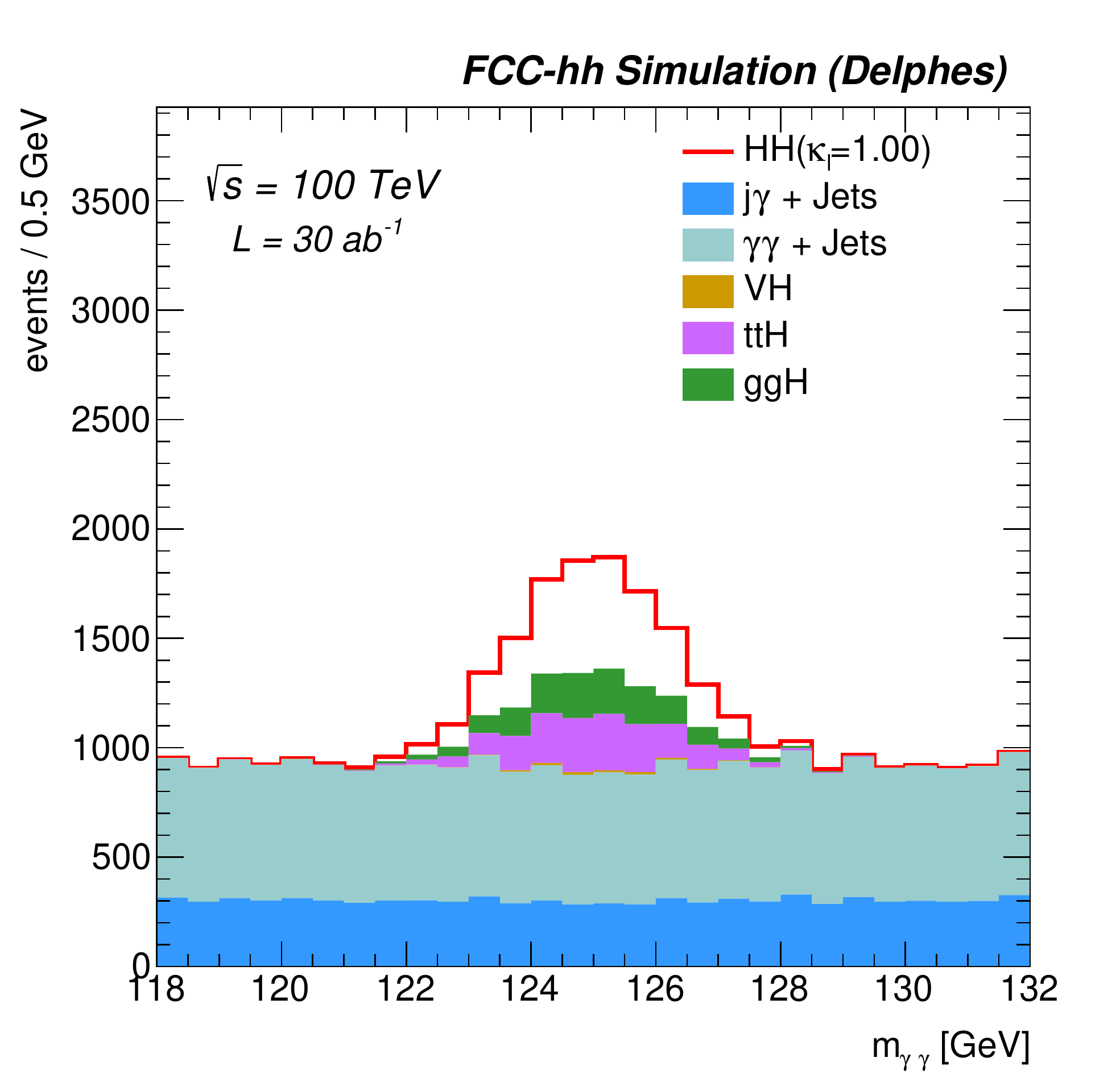}
  \caption{Left: Di-jet system invariant mass spectrum before applying the di-jet invariant mass selection. Di-Higgs (centre) and di-photon (right) candidates invariant mass spectra after applying all selection criteria.}
  \label{hhbbaa_spectra}
\end{figure}

\subparagraph{Event selection and signal extraction}

Events are required to contain at least two isolated photons and two \btagged jets. Jets are clustered using particle-flow candidates with the anti-k$_\mathrm{T}$ algorithm with radius parameter R=0.4. We required $\pT(\gamma,\rm{ b}) > 30$~GeV and $|\eta(\gamma,\rm{b})| < 3.0$. The Higgs candidates are formed from the two jets and photons with highest $\pT(\gamma,\rm b)$. The leading photon and \bjet are required to have $\pT(\gamma,\rm b) > 60$ GeV, and the di-photon and di-jet pairs $\pT(\gamma\gamma,\bb) > 125$ GeV. In order to suppress the $\ttbar\rm{H}$ background, we veto leptons with $\pT(\ell) > 25$ GeV and $|\eta(\ell)| < 3.0$ and require $\Delta R_{\bb} < 2.0$. The jet pair $\mbb$ is shown in Fig.~\ref{hhbbaa_spectra} (left). Finally, we apply a window cut on the invariant mass of the \bb pair $100 <\mbb < 130$ GeV.
The signal extraction is performed via a two dimensional likelihood fit over the the photon pair and the Higgs pair invariant masses, \myy and $m_{\gamma\gamma \bb}$, shown in Fig.~\ref{hhbbaa_spectra} (centre) and (right). At the LHC the 2D distribution ($\mbb$, $\myy$) is fitted to maximally discriminate against background and optimise the precision on the signal strength (see Sec.~\ref{subsubsec:bbaa_extr}). The strategy differs at the FCC-hh where differential information on the $\mhh$ distribution becomes accessible. The signal shape is parameterised by a Gaussian and the sum of a Landau and an exponential distribution respectively. 

\begin{figure}
  \centering
  \includegraphics[width=0.45\columnwidth]{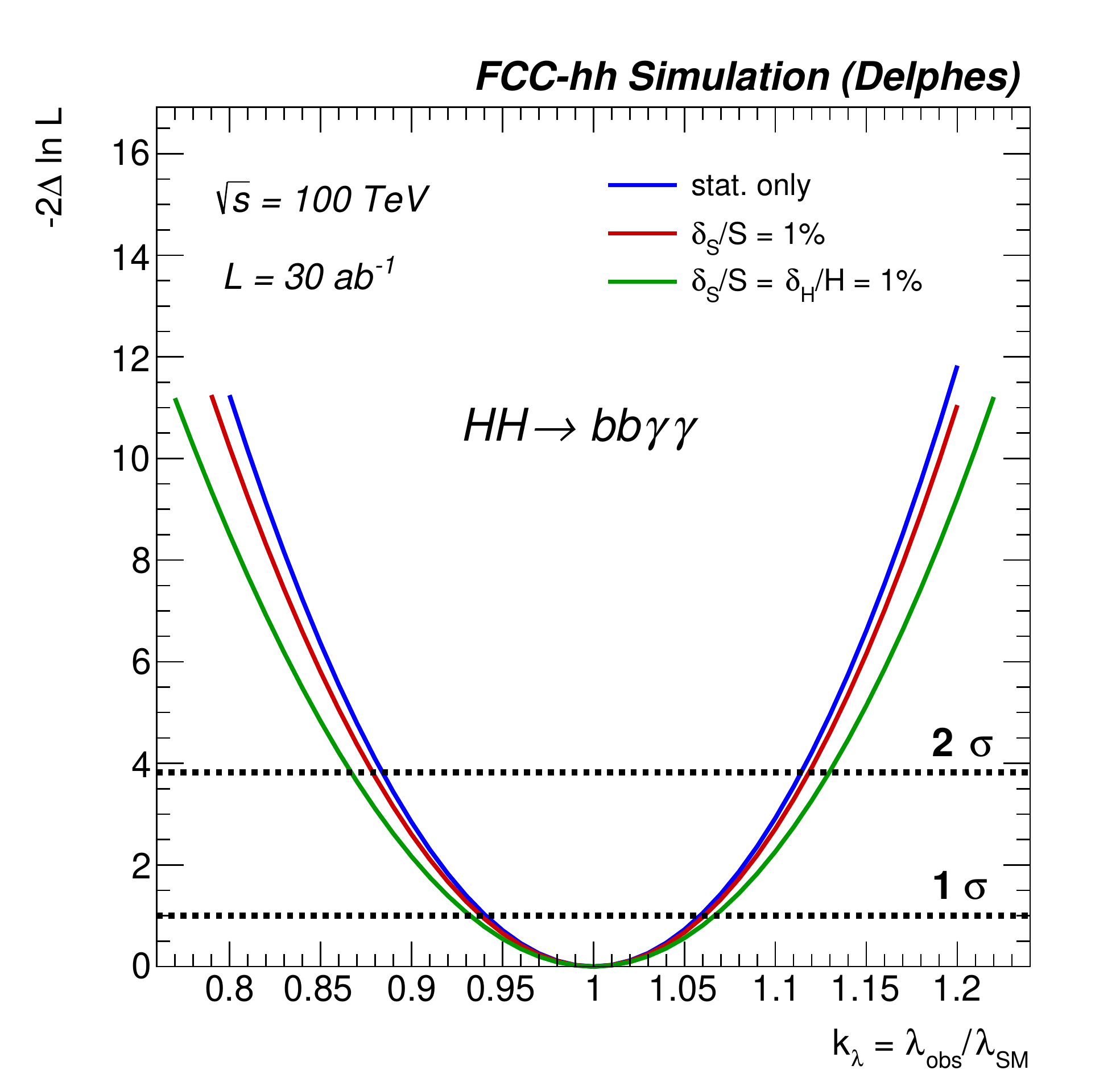}
  \includegraphics[width=0.45\columnwidth]{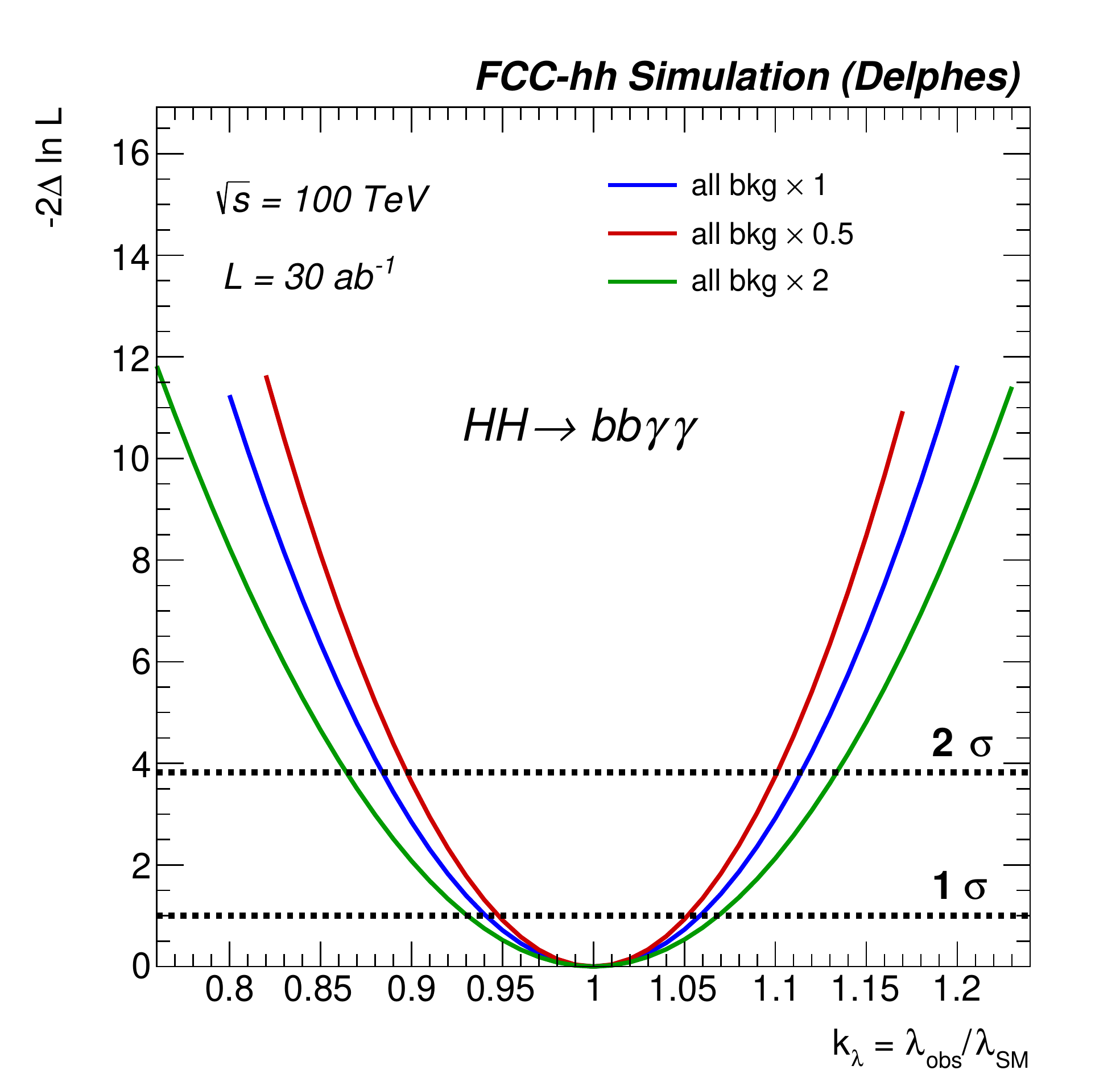}
  \caption{Expected precision on the Higgs self-coupling modifier $\kl$ with no systematic uncertainties (only statistical), 1\% signal uncertainty, 1\% signal uncertainty together with 1\% uncertainty on the Higgs backgrounds (left) and assuming respectively $\times 1$, $\times 2$, $\times 0.5$ background yields (right).)}
  \label{hhbbaa_kl_main}
\end{figure}

\begin{figure}
  \centering
  \includegraphics[width=0.32\columnwidth]{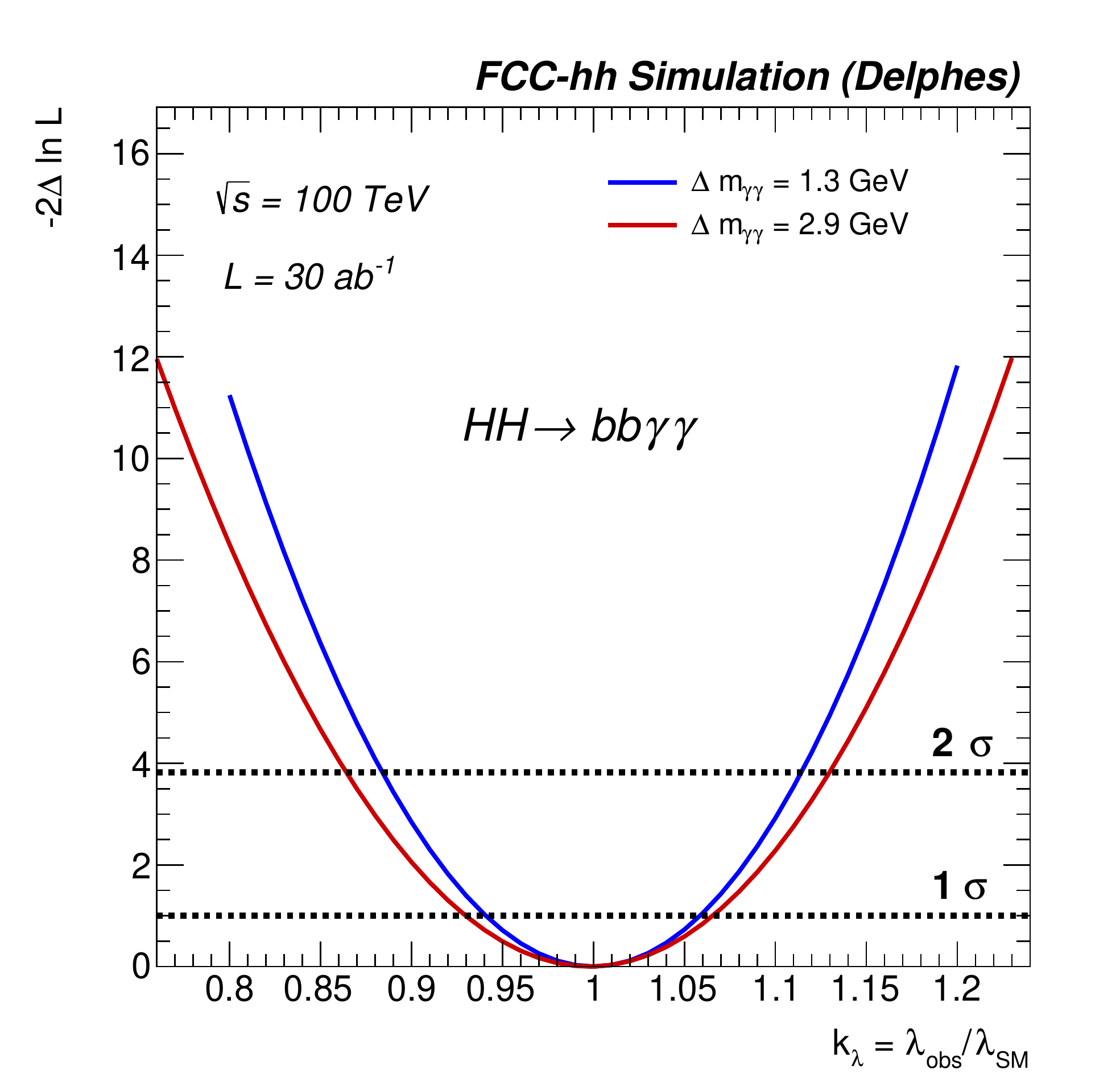}
  \includegraphics[width=0.32\columnwidth]{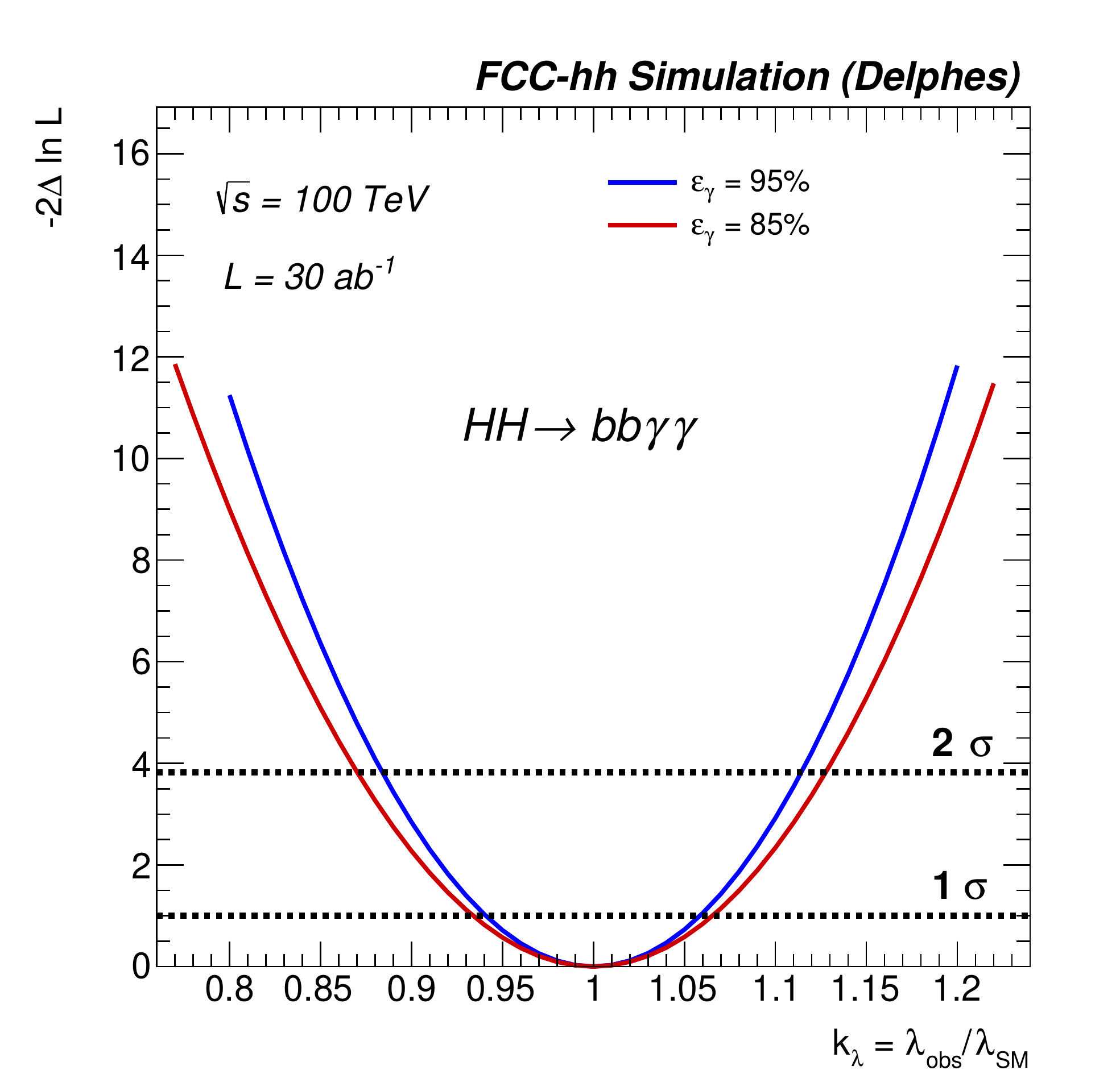}
  \includegraphics[width=0.32\columnwidth]{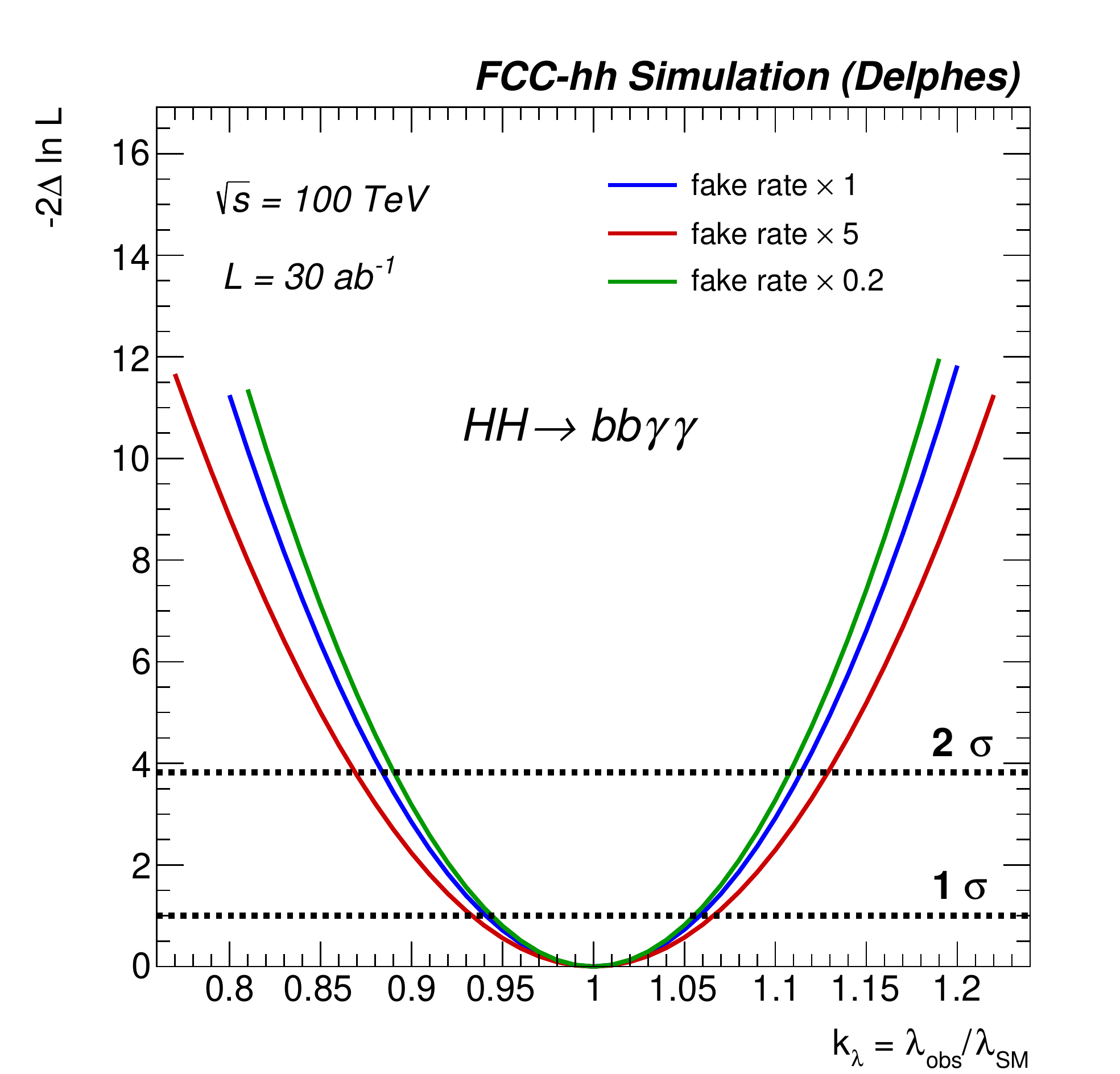}
  \caption{Expected precision on the Higgs self-coupling modifier $\kl$ obtained by varying the photon reconstruction performance. Left: Comparison of two scenarios with nominal ($\Delta \myy=1.3$~GeV) and degraded ($\Delta \myy=2.9$~GeV) energy resolution. Centre: Comparison of two scenarios with nominal ($\epsilon_{\gamma}$~=~95\%) and degraded ($\epsilon_{\gamma}$~=~85\%) photon reconstruction efficiency. Right: Comparison of three scenarios with nominal, degraded ($\times 5$) and improved ($\times 0.2$) photon mis-tag rate.}
  \label{hhbbaa_kl_syst}
\end{figure}

\subparagraph{Results and discussion}
\label{hhbbaa_results}
The negative log-likelihood (NLL) distribution for the parameter $\kl$ with respect to the best-fit value obtained for varying systematic effects, background normalisations and detector assumptions is shown in Figures~\ref{hhbbaa_kl_main} and~\ref{hhbbaa_kl_syst}. The 1$\sigma$ and 2$\sigma$ lines correspond to the 68\% and 95\% confidence levels (CL) respectively.

Figure~\ref{hhbbaa_kl_main} (left) shows the sensitivity obtained with different assumptions about the uncertainties. With only the statistical uncertainty (blue curve), we find $\delta \kl =$~5.5\%. When a 1\% systematic uncertainty on the signal normalisation is included (red curve) the expected precision decreases to $\delta \kl =$~6\%. The signal normalisation includes both theoretical uncertainties on the production cross section as well as the uncertainty on the integrated luminosity. An additional uncertainty of 1\% on the single Higgs backgrounds normalisation (green curve) is shown under the assumption that the QCD background can be extrapolated from a control sample defined by $|\maa - \mH| > 10$~GeV, with high statistics into the signal region. For the single Higgs background defining such a control sample is more challenging and we therefore assume an uncertainty of 1\% on the normalisation, motivated by expected precision on these processes at the FCC-hh~\cite{Plehn:2015cta}. In this scenario we find an expected precision $\delta \kl =$~6.5\%. Figure~\ref{hhbbaa_kl_main} (right) shows how the precision is affected by varying the overall background yields by factors of 2 and 0.5 and find an impact on the overall $\kl$ precision of $\approx \pm 1\%$.

Figure~\ref{hhbbaa_kl_syst} shows the impact of detector performance related assumptions on the sensitivity. Figure~\ref{hhbbaa_kl_syst} (left) shows the impact of degrading the energy resolution of the electromagnetic calorimeter, which affects the $\delta \maa$ resolution. Figure~\ref{hhbbaa_kl_syst} (centre) shows the impact of varying the photon reconstruction efficiency and Fig.~\ref{hhbbaa_kl_syst} (right) shows the impact of varying the jet-to-photon fake rate. Each of these scenarios degrades the precision on the self-coupling by 1-2\%. These less optimistic performance assumptions roughly correspond to the expected performance of the ATLAS and CMS detectors at HL-LHC (see Chapter~\ref{chap:hl-lhc}).

To summarise, within the stated assumptions on the expected performance of the FCC-hh detector, a precision on the Higgs self-coupling of $\delta \kl =$~5\% in the \hhbbyy channel can be achieved.

\subsection[$\hhbbZZ$]{$\hhbbZZ$ \\ \contrib{L.~Borgonovi, E.~Fontanesi}}

The large Higgs pair production cross section at 100 TeV allows for rare but cleaner final states to become accessible. One example is the \hhbbZZ\ decay channel (where $\ell = e^{\pm}$, $\mu^{\pm}$). This channel is not accessible at the HL-LHC. Despite a small
cross section at the FCC-hh ($\sigma_{\bb 4\ell}=178$~ab), the presence of four leptons in association with two \bjets leads to a very clean final state topology allowing to maintain a rather good signal selection efficiency while controlling the background. The main backgrounds processes are $\ttbar(\bb)\rm H(4\ell)$, gg$(\rm H)+\bb$, Z$(\bb)\rm H(4\ell)$ and $\ttbar Z(\ell\ell)$, followed by minor negligible contributions such as $4\ell+\bb$ continuum, $\ttbar(\rm \bar{b}\ell\nu_{\ell}\rm b\ell\nu_{\ell}) \rm H(\ell\ell)$ and $\ttbar \rm ZZ(4\ell)$.
 
The $\ttbar H$, $gg(H)+\bb$, $Z(\bb)H$ and $\ttbar Z(\ell\ell)$ background samples were generated at LO and higher order radiative corrections were accounted for by applying K-factors of $K(\ttbar H)=1.22$, $K(ggH)=3.2$ and $K(ZH)=1.1$ ~\cite{Contino:2016spe}. The contribution of the $4\ell$+jets ($ZZ^{*}, Z^{*}Z^{*}, ZZ$) continuum is evaluated using a $\ell\ell\ell\ell jj$ ($\ell=e^{\pm},\mu^{\pm}$) sample, generated with the four leptons invariant mass in the range [100, 150]~GeV and only heavy flavour partons ($b$/$c$). This background contribution was found to be negligible. The cross sections are summarised in \refta{tab:Cross section}.

\begin{table}[h]
\centering
\renewcommand\arraystretch{1.5}
\begin{tabular}{c|c|c|c|c}
\hline
$\hh \to \bb {ZZ}(4\ell)$ &$\ttbar H \to \bb 4\ell$ & $gg(H)+\bb \to \bb 4\ell$ & $ZH \to \bb 4\ell$ & $\ttbar Z \to \bb 4\ell$  \\
\hline
0.178  &   4.013   & 0.369   & 0.071   & 2594    \\
\hline
\end{tabular}
\caption{Cross section (fb) times branching ratio for the signal and the background processes~\cite{Contino:2016spe}.}
\label{tab:Cross section}
\end{table}

\subparagraph{Event Selection}
Events are required to have exactly four identified and isolated muons (electrons) with $\pT>5$($7$)~GeV and $|\eta|<4.0$. $Z$ boson candidates are formed from pairs of opposite-charge
leptons ($\ell^+\ell^-$). At least two di-lepton pairs are required. The $Z$ candidate with the invariant mass closest to the nominal $Z$ mass is denoted as $Z_1$; then, among the other opposite-sign lepton pairs, the one with the highest $\pT$ is labelled as $Z_2$.
$Z$ candidates must pass a set of kinematic requirements
that improve the sensitivity to the Higgs boson decay: the $Z_1$ and $Z_2$ invariant masses have to be in the [40, 120]~GeV and [12, 120]~GeV ranges, respectively.
At least one lepton is required to have $\pT>20$~GeV and a second is required to have $\pT>10$~GeV.
A minimum angular separation between two leptons is required to be $\Delta R(\ell_i, \ell_j)>0.02$.
The four leptons invariant mass, $m_{4\ell}$, is requested to be in the range $120 < m_{4\ell}< 130$~GeV.

At least two identified \bjets, reconstructed with the anti-k$_\mathrm{T}$ algorithm inside a cone of radius $R=0.4$, are required.
Their invariant mass is required to be in the range $80 < \mbb < 130$~GeV and the angular distance between the two \bjets has to be $0.5 < \Delta R_{\bb }<$ 2. These cuts are particularly effective to reject the \ttbar\H background.

\begin{figure}
\centering
\includegraphics[width=.32\textwidth]{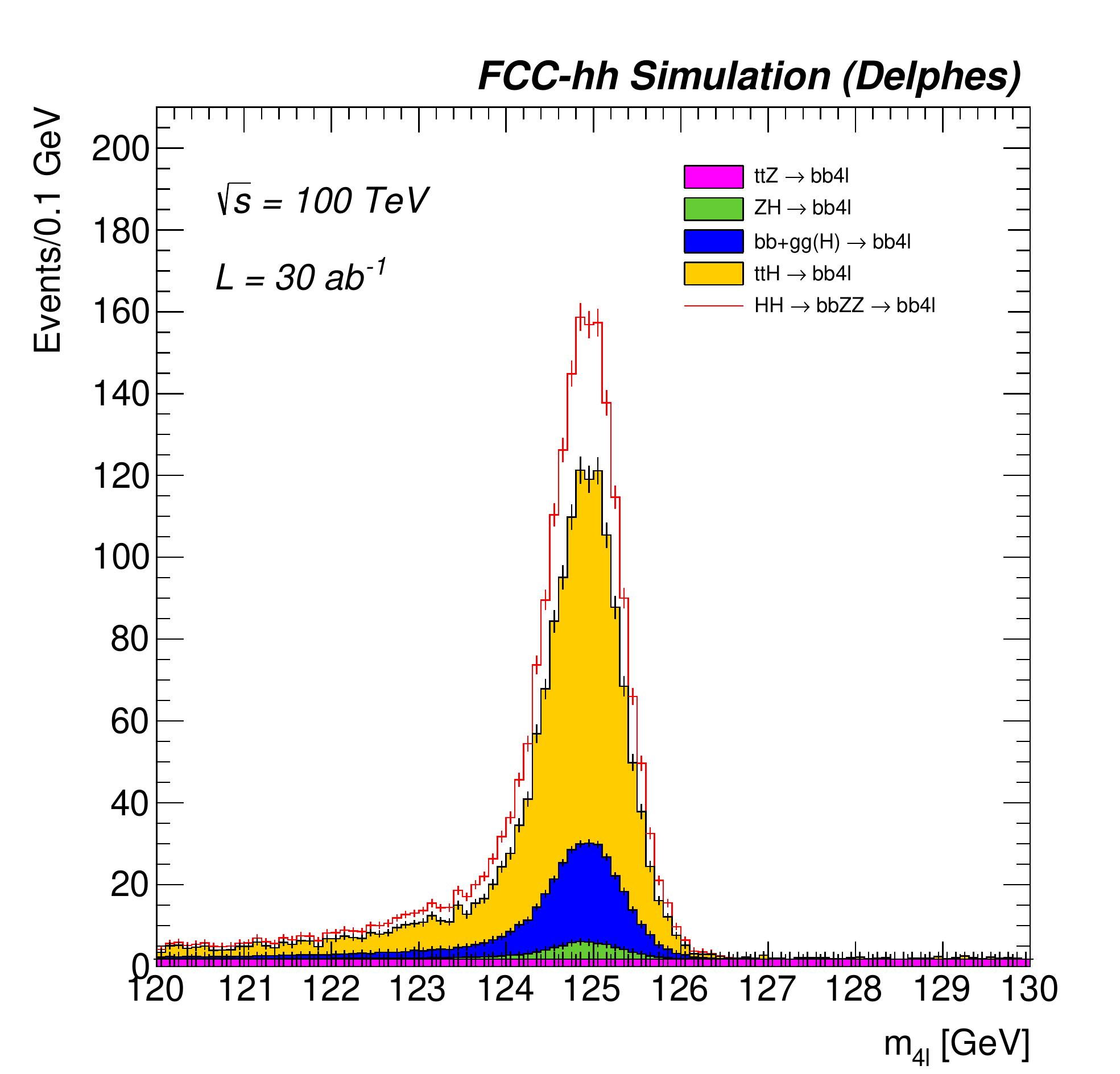}
\includegraphics[width=.32\textwidth]{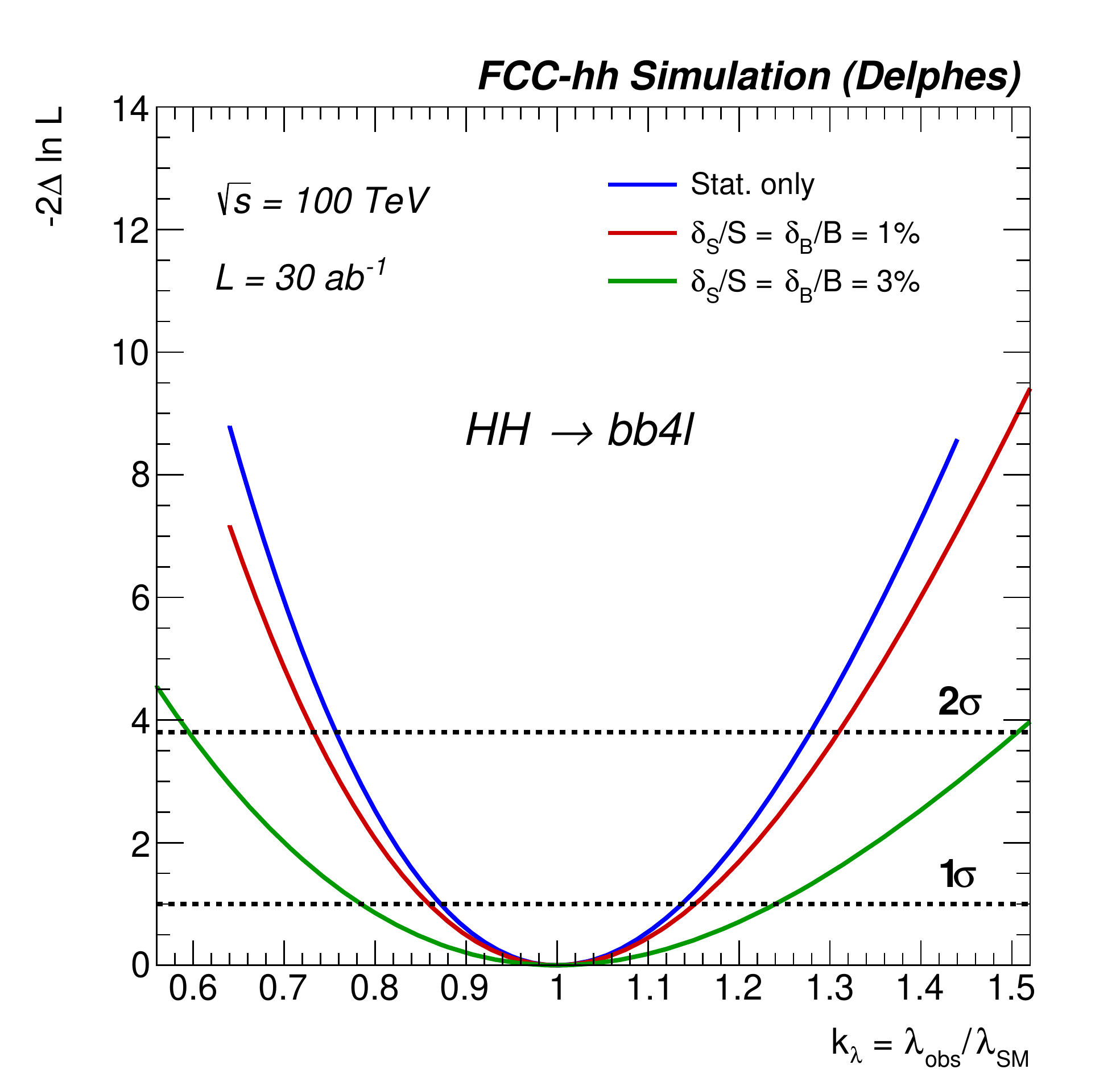}
\includegraphics[width=.32\textwidth]{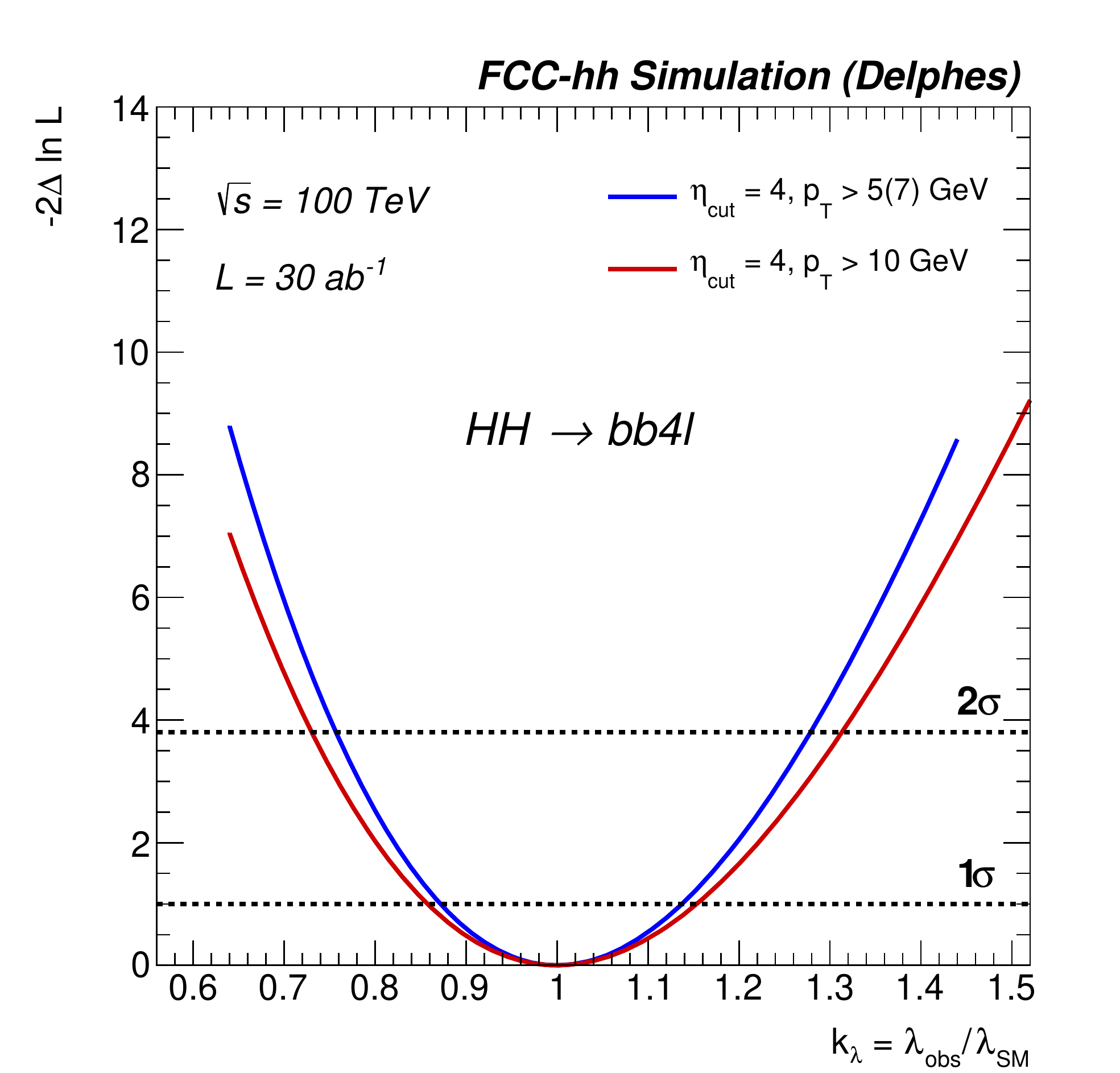}
\caption{Left: Distribution of the four leptons invariant mass for the \hhbbZZ signal and all the analyzed background processes after the full selection for 30 ab$^{-1}$. Centre: Expected precision on the Higgs self-coupling. Right: Comparison of two scenarios (without systematic uncertainties) with a cut on muon (electron) \pT larger than 5 (7)~GeV and 10 (10)~GeV.}
\label{fig:bb4l}
\end{figure}

\subparagraph{Results}
The invariant mass spectrum of the four leptons after the full event selection is shown in Fig.~\ref{fig:bb4l} (left). The NLL on the self-coupling modifier \klambda is shown in Fig. \ref{fig:bb4l}, (centre), for three different assumptions for the systematic uncertainties:
\begin{enumerate}[noitemsep]
\item Statistical uncertainties only
\item 1$\%$ systematic uncertainty on signal and background: $ \frac{\Delta S} {S} = \frac{\Delta B} {B} = 1 \%$
\item 3$\%$ systematic uncertainty on signal and background: $ \frac{\Delta S} {S} = \frac{\Delta B} {B} = 3 \%$
\end{enumerate}

The expected precision on the Higgs self-coupling modifier \klambda without systematic uncertainties is 14$\%$ at 68$\%$ CL, while when assuming a 1$\%$ systematic uncertainty on the signal and the backgrounds the precision on the measurement of \klambda becomes 15$\%$ while with a 3$\%$ systematic uncertainty it decreases to 24$\%$.
Figure \ref{fig:bb4l} (right) shows how the precision on the self-coupling is affected by the variation of the detector configuration (for example, assuming a larger tracker and/or higher magnetic field and consequently a minimum \pT for muons and electrons of 10~GeV). The precision on \klambda degrades from 14$\%$ to 15$\%$ at 68$\%$ CL, considering statistical uncertainties only.

\subsection[$HH \rightarrow$ \bbbb+jet]{$HH \rightarrow$ \bbbb+jet \\ \contrib{G.~Ortona, M.~Selvaggi}}
\label{subsec:bbbbj}

The fully hadronic channel has a high rate given the large Higgs branching fraction to a $\bb$ pair, but the overwhelming multi-jet background makes this measurement very difficult. This background can be reduced by requiring the Higgs to be boosted such that the decay products are contained inside a single, large-radius jet. A boosted configuration in which the both Higgs bosons have a large boost and recoil against each other can be effective in terms of background rejection but it provide low sensitivity to the Higgs self-coupling since the di-Higgs rate dependence on the trilinear largely originates from configurations with low \mhh. Following the approach in~\cite{Banerjee:2018yxy}, we study the configuration where the Higgs pair recoils against one or more jets, forcing the pair to have a small invariant mass. The main backgrounds include at least four \bjets, where the two $\bb$ pairs come from either strong production (QCD), mainly from $g \to \bb$ splittings, either QCD and electroweak production (QCD+EWK), e.g. $Z\bb$, or pure EWK production, e.g. $ZH$ or $ZZ$.

The signal sample consists of hh+jet and was generated taking into account the full top mass dependence at leading order (LO) with the jet $\pT^{\text{jet}}$ (or equivalently the di-Higgs \pTHH), $\pT > 200$~GeV, accounting for the full top mass. Higher order QCD corrections are accounted for with a K-factor $K=$1.95 applied to the signal samples~\cite{Banerjee:2018yxy}, leading to $\sigma_{\hh j} = 38$~fb for $\pT^{\text{jet}} > 200$~GeV and $\kl$=1. The LO cross sections used for the backgrounds are computed with $\pT^{\text{jet}} > 200$~GeV and are $\sigma_{\bb\bb j}(\text{QCD}) = 443.1$~pb, $\sigma_{\bb\bb j}(\text{QCD+EWK}) = 6.2$~pb and $\sigma_{\bb\bb j}(\text{EWK}) = 72$~fb.

\begin{figure}
  \centering
  \includegraphics[width=0.45\columnwidth]{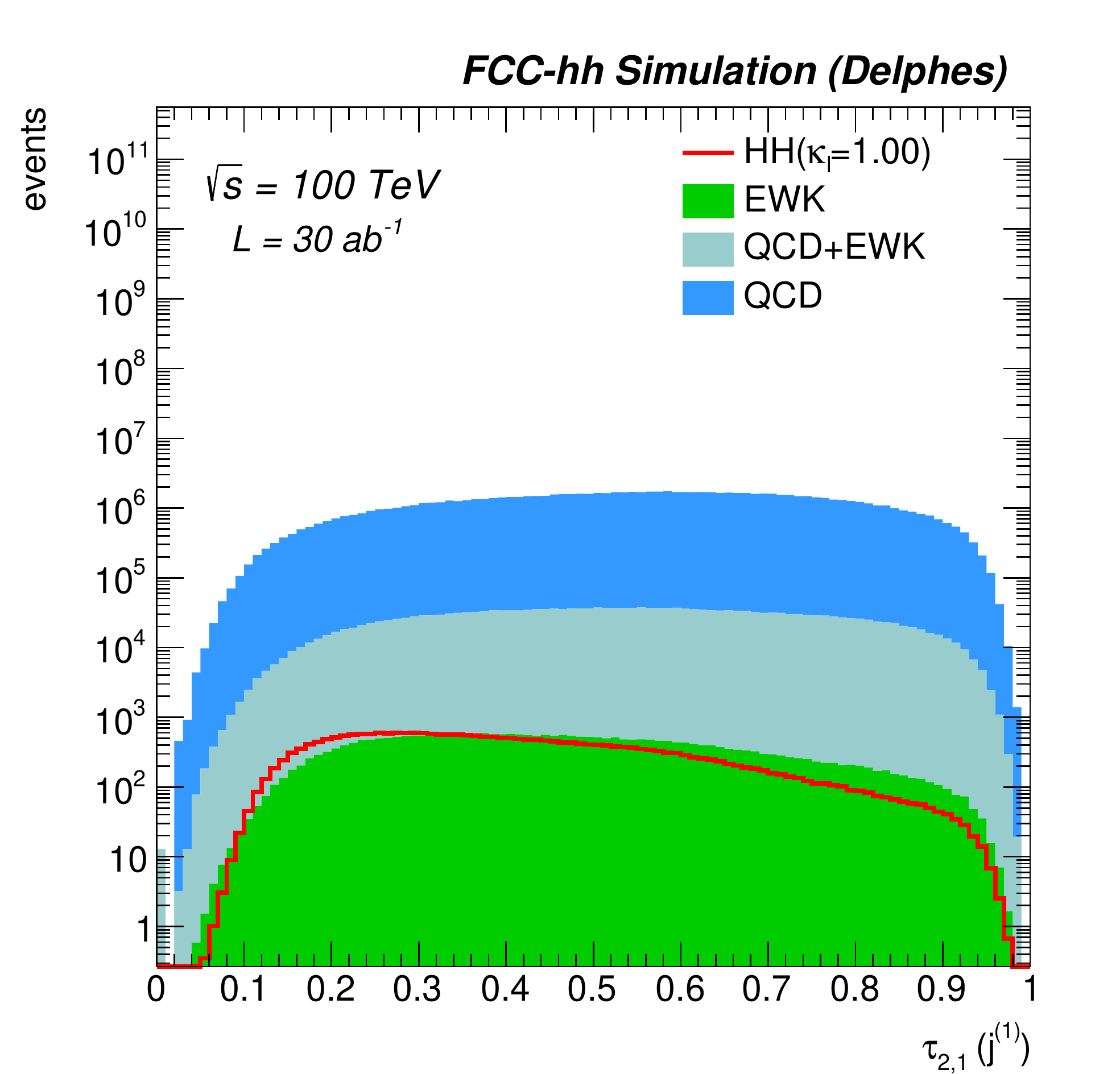}
  \includegraphics[width=0.45\columnwidth]{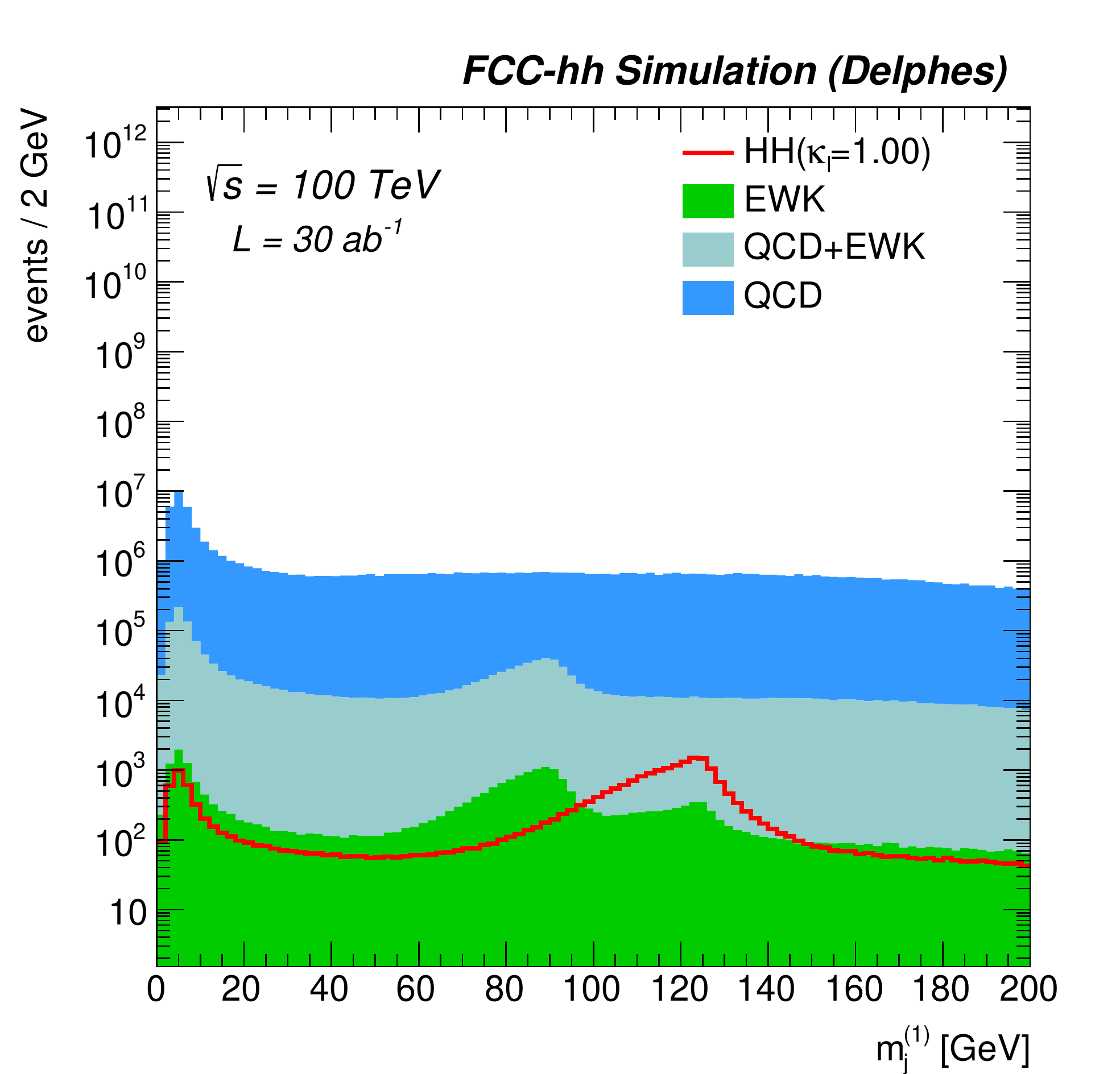}
  \caption{2-to-1 ($\tau_{2,1}$) subjettiness ratio (left) and soft-drop mass (left) spectra of the leading Higgs large-radius jet candidate.}
  \label{hhboosted_1}
\end{figure}

\subparagraph{Event selection and signal extraction}

Jets are clustered using particle-flow candidates with the anti-k$_\mathrm{T}$ algorithm with a large parameter $R=0.8$. The large cone size is chosen such that a large fraction of the Higgs decay products will be included in the jet, hence the denomination ``fat-jets''. Events are first pre-selected by requiring at least two central fat jets that contain at least two $b$-subjets. We assume a conservative $70\%$ \btagging efficiency. The fat-jets are selected if $\pT^{j} > 300$ GeV and $|\eta^{j}| < 2.5$. The two highest momentum double $b$-tagged fat-jets constitute our Higgs candidates. We further ask the di-jet pair to be sufficiently boosted, $\pT^{jj} > 250$ GeV, and the leading jet to have a $\pT^{j_1} > 400$ GeV. The b-tagging performance inside boosted jets is assumed to be equal to that of the resolved case. This is motivated by the relatively small boost of the Higgs fat-jets. The two fat-jets must have a small opening angle $\Delta R(j_1,j_2) < 3.0$. Finally, given that QCD splittings are characterized by a large momentum imbalance in the daughter partons, we require a small momentum imbalance $(\pT^{j_1} - \pT^{j_2})/\pT^{jj} < 0.9$.

\begin{figure}
  \centering
  \includegraphics[width=0.45\columnwidth]{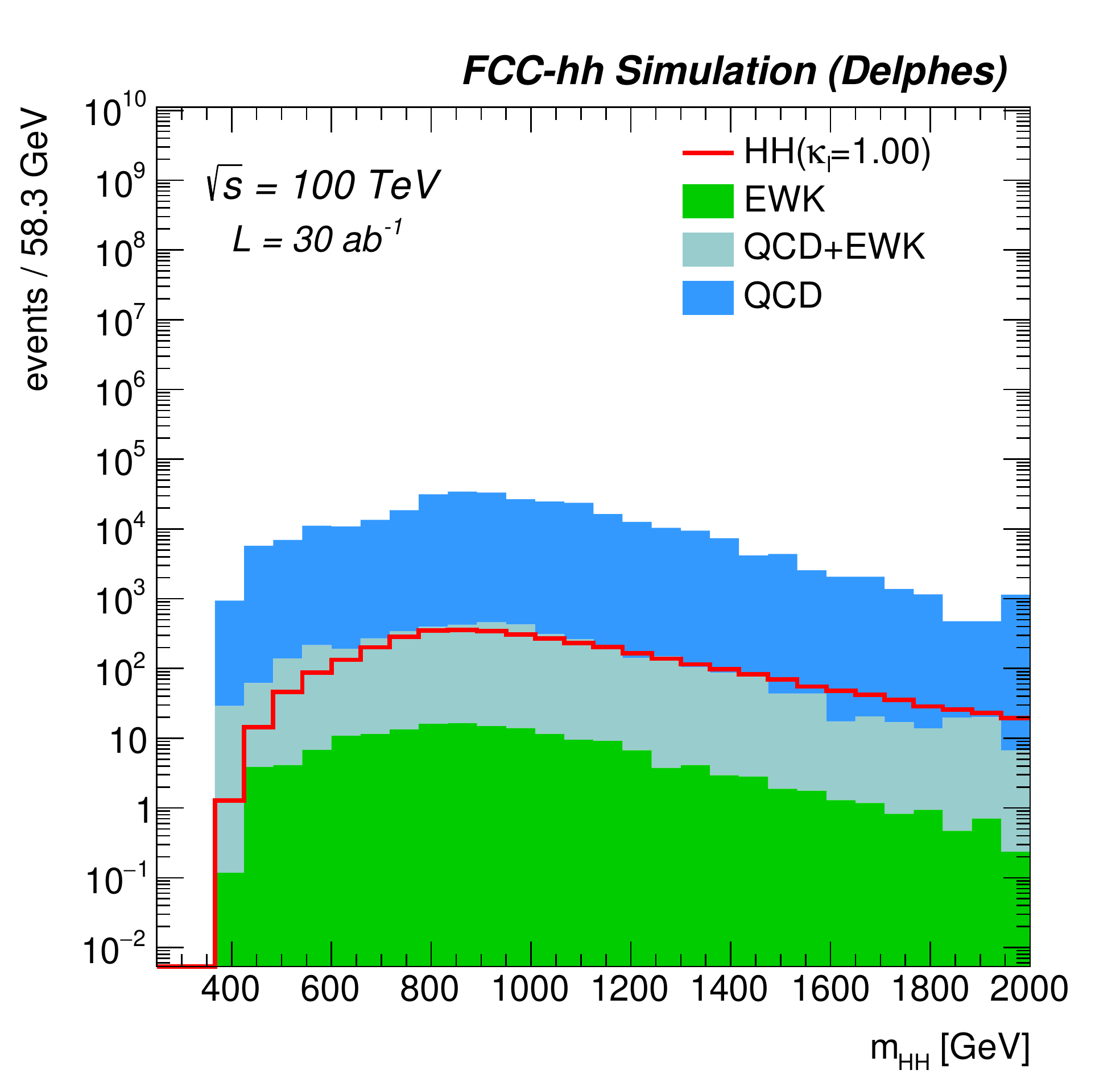}
  \includegraphics[width=0.45\columnwidth]{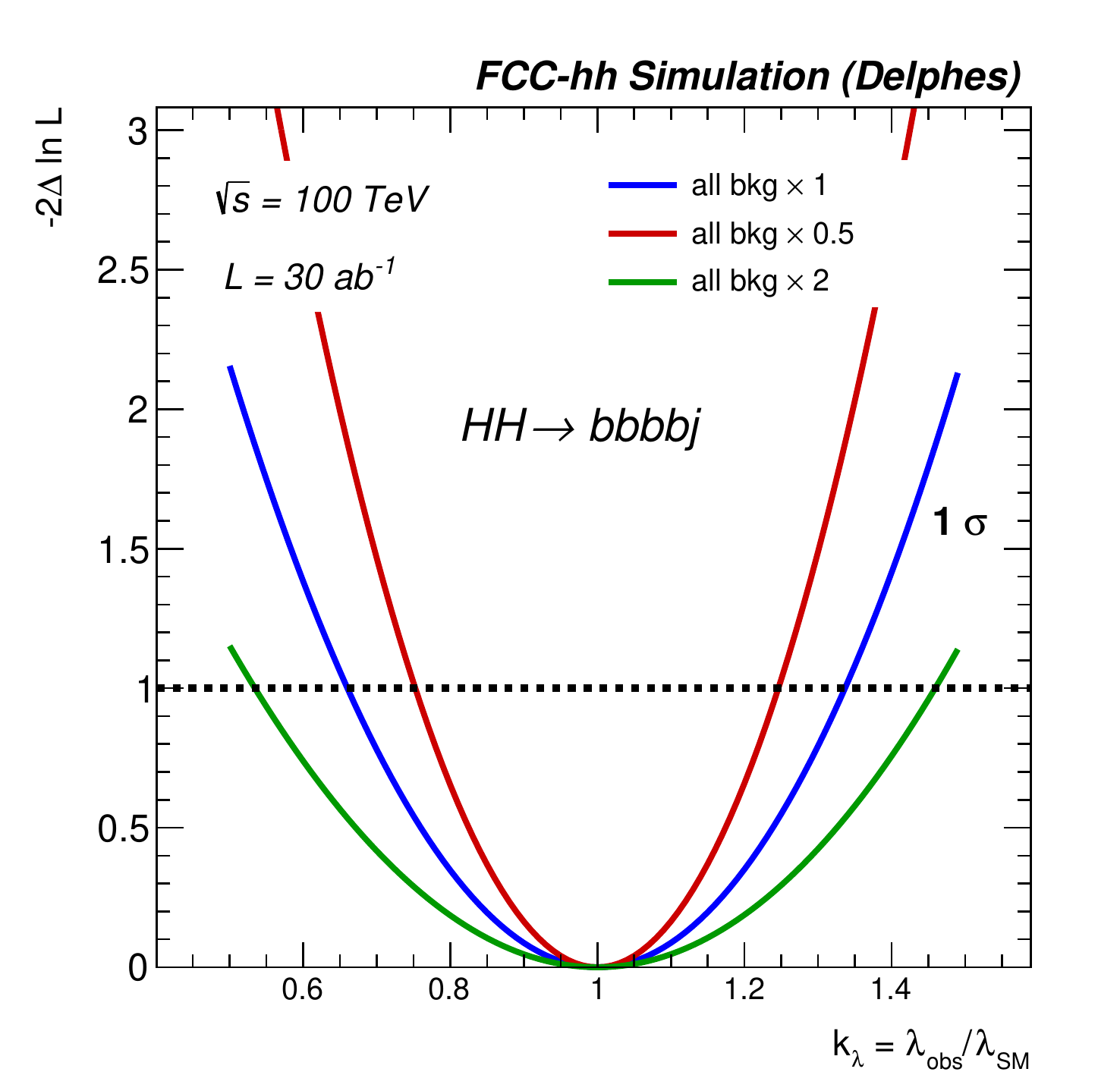}
  \caption{Left: Invariant mass spectrum of the di-Higgs pair constructed from the two large-radius jet Higgs candidates after the full event selection. Right: Expected precision on the Higgs self-coupling modifier $\kl$ assuming respectively $\times 1$, $\times 2$, $\times 0.5$ the nominal background yields.)}
  \label{hhboosted_2}
\end{figure}

Higgs jets are identified with standard boosted topologies techniques introduced in Chapter~\ref{sec:hbbbosted}. The N-subjettiness ratio $\tau_{2,1}$ observable~\cite{Thaler:2010tr} is shown in Fig.~\ref{hhboosted_1} (left) and the soft-drop mass $m_{SD}$ is shown in Fig.~\ref{hhboosted_1} (right). Higgs jets are tagged by selecting jets with $\tau_{2,1} < 0.35$ and $100 < m_{SD} < 130$ GeV, which yields signal tagging efficiency of 6\% and a background mis-identification rate of 0.1\%. Each of the two fat jets is required to be tagged. The signal extraction is performed via a one-dimensional likelihood fit on the di-Higgs mass observable \mhh, shown in Fig.~\ref{hhboosted_2} (left).

\subparagraph{Results and discussion}
\label{hhbosted_results}
The negative log-likelihood (NLL) distribution of the parameter $\kl$ is shown in Fig.~\ref{hhboosted_2} (right).
For the nominal detector and background yield assumptions we find an expected precision of the self-coupling of $\delta \kl =$~30\%. The uncertainty on the QCD background yield is parameterised by varying the overall normalisation by factors of 2 and 0.5 yielding an impact of the overall $\kl$ precision by $\approx \pm 10\%$.

This measurement can be improved by extending the analysis in the semi-resolved phase space region, where one Higgs is boosted and forms a fat-jet and the other is resolved, and into the fully resolved region with four \bjets in the final state.

\subsection[$HH\rightarrow\bbWW$]{$HH\rightarrow\bbWW$ \\ \contrib{B.~Di Micco}}



For the \bbWW decay mode, only the channel where one W boson decays hadronically and the other leptonically is considered. The dominant backgrounds are \ttbar
and multi-jet background, with smaller contributions from Drell-Yan and single top-quark production. Events are required to meet the following requirements: $\pT(WW)>150$~GeV, an invariant mass of the two \bjets system of $80 < \mbb < 180$~GeV and an angular distance between the two \bjets system of $\Delta R_{\bb}<$2.0.

The signal selection is optimised using a BDT. The input variables used by the BDT are the leptons, jets and neutrino 4-momenta as well as the azimuthal angular distance between various objects. The BDT is trained to discriminate the signal from the dominant background \ttbar. The event selection on the BDT output score has been optimised to ensure a high $S/\sqrt{B}$ ratio (where S is the number of signal events and B the number of background events after the full event selection).

\begin{figure}
	\begin{center}
		\includegraphics*[width=0.4\textwidth]{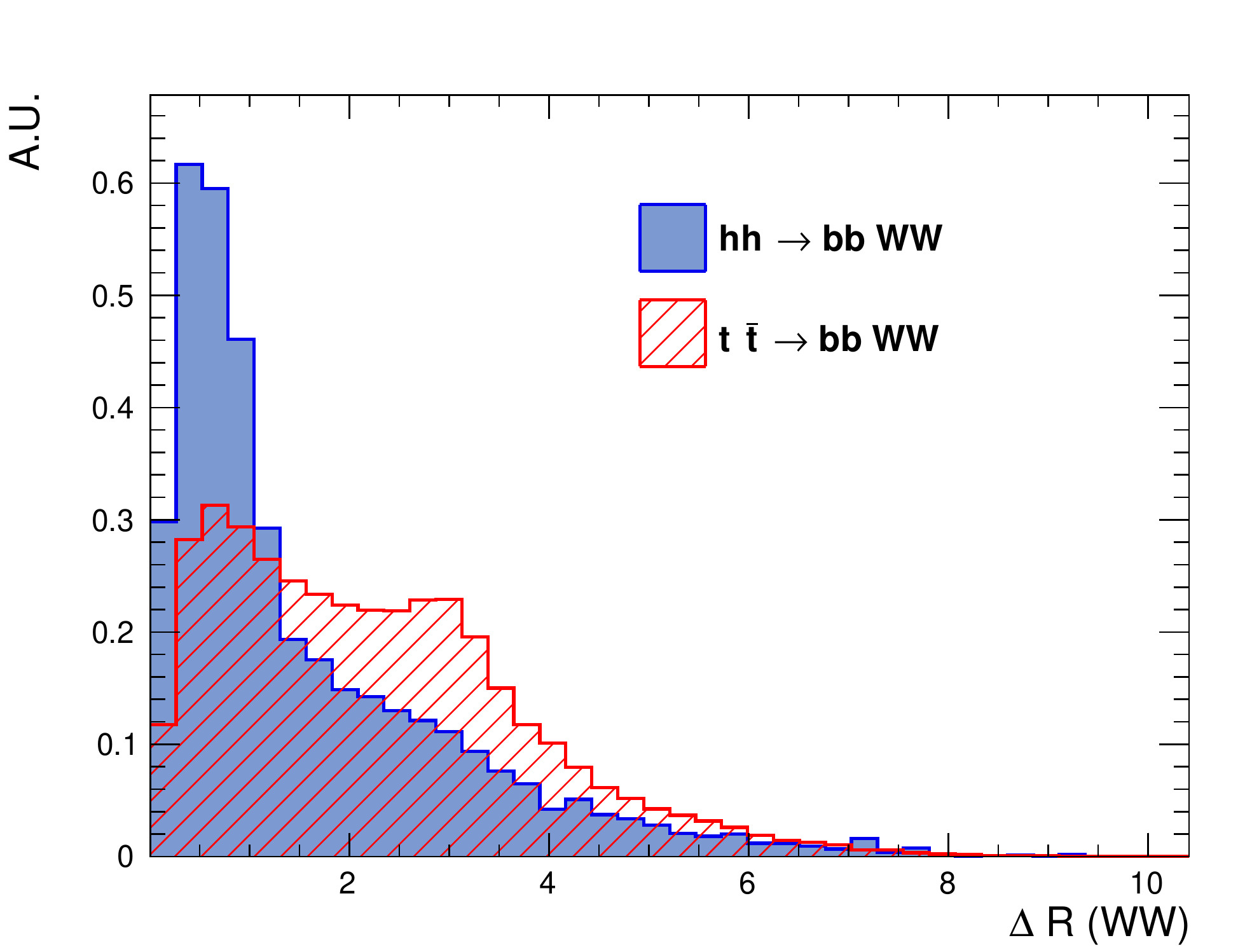}
        \includegraphics*[width=0.4\textwidth]{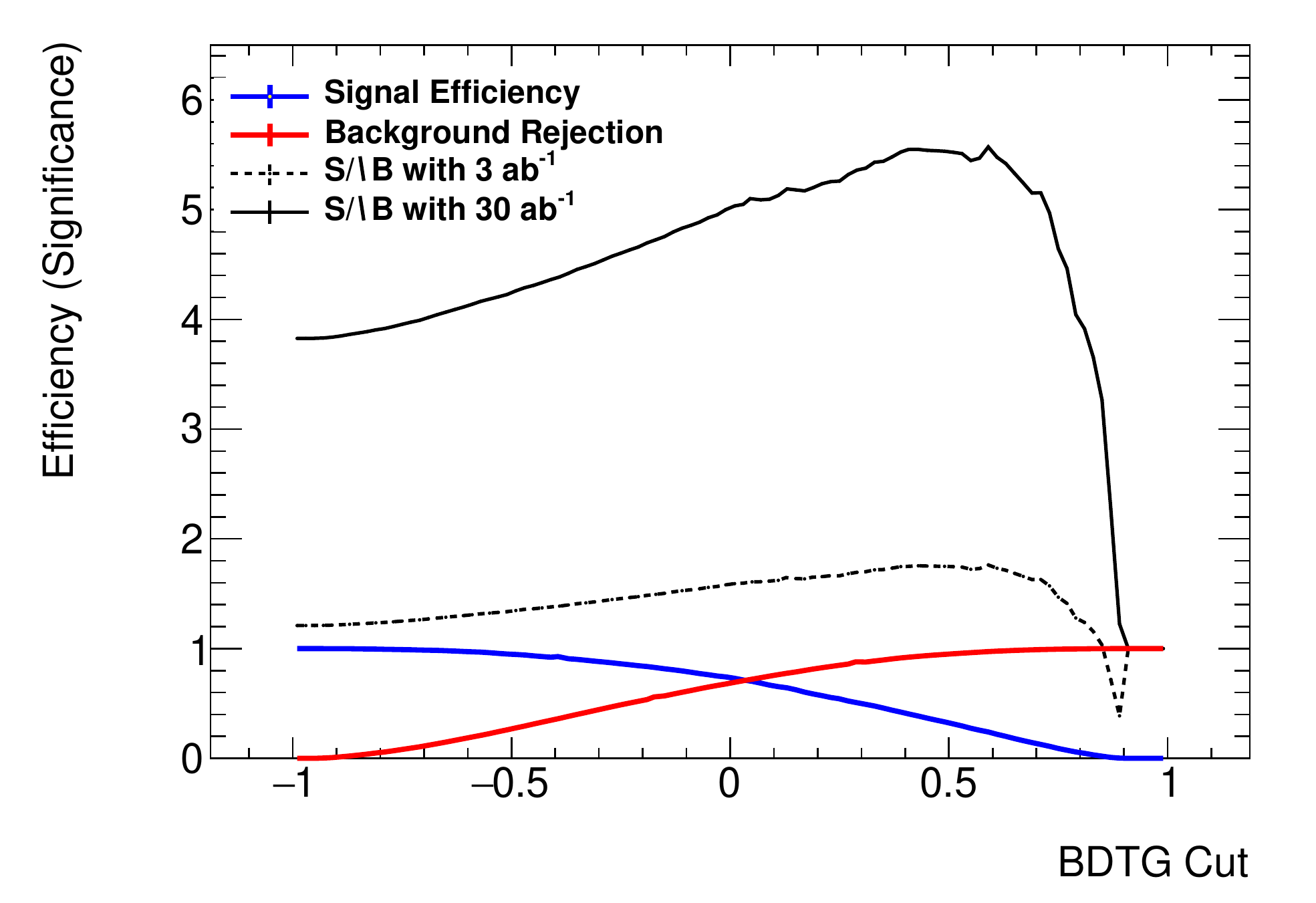}

	\end{center}
	\caption{\label{fig:bbWW_input} Left: The distribution of the most discriminant variables used in the BDT training for discriminating signal and background samples: the $\Delta R$ between the two $W's$. Right: The BDT efficiency and significance as a function of the applied cut on the BDT response for two reference integrated luminosity values: 3 ab$^{-1}$ and 30 ab$^{-1}$.}
\end{figure}

An example of an input distribution used in the BDT is shown in Fig.~\ref{fig:bbWW_input} (a), which is the angular separation between the two W bosons. The output BDT distribution for the signal and background is shown in Fig.~\ref{fig:bbWW_input} (b). With \intlumifcc, a significance of 5$\sigma$ can be achieved using the \bbWW decay mode, corresponding to a precision of $\delta \kl =$~40\%

\subsection{Summary of 100 TeV studies}
\label{sec:summary_HH100tev}

Reference~\cite{Banerjee:2018yxy} proposed using a boosted \hh final state to enhance the self-coupling sensitivity in the case of the \bbtt final state, following the approach discussed in Sec.~\ref{subsec:bbbbj}. A precision $\delta \klambda = 8\%$ can obtained at 68\%CL in this decay mode.

Reference~\cite{Goncalves:2018yva} performs a kinematic analysis of various \hh distributions in the \bbyy final state, considering quantities such as the invariant mass \mhh, the Higgs \pT and various angular correlations. The projected 1$\sigma$ sensitivity at 100~TeV (\intlumifcc) is found to be $\delta \klambda =$ 5\%, consistent with the results of the FCC-hh detector performance study, and with previous studies found in the literature~\cite{Contino:2016spe,Chang:2018uwu}.

A summary of the target precision in the measurement of $\klambda$ is given in \refta{tab:kappalambda}. Within the stated assumptions on the expected performance of the FCC-hh detector, a precision target on the Higgs self-coupling of $\delta \kl =$~5\% appears achievable, by exploiting several techniques and decay modes, and assuming the future theoretical progress in modelling signals and backgrounds.

\begin{table}[th]
\renewcommand{\arraystretch}{1.5}
 \begin{center}
   \begin{tabular}{l|c|c|c|c|c}
  \hline
 & \bbyy
 & \bbtt
 & \bbzz($4\ell$)
 & \bbWW(2j$\ell\nu$)
 & \bbbb+jet   \\ \hline
 $\delta\klambda$
 & 6\%
 & 8\%
 & 14\%
 & 40\%
 & 30\%
 \\
\hline 
\end{tabular}
\caption{\label{tab:kappalambda}
Precision of the direct Higgs self-coupling measurement in $gg\to$\hh production at \sqrtsfcc\ with \intlumifcc\ for various decay modes.}
\end{center}
\end{table}

\section{Other Probes of Multi-Linear Higgs Interactions}
\label{sec:other_probes}

\paragraph{The quartic coupling}
\contrib{F.~Maltoni, D.~Pagani,  A.~Shivaji, X.~Zhao}

\begin{figure}[tb]
\centering
\includegraphics[width=0.8 \textwidth]{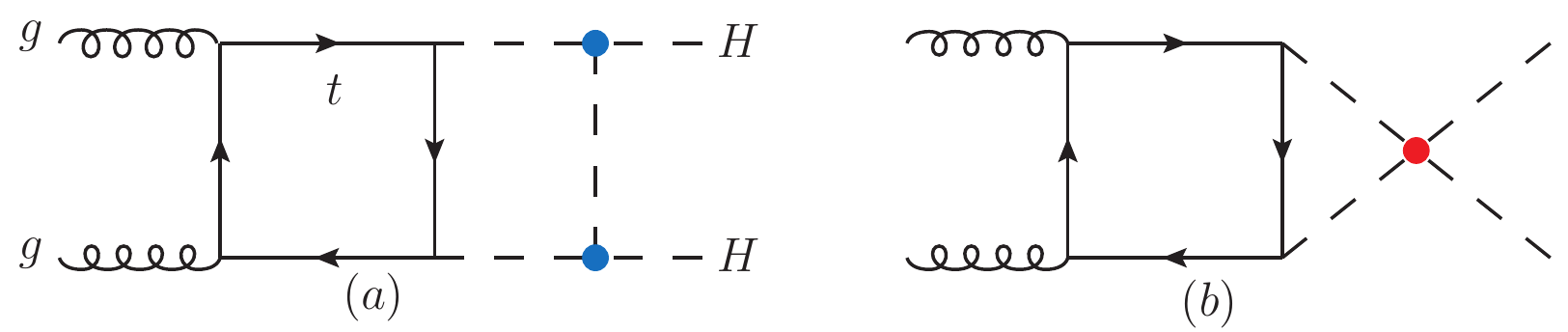}
\caption{Examples of two-loop diagrams contributing to double Higgs production.}
\label{fig:HH1L}
\end{figure}

At hadron colliders, di-Higgs boson production provides a direct access to the Higgs cubic self-coupling while the Higgs quartic self-coupling can be in principle directly probed through triple Higgs production \cite{Plehn:2005nk,Fuks:2015hna,Chen:2015gva,Fuks:2017zkg,Kilian:2017nio,Papaefstathiou:2019ofh}.
On the other hand, also di-Higgs production is sensitive to the Higgs quartic coupling via EW corrections; its measurement can thus provide an alternative way to constrain the quartic coupling indirectly (see Fig.~\ref{fig:HH1L}). The combined constraints that can be achieved at a future 100 TeV collider on the trilinear and quartic coupling for the case of gluon-gluon fusion, based on the results of Ref.~\cite{Borowka:2018pxx} is presented in what follows~\footnote{A similar study has also appeared in Ref.~\cite{Bizon:2018syu}. Differences among these two studies are commented in Ref.~\cite{Borowka:2018pxx}.}. This study relies on the theoretical framework (in particular the renormalization procedure) introduced in~\cite{Maltoni:2018ttu} and extends the idea of probing the trilinear Higgs self coupling ($\lambda_3$) via precise single Higgs measurement~\cite{McCullough:2013rea, Gorbahn:2016uoy,Degrassi:2016wml, Bizon:2016wgr, Maltoni:2017ims} to the case of the quartic ($\lambda_4$) and di-Higgs production.

We consider the loop corrections to di-Higgs production through gluon-gluon fusion in the EFT framework, taking into account both the $c_6$ and $c_8$ dependence in order to independently vary the  cubic and quartic self-couplings. Indeed, in our framework, $c_6$ and $c_8$ can be directly linked to the self couplings via the relations $\kappa_3\equiv\frac{\lambda_3}{\lambda_3^{\rm SM}} =  1+ c_6$  and $\kappa_4\equiv\frac{\lambda_4}{\lambda_4^{\rm SM}} =  1+ 6 c_6 + c_8 $ (see Equation~\ref{VNP})~\footnote{It should to note that $\kappa_3$ and $\kappa_4$ can also be affected by the $c_H$ coefficient as in Equation~\ref{eq:klrel} and neglected here.}.
Our phenomenological predictions are based on the following approximation $\sigma^{\rm pheno}_{\rm NLO}$ for the cross section:
\begin{align}
 \sigma^{\rm pheno}_{\rm NLO}  =& \sigma_{\rm LO} + \Delta\sigma_{c_6}+ \Delta\sigma_{ c_8}\,
\end{align}

where the quantity $\sigma_{\textrm{LO}}$ is the LO cross section, and $\Delta\sigma_{c_8}$ captures all $c_8$ contribution at NLO and $\Delta\sigma_{ c_6}$ corresponds to the leading NLO contribution from $c_6$ for large values of $c_6$, or equivalently, large values of $\lambda_3$.

We consider here only the case of 100 TeV; results for HL-LHC can be found in ref.~\cite{Borowka:2018pxx}. The signal extraction is performed in the \bbyy channel  assuming an integrated luminosity \intlumifcc\ via a fit on the \mhh distribution, following the approach described in Ref.~\cite{Azatov:2015oxa}.

\begin{figure}[tb]
	\centering
	\includegraphics[width=0.60\textwidth]{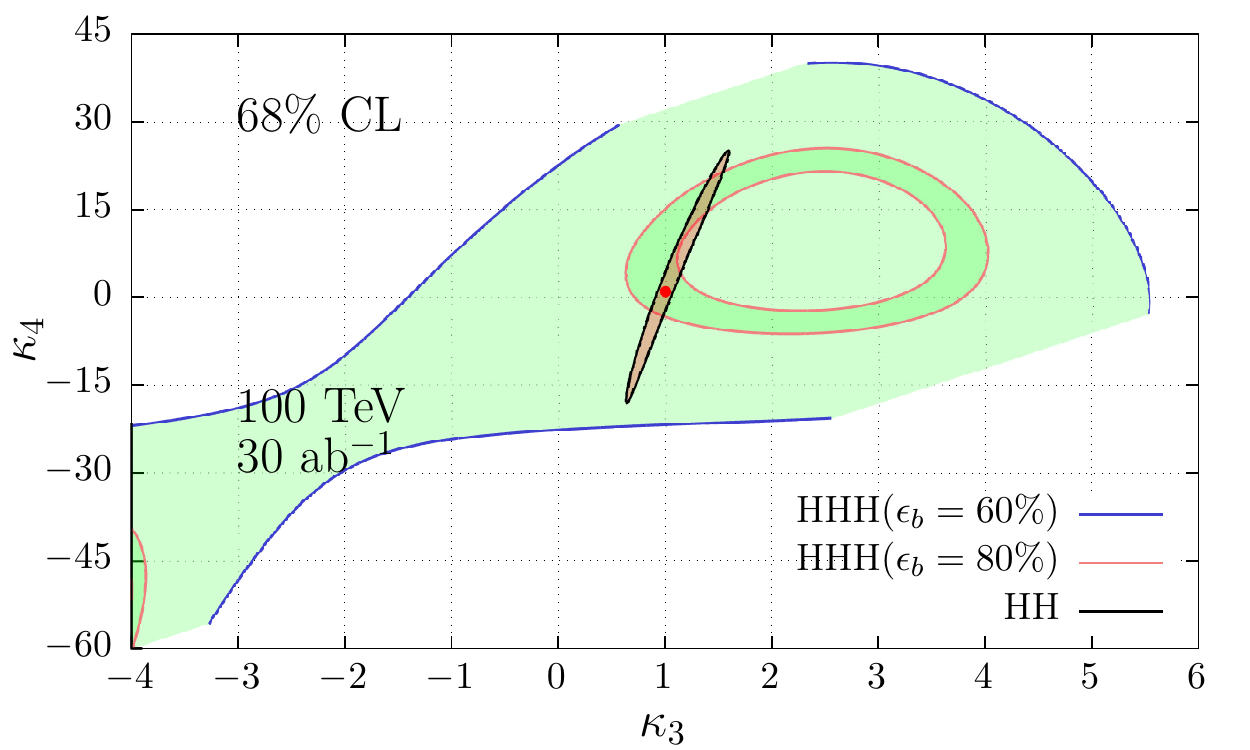}
	\caption{Comparison between expected $1\sigma$-bounds in the ($\kappa_3,\kappa_4$) plane obtained from double Higgs (indirect) and triple Higgs (direct) production. \label{fig:triple}}
\end{figure}

In Fig.~\ref{fig:triple}, we show the constraints that can be obtained on $\kappa_3$ and $\kappa_4$ (assuming the SM case) employing the indirect method described here. For comparison we also include what can be achieved via direct triple Higgs production. For the latter, we have followed the approach presented in Ref.~\cite{Papaefstathiou:2015paa, Contino:2016spe} based on the $ \bbbb \gamma\gamma$ signature, assuming an optimistic (80\%) and a conservative (60\%) scenarios on the $b$-tagging efficiency. As can be seen in Fig.~\ref{fig:triple}, the bounds obtained from triple Higgs production strongly depend on the $b$-tagging efficiency.

Within the conservative scenario the bounds obtained from double Higgs are stronger than those obtained with the conservative assumption in the triple Higgs analysis. In particular, for the Higgs quartic interaction we find that:
\begin{align}
\kappa_4 \in \left[-2.3, 4.3\right] \,\,\, \mathrm{at \,\, 68\% CL}
\end{align}
We stress however that double and triple Higgs production provide complementary constraints and their combination can be used to improve the bounds on the $(\kappa_3,\kappa_4)$ plane.

\paragraph{The $HHVV$ coupling}

Given the rates shown in \refta{table:xsec2-future}, the next process of interest for the production of Higgs pairs is vector boson fusion.
A study of this process focusing on the sensitivity to higher-dimension operators at the HL-LHC, was presented in Ref.~\cite{Bishara:2016kjn} and already discussed in Sec.~\ref{sec_exp_2dot9}. We simply remind here that the \mhh differential observable is a powerful probe of the gauge structure of the Higgs sector. The \mhh distribution is reconstructed in the \hhbbbb final state.

\begin{figure}[tb]
\begin{center}
\includegraphics*[width=0.48\textwidth]{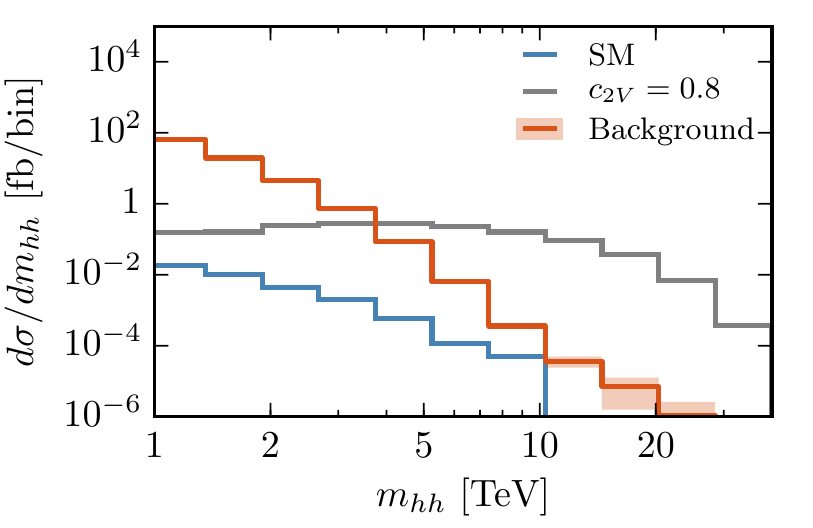}
\includegraphics*[width=0.48\textwidth]{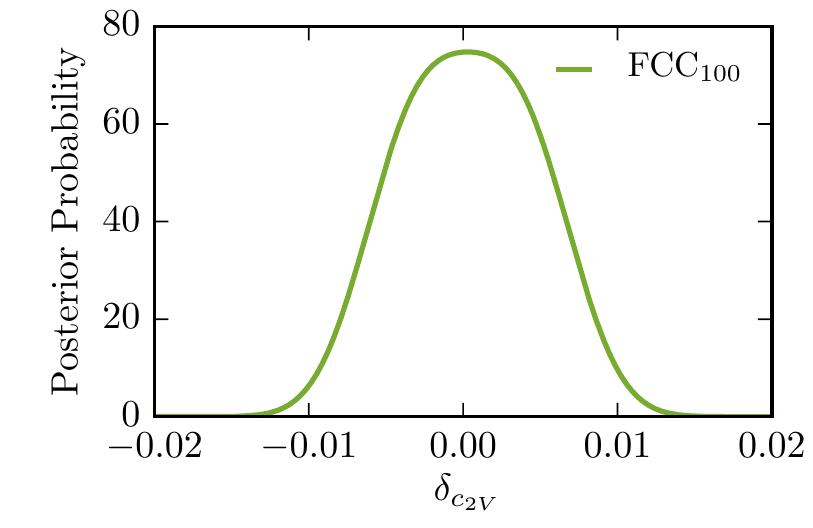}
\end{center}
\vspace*{-0.5cm}
\caption{\label{fig:vbfHH2} Distribution of the $m_{HH}$ (left) and posterior probability on the determination of $\delta_{c_{2V}}$ at the FCC-hh (see Equation~\ref{eq:delta_c}). }
\end{figure}

Boosted-jet tagging techniques -- justified by the high \pT of the Higgs bosons in the relevant kinematic region -- have been applied to minimise the dominant backgrounds (4b, 2b2j, t\={t}2j, Hjj). An example of the impact of $\delta_{c_{2V}} = c_{2V}^2-c_V \ne 0$ (as defined in Sec.~\ref{sec_exp_2dot9}, Eq.~\ref{eq:delta_c} as $\delta_c$) is shown in Fig.~\ref{fig:vbfHH2}, where the di-Higgs mass spectrum in the SM and in a $c_V=1$, $c_{2V}=0.8$ scenario are compared to the expected backgrounds (in the parton-level simulation). A study of the shape of the mass distribution results in the probability density distribution shown in the right plot of Fig.~\ref{fig:vbfHH2}. Several robustness tests have been performed, including assigning large uncertainties on the background rates. The signal itself is already known with a precision at level of few percent (see \refta{table:xsec2-future}). Since $c_V$ will be measured with a few per-mille  precision at FCC-ee (independently of whether it agrees or not with the SM), and given that the cubic Higgs self-coupling contribution is suppressed at the multi-TeV mass values considered in this analysis, the constraints on $\delta_{c_{2V}}$ at FCC-hh will translate directly into a constraint on $c_{2V}$ of better than $\pm 1\%$ which constitutes a large improvement compared to the 40\% precision that can be obtained at the HL-LHC.  

\paragraph{The \ttbar\hh coupling}
\contrib{S.~Banerjee, F.~Krauss, M.~Spannowsky}

\begin{figure}[tb]
\centering
\includegraphics[scale=0.4]{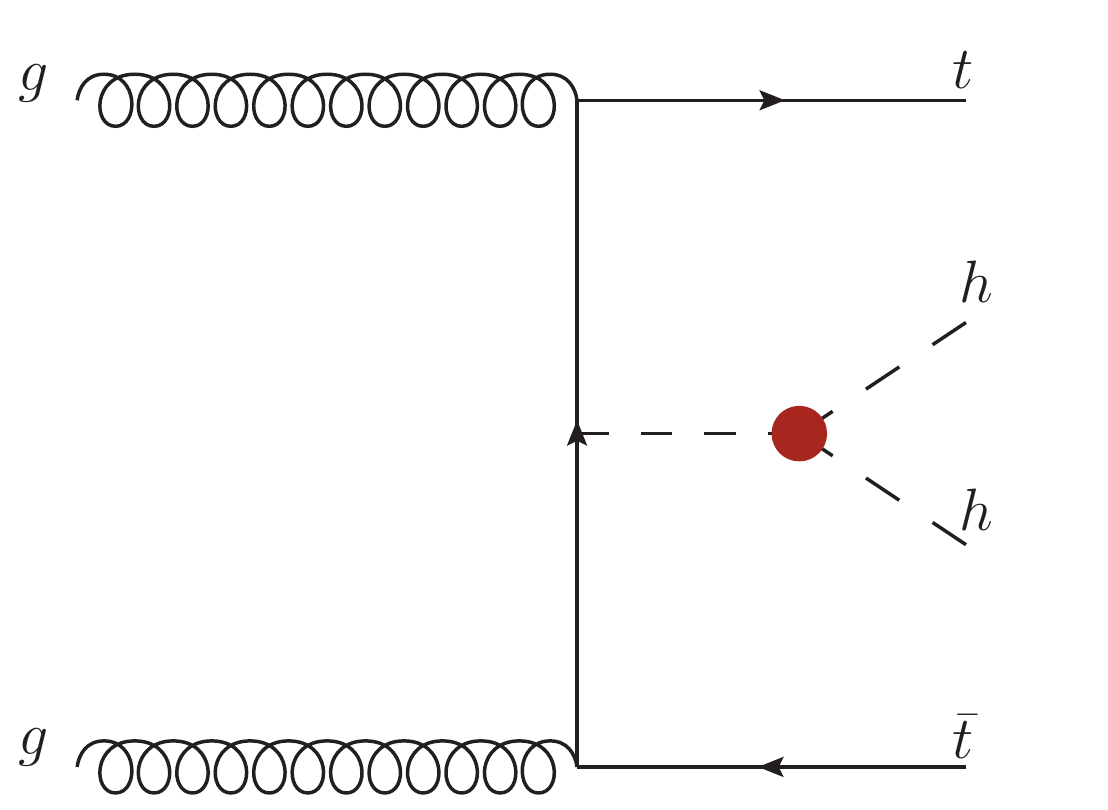}~~~\includegraphics[scale=0.4]{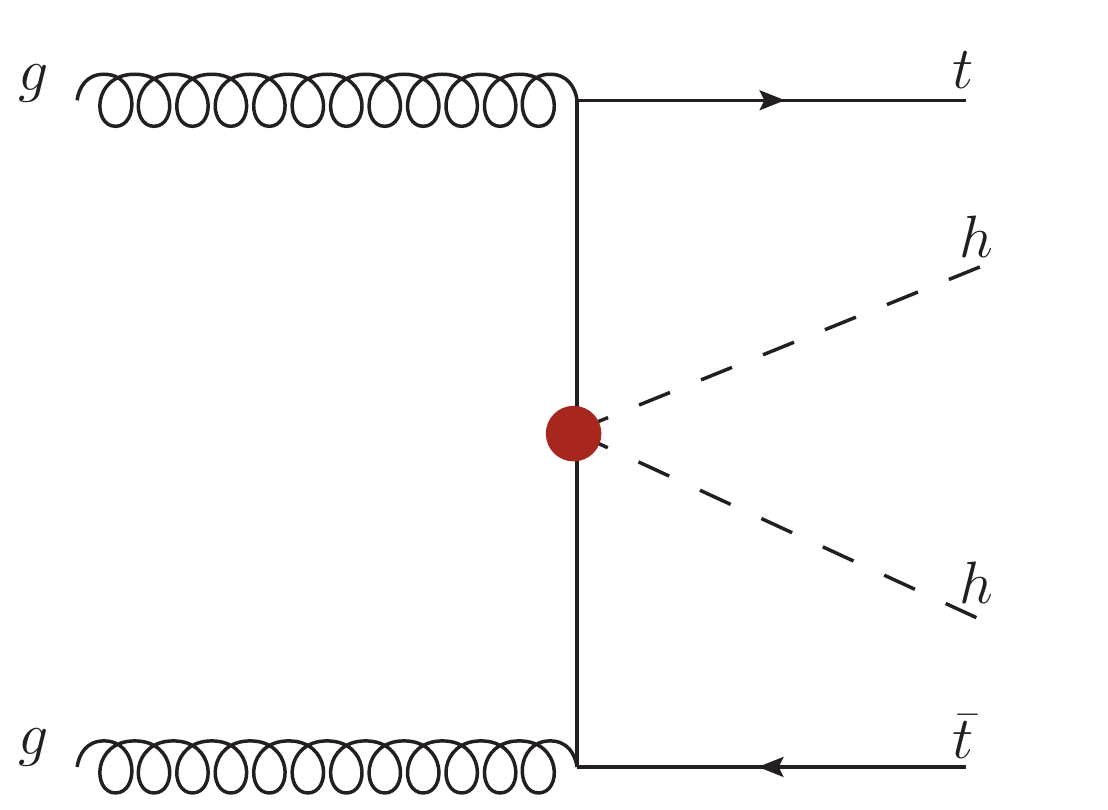}~~~\includegraphics[scale=0.4]{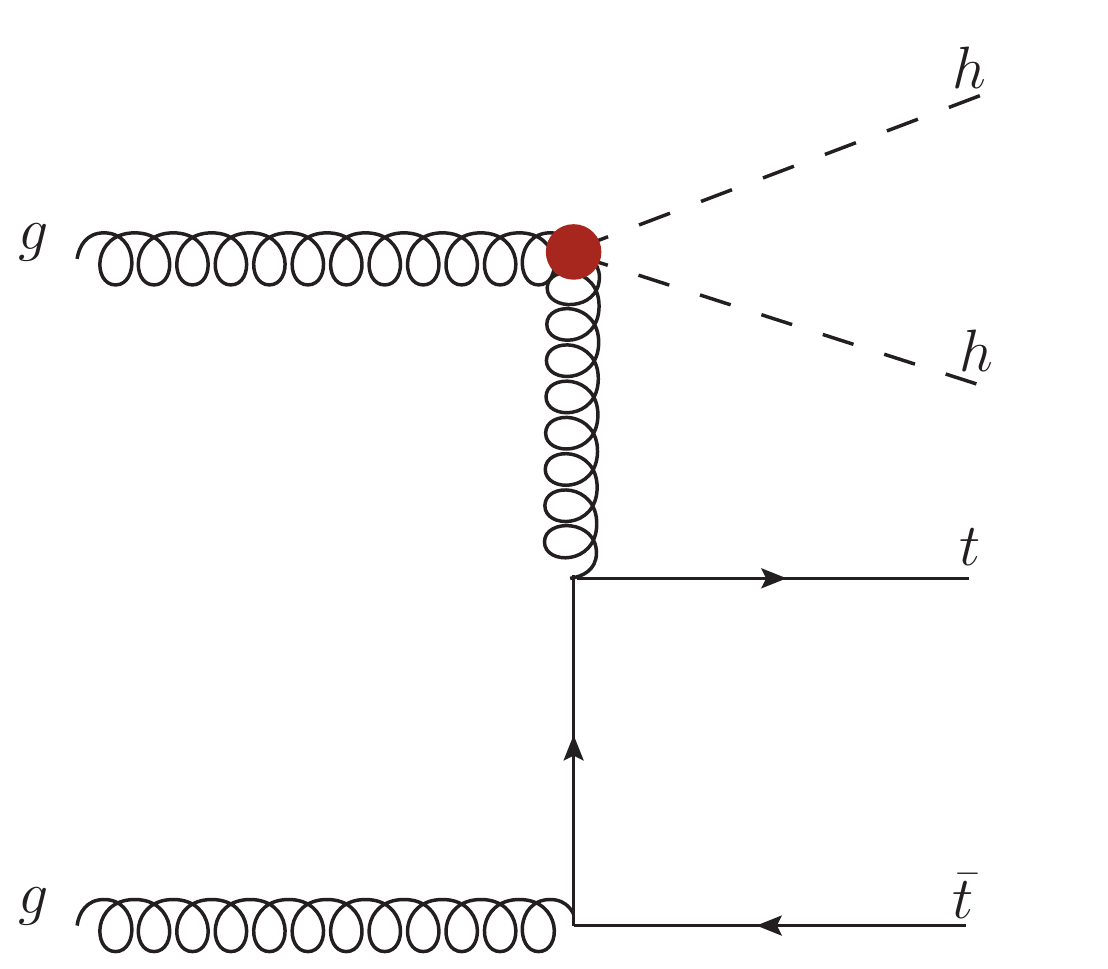}
\caption{Feynman diagrams showing the impact of the three effective vertices, \textit{i.e.}, $HHH$, \ttbar\hh and $gg$\hh.}
\label{fig:Feynman}
\end{figure}

    The \ttbar\hh four-point vertex is often neglected as it is not present in the SM. However, this vertex is accessible upon including the dimension 6 operator~\cite{Englert:2014uqa, Liu:2014rva, He:2015spf, Banerjee:2019jys}. This vertex arises for example when considering the non-linear realisation of the $SU(2) \times U(1)$ symmetry~\cite{Feruglio:1992wf, Bagger:1993zf, Koulovassilopoulos:1993pw, Buchalla:2012qq, Buchalla:2013rka}, \textit{i.e.}, defined in Sec.~\ref{sec:heft}, Eq.~\ref{eq:ewchl}. While in a linearised realisation the $\ttbar \rm{H}$ and the \ttbar\hh vertices ($c_t$ and $c_{tt}$) are correlated, one may probe these couplings independently in purview of the non-linear EFT formalism. In Fig.~\ref{fig:Feynman} we show the various possible vertex deformations for the $pp \to \ttbar\hh $ channel.

\begin{figure}[tb]
\centering
\includegraphics[scale=0.4]{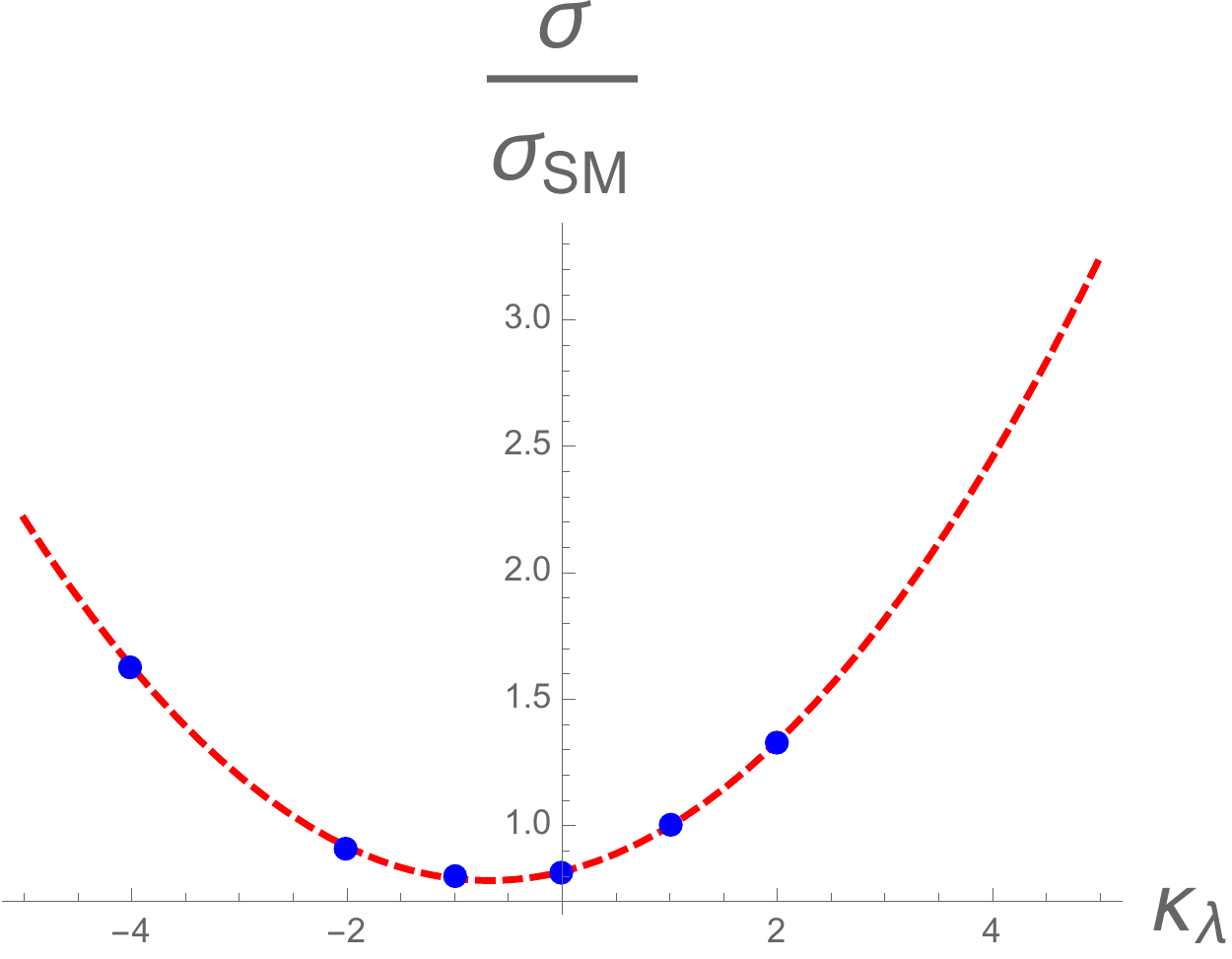}~~~~\includegraphics[scale=0.35]{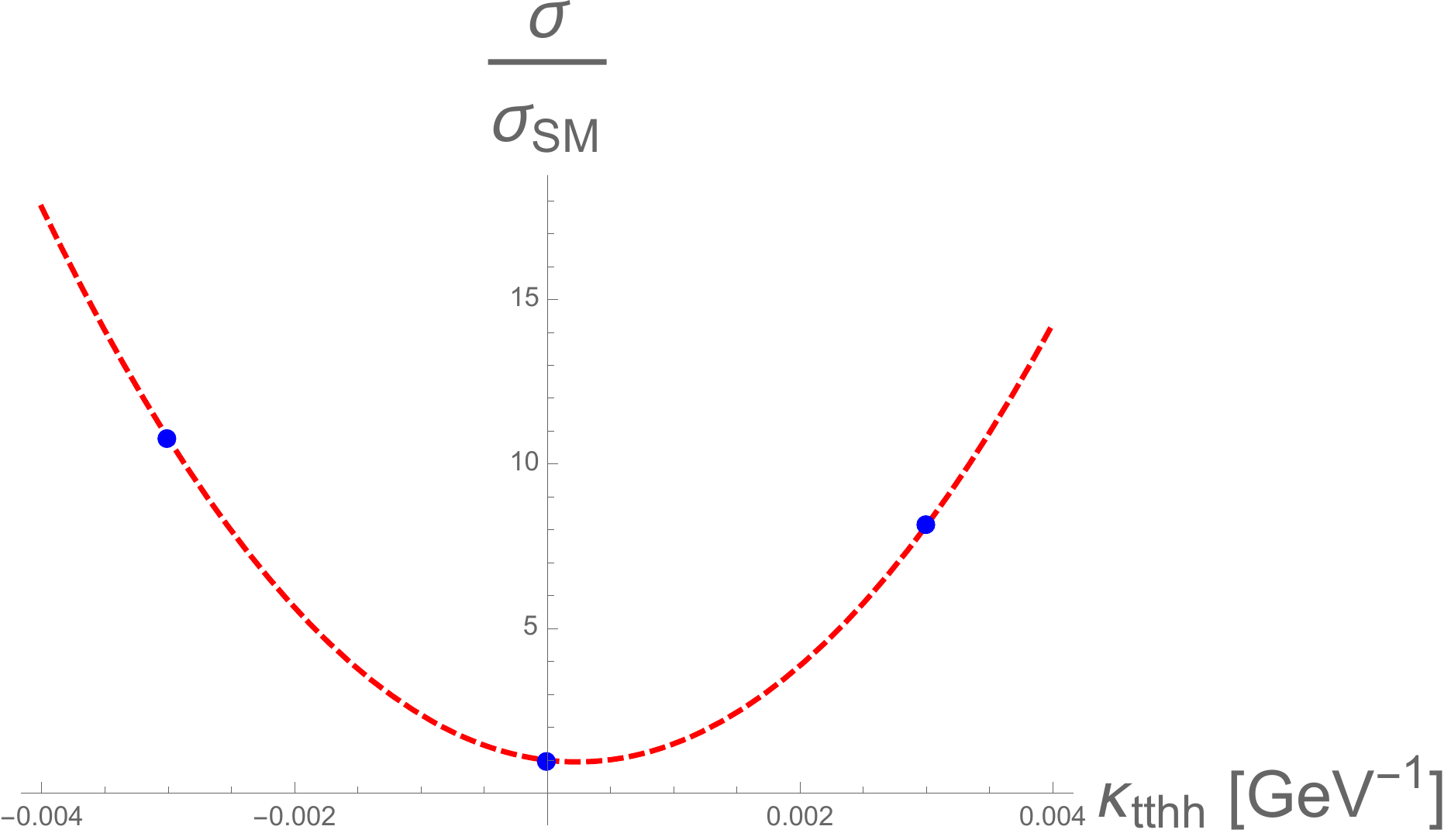}
\caption{$\sigma/\sigma_{\textrm{SM}}$ as a function of $\klambda$ (left) and $\kappa_{\ttbar\hh}$ [GeV$^{-1}$] (right), where $\kappa_{\ttbar\hh} = - m_t c_{tt} / v^2$.}
\label{fig:signalXS}
\end{figure}

In this section, without varying the $G^{a}_{\mu \nu } G^{\mu \nu}_a h h$, $HHH$ and \ttbar$\rm{H}$ vertices we want to see how far the \ttbar\hh coupling can be constrained at \sqrtsfcc. Unlike the other di-Higgs processes, this channel shows a growth in cross section for $|c_{tt}| > 0$ or for $|\klambda| > 0$ (see Fig.~\ref{fig:signalXS}). The \ttbar\hh cross section increases by a factor of $\sim 75$ upon going from the 14 TeV to the 100 TeV machine (see \refta{table:xsec2-future}) .

Here we consider the final state comprised of six \btagged jets, one isolated lepton ($e,\mu$), at least two light jets and $\slashed{E}_T$. We employ the technique outlined in Refs.~\cite{Englert:2014uqa, Banerjee:2019jys} to reconstruct the two Higgs bosons and the hadronic top. There are several backgrounds at play, \textit{i.e.}, $\ttbar ZZ$, $\ttbar HZ$ ($b$-quarks coming from $Z$/$H$ decays), $\ttbar H \bb$, $\ttbar Z \bb$, $\ttbar\bb\bb$ ($b$-quarks produced through gluon splitting in QCD) and $W$ plus four \bjets where $W$ decays leptonically. Besides, there are fake sub-dominant backgrounds, \textit{e.g.}, $\ttbar c\bar{c}c\bar{c}$ and $W^{\pm}c\bar{c}c\bar{c}$ ($c$ mis-identified as $b$) or $\ttbar\bb$, $\ttbar H$, $\ttbar Z$, and W$^{\pm}\bb$ associated with light jets. For the various scale choices and the details of the analysis, we refer the readers to Ref.~\cite{Banerjee:2019jys}. A log-likelihood CLs hypothesis test considering the SM as the null hypothesis, assuming $\klambda = c_t = 1$, and zero systematic uncertainties gives at 68\% CL:

\begin{eqnarray}
-0.24 \; \textrm{TeV}^{-1} < \kappa_{t\bar{t}hh} < 0.60 \; \textrm{TeV}^{-1} \; \; \; \; \textrm{30/ab}.
\end{eqnarray}

where $\kappa_{\ttbar\hh} = - m_t c_{tt} / v^2$.

%% file: HHFuture/intro_future.tex
\contrib{M.~E.~Peskin, M. Selvaggi}

\noindent Future colliders will play a key role in the measurement of the Higgs self-coupling and the investigation of the nature of the Higgs potential. There are two methods to probe the Higgs self-coupling, using the measurement of $HH$ production --  requiring parton centre of mass energy sufficiently far above the di-Higgs production threshold to provide a useful event sample -- and using a global fit to single Higgs measurements -- requiring a high level of control over other new physics effects that contribute to the relevant observables. 

In the past Chapters, we have evaluated how these methods can be used at proposed future hadron and $e^{+}e^{-}$ colliders. To summarise the capabilities of the various colliders we adopt the scheme introduced in Sec.~\ref{sec:BSMsummary}. The estimates of the precision on the self-coupling achievable at the various future colliders is presented in  Table~\ref{tab:future_comp}.

\begin{table}[tb]
\centering
{\small
\begin{tabular}{ c  c  c  c }
\hline 
collider       &  single-$H$ &  $HH$   & combined   \\
\hline
HL-LHC         &  100-200\%    & 50\%    & 50\%   \\  \hline
CEPC$_{240} $          & 49\%    & $-$    &49\%  \\
ILC$_{250}$    &  49\%    & $-$   &49\%    \\
ILC$_{500}$    &  38\%      & 27\%   &22\%  \\
ILC$_{1000}$    &  36\%    & 10\%    &10\%  \\
CLIC$_{380}$   & 50\%     & $-$    &50\%  \\
CLIC$_{1500}$  &  49\%   & 36\%  & 29\%  \\
CLIC$_{3000}$  &  49\%  & 9\% &  9\% \\
FCC-ee  &  33\%   & $-$    &33\%   \\
FCC-ee (4 IPs) &  24\%   & $-$    &24\%  \\  \hline
HE-LHC         &   -   & 15\%    & 15\%   \\
FCC-hh         &  -  & 5\%  & 5\%  \\
\hline
\end{tabular}
}
\caption{\label{tab:h3} Sensitivity at 68\% probability on the Higgs cubic self-coupling at the various future colliders, as discussed in Chapters 8--10. Values for single Higgs determinations below the first line are taken from \cite{deBlas:2019rxi}.  These values are quote here as combined with an independent determination of the self-coupling with uncertainty 50\% from the HL-LHC.  Please see the discussion in the text on the interpretation of this table. }
\label{tab:future_comp}
\end{table}

By the end of Run 3 in 2024, the LHC will have collected, by combining the ATLAS and CMS dataset, around 600\ifb of integrated luminosity. Naive extrapolations of current results~\cite{Sirunyan:2018two,Aad:2019uzh} (see Table \ref{exp-summary-table}) indicate that double Higgs production as predicted by the SM will not be observed even with the Run 3 dataset. Assuming current detector performance, it will be possible to set an upper limit on the di-Higgs production cross-section of 1-3 times the SM value at 95 \% CL at best. According to our convention, such sensitivity would qualify as a \emph{bronze} type measurement (see sec.~\ref{sec:BSMsummary}). A measurement of the Higgs self-coupling is thus out of reach of Run 3 and requires either a larger dataset, or/and a higher collision energy. 

The HL-LHC will collide protons at 14 TeV (which constitutes a moderate although non-negligible increase in centre of mass energy with respect to 13 TeV at current LHC), and is expected to produce an integrated luminosity of 3~$\mathrm{ab}^{-1}$ per interaction point.

Such a large increase in the luminosity will allow for the milestone observation of double Higgs production at 5$\sigma$.  This would correspond to observation at the 95\%CL that the Higgs self-coupling is nonzero.   Still,  the corresponding precision on the Higgs self-coupling will be only of order 50\%, barely approaching the~\emph{silver} level of precision. This measurement will be largely driven by the measurement of $HH$ production.  The projections for individual decay channels and their combination, including indirect self-coupling constraints from single Higgs production have been summarised in Chapter~\ref{chap:hl-lhc}. 

The goal for future machines beyond the HL-LHC should be to probe the Higgs potential quantitatively.  This requires at least \emph{gold} quality precision for the self-coupling parameter. Such level of precision is achievable through the measurement of  $HH$ production at the highest energy lepton machines (ILC$_{1000}$ or CLIC$_{3000}$) and hadron machines (FCC-hh). 

The proposed $e^+e^-$ Higgs factories---CEPC, ILC, CLIC, and FCC-ee---can access the Higgs self-coupling through analysis of single Higgs measurements.  This relies on the fact that these colliders will measure a large number of individual single Higgs reactions with high precision, allowing a highly model-independent analysis of possible new physics contributions.   It will  be important to have data at two different CM energies to reach the silver level of precision. This requires reaching the second stage of a staged run plan: 365~GeV for FCC-ee, 500~GeV for ILC, 1.5~TeV for CLIC.  Running beyond 240~GeV is not in the  CEPC baseline plan. It should be  added to achieve a competitive result.  For FCC-ee, running with 4 IPs has been considered to increase the data set and reach a precision of 24\% on the Higgs self-coupling.   All of these points have been 
reviewed in Chapter~\ref{chap:epem}.

In Chapter~\ref{chap:helhc-fcchh} we have reviewed the prospects for future energy hadron colliders beyond LHC, in particular, the High Energy LHC (27 TeV) and the FCC-hh (100 TeV).  These machines are also planned to  produce higher luminosities than the HL-LHC.  The studies reported in Chapter~\ref{chap:helhc-fcchh} have indicated that respectively 5\% (FCC-hh) and 15\% (HE-LHC) precision on the Higgs self-coupling are within reach at those machines, based on the  method of measuring the $HH$ production cross section.

Some caution is necessary in directly comparing the numbers given in Table~\ref{tab:h3}.
The values for the single Higgs method given in the lines below HL-LHC are combined with the HL-LHC projected error of 50\%~ \cite{deBlas:2019rxi}.  Thus, only values well below 50\% represent a significant improvement.  The various estimates in the table are computed using different assumptions on the inclusion of SMEFT parameters representing other new physics effects.  We have tried to clarify these in the discussions of the individual analyses. In particular, many of the numbers from $HH$ production  are derived from fits including the single parameter $\kappa_\lambda$ only. At $e^+e^-$ colliders it is more straightforward to simulate the relevant backgrounds, but we have less experience with the high-energy regime studied here.  The uncertainties in the direct determinations at $e^+e^-$ colliders are computed using full-simulation analyses based on current analysis methods.   These have much room for improvement when the actual data is available.  The analyses at hadron colliders are based on estimates of the achievable  detector performance in the presence of very high pileup.  These are extrapolations, but the estimates are consistent with the improvements in analysis methods that we have seen already at the LHC. 

Despite the uncertainties, it is clear that the highest-energy $e^+e^-$ and hadron colliders can achieve the {\it gold} level of precision  set out in Sec.~\ref{sec:BSMsummary}.  With new resources, and with patient improvement of the experimental state of the art, we will achieve an excellent understanding of the underlying physics of spontaneous symmetry breaking generated by the Higgs boson.

%% file: acknowledgements.tex
The editors would like to thank all the authors for their contribution to this report, and in particular to Christophe Grojean and Michael Peskin for fruitful discussions, and to Marc Oscherson for carefully reading the document.
This research was supported in part by the COST Action CA16201 (“Particleface") of the European Union.
Monika Blanke, Simon Kast, Susanne Westhoff and Jos\'e Zurita are supported in part by the Heidelberg Karlsruhe Strategic Partnership (HEiKA) through the research bridge ``Particle Physics, Astroparticle Physics and Cosmology (PAC)''.
The research of Monika Blanke and Susanne Westhoff is supported in part by the Deutsche Forschungsgemeinschaft (DFG, German Research Foundation) under grant  396021762 -- TRR 257 ``Particle Physics Phenomenology after the Higgs Discovery''.
Simon Kast acknowledges  the  support  of the DFG-funded Doctoral School ``Karlsruhe School of Elementary and Astroparticle Physics: Science and Technology'' (KSETA).
Susanne Westhoff is supported in part by the Carl Zeiss foundation through an endowed junior professorship (\emph{Junior-Stiftungsprofessur}).
The work of Gerhard Buchalla has been supported in part by the
Deutsche Forschungsgemeinschaft (DFG) under grant BU 1391/2-2
(project number 261324988) and the DFG Cluster Exc 2094 "ORIGINS''.
Sally Dawson is supported by the U.S.~Department of Energy under Grant Contract DE-SC0012704.
Christoph Englert is supported by the STFC under grant ST/P000746/1.
The research of Arnaud Ferrari is supported by Vetenskapsrådet (DNR 2015-03942). 
Pier Paolo Giardino is supported by the Spanish Research Agency (Agencia Estatal de Investigacion) through the contract FPA2016-78022-P and IFT Centro de Excelencia Severo Ochoa under grant SEV-2016-0597.
The work of Seraina Glaus is supported by the Swiss National Science Foundation (SNF).
Ramona Gr\"ober is supported by the ``Berliner Chancengleichheitsprogramm''.
Christophe Grojean and Tim Stefaniak are supported by Deutsche Forschungsgemeinschaft (DFG, German Research Foundation) under Germany‘s Excellence Strategy -- EXC 2121 ``Quantum Universe'' -- 390833306.
The research of Stefania Gori is supported by the National Science Foundation under the CAREER grant PHY-1915852.
Peisi Huang is supported in part by United States National Science Foundation under grant PHY-1820891, and the National Science Foundation supported Nebraska EPSCoR program under grant number OIA-1557417.
Michael Kagan is supported by the US Department of Energy (DOE) under grant DE-AC02-76SF00515 and by the SLAC Panofsky Fellowship.
Jonathan Kozaczuk is supported by NSF grant PHY-1719642. 
Ian M. Lewis is supported in part by United  States  Department  of Energy grant number DE-SC0017988.
Heather E.~Logan is supported by the Natural Sciences and Engineering Research Council of Canada and by the H2020-MSCA-RISE-2014 grant no.~645722 (NonMinimalHiggs).
Andrew J. Long is supported at the University of Michigan by the US Department of Energy under grant DE-SC0007859.
The work of Davide Pagani has been supported by the Alexander von Humboldt Foundation, in the framework of the Sofja Kovalevskaja Award Project ``Event Simulation for the Large Hadron Collider at High Precision'', and by the Deutsche Forschungsgemeinschaft (DFG) through the Collaborative Research Centre SFB1258.
Michael Peskin is supported by the US Department of Energy (DOE) under
grant DE-AC02-76SF00515.
Tania Robens is supported by the European Union through the European Regional Development Fund - the Competitiveness and Cohesion Operational Programme (KK.01.1.1.06).
Nausheen R.~Shah is supported by Wayne State University and by the United States Department of Energy under grant number DE-SC0007983.
Ambresh Shivaji very much appreciates the support from MOVE-IN Louvain incoming postdoctoral fellowship co-funded by the Marie Curie Actions of the European Commission.
For Suyog Shrestha, the work was partially supported by the U.S. Department of Energy through the grant DE-SC0011726.
Kuver Sinha is supported in part by United  States  Department  of Energy grant number DE-SC0009956.
Matthew Sullivan is supported by the Kansas EPSCoR grant program.
Maximilian Swiatlowski is supported by the University of Chicago and through the NSF grant PHY-1707981.
Rafael Teixeira De Lima is supported by the SLAC Panofsky Fellowship.
Caterina Vernieri is supported by the SLAC Panofsky Fellowship.